\documentclass[a4paper]{article}
\usepackage{icrc2015}
\usepackage[english]{babel}
\usepackage{amsmath}
\usepackage{amssymb}
\usepackage{threeparttable}
\usepackage[usenames,dvipsnames]{color}
\usepackage[rightcaption,outercaption]{sidecap}
\usepackage{graphicx}   
\usepackage{cite}
\addto\captionsenglish{
  \renewcommand{\contentsname}%
    {\centerline{{\LARGE \textcolor{blue}{List of Contributions}}}%
    \vskip 0.25cm%
    \hrule%
    }%
}

\usepackage[numbers,sort&compress]{natbib}
\usepackage{mdwlist}
\usepackage{caption}
\usepackage{subcaption}
\usepackage{wrapfig}
\usepackage[quotient-mode=fraction]{siunitx}
\usepackage{booktabs}
\usepackage{multirow}
\usepackage[percent]{overpic}  

\newcommand{\unitx}{\,\mathrm}

\newcommand{\comment}[1]{}

\newcommand{\ant}{\textsc{Antares}}
\newcommand{\kmnet}{{\sc KM3NeT}}
\definecolor{bianco}{rgb}{1.00,1.00,1.00}
\definecolor{color01}{rgb}{0.00,0.00,0.00}

\newcounter{IdContrib}
\newcommand{\id}[1]{\refstepcounter{IdContrib}\label{#1}}

\begin{document}

\shorttitle{}

\begin{onecolumn}

\begin{center}
%

{
{\Huge \bf%

\textcolor{blue}{The {\sc Antares} Collaboration} } \\
 \vskip 0.25cm
\hrule
\vskip 0.5cm
 {\Large Contributions to the\\
 34$^{th}$  International Cosmic Ray Conference (ICRC 2015) \\
 \vskip 0.25cm
The Hague, The Netherlands\\
 July 2015}
  \vskip 0.25cm
\hrule
 }
 \end {center}
  \vspace{2cm}
\centerline{\includegraphics[width=0.15\textwidth]{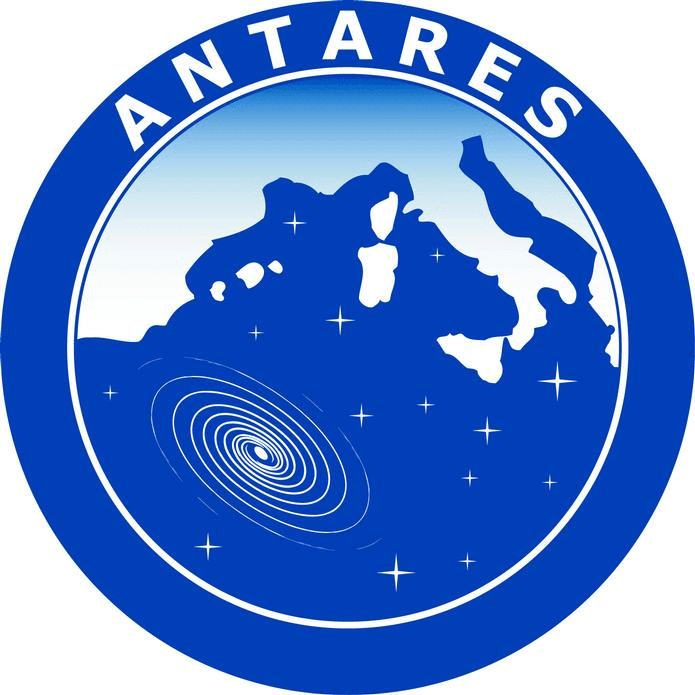}}
 \vspace{2cm}
 \begin{center}
 \textcolor{blue}{{\LARGE \bf Abstract}}
 
  \end{center} 
{\large
The ANTARES detector, completed in 2008, is the largest neutrino telescope in the Northern hemisphere. 
Located at a depth of 2.5 km in the Mediterranean Sea, 40 km off the Toulon shore, its main goal is the 
search for astrophysical high energy neutrinos. 
In this paper we collect 22 contributions of the ANTARES collaboration to the 34th International Cosmic Ray Conference (ICRC 2015). 
The scientific output is very rich and the contributions included in these proceedings cover the main physics results,
ranging from steady point sources, diffuse searches, multi-messenger analyses to exotic physics.

 }
 \vskip 0.125cm
\hrule

\vskip 2cm
\begin{center}
 \textcolor{blue}{{\Large \bf Acknowledgements}}
   \end{center} 

\noindent  
The authors acknowledge the financial support of the funding agencies:
Centre National de la Recherche Scientifique (CNRS), Commissariat \`a
l'\'ener\-gie atomique et aux \'energies alternatives (CEA),
Commission Europ\'eenne (FEDER fund and Marie Curie Program), R\'egion
\^Ile-de-France (DIM-ACAV) R\'egion Alsace (contrat CPER), R\'egion
Provence-Alpes-C\^ote d'Azur, D\'e\-par\-tement du Var and Ville de La
Seyne-sur-Mer, France; Bundesministerium f\"ur Bildung und Forschung
(BMBF), Germany; Istituto Nazionale di Fisica Nucleare (INFN), Italy;
Stichting voor Fundamenteel Onderzoek der Materie (FOM), Nederlandse
organisatie voor Wetenschappelijk Onderzoek (NWO), the Netherlands;
Council of the President of the Russian Federation for young
scientists and leading scientific schools supporting grants, Russia;
National Authority for Scientific Research (ANCS), Romania; 
Mi\-nis\-te\-rio de Econom\'{\i}a y Competitividad (MINECO), Prometeo 
and Grisol\'{\i}a programs of Generalitat Valenciana and MultiDark, 
Spain; Agence de  l'Oriental and CNRST, Morocco. We also acknowledge 
the technical support of Ifremer, AIM and Foselev Marine for the sea 
operation and the CC-IN2P3 for the computing facilities.
\end{onecolumn} 

\newpage
\null
\newpage
\shorttitle{{\sc Antares} Contributions to ICRC2015}
\begin{center}
{\large \bf \textcolor{blue}{The {\sc Antares} Collaboration}
\phantomsection
 \addcontentsline{toc}{part}{The {\sc Antares} Collaboration, list of authors}}
\end{center}


\authors[{
\noindent
S.~Adri\'an-Mart\'inez$^a$,
M.~Ageron$^g$,
A.~Albert$^b$, 
M.~Andr\'e$^c$, 
G.~Anton$^e$, 
M.~Ardid$^a$, 
J.-J.~Aubert$^g$,
B.~Baret$^h$,
J.~Barrios-Mart\'{\i}$^i$,
S.~Basa$^j$,
V.~Bertin$^g$,
S.~Biagi$^{k,l}$,
R.~Bormuth$^{g,ak}$,
M.C.~Bouwhuis$^f$,
R. Bruijn$^{f,ac}$,
J.~Brunner$^g$,
J.~Busto$^g$,
A.~Capone$^{m,n}$,
L.~Caramete$^o$,
J.~Carr$^g$,
T.~Chiarusi$^k$,
M.~Circella$^r$,
A.~Coleiro$^h$,
R.~Coniglione$^v$,
H.~Costantini$^g$,
P.~Coyle$^g$,
A.~Creusot$^h$,
I.~Dekeyser$^s$,
A.~Deschamps$^q$,
G.~De~Bonis$^{m,n}$,
{C.~Distefano$^v$,
C.~Donzaud$^{h,w}$,
D.~Dornic$^g$,
D.~Drouhin$^b$,
A.~Dumas$^p$,
T.~Eberl$^e$,
I.~El Bojaddaini$^{am}$,
D.~Els\"asser$^y$,
A.~Enzenh\"ofer$^e$,
S
K.~Fehn$^e$,
I.~Felis$^a$,
P.~Fermani$^{m,n}$,
L.A.~Fusco$^{k,l}$,
S.~Galat\`a$^h$,
P.~Gay$^p$,
S.~Gei{\ss}els\"oder$^e$,
K.~Geyer$^e$,
V.~Giordano$^z$,
A.~Gleixner$^e$,
H.~Glotin$^{an}$,
R.~Gracia-Ruiz$^h$,
K.~Graf$^e$,
S.~Hallmann$^e$,
H.~van~Haren$^{aa}$,
A.J.~Heijboer$^f$,
Y.~Hello$^q$,
J.J.~Hern\'andez-Rey$^i$,
J.~H\"o{\ss}l$^e$,
J.~Hofest\"adt$^e$,
C.~Hugon$^d$,
C.W~James$^e$,
M.~de~Jong$^f$,
M.~Kadler$^y$,
O.~Kalekin$^e$,
U.~Katz$^e$,
D.~Kie{\ss}ling$^e$,
P.~Kooijman$^{f,ab,ac}$,
A.~Kouchner$^h$,
M.~Kreter$^y$,
I.~Kreykenbohm$^{ad}$,
V.~Kulikovskiy$^{d,ae}$,
C.~Lachaud$^h$,
D. ~Lef\`evre$^s$,
E.~Leonora$^{z,af}$,
S.~Loucatos$^{ah}$,
M.~Marcelin$^j$,
A.~Margiotta$^{k,l}$,
A.~Marinelli$^{ao,ap}$,
J.A.~Mart\'inez-Mora$^a$,
A.~Mathieu$^g$,
T.~Michael$^f$,
P.~Migliozzi$^t$,
A.~Moussa$^{am}$,
L.~Moscoso$^{h,\dagger}$,
C.~Mueller$^y$,
E.~Nezri$^j$,
G.E.~P\u{a}v\u{a}la\c{s}$^o$,
P.~Payre$^{g,\dagger}$,
C.~Pellegrino$^{k,l}$,
C.~Perrina$^{m,n}$,
P.~Piattelli$^v$,
V.~Popa$^o$,
T.~Pradier$^{ai}$,
C.~Racca$^{b}$,
G.~Riccobene$^v$,
K.~Roensch$^e$,
M.~Salda\~{n}a$^a$,
D.F.E.~Samtleben$^{f,ak}$,
A.~S{\'a}nchez-Losa$^i$,
M.~Sanguineti$^{d,al}$,
P.~Sapienza$^v$,
J.~Schmid$^e$,
J.~Schnabel$^e$,
F.~Sch\"ussler$^{ah}$,
T.~Seitz$^e$,
C.~Sieger$^e$,
M.~Spurio$^{k,l}$,
J.J.M.~Steijger$^f$,
Th.~Stolarczyk$^{ah}$,
M.~Taiuti$^{d,al}$,
C.~Tamburini$^s$,
A.~Trovato$^v$,
M.~Tselengidou$^e$,
D.~Turpin$^g$,
C.~T\"onnis$^i$,
B.~Vallage$^{ah}$,
C.~Vall\'ee$^g$,
V.~Van~Elewyck$^h$,
E.~Visser$^f$,
D.~Vivolo$^{t,u}$,
S.~Wagner$^e$,
J.~Wilms$^{ad}$,
J.D.~Zornoza$^i$,
J.~Z\'u\~{n}iga$^i$}


\afiliations[{\noindent\scriptsize{$^a$ Institut d'Investigaci\'o per a la Gesti\'o Integrada de les Zones Costaneres (IGIC) - Universitat Polit\`ecnica de Val\`encia. C/  Paranimf 1 , 46730 Gandia, Spain\\}}
\afiliations[{\scriptsize{$^b$ GRPHE -Universit\'e de Haute Alsace \& Institut universitaire de technologie de Colmar, 34 rue du Grillenbreit BP 50568 - 68008 Colmar, France\\}}
\afiliations[{\scriptsize{$^c$ Technical University of Catalonia, Laboratory of Applied Bioacoustics, Rambla Exposici\'o,08800 Vilanova i la Geltr\'u,Barcelona, Spain\\}}
\afiliations[{\scriptsize{$^d$ INFN - Sezione di Genova, Via Dodecaneso 33, 16146 Genova, Italy\\}}
\afiliations[{\scriptsize{$^e$Friedrich-Alexander-Universit\"at Erlangen-N\"urnberg, Erlangen Centre for Astroparticle Physics, Erwin-Rommel-Str. 1, 91058 Erlangen, Germany\\}}
\afiliations[{\scriptsize{$^f$Nikhef, Science Park,  Amsterdam, The Netherlands\\}}
\afiliations[{\scriptsize{$^g$Aix Marseille Universit\'e, CNRS/IN2P3, CPPM UMR 7346, 13288, Marseille, France\\}}
\afiliations[{\scriptsize{$^h$APC, Universit\'e Paris Diderot, CNRS/IN2P3, CEA/IRFU, Observatoire de Paris, Sorbonne Paris Cit\'e, 75205 Paris, France\\}}
\afiliations[{\scriptsize{$^i$IFIC - Instituto de F\'isica Corpuscular, c/ Catedra\'atico Jos\'e Beltr\'an, 2 E-46980 Paterna, Valencia (Spain)\\}}
\afiliations[{\scriptsize{$^j$LAM - Laboratoire d'Astrophysique de Marseille, P\^ole de l'\'Etoile Site de Ch\^ateau-Gombert, rue Fr\'ed\'eric Joliot-Curie 38,  13388 Marseille Cedex 13, France\\}}
\afiliations[{\scriptsize{$^k$INFN - Sezione di Bologna, Viale Berti-Pichat 6/2, 40127 Bologna, Italy\\}}
\afiliations[{\scriptsize{$l$Dipartimento di Fisica dell'Universit\`a, Viale Berti Pichat 6/2, 40127 Bologna, Italy\\}}
\afiliations[{\scriptsize{$^m$INFN -Sezione di Roma, P.le Aldo Moro 2, 00185 Roma, Italy\\}}
\afiliations[{\scriptsize{$^n$Dipartimento di Fisica dell'Universit\`a La Sapienza, P.le Aldo Moro 2, 00185 Roma, Italy\\}}
\afiliations[{\scriptsize{$^o$Institute for Space Sciences, R-77125 Bucharest, M\u{a}gurele, Romania\\}}
\afiliations[{\scriptsize{$^p$Laboratoire de Physique Corpusculaire, Clermont Universit\'e, Universit\'e Blaise Pascal, CNRS/IN2P3, BP 10448, F-63000 Clermont-Ferrand, France\\}}
\afiliations[{\scriptsize{$^q$G\'eoazur, Universit\'e Nice Sophia-Antipolis, CNRS, IRD, Observatoire de la C\^ote d'Azur, Sophia Antipolis, France \\}}
\afiliations[{\scriptsize{$^r$INFN - Sezione di Bari, Via E. Orabona 4, 70126 Bari, Italy\\}}
\afiliations[{\scriptsize{$^s$Aix Marseille Universit\'e, CNRS/INSU, IRD, Mediterranean Institute of Oceanography (MIO), UM 110, Marseille, France ; Universit\'e de Toulon, CNRS, IRD, Mediterranean Institute of Oceanography (MIO), UM 110, La Garde, France\\}}
\afiliations[{\scriptsize{$^t$INFN -Sezione di Napoli, Via Cintia 80126 Napoli, Italy\\}}
\afiliations[{\scriptsize{$^u$Dipartimento di Fisica dell'Universit\`a Federico II di Napoli, Via Cintia 80126, Napoli, Italy\\}}
\afiliations[{\scriptsize{$^v$INFN - Laboratori Nazionali del Sud (LNS), Via S. Sofia 62, 95123 Catania, Italy\\}}
\afiliations[{\scriptsize{$^w$Univ. Paris-Sud , 91405 Orsay Cedex, France\\}}
\afiliations[{\scriptsize{$^y$Institut f\"ur Theoretische Physik und Astrophysik, Universit\"at W\"urzburg, Emil-Fischer Str. 31, 97074 W\"urzburg, Germany\\}}
\afiliations[{\scriptsize{$^z$INFN - Sezione di Catania, Viale Andrea Doria 6, 95125 Catania, Italy\\}}
\afiliations[{\scriptsize{$^{aa}$Royal Netherlands Institute for Sea Research (NIOZ), Landsdiep 4,1797 SZ 't Horntje (Texel), The Netherlands\\}}
\afiliations[{\scriptsize{$^{ab}$Universiteit Utrecht, Faculteit Betawetenschappen, Princetonplein 5, 3584 CC Utrecht, The Netherlands\\}}
\afiliations[{\scriptsize{$^{ac}$Universiteit van Amsterdam, Instituut voor Hoge-Energie Fysica, Science Park 105, 1098 XG Amsterdam, The Netherlands\\}}
\afiliations[{\scriptsize{$^{ad}$Dr. Remeis-Sternwarte and ECAP, Universit\"at Erlangen-N\"urnberg,  Sternwartstr. 7, 96049 Bamberg, Germany\\}}
\afiliations[{\scriptsize{$^{ae}$Moscow State University,Skobeltsyn Institute of Nuclear Physics,Leninskie gory, 119991 Moscow, Russia\\}}
\afiliations[{\scriptsize{$^{af}$Dipartimento di Fisica ed Astronomia dell'Universit\`a, Viale Andrea Doria 6, 95125 Catania, Italy\\}}
\afiliations[{\scriptsize{$^{ah}$Direction des Sciences de la Mati\`ere - Institut de recherche sur les lois fondamentales de l'Univers - Service de Physique des Particules, CEA Saclay, 91191 Gif-sur-Yvette Cedex, France\\}}
\afiliations[{\scriptsize{$^{ai}$Universit\'e de Strasbourg, IPHC, 23 rue Becquerel 67087 Strasbourg, France
CNRS, UMR7178, 67087 Strasbourg, France\\}}
\afiliations[{\scriptsize{$^{ak}$Universiteit Leiden, Leids Instituut voor Onderzoek in Natuurkunde, 2333 CA Leiden, The Netherlands\\}}
\afiliations[{\scriptsize{$^{al}$Dipartimento di Fisica dell'Universit\`a, Via Dodecaneso 33, 16146 Genova, Italy\\}}
\afiliations[{\scriptsize{$^{am}$University Mohammed I, Laboratory of Physics of Matter and Radiations, B.P.717, Oujda 6000, Morocco\\}}
\afiliations[{\scriptsize{$^{an}$
LSIS, Aix Marseille Universit\'e CNRS ENSAM LSIS UMR 7296 13397 Marseille, France ; Universit\'e de Toulon CNRS LSIS UMR 7296 83957 La Garde, France ; Institut Universitaire de France, 75005 Paris, France\\}}
\afiliations[{\scriptsize{$^{ao}$INFN - Sezione di Pisa, Largo B. Pontecorvo 3, 56127 Pisa, Italy\\}}
\afiliations[{\scriptsize{$^{ap}$Dipartimento di Fisica dell'Universit\`a, Largo B. Pontecorvo 3, 56127 Pisa, Italy\\}}

\afiliations[{\scriptsize{$^{\dagger}$Deceased\\}}



%
%

\newpage
\null
\newpage
\setcounter{tocdepth}{0}
\vspace*{0.4cm}

\tableofcontents

\newpage



\newpage
\id{id_cj}
\addcontentsline{toc}{part}{\textcolor{blue}{\arabic{IdContrib} - {\sl Clancy W. James - Highlight Talk} : Highlights from \ant, and prospects for \kmnet}}

\title{\arabic{IdContrib} - Highlights from ANTARES, and prospects for KM3NeT}

\shorttitle{ANTARES highlights and KM3NeT prospects}

\authors{Clancy~W.~James, on behalf of the ANTARES and KM3NeT Collaborations}
\afiliations{        ECAP, University of Erlangen-Nuremberg}
\email{clancy.james@physik.uni-erlangen.de}


\abstract{The ANTARES experiment has been running in its final configuration since 2008. It is the largest neutrino telescope in the Northern hemisphere. After the discovery of a cosmic neutrino diffuse flux by the IceCube detector, the search for its origin has become a key mission in high-energy astrophysics. Particularly interesting is the indication (although not significant with the present IceCube statistics) of an excess of signal events from the Southern sky region.

The ANTARES sensitivity is large enough to constrain the origin of the IceCube excess from regions extended up to 0.2 sr in the Southern sky. Assuming different spectral indices for the energy spectrum of neutrino emitters, the Southern sky and in particular central regions of our Galaxy are studied searching for point-like objects, for extended regions of emission, and for signal from transient objects selected through multimessenger observations. For the first time, cascade events are used for these searches, using a new method with $3^{\circ}$ angular resolution.

ANTARES has also provided results on
searches for rare particles (such as magnetic monopoles and nuclearites in the cosmic radiation), and multi-messenger studies of the sky in combination with different experiments.
Of particular note are the searches for Dark Matter: the limits obtained for the spin-dependent WIMP-nucleon cross section overcome that of existing direct-detection experiments.

The contribution concludes with an outlook to the next-generation experiment KM3NeT, which is already under construction. KM3NeT will consist of two components: ORCA, optimised for measuring atmospheric neutrino oscillation parameters in the few-GeV range; and ARCA, for studying astrophysical neutrinos at higher energies. The status of KM3NeT will be summarised and the resulting prospects for ORCA and ARCA discussed.
}
%

\newcommand*\aap{A\&A}
\let\astap=\aap
\newcommand*\aapr{A\&A~Rev.}
\newcommand*\aaps{A\&AS}
\newcommand*\actaa{Acta Astron.}
\newcommand*\aj{AJ}
\newcommand*\ao{Appl.~Opt.}
\let\applopt\ao
\newcommand*\apj{ApJ}
\newcommand*\apjl{ApJ}
\let\apjlett\apjl
\newcommand*\apjs{ApJS}
\let\apjsupp\apjs
\newcommand*\aplett{Astrophys.~Lett.}
\newcommand*\apspr{Astrophys.~Space~Phys.~Res.}
\newcommand*\apss{Ap\&SS}
\newcommand*\araa{ARA\&A}
\newcommand*\azh{AZh}
\newcommand*\baas{BAAS}
\newcommand*\bac{Bull. astr. Inst. Czechosl.}
\newcommand*\bain{Bull.~Astron.~Inst.~Netherlands}
\newcommand*\caa{Chinese Astron. Astrophys.}
\newcommand*\cjaa{Chinese J. Astron. Astrophys.}
\newcommand*\fcp{Fund.~Cosmic~Phys.}
\newcommand*\gca{Geochim.~Cosmochim.~Acta}
\newcommand*\grl{Geophys.~Res.~Lett.}
\newcommand*\iaucirc{IAU~Circ.}
\newcommand*\icarus{Icarus}
\newcommand*\jcap{J. Cosmology Astropart. Phys.}
\newcommand*\jcp{J.~Chem.~Phys.}
\newcommand*\jgr{J.~Geophys.~Res.}
\newcommand*\jqsrt{J.~Quant.~Spec.~Radiat.~Transf.}
\newcommand*\jrasc{JRASC}
\newcommand*\memras{MmRAS}
\newcommand*\memsai{Mem.~Soc.~Astron.~Italiana}
\newcommand*\mnras{MNRAS}
\newcommand*\na{New A}
\newcommand*\nar{New A Rev.}
\newcommand*\nat{Nature}
\newcommand*\nphysa{Nucl.~Phys.~A}
\newcommand*\pasa{PASA}
\newcommand*\pasj{PASJ}
\newcommand*\pasp{PASP}
\newcommand*\physrep{Phys.~Rep.}
\newcommand*\physscr{Phys.~Scr}
\newcommand*\planss{Planet.~Space~Sci.}
\newcommand*\pra{Phys.~Rev.~A}
\newcommand*\prb{Phys.~Rev.~B}
\newcommand*\prc{Phys.~Rev.~C}
\newcommand*\prd{Phys.~Rev.~D}
\newcommand*\pre{Phys.~Rev.~E}
\newcommand*\prl{Phys.~Rev.~Lett.}
\newcommand*\procspie{Proc.~SPIE}
\newcommand*\qjras{QJRAS}
\newcommand*\rmxaa{Rev. Mexicana Astron. Astrofis.}
\newcommand*\skytel{S\&T}
\newcommand*\solphys{Sol.~Phys.}
\newcommand*\sovast{Soviet~Ast.}
\newcommand*\ssr{Space~Sci.~Rev.}
\newcommand*\zap{ZAp}

\maketitle

\section{Introduction}

The underwater neutrino telescope ANTARES has been operating in its final configuration since $2008$. Anchored to the seabed at a depth of $2.5$~km, and located $40$~km off the coast of Toulon, France, it is the largest neutrino telescope in the Northern Hemisphere. Consisting of an array of $885$ $10''$ photomultiplier tubes covering an instrumented volume of approximately $0.01$~km$^3$,
it is designed primarily to search for $E \gtrsim 100$~GeV muons resulting from the charged-current interactions of $\nu_{\mu}$ in the vicinity of the detector.

Highlights from a wide range of analyses using ANTARES data are reported here. These include several measurements which are used to constrain both point-like (Sec.\ \ref{sec:point_sources}) and extended (Sec. \ref{sec:diffuse}) origins of the astrophysical flux observed by IceCube \cite{2013Sci...342E...1I,2014PhRvL.113j1101A,2015ApJ...809...98A}, and a new cascade reconstruction method which, due to its high angular resolution, for the first time allows a point-source search with cascade events (Sec.\ \ref{sec:cascades}). Updated limits on dark matter are also given in Sec.\ \ref{sec:dark_matter}).

ANTARES is planned to cease operation in $2017$. At the same time, Phase $1$ of the next-generation instrument KM3NeT will be completed. With a flexible block design, KM3NeT will be deployed in both a compact configuration (`ORCA') to study neutrino oscillations and the neutrino mass hierarchy, and a sparser configuration (`ARCA') for performing high-energy neutrino astronomy. The status of KM3NeT deployment, and the prospects for ARCA and ORCA during Stage $2$ of KM3NeT, is given in Sec.\ \ref{sec:km3net}.


\section{Searches for astrophysical neutrino point sources}
\label{sec:point_sources}

The main channel by which ANTARES searches for astrophysical point-like sources of neutrinos is by searching for an excess of energetic $\mu$ from the interactions of $\nu_{\mu}$ in the vicinity of the detector. The high rate of downgoing $\mu$ from the interactions of cosmic rays (CR) in the Earth's atmosphere restricts such searches to events coming from below, or only a few degrees above, the horizon. The primary background to such searches then becomes the flux of atmospheric $\nu_{\mu}$, and those few atmospheric $\mu$ events mis-reconstructed as up-going. The long scattering length of blue light in seawater provides an excellent directional resolution on the $\nu_{\mu}$ primary of $0.38^{\circ}$ for an $E^{-2}$ source \cite{2014ApJ...786L...5A}, which is tested using the Moon shadow (M.~Sanguineti, ICRC2015 1138). This allows a very strong suppression of both backgrounds, and a correspondingly good sensitivity to neutrino sources from the Southern Hemisphere. Its ability to probe the origin of the IceCube astrophysical flux is best-characterised through the joint analysis described below. Throughout, flavour-uniform spectra are assumed, consistent with observations \cite{2015ApJ...809...98A}.

\subsection{Joint analysis with IceCube}
\label{sec:joint_analysis}

\begin{figure*}
\includegraphics[width=0.48\textwidth]{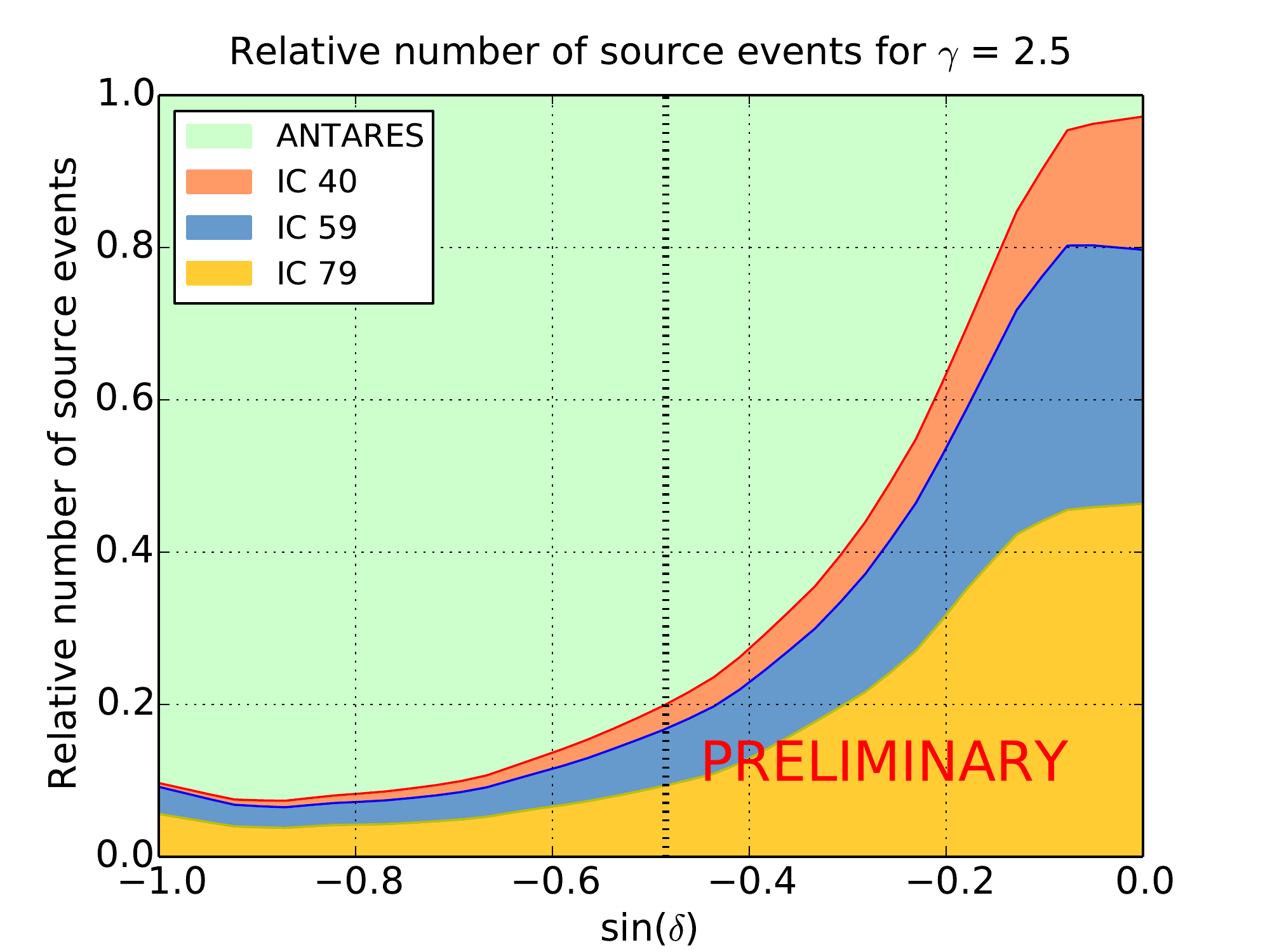} \includegraphics[width=0.48\textwidth]{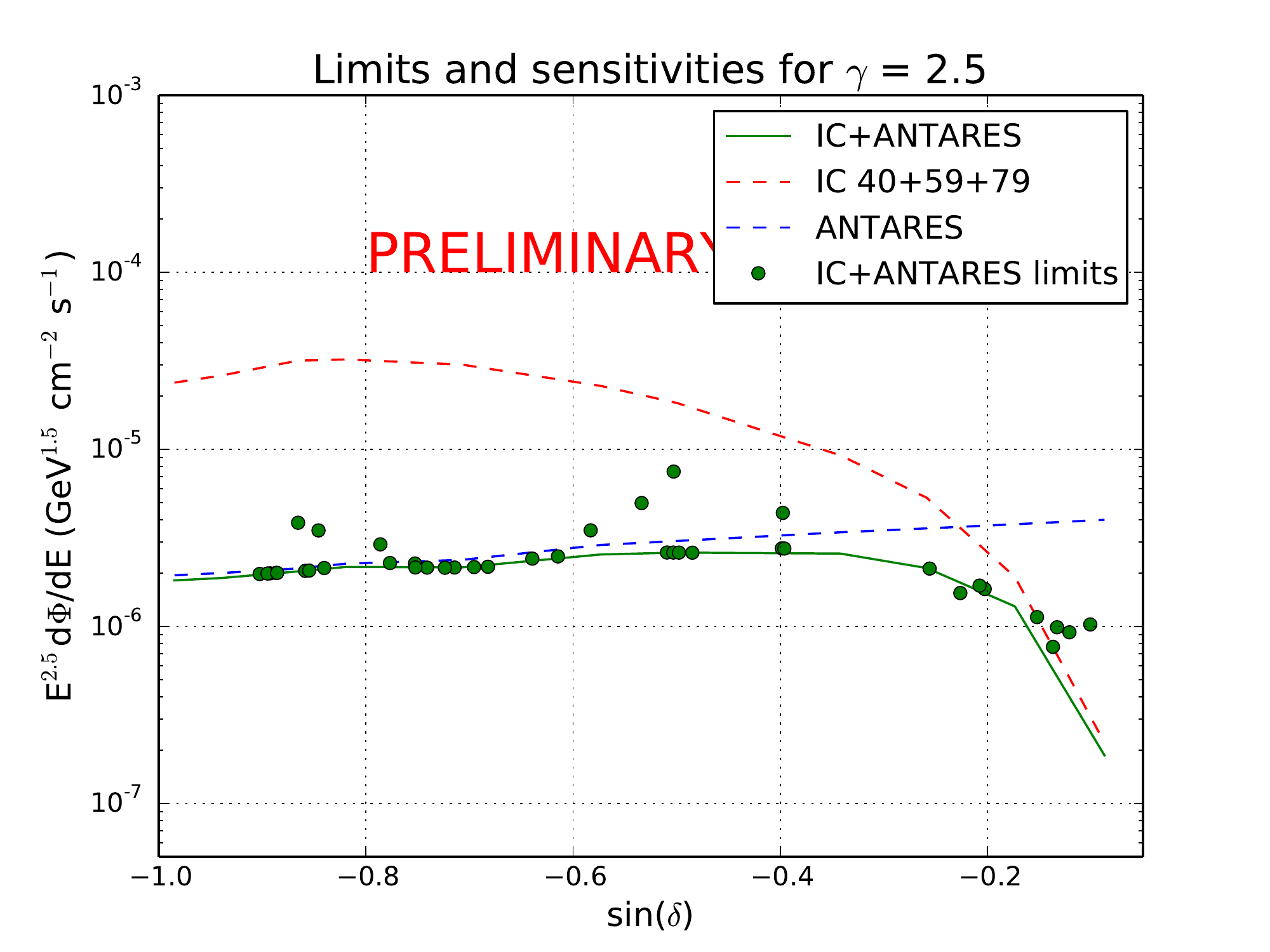}
\caption{Left: Fractional contributions of each data set to the total number of signal events passing cuts in the joint ANTARES--IceCube analysis (Barrios-Mart\'i \& {Finley}, ICRC2015 1076), for sources with an $E^{-2.5}$ spectrum, as a function of declination $\delta$. Right: Sensitivities (lines) and limits (dots) to an $E^{-2.5}$ flux with no cutoff, using ANTARES (blue), IceCube (red), and combined (green) data, as a function of $\delta$.} \label{fig:antares_icecube}
\end{figure*}


A joint analysis using ANTARES 
and IceCube
data is detailed in {Barrios-Mart\'i} \& {Finley} (ICRC2015 1076). The fractional number of source events expected to be present in each data set is shown in Fig.\ \ref{fig:antares_icecube} (left) for an $E^{-2.5}$ spectra, the current best-fit to the IceCube flux. The fraction of events contributed by the ANTARES sample is greater for $\delta \lesssim 15^{\circ}$, where ANTARES is more sensitive to low-energy upcoming muon tracks, while IceCube requires high-energy events to distinguish them from the down-going muon background. The sensitivity is also a function of the background rates, and angular and energy resolutions, which are not shown.



The results of the combined search are shown in Fig.\ \ref{fig:antares_icecube} (right), for an $E^{-2.5}$ source spectrum. No significant cluster is found, with the most significant source on the candidate list being 3C~279, with a pre-trial $p$-value of $0.05$. Over the entire Southern sky, the combined analysis improves on the results from both experiments, indicating the complementarity of the two instruments.

\subsection{Limits on point-source origins of the HESE}
\label{sec:point_source_hese}


\begin{SCfigure*}
\includegraphics[width=0.48\textwidth]{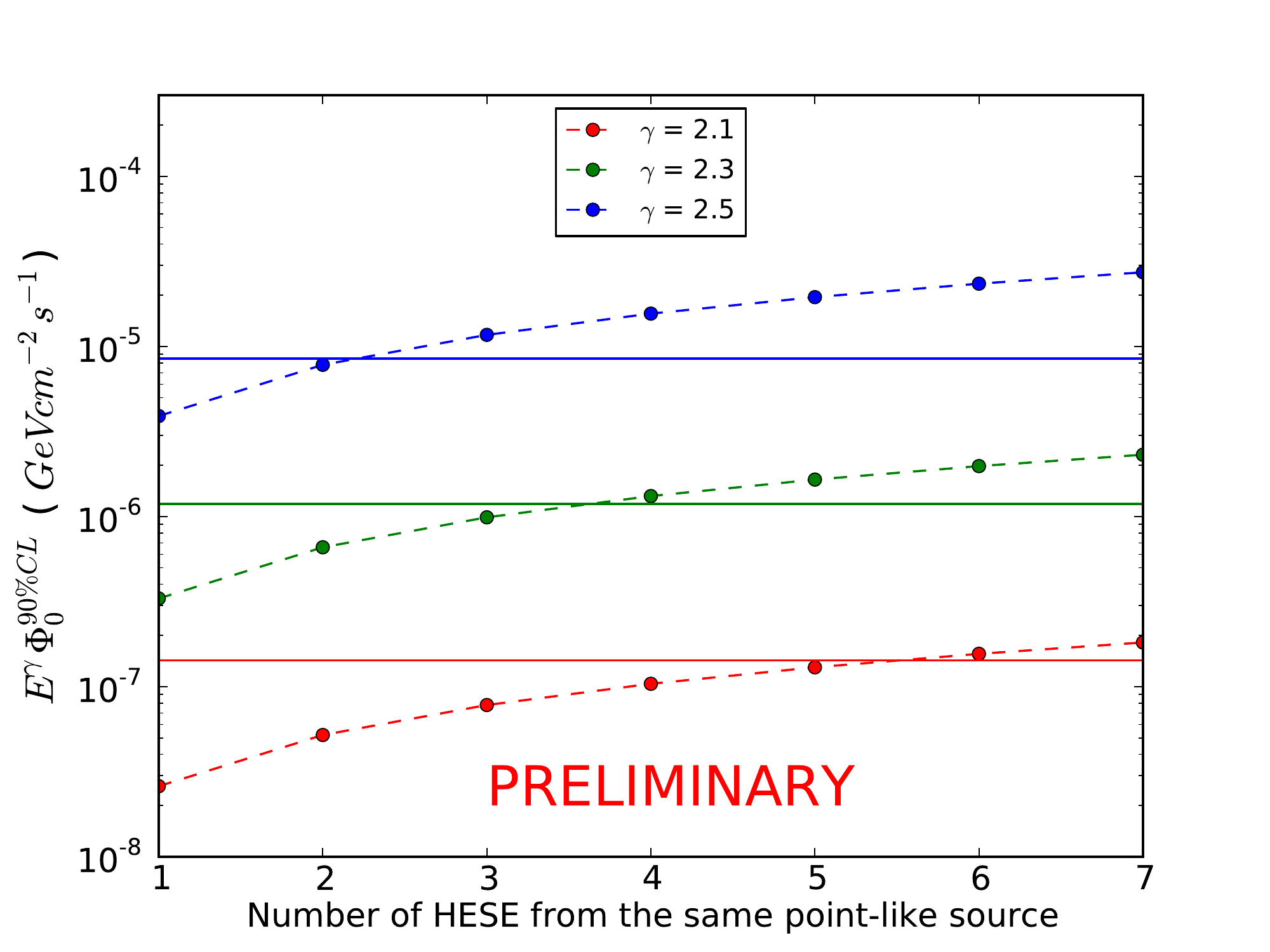}
\caption{ANTARES limits (solid lines) at 90\% C.L.\ on the contribution of point-like sources to the IceCube HESE sample \cite{2014PhRvL.113j1101A} for different spectral indices, shown for a source at $\delta = -29^{\circ}$. (J.~{Barrios-Mart\'i}, ICRC2015 1077). These are compared to (dashed lines) the flux required to produce a given expected number of HESE \cite{Spurio_GC}. The result is similar for other declinations around the Galactic Centre.} \label{fig:icecube_pointlike_results}
\end{SCfigure*}


It has been proposed \cite{2014APh....57...39G} that the cluster of IceCube events seen in Ref.\ \cite{2014PhRvL.113j1101A} could be due to a single point-like source, which is not detectable due to the low angular resolution. The non-detection of an ANTARES point-like source in this region, as reported by J.~{Barrios-Mart\'i} (ICRC2015 1077), limits the flux of such a source as a function of spectral index, shown by the solid lines and y-axis of Fig.\ \ref{fig:icecube_pointlike_results}. The flux required to produce a given number of events in the HESE analysis (x-axis) is also shown. The range where the latter is greater than the former rules out a corresponding contribution from any single point-like source with that spectral index at $90$\% confidence level (C.L.). 

The result above is particularly relevant because the current best-fit spectrum (between $25$~TeV and $2.8$~PeV) of the IceCube flux has a spectral index of $-2.50 \pm 0.09$ \cite{2015ApJ...809...98A}. ANTARES can thus rule out any single point-source of neutrinos in the region of the Galactic Centre with spectral index of $-2.5$ as having a flux corresponding to more than $2$~HESE.



\subsection{Flares from AGN and X-ray binaries}
\label{sec:agn_flares}

AGN have long been proposed as a source of high-energy cosmic rays and, hence, neutrinos \cite{1995PhR...258..173G}. Blazars, being active galactic nuclei with jets pointed towards the line-of-sight, exhibit bright flares which dominate the extragalactic $\gamma$-ray sky observed by \emph{Fermi}-LAT \cite{2009ApJ...697.1071A}.

Using multi-wavelength observations, several bright blazars have been reported by the TANAMI collaboration \cite{2014A&A...566L...7K} to lie within the $50$\% error bounds of the reconstructed arrival directions of the PeV-scale events IC~14 and IC~20 observed by IceCube \cite{2014PhRvL.113j1101A}. As discussed by Kadler et~al.\ (ICRC2015 1090), ANTARES observes signal-like events from the two brightest blazars, both in the field of IC~20 \cite{2015A&A...576L...8A}, although this is also consistent with background fluctuations. A lack of such events from the field of IC~14 excludes a neutrino spectrum softer than $E^{-2.4}$ as being responsible for this event. The highest-energy `Big Bird' event (IC~35) was detected during an extremely bright flare from the blazar PKS B1424-418, which lies within the $50$\% error region of the IC~35 arrival direction. ANTARES finds only one event within $5^{\circ}$ of this source during the flaring period, whereas approximately three would be expected from random background fluctuations alone.

\begin{SCfigure*}
\includegraphics[width=0.55\textwidth]{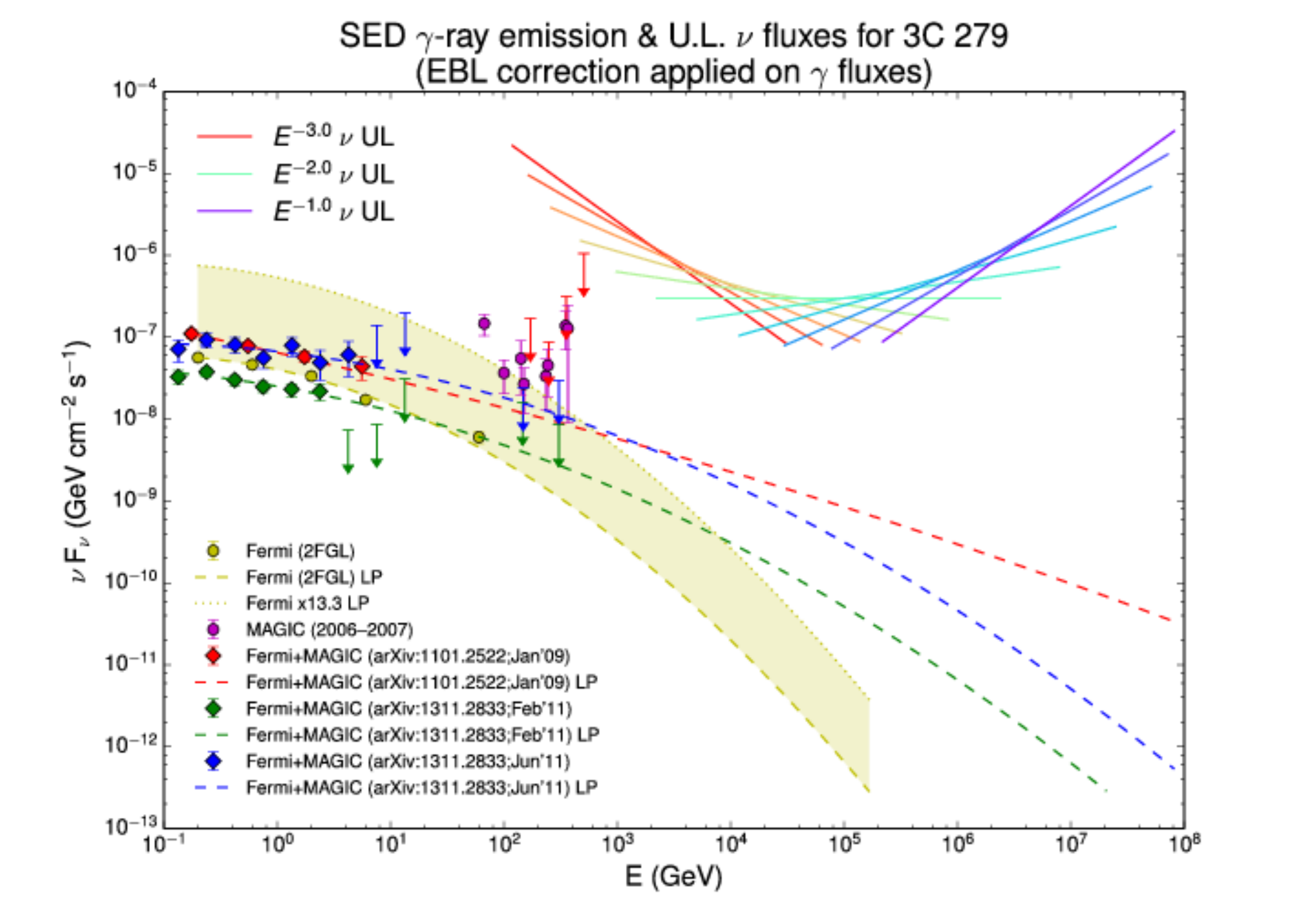}
\caption{Result from time-dependent analyses using gamma-ray data. Limits on the neutrino flux from the blazar 3C~279 as a function of spectral index (solid lines; {S\'anchez-Losa} \& {Dornic}, ICRC2015 1075), compared to the observed (points) and extrapolated (dashed lines) gamma-ray spectra observed by \emph{Fermi} and IACTs.
 } \label{fig:flaring_sources}
\end{SCfigure*}

In another analysis ({S\'anchez-Losa} \& {Dornic}, ICRC2015 1075), ANTARES targets a sample of $41$ blazar flares observed by \emph{Fermi} LAT and $7$ by the IACTs H.E.S.S., MAGIC, and VERITAS. The lowest pre-trial p-value of $3.3$\% was found for the blazar 3C~279, which comes from the coincidence of one event with a 2008 flare previously reported by  Ref.\ \cite{2012APh....36..204A}. However, the post-trial p-value is not significant. The resulting limits are given in Fig.\ \ref{fig:flaring_sources}.


Similar methods were also used to search for neutrino emission during the flares from galactic x-ray binaries ({Dornic} \& {S\'anchez-Losa}, ICRC2015 1046).
A total of $34$ x-ray- and $\gamma$-ray-selected binaries were studied, with no significant detections, allowing some of the more optimistic models for hadronic acceleration in these sources to be rejected at $90$\% C.L..

\subsection{Gamma-ray bursts}
\label{sec:GRBs}

\begin{SCfigure*}
\includegraphics[clip=true,trim={0 8cm 0 8.5cm}, width=0.49\textwidth]{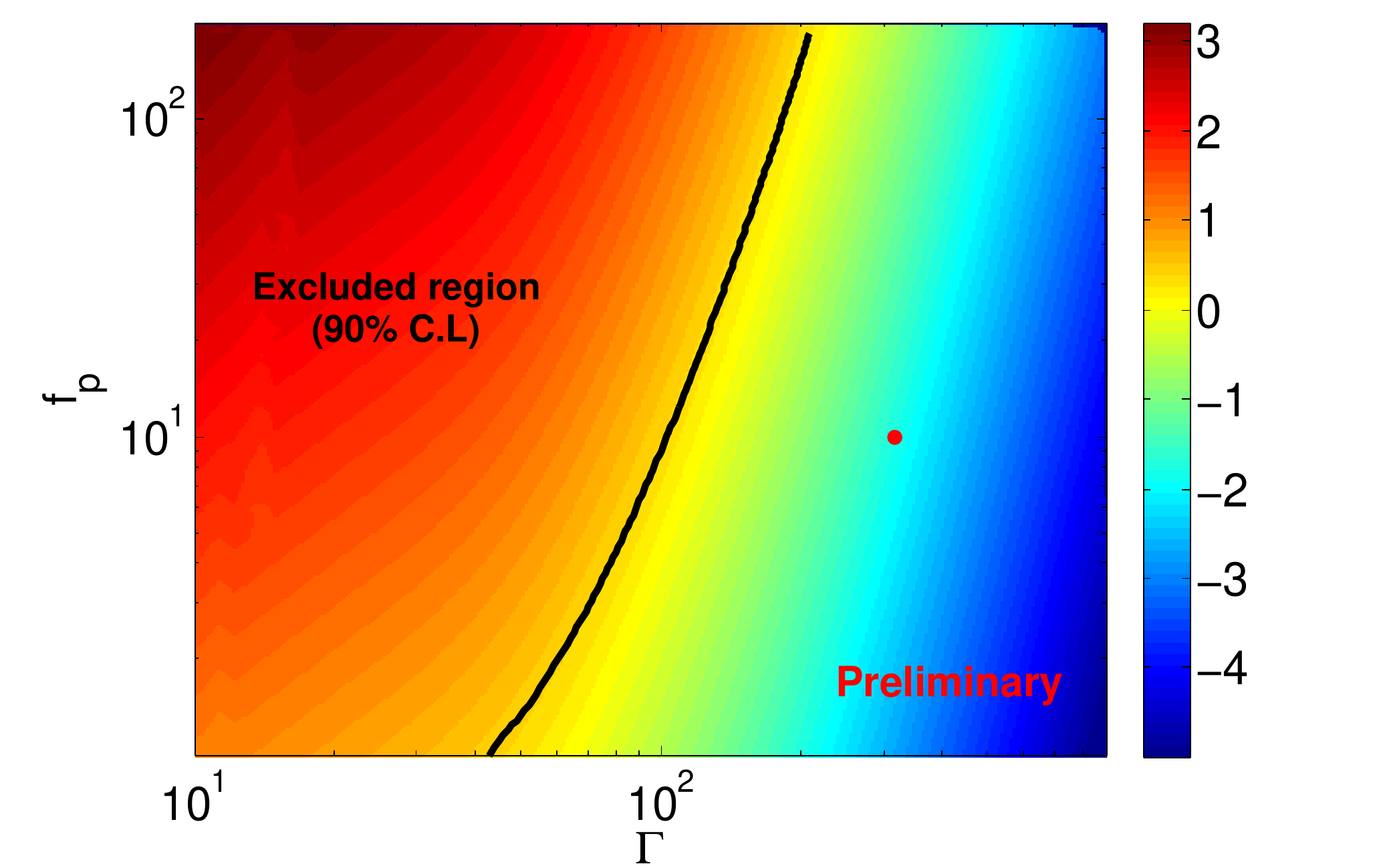}
\caption{Range of jet $\Gamma$-factor and baryonic loading $f_p$ excluded by ANTARES in the case of GRB110918A using the NeuCosmA model of Ref.\ \cite{2010ApJ...721..630H}, as described by Schmid \& Turpin (ICRC2015 1057). The assumed values of $\Gamma = 316$ and $f_p=10$ are shown by the red point, while the colour-coding gives the expected number of observable neutrinos. The predicted $\nu$ emission scales with $\Gamma^{-5}$ and linearly with $f_p$.} \label{fig:grb_update_fig2_left}
\end{SCfigure*}

Long-duration gamma-ray bursts (GRBs) have been proposed as a source of the highest-energy cosmic rays \cite{1997PhRvL..78.2292W}. ANTARES searches for a neutrino flux from GRBs considering two methods of modelling emission processes: the NeuCosmA description of Ref.\ \cite{2010ApJ...721..630H}, and the `photospheric' model of Ref.\ \cite{2012JCAP...11..058G}.
In each case, the expected signal is simulated on a burst-by-burst basis, and the detector response and background are modelled using the exact oceanic conditions at the time of the burst. The ANTARES analysis using the NeuCosmA model was developed and applied to a sample of 296 bursts in Ref.\ \cite{2013A&A...559A...9A}, with no coincident neutrino events detected. Since then, one especially powerful burst GRB110918A, and the nearby burst GRB130427A, have been identified as promising candidates for neutrino detection, and studied in detail by Schmid \& Turpin (ICRC2015 1057). No coincident events are observed from either burst, with limits set on the bulk gamma-factor and baryonic loading of the jet, as shown in Fig.\ \ref{fig:grb_update_fig2_left}. 

A search using the photospheric models is developed by M.~{Sanguineti} (ICRC2015 1068), and will shortly be unblinded.
The GRB search methods are also being extended to test Lorentz invariance violation (Schmid \& Turpin, ICRC2015 1057), which would delay the arrival times of TeV neutrinos compared to GeV photons.

\subsection{Optical and X-ray follow-up}
\label{sec:tatoo}

\begin{SCfigure*}
\includegraphics[width=0.4\textwidth]{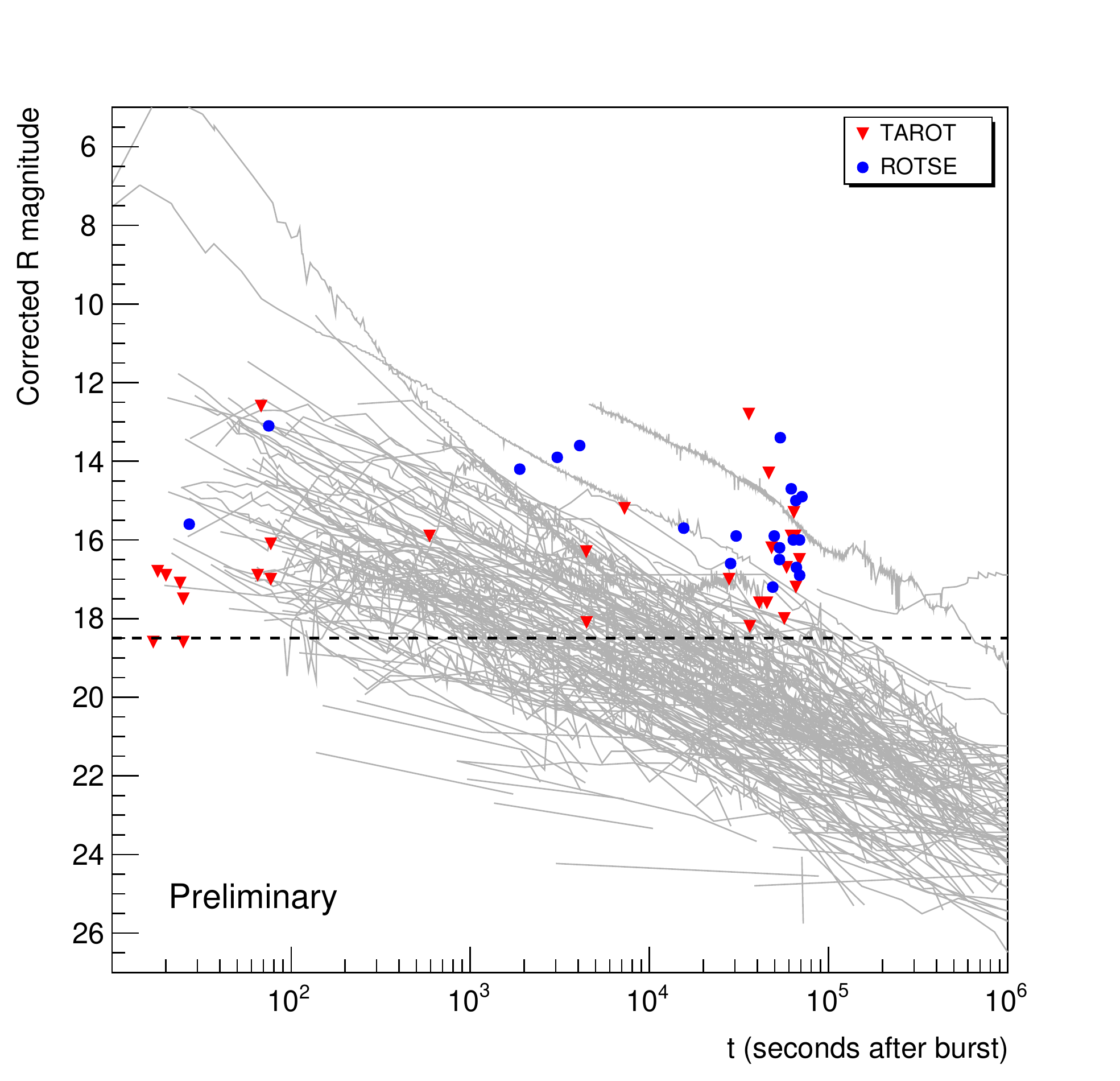}
\caption{Points: limiting magnitudes and delay times of optical follow-up observations to ANTARES alerts with ROTSE and TAROT  (A.~{Mathieu}, ICRC2015 1093) compared to (grey lines) the light-curves from measured GRBs.} \label{fig:tatoo_optical_fig1_left}
\end{SCfigure*}

The TAToO (telescopes--ANTARES target-of-opportunity) program \cite{2012APh....35..530A} performs near-real-time reconstruction of muon-track events. If a sufficiently high energy event is reconstructed as coming from below the horizon (i.e.\ those events most likely to be of astrophysical origin), an alert message is generated to trigger robotic optical telescopes, and, with a higher threshold, the \emph{Swift}-XRT. The very short alert-generation time (a few seconds) and half-sky simultaneous coverage of ANTARES makes it ideal for detecting transient signals, and optical and x-ray follow up observations have been initiated within 20~s and one hour respectively.
 
Result from $42$ optical and $7$ x-ray alerts are reported by A.~{Mathieu} (ICRC2015 1093). While no associated transient event was detected, this non-observation can be used to place limits on the astrophysical origin of the detected neutrinos (A.~{Mathieu}, ICRC2015 1093), as shown in Fig.\ \ref{fig:tatoo_optical_fig1_left}.
The steep fall-off of the light-curves emphasises the need for a rapid alert generation and follow-up: observations within one minute can rule out a GRB origin with high confidence, while those after one day would be unlikely to detect even a bright GRB.

\section{Cascades}
\label{sec:cascades}


The effective area of neutrino telescopes such as ANTARES and IceCube to cascade events (neutral-current (NC) interactions, and $\nu_{\mu}$ and $\nu_{\tau}$ charged-current (CC) interactions) is generally lower than to $\nu_{\mu}$ CC interactions, due to the very long range of the outgoing $\mu$. Additionally, the angular resolution to through-going $\mu$ is superior. However, the cascade channel has several advantages: neutrino events are more easily distinguished from the background of atmospheric muons, allowing both up- and down-going events to be studied; and the energy deposited in the detector is more-strongly correlated with the energy of the neutrino primary.  It was these latter advantages that allowed the diffuse cosmic neutrino flux detected by IceCube to be first observed in the cascade channel \cite{2013Sci...342E...1I}.


Cascade event identification and reconstruction has been in development in ANTARES for several years, and its application in a search for a diffuse flux is reported below. The most important development however has been a new cascade reconstruction algorithm with an unprecedented angular resolution, of typically $3^{\circ}$ accuracy, which for the first time enables a point-source search using the cascade channel.

\subsection{Cascade reconstruction in ANTARES}


\begin{figure*}
\includegraphics[width=0.5\textwidth]{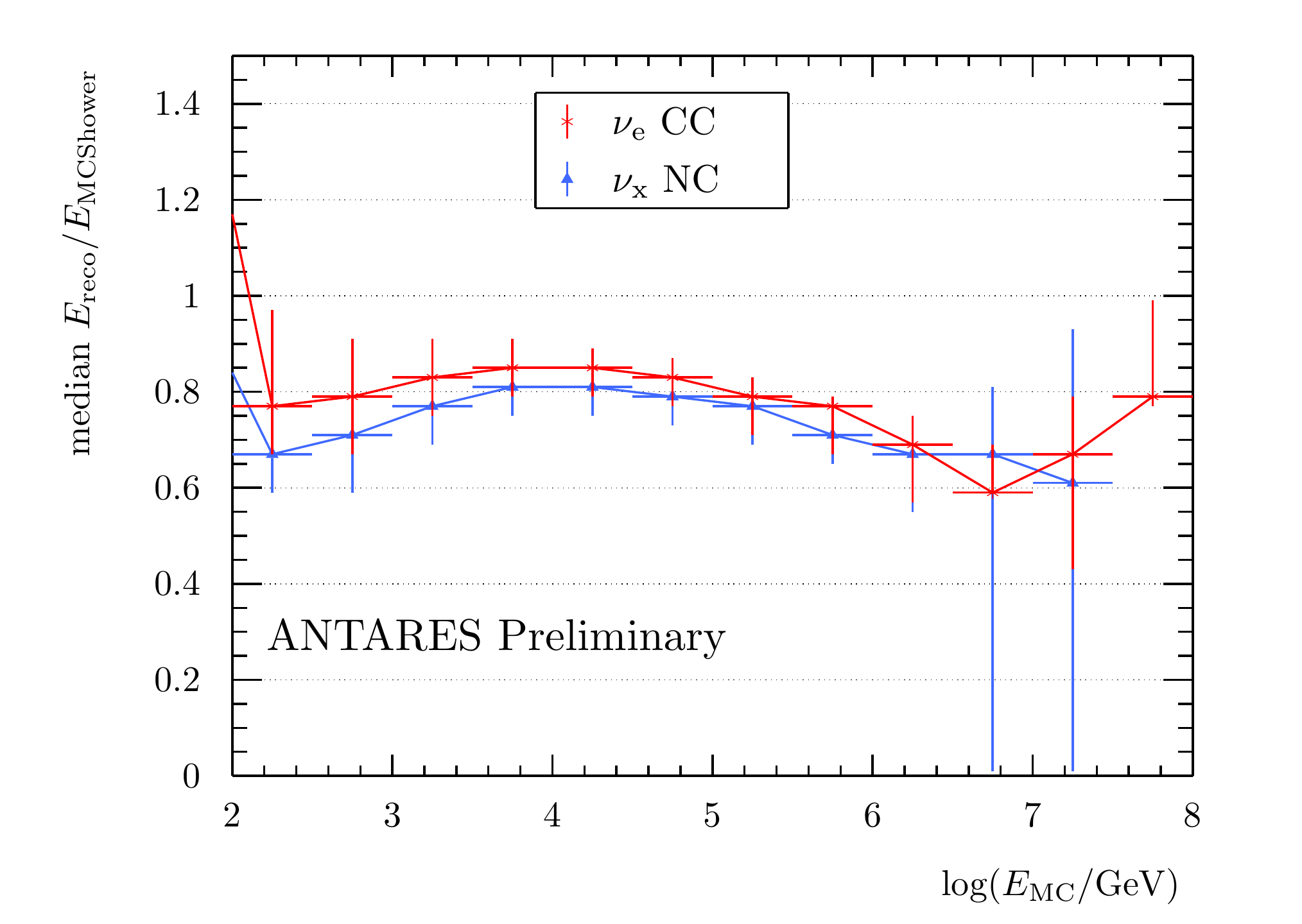} \includegraphics[width=0.5\textwidth]{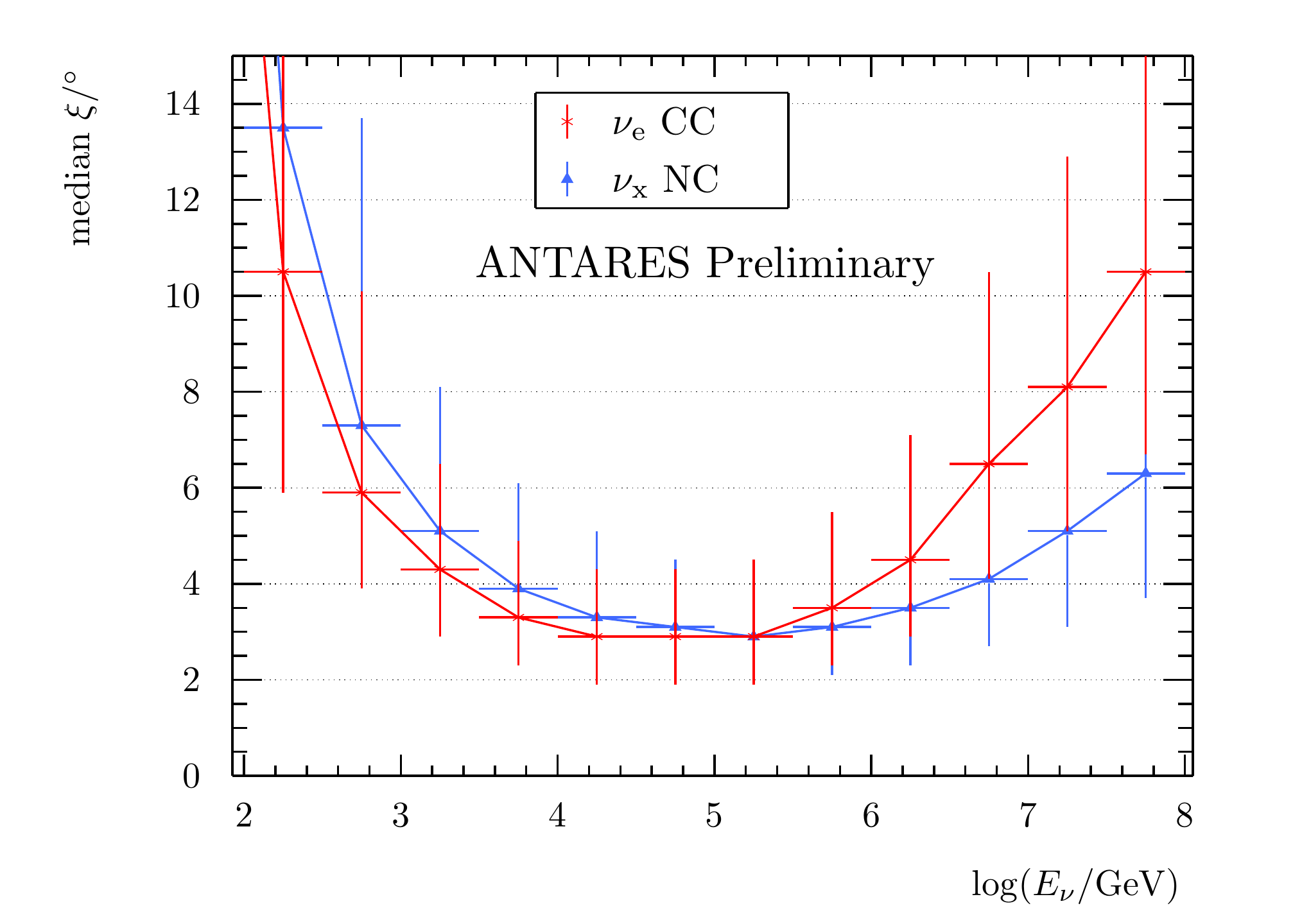}
\caption{Energy (left) and angular (right) resolutions for $\nu_{\mu}$ and $\nu_e$ NC (blue), and $\nu_e$ CC (red), events with ANTARES (T.~{Michael}, ICRC2015 1078).} \label{fig:cascade_res}
\end{figure*}

The most recent cascade-reconstruction algorithm developed for ANTARES, reported by T.~{Michael} (ICRC2015 1078), is termed `Tantra'. Its performance is shown in Fig.\ \ref{fig:cascade_res}. Over the approximate $10$-$300$~TeV range, arrival directions are reconstructed with a median angular error of $3^{\circ}$, and a resolution on deposited energy of $5$\% (the offset from $E_{\rm reco} / E_{\rm MC} = 1$ is easily corrected for), although the latter is limited by the total ANTARES systematic energy uncertainty of approximately $10$\%.  Below $10$~TeV, the resolutions worsen due to a decreasing number of photons being detected, while above $300$~TeV, the events begin to saturate the detector. Over the entire $100$~GeV to $100$~PeV range, the median angular resolution improves on the IceCube resolution for purely shower-like high-energy starting events (i.e.\ those without an outgoing muon) \cite{2014PhRvL.113j1101A}.

\subsection{Point-source search including cascades}
\label{sec:point_source_cascades}

\begin{figure*}
\includegraphics[width=0.48\textwidth]{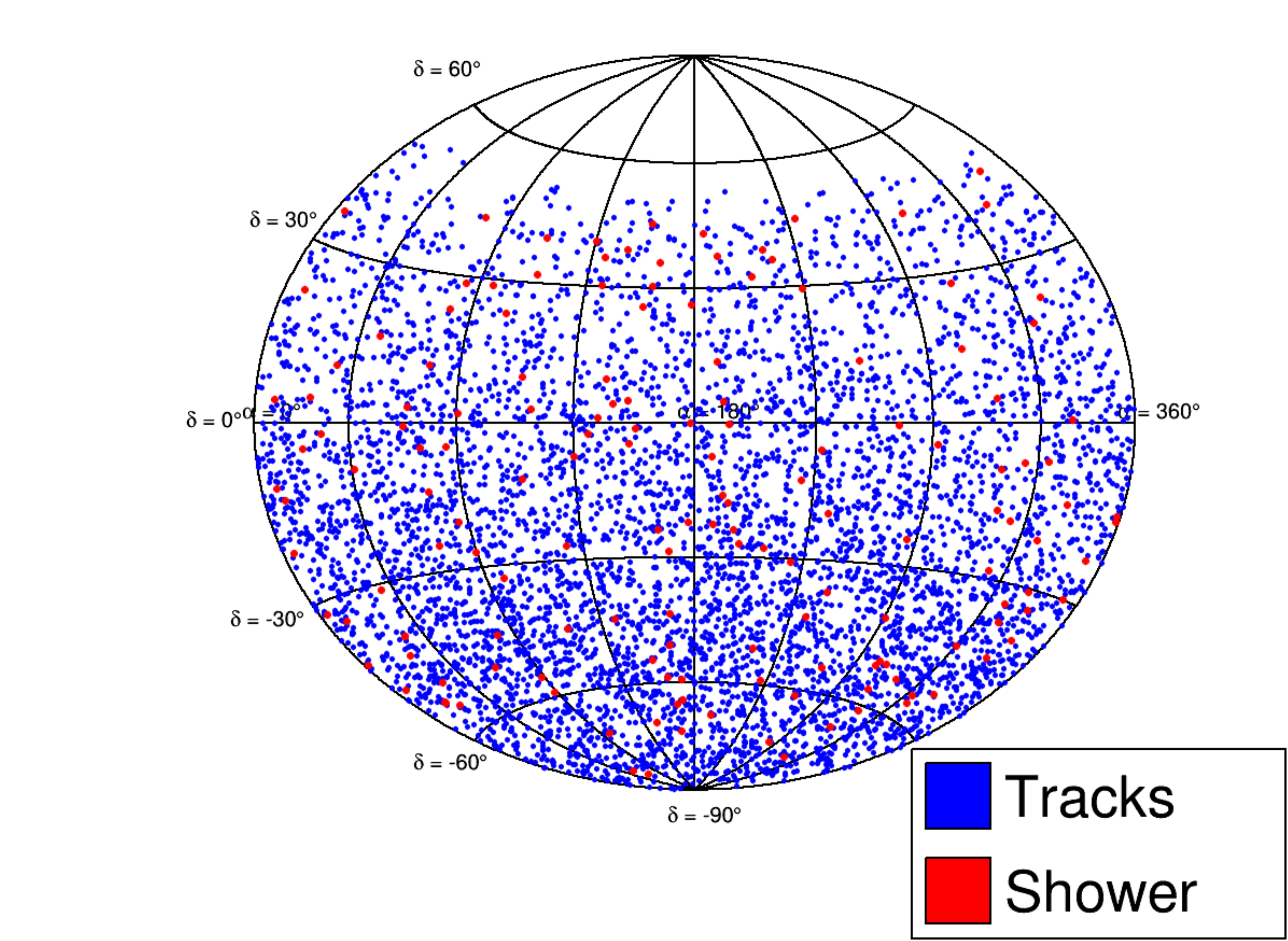} \includegraphics[width=0.48\textwidth]{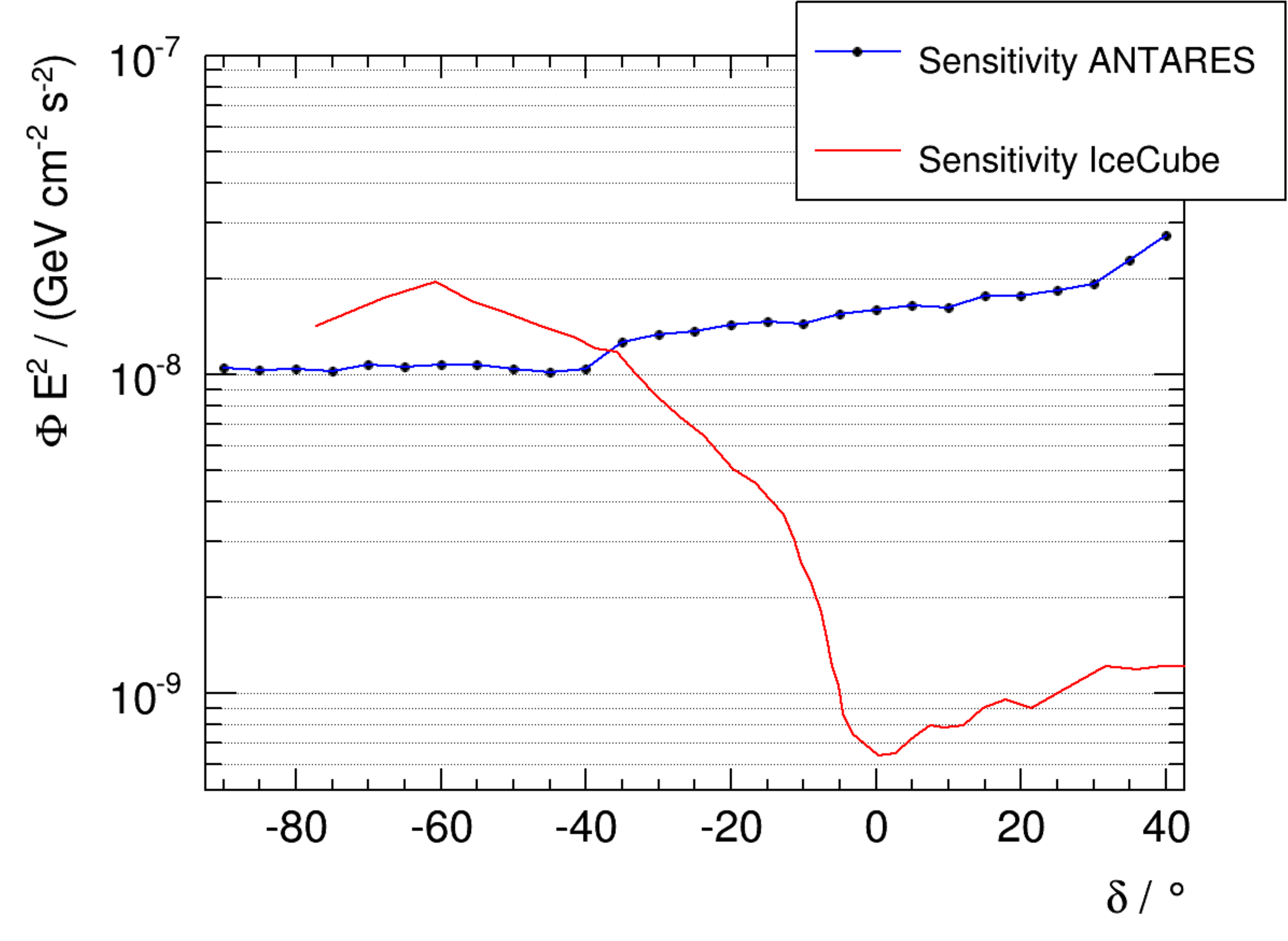}
\caption{Left: The arrival directions of events used in the ANTARES all-sky point-source-search sample. Right: sensitivity of the ANTARES targeted search to flavour-uniform neutrino point sources with $E^{-2}$ spectra in terms of flux per flavour, compared to the IceCube results from Ref.\ \cite{2015ApJ...809...98A}.} \label{fig:point_source_results}
\end{figure*}

A combined point-source search using both muon-track and cascade events has been performed using $1622$ days of effective livetime from $2007$ to $2013$ (T.~{Michael}, ICRC2015 1078). After cuts, the sample consisted of $6261$ muon-track events, and $156$ cascade events, with an estimated contamination of $10$\% mis-reconstructed atmospheric muons in each.

The resulting skymap is shown in Fig.\ \ref{fig:point_source_results} (left). An untargeted point-source search, a search over a list of pre-specified candidates, and a search using the origins of the IceCube events reported in Ref.\ \cite{2014PhRvL.113j1101A} were applied to this data. No significant excess was observed. The resulting limits on point-like sources are given in Fig.\ \ref{fig:point_source_results} (right). While the atmospheric background produces predominantly muon-track events, an $E^{-2}$ point source with a flavour-uniform flux would be expected to produce a cascade-to-track ratio of $3$:$10$, significantly increasing the sensitivity of the search. Thus the achieved search sensitivity was approximately $10^{-8}$ GeV$^{-1}$ cm$^{-2}$ s$^{-1}$ for $\delta < -40^{\circ}$.

\subsection{Diffue flux search}
\label{sec:diffuse_search}

A diffuse flux search in ANTARES has been developed that makes optimal uses of both muon-track and cascade events (Schnabel \& {Hallmann}, ICRC2015 1065).
Since any explicit selection of muon-like and cascade-like events inevitably discards events with topologies falling between the two classes, no such selection was made.

The procedure was first optimised for, and applied to, the $913$ days of effective livetime between $2007$ and $2013$ exhibiting the best data-taking conditions (mostly low bioluminescent activity). The expected number of events from the standard and prompt atmospheric background \cite{2007PhRvD..75d3006H,2008PhRvD..78d3005E} was $9.5 \pm 2.5$, composed of $5.5$ $\nu_{\mu}$ CC, $1$ atmospheric $\mu$, and $2.9$ $\nu$ NC and $\nu_e$ events. The expectation from the IceCube neutrino flux reported by Ref.\ \cite{2015ApJ...809...98A}\footnote{Flux-per-flavour of $\Phi = 2.23 \times 10^{-18} (E / {\rm 1 TeV})^{-2.5}$ GeV$^{-1}$ cm$^{-2}$ s$^{-1}$ sr$^{-1}$} was $5.0 \pm 1.1$ events.

\begin{figure*}
\includegraphics[width=\textwidth]{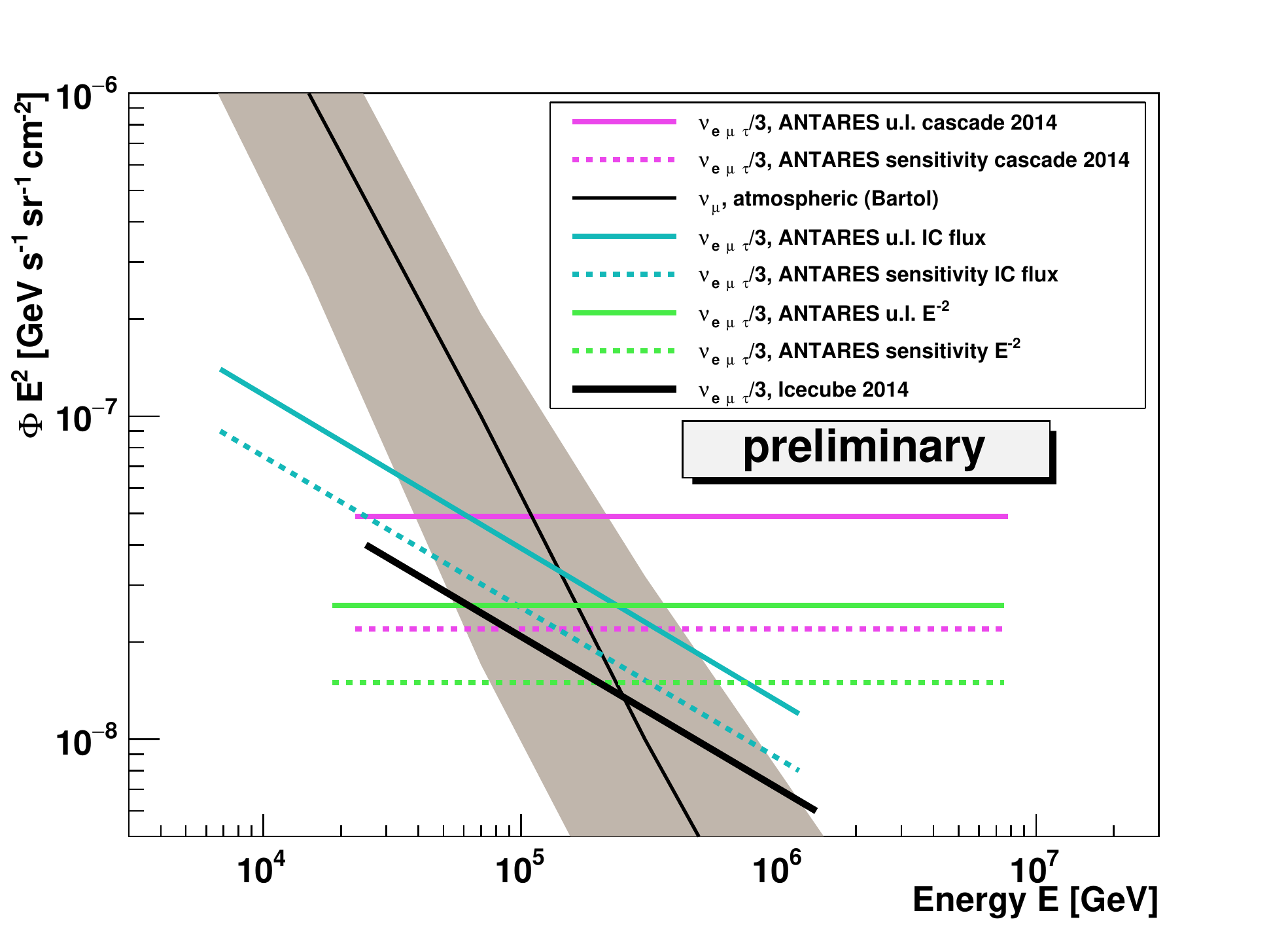}
\caption{ANTARES sensitivity to (dotted), and limits on (solid), a diffuse astrophysical neutrino flux (Schnabel \& {Hallmann}, ICRC2015 1065), showing (pink) the previous ANTARES limit \cite{2011PhLB..696...16A}, (green) this work, and (blue) the flux observed by IceCube \cite{2015ApJ...809...98A}. This is compared to the conventional atmospheric background flux (black) \cite{2007PhRvD..75d3006H}, with associated error (grey shading).} \label{fig:diffuse_limits}
\end{figure*}

After unblinding, $12$ events passed the selection cuts --- consistent with both background-only, and background and IceCube diffuse flux expectations. The resulting limits on an $E^{-2}$ flux are given in Fig.\ref{fig:diffuse_limits}.

\section{Extended source searches}
\label{sec:diffuse}

In addition to the numerous point-like candidate neutrino sources targeted in Secs. \ref{sec:point_sources} and \ref{sec:cascades}, several extended regions have been proposed as hadronic acceleration sites. ANTARES searches for an excess neutrino flux from these regions using 'on-zones' defined by specific templates, which are compared to 'off-zones' of exactly the same size and shape, but offset in right ascension. Thus the off-source regions give an unbiased estimate of the background in the source region in a way that is independent of simulations. Results for the Fermi Bubbles, Galactic plane, and the IceCube cluster are described below.

\subsection{Fermi bubbles}
\label{sec:fermi_bubbles}

The Fermi Bubbles \cite{2010ApJ...724.1044S} are giant regions of $\gamma$-ray emission extending out of the galactic centre, and are proposed hadronic acceleration site \cite{2015PhRvD..92b1301L}, with neutrinos expected from $p$--$p$ collisions. A first search in ANTARES data from 2008--2011 for emission from these regions was presented by Ref.\ \cite{2014EPJC...74.2701A} --- here, an update is presented using 2012--2013 data.

\begin{figure*}[h!]
\centerline{\includegraphics[trim={2cm 0 0 0},clip=true, width=0.5\textwidth]{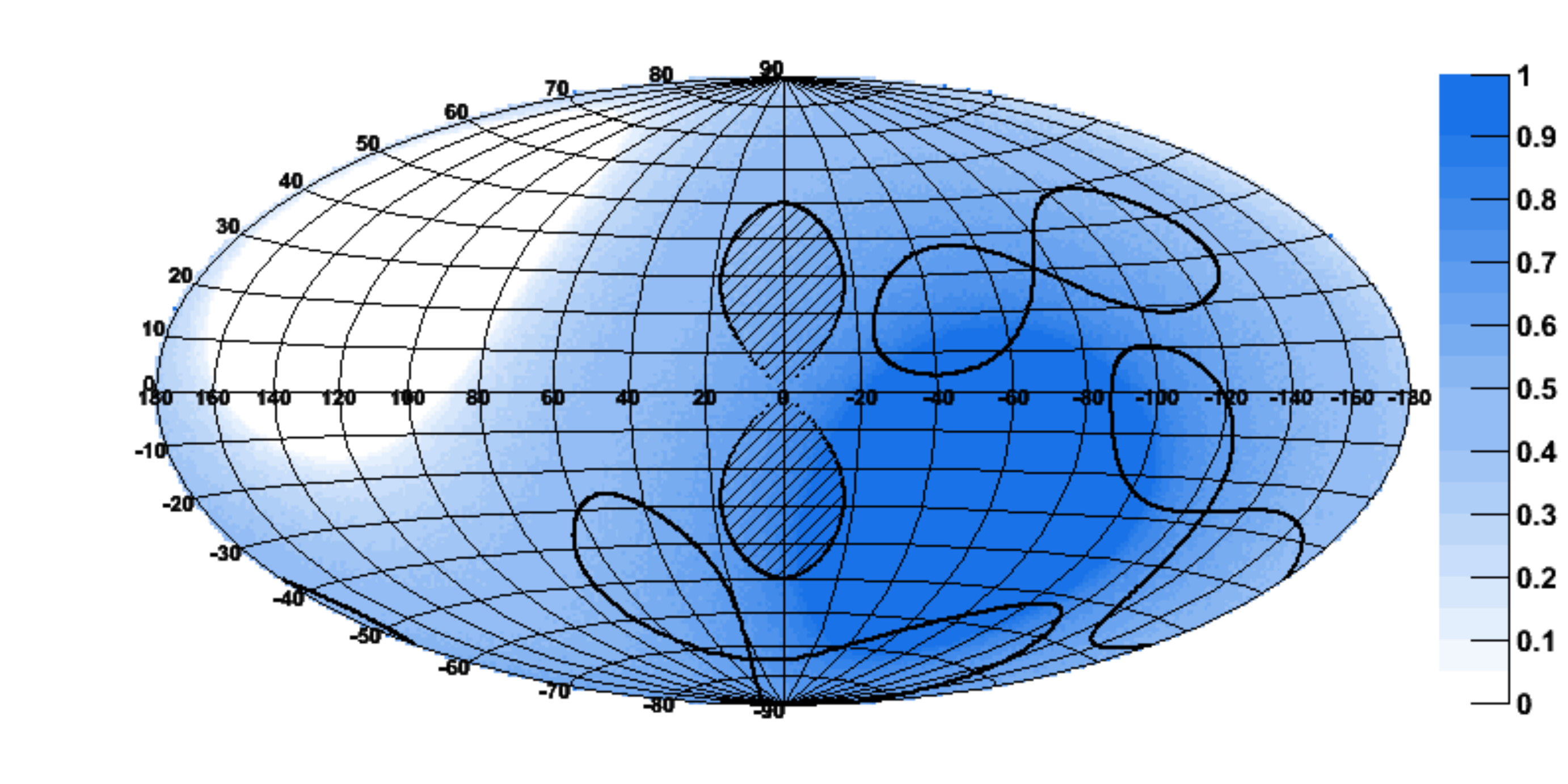}\includegraphics[width=0.45\textwidth]{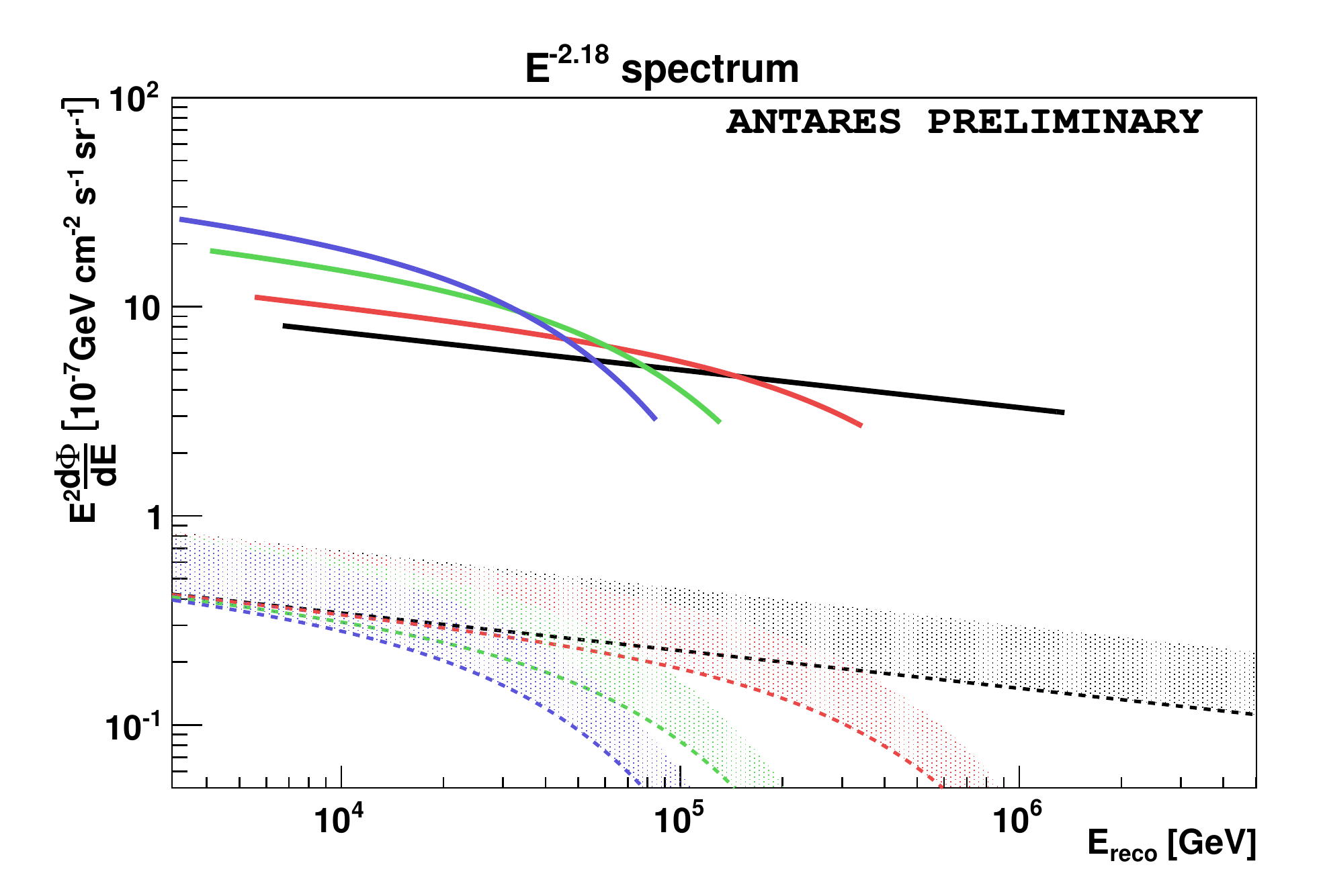}}
\caption{Left: on- and off-zone search regions for the Fermi Bubble search of S.~{Hallmann} (ICRC2015 1059), compared to the ANTARES visibility (blue shading). Right: $90$\% C.L. upper limits (lines) on the neutrino flux from the Fermi Bubbles, compared to (shaded regions) expectations \cite{2015PhRvD..92b1301L} for different spectral shapes.} \label{fig:fermi_bubbles_fig2_fig5l}
\end{figure*}

The on- and off-zone regions used in the Fermi Bubble analysis are shown in Fig.\ \ref{fig:fermi_bubbles_fig2_fig5l} (left). Flavour-uniform $E^{-2}$ and $E^{-2.18}$ neutrino fluxes are assumed, where the latter is motivated by the best-fit proton spectrum of $E^{-2.25}$ reported by Ref.\ \cite{2015PhRvD..92b1301L}.
Exponential cut-offs at energies of $500$, $100$, and $50$ TeV are also tested.

A slight excess is found in the source region, corresponding to a $1.9~\sigma$ significance. The corresponding upper limits on an $E^{-2.18}$ neutrino flux are compared in Fig.\ \ref{fig:fermi_bubbles_fig2_fig5l} (right) to the expectations from Ref.\ \cite{2015PhRvD..92b1301L}.

\subsection{Galactic plane}
\label{sec:gal_plane}

Cosmic rays in our galaxy will collide with the interstellar medium to produce pions and, hence, neutrinos. Direct evidence for these processes comes from observations by \emph{Fermi}-LAT \cite{2012ApJ...750....3A} of the diffuse galactic $\gamma$-ray background. It is also interesting that the number of IceCube high energy starting events (HESE) in the $E>100$~TeV range \cite{2014PhRvL.113j1101A} with angular reconstructions consistent with this region corresponds to a flux consistent with that observed in $\gamma$ rays \cite{2014PhRvD..89j3002N}, as shown in Fig.\ \ref{fig:galactic_plane_fluxes}. The large uncertainty in the arrival directions of cascade-like HESE, and their low number, makes this comparison difficult however. More-detailed simulations of the expected neutrino flux are given in Refs.\ \cite{2015arXiv150400227G}.

\begin{figure*}
\includegraphics[width=\textwidth]{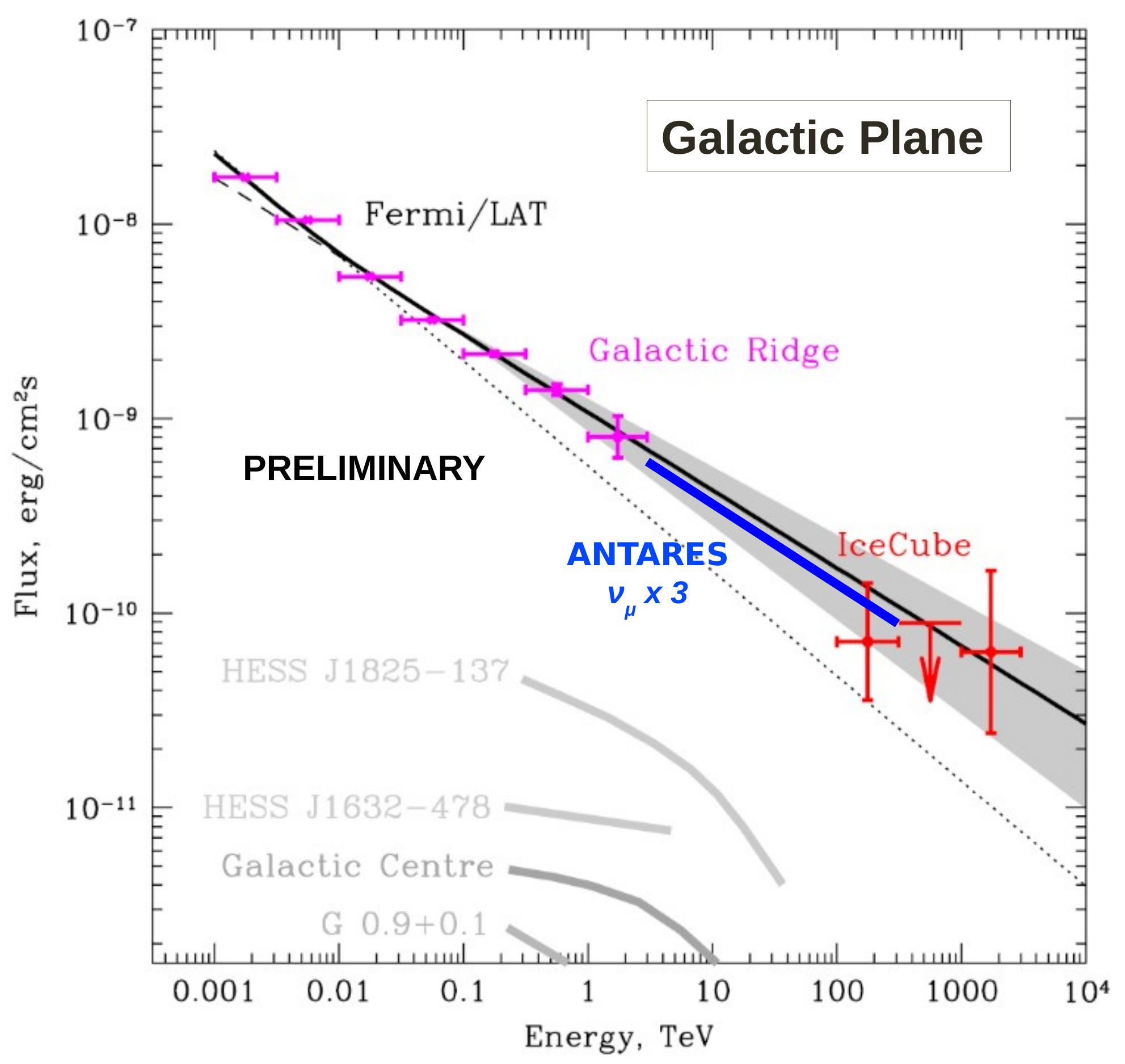}
\caption{\emph{Fermi}-LAT-detected gamma-ray flux from the Galactic Ridge (pink), and IceCube events consistent with this region (red), as computed in Ref.\ \cite{2014PhRvD..89j3002N}, compared to the ANTARES all-flavour flux limit (blue) (L.~{Fusco}, ICRC2015 1055) and gamma-ray fluxes from various other galactic sources (grey) \cite{2015arXiv150400227G}.} \label{fig:galactic_plane_fluxes}
\end{figure*}

ANTARES' northern latitude is ideally suited to studying the expected neutrino flux from the inner galactic plane, and a search has been performed searching in the regions of galactic longitude $|l| < 40^{\circ}$ and latitude $|b| < 3^{\circ}$, as reported by L.~{Fusco} (ICRC2015 1055). The search used nine off-zones and one on-zone, and found no excess in the on-zone region (one event compared to an average of $2.5$ for the off-zones). The resulting limits are shown in Fig.\ \ref{fig:galactic_plane_fluxes}. In particular, the hypothesis of a 1--1 relation between the $\gamma$-ray and neutrino flux from the Galactic Ridge is ruled out at $90$\% confidence, showing that ANTARES is already testing the well-established multimessenger $\gamma$--$\nu$--CR paradigm in our galaxy. The limits cannot rule out however models from more-detailed simulations of galactic cosmic-ray propagation.

\subsection{IceCube cluster}
\label{sec:icecube_cluster}

The same search techniques employed in the galactic plane search were used to probe the origin of the cluster of IceCube events seen in Ref.\ \cite{2014PhRvL.113j1101A}. The analysis (L.~{Fusco}, ICRC2015 1055) used twelve off-zones and one on-zone to search for an excess of events. One event passing selection cuts is observed in both the on-zone and the average off-zone, i.e.\ no excess (significant or otherwise) is observed. Resulting limits on the maximum number of HESE produced by a source with different spectral indices are presented in Fig.\ \ref{fig:icecube_cluster_results}, calculated analogously to the point-source search of Sec.\ \ref{sec:point_sources} and Fig.\ \ref{fig:icecube_pointlike_results}.
For the best-fit IceCube diffuse spectral index $\Gamma = 2.5$ \cite{2015ApJ...809...98A}, ANTARES rejects at $90$\% confidence a flux from this region expected to produce three or more of the IceCube events in the cluster. This extends the results of Ref.\ \cite{2014ApJ...786L...5A} and J.~{Barrios-Mart\'i} (ICRC2015 1077) for this region, which limit the existence of point-like and mildly extended sources in this region.

\begin{figure*}
\includegraphics[width=\textwidth]{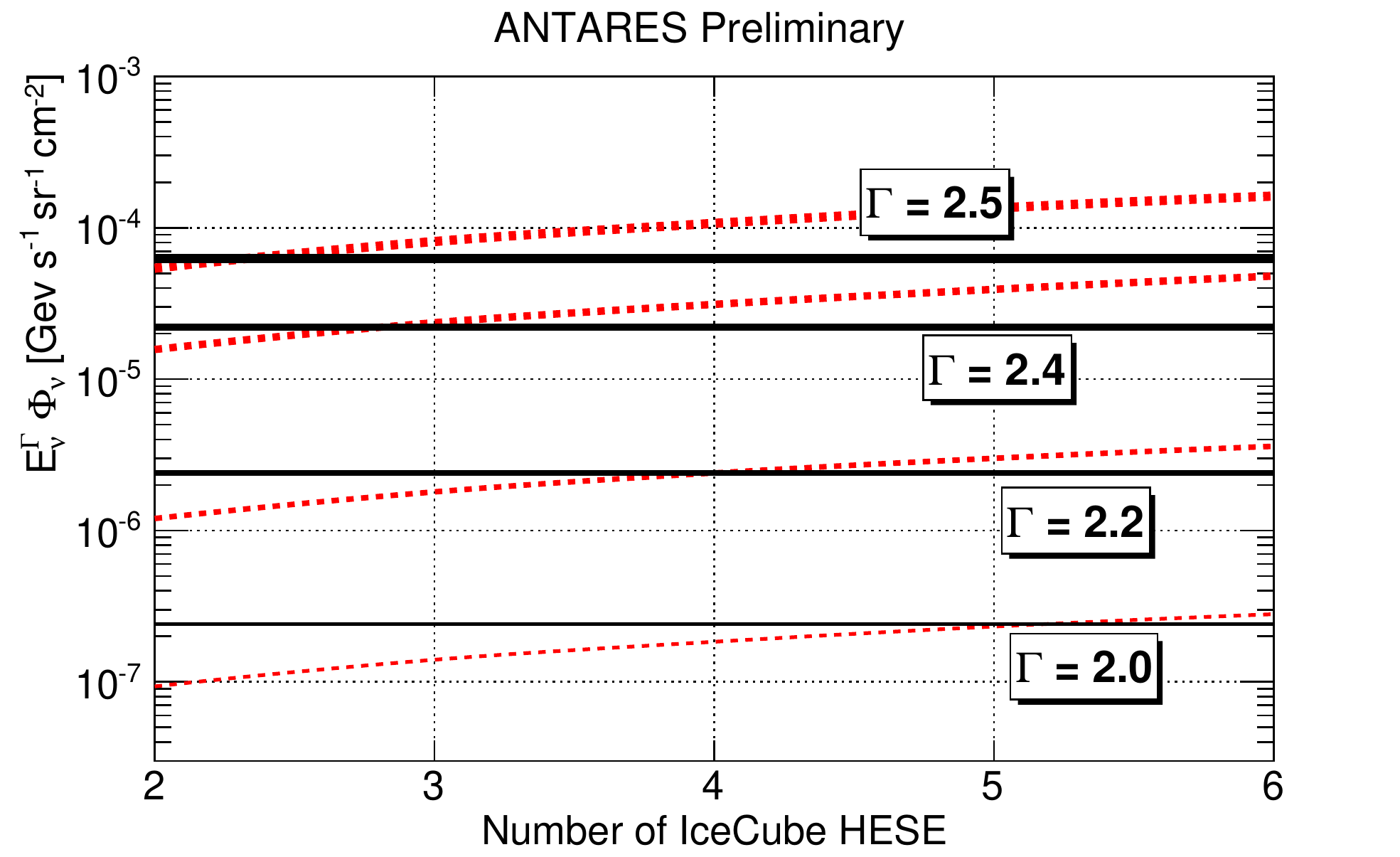}
\caption{ANTARES upper limits at $90$\% C.L.\ (black) on a flavour-uniform neutrino flux from the IceCube cluster region as a function of the spectral index $\Gamma$, compared to (red) the flux required to produce an expected number of events in the IceCube HESE analysis \cite{Spurio_GC}. The maximum number of IceCube events allowed at $90$\% C.L.\ is indicated by the crossing points of the red and black lines for a given spectral index. See L.~{Fusco} (ICRC2015 1055) for details.} \label{fig:icecube_cluster_results}
\end{figure*}

\subsection{Model-independent searches}

It is possible that as-yet unknown sources or source populations produce a significant neutrino flux.
Two techniques have been used by ANTARES to perform the most general searches possible. A two-point autocorrelation analysis is performed by R.~{Gracia Ruiz} (ICRC2015 1074), searching for an excess of clustering on angular scales up to $60^{\circ}$. A small ($2.2 \sigma$) excess is found at angular scales of less than $0.5^{\circ}$, i.e.\ within the reconstruction accuracy of the detector, though this is not statistically significant.

In S.~{Gei{\ss}els{\"o}der} (ICRC2015 1054), a search for individual sources of arbitrary shape and size is presented. The algorithm searches for local clustering, and identifies regions with an excess of events. This procedure identified a very large structure of unusual shape containing the galactic centre region, with a post-trial p-value of $2.5\sigma$ based on simulations and data-scrambling.
A detailed analysis of possible systematic effects has not identified any reason for such a fluctuation, and the correct interpretation of this result remains an open question.

\section{Dark matter and Exotics}
\label{sec:dark_matter}

ANTARES can place limits on different WIMP dark-matter scenarios by limiting the neutrino flux expected from WIMP interactions in the Sun, Earth, Galactic Centre, and dwarf galaxies. Since the expected dark-matter density tends to be strongly peaked near the centres of these objects, and ANTARES has an excellent angular resolution, competitive limits can be set in the $E_{\rm WIMP} \gtrsim 50$~GeV range where ANTARES is sensitive.

\begin{figure*}
\centerline{\includegraphics[width=0.49\textwidth]{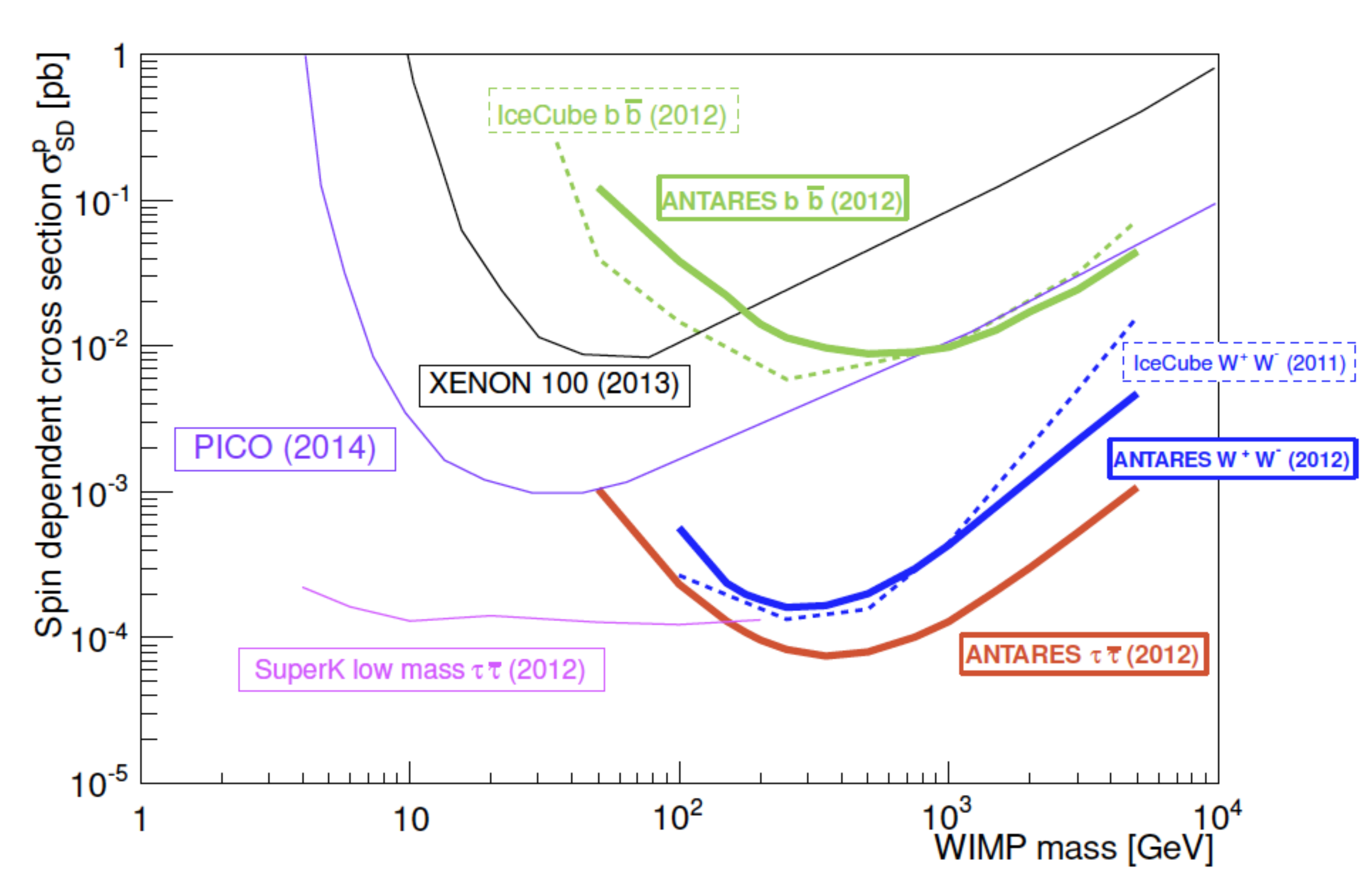} \includegraphics[width=0.45\textwidth]{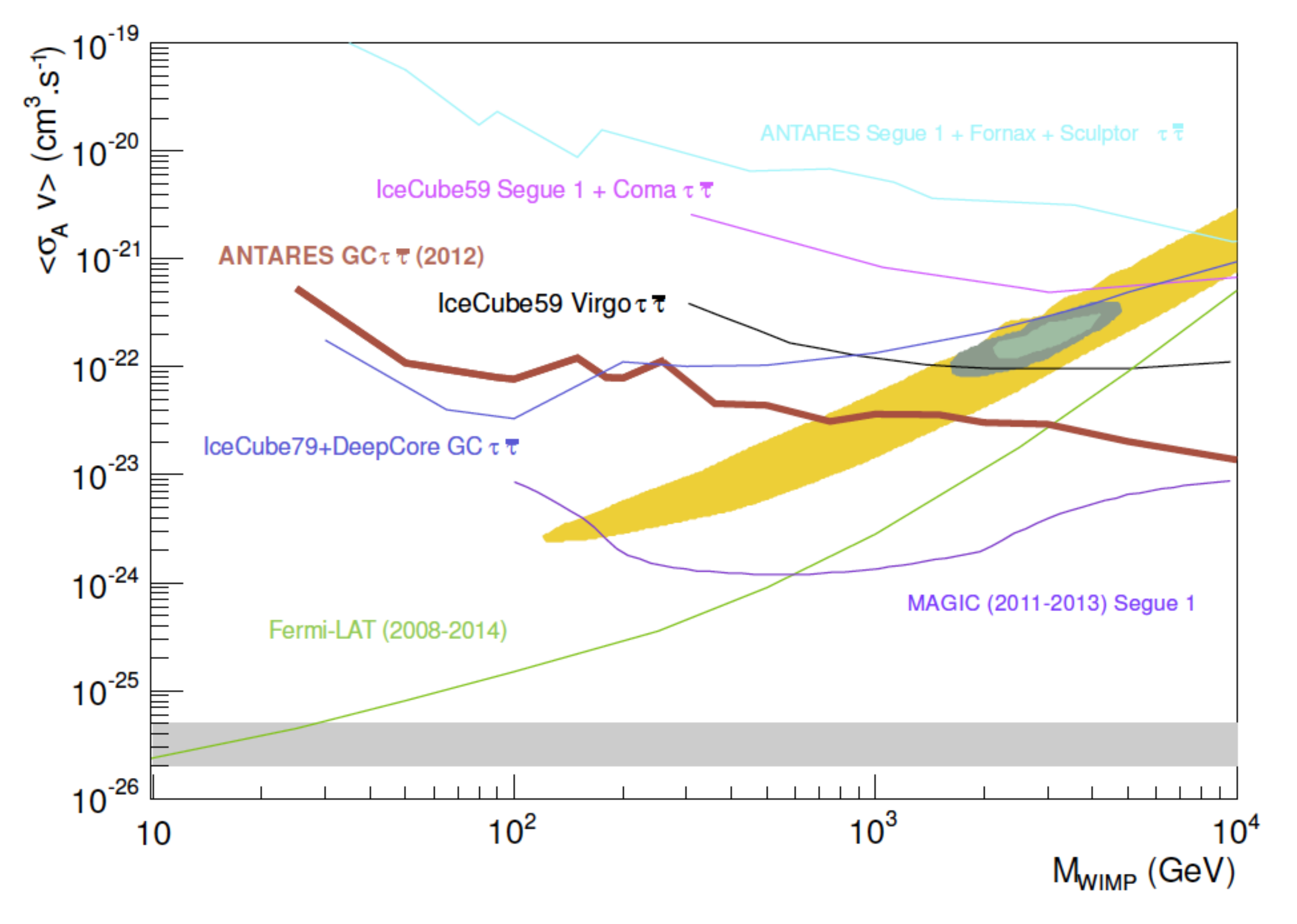} }
\caption{ANTARES limits $\sigma^p_{\rm SD}$ from the Sun (left) and on $\left< \sigma_A v \right>$ from the Galactic Centre (right) as a function of the WIMP mass. See C.~{T{\"o}nnis} (ICRC2015 1207) for details and associated references.} \label{fig:wimp_limits}
\end{figure*}

Limits on the spin-dependent (WIMP-proton) interaction cross section $\sigma^p_{\rm SD}$ from ANTARES observations of the Sun (left), and on the WIMP-WIMP velocity-averaged self-annihilation cross section $\sigma_A v$ from the Galactic Centre (right) using the $\tau \bar{\tau}$ channel are given in Fig.\ \ref{fig:wimp_limits}, and are described in further detail by C.~{T{\"o}nnis} (ICRC2015 1207).

Dark-matter analyses by ANTARES also includes a search for a WIMP signature from the centre of the Earth ({Gleixner} \& {T{\"o}nnis}, ICRC2015 1110), and a test of secluded dark-matter models in the Sun ({Ardid} \& {T{\"o}nnis}, ICRC2015 1212).

ANTARES also places limits on beyond-the-standard-model physics, with searches for magnetic monopoles and nuclearites. Updates to existing limits are presented in Ref.\ {El Bojaddaini} \& {P{\u{a}}v{\u{a}}la{\c{s}}} (ICRC2015 1060)  and G.~{P{\u{a}}v{\u{a}}la{\c{s}}} (ICRC2015 1060) respectively.

\section{KM3NeT -- ARCA and ORCA}
\label{sec:km3net}

KM3NeT ({\tt www.km3net.org}) is a multi-site deep-sea research infrastructure. Two components are described here: ARCA (Astrophysical Research with Cosmics in the Abyss), a neutrino telescope for performing high-energy neutrino astronomy
(P.~Piattelli, ICRC2015 1158); and ORCA (Oscillations Research with Cosmics in the Abyss), to study neutrino oscillations parameters
and resolve the neutrino mass hierarchy (J.~Brunner, ICRC2015 1140).

\begin{figure*}
\begin{center}
\raisebox{-0.5\totalheight}{  \includegraphics[width=0.37\textwidth]{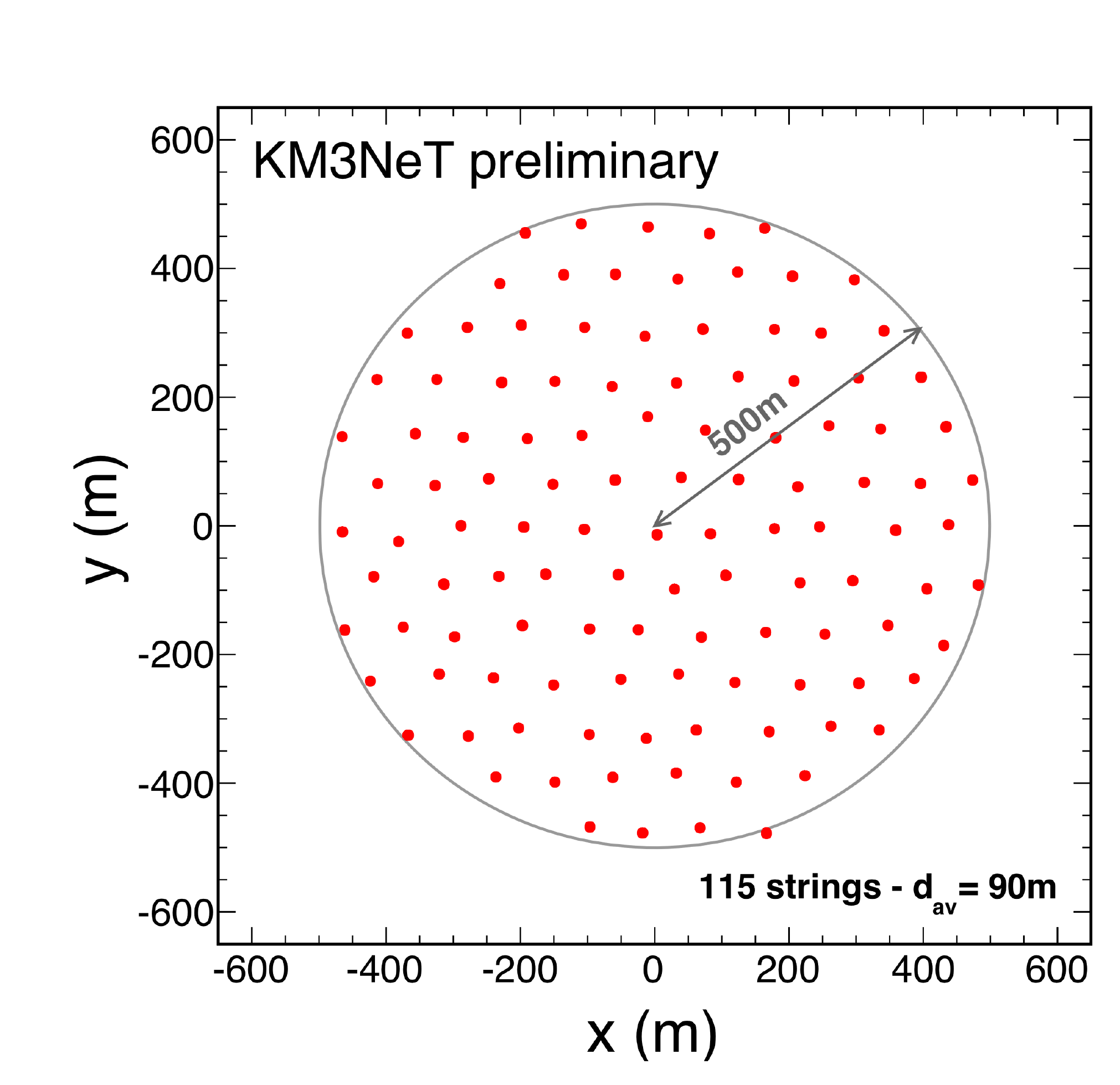} } \raisebox{-0.5\totalheight}{  \includegraphics[width=0.35\textwidth]{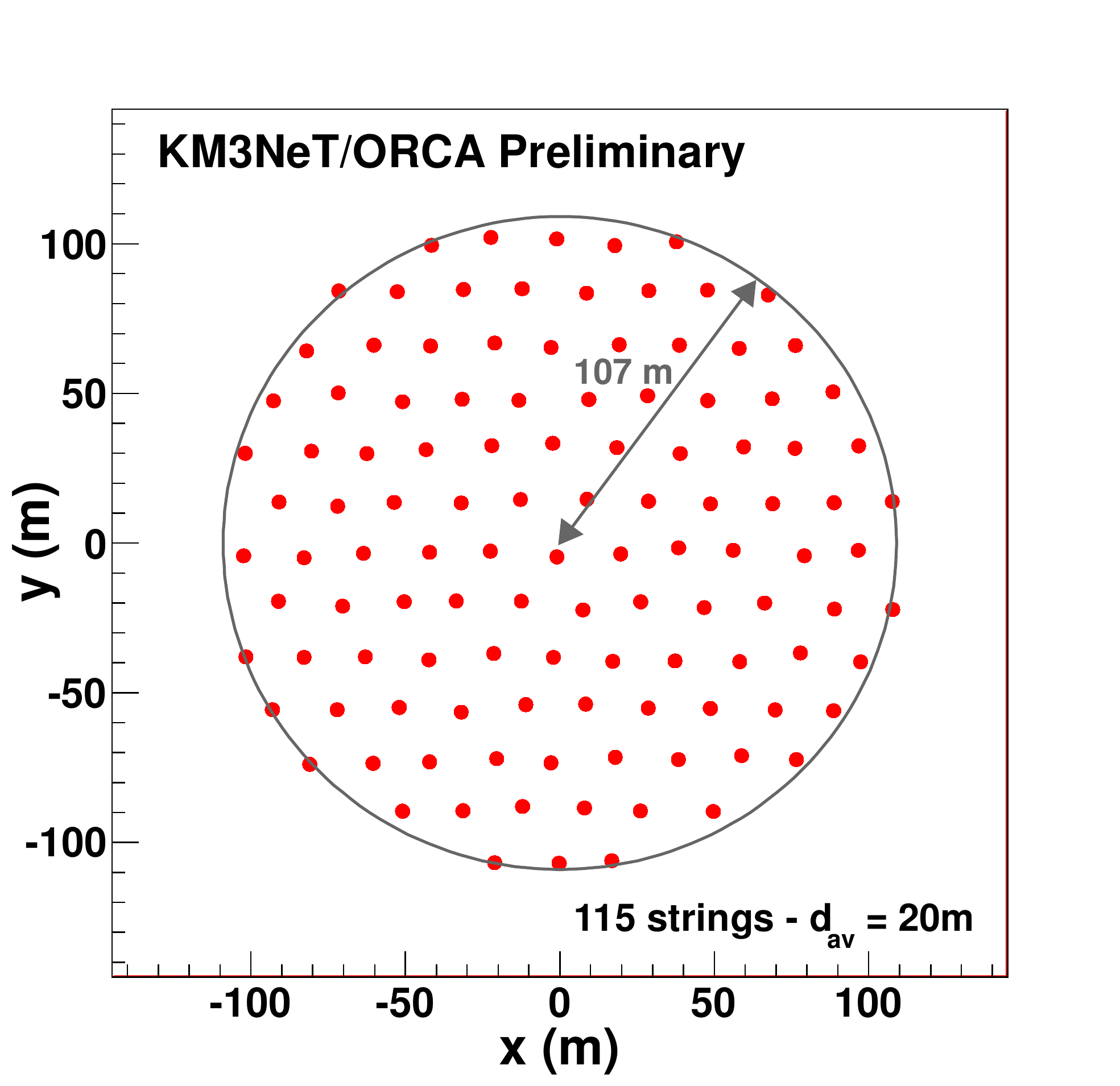} }
\raisebox{-0.5\totalheight}{  \includegraphics[width=0.17\textwidth]{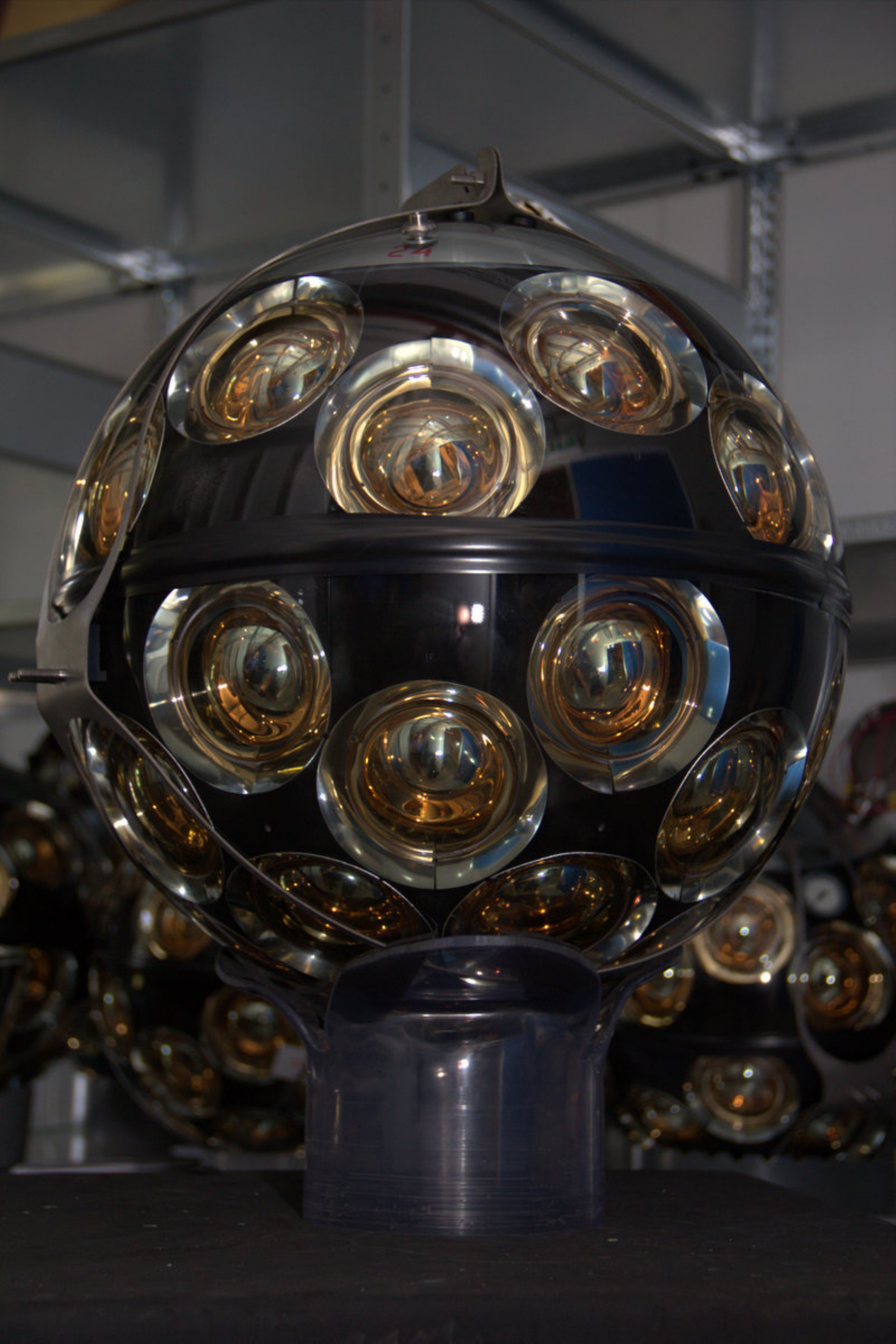} }
\end{center}
\caption{Preliminary seafloor layouts of the 115 detection units (DUs) in the ARCA (left) and ORCA (centre) blocks. Right: photograph of a KM3NeT DOM immediately after assembly.
} \label{fig:block}
\end{figure*}

ARCA will consist of two detection `blocks', each consisting of $115$ vertical detection units (DUs) with $18$ multi-PMT digital optical modules (DOMs) with $31$ photomultiplier tubes (PMTs) per DOM. A sketch of the ARCA block layout is given in Fig.\ \ref{fig:block} (left). Both blocks will be deployed $10$~km apart at the KM3NeT Italian site (shore station at Capo Passero), with seafloor depth $3500$~m, during Phase $2$ of deployment. The KM3NeT-It site has been extensively studied in the context of the NEMO experiment (see e.g.\ Ref.\ \cite{2015APh....66....1A}). ARCA is envisaged to be extended to a total of six blocks over multiple sites during Phase $3$.

ORCA will consist of a single block with the same number of DUs and DOMs, but in a denser configuration (Fig.\ \ref{fig:block}, middle). It will be fully deployed during Phase $2$ at the KM3NeT France site (seafloor depth $2475$~m), $10$~km East of the current ANTARES detector, with shore station at Lyon, France.

\subsection{KM3NeT Phase 1: status}

KM3NeT has completed its initial design and technical verification, and is currently in Phase $1$ of production and deployment. Procedures for PMT testing ({Mollo} \& {Piattelli}, ICRC2015 1159) and DU deployment (P.~{Kooijman}, ICRC2015 1173) are in place, and timing (M.~{Bouwhuis}, ICRC2015 1170), acoustic positional (S.~{Viola} et~al., ICRC2015 1169), and environmental (van Elewyck, Keller \& Lindsey Clark, ICRC2015) calibration devices have been developed. A data-acquisition system (Biagi et al., ICRC2015 1172) based on the ``all-data-to-shore'' philosophy is in place at the KM3NeT-It site, while the main electro-optical cable and junction box have been deployed at KM3NeT-Fr.



Several stages of prototype DOMs have been deployed and tested, from an initial prototype DOM deployed at the ANTARES site in 2013, to a prototype detection unit with three DOMs at KM3NeT-It in 2014 (Biagi, Creusot, \& Bormuth, ICRC2015 1164). The final design of the KM3NeT DOM is reported by {Bruijn} \& {van Eijk} (ICRC2015 1157), and is the technology upon which both ORCA and ARCA is based.

\begin{figure*}
\begin{center}
\centerline{\includegraphics[width=0.32\textwidth]{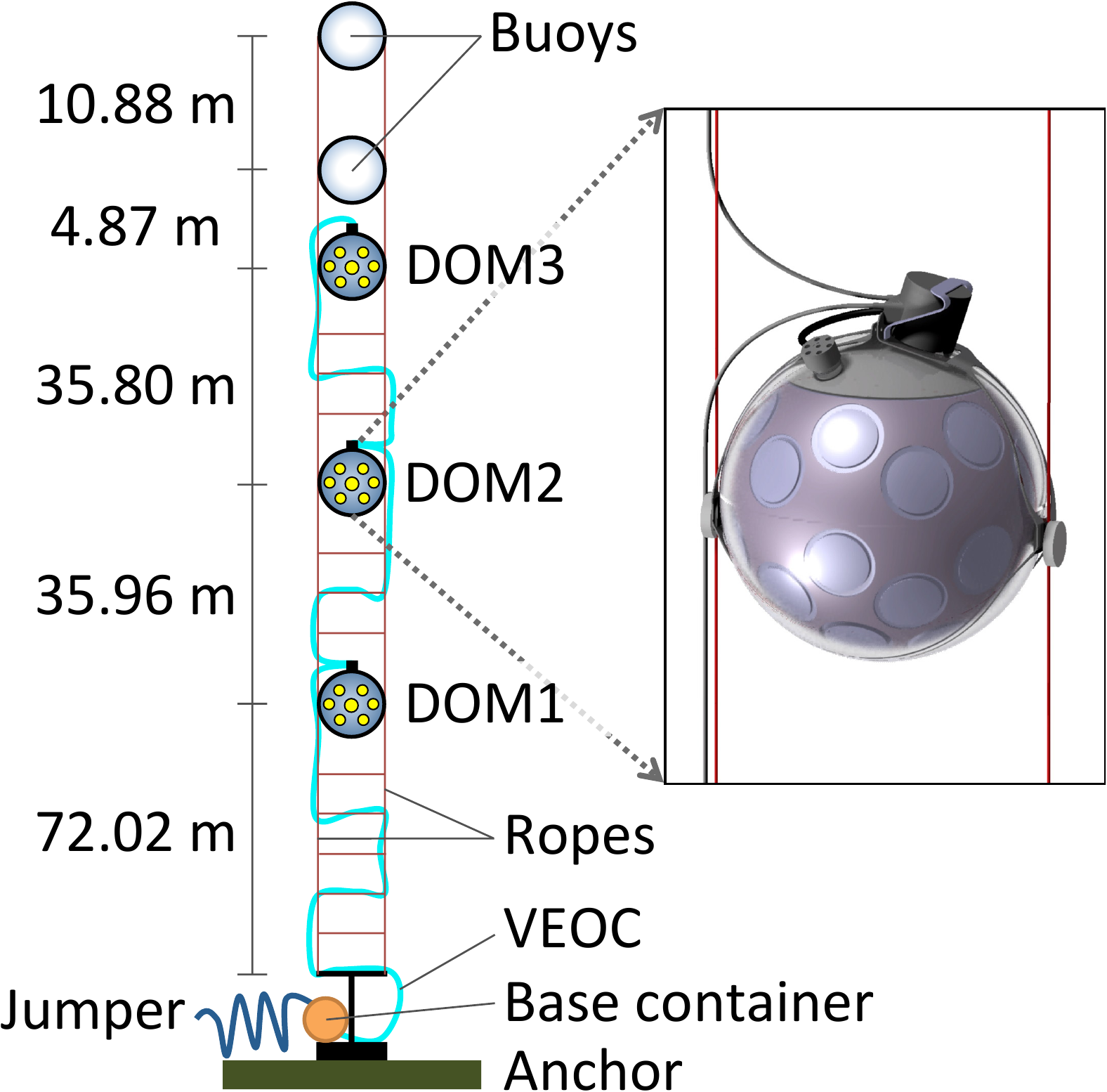} \includegraphics[width=0.55\textwidth]{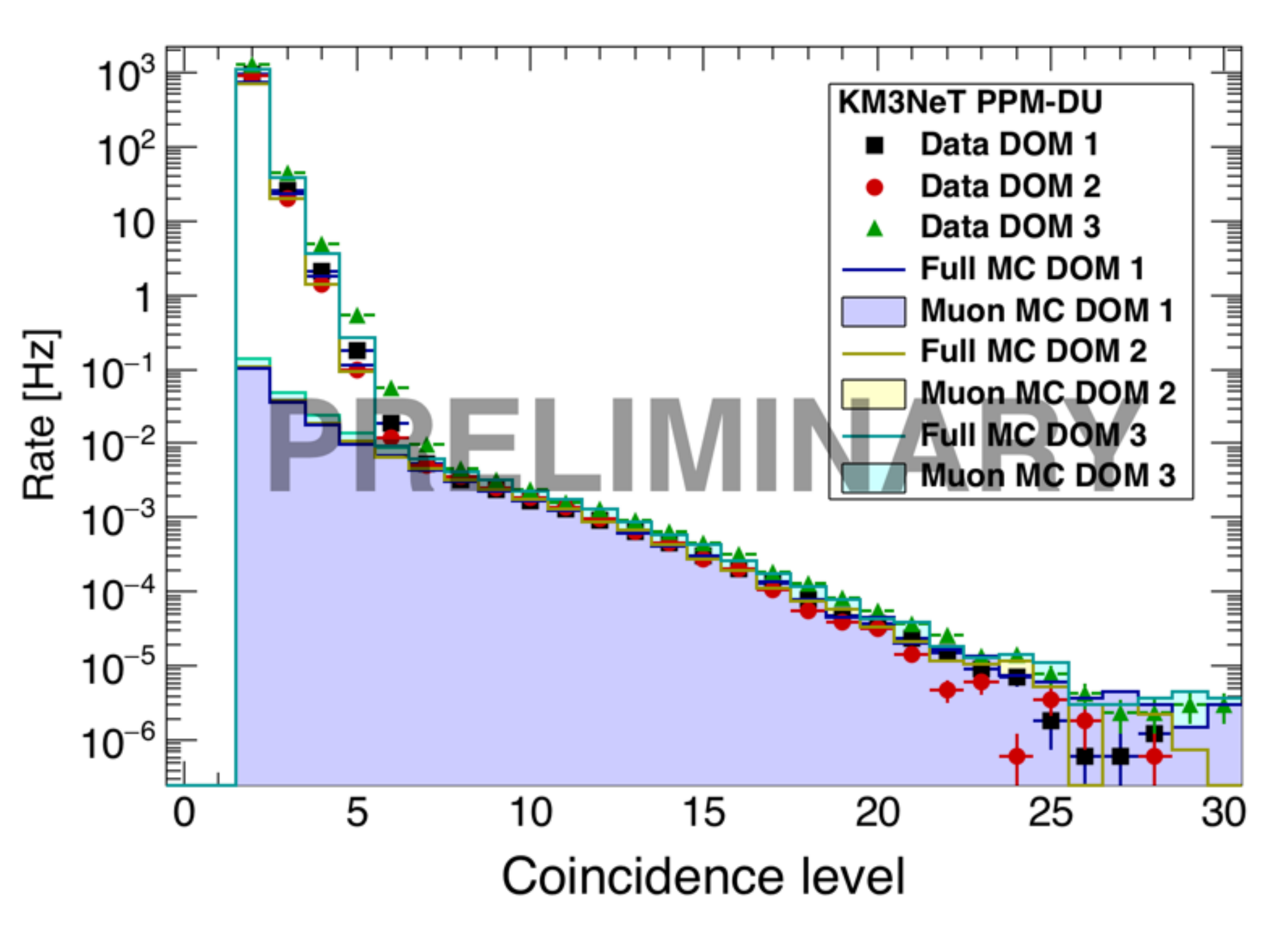} }
\end{center}
\caption{(From Biagi, Creusot, \& Bormuth, ICRC2015 1164) Left: schematic diagram of the KM3NeT prototype detection unit deployed at KM3NeT-It, and Right: comparison of coincidence rates from data compared to Monte Carlo simulation. The second figure is an update to Fig.\ 4(a) from Biagi, Creusot, \& Bormuth, after the cause of the excess of many-fold coincidences in simulations was discovered.} \label{fig:ppmdu}
\end{figure*}

A photograph of a KM3NeT DOM is given in Fig.\ \ref{fig:block} (right). The total effective area of the entire DOM
 is comparable to an ANTARES storey of three $10''$ PMTs, but with a much more uniform angular coverage. Having many small co-located PMTs has several other advantages, including a large effective dynamic range, and the ability to calibrate on multi-fold coincidences from potassium 40 decays. Detailed GEANT simulations of the DOM are described in C.~Hugon, ICRC2015 1106, and these are used as input to Monte Carlo simulations of the response of the prototype detection unit to background light and the atmospheric muon flux. Fig.\ \ref{fig:ppmdu} compares the results with data: it is evident that over the entire range, the prototypes are behaving as expected, and are well-modelled by the simulations.

The first full KM3NeT (ARCA) DU has recently been assembled and tested on-shore (A.~{Creusot}, ICRC2015 1154), and is currently awaiting deployment at KM3NeT-It. KM3NeT Phase $1$, which has now begun, will build and deploy $31$ ARCA-scale DUs at KM3NeT-It, and 7 ORCA-scale DUs at KM3NeT-Fr, during $2015$--$2017$. The rest of this contribution outlines the expected science potential of the Phase 2 instruments ARCA and ORCA, which are scheduled for completion as early as $2020$.

\subsection{ARCA}
\label{sec:arca}

The main goal of ARCA is to perform high-energy neutrino astronomy. The total instrumented volume in Phase $2$ will be comparable to that of the IceCube detector. It's northern latitude and excellent angular resolution will give it a superior sensitivity to southern and point-like sources, and hence ARCA will be ideally suited to identifying prospective Galactic sources of cosmic-ray acceleration, e.g.\ young supernova remnants such as RXJ~1713 \cite{2006PhRvD..74c4018K}, and pulsar wind nebula such as Vela~X \cite{2006A&A...448L..43A}. Further details of ARCA are given by P.~Piattelli (ICRC2015 1158).

\begin{figure*}
\begin{center}
\centerline{\includegraphics[width=0.4\textwidth]{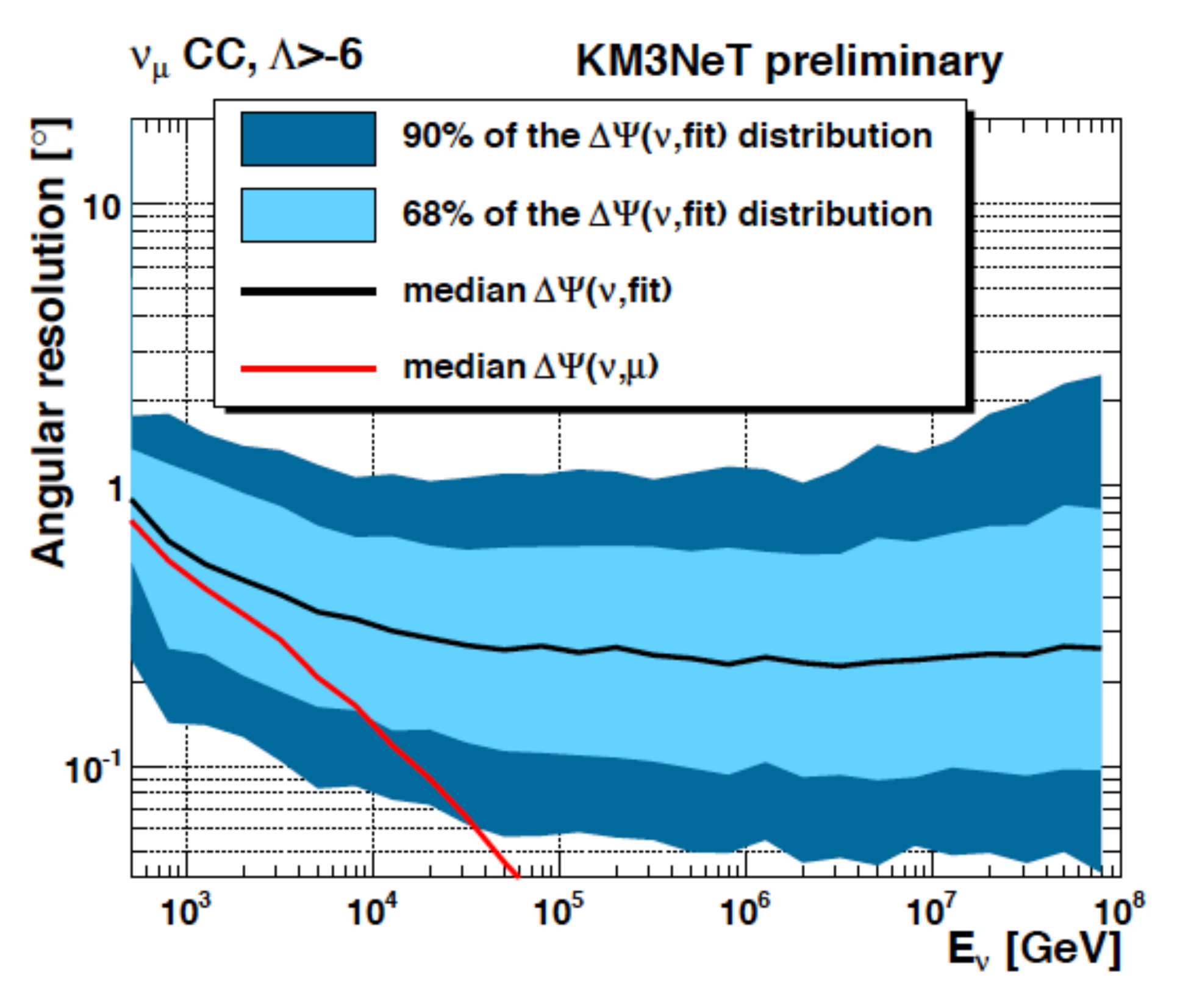} \includegraphics[width=0.45\textwidth]{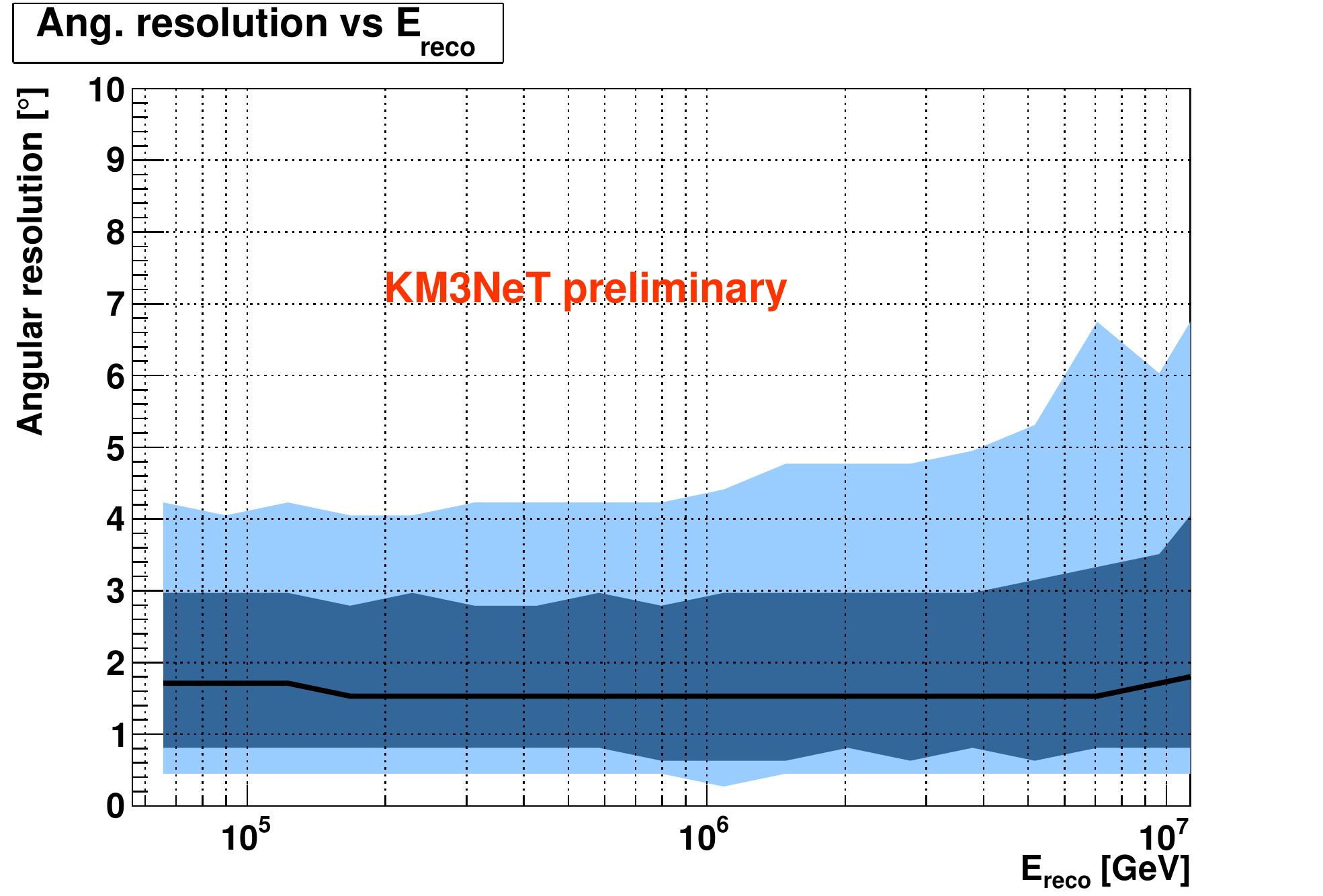}}
\end{center}
\caption{Angular resolutions of ARCA to (left) $\nu_{\mu}$ and (right) $\nu_e$ CC events, showing the median (black lines), and $68$\% (inner shading) and $95\%$ (outer shading) ranges, as a function of the neutrino energy.} \label{fig:arca_ang_res}
\end{figure*}

The expected reconstruction accuracies of muon track and cascade events in ARCA are described in {Trovato}, {Drakopoulou} \& P.~{Sapienza} (ICRC2015 1114) and D.~{Stransky} et~al.\ (ICRC2015 1108) respectively. The angular resolutions are shown in Fig.\ \ref{fig:arca_ang_res} after basic quality cuts. For the energy range above $30$~TeV, the resolution is approximately $0.25^{\circ}$ and $1.5^{\circ}$ for $\nu_{\mu}$ and $\nu_e$ CC events respectively.

\begin{figure*}
\begin{center}
\centerline{\includegraphics[width=0.48\textwidth]{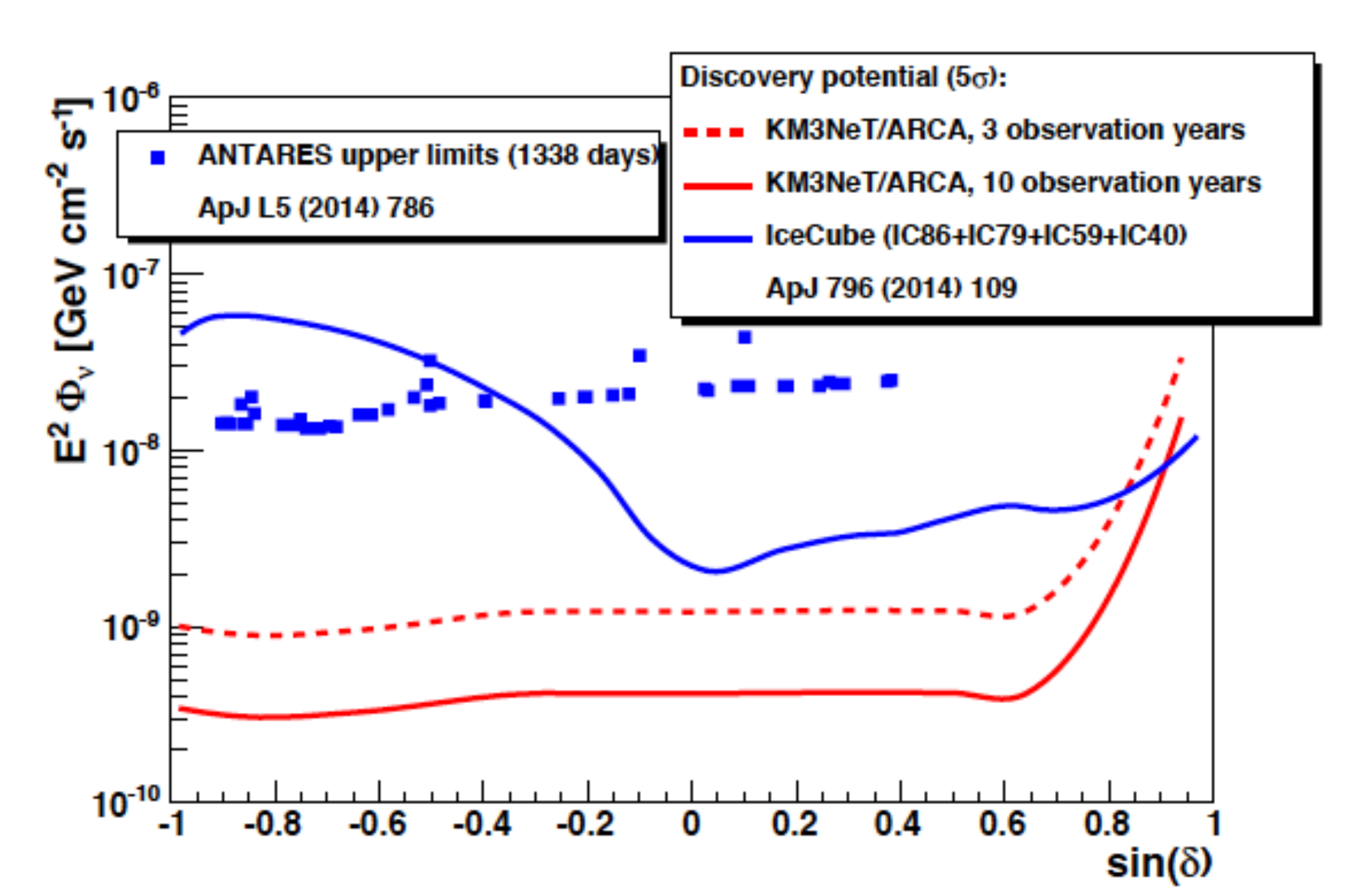}  \includegraphics[width=0.45\textwidth]{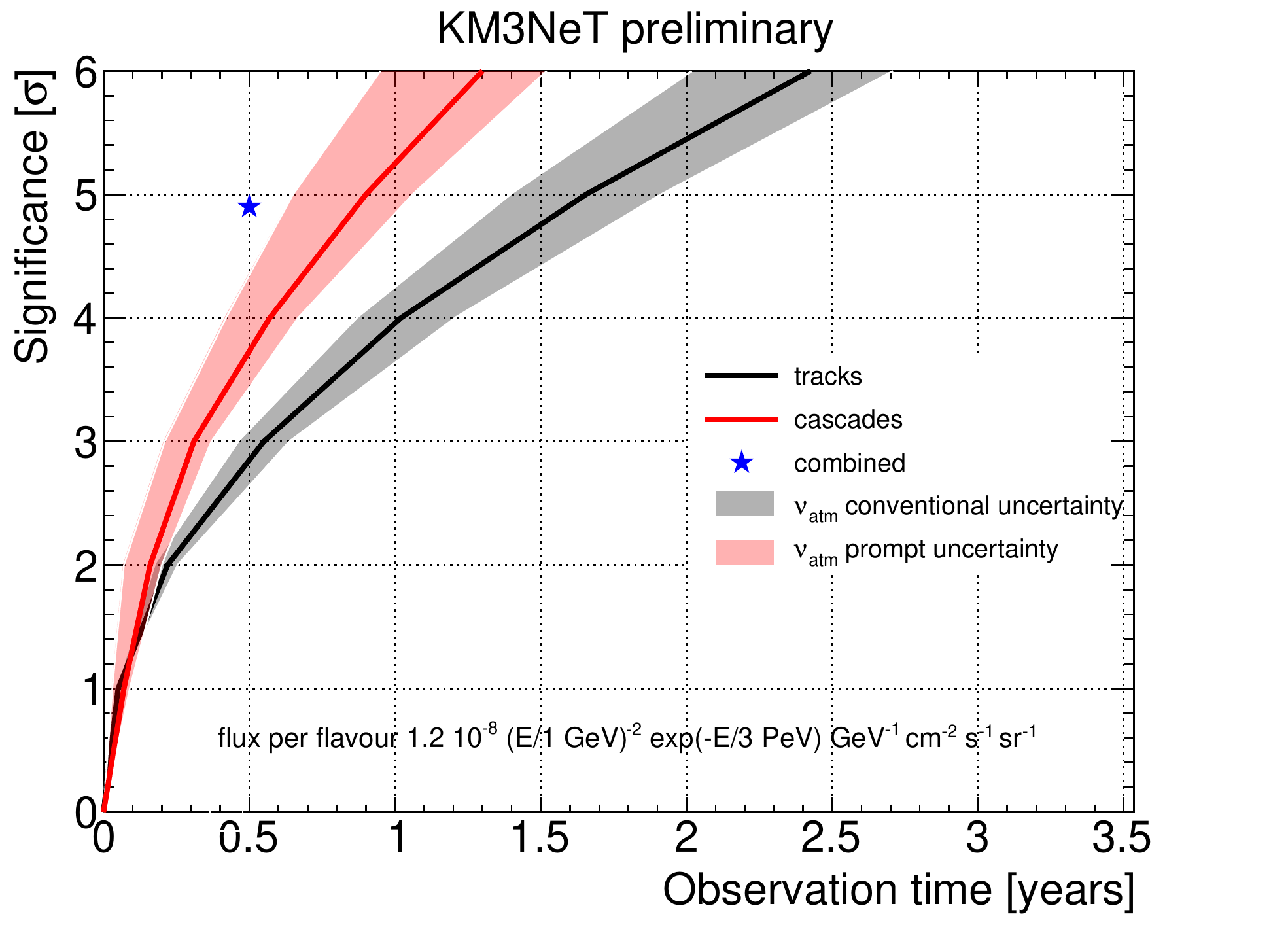}}
\end{center}
\caption{Left: ARCA $5~\sigma$ discovery flux to $E^{-2}$ point-like sources of neutrinos after $3$ and $10$ years of operation ({Trovato} \& {Barrios-Mart{\'i}}, ICRC2015 1113), and Right: expected ARCA detection significance ({Stransky}, {Coniglione} \& {Fusco}, ICRC2015 1107) as a function of time to the diffuse neutrino flux shown (c.f.\ Ref.\ \cite{2013Sci...342E...1I}).} \label{fig:arca_sensitivity}
\end{figure*}

The expected sensitivity of ARCA to astrophysical neutrino fluxes has been characterised by the sensitivity to point-like ({Trovato} \& {Barrios-Mart{\'i}}, ICRC2015 1113) and diffuse ({Stransky}, {Coniglione} \& {Fusco}, ICRC2015 1107) sources in both the track and cascade channels. Several methods to discriminate against the atmospheric muon background have been developed, including the `self-veto' effect \cite{2009PhRvD..79d3009S} on downgoing atmospheric neutrinos ({Heid}, {James} \& {Pikounis}, ICRC2015 1067). The estimated flux from generic $E^{-2}$ point-like sources required for a $5~\sigma$ discovery as a function of their declination is compared to the discovery flux of IceCube, and limits from ANTARES, in Fig.\ \ref{fig:arca_sensitivity} (left). Fig.\ \ref{fig:arca_sensitivity} (right) shows the expected significance to a diffuse flux in both the track and cascade channels, as well as a preliminary estimate of their combined sensitivity.

\subsection{ORCA}
\label{sec:orca}

ORCA (Oscillations Research with Cosmic in the Abyss) will be a KM3NeT block in a dense configuration at the KM3NeT-Fr site near Toulon, and is described in detail by J.~{Brunner} (ICRC2015 1140). As pointed out by Ref.\ \cite{2013JHEP...02..082A}, the effects of the specific values of neutrino oscillation parameters --- particular $\Delta M_{23}$, $\theta_{23}$, $\delta_{\rm CP}$, and the neutrino mass hierarchy (NMH) itself --- imprint themselves on the atmospheric neutrino flux in the few-GeV range. The goal of ORCA therefore is to study neutrino interactions in this range, and measure the zenith-angle and energy-dependence of the interaction rate for different interaction types --- in particular, $\nu_e / \bar{\nu}_e$ and $\nu_{\mu} / \bar{\nu}_{\mu}$ CC interactions.

\begin{figure*}
\begin{center}
\includegraphics[width=0.4\textwidth]{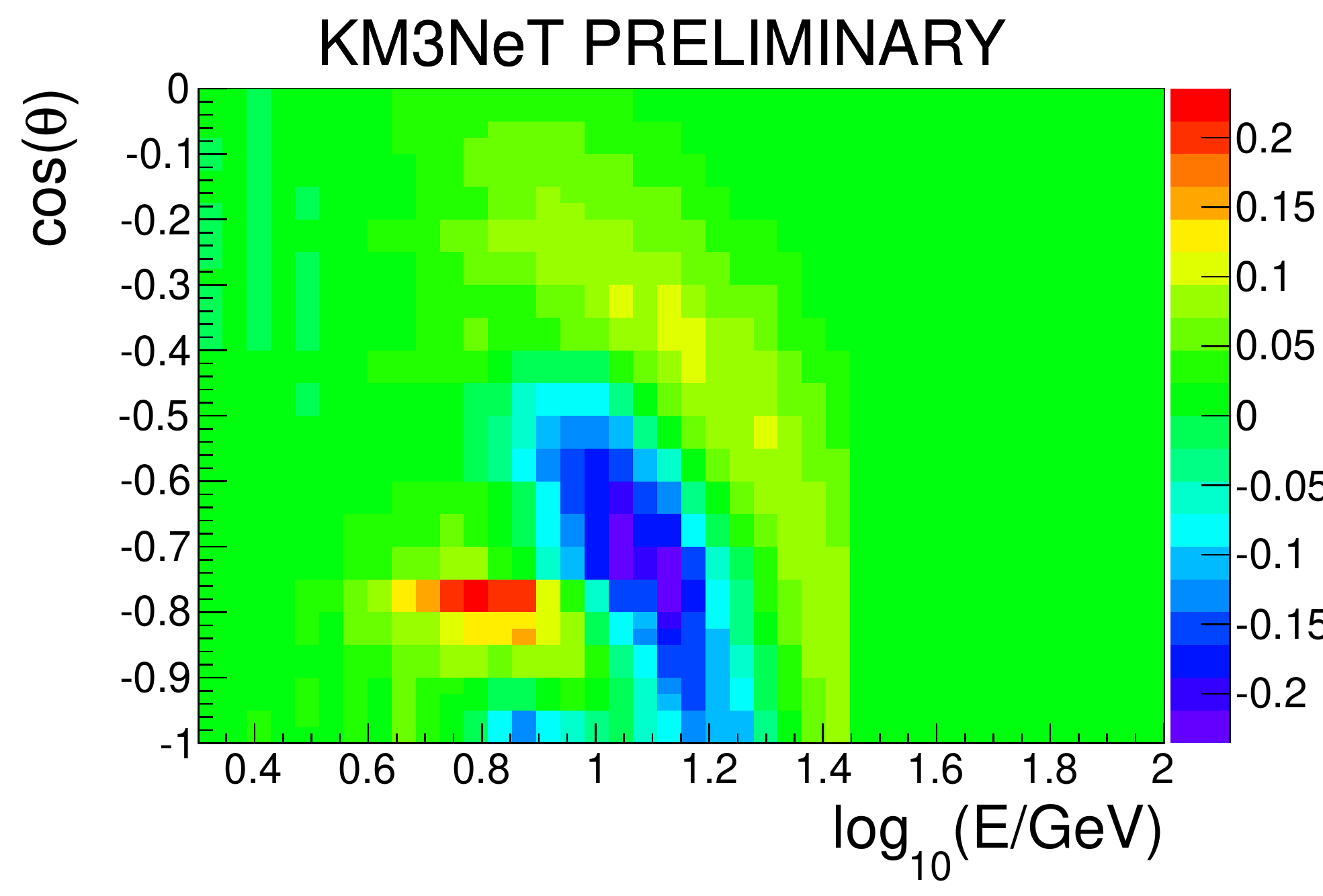} \includegraphics[width=0.4\textwidth]{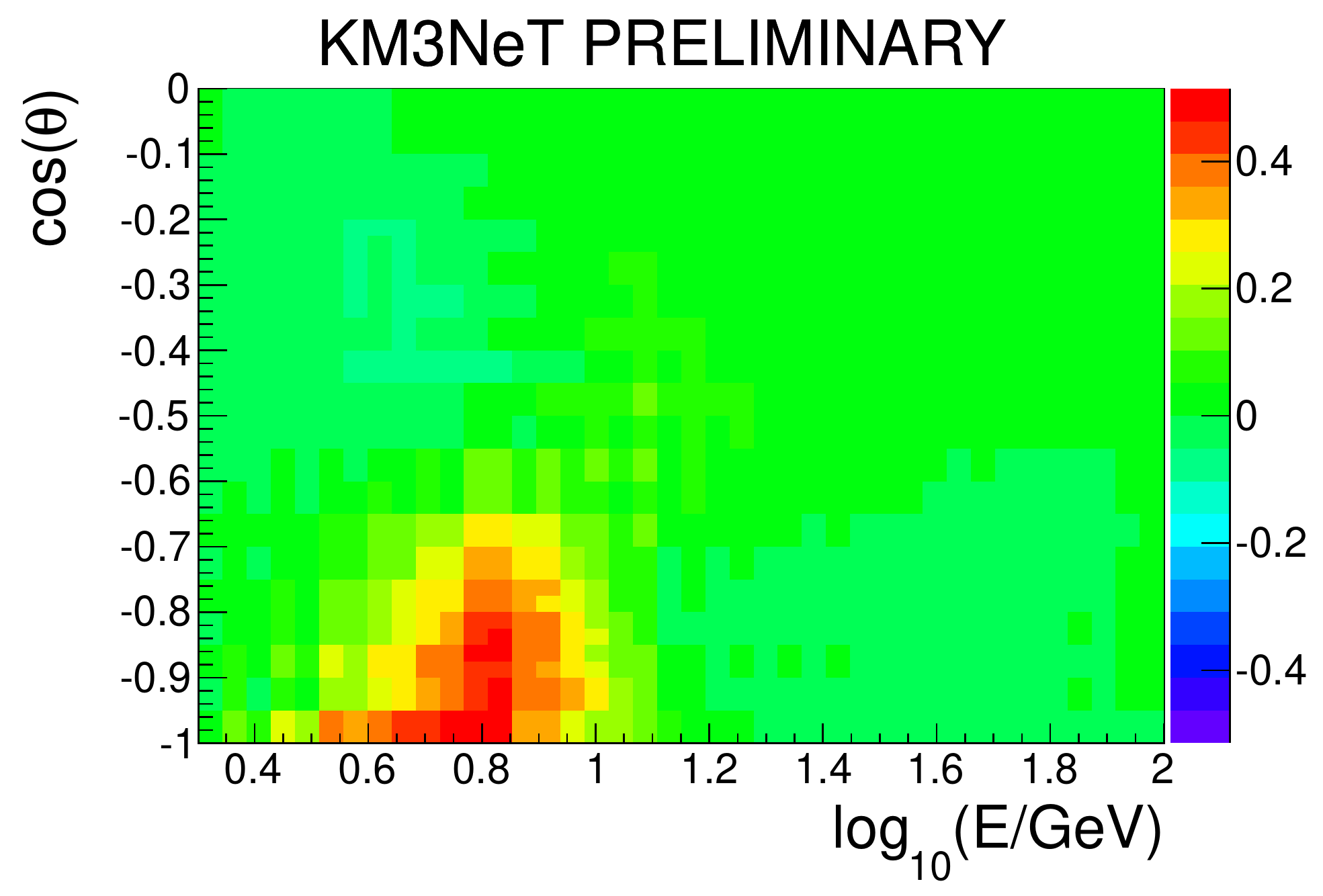}
\end{center}
\caption{Mass hierarchy sensitivity ($|N_{\rm NH} - N_{\rm IH}| / \sqrt{N_{\rm NH}}$, where $N$ is the number of events per bin after one year) showing the expected relative fluctuations in $\nu_{\mu}$~CC (left) and $\nu_e$~CC (right), taking the resolution of ORCA into account (M.~Jongen, ICRC2015 1092).} \label{fig:orca_resolutions}
\end{figure*}

Extensive studies have been carried out on the ability of the ORCA baseline detector (Fig.\ \ref{fig:block}, middle) to resolve the NMH, with intrinsic limits on reconstruction accuracy given by {Hofest{\"a}dt} and {James} (ICRC2015 1084). The ability of ORCA to reconstruct track- and cascade-like events is described by S.~{Galat{\'a}} (ICRC2015 1102) and J.~{Hofest{\"a}dt} (ICRC2015 1083) respectively, and atmospheric muon rejection is detailed in L.A.~Fusco (ICRC2015 1072). Including these resolutions gives the hierarchy signature shown in Fig.\ \ref{fig:orca_resolutions}.

\begin{SCfigure*}
\includegraphics[width=0.5\textwidth]{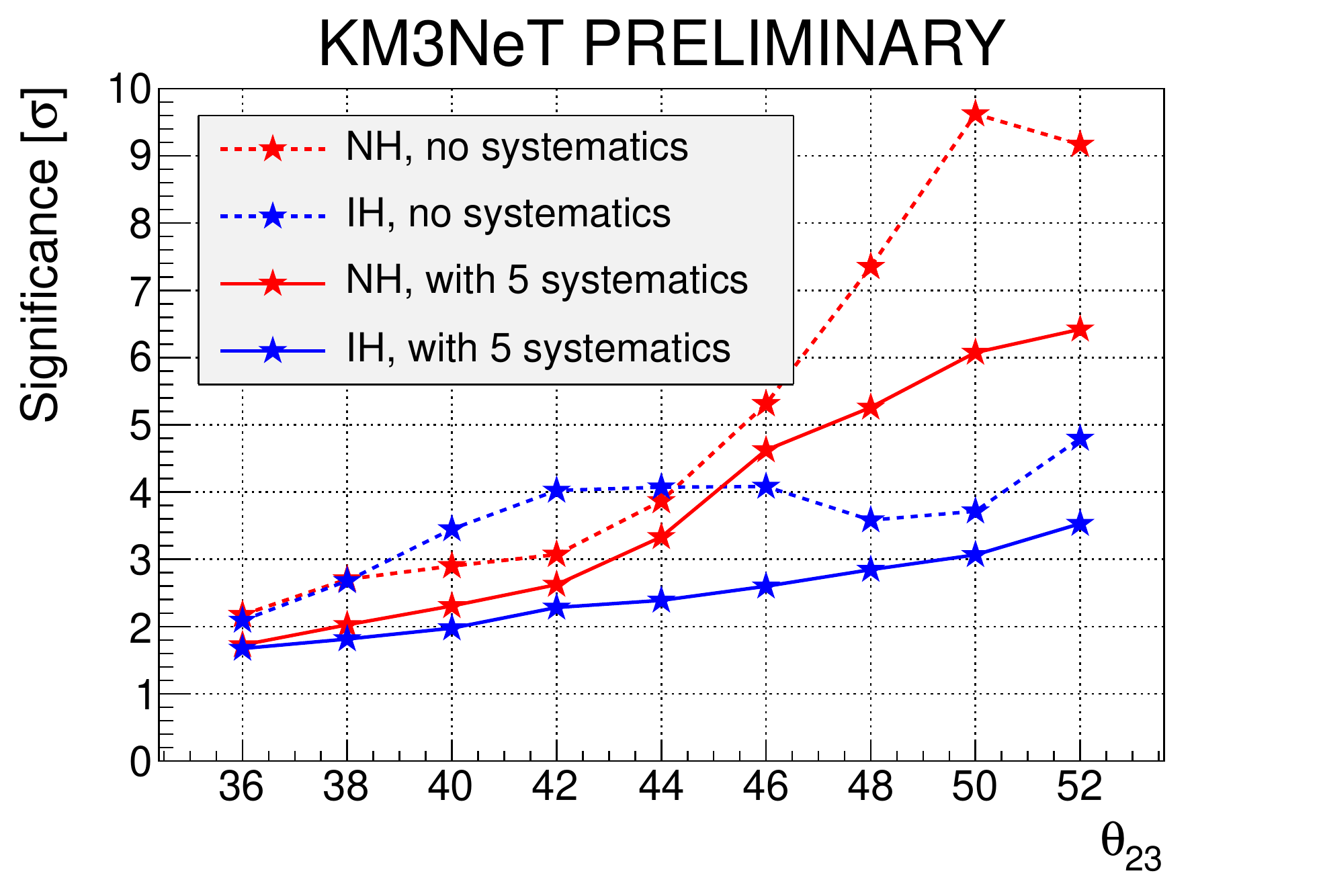}
\caption{Expected sensitivity of ORCA to the neutrino mass hierarchy after $3$~years of operation, both fitting for $\delta_{\rm CP}$ and allowing it to take all values, and setting $\delta_{\rm CP}=0$ (M.~Jongen, ICRC2015 1092).} \label{fig:orca_nmh_sensitivity}
\end{SCfigure*}

The expected sensitivity of ORCA to the NMH after $3$~years is given in Fig.\ \ref{fig:orca_nmh_sensitivity} as a function of the experimental lifetime. The calculation, described in detail by M.~Jongen (ICRC2015 1092), includes fitting for five systematic `nuisance' parameters, as well as $\theta_{23}$, and also indicate the effects of artificially setting $\delta_{\rm CP}=0$ vs.\ including this in the fit. The ability to differentiate between hierarchies is dependent in particular upon the true value of $\theta_{23}$ and, to a lesser extent, $\delta_{\rm CP}$, whereas commonly $\delta_{\rm CP}=0$ is assumed. A significance of $3 \sigma$ is expected after three years.


\section{Conclusion}
\label{sec:conclusion}

The ANTARES neutrino telescope has proved itself to be a highly successful instrument for performing a wide range of physics analyses. In particular, its excellent angular resolution on both muon-track and cascade events, facilitated by the optical properties of deep-sea water, is well-suited to studying point-like sources of neutrinos. This capability has come to the fore now that an astrophysical neutrino flux has been detected by IceCube, and the key question now is: what produces it? ANTARES has been able to limit a wide range of source scenarios, from galactic plane emission to blazars, and has performed the first point-source search using cascade events.

A new era in neutrino astronomy will begin in 2017, with the decommissioning of ANTARES, and the completion of KM3NeT Phase $1$.
The unique design of KM3NeT multi-PMT optical modules is expected to allow a very high resolution of neutrino interactions. The KM3NeT ORCA block will study the atmospheric flux in the $1$--$20$~GeV range, and is expected in Phase $2$ to determine the neutrino mass hierarchy to $3~\sigma$ significance in three years. In a sparser configuration at KM3NeT-It, ARCA in Phase $2$ will be a similarly sized instrument to IceCube, but have a much-improved angular resolution. Eventually to reach $6$ blocks during Phase $3$, ARCA is optimised to study galactic sources of hadronic acceleration, and will study the astrophysical neutrino flux in unprecedented detail.

\let\OLDthebibliography\thebibliography
\renewcommand\thebibliography[1]{
  \OLDthebibliography{#1}
  \setlength{\parskip}{0pt}
  \setlength{\itemsep}{0pt plus 0.3ex}
}







\setcounter{figure}{0}
\setcounter{table}{0}
\setcounter{footnote}{0}
\setcounter{section}{0}
\setcounter{equation}{0}

%
%
\newpage
\addcontentsline{toc}{part}{{\sc Search for Point Sources and Diffuse Fluxes}%
\vspace{-0.5cm}
}
\id{id_barriosmarti}
\addcontentsline{toc}{part}{\textcolor{blue}{\arabic{IdContrib} - {\sl J. Barrios-Mart\'i} : Limits on point-like sources with different spectral indexes around the Galactic Centre using the ANTARES neutrino telescope}%
}

\title{\arabic{IdContrib} - Limits on point-like sources with different
spectral indexes around the Galactic Centre
using the ANTARES neutrino telescope}

\shorttitle{\arabic{IdContrib} - Point-source limits around the Galactic Centre}

\authors{J.~Barrios-Mart\'i}
      \afiliations{Instituto de F\'isica Corpuscular, IFIC (UV-CSIC), Parque Cient\'ifico, C/Catedr\'atico Jos\'e Beltr\'an 2, E-46980 Paterna, Spain}
\email{javier.barrios@ific.uv.es}



\abstract{Motivated by an accumulation of events close to the Galactic Centre in the High Energy
Starting Events (HESE) reported by the IceCube Collaboration, a search for point-like
sources up to an extension of a few degrees in a wide region around the Galactic center has
been performed using the ANTARES neutrino telescope. Different spectral indexes for the
energy spectra of the sources, in addition to the default value of $\gamma=2.0$, have been
tested. Upper limits on the flux normalization as a function $\gamma$ have been set.}

%
\maketitle

\section{Introduction}
\label{sec:Introduction}

The IceCube collaboration reported an excess of high energy neutrinos which cannot be explained by the expected contribution of atmospheric muons and neutrinos \cite{IceCube1}, \cite{IceCube2}. An accumulation of events is seen in the surroundings of the Galactic Centre. The point with the lowest p-value was found at equatorial coordinates of ($\alpha$,$\delta$) = ($-$79$^\circ$, $-$23$^\circ$). Although the significance is not enough to identify a point-source, some authors have considered this accumulation could come from a single point-source \cite{MCGonz}, with an expected flux normalisation of $\Phi_0 = 6 \times 10^{-8}$  \textrm{GeVcm$^{-2}$s$^{-1}$}. Triggered by this hypothesis, a search for $E^{-2}$ point-sources around the Galactic Center was performed in the last ANTARES\cite{ANT-INST} point-source analysis \cite{PS-ANTARES}, with no significant results. \\

Although the expected energy spectrum for neutrino sources is not completely unkwnown, there is uncertainty on the spectral index. The last HESE analysis results \cite{IC25} show an expected index of $\gamma = $ 2.50 $\pm$ 0.09.  Herewith, an update of the results of the previous ANTARES analysis around the Galactic Center for different energy spectra (from $\gamma$ = 2.0 to 2.5) is presented. The data sample for this analysis is described in Section \ref{sec:sample}. The performance of the ANTARES telescope for different energy spectra is shown in Section \ref{sec:performance}. The procedure of this analysis is explained in Section \ref{sec:search}, with the results on Section \ref{sec:results}.

\section{Data sample}
\label{sec:sample}

The same data sample as in the last published ANTARES point-source analysis is used. The data was collected between January 29, 2007 until December 31, 2012, with a total livetime of 1338 days. \\

The events in the data sample consist of muon-neutrino source candidates, which are selected following a blind procedure on pseudo-experiments. The selection of the events was tuned to minimise the neutrino flux required for a 5$\sigma$ discovery in 50\% of the experiments for an $E^{-2}$ spectrum. This minimisation was performed by considering different cuts on three parameters of the events: the quality of the track fit, $\Lambda$; the angular error estimate, $\beta$; and the zenith angle, $\theta$. The values of these parameters are obtained by the track reconstruction of the neutrino events, which uses a maximum likelihood (ML) method \cite{AartThesis}. The reconstruction is based on a multi-step algorithm to fit the direction of the reconstructed muon by means of a maximisation in the likelihood of the reconstruction. The angular error estimate is later extracted from the estimated uncertainty on the zenith and azimuth angles obtained from the covariance matrix. \\


A total number of 5516 events are selected for the whole sky in the final sample, with an estimated contamination of mis-reconstructed atmospheric of 10\%.

\section{Expected number of events for different energy spectra}
\label{sec:performance}

The number of expected signal events which can be detected varies depending on the considered source spectra. For this analysis, power-law spectra of $d\Phi/dE_{\nu} = \Phi_0 (E_{\nu}/GeV)^{-\gamma}$ are assumed. By considering the effective area of the telescope, $A_{eff}$, given the neutrino energy, E$_{\nu}$, and the declination of the source, $\delta$, it is possible to estimate this number as

\begin{equation}\label{eq:acc}
N(\delta, \gamma) = \int dt \int dE_\nu A_{eff}(E_\nu,\delta) \Phi_{0} \left( \frac{E_\nu}{GeV} \right)^{-\gamma} \; ,
\end{equation}

\noindent where the time integration ranges for the whole lifetime of 1338 days. The expected number of signal events for a normalization flux of $\Phi_0$ = 10$^{-8}$ \textrm{GeV$^{-1}$cm$^{-1}$s$^{-1}$} and spectral indices between 2.0 and 2.5 in steps of 0.1 are shown in Figure \ref{fig:Nevents}.

\begin{figure}[!ht]
	\centering
	\includegraphics[width=.6\textwidth]{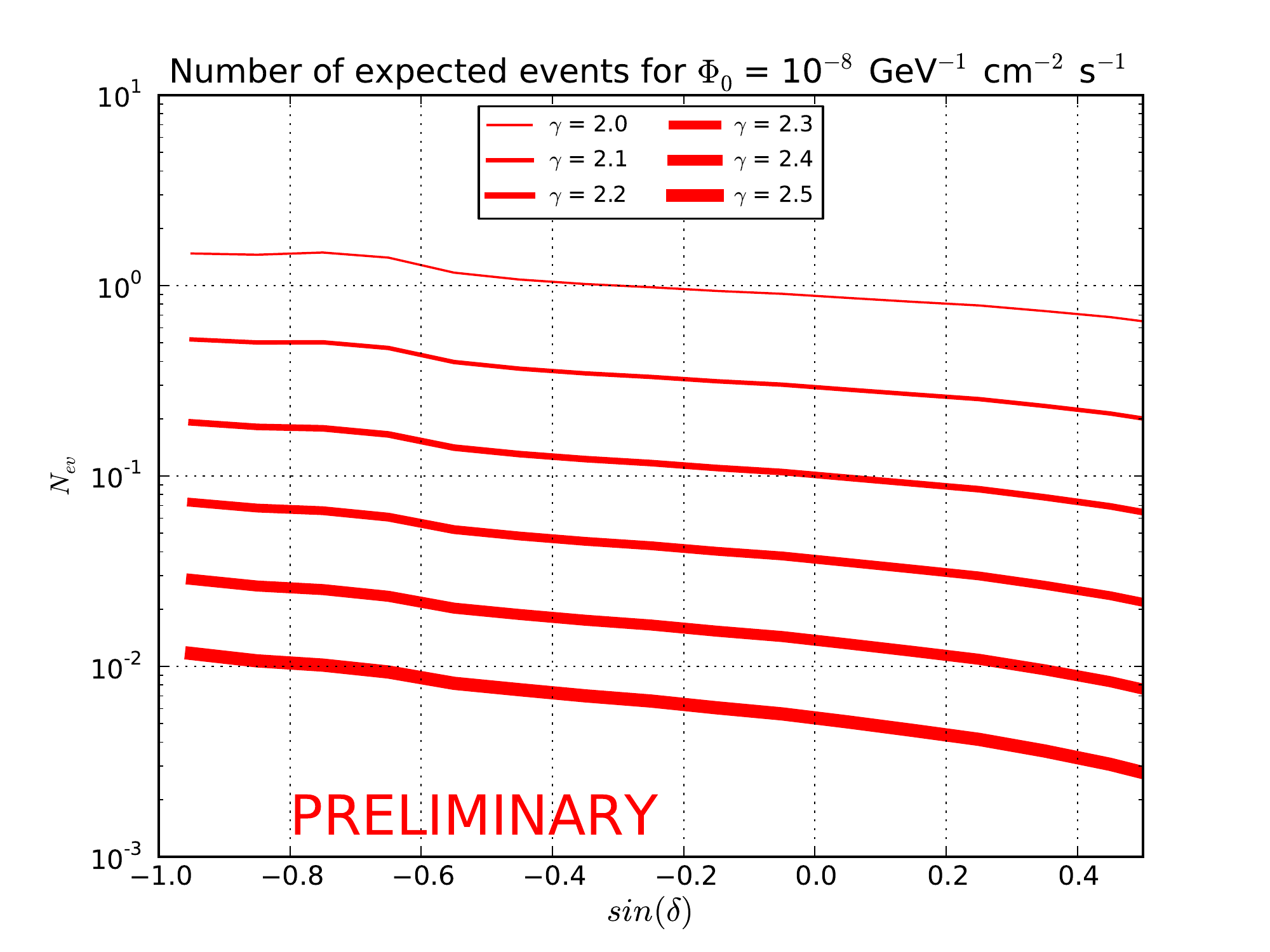}
	\caption{Number of expected signal events as a function of the declination, $\delta$, for energy source spectra between 2.0 and 2.5 in steps of 0.1. A normalization flux of $\Phi_0$ = 10$^{-8}$ \textrm{GeV$^{-1}$cm$^{-1}$s$^{-1}$} has been assumed in all cases.}
	\label{fig:Nevents}
\end{figure}

\section{Search Method}
\label{sec:search}

A search of signal events is performed by means of a maximum-likelihood estimation. This likelihood describes the data in terms of  signal and background probability density functions (PDFs). This likelihood is described as,

\begin{equation}
	\log L(n_s) = \sum_i \log \left[ \frac{n_s}{N}S_i + \left(1-\frac{n_s}{N}\right)B_i \right] ,
\end{equation}

\noindent where $n_s$ indicates the fitted number of signal source events, $N$ is the total number of events in the sample, and $B_i$ and $S_i$ are the background and signal PDFs for the $i$th event, respectively. In order to describe the signal and background PDFs, the information of the number of hits, $\mathcal{N}^{hits}$, angular error estimate, $\beta$, and position in equatorial coordinates, $\vec{x}_{s} = (\alpha, \delta)$, is considered. The signal PDF is described as

\begin{equation}
S_i = \frac{1}{2\pi\beta_i^2} \exp\left(-\frac{\psi_i(\vec{x}_s)^2}{2\beta_i^2}\right) P_{s}(\mathcal{N}^{hits}_i, \beta_i|\gamma) ,
\end{equation}

\noindent where $P_{s}(\mathcal{N}^{hits}_i, \beta_i|\gamma)$ indicates the probability for the $i$th event to be reconstructed as signal given a number of hits of $\mathcal{N}^{hits}_i$ and an angular error estimate of $\beta_i$ for a spectral index of $\gamma$, and $\psi_i(\vec{x}_s)^2$ represents the angular distance to the assumed source direction, $\vec{x}_s$. The distribution of $P_{s}(\mathcal{N}^{hits}, \beta|\gamma)$ is obtained from simulated events, and it depends on the assumed energy spectra, $\gamma$. \\

The background PDF is defined as 

\begin{equation}
B_i = \frac{B(\delta_i)}{2\pi} P_{b}(\mathcal{N}^{hits}_i, \beta_i) ,
\end{equation}

\noindent where $B(\delta_i)$ is the probability for an event to be background given its declination, and  $P_{b}(\mathcal{N}^{hits}_i, \beta_i)$ is the probability for an to be reconstructed as background with an angular error estimate of $\beta_i$ and a number of hits $\mathcal{N}^{hits}_i$. The distribution $B(\delta_i)$ is obtained from the background rate of events from the data sample. The $P_{b}(\mathcal{N}^{hits}_i, \beta_i)$ is obtained also from the information in the data. \\

In order to determine the significance of any cluster, the test statistic, TS, is defined as  TS = $\log L(n_s) - \log L(n_s = 0)$. $L_{b}$ indicates the value of the likelihood where only background events are expected. Larger values of the TS designate a smaller probability of the cluster to be generated from only atmospheric events. \\

In order to take into account the large uncertainty of the angular error estimates of the IceCube events around the Galactic Center, a search around a region of 20$^\circ$ around the proposed location ($\alpha$, $\delta$) = ($-$79$^\circ$ $-$23$^\circ$) is performed. For this purpose, the TS is evaluated in steps of 1$^\circ$ $\times$ 1$^\circ$, while leaving the expected source position, $\vec{x}_s$, as a free parameter within these boundaries. In order to estimate the limits, 7 different source declinations were considered in the simulations.


\section{Results}
\label{sec:results}

No significant cluster has been found in the defined area around the Galactic Centre.  Figure \ref{fig:limits_old} shows the results presented in the last ANTARES point-source analysis, where different source extensions were considered (point source, 0.5$^\circ$, 1$^\circ$ and 2$^\circ$). 90\% C.L. upper limits on the flux normalisation, $\Phi_0$, for the different assumed source spectra and as function of the declination can be seen in figure \ref{fig:limits}. Figure \ref{fig:limits_spec} shows the limits for a declination of $\delta$ = -29$^\circ$. In this figure, the expected flux normalisation from the hypothetical source depending on the number of HESE events which would be originated in this source is also considered. These values have been obtained from \cite{Maurizio}. A point-like source with values of the spectral index closer to 2.5 are more disfavoured than for values closer to 2.0.  

\begin{figure}[!ht]
	\centering
	\includegraphics[width=.55\textwidth]{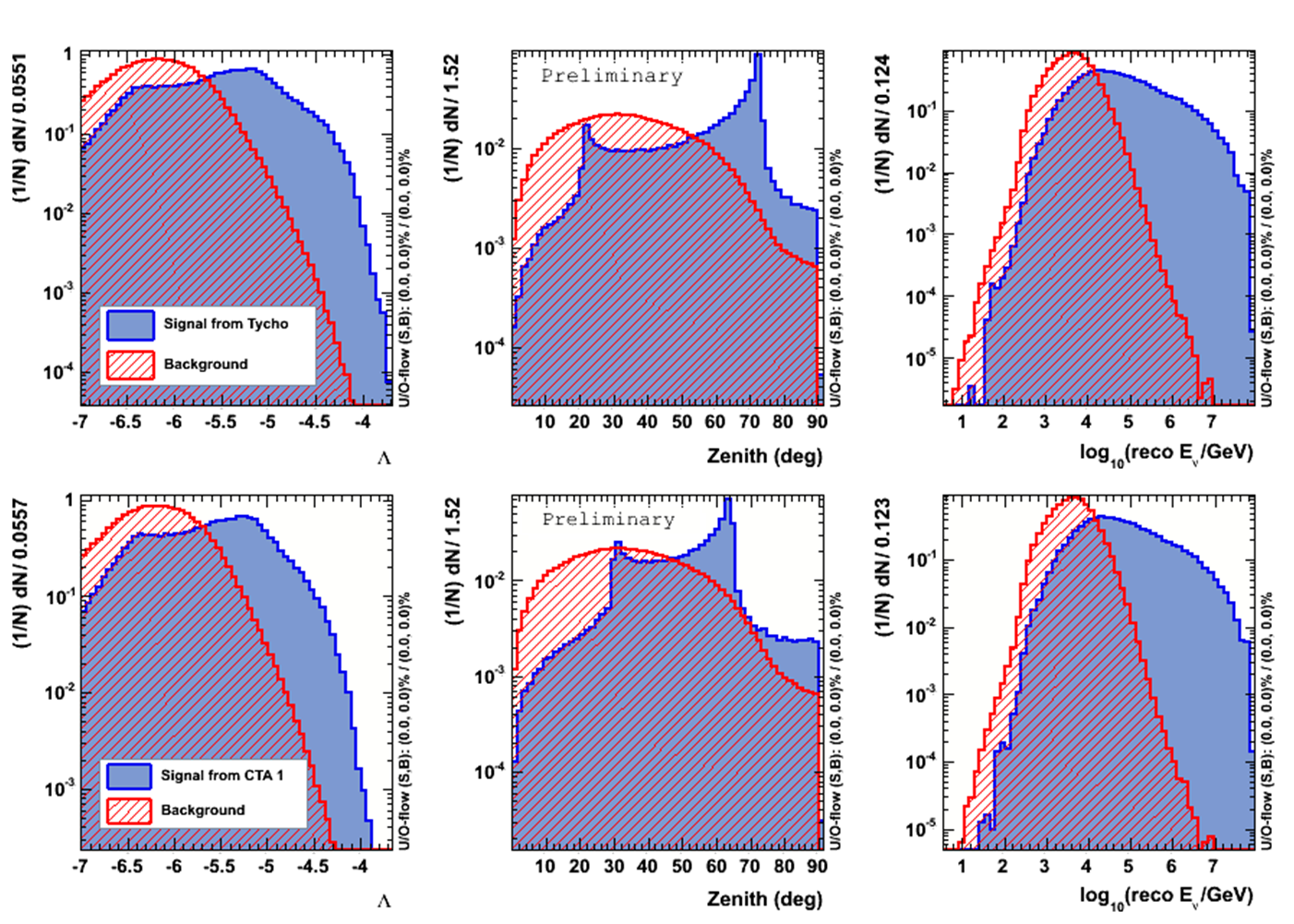}
	\caption{90\% C.L. upper limits for a point-source and for source extensions of 0.5$^\circ$, 1$^\circ$ and 2$^\circ$ as a function of the declination. }
	\label{fig:limits_old}
\end{figure}

\begin{figure}[!ht]
	\centering
	\includegraphics[width=.55\textwidth]{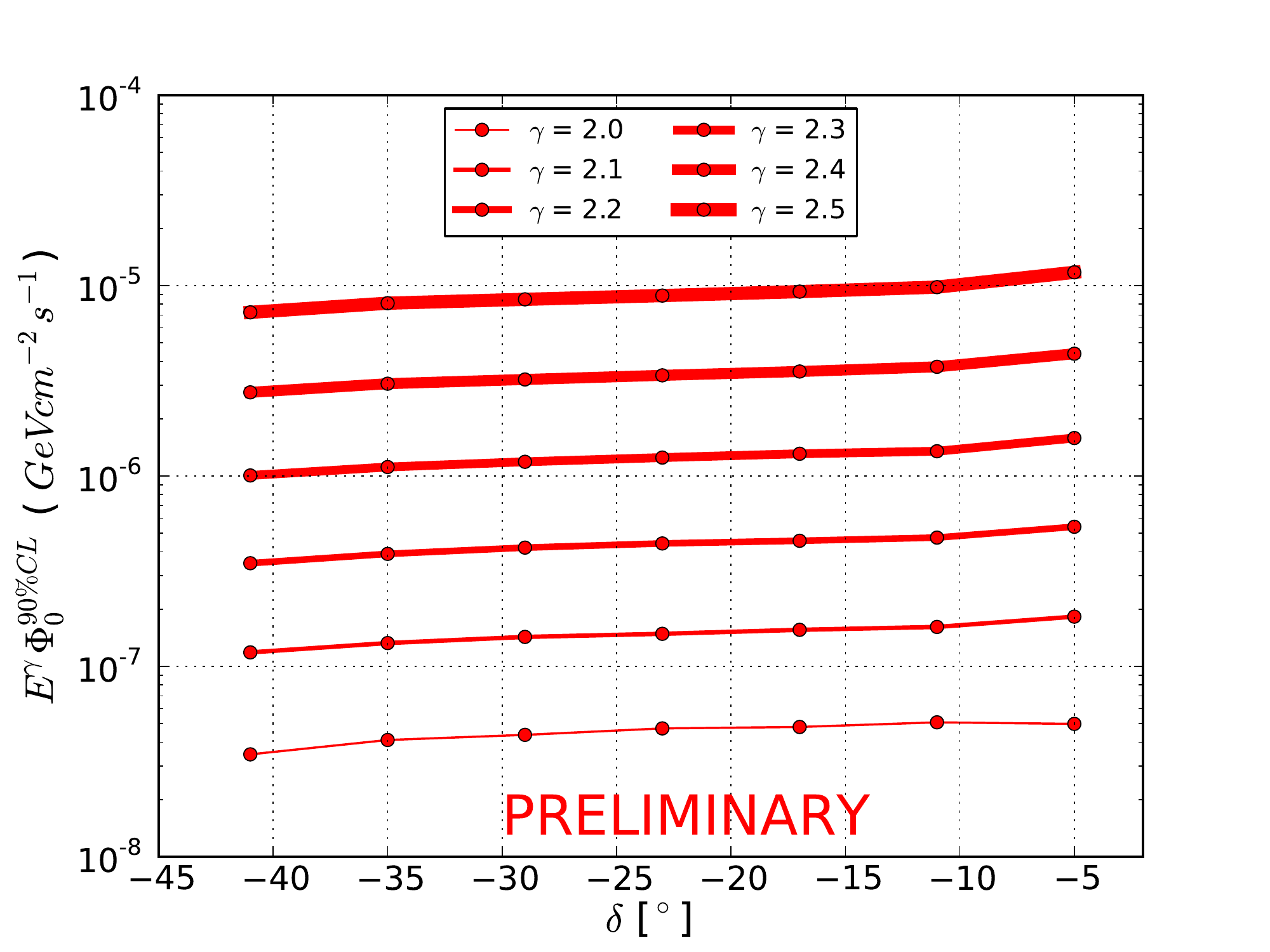}
	\caption{90\% C.L. upper limits for source spectra between 2.0 and 2.5  as a function of the declination of the source. }
	\label{fig:limits}
\end{figure}

\begin{figure}[!ht]
	\centering
	\includegraphics[width=.49\textwidth]{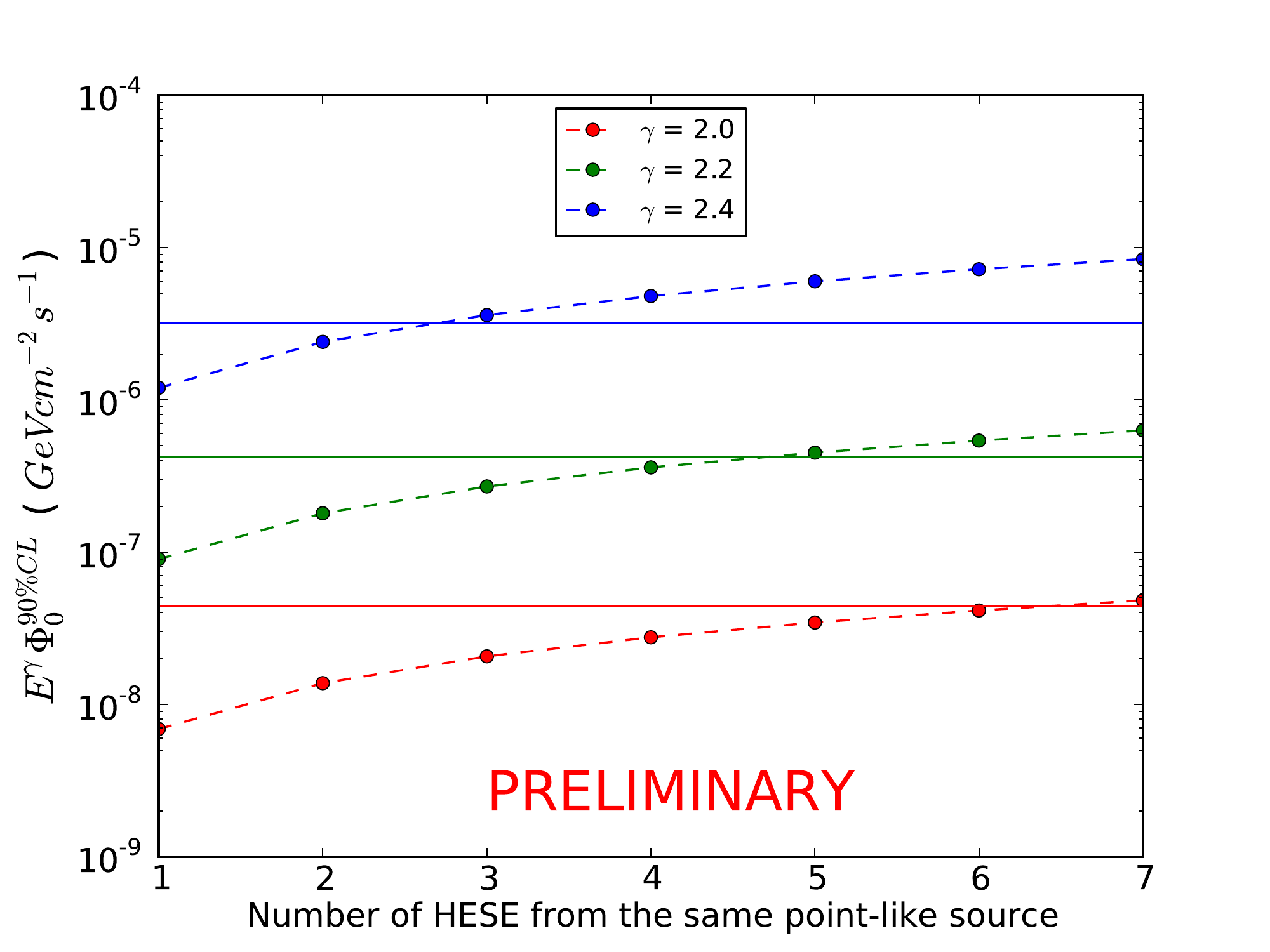}
	\includegraphics[width=.49\textwidth]{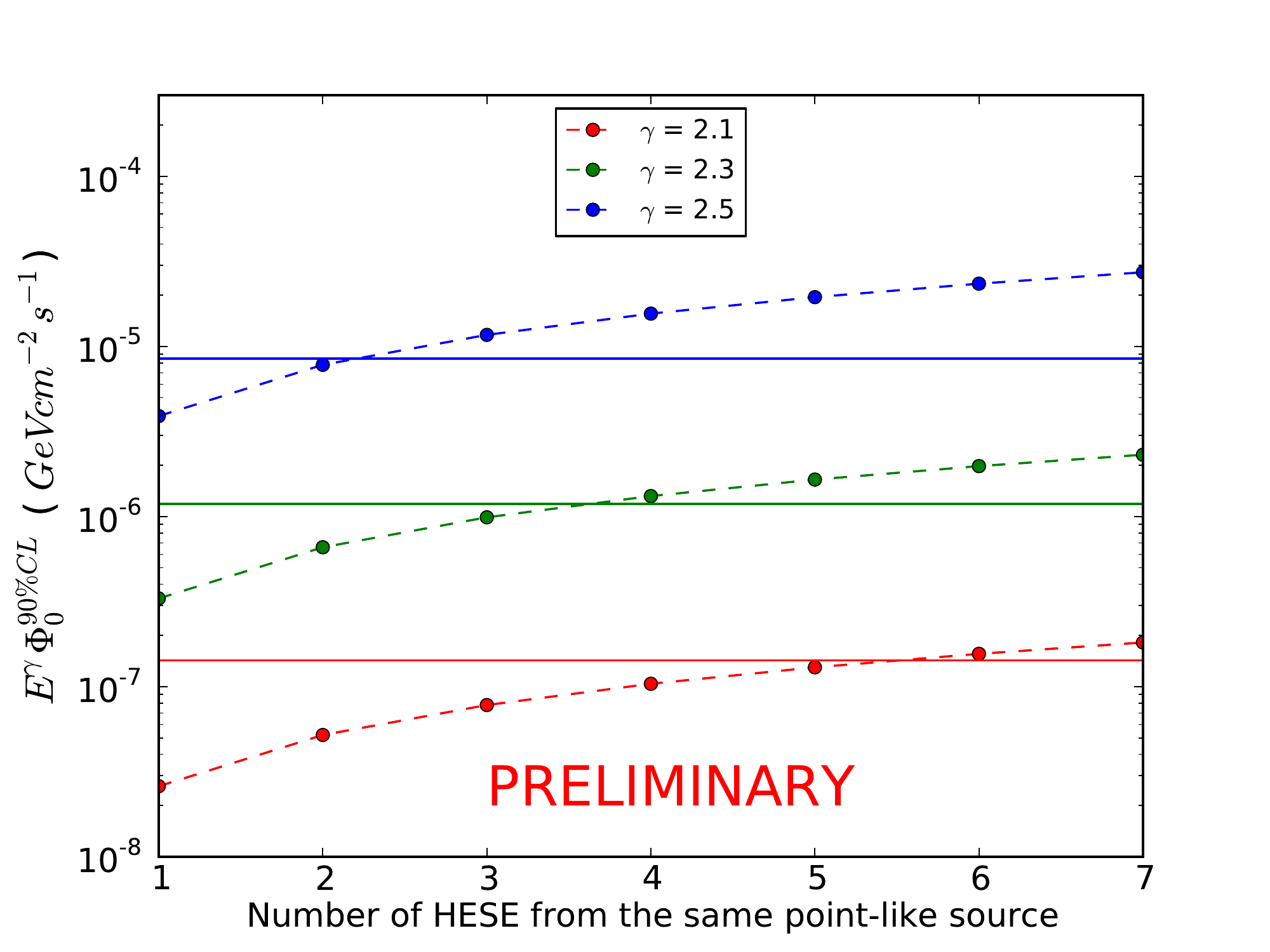}
	\caption{Solid lines: 90\% C.L. upper limits for source spectra between 2.0 and 2.5 and a source declination of $\delta =$ --29$^\circ$. The figure on the left contains the values for $\gamma$ = 2.0, 2.2, and 2.4, whereas the figure on the right contains the ones for $\gamma$ = 2.1, 2.3 and 2.5. Dashed lines: expected flux normalisation of the proposed source as a function of the number of HESE events coming from this source. Values above the solid lines are disfavoured with a confidence level larger than 90\%. }
	\label{fig:limits_spec}
\end{figure}

\section{Conclusions}
\label{sec:conclusions}

A point-source search around the Galactic Centre for spectral indices between 2.0 and 2.5 with the ANTARES neutrino telescope has been performed. No significant cluster has been found, and 90\% C.L. upper limits have been set. According to these limits, point-like sources with softer spectra are more disfavoured to explain the accumulation of events observed by IceCube.

\section*{Acknoledgements}
The authors acknowledge the support of the Spanish MINECO through project FPA2012-37528-C02-01, the MultiDark Consolider Project CSD2009-00064, the Generalitat Valenciana via Prometeo-II/2014/079, and of Universitat de Val\`{e}ncia, Atracci\'{o} de Talent.


\setcounter{figure}{0}
\setcounter{table}{0}
\setcounter{footnote}{0}
\setcounter{section}{0}
\setcounter{equation}{0}
\newpage
\id{id_coleiro}
\addcontentsline{toc}{part}{\textcolor{blue}{\arabic{IdContrib} - {\sl A. Coleiro} : Transient neutrino emission from the Galactic Center studied by ANTARES}%
}

\title{\arabic{IdContrib} - Transient neutrino emission from the Galactic Center studied by ANTARES}

\shorttitle{\arabic{IdContrib} - Transient neutrino emission studied by ANTARES}

\authors{Alexis Coleiro}
        \afiliations{APC, Universit\'e Paris Diderot, CNRS/IN2P3, CEA/Irfu, Observatoire de Paris, Sorbonne Paris Cit\'e, 10 rue Alice Domon et L\'eonie Duquet, 75205 Paris Cedex 13, France.}
\email{coleiro@apc.univ-paris7.fr}


\abstract{We present a search for ANTARES neutrino events in temporal coincidence with IceCube High-Energy Starting Events (HESE), between May 2010 and December 2012. This study uses a two-point correlation function and focuses on HESE located within 45$^\circ$ from the Galactic Center (GC). This approach is sensitive to transient emission and requires neither prior on the burst timing structure nor on the electromagnetic emission. Therefore, it provides an effective way to acquire information on the possible origin of the IceCube astrophysical signal from transient sources.
}
%

\maketitle

\section{Introduction}

High energy neutrinos are expected to be produced in sources of cosmic rays (active galactic nuclei, X-ray binaries, supernova remnant, etc.). Since they are neutral, weakly interacting and traveling straight from their source without suffering from absorption, neutrinos are unique messengers to further understand the particle acceleration processes in such astrophysical sources.

Time integrated analyses suffer from a high background of both atmospheric muons and neutrinos. When dealing with transient emission, this background can be significantly reduced using a time-dependent approach that usually consists in searching for astrophysical neutrinos in smaller time windows around flares (see e.g. \cite{astropart_blazars} and \cite{JHEA_muqua}). Here, we propose a model-independent approach based on the timing properties of both the ANTARES and IceCube data samples.

The IceCube collaboration announced recently (\cite{icecube1} , \cite{icecube2}) the discovery of the first extraterrestrial very-high energy neutrinos in the energy range from 30 TeV to 2 PeV. Nine of these so-called High Energy Starting Events (HESE), occurring between June 2010 and January 2013, are positionally consistent with the Galactic Center (GC). Moreover, it was pointed out recently that two of these IceCube HESE occurred within 1 day of each other with a p-value of 1.6\% \cite{lighthouse}. Consequently, this was interpreted as possibly the signature of a transient point source of very-high energy neutrinos in this part of the sky.

In order to search for neutrino flares in the Galactic center region, we perform a time correlation study between these nine IceCube HESE and the ANTARES dataset. This approach, requires neither prior on the burst timing structure nor on the potential electromagnetic emission. Therefore, it provides an effective way for further shedding light on the IceCube astrophysical signal possibly emitted by transient astrophysical sources. In Section \ref{sample_dat}, we first report the ANTARES and IceCube data samples, focusing on the ANTARES quality cuts optimization. Section \ref{methodo} presents both the approach used in this analysis and the related discovery potential. Preliminary results are provided in Section \ref{results_section}.

\section{Event samples}\label{sample_dat}
\subsection{ANTARES data selection}

The considered dataset was collected by the ANTARES neutrino telescope between May 01 2010 and November 30 2012. The event selection criterion has been optimized through Monte-Carlo simulations to reach a constant neutrino candidate rate over time. In the following, we assume that the run-by-run variations of the ANTARES data stream are mainly due to the evolution of the data taking conditions in the sea. To accurately take into account these variations, we use the mean counting rate of the optical modules (\texttt{MR}) as a good measure of the real conditions in the sea. The dataset has been divided into five ranges of \texttt{MR} and the reconstruction quality parameter $\Lambda$ was optimized separately for each of these sub-samples to reach a constant neutrino candidate rate. Consequently, the same quality cut is used for runs subject to similar data taking conditions. The neutrino candidate rate was chosen according to a Model Discovery Potential (MDP - see \cite{HillsMDP}) optimization, varying the cut on $\Lambda$ in the sub-sample of \texttt{MR} related to the best data taking conditions (50 kHz $< $ \texttt{MR} $<$ 100 kHz). The expected signal used in the MDP optimization was defined as the last estimation of the diffuse IceCube astrophysical neutrino spectrum: $\Phi_{\nu} = 2.06^{+0.4}_{-0.3}\times10^{-18}\left(E_{\nu}/10^5\mathrm{GeV}\right)^{-2.46\pm0.12}$  (see \cite{icecube3}). From the optimized $\Lambda$ cut, the corresponding neutrino candidate rate was computed. For the other ranges of \texttt{MR}, we computed the $\Lambda$ cut that enable to reach the previously defined neutrino candidate rate. We point out that the $\Lambda$ cut obtained for each range of \texttt{MR} (see Table \ref{lambda_val}) is close to the quality cut that would have been set by a separate MDP optimization in each subsample. This approach leads to a final sample consisting of 4337 events.

\begin{table}[h!]
\centering
\begin{tabular}{|c|c|}
  \hline
    \texttt{MR} range (kHz) & $\Lambda$ cut \\
  \hline
  50 -- 99 & -5.2\\
  100 -- 199 & -5.4 \\
  200 -- 398 & -5.5\\
  399 -- 794 & -5.6\\
  795 -- 1585 & -5.7\\
  \hline
  \end{tabular}
  \caption{$\Lambda$ cuts for the 5 bins of \texttt{MR}.}
  \label{lambda_val}
\end{table}

\subsection{IceCube data selection}
The aim of this analysis is primary to constrain the potential transient origin of the IceCube astrophysical signal close to the GC.  Among the three-year (988 days) HESE dataset, consisting of thirty-seven events (see \cite{icecube2}), nine of them are located within 45$^\circ$ from the GC, of which eight occur between May 2010 and November 2012. For overlap with the considered ANTARES data sample, we will thus only consider these eight events listed in Table \ref{ICtable_t}. 
\begin{table}
\tiny
\centering
\begin{tabular}{l c c c c c c}
  \hline
    HESE ID & Date (MJD) & Energy (TeV) & RA (Deg) & Dec (Deg) & Angular error (Deg) & Distance from GC (Deg) \\
  \hline
  \hline
 2 & 55351.4659661 & 117 & 282.6 & -28 & 25.4 & 14.6\\
 12 & 55739.4411232 & 104 & 296.1 & -52.8 & 9.8 & 32.5\\
 14 & 55782.5161911 & 1040 & 265.6 & -27.9 & 13.2 & 1.2\\
 15 & 55783.1854223 & 57.5 & 287.3 & -49.7 & 19.7 & 26.3\\
 22 & 55941.9757813 & 219.5 & 293.7 & -22.1 & 12.1 & 25.9\\
 24 & 55950.8474912 & 30.5 & 282.2 & -15.1 & 15.5 & 20.4\\
 25 & 55966.7422488 & 33.5 & 286.0 & -14.5 & 46.3 & 23.5\\
 33 & 56221.3424023 & 385 & 292.5 & 7.8 & 13.5 & 44.8\\
 36 & 56308.1642740 & 28.9 & 257.7 & -3.0 & 11.7 & 27.2\\
  \hline
  
\end{tabular}
\caption{Properties of the IceCube HESE considered in the analysis.}
\label{ICtable_t}
\end{table}

\section{Search methodology}\label{methodo}
\subsection{The algorithm: Two-point correlation function}
In this analysis, we extend the two-point correlation function, which is commonly used to detect spatial clustering (see e.g. \cite{JCAP_clustering}) to the time domain. Thus, we consider the two-point cumulative distribution defined as:
\begin{equation}
\mathcal{N}(\Delta t) = \sum_{i=1}^{N_{IC}}\sum_{j=1}^{N_{ANT}}\omega_{ij}\left[1-H\left(\Delta t_{ij}-\Delta t\right)\right]
\label{cum_correl}
\end{equation}
where $H$ is the Heaviside function depending on the absolute value of the temporal distance $\Delta t$. $\omega_{ij}$ are the weights assigned to each couple of IceCube (IC) and ANTARES (ANT) events, named respectively $i$ and $j$. Each weight $\omega_{ij}$ is computed according to a normal distribution centered on the IceCube HESE and defined as:
\begin{equation}
\omega_{ij} = \mathrm{exp}\left(\frac{-\Delta\Omega_{ij}^2}{2\sigma_i^2}\right)
\label{weight_eq}
\end{equation}
where $\Delta\Omega_{ij}$ is the angular distance between each couple of IceCube and  ANTARES events, and $\sigma_i$ is the standard deviation of the $i^\mathrm{th}$ IceCube HESE angular error distribution, computed from its median angular error provided in Table \ref{ICtable_t} (see also \cite{icecube2}). In this definition, the angular error of the ANTARES events is neglected as it is much smaller than the one of the IceCube HESE. The temporal binning $\Delta t$ is set equal to 0.01 day between 0 and 10 days and equal to 1 day between 10 and 1000 days. 

\subsection{Background estimation}\label{bkg_estimate}
To detect correlated structures in the dataset, a reference cumulative correlation distribution (considered as the background distribution) has been built by generating $10^4$ pseudo-experiments. 

\subsubsection{Time generation}\label{time_gene}

The event times are randomized following the procedure described hereafter, assuming approximately constant neutrino candidate rates for different periods.

\begin{figure}[h!]
\begin{center}
\includegraphics[width=0.5\linewidth]{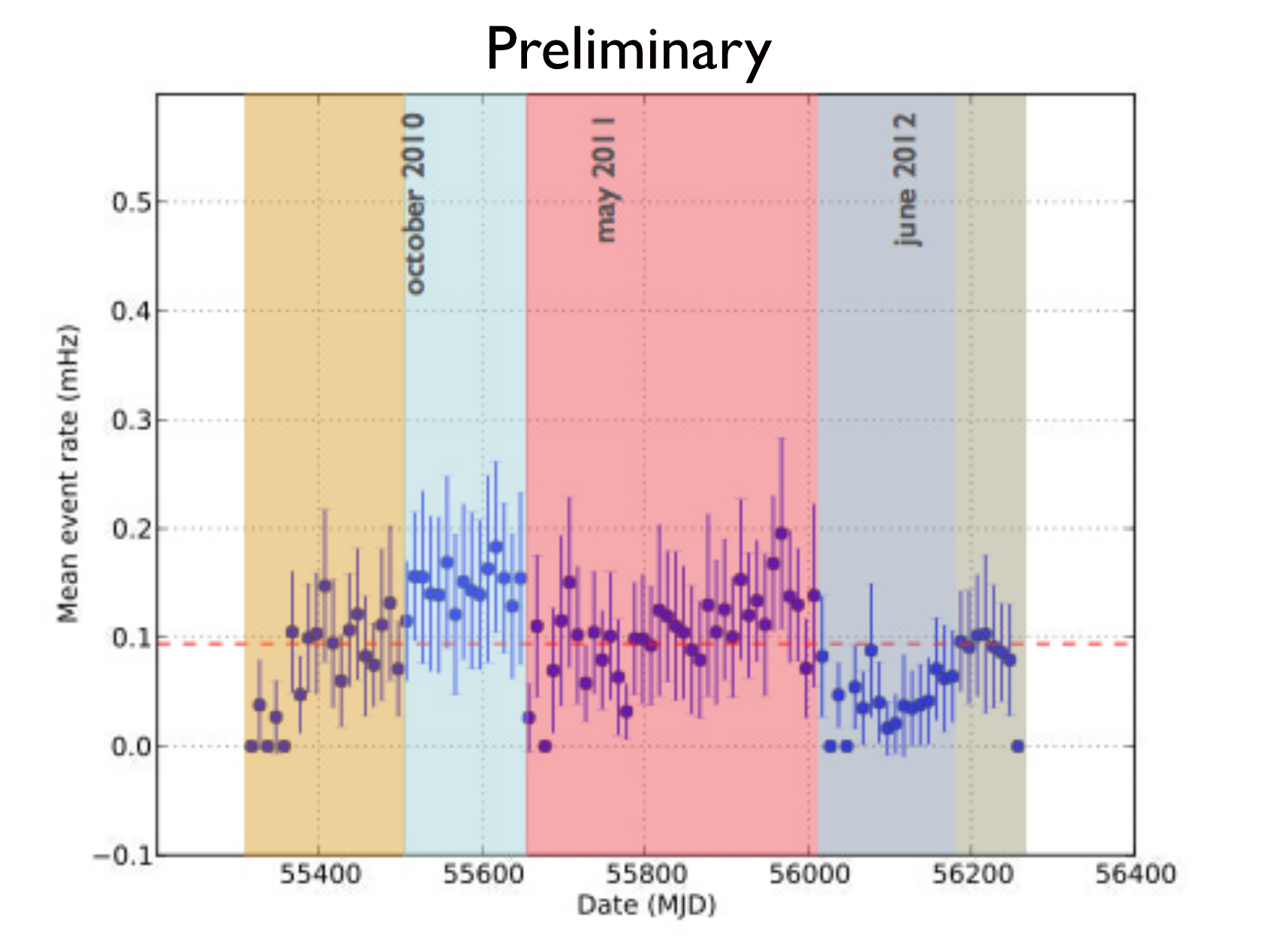}
\caption{Mean neutrino candidate rate per bins of 10 days for the final dataset. The 5 time periods defined to generate the pseudo-experiments are indicated by the 5 colored areas.}
\label{4periods}
\end{center}
\end{figure}

The evolution of the mean neutrino candidate rate per bins of ten days is shown in Figure \ref{4periods}. From this plot, we  define five sub-periods characterized by a small evolution of the mean neutrino candidate rate, which roughly follow the seasonal variation of the ANTARES data stream. For each of them, we define a mean neutrino candidate rate as:
\begin{equation}
\overline{\mathrm{event\ rate}} = \frac{\sum_{k} \mathrm{number\ of\ neutrino\ candidates\ over\ run\ }k}{\sum_{k} \mathrm{run\ duration}_{k}}
\label{meanevtrateeq}
\end{equation}
where $k$ refers to each run included in the given sub-period.

Every sub-time period is now treated separately. For each run $k$ belonging to one given sub-time period, we draw a number of neutrino candidates from a poissonian distribution of mean $\mu$ given by:

\begin{equation}
\mu = \overline{\mathrm{event\ rate}} \times \mathrm{run\ duration}_{k}
\end{equation}
where $\overline{\mathrm{event\ rate}}$ is defined in Equation \ref{meanevtrateeq}.
Finally, we draw the event time from an uniform distribution between the run start and the run stop dates of the given run.

\subsubsection{Local coordinates generation}\label{coord_gene}
The local coordinates (Azimuth; Zenith) of each event are generated from the 2D distribution of the local coordinates related to each of the five time periods defined above. Corresponding equatorial coordinates are computed from the event local coordinates, the event time and the ANTARES detector location.\\

This randomization process is performed $10^4$ times. Each pseudo-experiment is then analyzed in exactly the same way as the data to derive the normalized cumulative distribution function. All the pseudo-experiment cumulative distributions are finally averaged to compute a background estimation. Figure \ref{cum_distri} shows the resulting two-point cumulative distribution for the background.

\begin{figure}[h!]
\begin{center}
\includegraphics[width=7cm]{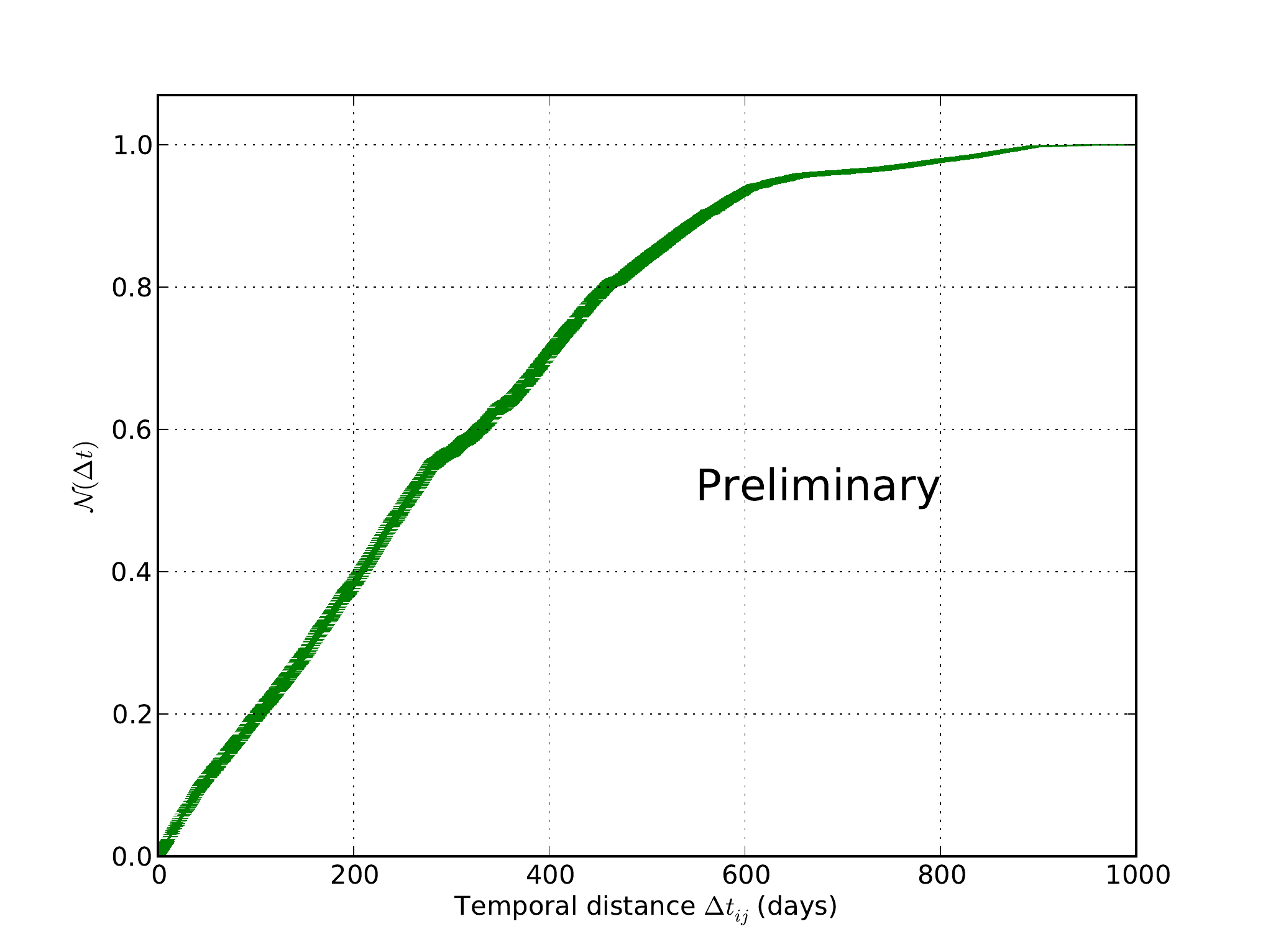}\hfill
\caption{Cumulative two-point distribution function in the background only hypothesis (average of 10$^4$ pseudo-experiments). The green band indicates the standard deviation of the 10$^4$ pseudo-experiments in each bin.}\label{cum_distri}
\end{center}
\end{figure}

\subsection{Test Statistic}

In order to detect a possible timing correlation between ANTARES events and IceCube HESE, the cumulative two-point distribution of the data is compared, bin-by-bin, with the reference distribution built according to Section \ref{bkg_estimate}. Time correlation features will thus appear as differences between these two distributions. To test the significance of such a correlation, a test statistics (TS) is defined by maximizing a value over all the timescales $\Delta t_i$, as given by Equation \ref{TStat}: 
\begin{equation}
\mathrm{TS}=\mathrm{max}_{\Delta t_{i}}\left[\frac{\mathrm{N_{on}}-\mathrm{N_{off}}}{\sigma}\right]
\label{TStat}
\end{equation}
where N$_\mathrm{on}$ and N$_\mathrm{off}$ correspond to the value of the cumulative distribution function (see equation \ref{cum_correl}) for a given bin $\Delta t_i$, computed respectively for the data and for the background. The denominator $\sigma$ is taken equal to the standard deviation of the poissonian  pseudo-experiments distribution in each bin. 
To avoid a divergence of the test statistic (due to small $\sigma$ values related to large $\Delta t_i$) and to limit the number of trials, we scan only up to 10 days.

\subsection{Discovery potential}
In order to estimate the detection power of the analysis, we perform pseudo-experiments in which one or more event(s) is (are) replaced with one or more signal event(s) that would have come from a transient astrophysical source.

We first choose randomly one of the IceCube HESE around which the generated signal event will occur. The neutrino flare is described by a gaussian distribution of mean $\mu_{flare}$ equal to the IceCube event time and standard deviation $\sigma_{flare}$, that will be considered as an estimate of the characteristic flare duration. Thus, the generated signal event time is drawn from this probability distribution, while the angular distance between the generated signal event and the IceCube HESE, $\Delta\Omega_{ij}$,  is drawn from a normal probability distribution function defined as:
\begin{equation}
\mathrm{pdf}\left(\Delta\Omega_{ij}\right) = \mathrm{sin}\left(\Delta\Omega_{ij}\right)\times\mathrm{exp}\left(\frac{-\Delta\Omega_{ij}^2}{2\sigma_i^2}\right)
\end{equation}
where $\sigma_i$ is defined as in Equation \ref{weight_eq}.
Finally, the equatorial coordinates of the generated event are drawn randomly from all the points located $\Delta\Omega_{ij}$ degrees from the $i^\mathrm{th}$ IceCube HESE. 
Based on the arrival time of this generated event, its equatorial coordinates are then translated into local coordinates. If these local coordinates belongs to the ANTARES visibility map (if the Zenith is $\geq \pi/2$), the event is included to the generated events list. Elsewhere, it is not taken into account.

The probability for a 3$\sigma$ effect is finally provided with respect to the different combinations of flare duration and number of generated signal events (see Figure \ref{disco_potential}). We point out that the number of generated signal events corresponds to the total number of events generated around the eight IceCube HESE. Moreover, one has to keep in mind that the region of the sky around the GC is visible $\sim$70\% of the time by the ANTARES telescope. This means that only $\sim$70\% of the total generated signal events might be observed by ANTARES (this is particularly true for the longest flares). Thus, the effective number of ANTARES signal events per IceCube HESE is provided in the upper part of Figure \ref{disco_potential}.

\begin{figure}[h!]
\begin{center}
\includegraphics[width=0.7\linewidth]{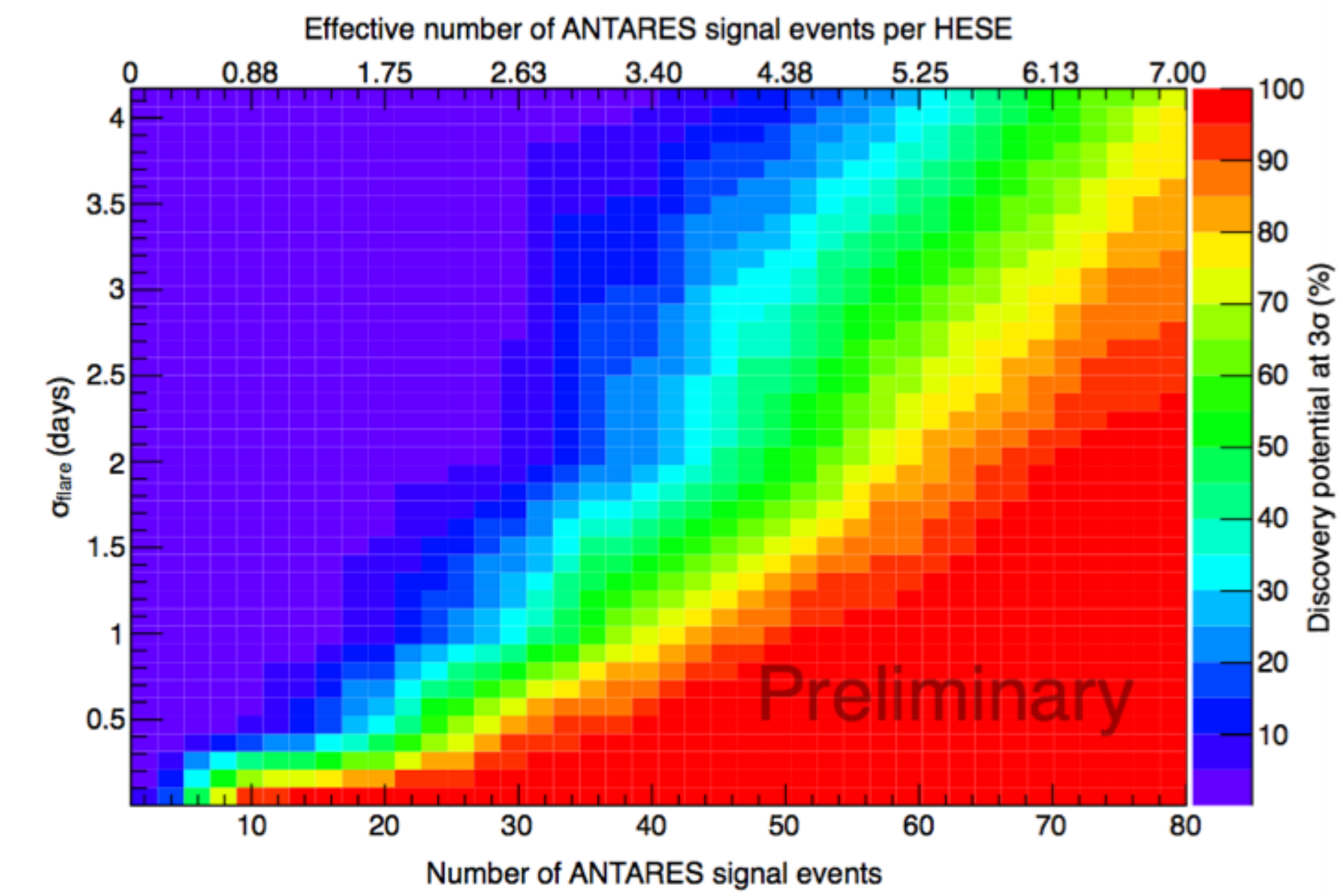}
\caption{Discovery potential at 3$\sigma$ with respect to the flare duration ($\sigma_{flare}$) and the total number of signal events. The effective number of signal events (see text) is also given in the upper part.}
\label{disco_potential}
\end{center}
\end{figure}

\section{Results} \label{results_section}

Figure \ref{TSbkg} (left panel) shows the distribution of the test statistic derived for 10$^4$ pseudo-experiments by comparing the cumulative two-point distribution of each pseudo-experiment with the cumulative reference distribution built according to Section \ref{bkg_estimate}.
A test statistic value of 0.085 would correspond to a 3$\sigma$ effect. For the unblinded ANTARES dataset, a test statistic value of 0.027 is found. The probability of obtaining a TS value at least as large as this in the background only hypothesis is equal to 35\%, which corresponds to a $\sim$0.9$\sigma$ effect. Thus, no significant correlation has been found. The comparison between the data and the background (see Figure \ref{TSbkg}, right panel) shows that the largest deviation between the two-point cumulative distributions corresponds to a time scale of 6.1 days.

Consequently, we set upper limits on such a time correlation. Figure \ref{uplimit} provides the 90\% confidence level upper limit on the number of ANTARES events temporally correlated with the IceCube HESE as a function of the flare duration. The blue area indicates the region excluded at a 90\% confidence level. Thus, for flares shorter than $\sim$1 hour, we can exclude the fact that at least two ANTARES events are temporally correlated with any of the 8 IceCube HESE. For larger flares, the total number of ANTARES events needed to detect a significant time correlation increases as the flare duration increases. Thus, for instance, considering a 1-day flare, one can exclude that fifteen ANTARES neutrino events arrive in correlation with any of the IceCube HESE (which corresponds roughly to 1.3 event per IceCube HESE) . 

\begin{figure}[h!]
\begin{center}
\includegraphics[width=0.5\textwidth]{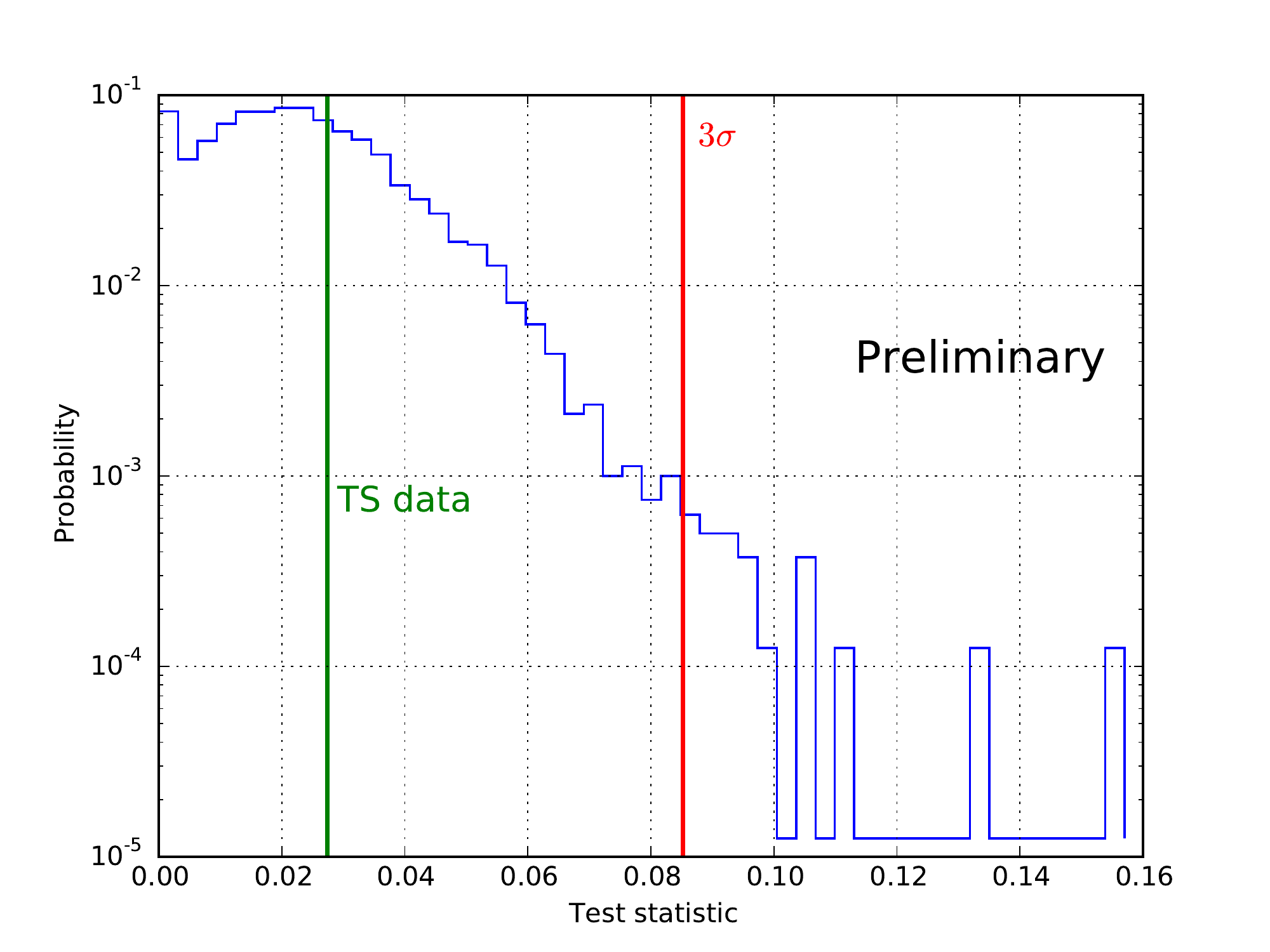}\hfill
\includegraphics[width=0.5\textwidth]{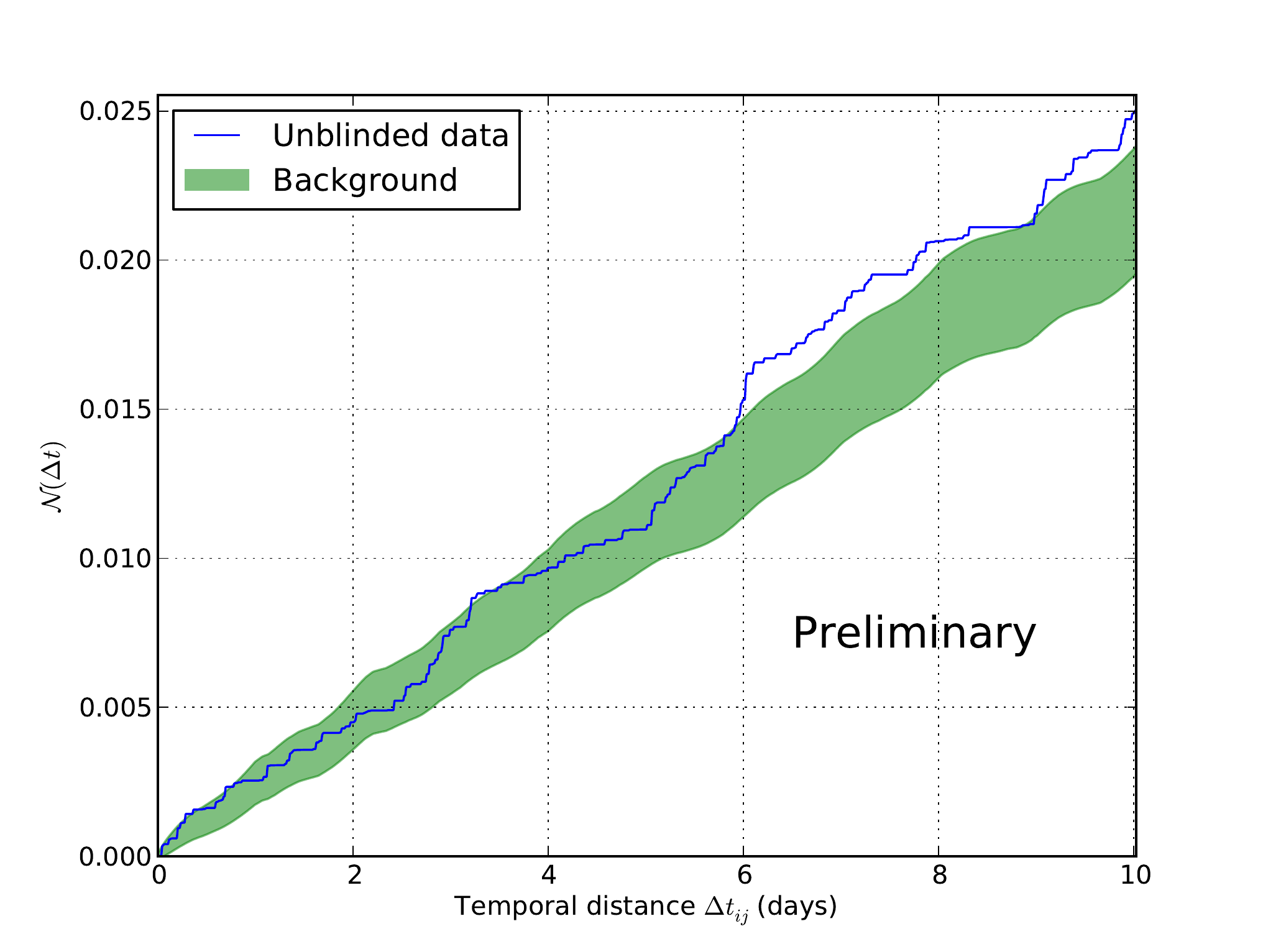}
\caption{\textbf{Left:} Distribution of the test statistic (TS) in the background only hypothesis. The 3$\sigma$ probability is given by the red line while the TS value obtained for the ANTARES unblinded dataset is represented by the green vertical line. \textbf{Right:} Cumulative two-point distribution over 10 days. The green area corresponds to the standard deviation of each bin of the reference cumulative distribution (built according to Section \ref{bkg_estimate}), while the two-point cumulative distribution for the unblinded dataset is represented in blue.}
 \label{TSbkg} 
\end{center}
\end{figure}

\begin{figure}[h!]
\begin{center}
\includegraphics[width=0.8\linewidth]{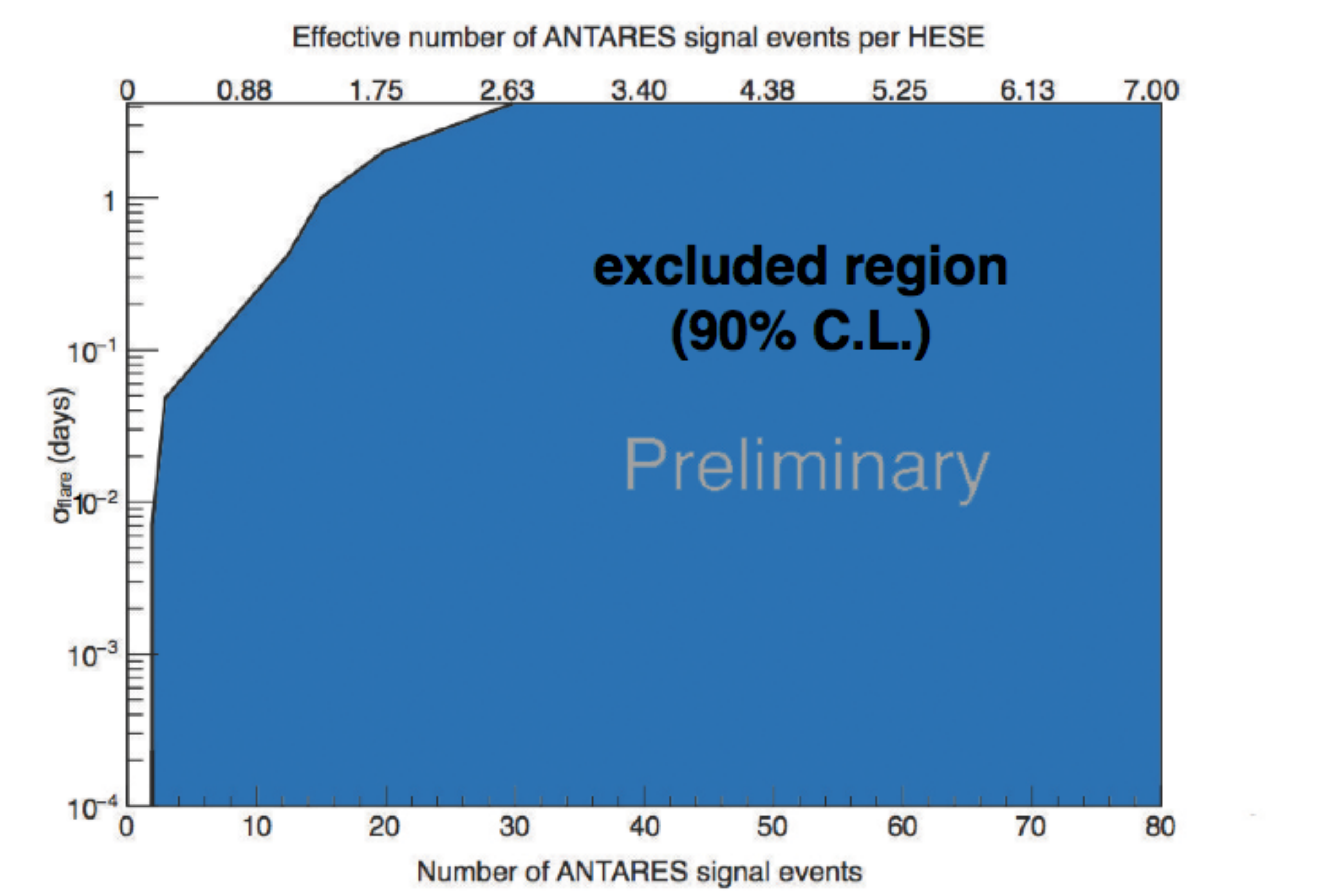}
\caption{90\% confidence level upper limit on the number of ANTARES events temporally correlated with the IceCube HESE as a function of the flare duration. Blue area indicates the region excluded at a 90\% confidence level.}
 \label{uplimit} 
\end{center}
\end{figure}

\section{Conclusion}
We have described a time-correlation analysis based on the IceCube HESE positionally consistent with the GC and the ANTARES events recorder in the period from May 2010 and November 2012 to look for the possible signature of transient neutrino emission in this region of the sky. The discovery potential was evaluated for different flare durations and numbers of signal events. No significant time correlation was found between the two samples.


\setcounter{figure}{0}
\setcounter{table}{0}
\setcounter{footnote}{0}
\setcounter{section}{0}
\setcounter{equation}{0}
\newpage
\id{id_lafusco}
\addcontentsline{toc}{part}{\textcolor{blue}{\arabic{IdContrib} - {\sl L. A. Fusco} : Search for an enhanced emission of neutrinos from the Southern Sky with the ANTARES telescope}%
}
\noindent
\title{\arabic{IdContrib} - Search for an enhanced emission of neutrinos from the Southern Sky with the ANTARES telescope }

\shorttitle{\arabic{IdContrib} - Southern Sky neutrinos with ANTARES}

\authors{Luigi Antonio Fusco}
\afiliations{Dipartimento di Fisica e Astronomia dell'Universit\`a di Bologna, Viale Berti-Pichat 6/2, 40127, Bologna, Italy.\\
  INFN -- Sezione di Bologna, Viale Berti-Pichat 6/2, 40127, Bologna, Italy}
  \email{lfusco@bo.infn.it}

\abstract{Compelling evidence of the existence of cosmic neutrinos has been reported by the IceCube collaboration. Some features of this signal could be explained by a Northern/Southern sky asymmetry of the flux. This possible asymmetry could be related to the presence of the bulk of our Galaxy in the Southern sky.

The ANTARES detector, located in the Mediterranean Sea, is currently the largest operating under-water neutrino telescope. Its effective area and good exposure to the Southern sky allows to constrain an enhanced muon neutrino emission from extended sources.

  Two signal regions are defined: one around the largest accumulation of events from the IceCube High Energy Starting Events and one surrounding the Galactic Plane area; the background from atmospheric events is estimated looking at data from off-zones for which ANTARES has the same exposure as for the signal region. The ANTARES sensitivity to such a flux has been computed and the results of the analysis of data from 2007 to 2013 are presented.}


\maketitle

\section{Introduction} \label{sec:intro}

The scientific goal of the ANTARES neutrino telescope \cite{bib:ANTARES_ICRC} is to detect high energy neutrinos of cosmic origin. The IceCube Collaboration has announced the observation of a cosmic neutrino signal in the High Energy Starting Events (HESE) analysis \cite{bib:IC2y} of two years of data with the complete detector. The purely atmospheric expectation is rejected at a level of 5.7$\sigma$ with the inclusion of a third year of data \cite{bib:IC3y}. 

The IceCube neutrino signal is reported to be compatible with a cosmic flux which is equally distributed in the three neutrino flavours \cite{bib:IC_flavour} and isotropic in the neutrino arrival direction. However, some possible enhancement of the neutrino flux from the Southern with respect to the Northern hemisphere has been underlined for example in \cite{bib:maurizio}. The low energy extension of the IceCube HESE analysis \cite{bib:IC_25T} shows a steepening of the energy spectrum of the cosmic signal and more events in the low energy part of the signal. Even if the IceCube detector is much larger in volume, the effective area of ANTARES in the region between 20 and 50 TeV in the Southern Sky is similar to that of IceCube: some IceCube-like signal events could be observed by ANTARES, too.

Because of the nature of the IceCube analysis, based on vetoing techniques to detect downward-going events \cite{bib:ICveto}, the signal is mainly distributed in the shower channel: the directional resolution for these events is poor and the signal appears as an all sky flux. Since ANTARES is located in the Mediterranean Sea, the Southern sky is accessible to the detector in up-going tracks, for which an extremely good angular resolution can be achieved. For this reason a diffuse flux in the shower channel for IceCube might appear as an ensemble of individual point sources in ANTARES or a region with an enhanced diffuse emission. The first possibility is addressed in \cite{bib:PS_ICRC}, while the latter will be presented in this contribution.

\section{Neutrinos from the Milky Way} \label{sec:gal_nu}

Neutrinos can be produced close to galactic Cosmic Rays (CR) accelerators such as supernova remnants when high energy protons or nuclei interact with the surrounding matter. A large amount of pions is produced, and the number of produced $\pi^0$ is equal to the sum of $\pi^+$ and $\pi^-$. While $\pi^0$ immediately decay into a pair of $\gamma$-rays, $\pi^\pm$ mesons decay into muons and muon neutrinos. As a consequence, the expected neutrino flux is equal to the $\gamma$ flux of hadronic origin. The energy spectrum of this flux follows the CR spectrum at the acceleration site - E$^{-\Gamma}$ with $\Gamma$ = 2.0 for Fermi acceleration scenarios \cite{bib:fermiacc} - since the decay usually takes place before the primary particle can interact.

Another neutrino contribution from the Milky Way is expected from CRs propagating in the inner region of our Galaxy. A CR can interact with the dense environment, producing $\gamma$-rays and neutrinos. Data from the Fermi/LAT detector provide the best observation of this diffuse $\gamma$ flux in the Galactic Plane \cite{bib:fermigal}, though no observation of the neutrino counterpart is available.

Assuming that a certain fraction of the observed diffuse $\gamma$ flux in the central region of the Galaxy originates from hadronic mechanisms, the neutrino yield from CR propagation can be calculated. Different predictions are available such as the ones in \cite{bib:neronov, bib:fresh, bib:antonio}. Each of these provides a different description of the expected neutrino flux, with an overall normalisation that can vary by one order of magnitude. A rather hard spectrum is expected, described, at least in part of the energy range, by a broken power law with spectral index $\Gamma\sim2.4\div2.5$.

\section{The IceCube signal} \label{sec:ICsig}

The IceCube cosmic neutrino signal was initially \cite{bib:IC3y} fitted by a broken power law spectrum with spectral index $\Gamma$ equal to 2:
\begin{equation}
  E^{2}\frac{d\Phi}{dE} = (0.95\pm 0.3)\times10^{-8}\;\; GeV\,s^{-1}\,sr^{-1}\,cm^{-2}.
  \label{eq:ic_spectrum_2}
\end{equation}
while also an E$^{-2.3}$ spectrum was reported to be compatible with the observed signal, mainly because of the absence of events above 2 PeV and the lack of an enhanced neutrino detection in the region of the Glashow resonance.

The further extension of the IceCube data set and refinements in the analysis for the observation of lower energy neutrinos resulted in a steepening of the energy spectrum; in the most recent publication \cite{bib:IC_25T} the best fit is reported to be:
\begin{equation}
  \frac{d\Phi}{dE} = \left(2.06 ^{+0.4}_{-0.3} \right) \times 10^{-18} \left(\frac{E}{100\; TeV}\right)^{2.46\pm0.12}\;\; GeV^{-1}\,s^{-1}\,sr^{-1}\,cm^{-2}
  \label{eq:ic_spectrum_flav}
\end{equation}
for 25 TeV < E$_{\nu}$ < 1.4 PeV, rejecting the $E^{-2}$ hypothesis with a significance of 3$\sigma$.

This flux is observed as an all sky flux by IceCube. This can be related to the bias in efficiency of the analysis to shower events, for which IceCube has a rather bad angular resolution. However, an accumulation of shower events is present in the IceCube sky map of \cite{bib:IC3y} even if it is not significant because of the poor directional reconstruction. This accumulation could point towards a diffuse emission region of cosmic neutrinos. In addition, as the Southern sky contains (most) of the Galactic Plane, the soft spectrum derived from IceCube measurement and an enhancement of the neutrino flux in the Southern sky hint towards a possible neutrino emission from the Galactic Plane.

\section{Data analysis} \label{sec:analisi}

Given the IceCube effective area, the cosmic flux which can produce a certain number of events from a region of the sky with angular size $\Omega \sim 0.1-0.2$ sr can be computed as a function of the signal spectral index \cite{bib:maurizio}. 

Two possible signal regions have been chosen for this analysis. The first one corresponds to a 10$^\circ$ circular region around the largest accumulation of IceCube signal events (IceCube hotspot). The centre of this circular area is at l = 18$^\circ$ and b = -9$^\circ$ in galactic coordinates. This position has been computed averaging the position of the HESE events weighted according their energy and angular uncertainty.

The second signal region is chosen to represent the Galactic Plane area. A rectangular region having | l | < 40$^\circ$ and | b | < 3$^\circ$ is selected since it encloses the central part of the Fermi/LAT diffuse galactic plane flux.

In any case, as the background per unit solid angle can be considered rather similar for regions that have similar exposures and since the sensitivity depends on the intensity of the background flux in terms of number of events per steradians, the results are independent on the choice of the signal region.

Data collected by the ANTARES neutrino telescope from 2007 to 2013 have been considered for this analysis. 

\subsection{Atmospheric background} \label{sec:atmo}

The main background in the search for cosmic neutrinos is given by downward-going atmospheric muons reaching the detector; these events are simulated using the MUPAGE software \cite{bib:mupage}. Since only neutrinos can traverse the Earth, neutrino telescopes look at upward going events to reject this background. Wrongly reconstructed atmospheric muons, mimicking upgoing neutrino events can be rejected by a selection on the track quality parameter $\Lambda$ and the estimate for the angular error $\beta$.

An irreducible background comes from atmospheric neutrinos coming from the decay of short-lived particles in extensive air showers. The \textit{conventional} component, coming from the decay of pions and kaons, is described by the \textit{Honda et al.} flux \cite{bib:honda}: in general this conventional flux can be described with a broken power law energy spectrum, with spectral index $\Gamma$ asymptotically going to 3.7. 

A \textit{prompt} component is expected to come from charmed hadrons, which decay in a much shorter time and give a harder neutrino energy spectrum ($\Gamma\sim$2.7). The \textit{Enberg et al.} \cite{bib:enberg} model is used in this work to parametrise the \textit{prompt} component. A measurement of the atmospheric neutrino energy spectrum using ANTARES data has been performed and the results are shown in \cite{bib:atmonu}. 

Since signal and background spectra can be described by power laws with different indices, a cut on the reconstructed energy \cite{bib:jutta} can provide a selection for cosmic neutrinos, since at high energy the signal flux is naturally enhanced.

\subsection{Background estimation from data} \label{sec:ON}

A signal from a Southern Sky region is searched for by comparing the number of selected events from the chosen on-zones to that of similar regions with no expected signal (off-zones). This choice avoids simulation related biases in the estimation of the signal intensity after event selection.

Off-zones are defined as fixed regions in equatorial coordinates which have identical size and shape as the on-zone but have no overlap with it. In local coordinates, off-zones have the same sidereal-day periodicity as the on-zone and span the same fraction of the sky, but with some fixed delay in time. This approach has been already used in ANTARES to search for events from the Fermi bubbles \cite{bib:FB, bib:FB_ICRC} and in a previous analysis of the Galactic Plane region \cite{bib:erwin}.

Figure \ref{fig:ONOFF} reports the position of the signal and background regions in galactic coordinates; also the position of the Fermi Bubbles is shown. Off-zones are shifted in the sky to avoid any overlap with the Fermi bubbles, so that none of the possible signal events from these areas enters in the background estimation. In the case of the IceCube Hotspot, up to 12 off-zones can be used for background estimation, while for the Galactic Plane selection the maximum is 9 off-zones.

\begin{figure}
 \centering
  \begin{subfigure}[b]{0.49\textwidth}
   \includegraphics[width=\textwidth]{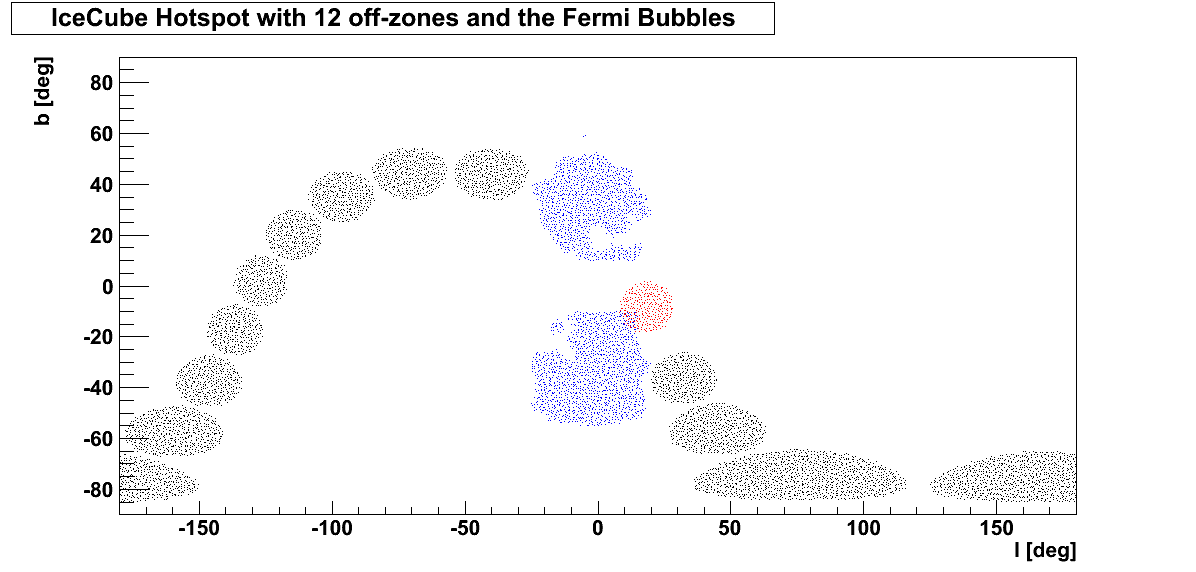}
   \caption{IceCube hotspot}
  \end{subfigure}
  \begin{subfigure}[b]{0.49\textwidth}
   \includegraphics[width=\textwidth]{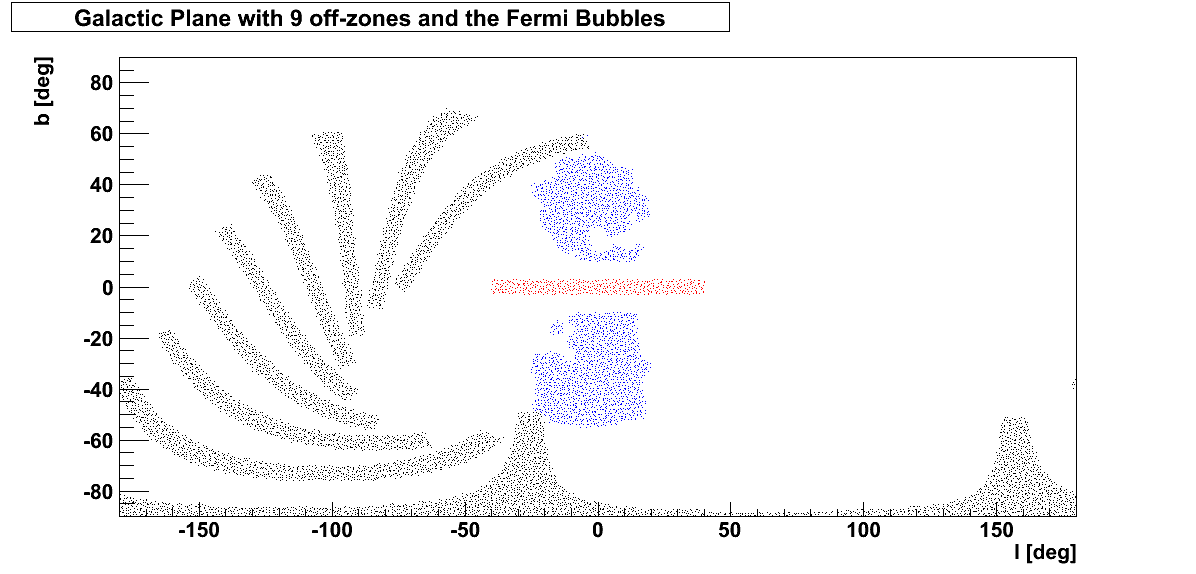}
   \caption{Galactic plane}
  \end{subfigure}
 \caption{Signal (red dots) and background (black dots) regions for the on/off-zones of the analysis of: a) the IceCube hotspot; b) the Galactic Plane region. Also shown the shape of the Fermi bubbles (blue dots) as used in the analysis of \cite{bib:FB_ICRC}.}
                      \label{fig:ONOFF}
\end{figure}

While data from the signal regions are blinded until the event selection has not been completely defined, off-zones can be used to estimate the agreement between data and Monte Carlo as well as the relative agreement between data and data from different off-zones. No anomalous behaviour has been found in the analysed data set. 

\subsection{Cut optimisation} \label{sec:cut}

The optimisation of the event selection to enhance the possible cosmic signal against the atmospheric background is made on the basis of track quality parameters to reject wrongly reconstructed atmospheric muons and on the energy estimation to select cosmic neutrinos over the atmospheric background. The optimal selection cut, which maximises the sensitivity, is computed on the basis of the Model Rejection Factor (MRF) procedure \cite{bib:mrf} for the $\Gamma$ = 2.4 and $\Gamma$ = 2.5 hypotheses.

The optimal selection cut is:
\begin{equation}
  \Lambda > -5.0, \;\; \beta < 0.5^\circ, \;\; E_{ANN} > 10\; TeV
  \label{eq:cut}
\end{equation}
where $\Lambda$ and $\beta$ are the track quality parameters described in section \ref{sec:atmo}, while E$_{ANN}$ is the energy estimator from the Artificial Neural Network algorithm presented in \cite{bib:jutta}. 

Considering a signal flux with an energy spectrum $\sim$E$^{-2.4}$ (E$^{-2.5}$) the 90\% confidence level sensitivity is 2.0 (6.0) 10$^{-5}$ GeV$^{-1}$cm$^{-2}$s$^{-1}$sr$^{-1}$. For comparison, assuming an E$^{-2.4}$ spectrum, the normalisation of an IceCube-like flux producing 2 or more events from a region in the sky of 0.1 sr is larger than 2.0 10$^{-5}$ GeV$^{-1}$cm$^{-2}$s$^{-1}$sr$^{-1}$ and any of these scenarios can be rejected. Figure \ref{fig:sens} compares the obtained sensitivity in the Galactic Plane region to the expected neutrino flux from CR propagation of \cite{bib:neronov} and \cite{bib:antonio}. This sensitivity holds in the energy range 3 TeV - 300 TeV, which contains the central 90\% of the expected signal.

\begin{figure}
 \centering
  \begin{subfigure}[b]{0.48\textwidth}
   \includegraphics[width=\textwidth]{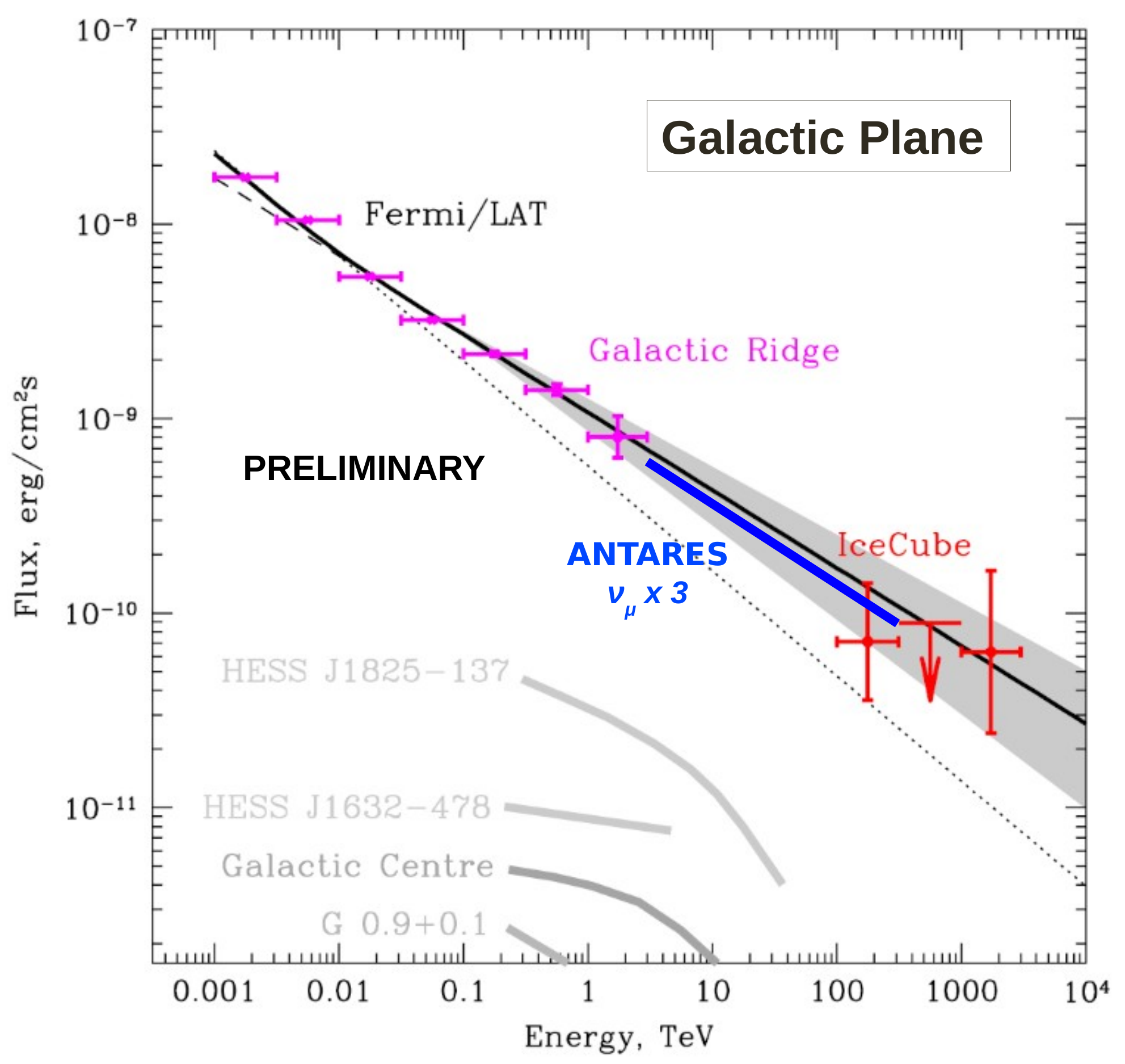}
  \end{subfigure}
  \begin{subfigure}[b]{0.48\textwidth}
   \includegraphics[width=\textwidth]{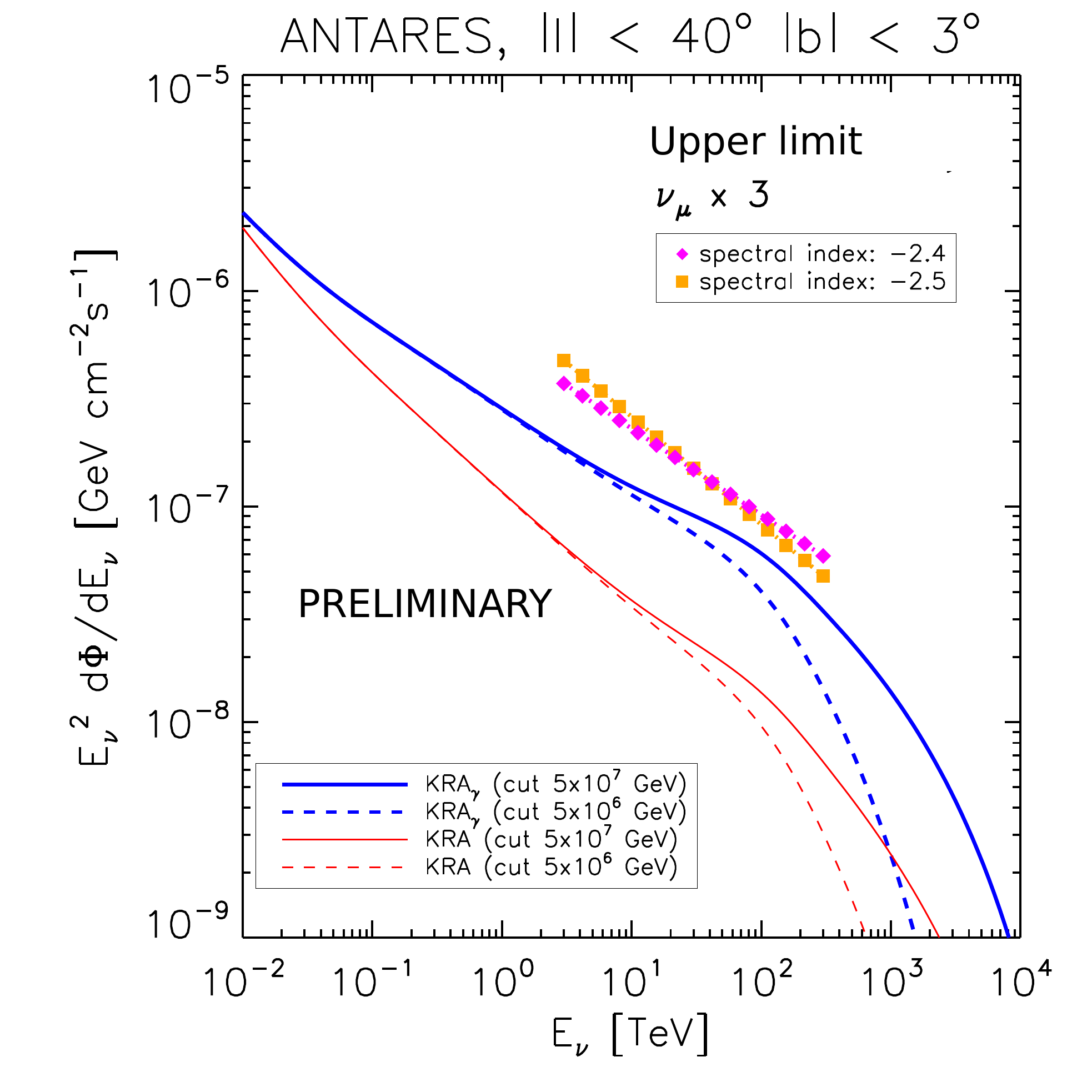}
  \end{subfigure}
 \caption{a) ANTARES sensitivity for the Galactic plane region assuming neutrino spectral index $\Gamma$ = 2.4 (blue line) compared to theoretical expectations and experimental data from Fermi/LAT and IceCube as computed in \cite{bib:neronov}; b) 90\% C.L. upper limit for signal spectrum E$^{-2.4}$ (magenta dots) and E$^{-2.5}$ (orange dots) for the null observation of this analysis compared to the expected neutrino flux from the simulations presented in \cite{bib:antonio, bib:antonio_icrc}.}
                      \label{fig:sens}
\end{figure}


\section{Results} \label{sec:results}

As far as the IceCube HESE hotspot is concerned, the average number of events coming from the chosen off-zones passing the signal selection criteria is 1.0 over the entire period. One event is also observed from the signal region and the measurement is perfectly compatible with the background only expectations. For the Galactic Plane 2.5 events are observed on average for the background regions and one is detected from the on-zone after the final selection. An underfluctuation of the background is thus present in the signal region. The reconstructed energy distributions of events for the on and off-zones are reported in figure \ref{fig:Endistr}.


\begin{figure}
 \centering
  \begin{subfigure}[b]{0.48\textwidth}
   \includegraphics[width=\textwidth]{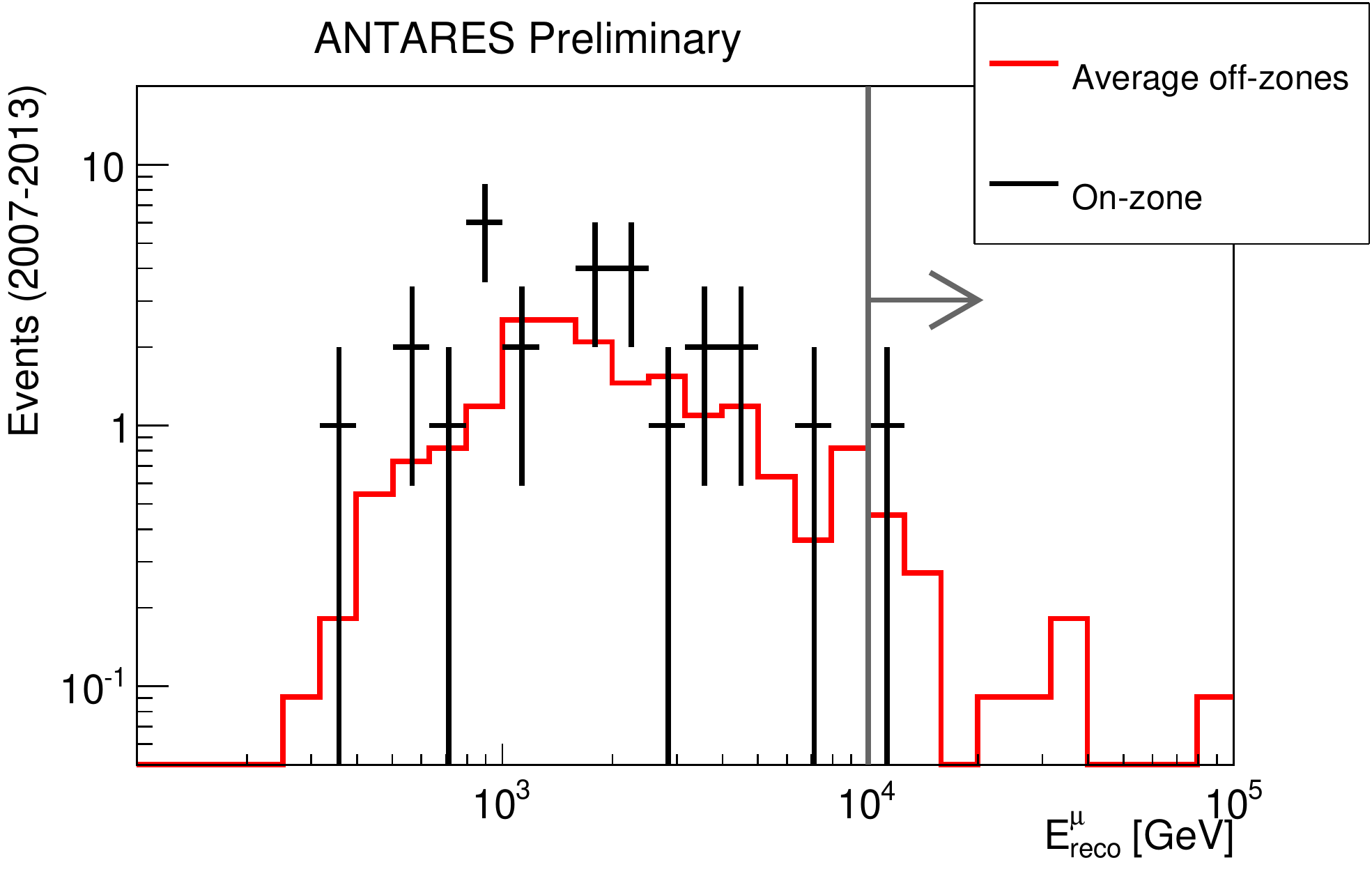}
   \caption{IceCube HESE hotspot}
  \end{subfigure}
  \begin{subfigure}[b]{0.48\textwidth}
   \includegraphics[width=\textwidth]{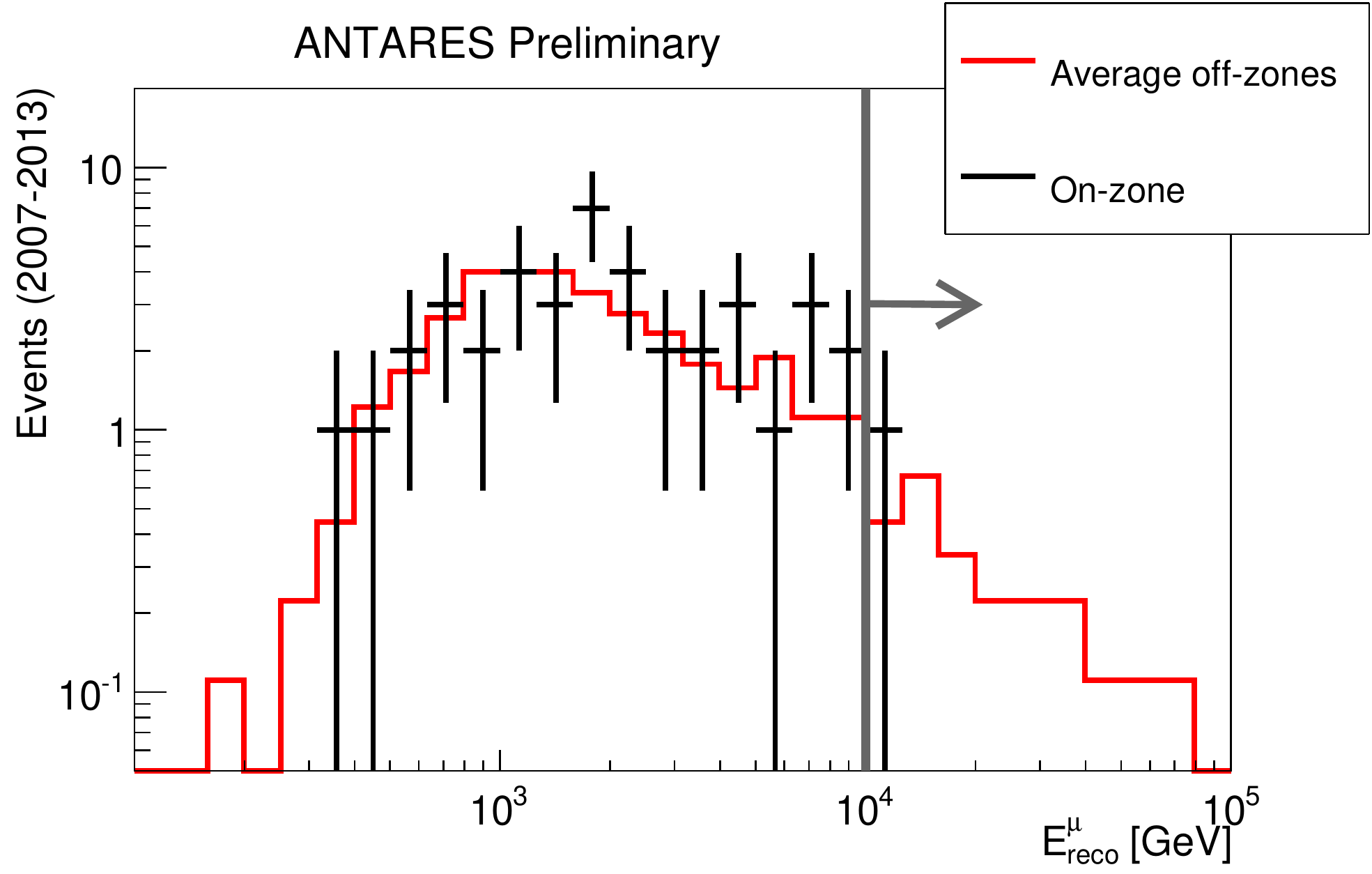}
      \caption{Galactic Plane}
  \end{subfigure}
 \caption{Reconstructed energy distribution of events in the signal (black crosses) and background (red line) regions for a) the IceCube hotspot region, b) Galactic Plane. The gray line shows the energy selection cut applied in the procedure. No significant excess is observed at high energy.}
                       \label{fig:Endistr}
\end{figure}

Neither the hotspot region nor the Galactic Plane area present an excess of events with respect to the background only evaluation. For this reason an upper limit at 90\% confidence level on the signal flux can be set and it corresponds to the ANTARES sensitivity for the analysed spectral indexes. The obtained upper limits are reported in figure \ref{fig:limitsIC} compared to the expected neutrino flux that would induce a certain number of IceCube HESE events. These fluxes have been computed on the basis of the effective areas reported in \cite{bib:IC3y}. Any model producing more than 3 events in the IceCube HESE sample from the Galactic Plane region is excluded at 90\% confidence level for a spectral index larger than 2.4. The upper limits for harder spectral assumptions with the same selection criteria are also shown.

\begin{figure}
\centering
\includegraphics[width=0.8\textwidth]{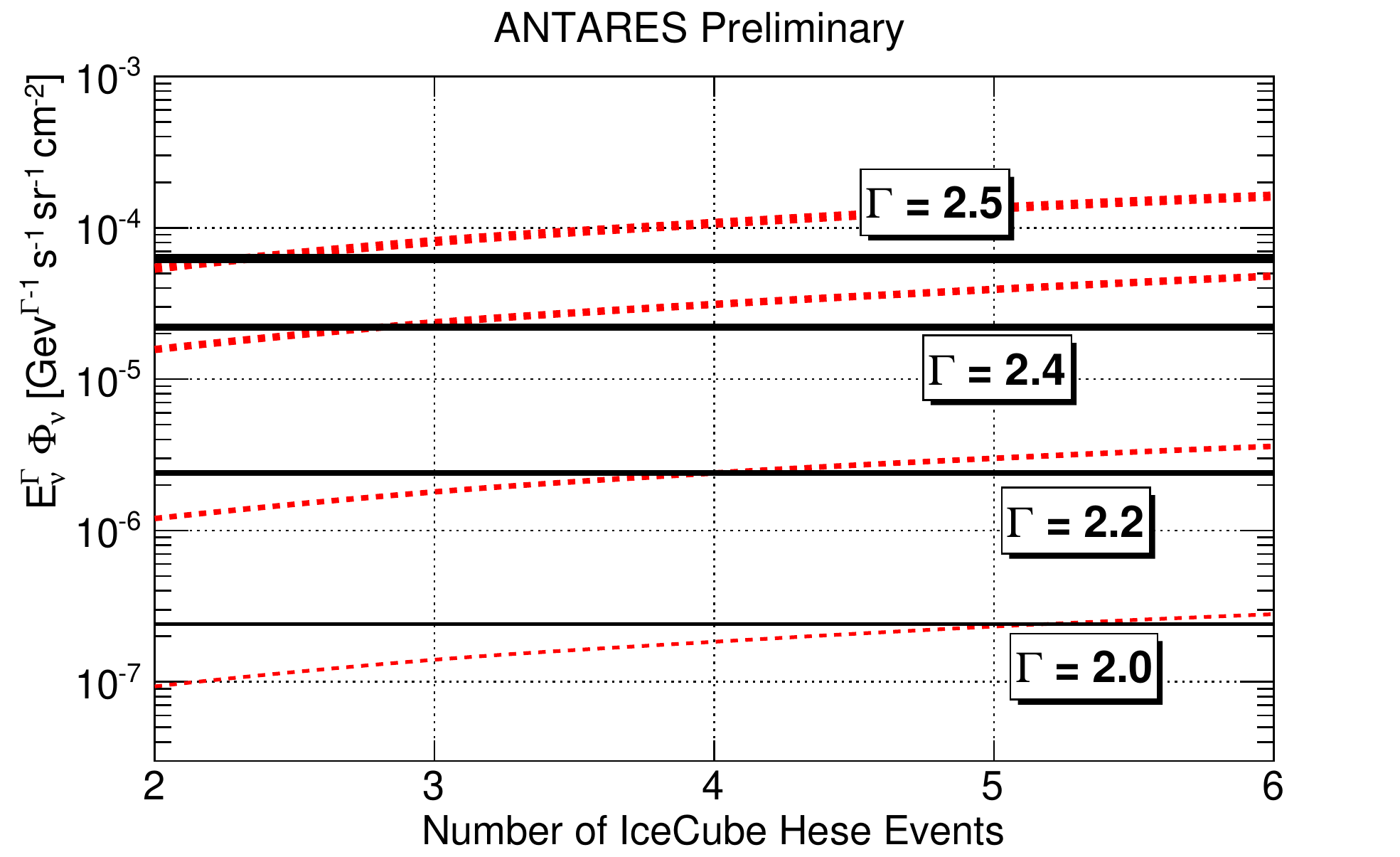}
\caption{Upper limits coming from the null observation in the Galactic Plane region compared to the expected flux producing a certain number of IceCube HESE events. Selection cuts are optimised for $\Gamma$ = 2.4 and 2.5, while the limits for harder spectral index are computed with non optimal selection.}
\label{fig:limitsIC}
\end{figure}

The lack of signal events in ANTARES data significantly constrains the possible Galactic origin of the IceCube Southern sky excess. Adding cascade events in the analysis is foreseen. For these events the angular resolution of ANTARES is much better than that of IceCube \cite{bib:ICRC_show} and the combined search in track and showers can improve the possiblity to observe such a diffuse flux from our Galaxy and to explain the Southern sky excess in IceCube data.


\setcounter{figure}{0}
\setcounter{table}{0}
\setcounter{footnote}{0}
\setcounter{section}{0}
\setcounter{equation}{0}

\newpage
\id{id_geiselsoder}
\addcontentsline{toc}{part}{\textcolor{blue}{\arabic{IdContrib} - {\sl R. Gei{\ss}els\"oder} : Model-independent search for neutrino anisotropies with the ANTARES neutrino telescope
}%
}
%

\title{\arabic{IdContrib} - Model-independent search for neutrino anisotropies with the ANTARES neutrino telescope}

\shorttitle{\arabic{IdContrib} - Model-independent multiscale anisotropy search}

\authors{Stefan Gei{\ss}els{\"o}der}
       \afiliations{ Friedrich-Alexander University of Erlangen-N{\"u}rnberg - Erlangen Centre for Astroparticle Physics}
\email{stefan.geisselsoeder@fau.de}

\abstract{ANTARES is the largest operational neutrino telescope in the Northern Hemisphere, 
located in the Mediterranean Sea at a depth of 2475 meter.
The direction and energy of the observed particles are reconstructed from the 
time and amplitude information recorded by the photomultiplier tubes. 
The collected set of reconstructed events can be analyzed with respect to the spatial, temporal and energy distribution. 

The approach shown in this contribution focuses on the spatial distribution, searching unbiasedly for a significant 
excess of neutrinos with an arbitrary size and shape from any direction in the sky. 
Techniques originating from the domain of pattern recognition and image processing are used. 
In contrast to a dedicated search for a specific neutrino emission model this approach is sensitive to a wide range of 
possible source structures. 
The result of this method applied to the ANTARES data are presented. 
}

%
\maketitle
\section{Neutrino astronomy with ANTARES}

Neutrinos are able to traverse through dense matter and are not deflected by galactic or extragalactic magnetic fields. 
While these properties make them favorable for astronomy, their detection becomes more complex and requires large volumes. 
ANTARES \cite{antares:antares} is the largest operational neutrino telescope in the Northern Hemisphere
, providing a good view on the Galactic Center through the Earth. 
Despite many dedicated searches \cite{antares:ps} \cite{antares:grb} \cite{antares:GC} \cite{antares:fermi} 
focusing on promising candidates for cosmic neutrino sources, 
no source has been identified statistically significant yet. 

\section{Model-independent multiscale source search}
\label{sec:multiscale}

The model-independent multiscale source search presented here tries to identify regions of 
arbitrary position, size, shape and internal neutrino distribution 
in which an excess of neutrino events with respect to the background expectation has been observed. 
In contrast to the testing of preselected hypotheses, an unbiased approach can also detect 
unexpected structures. 
The main drawbacks are higher trial factors than in a dedicated search and possibly a less straight forward interpretation of the result. Since there is no physical model involved, 
any kind of deviation, for instance uncompensated systematic effects, could be detected, but nevertheless this would be a valuable result. 

A discrete spherical grid with 165016 gridpoints, 
corresponding to a spacing of $\approx$ 0.5$^{\circ}$, 
is used to evaluate the directions of the measured neutrinos. 
Figure \ref{fig:sourceSetup} 
shows such a spherical grid with gridpoints in blue and random neutrino events in white. 
In the example shown in Figure \ref{fig:sourceSetup}, 
random events with two artificial point sources with 12 and 18 events  
at a declination of -70$^{\circ}$ have been added to demonstrate the analysis method. 

The search evaluates scales from 0.0$^{\circ}$ up to 90.0$^{\circ}$ in steps of 0.5$^{\circ}$. 
It starts by counting the number of neutrinos located in 
a ring around each gridpoint with a radius corresponding to the current search scale. 
The result of this evaluation is one number for each gridpoint in each scale. 
The counting is visualized in Figure \ref{fig:searchRingScheme}. 
The results for three scales can be seen in Figure \ref{fig:counting}. 


The next step is to calculate the Poisson probability for each observed value. 
The number of neutrinos $n$ around each gridpoint has been counted and the expected mean number $\lambda$ is estimated from scrambled data. 
With this information the Poisson probability $P(n)$ can be computed.
For technical reasons the Poisson probabilities of each gridpoint 
are then processed as described in formula \ref{equ:poissonAlter}.
\begin{equation}
\label{equ:poissonAlter}
R = log_{10}(\frac{1}{P(x \ge n)})
\end{equation}

The effect of the computations of this step on the search spheres is shown in Figure \ref{fig:countingToPoisson}. 
Potential source regions containing more 
neutrinos should be linked to higher values on these spheres, 
low values can be assumed not to be linked to detectable sources. 
Separating background from potentially relevant information is called segmentation. 
In this search this is done by the application of a threshold to all R values. 
The threshold is derived from the histogram of all observed R values. 
It it set to the beginning of the tail of the distribution. 
The procedure is visualized in Figure \ref{fig:threshold}. 
\newpage

\begin{figure}[h!]
\begin{center}
\includegraphics[width=0.3\textwidth]{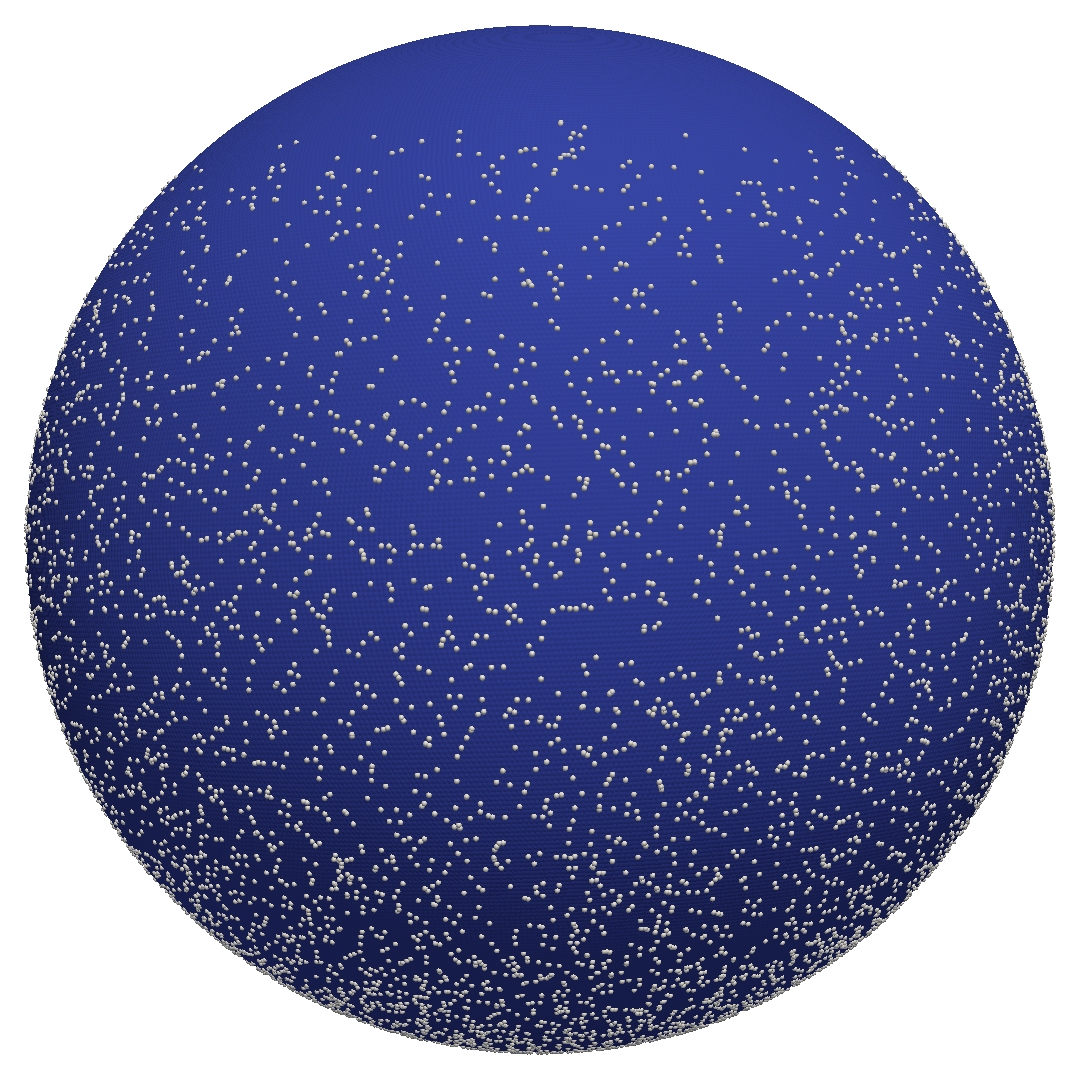}
\hfill
\includegraphics[width=0.3\textwidth]{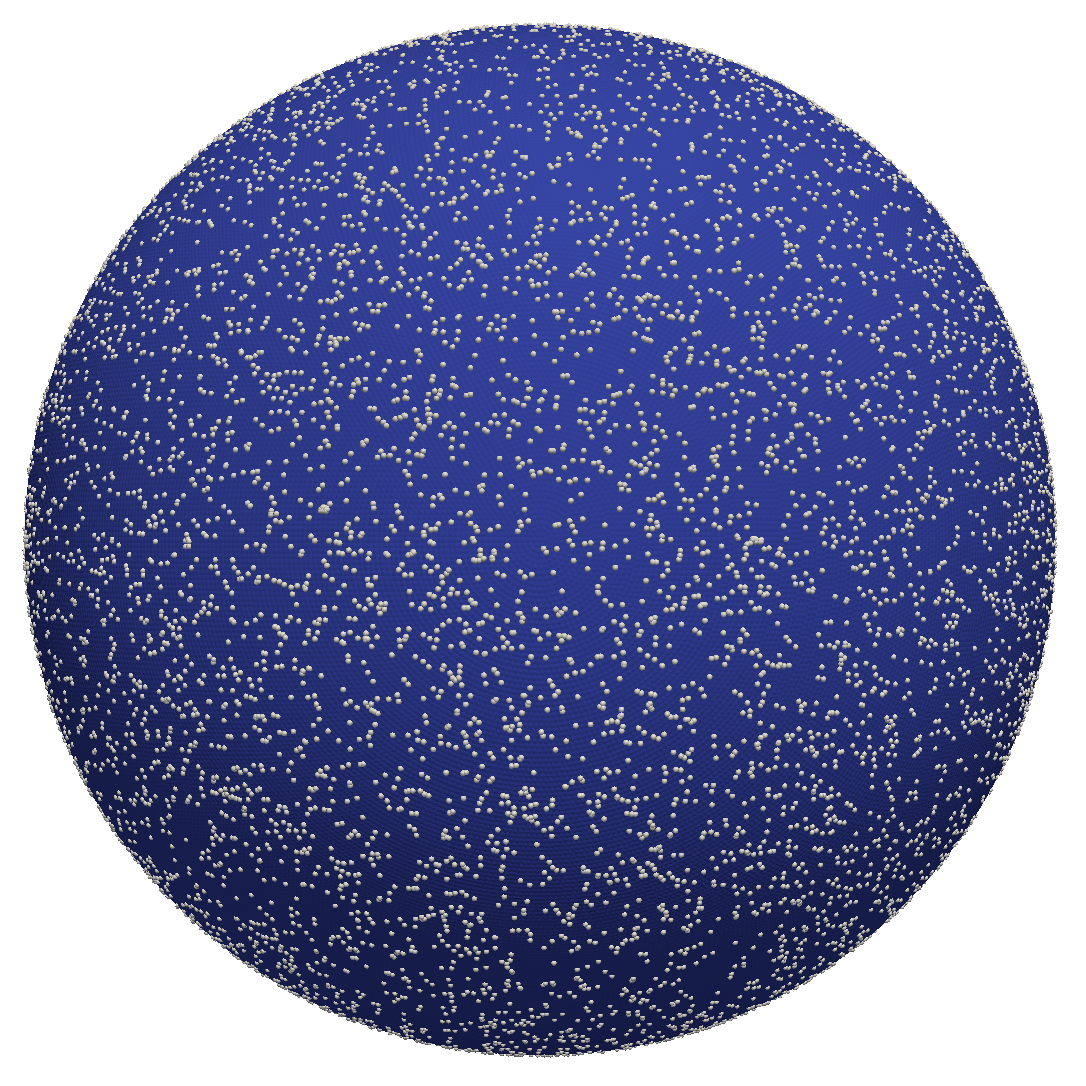}
\hfill
\includegraphics[width=0.3\textwidth]{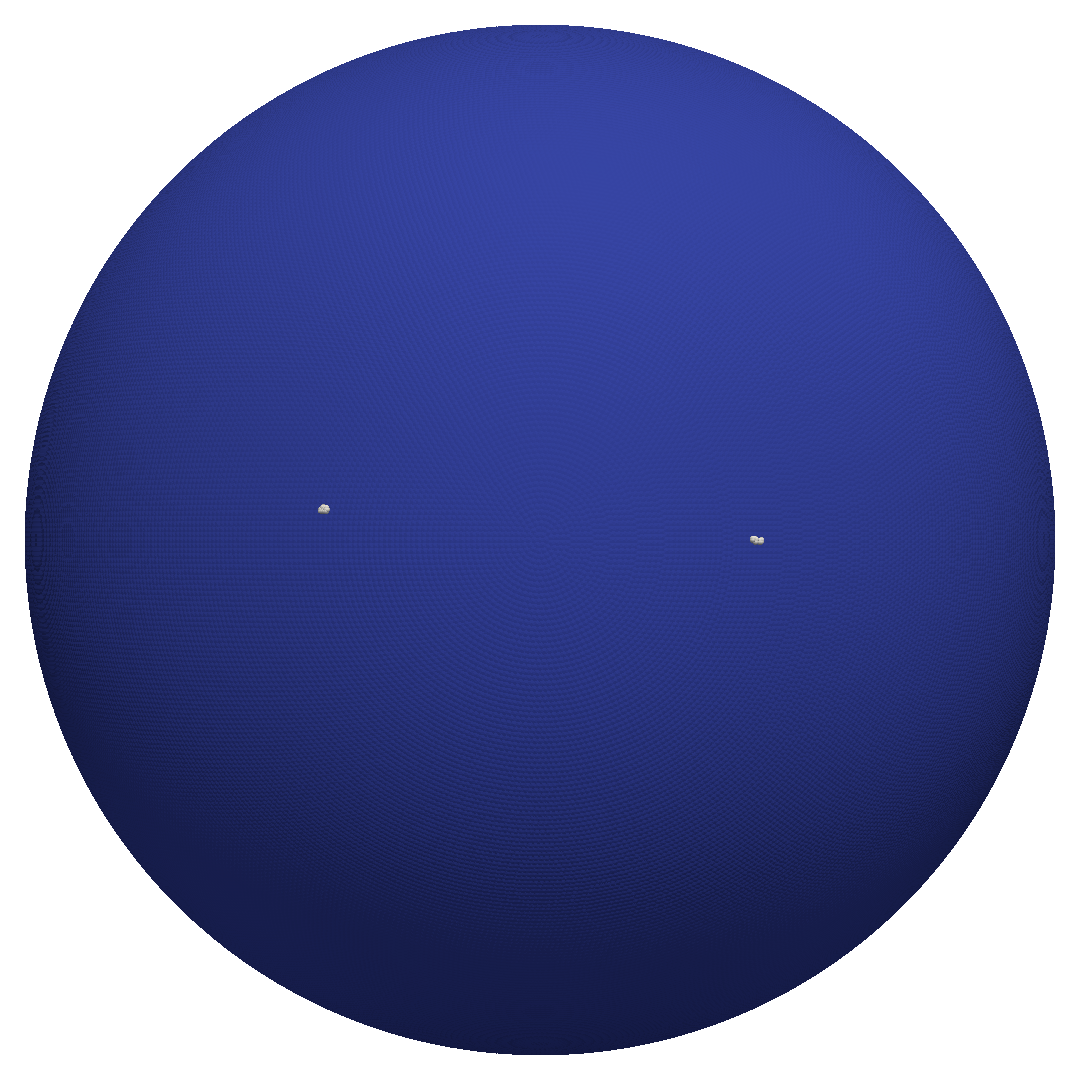}
\end{center}
\caption{
Left: A spherical grid with 12000 random events and two point-like sources. 
The gridpoints are rendered with a radius of about 0.5$^{\circ}$, hence the overlap and form a closed sphere. 
View on the equator (declination of 0$^{\circ}$). 
Middle: View from below to the south pole (declination of -90$^{\circ}$). 
Right: The same setup displayed without the random events.
Since the sphere is a three dimensional model, only the part facing the observer is visible.}
\label{fig:sourceSetup}
\end{figure}
\begin{figure}[h!]
\begin{center}
\includegraphics[width=0.3\textwidth]{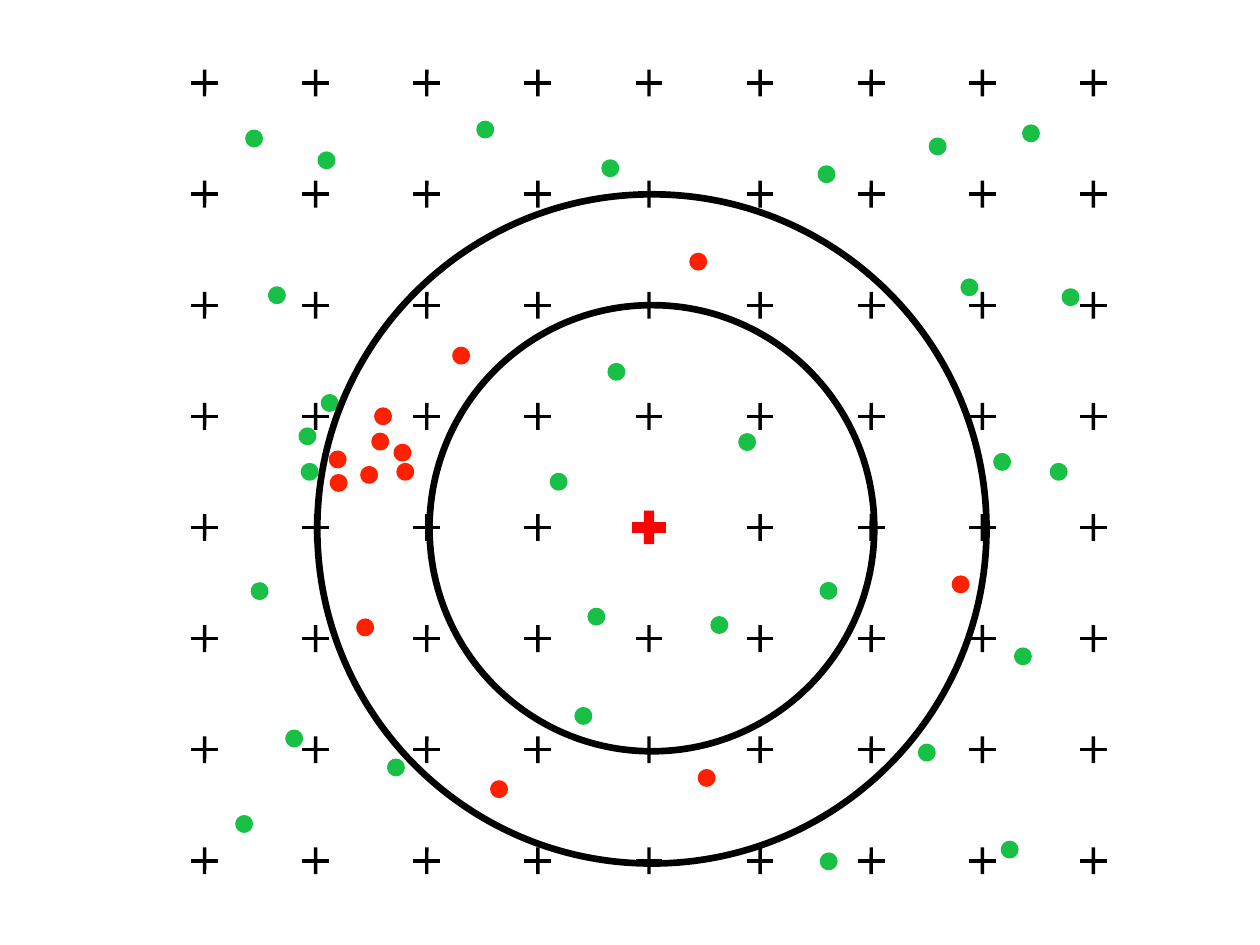}
\caption{Scheme of the neutrino counting. Crosses mark the gridpoints with a distance of 0.5$^{\circ}$ between them. 
Green and red dots are neutrinos. The red cross is the gridpoint that is being evaluated. 
The current search scale is between the black circles. It is 1.0$^{\circ}$ (inner circle) 
to 1.5$^{\circ}$ (outer circle) in this example. Neutrinos which are found for the current search 
scale at the current searchpoint are shown in red. 
The result of the evaluation of this scale at the red gridpoint is 13.}
\label{fig:searchRingScheme}
\end{center}
\end{figure}

\begin{figure}[h!]
\includegraphics[width=0.25\textwidth]{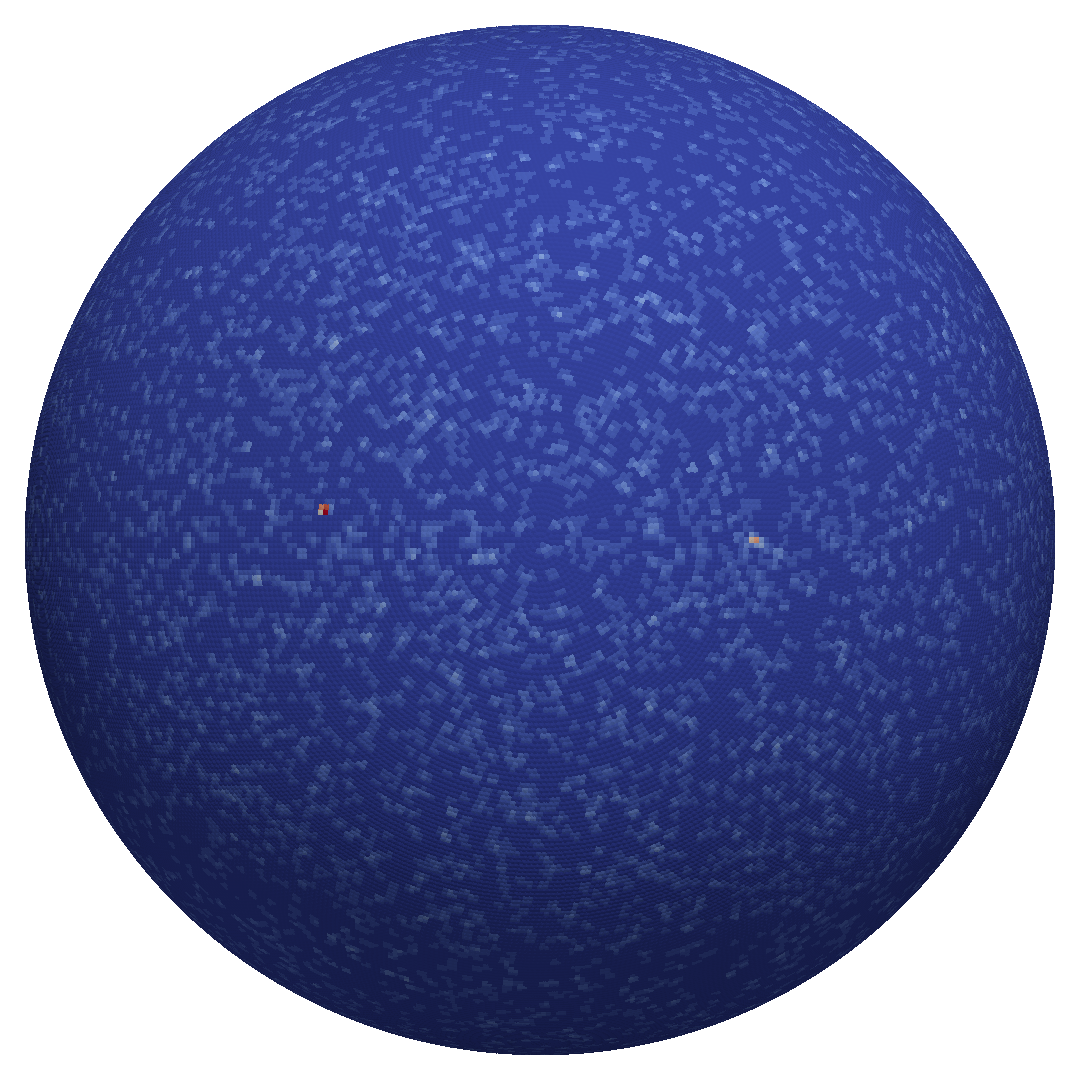}
\hfill
\includegraphics[width=0.25\textwidth]{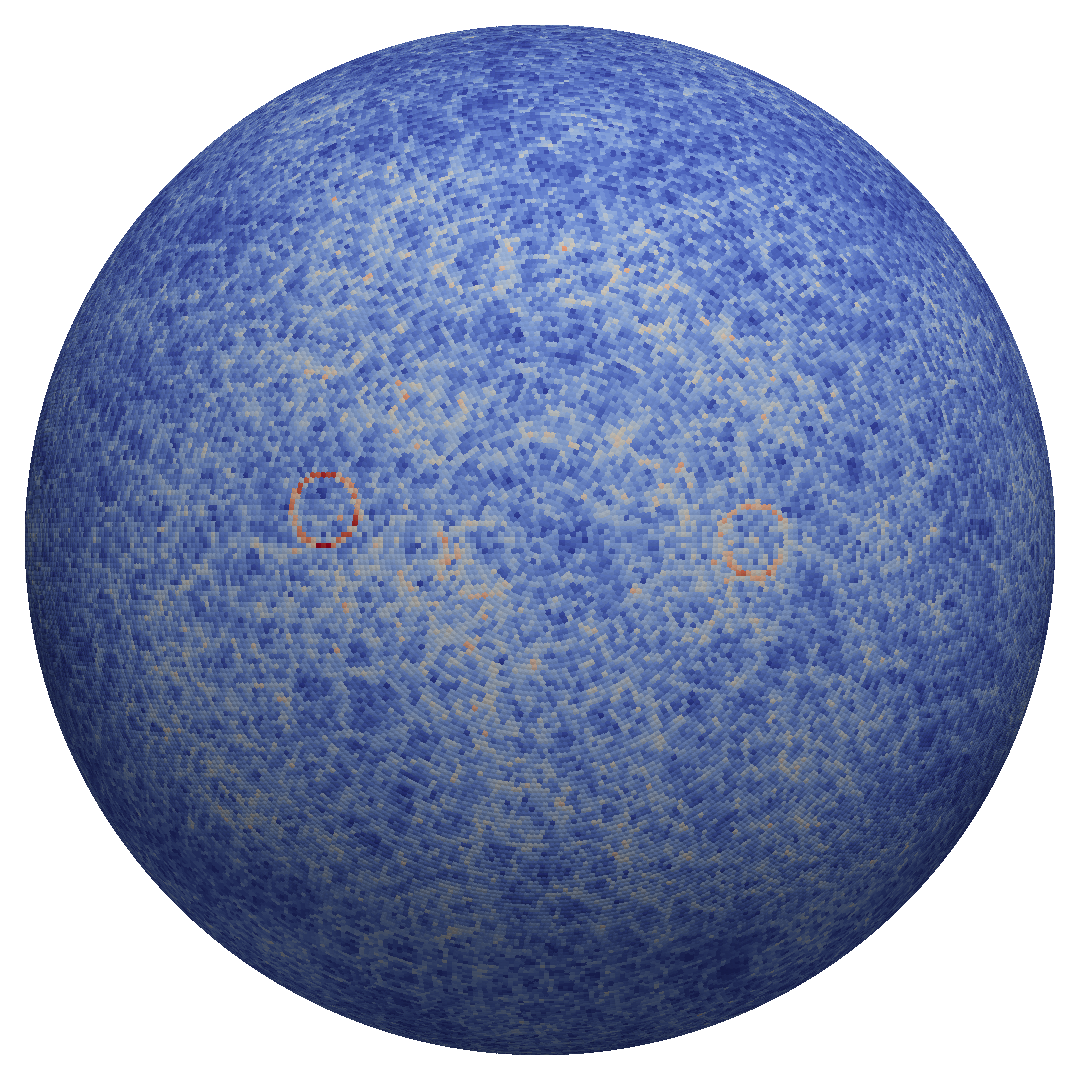}
\hfill
\includegraphics[width=0.25\textwidth]{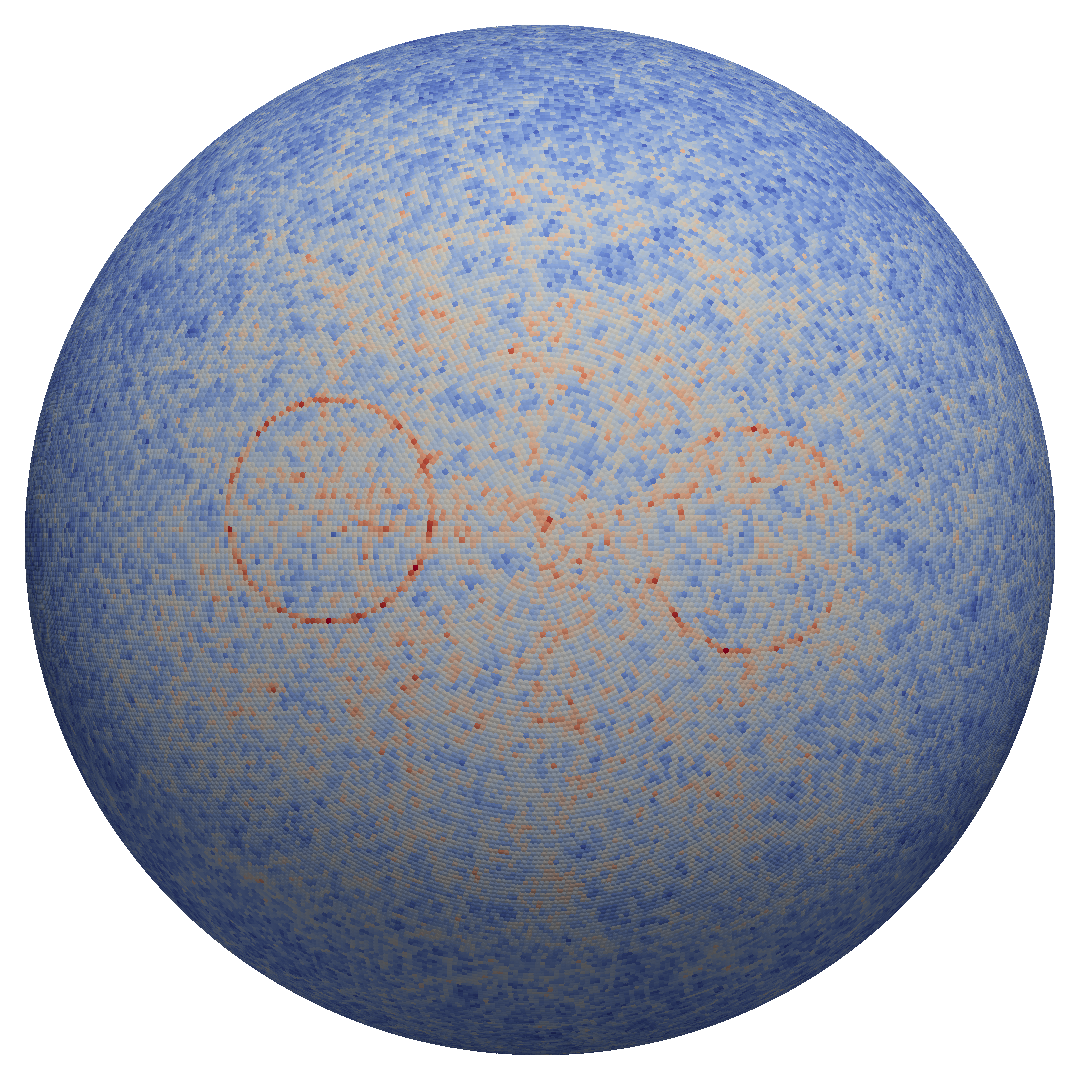}
\caption{a) The spherical search grid with the number of counted events 
in a circle between 0.0$^{\circ}$ and 0.5$^{\circ}$ around each gridpoint. 
b) Number of events between 3.0$^{\circ}$ and 3.5$^{\circ}$. c) Number of events between 10.0$^{\circ}$ and 10.5$^{\circ}$.
The color scale is readjusted between the different scales.}
\label{fig:counting}
\end{figure}

\newpage
\begin{figure}[h!]
\includegraphics[width=0.3\textwidth]{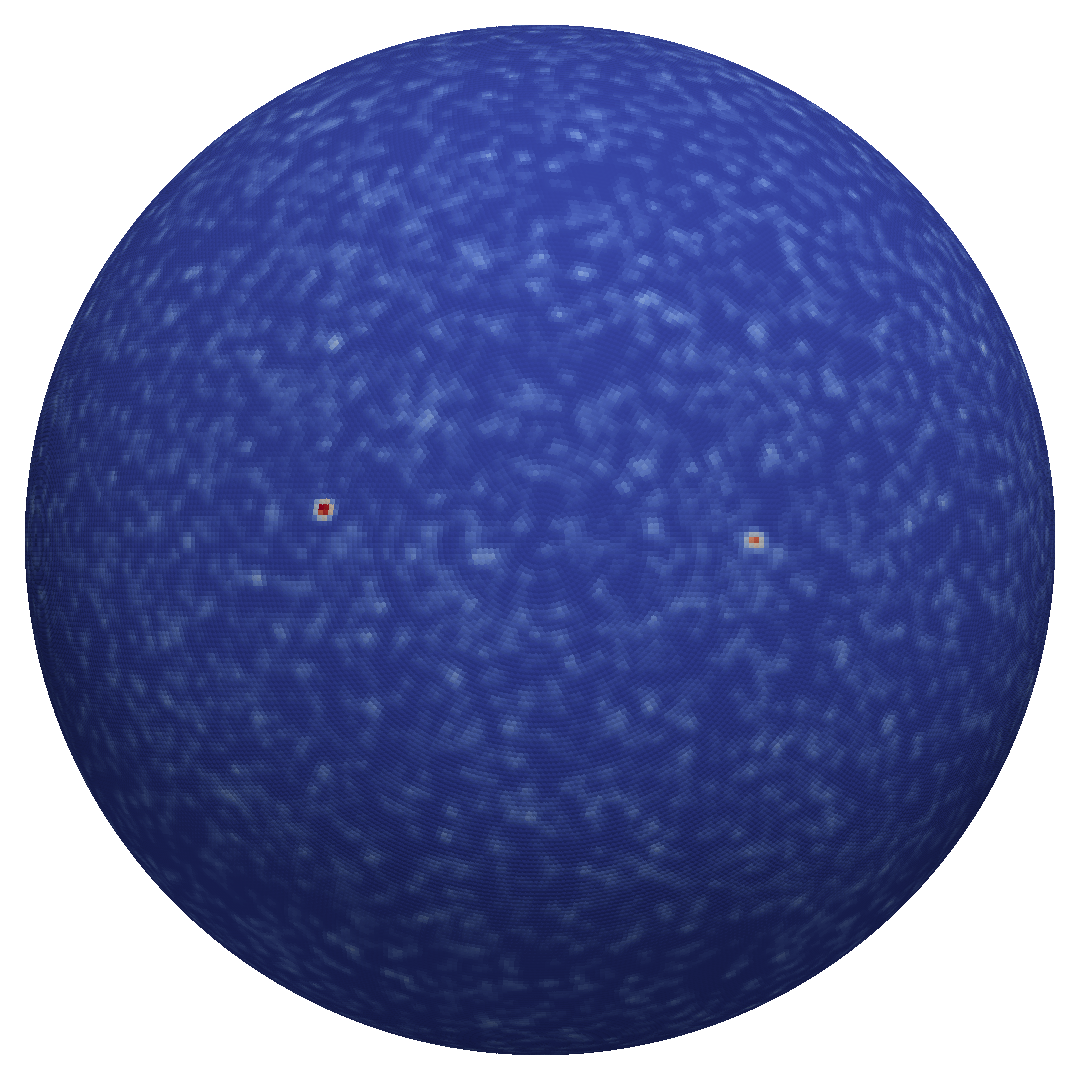}
\hfill
\includegraphics[width=0.3\textwidth]{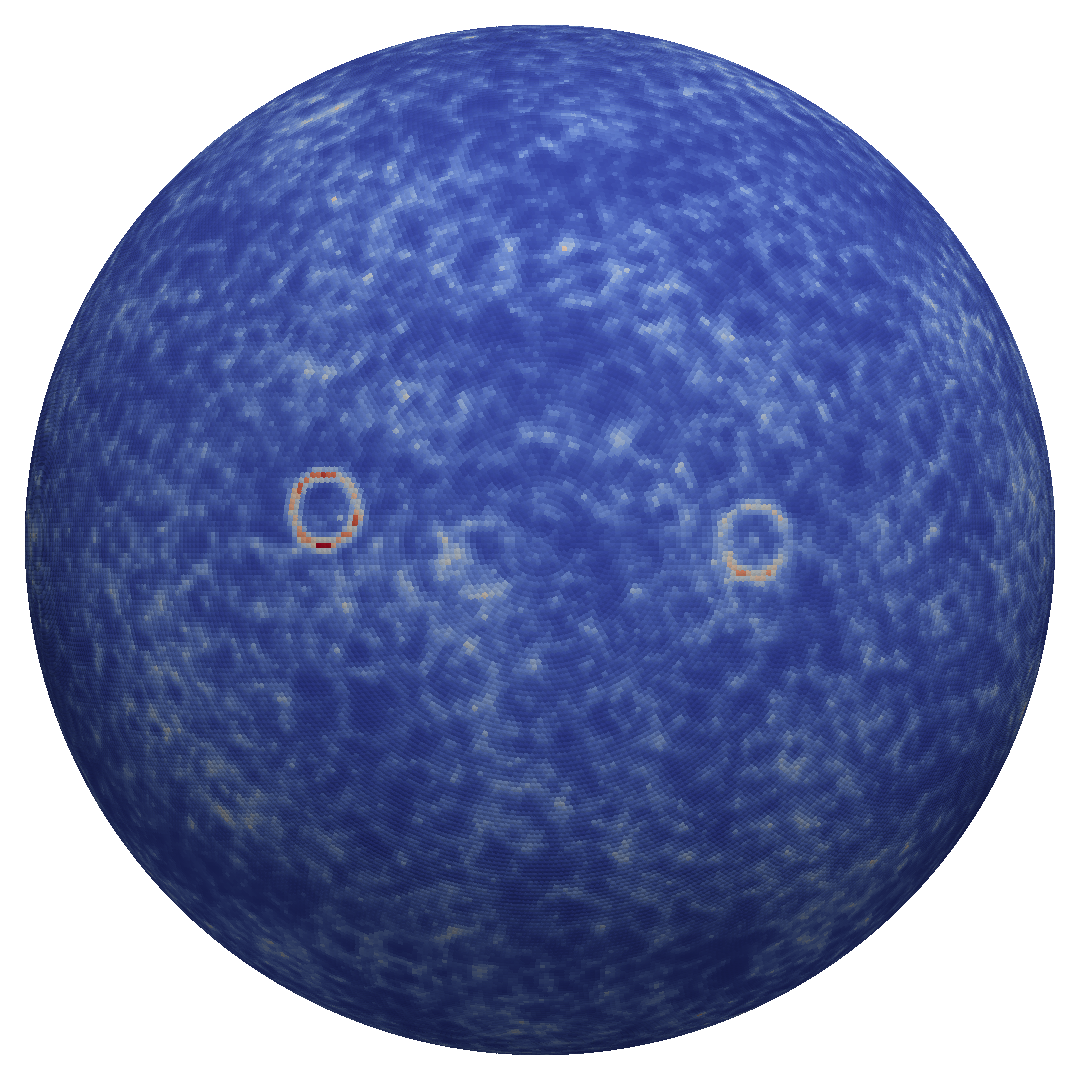}
\hfill
\includegraphics[width=0.3\textwidth]{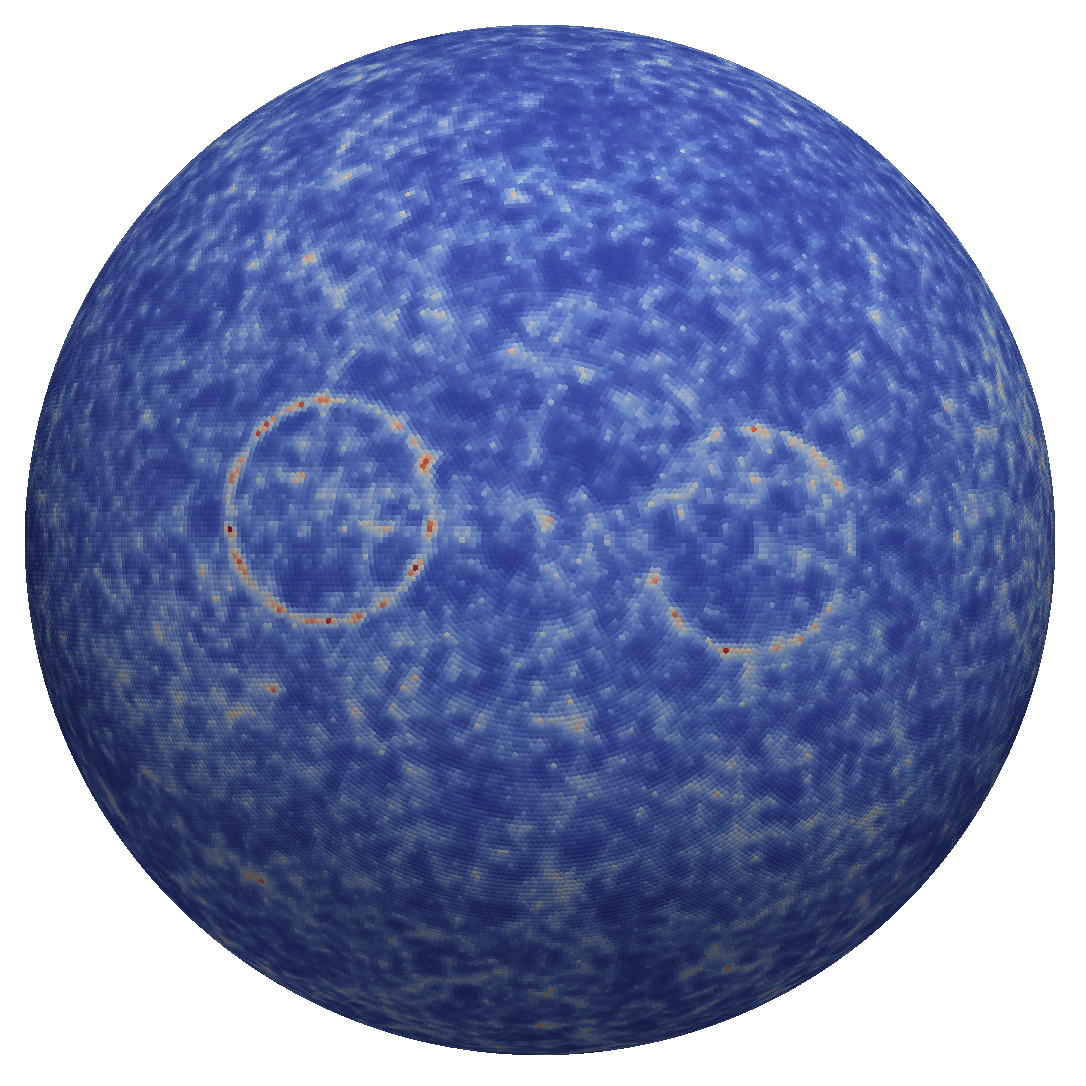}
\caption{Left: The spherical grid for the scale 0.0$^{\circ}$ to 0.5$^{\circ}$ after computation of the Poisson probabilities. 
Middle: 3.0$^{\circ}$ to 3.5$^{\circ}$. 
Right: 10.0$^{\circ}$ to 10.5$^{\circ}$. 
}
\label{fig:countingToPoisson}
\end{figure}
\begin{figure}[h!]
\begin{center}
\includegraphics[width=0.4\textwidth]{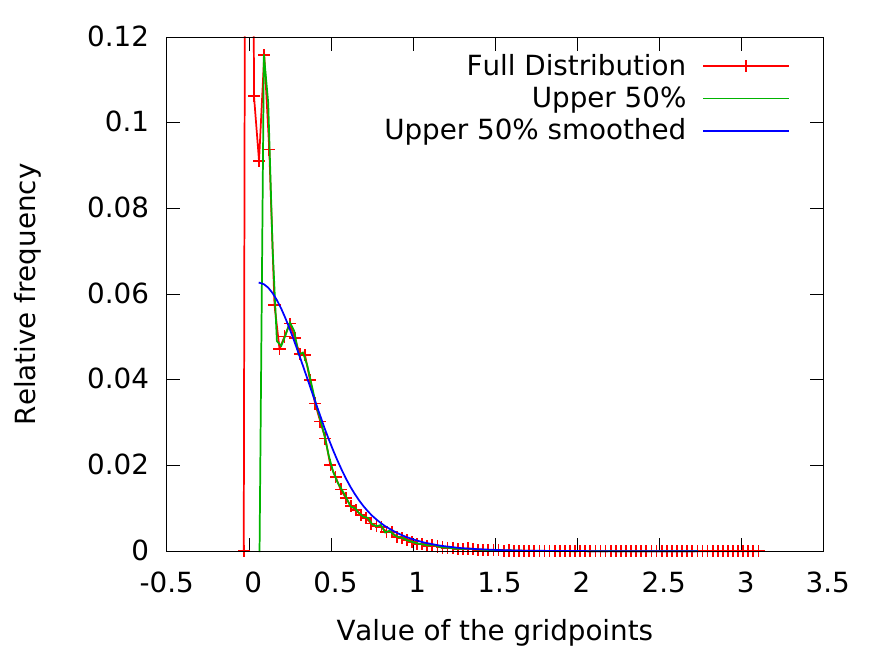}
\includegraphics[width=0.4\textwidth]{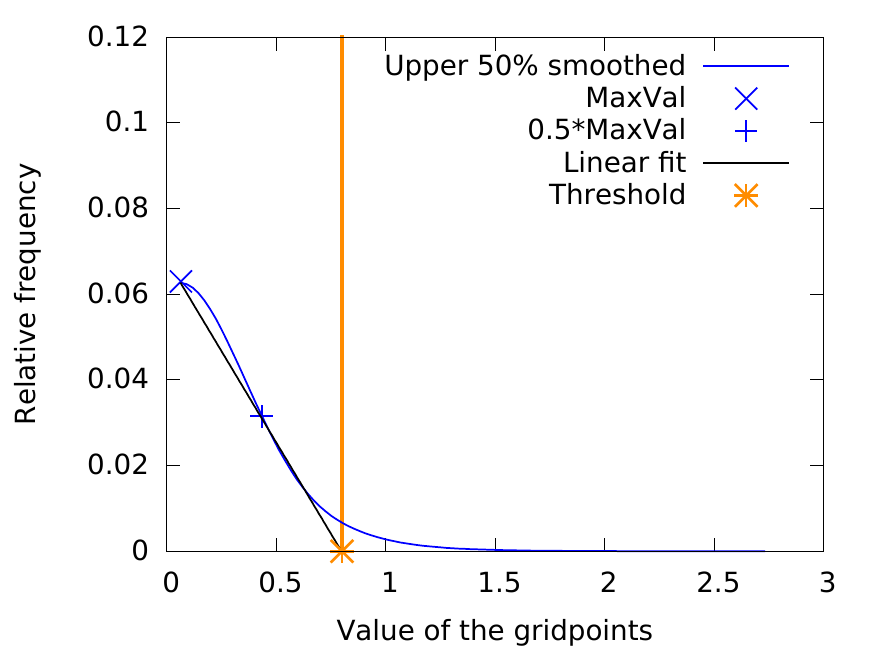}
\end{center}
\caption{Left: The histogram of R values in red and various intermediate steps 
of the threshold computation. Right: Various intermediate steps of the threshold 
computation and the final threshold in orange.}
\label{fig:threshold}
\end{figure}
\begin{figure}[h!]
\includegraphics[width=0.3\textwidth]{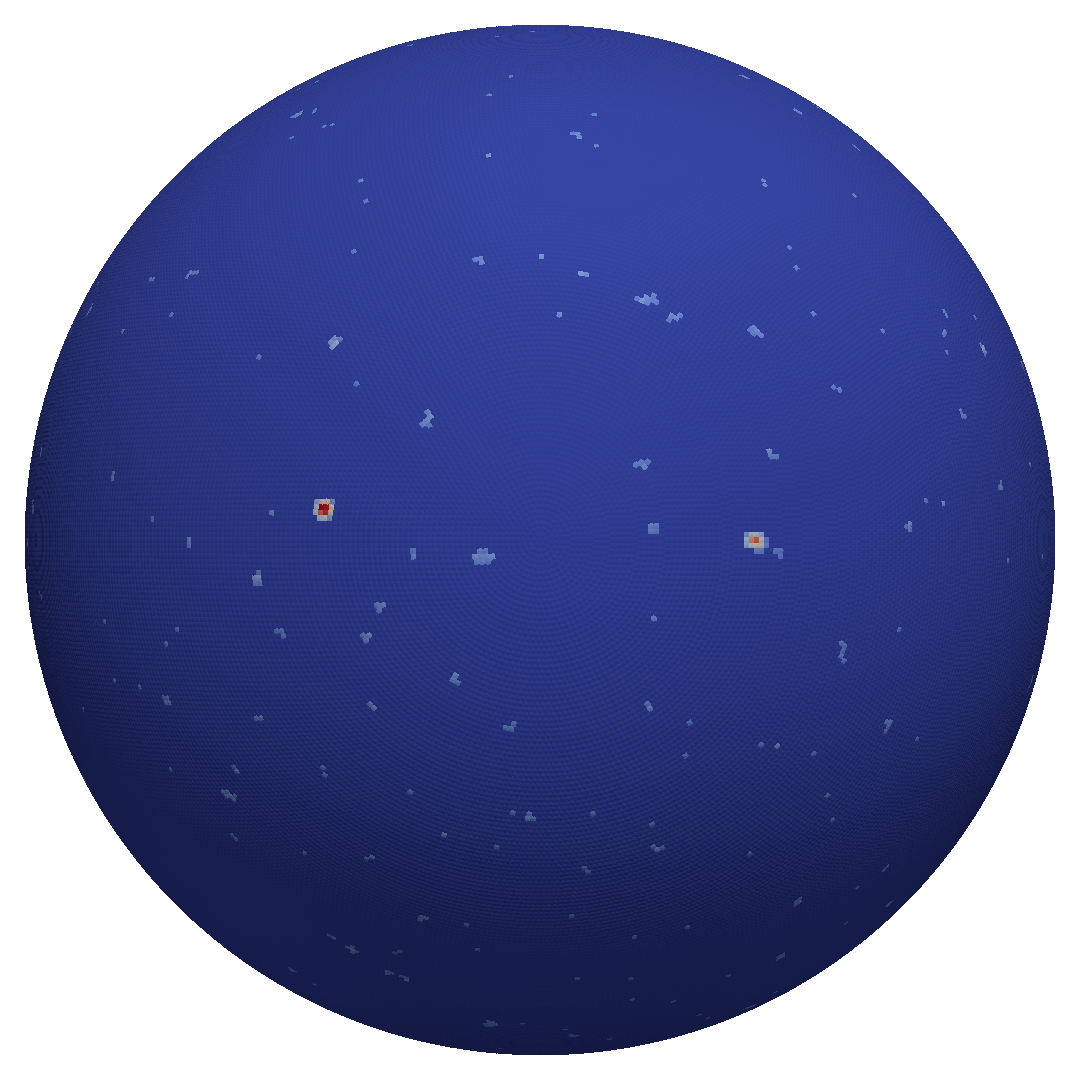}
\hfill
\includegraphics[width=0.3\textwidth]{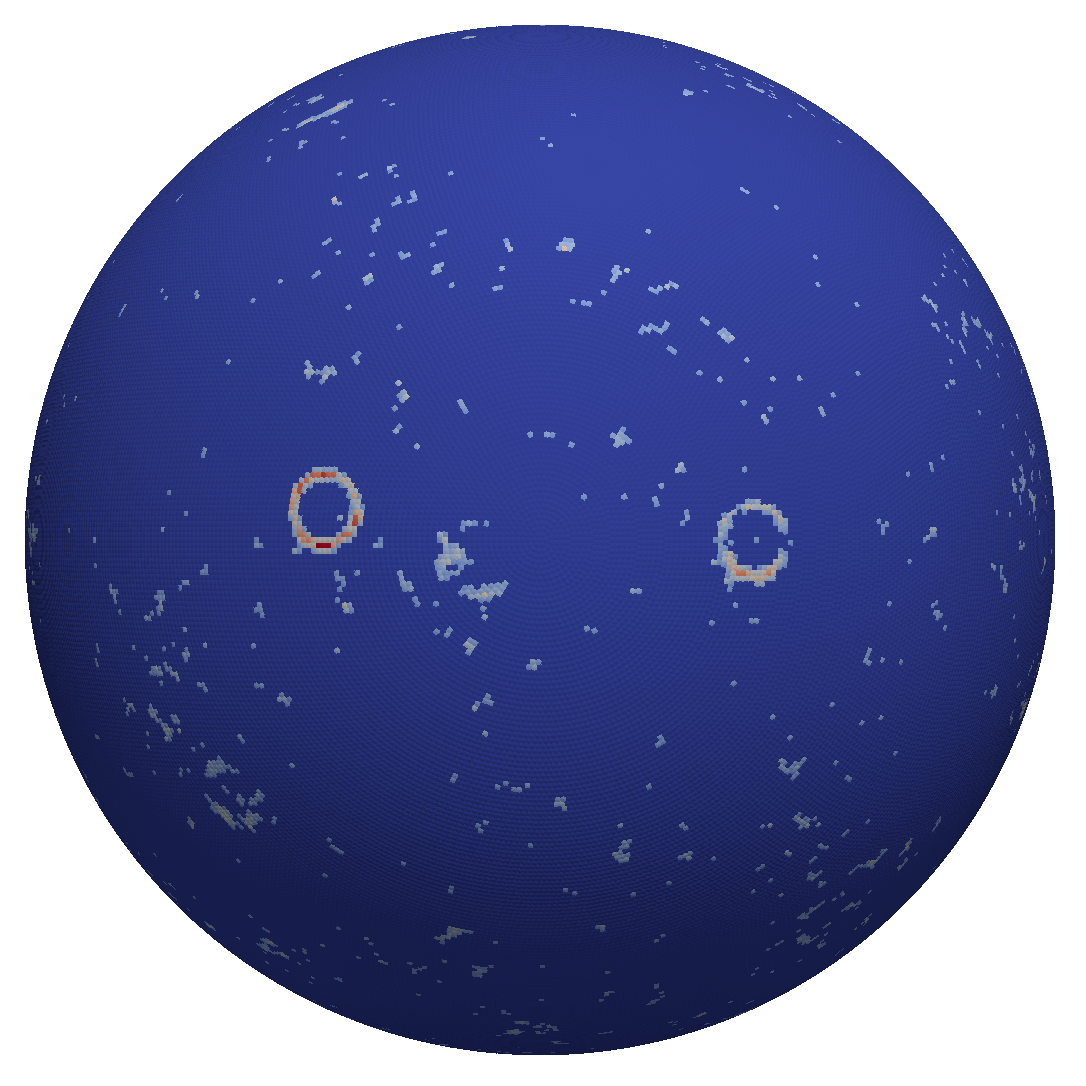}
\hfill
\includegraphics[width=0.3\textwidth]{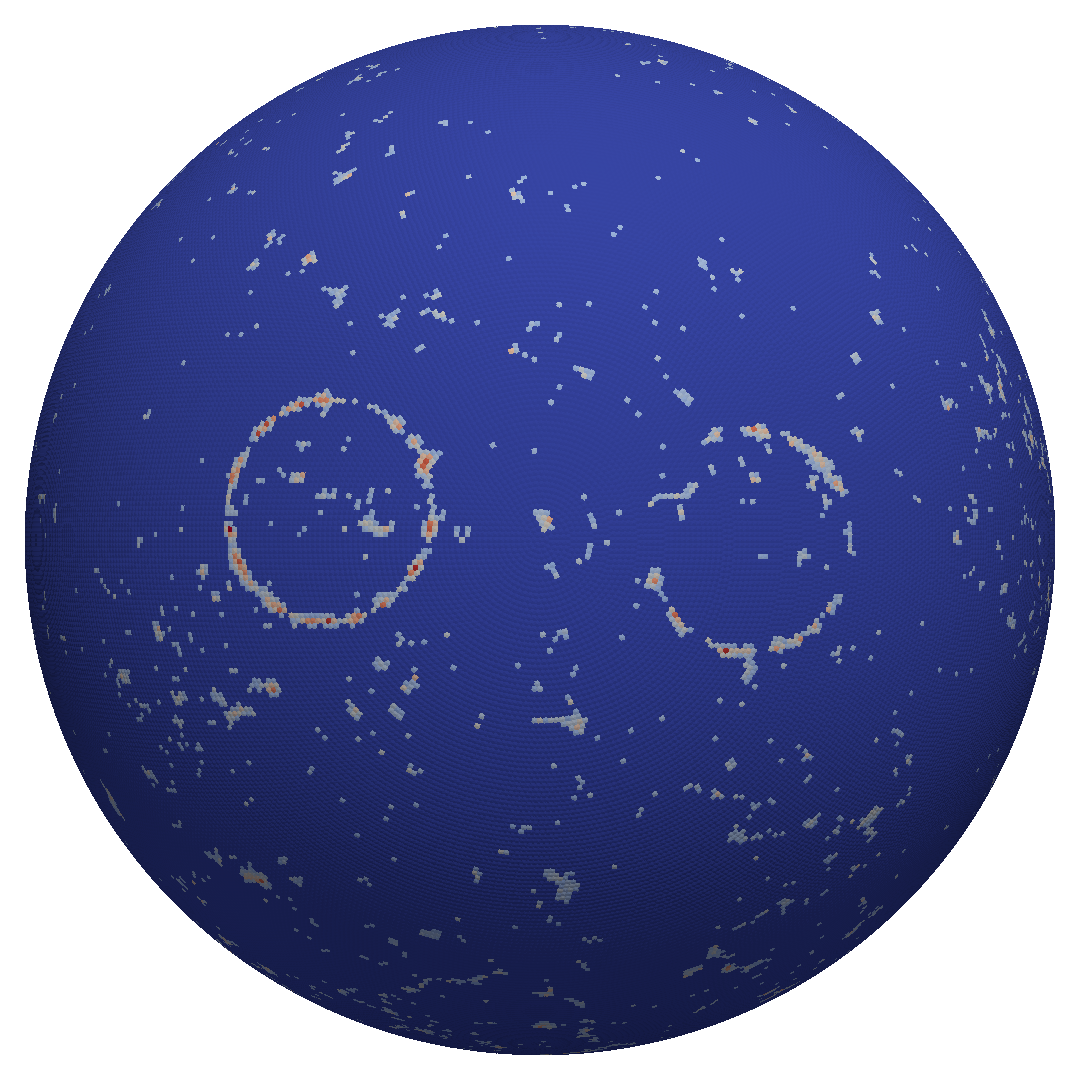}
\caption{Left: The spherical grid for the scale 0.0$^{\circ}$ to 0.5$^{\circ}$ after segmentation. 
Middle: 3.0$^{\circ}$ to 3.5$^{\circ}$. 
Right: 10.0$^{\circ}$ to 10.5$^{\circ}$.}
\label{fig:PoissonToSegment}
\end{figure}

\newpage

The result after the segmentation is shown in Figure \ref{fig:PoissonToSegment}. 

The next step is to reconstruct the original location of the neutrinos that caused the detected overfluctuations. 
For the search sphere with a search distance between 0.0$^{\circ}$ 
and 0.5$^{\circ}$ nothing changes, since the neutrinos have been counted at the location where the information is stored. 
For all other scales $d > 0.5^{\circ}$, the information stored at a gridpoint 
originated from counting neutrinos that are $d$ degrees away.
To achieve the remapping of the information to the original location a second grid is initialized with 0.0 values.
For each gridpoint $p$ in the original grid, the set $p_{d}$ of all gridpoints at a distance $d$ around it is computed.
For each of the gridpoints within this set, the mean contribution of a gridpoint at this distance to the observed value 
$\frac{R_{p}}{size(p_{d})}$
is added (in the new grid). 
Afterwards the new grid contains the corresponding fractions of the overfluctuations mapped back to their origin 
and this grid is used from there on.
The result of these computations is shown in Figure \ref{fig:SegmentToRemap}. 
The information where the neutrino distribution had a higher density is automatically encoded in the pattern how 
the remapped circles around the old gridpoints overlap in the new grid, see middle and right of Figure \ref{fig:SegmentToRemap}. 
\begin{figure}[ht]
\includegraphics[width=0.32\textwidth]{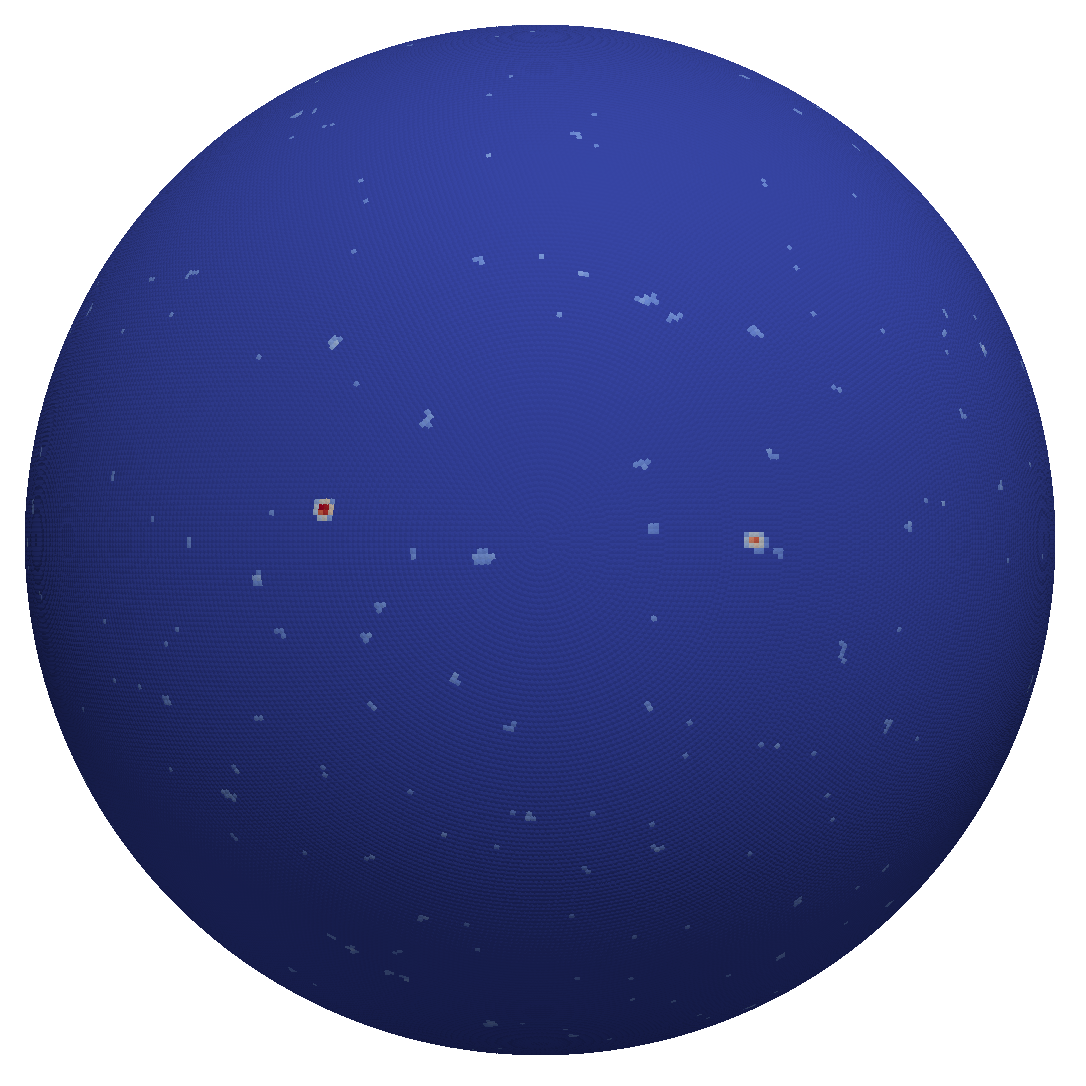}
\hfill
\includegraphics[width=0.32\textwidth]{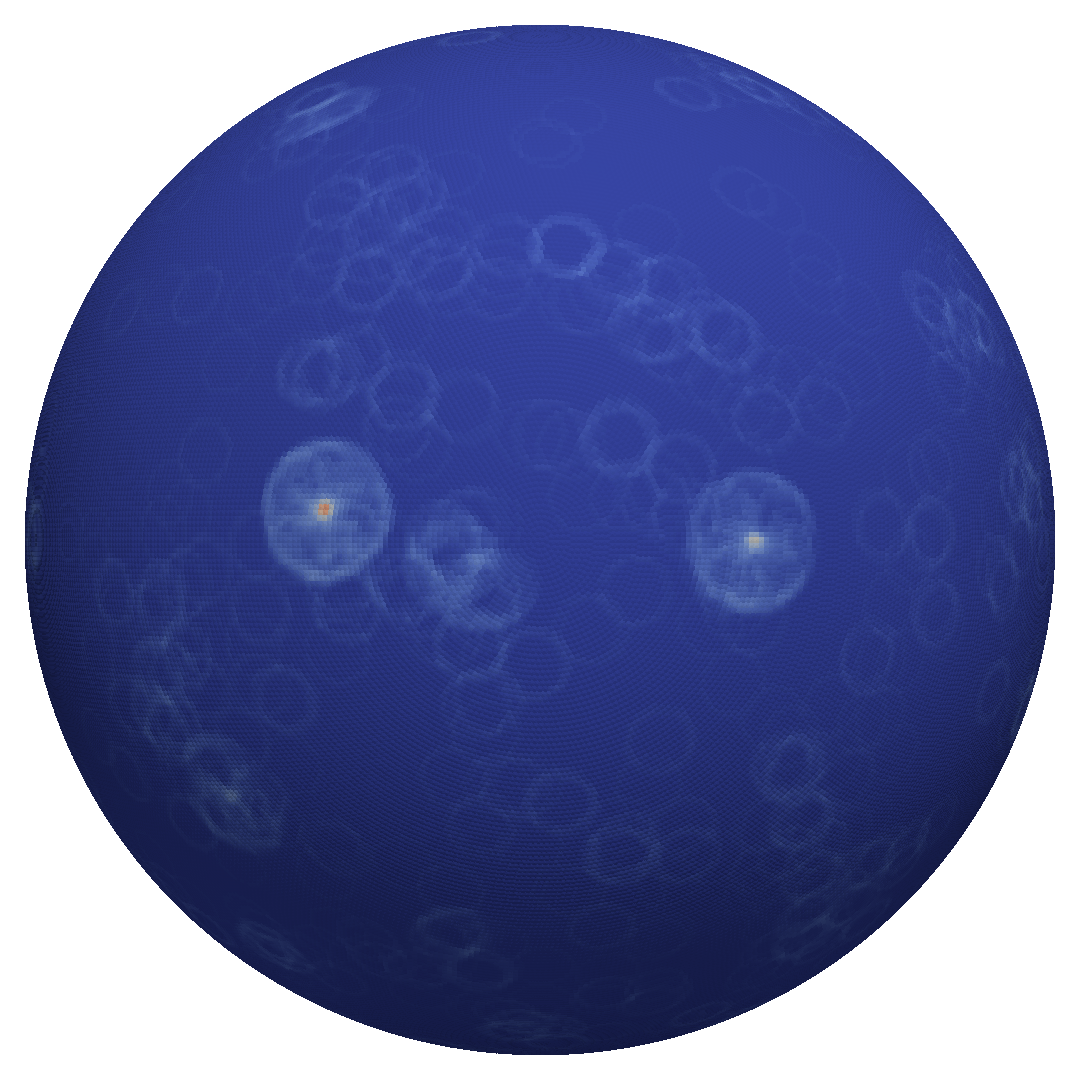}
\hfill
\includegraphics[width=0.32\textwidth]{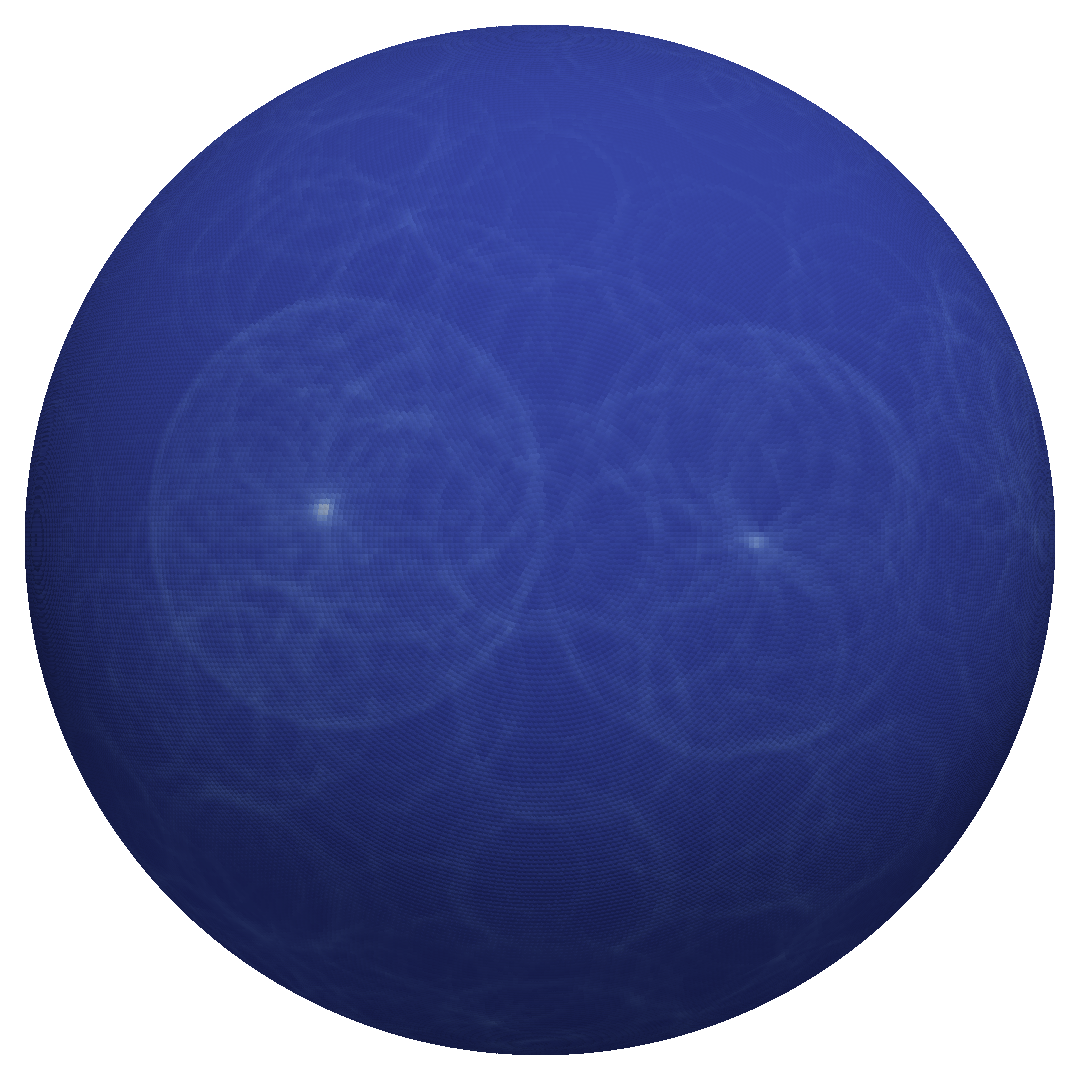}
\caption{Left: The spherical grid for the scale 0.0$^{\circ}$ to 0.5$^{\circ}$ after the remapping. 
Middle: 3.0$^{\circ}$ to 3.5$^{\circ}$. 
Right: 10.0$^{\circ}$ to 10.5$^{\circ}$.}
\label{fig:SegmentToRemap}
\end{figure}
In order to evaluate the 180 different search scales they have to be combined in some way.
The best successfully developed robust solution turned out to be taking the sum of all scales. 
It should be noted that there is the potential to exploit the available information better, 
for instance by an individual evaluation of the 180 spheres 
and a more sophisticated combination of the information derived from each. 
The result of the summation can be seen in Figure \ref{fig:sumSphere}. 

\begin{figure}[ht]
 \begin{center}
   \includegraphics[width=0.3\textwidth]{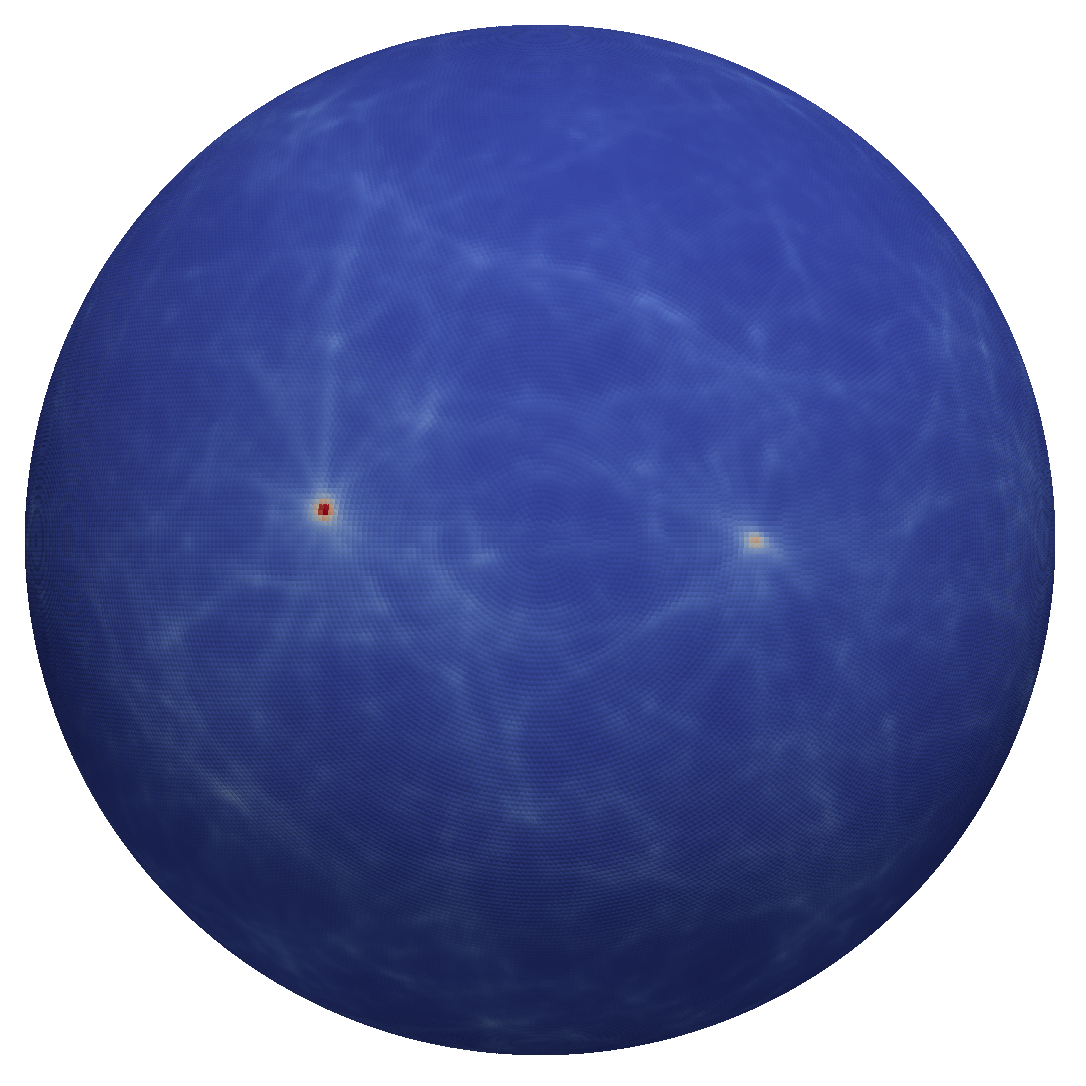}
 \end{center}
\caption{The sum of all 180 search scales. \bigskip}
\label{fig:sumSphere}
\end{figure}
Identifying connected regions with high values on the final sphere, 
which could be linked to possible neutrino source morphologies and strengths, 
is again achieved by a segmentation as already described for the individual scales. 
The same procedure is used, but this time with the 
additional option to obtain different thresholds by scaling the distance between the previous minimum value $x_{min\_old}$ 
and $\theta$, using a factor $\alpha$. The new threshold is then given by equation \ref{equ:thresholdAlpha}.
\begin{equation}
\label{equ:thresholdAlpha}
\theta_{final} = (\alpha \cdot x_{min\_old} + (1-\alpha) \cdot \theta)
\end{equation}
The effect of different thresholds for segmentation is shown in Figure \ref{fig:segmentationComparison}. 
High positive values for $\alpha$ allow larger extended source structures to be found, hard negative cuts only preserve the high peaks. 
Multiple values pronounce different aspects of the obtained result, 
but on the other hand they also increase the trial factor for the final result. 
By heuristic optimization based on a variety of simulated sources, the values for the 
segmentation have been fixed to $\alpha$ = 0.25 and $\alpha$ = -0.11, 
since a single value cannot cover the targeted range of sources. 

The gridpoints of the resulting segmented grid are checked for connectedness. 
A connected group of gridpoints is called a cluster. 
The next step is to distinguish potentially significant clusters from random accumulations. 
To achieve this one needs to know the probability how likely a cluster could have been generated by random events. 
This probability could in theory be determined by pseudo-experiments using scrambled data. 
A specific cluster shape, position and composition is unlikely to be reproduced, therefore the analysis
must rely on more generic attributes to evaluate the significance of a cluster. 
For instance one can compute the probability for a cluster of the same 
size or larger, with size measured by the number of gridpoints. 
Only considering size for a relevance measurement is not sensitive to smaller or even point-like sources. 
A better metric to find these is for example the maximal value of any gridpoint in the cluster. 
Many others metrics have been tested, each with different sensitivities to different source characteristics. 
But if many relevance metrics are used, also a large trial factor has to be considered. 
Since this search is not intended for a specific source model, 
the optimized selection had to be done heuristically with a multitude of simulated sources. 
To specifically detect point sources one would use e.g. the maximal value within a cluster. 
Justified by the fact that ANTARES has already conducted specialized searches for promising small and point-like sources, 
the metric size in gridpoints $N$ has been chosen, performing best for large, extended source morphologies. 
Due to the increased trial factor that comes with more metrics, the second best, 
the mean value of the $\sqrt{N}$ highest pixels within a cluster, is not included. 
The significance for each cluster is then derived from pseudo-experiments with scrambled data. 

\begin{figure}[hb]
\includegraphics[width=0.32\textwidth]{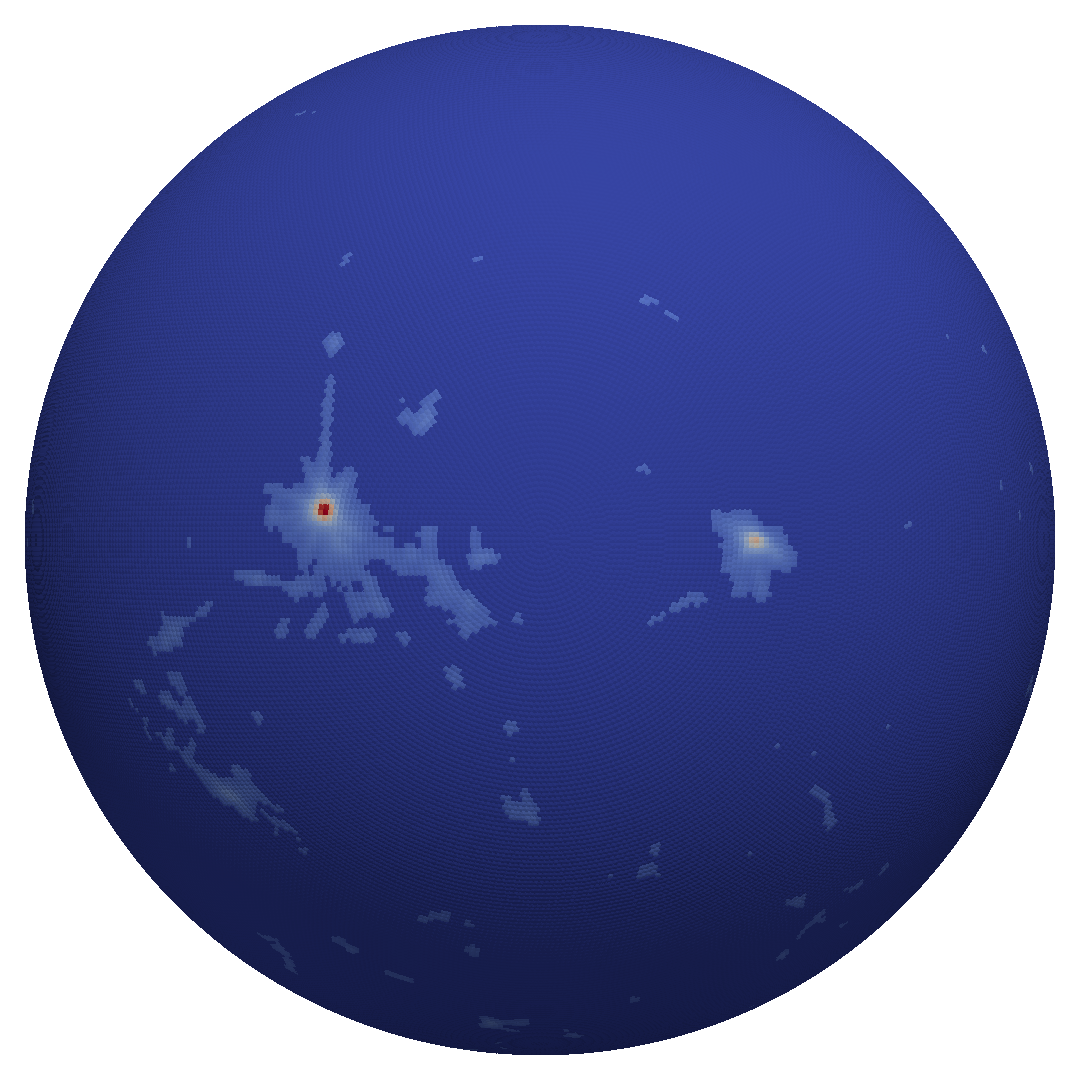}
\hfill
\includegraphics[width=0.32\textwidth]{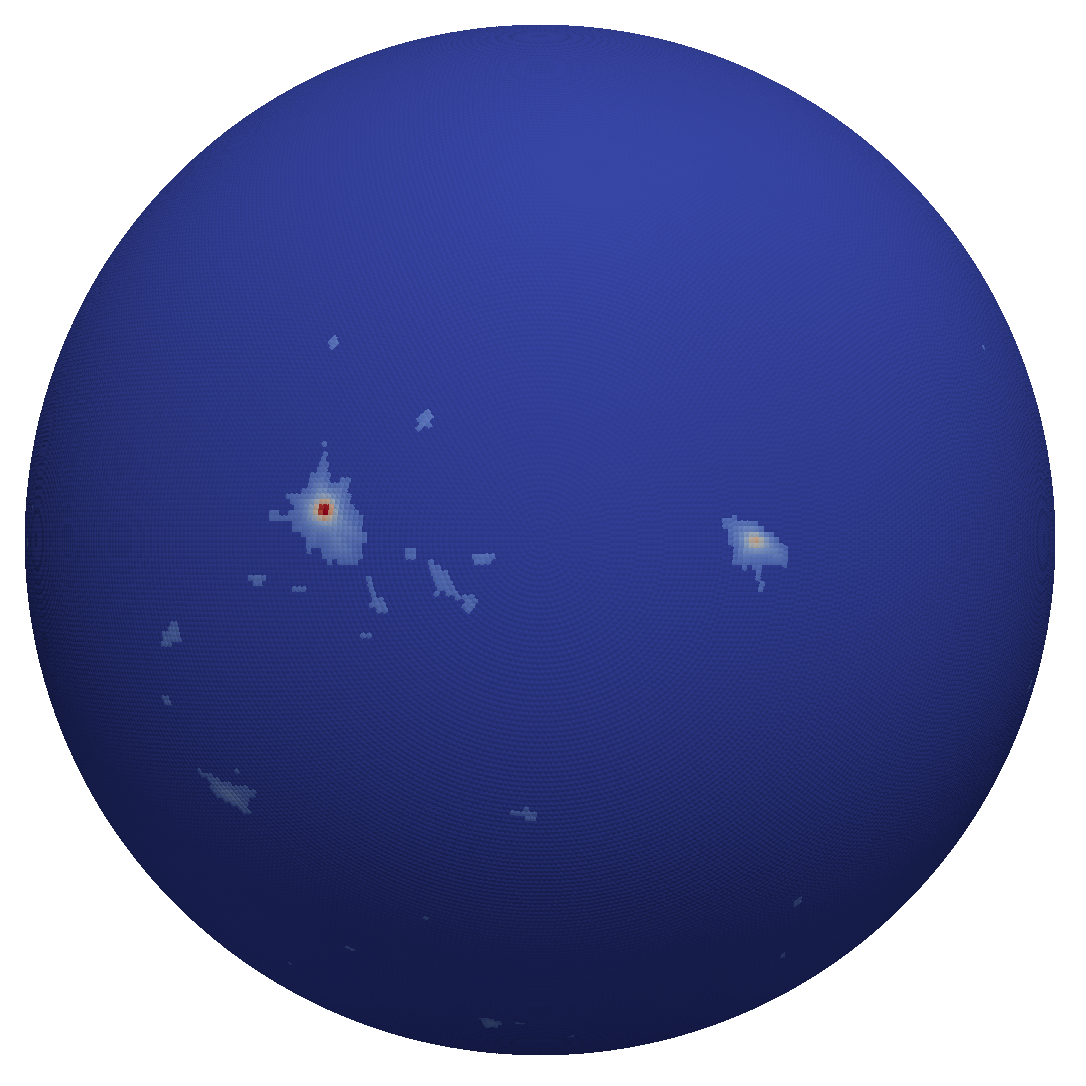}
\hfill
\includegraphics[width=0.32\textwidth]{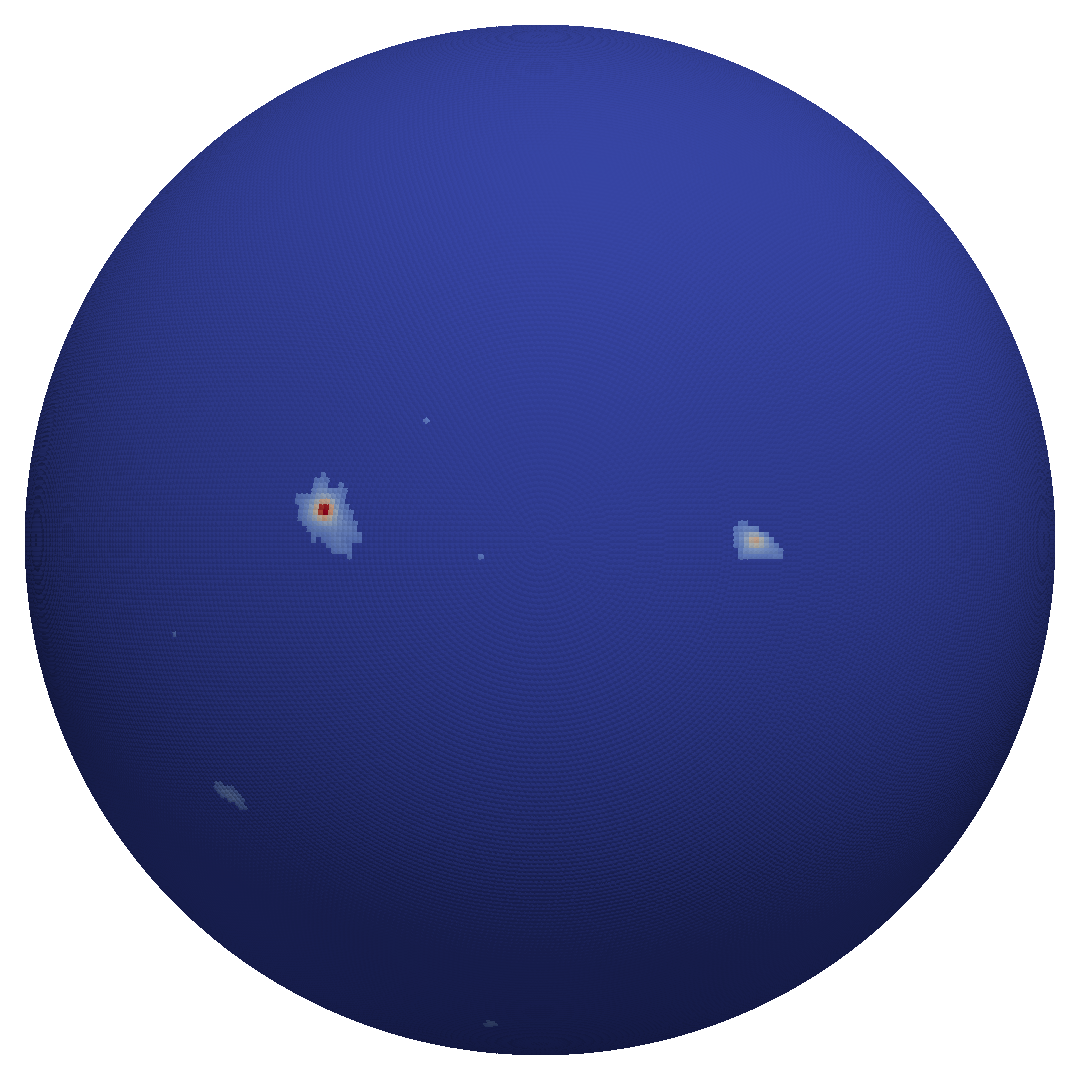}
\caption{The effect of different $\alpha$ values on the segmentation. 
Left: +0.25, Middle: +0.05, Right: -0.1.}
\label{fig:segmentationComparison}
\end{figure}

\section{Selectfit}
\label{sec:selectfit}

A new method for the direction reconstruction of events detected 
by the ANTARES neutrino telescope called "SelectFit" is introduced here.
Instead of using one reconstruction algorithm for the whole sample, Selectfit combines the results of multiple available 
reconstruction algorithms, trying to select the reconstruction algorithm for each event, 
which gives the most precise result. 
This selection is done by a machine learning technique called ``Random decision forest'' (RDF) \cite{ho:RDFs}. 
Selectfit decreases the angular reconstruction uncertainty of a sample of neutrinos or 
allows to increase the sample size for a fixed angular uncertainty, 
as illustrated in Figure \ref{fig:selectfit}.
Since a search for extended objects does not need the same angular precision, 
which is required for a point source search, less strict quality cuts for the reconstruction can be applied. 
Together these two aspects increase the number of neutrino events 
in this analysis compared to the standard ANTARES point-source search. 

\begin{figure}[h]
 \begin{center}
   \includegraphics[width=0.5\textwidth]{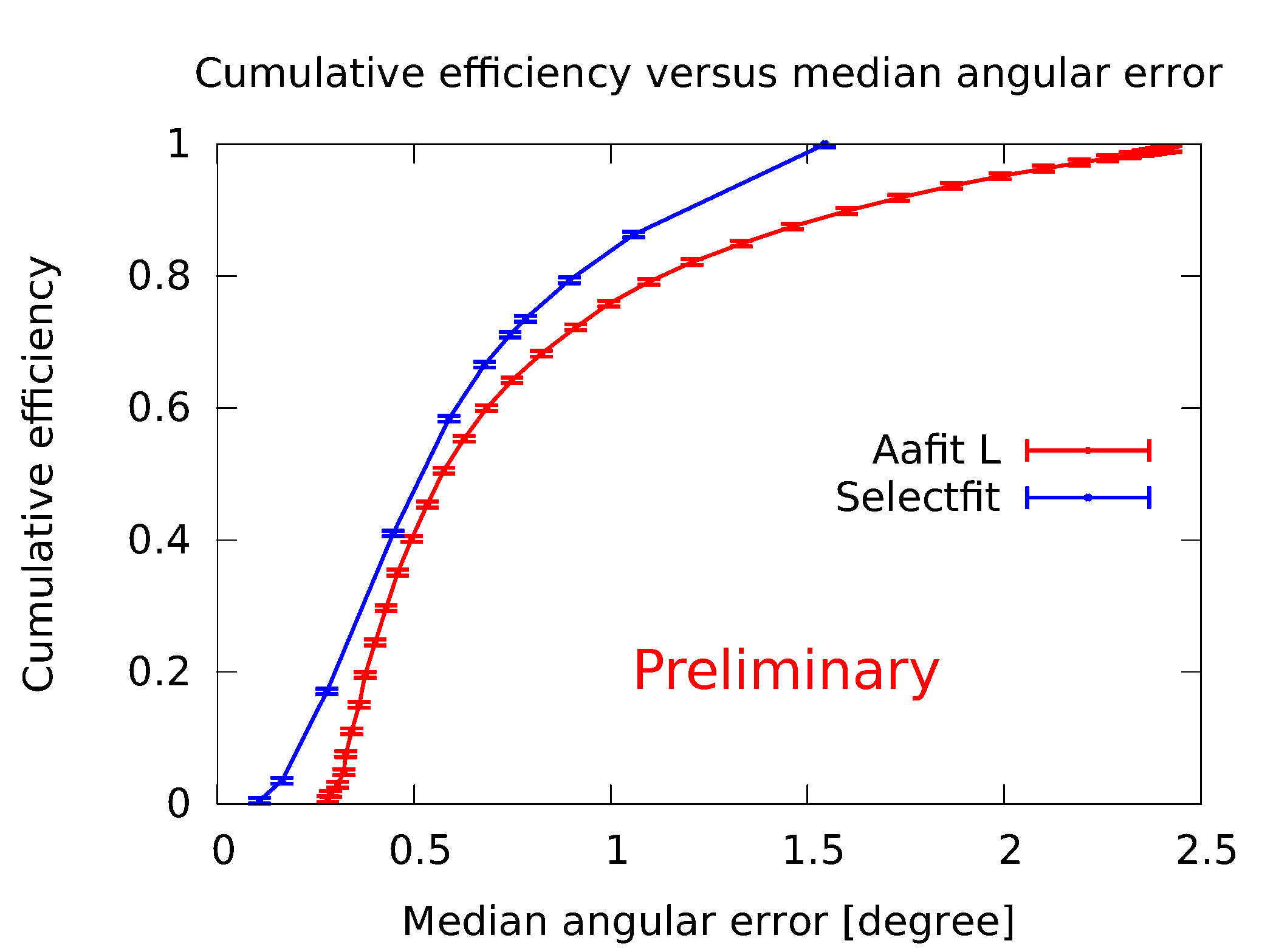}
 \end{center}
\caption{Comparison of the performance of the best individual direction reconstruction algorithm ``Aafit'' and ``Selectfit'',
the introduced method to efficiently combine multiple reconstruction algorithms.}
\label{fig:selectfit}
\end{figure}


\section{Results}

The unblinding of ANTARES data from 2007 to 2012 resulted in 13283 neutrino event candidates. 
The analysis of these events with the method explained in chapter \ref{sec:multiscale} 
yielded the preliminary results shown in Figure \ref{fig:resultANTARES}. 
Using the harder segmentation threshold $\alpha$ = -0.11 no cluster with a significance above 0.8$\sigma$ has been found. 
With $\alpha$ = 0.25 a very large structure is found.
A wide range of checks for systematic effects that could possibly influence the result has been performed, 
including for instance the small effect of time variations in the data taking efficiency on the event distributions. 
After accounting for all known systematic effects, 
the large structure has a post-trial significance of 2.5$\sigma$. 
It contains the galactic center, which is located in the center of skymaps in galactic coordinates.
More details how these structures have formed can be seen in Figure \ref{fig:detailedANTARES}, 
which shows the result of the summation of all scales before the segmentations. 
One has to keep in mind that the exact borders of these structures are certainly influenced by random fluctuation. 
\begin{figure}[h!]
\begin{center}
\includegraphics[width=0.425\textwidth]{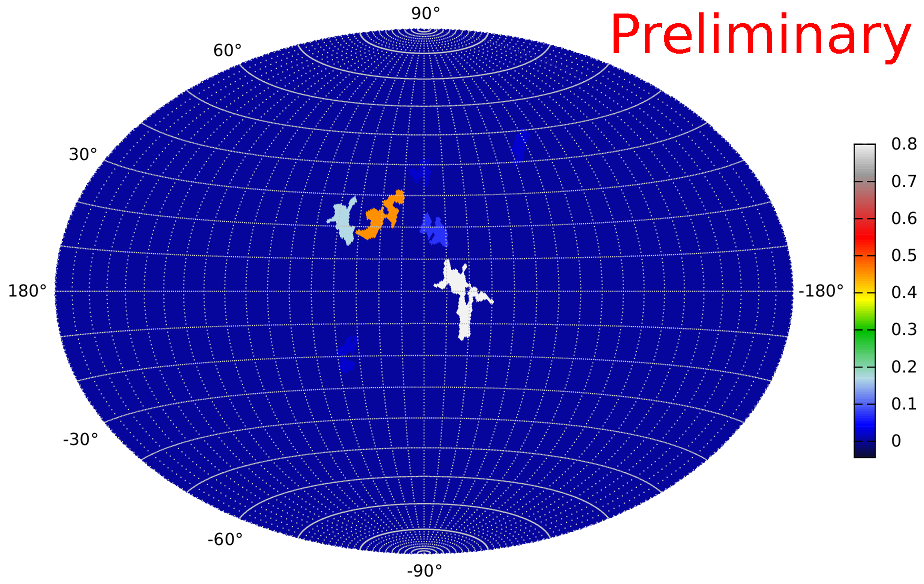}
\includegraphics[width=0.425\textwidth]{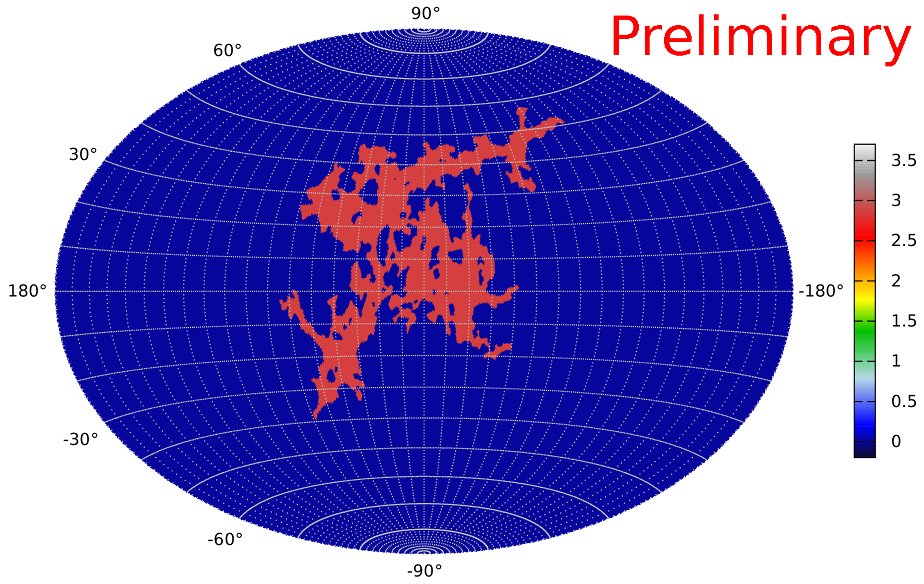}
\end{center}
\caption{Left: The result on ANTARES data with segmentation using $\alpha$ = -0.11 in galactic coordinates. 
Right: The result with $\alpha$ = +0.25. The color code of the clusters shows the significance in $\sigma$. 
}
\label{fig:resultANTARES}
\end{figure}

\begin{figure}[h!]
\begin{center}
\includegraphics[width=0.425\textwidth]{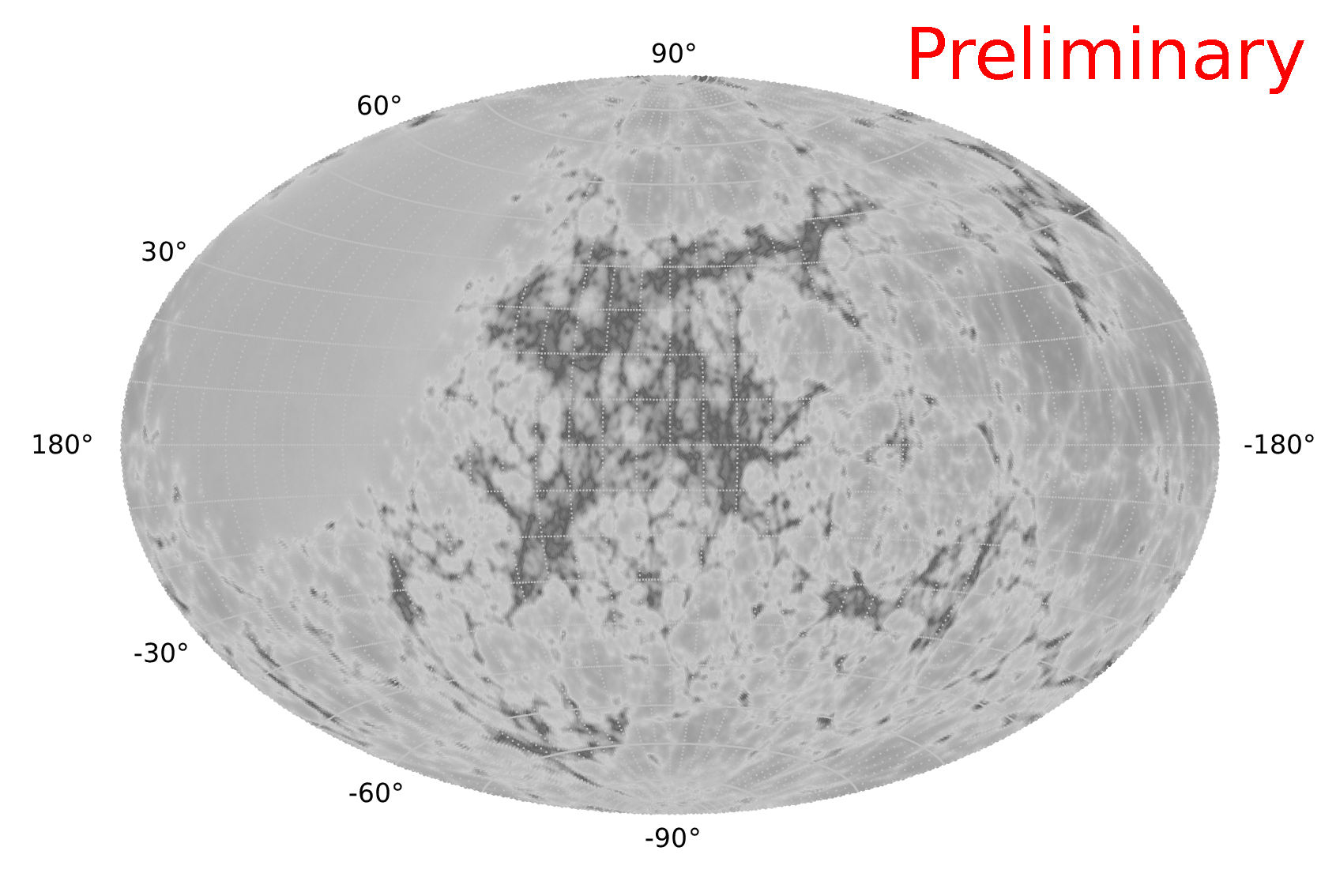}
\includegraphics[width=0.425\textwidth]{galacticFineGreyLightness.png}
\end{center}
\caption{Left: The detailed structure behind the result on ANTARES data 
in galactic coordinates before the segmentations in arbitrary units. 
Right: An example for the detailed structure for a random dataset with the same color code for comparison. 
The observed intensity of the overfluctuation is common, the clustering of their locations is rare.}
\label{fig:detailedANTARES}
\end{figure}

\newpage
\section{Conclusions}
We have devised a new analysis method that is able to detect sources 
of arbitrary location and morphology without relying on assumptions on the source properties. 
Applied to ANTARES data this method found a large structure with a post-trial significance of 2.5$\sigma$. 
This preliminary result is consistent with a random background fluctuation. 

Even though an unspecific, model-independent analysis is unlikely to obtain a result 
with high statistical significance due to the high trial factors, 
these searches provide a good way to become aware of the most interesting structures in data, 
which then may be worth further, more specific investigations.


\setcounter{figure}{0}
\setcounter{table}{0}
\setcounter{footnote}{0}
\setcounter{section}{0}
\setcounter{equation}{0}

\newpage
\id{id_gracia}
\addcontentsline{toc}{part}{\textcolor{blue}{\arabic{IdContrib} - {\sl R. Gracia} : Search for signal emission from unresolved point sources with the ANTARES neutrino telescope
}%
}

\title{\arabic{IdContrib} - Search for signal emission from unresolved point sources with the ANTARES neutrino telescope}

\shorttitle{\arabic{IdContrib} - Search for unresolved point sources}

\authors{Rodrigo Gracia Ruiz} 
\afiliations{        APC, Universit\'e Paris Diderot, CNRS/IN2P3, CEA/Irfu, Observatoire de Paris, Sorbonne Paris Cit\'e, 10 rue Alice Domon et L\'eonie Duquet, 7205 Paris Cedex 13, France}
\email{rgracia@in2p3.fr}


\abstract{A two point correlation analysis is used to search for inhomogeneities in the arrival directions of the high energy muon neutrino candidates detected by the ANTARES neutrino telescope. This approach is complementary to a point source likelihood-based search, which is mainly sensitive to point like sources and not to collective effects. We present the results of a search based on this two-point correlation method performed on ANTARES 2007-2012 data, providing constraints on models of a population of point sources too faint to be detected by the searches for point like sources.}

%
%
\maketitle
\section{Introduction}
The origin of cosmic rays (CR) is still an open question that can only be addressed by identifying their sources and the physical mechanisms by which they are accelerated up to energies of the order of $\sim 10^{20}\mathrm{eV}$. The magnetic fields in the galactic and intergalactic space deflect the CRs during their propagation, making it difficult to resolve their source's positions by measuring their arrival direction. Neutrinos are believed to be produced in hadronic processes in the CR accelerators. The fact that they are electrically neutral and weakly interacting particles make of neutrinos good candidates to determine unambiguously the position of the CR accelerators. 

Neutrino telescopes aim at detecting the Cherenkov light emitted by charged leptons resulting from the interaction of astrophysical neutrinos with the matter surrounding the instrumented volume. The good angular resolution (below $0.5^o$) achieved with the ANTARES neutrino telescope for muon tracks allows for the search of small scale anisotropies (eg point sources) as well as large scale structures. In the following a model independent search is presented based on a modified two point correlation function. The results are interpreted in terms of upper limits on the population of unresolved point-like sources.

\subsection{Motivation}
The interaction of high energy cosmic rays with the Earth's atmosphere induces air showers in which among other particles, muons and neutrinos are present. These so called atmospheric muons and atmospheric neutrinos constitute the two main sources of background for the ANTARES detector. 
Given that the Earth is opaque to all particles with the exception of neutrinos, because they interact weakly with matter, the atmospheric muon background can be reduced by selecting only those events that are reconstructed with an upwards direction with respect to the ANTARES neutrino telescope. Nevertheless, some muon tracks coming from above can be reconstructed as up going. The amount of wrongly reconstructed muons can be reduced by means of quality cuts in the muon tracks reconstruction parameters.

Atmospheric neutrinos can traverse the Earth and produce muon tracks that will remain as an irreducible source of background. 

The challenge of the statistical analyses carried out within the ANTARES collaboration is to unmask those events with astrophysical origin, hidden within a background dominated ensemble of isotropically reconstructed events. One way of looking for an astrophysical signal, is to look for clustering in the arrival directions of the reconstructed events. The autocorrelation analysis is a way of looking for spatial clustering in discrete data ensembles. An improved autocorrelation analysis was presented in \cite{fabian} and applied to the neutrino candidates detected by the ANTARES neutrino telescope during its first three years of data taking. In the present analysis the method is applied to five years of data and used to search for a signal coming from sources that are too faint to be detected by other statistical analyses such as the ones relying on a likelihood-based method \cite{psources}.  In absence of such a signal we will set upper limits on the neutrino fluxes.
 
\section{The autocorrelation analysis}
\subsection{The method}\label{method}
The autocorrelation analysis allows to find inhomogeneities within a discrete data set by studying the two point correlation distribution, which is defined as the distribution of the number of pairs of events as a function of their mutual angular distance $\Delta \Omega$. As it was shown in \cite{fabian} and \cite{TheANTARES:2013iwa}, weights based on some energy estimator $\bar{E}$ can be applied to the events in order to discriminate between astrophysical neutrinos, which dominate at lower energies, and cosmic neutrinos, which spectral distribution in energy is harder. This behaviour is shown in fig.\ref{jits} for simulated events.
 
Formally, the cumulative autocorrelation distribution can be defined as
\begin{equation}
\mathcal{N}_{n_{Hit}}(\Delta \Omega) = \sum\limits_{i=1}^{N}\sum\limits_{j=i+1}^{N}\omega _{ij}\cdot \left[1-H(\Delta \Omega _{ij}-\Delta \Omega)\right],
\end{equation}
where $H$ is the Heaviside step function and $\omega _{ij} = \omega _{i}\cdot \omega _{j}$ are weights assigned to the couple of events $i$ and $j$. Each of the individual weights $\omega _i$ is defined as
\begin{equation}
\omega (\bar{E}_i) = \int\limits_{0}^{\bar{E}_i}f(\bar{E})d\bar{E}
\end{equation}
$f(\bar{E})$ is the normalized distribution of the energy estimator, which can be obtained from Monte Carlo simulations. As it is shown in \cite{fabian}, the selected choice for the energy estimator is the number of hits used in the event reconstruction, $n_{hit}$. Figure (\ref{jits}) shows a comparison between the $n_{hit}$ distribution for atmospheric and astrophysical neutrinos.

\begin{figure}
\centering
\includegraphics[scale=0.5]{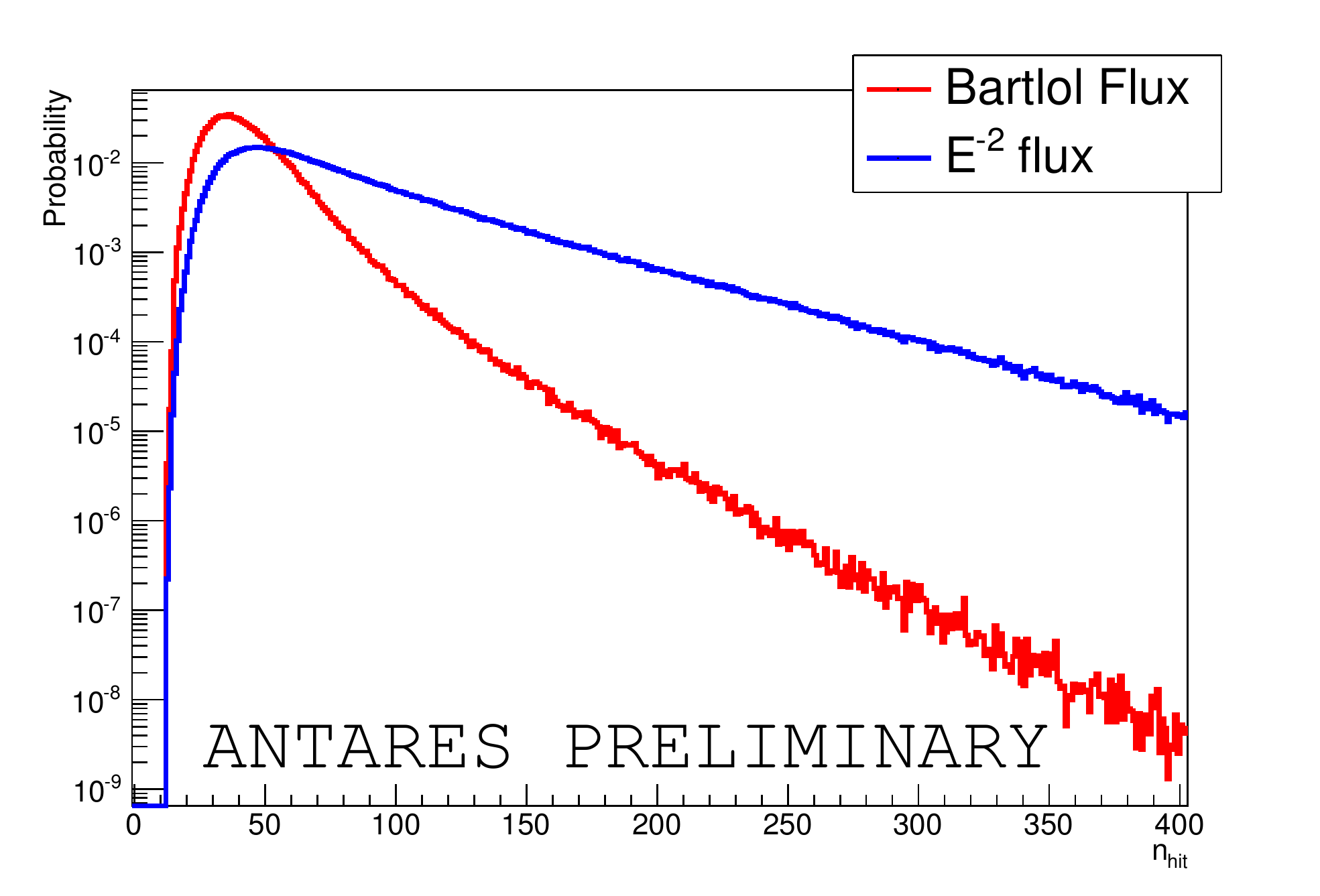}
\caption{Distributions of the $n_{hit}$ energy estimator for simulated atmospheric neutrinos, following a Bartol flux ($\sim E^{-3.7}$), and simulated cosmic neutrinos following an $E^{-2}$ spectrum.}
\label{jits}
\end{figure}

A comparison of the autocorrelation function resulting from measured events with the one corresponding to an isotropic sample will allow to detect possible clusters of events if a significant excess with respect to the isotropy is present in the data.
\subsection{The ANTARES data set}\label{dataset}
For the present analysis, a data set recorded by the ANTARES neutrino telescope between 2007 and 2012 has been used. The sample contains 5243 neutrino candidates that satisfy selection criteria optimized in order to obtain the best average upper limit on the flux of neutrino coming from point like sources and extends the dataset of the previous analysis \cite{fabian} by about 50$\%$. These selection criteria consist in a cut on the reconstructed zenith angle $\theta > 90º$, a cut on the angular uncertainty in the track reconstruction $\beta < 1º$, and a cut on the reconstruction quality parameter $\Lambda > -5.2$  
\subsection{The reference autocorrelation distribution}\label{isotropicset}
The reference autocorrelation distribution has been built as the average of the autocorrelation distributions derived from about $10^4$ isotropic data sets. Each of them was obtained by keeping the local coordinates of one neutrino candidate, but assigning it a time randomly selected from another event. This process allows to build an isotropic sky map with the same coverage as the original data set, and in which the non uniformity in the data taking conditions is taken on account. 
\subsection{Statistical comparison between the data and the reference distributions}
Fig.\ref{cumulative} shows the cumulative autocorrelation distributions described in section \ref{method}. The statistical comparison between both distributions is based on an hypothesis test in which the test statistics (TS) will be given by the maximum of a quantity computed for each angular scale, that measures the difference between both distributions:

\begin{equation}
TS = \smash{\displaystyle\max}\left\lbrace \left(\frac{\mathcal{N}_{n_{Hit}}^{data} - \mathcal{N}_{n_{Hit}}^{iso}}{\sigma} \right) \right\rbrace _{\Delta \Omega _i}
\end{equation}
where $\sigma$ denotes the standard deviation for the isotropic distribution with respect to its mean.  

A distribution of the test statistics for background like ensembles will be built by comparing the autocorrelation distribution of about $10^4$ randomized sky maps with the isotropic one. This distribution will be used to compute the p-value as the probability of finding in a background like ensemble, the same value for the test statistics or a higher one than the corresponding to the data.    
\begin{figure}
\centering
\includegraphics[scale=0.5]{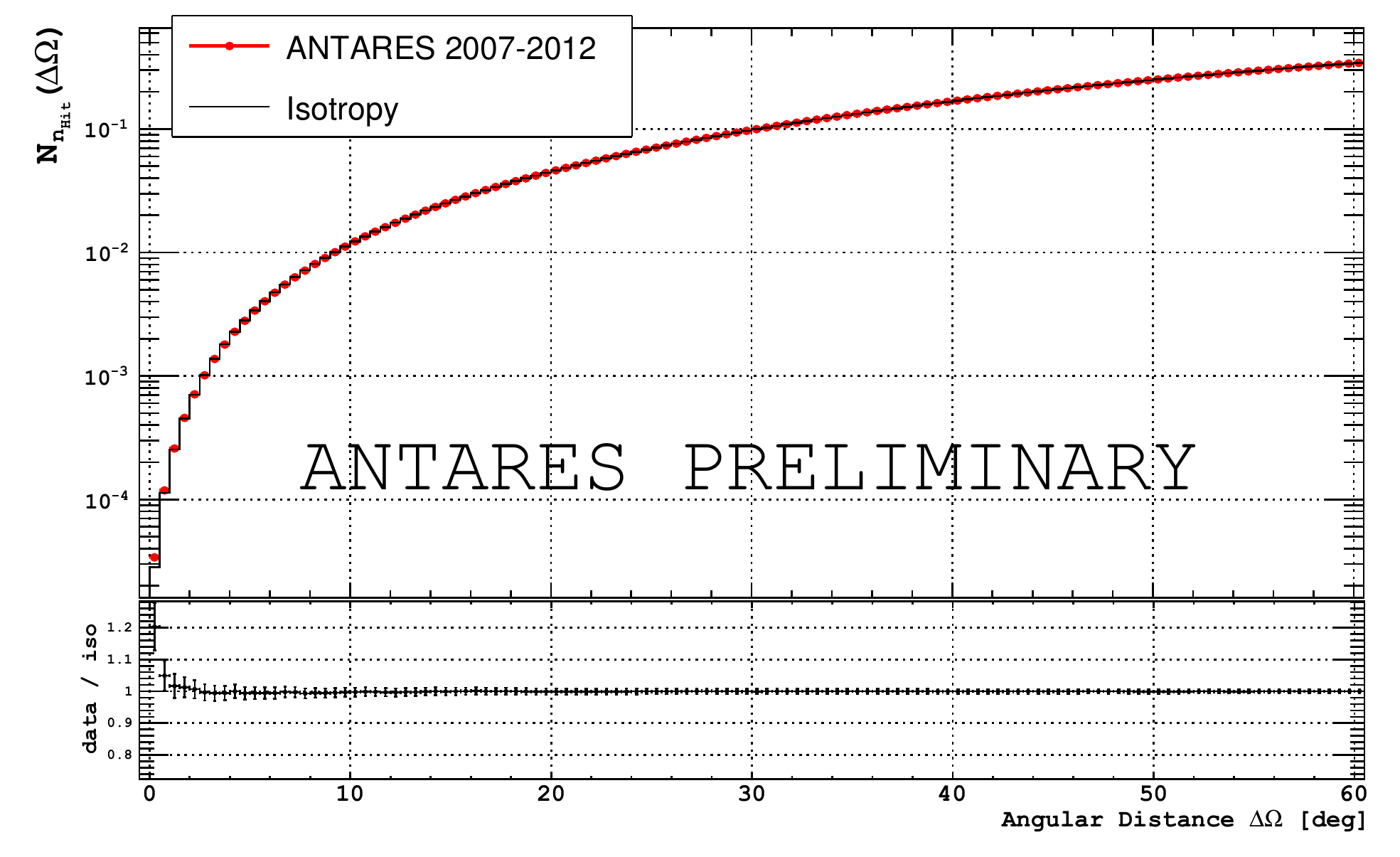}
\caption{Cumulative autocorrelation distributions for the 2007-2012 ANTARES data set (red points) and for the average isotropic ensemble (black line). }
\label{cumulative}
\end{figure}    

\subsection{Performance and sensitivity of the method}

The detection power of the autocorrelation method has been previously tested with background sky maps in which some of the events had been substituted by signal events that would have come from a single point source \cite{fabian}. The results, showed that a dedicated point source search analysis is slightly more efficient in the detection of single point sources than the autocorrelation method, but outperforms it as soon as more than one source is present. Here the detection power of this method is studied using signal coming from populations of point sources that are too faint to be detected by the dedicated point source search. 

\subsubsection{Model for sky maps with signal}\label{models}

In background like sky maps produced as described in section \ref{isotropicset}, a certain proportion of events has been removed from their positions and substituted by signal like events distributed in a simulated population of sources inspired by \cite{unresolved}, whose mean neutrino luminosities follow a power law  
\begin{equation}\label{luminosity distribution}
\frac{dN}{dS} \propto \cdot S^{-\gamma}
\end{equation}
between some limits $S_{min}$ and $S_{max}$, where $S$ is a dimensionless quantity corresponding to the  integrated mean luminosity expressed in number of detected neutrinos. $S_{max}$ has been fixed to the faintest point source compatible with the limit set by the search for point sources with ANTARES data \cite{psources}, and $S_{min}$ is a free parameter bounded above by $S_{max}$, which will characterize a source population. 

The luminosity function can be considered as a proxy of the power of the sources. Although the luminosity function for neutrino sources is unknown, one can assume that it follows the same general rules as the luminosity functions in x-rays or gamma rays. We can thus rely on known populations of point sources to choose the range of $\gamma$. Studies of galactic type sources like low-mass X-ray binaries in Centaurus A \cite{sources1} \cite{sources2} or in the Milky Way \cite{Gilfanov:2003th} yield a typical spectral index below $2.0$, and Fermi LAT studies show that the gamma ray luminosity function of AGNs is well described by a power law with $\gamma\sim 2.2$ \cite{sources3}. 

Therefore a signal sky map is characterized by the couple of parameters, $(\gamma, S_{min})$ and the proportion of injected signal events. $\gamma$ characterizes the type of objects that constitute the population. It drives the relative contributions of the different source luminosities to the total flux. For a given average detected flux, populations characterized by higher values of $\gamma$ will consist in a higher number of less luminous sources. $S_{min}$, independently of the type of sources, will characterize their effective average minimal detectable luminosity within the population. We will test the parameter space $1.8<\gamma<2.3$, and $0.025<S_{min}<1$.


The energy estimator of the signal events is generated from Monte Carlo simulations assuming that the signal events follow an $E^{-2}$ spectrum. Figure \ref{maps} shows an example of a sky map where the 0.5$\%$ of the background events have been replaced by events coming from sources distributed with a spectral index of $\gamma = 1.9$ and $S_{min} = 0.025$ 
\begin{figure}[top]
\centering
\includegraphics[scale=0.25]{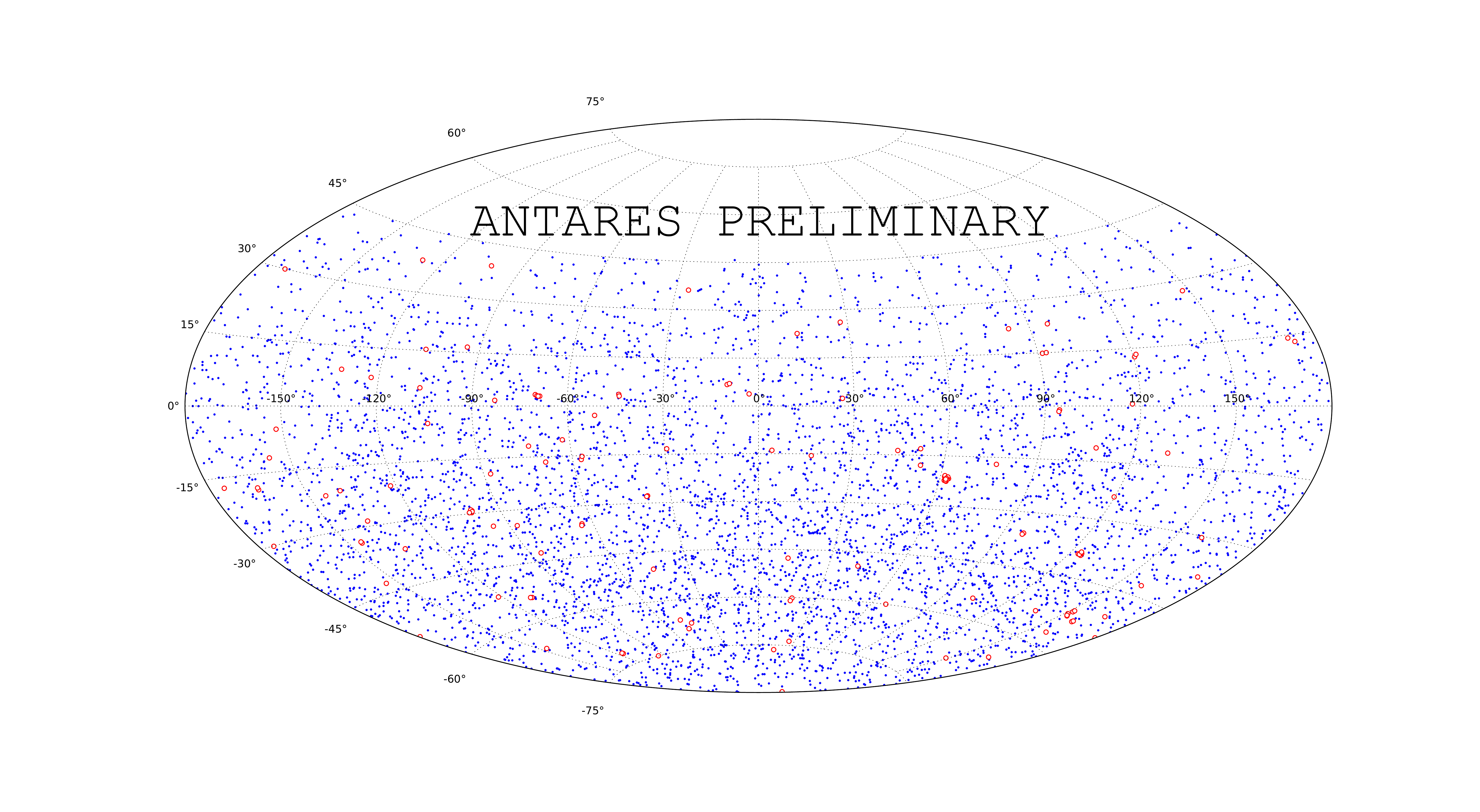}
\caption{Example of a sky map with signal coming from unresolved point sources. The events in blue come from background while the red events correspond to signal.}
\label{maps}
\end{figure}    
\subsubsection{Discovery potential of a population of unresolved point sources}
Following the autocorrelation method described in section \ref{method}, sky maps characterized by different couples $(\gamma, S_{min})$ can be analysed to determine the cumulative diffuse flux of cosmic neutrinos coming from  populations of unresolved point sources that would be detected at a 3$\sigma$ significance with a 90$\%$ probability. As the neutrino luminosity function for different kinds of sources is unknown, the spectral index was assumed to lay within the same range as the spectral index for the luminosity function in x-rays or in gamma rays for different kinds of sources \cite{unresolved}. The discovery potential as a function of the spectral index $\gamma$ and $S_{min}$ under the above conditions is shown in fig.\ref{sensiss}

\begin{figure}
\centering
\includegraphics[width=0.80\linewidth]{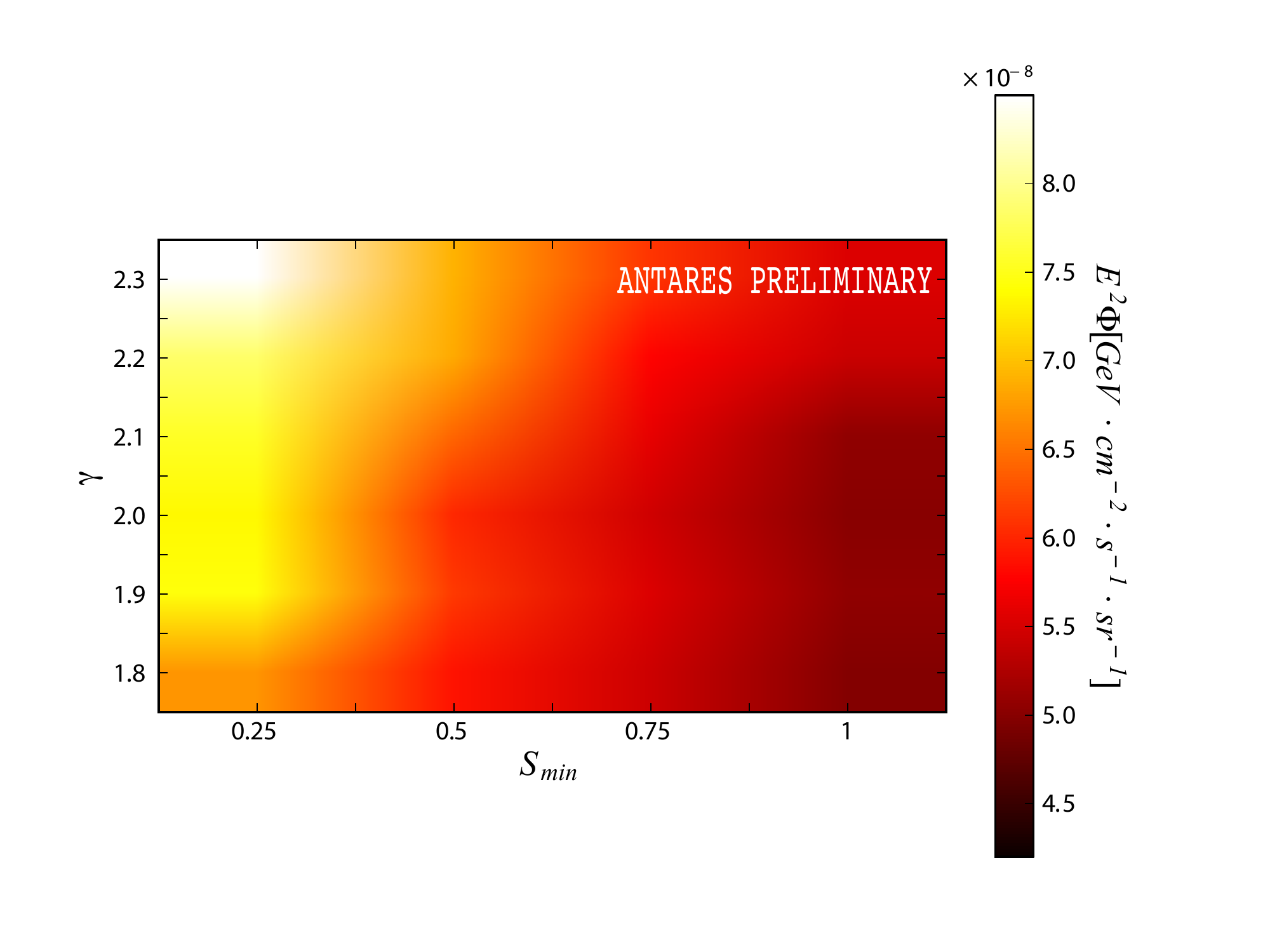}
\caption{Discovery potential at 3$\sigma$ significance with a 90$\%$ probability of the model presented in section \protect\ref{models} as a function of the model's parameters $(\gamma, S_{min})$.}
\label{sensiss}
\end{figure}

\begin{figure}
\centering
\includegraphics[width=0.80\linewidth]{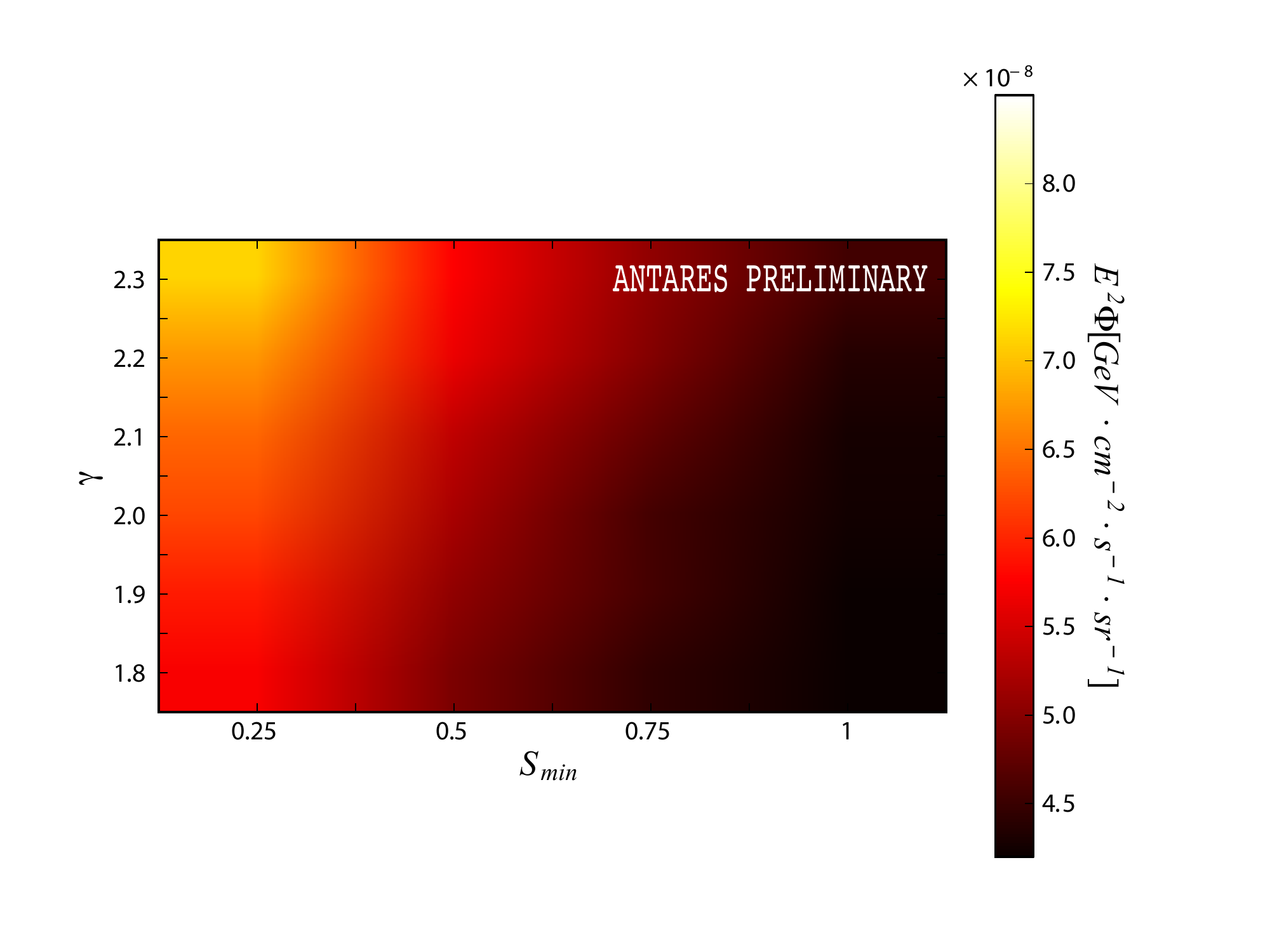}
\caption{90$\%$ confidence level upper limit on the cumulative diffuse flux as a function of the model's parameters $(\gamma, S_{min})$.}
\label{limitss}
\end{figure}

\subsection{Results and Discussion}
The autocorrelation method has been applied to the 5243 events measured by ANTARES from 2007 to 2012 and the result has been compared to the one expected from a purely isotropic sample. The comparison, showed that the largest difference between the distributions corresponds to an excess of the ANTARES neutrino candidates with respect to the isotropic ensemble, at scales $<0.5^o$. The statistical analysis leads to the conclusion that this corresponds to a $\sim 2.3 \sigma$ excess. In previous point source search analyses \cite{fabian} \cite{psources} a 2.2 $\sigma$ excess was found around ($\alpha$,$\delta$) = (313.20,-64.90). Removing events closer than $0.5^o$ from this point reduces the significance of the observed excess in the current analysis to 2.15 $\sigma$, and therefore we can conclude that they are not responsible for the observed deviation with respect to the background.

 We thus set upper limits on the cumulative diffuse flux of the model of unresolved point sources population presented above, as a function of the spectral index $\gamma$ and $S_{min}$, which are presented in figure \ref{limitss}. As expected from the definitions in section \ref{models}, the constraints are stronger for increasing $S_{min}$. One can also see that populations of sources with low values of $\gamma$ (similar to high energy galactic type sources as x-ray binaries) are more strongly constrained. These results, in addition and independently of the high visibility of the galactic region, confirm the latter as a favoured subject of study for the ANTARES neutrino telescope.     
 

\section{Summary}
The detection of astrophysical neutrinos would provide fundamental information about the location of CR sources. This is the aim of the ANTARES neutrino telescope. In this analysis an improved autocorrelation method was applied in order to search for clustering in the directions of the reconstructed neutrino candidates but not significant deviation from background was found. Upper limits for a neutrino flux coming from a population of point sources characterized by a two parameter model were set.


\setcounter{figure}{0}
\setcounter{table}{0}
\setcounter{footnote}{0}
\setcounter{section}{0}
\setcounter{equation}{0}

\newpage
\id{id_hallmann}
\addcontentsline{toc}{part}{\textcolor{blue}{\arabic{IdContrib} - {\sl S. Hallmann} : Search for a neutrino flux from the Fermi Bubbles with the ANTARES telescope
}%
}
%

\newcommand{\remark}[1]{{\color{blue}NB:\ #1}}
\newcommand{\todo}[1]{{\color{red} ToDo:\ #1}}
\newcommand{\antares}{{\textsc{Antares} }}

\title{\arabic{IdContrib} - Search for a neutrino flux from the Fermi Bubbles with the \antares telescope}

\shorttitle{\arabic{IdContrib} - Search for a neutrino flux from the Fermi Bubbles with the \antares telescope}

\authors{S. Hallmann}
        \afiliations{University of Erlangen-N\"urnberg, Erwin-Rommel-Str. 1, 
91058 Erlangen, Germany\\
		Erlangen Centre for Astroparticle Physics (ECAP)}
\email{steffen.hallmann@fau.de}


\abstract{
The Fermi Bubbles are two giant lobes of $\gamma$-ray emission above and below the Galactic Center. Whereas the origin of the observed $\gamma$-ray flux remains obscure, the measurement of a neutrino flux from the Fermi Bubbles could distinguish between leptonic and hadronic emission scenarios. Such a search for a neutrino signal from the Fermi Bubbles has been performed with the \antares neutrino telescope in the Mediterranean Sea using four years of data. The search has used charged current muon neutrino interactions, which produce muons with long tracks in the detector and therefore have an angular resolution of well below one degree. In the analysis, the background is determined from off-regions and compared to the number of events observed in the Fermi Bubble zone. The results of an update using data from 2012 and 2013 are presented. Since no statistically significant excess was found the new upper limits for six years of \antares data are presented.
}


%
%
\maketitle
\section{Introduction}

The Fermi-LAT experiment has revealed two giant lobes of $\gamma$-ray emission extending $7-\SI{8}{kpc}$ ($\approx 50^\circ$) above and below the Galactic Centre \cite{Su:2010qj}. These are commonly referred to as the Fermi Bubbles (FB).
Structures in spatial correlation with the FB have also been observed in X-rays \cite{Snowden97}, in the microwave band \cite{Dobler2012Apj} and radio-wave band \cite{Carretti2013}. 
To date the origin of the FB remains unknown. Several proposed models explaining the emission include hadronic mechanisms, in which the $\gamma$-rays together with a corresponding neutrino signal are produced by the collisions of cosmic-ray protons with interstellar matter \cite{Crocker:2010dg,2014MNRAS.444L..39L,2013ApJ...778L..20T}.  In contrast, models based on leptonic mechanisms or dark matter decay would yield less neutrino emission or none at all \cite{Su:2010qj,2014MNRAS.444L..39L,1104.3585,1102.5095}. The observation of a neutrino signal from the FB region would therefore give a unique possibility to discriminate between the different models.

A search for a signal from the Fermi Bubbles with the \antares neutrino detector with four years of data (2008--2011) has already set an upper limit on the neutrino flux \cite{AntaresFBTracks}. The analysis used off-zones with same visibility in the \antares detector to determine the background in the Fermi Bubbles' region. In the signal region a statistically insignificant excess of 1.2~$\sigma$ over the background was observed. In this proceeding, the result of an update on the existing analysis using two additional years of data taking (2012 \& 2013) is presented.


The \antares telescope~\cite{2011NIMPA.656...11A,clancyICRC} is a deep-sea neutrino detector located $\SI{40}{km}$ off Toulon (France) 
taking data in its final configuration since 2008.
In the search for a neutrino signal from the FB muons and neutrinos emerging from cosmic-ray interactions in the atmosphere constitute the main backgrounds. While the water overburden acts as a partial shield, the rate of atmospheric muons coming from above the detector still dominates over the neutrino signal. Signal searches reduce this background by looking only at events coming from below the detector. The cosmic signal is distinguished from atmospheric neutrinos by its harder energy spectrum. A cut on the reconstructed energy exploits this feature.

This analysis focusses on charged current interactions of muon neutrinos ($\nu_\mu+\overline{\nu}_\mu$). In this interaction channel a relativistic muon is produced and emits Cherenkov light along its path through the water. The direction is reconstructed by maximising a likelihood which fits the photon arrival times at the optical modules (hits) to the Cherenkov emission on the hypothesised muon track. This gives a median angular resolution on the neutrino direction of $0.46^\circ$ \cite{1207.3105}.

Thanks to the detector position at $43^\circ$ latitude in the northern hemisphere, \antares has an excellent visibility to the region around the Galactic Centre. Their position hence makes the Fermi Bubbles an ideal target to look for galactic neutrino emission.

\begin{figure}[h]
	\begin{minipage}[c]{.5\textwidth}
	\begin{center}
		\includegraphics[width=0.7\textwidth]{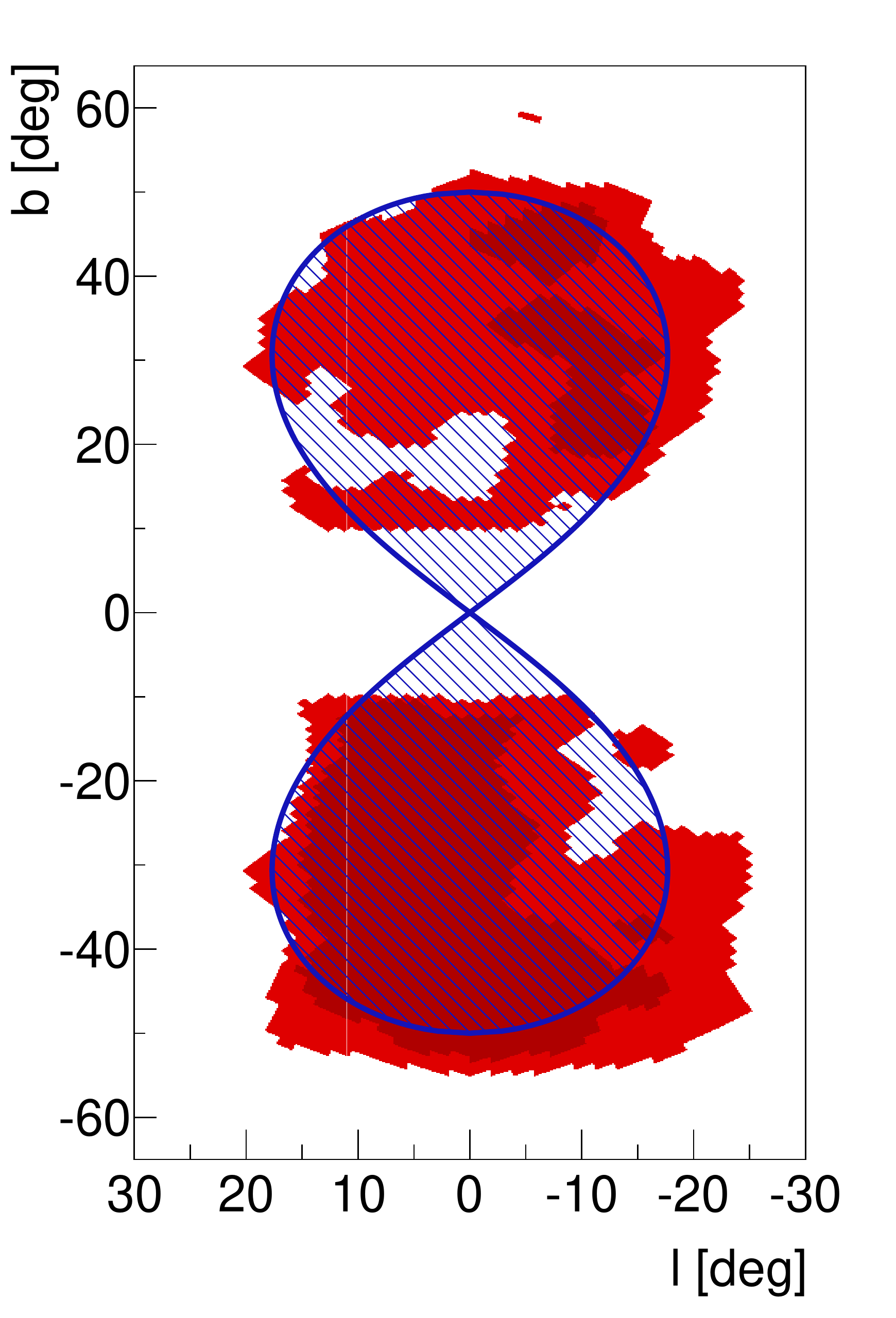}
	\end{center}
	\end{minipage}
	\begin{minipage}[c]{.5\textwidth}
	\caption{The geometric shape used in the analysis (shown in blue) has a good overlap with the shape of the FB structure found in Ref. \cite{Fermi-LAT:2014sfa} (indicated in red). Especially the 'cocoon' structure (dark red), which shows a higher $\gamma$ intensity, is well covered (91\%).}
	\label{fig:FBshape}
	\end{minipage}	
\end{figure}

\section{Spectrum of the expected neutrino flux from the Fermi Bubbles}\label{sec:SpectrumUpdate}
Fig.~\ref{fig:FBshape} shows the shape of the $\gamma$-ray lobes observed with Fermi-LAT. They show a relatively uniform $\gamma$-ray emission over the whole region~\cite{Su:2010qj}. Ref.~\cite{Su:2010qj} measured a hard $\gamma$-ray spectrum without visible cutoff compatible with a power-law $E^{-\alpha}$ with spectral index $\alpha=2$, and a corresponding flux of
\begin{equation}
E^2\frac{\mathrm{d}\Phi_\gamma}{\mathrm{d}E} \approx 3-\SI{6e-7}{GeV cm^{-2} s^{-1} sr^{-1}}.
\label{eq:flux200}
\end{equation}
A more recent study by Fermi-LAT on the spectrum of the $\gamma$-rays prefers steeper spectra or low cutoff energies. Using the $\gamma$-flux parametrisation from the \textsc{Sybill}-code from Ref.~\cite{astro0606058}, it is shown in Ref. \cite{1504.07033} that an $E^{-2.25}$ proton spectrum can produce a $\gamma$-flux that fits the Fermi-LAT data well. A power-law fit to this parametrisation at energies beyond $\SI{10}{GeV}$ yields a spectral index of $\alpha=2.18$ and a $\gamma$-flux~(c.f. \cite[Fig.~2]{1504.07033})
\begin{equation}
E^{2.18}\frac{\mathrm{d}\Phi_\gamma}{\mathrm{d}E} \approx 0.5-\SI{1.0e-6}{GeV^{1.18} cm^{-2} s^{-1} sr^{-1}}.
\label{eq:flux218}
\end{equation} 

In a purely hadronic emission scenario a $\gamma$-ray flux and a corresponding neutrino flux are generated by the decay of neutral and charged pions, which emerge from the interaction of cosmic-ray protons with the interstellar gas \cite{astro0606058}. 
At high energies the neutrino and $\gamma$-ray flux in this hadronic case differ only by a scaling factor $\xi(\alpha_\gamma)$~\cite{0807.4151},
\begin{equation}
\Phi_\nu(E) = \xi(\alpha_\gamma)\times\Phi_\gamma(E).
\end{equation}
The scaling depends on the spectral index of the $\gamma$-rays, $\alpha_\gamma$, and is  $\xi(\alpha_\gamma)\approx0.41\ (0.36)$ for an $E_\gamma^{-2}$ ($E_\gamma^{-2.18}$) spectrum~\cite{0807.4151}.

The Fermi satellite due to its limited size can only measure the photon spectrum to energies of some $\SI{100}{GeV}$. The spectrum and cutoff of the FB signal at higher energies is to date undetermined. Within our galaxy it is however assumed that protons can only be efficiently accelerated up to energies of 1--10~PeV \cite{Crocker:2010dg}. This will induce also a cutoff in the observed $\gamma$-ray and neutrino spectra. As a crude approximation 20\% of the proton energy is on average converted into charged pions. An equal distribution over the four daughters in pion decay yields
\begin{equation}
E_\nu^\mathrm{cutoff} = 0.05\times E_p^\mathrm{cutoff}
\label{eq:cutoff}
\end{equation}
for the neutrino cutoff, i.e. cutoffs in the range of $50-\SI{500}{TeV}$.
Combining Eq.~\ref{eq:flux200} with an exponential cutoff from Eq.~\ref{eq:cutoff} and taking into account the scaling factor yields the expected neutrino fluxes,
\begin{equation}
E^\alpha\frac{\mathrm{d}\Phi_{\nu_\mu+\overline{\nu}_\mu}}{\mathrm{d}E} = A^\alpha_\mathrm{model}\times\exp\left(-\frac{E}{E^\mathrm{cutoff}_\nu}\right),
\label{eq:modelFlux}
\end{equation}

\begin{equation}
A_\mathrm{model}^{2.0} = 1.2-\SI{2.4e-7}{GeV cm^{-2} s^{-1} sr^{-1}}\qquad\mathrm{for\ }\alpha=2.0,\phantom{\ } 
\end{equation}
and similarly for the flux assumption from Eq.~\ref{eq:flux218},
\begin{equation}
A_\mathrm{model}^{2.18} = 1.8-\SI{3.6e-7}{GeV^{1.18} cm^{-2} s^{-1} sr^{-1}}\qquad\mathrm{for\ }\alpha=2.18.
\end{equation}

\section{Event selection for the on-/off-zone analysis}\label{sec:eventReco}
For the analysis a preliminary event selection is applied on the data to reject badly reconstructed events and background: To reject most of the atmospheric muons only events reconstructed as up-going are selected. Events are kept, if the track fit algorithm used more than 10 hits. A cut on a parameter describing the angular error of the reconstruction, $\beta<1^\circ$, deselects events with misreconstructed directions. 
Shower-like events are identified by an alternative $\chi^2$-based fit algorithm. This algorithm assumes the hypothesis of a showering event signature ($\chi^2_\mathrm{point}$) and that of a muon track ($\chi^2_\mathrm{track}$). Events which are shower-like ($\chi^2_\mathrm{point}<\chi^2_\mathrm{track}$) are excluded from the analysis.  

The optimisation of the event selection is done on two parameters: The track fit quality $\Lambda$, and the reconstructed energy $E_\mathrm{reco}$. The cut on $\Lambda$ is mainly used to reject atmospheric muons. The energy estimate $E_\mathrm{reco}$ is determined by Artificial Neutral Networks. To produce these a machine learning algorithm was trained to derive an energy estimate~\cite{ANNenergy} from a set of variables, such as the number of detected photons and the total charge deposited on the optical modules. For $\SI{10}{TeV}$ muons the median resolution is 30\% on $\log_{10}(E_\mathrm{reco}[\mathrm{GeV}])$.

A blind strategy is adopted for the analysis in which the optimisation of the cuts on $\Lambda$ and $E_\mathrm{reco}$ is performed using simulated signal and background data only.

\begin{figure}
	\begin{minipage}[c]{.33\textwidth}
	\caption{Hammer equal-area map projection in galactic coordinates showing the on-zone and off-zones. The shaded area is the Fermi Bubbles region (on-zone). The three off-zones are shifted by 6, 12 and 18 hours in time. The colour scale represents the visibility of the sky at the ANTARES site ranging from $\SI{0}{h}$ (white) to $\SI{24}{h}$ (blue) per day. Figure taken from Ref.~\cite{inspireVladimirsFB}.}
	\label{fig:onoffzones}
	\end{minipage}	
	\begin{minipage}[c]{.67\textwidth}
	\begin{center}
		\includegraphics[width=\textwidth]{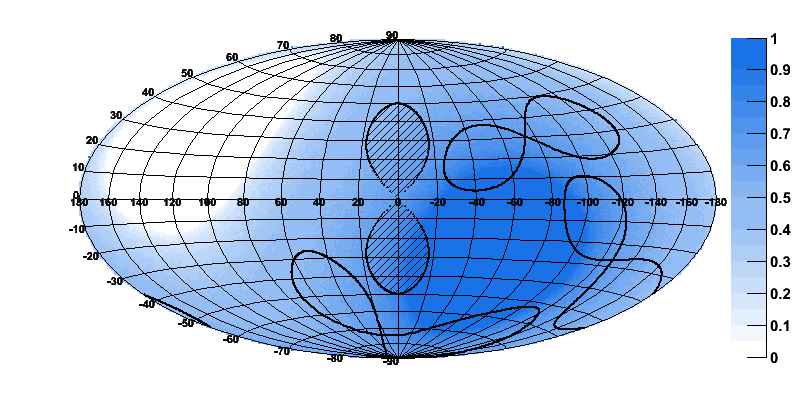}
	\end{center}
	\end{minipage}
\end{figure}

For the signal search the number of events originating from the combined region of the two FB lobes above and below the Galactic Centre (on-zone) is compared to the background observed in regions from which no signal is expected (off-zones). A simplified shape of the FB, which approximates the template area identified in Ref. \cite{Fermi-LAT:2014sfa}, is used for the analysis. The exact shape is illustrated in Fig.~\ref{fig:FBshape}.
The off-zones are chosen as fixed regions in galactic coordinates of identical shape and size as the on-zone. In the detector these shapes are observed with a time-shift of 1/4, 1/2 and 3/4 of a siderial day and therefore have the same visibility as the on-zone (see Fig.~\ref{fig:onoffzones}). Gaps in the data-taking and slight changes in the detector efficiencies can however lead to differences in the observed numbers of events in the on- and offzones. This effect was checked and found to be negligible. More specifically the numbers of events recorded in each of the off-zones were compared for various cut levels ($\Lambda^\mathrm{cut}$, $E_\mathrm{reco}^\mathrm{cut}$) and the differences were found to be within the statistical uncertainty. The approach of using on- and off-zones has also been used recently in a search for an enhanced neutrino emission from the southern sky~\cite{luigiHotspot}.
The distributions of the parameters used for the cut optimisation, $\Lambda$ and $E_\mathrm{reco}$, are shown in Fig.~\ref{fig:recoParams} for events coming from the off-zones with the preliminary event selection applied.

\begin{figure}[b!]
	\begin{center}
		\includegraphics[width=0.49\textwidth,height=.2\textheight]{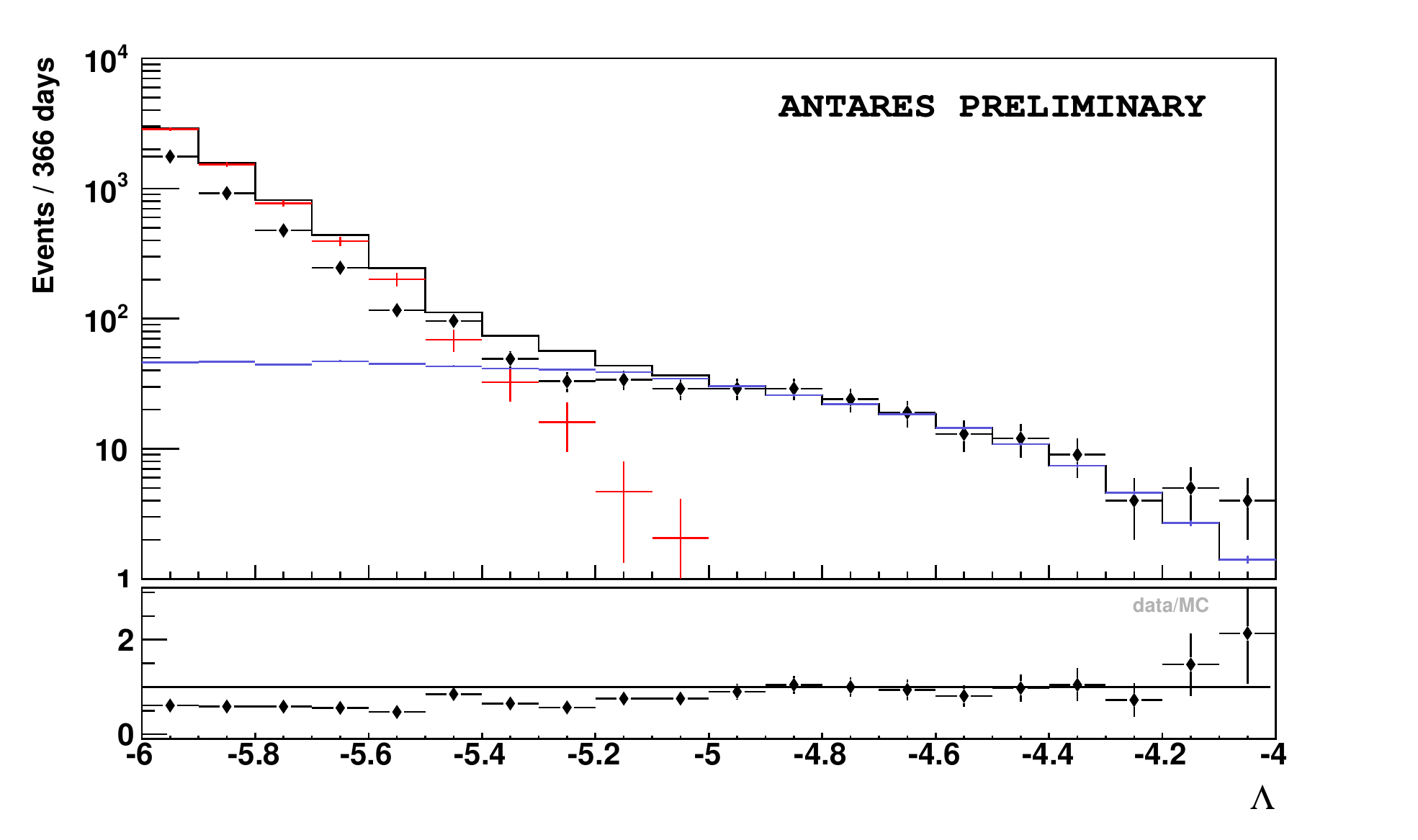}
		\hfill
		\includegraphics[width=0.49\textwidth,height=.2\textheight]{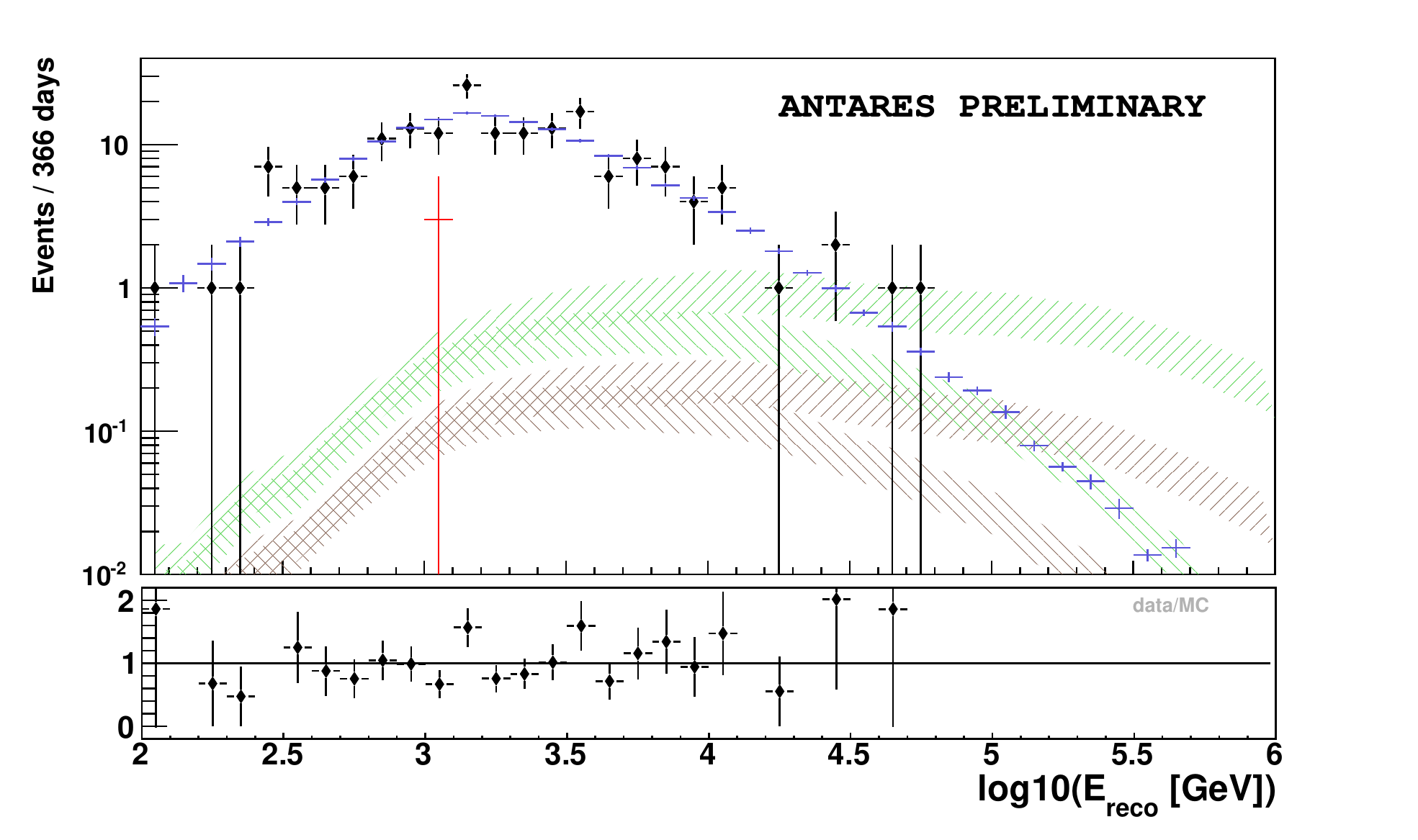}
	\end{center}
	\vspace{-\baselineskip}

	\caption{Off-zone distribution for measured (black points) and simulated events of the two reconstruction parameters used for optimisation of the signal sensitivity: On the left a transition of the main contribution from muons (red) to neutrinos (blue) is seen in the track fit quality parameter at $\Lambda\approx-5.2$. On the right the distribution of the reconstructed energy for $\Lambda>-5.1$ is compared to the distribution of simulated data. The signal flux (scaled up by a factor of 3 for easy comparison with the off-zones) for an $E^{-2}$ spectrum (green) and an $E^{-2.18}$ spectrum (brown) is also indicated for a 50~TeV cut-off and no cut-off . The preliminary event selection mentioned in the text has been used. A scaling factor within the systematic uncertainties of the Bartol model has been applied to the atmospheric neutrino flux to allow for better agreement between measured data and simulation.}
	\label{fig:recoParams}	
\end{figure}

At energies of 100~TeV and beyond the prompt neutrino flux from semi-leptonic decay of charmed particles might be a major contribution to the atmospheric neutrino background. This component is not present in the simulated data and the uncertainty on its flux is large. Due to the on- and off-zone approach this effect will however not alter the final result significantly.  

\section{Cut optimisation}\label{sec:cutOptim}
To determine the optimal cut values for the dataset used in the update, the result of the 4--year FB analysis needs to be taken into account. This first measurement has observed an average background of $n_{\mathrm{off},1} = 11$ in the off-zones and $n_{\mathrm{on},1}=16$ events in the on-zone.
The optimal cut values for the new data are obtained by minimising the average upper limit on the flux:
\begin{equation}
\overline{\Phi}_{90\%} = \Phi_\mathrm{\nu_\mu+\overline{\nu}_\mu}\frac{\overline{s}_{90\%}(b_2|n_{\mathrm{on},1}, n_{\mathrm{off},1})}{s_1+s_2}, 
\label{eq:mrf}
\end{equation}
where $s_1+s_2$ is the number of signal events simulated with the assumed neutrino flux $\Phi_\mathrm{\nu_\mu+\overline{\nu}_\mu}$ from Eq.~\ref{eq:modelFlux} in the whole data taking period used for the initial analysis ($s_1$) and the update ($s_2$). For a known number of simulated background events in the new dataset, $b_2$, signal upper limits with a 90\% confidence level, $\mu_{90\%}$, are calculated following the approach in Ref.~\cite{1998PhRvD..57.3873F} to obtain the upper limit
\begin{equation}
	\overline{s}_{90\%}(b_2|n_{\mathrm{on},1}, n_{\mathrm{off},1}) = \sum_{k = 0}^{\infty} \mu_{90\%}(k+n_{\mathrm{on},1}, b_2+n_{\mathrm{off},1})\times\mathrm{Poisson}(k|b_2),
\end{equation}
which is an average over all possible numbers of events $k$ observed in the on-zone weighted by their Poisson probability.
In the case of no discovery this best average upper limit represents the sensitivity of the \antares detector to the neutrino flux from the Fermi Bubbles~\cite{mrfTechnique}.

The sets of cuts ($\Lambda^\mathrm{cut}$, $E^\mathrm{cut}_\mathrm{reco}$) optimising the average upper limit on the neutrino flux given in Eq.~\ref{eq:mrf} and the respective flux normalisations are reported in Tab.~\ref{tab:TableCuts} for an $E^{-2}$ neutrino spectrum. For an $E^{-2.18}$ spectrum the corresponding values are also given.

\begin{table}[t!]
\caption{Resulting cut values ($\Lambda^\mathrm{cut}$, $\log_{10}\left(E_\mathrm{reco}^\mathrm{cut}[\mathrm{GeV}]\right)$) from the cut optimisation for an $E^{-2}$ ($E^{-2.18}$) neutrino spectrum on the left (right) with different cutoff energies. The average upper limits on the flux coefficient $\overline{A}_{90\%}^{\alpha}$ are given in units of $\SI{1e-7}{GeV^{(\alpha-1)} cm^{-2} s^{-1} sr^{-1}}$. In the last row, the cut values $\Lambda^\mathrm{cut}=-5.14$ and $\log_{10}\left(E_\mathrm{reco}^\mathrm{cut}[\mathrm{GeV}]\right)=4.03$ from the previous 4--year analysis have been applied for all cut-off energies.}
\label{tab:TableCuts}
\begin{center}
	\begin{tabular}{l|rrrr|rrrr}
	\multicolumn{1}{c}{}&\multicolumn{4}{c}{$E^{-2}$ neutrino spectrum:} & \multicolumn{4}{c}{$E^{-2.18}$ neutrino spectrum:} \\
	\toprule
	$E_\nu^\mathrm{cutoff}$	[TeV]& $\infty$	& 500	& 100	& 50 & $\infty$	& 500	& 100	& 50\\
	\midrule
	$\Lambda^\mathrm{cut}$ &      -5.34 & -5.16 & -5.16& -5.34 & -5.16 & -5.16 & -5.16 & -5.32 \\
	$\log_{10}\left(E_\mathrm{reco}^\mathrm{cut}[\mathrm{GeV}]\right)$& 4.04 & 3.78 & 3.64 & 3.52 & 3.68 & 3.64 & 3.44 & 3.36 \\		
	\midrule
	$\overline{A}_{90\%}^{\alpha}$ & 3.73 & 5.60 & 9.41 & 13.9 & 29.3 & 38.3 & 59.0 & 78.3\\
	$\overline{A}_{90\%}^{\alpha}$ (cuts from \cite{AntaresFBTracks}) & 3.78 & 5.74 & 10.0 & 15.5 & 30.0 & 40.2 & 65.3 & 91.3 \\
	\bottomrule
	\bottomrule
	\end{tabular}
\end{center}
\end{table}

\section{Results}\label{sec:Result}
The analysis used data taken in 2012 and 2013. In addition, two months of 2010 data which were not part of the 4--year analysis were added to the new analysis. Using only runs with low optical background from bioluminescence and runs with good data taking conditions the total lifetime of the additional dataset sums to 366~days (c.f. 806 days in the 4--year analysis). Since the sensitivity does not change significantly when using the cuts of the 4--year analysis, i.e. $\Lambda>-5.14$ and $\log_{10}(E_\mathrm{reco}[\mathrm{GeV}])>4.03$, these cuts are also applied to the unblinded new dataset. In the three off-zones 1, 2 and 3 events are observed and add to the 33 background events in the 4--year analysis. In the region of the Fermi Bubbles 6 events are detected in addition to the 16 events in the first analysis. In Fig.~\ref{fig:energiesOnOff} the energy distribution of the signal events in the on-zone is compared to the off-zones. Using the calculation from Ref.~\cite{LiMaSensitivity}, the observed excess in the signal region is $1.9\sigma$. The 90\% upper limits on the neutrino flux for the \antares data from 2008--2013 were calculated using the approach of Feldman\&Cousins~\cite{1998PhRvD..57.3873F} and are presented in Fig.~\ref{fig:limits}. At the moment of writing this proceeding, a dedicated study of the systematic error is still ongoing.

\begin{figure}[h!]
	\begin{minipage}[c]{.53\textwidth}
	\begin{center}
		\includegraphics[width=\textwidth]{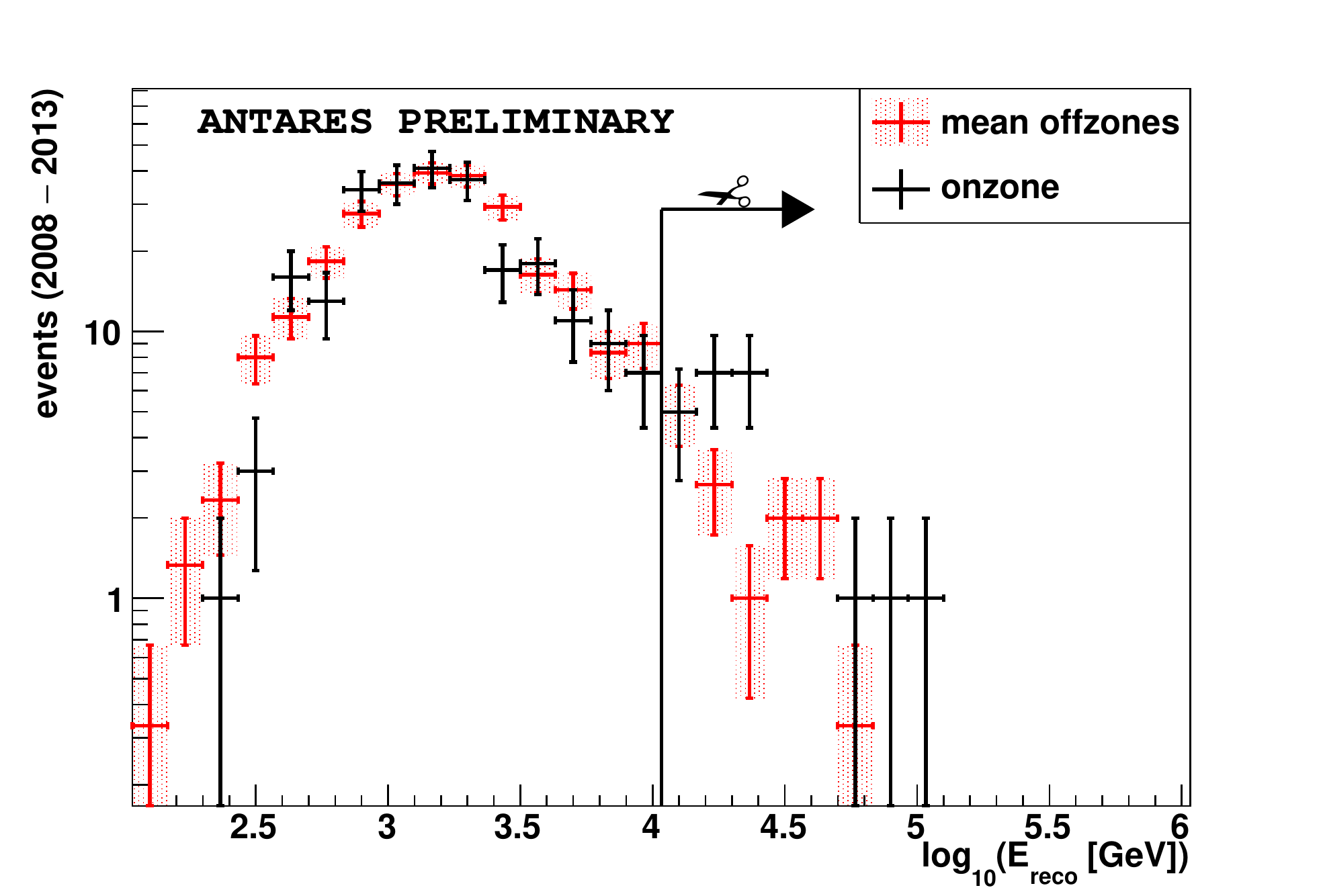}
	\end{center}
	\end{minipage}
	\begin{minipage}[c]{.45\textwidth}
	\caption{Distribution of the reconstructed energies for the six years of \antares data analysed with the preliminary event selection and the final cut on the reconstruction quality parameter, $\Lambda^\mathrm{cut}>-5.14$, applied. From comparison with the off-zones an excess in the on-zone can be seen at energies beyond the cut at $\log_{10}\left(E_\mathrm{reco} [\mathrm{GeV}] \right)>4.03$.}
	\label{fig:energiesOnOff}
	\end{minipage}	
\end{figure}
\begin{figure}[h!]
	\begin{center}
		\includegraphics[width=0.49\textwidth]{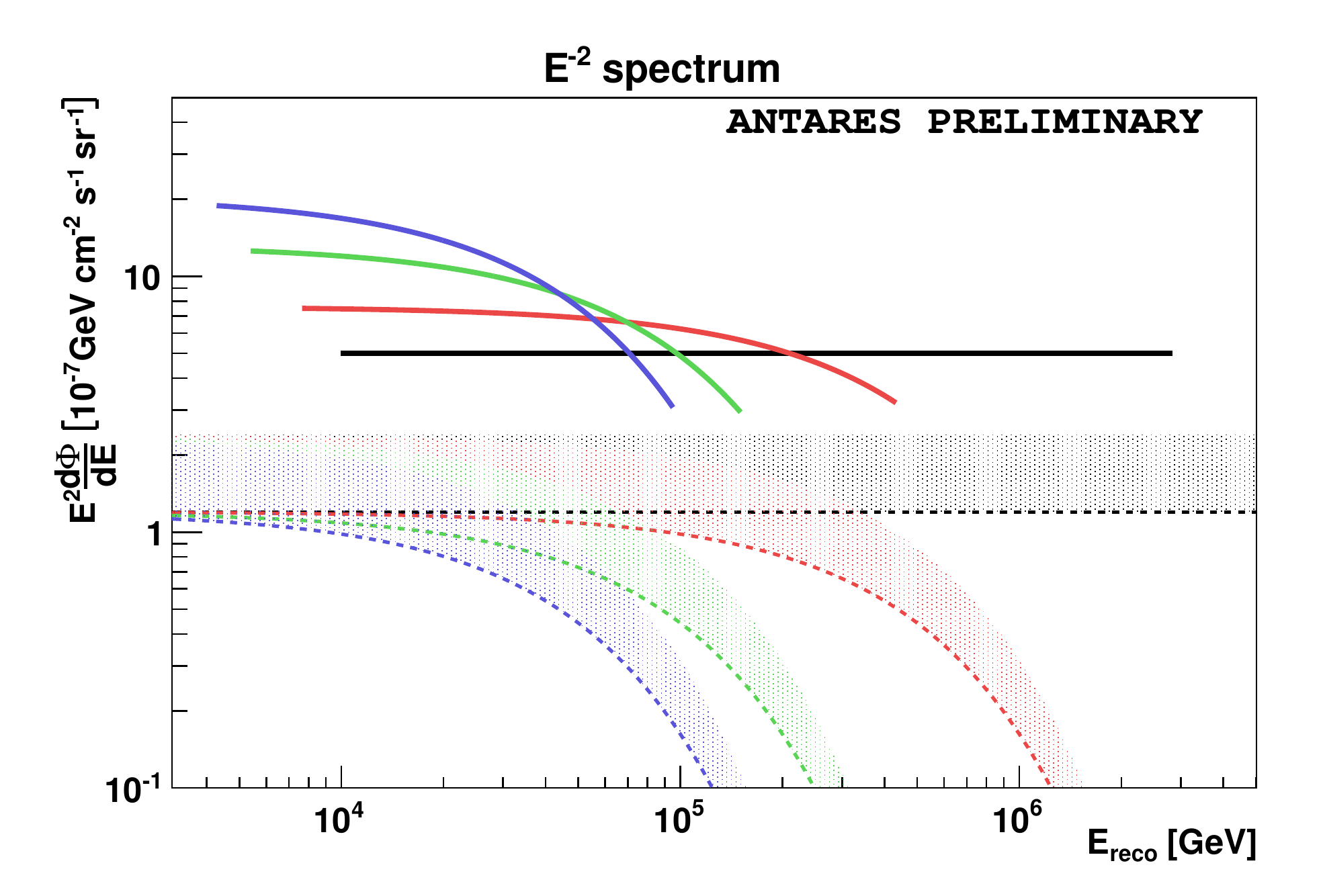}
		\hfill
		\includegraphics[width=0.49\textwidth]{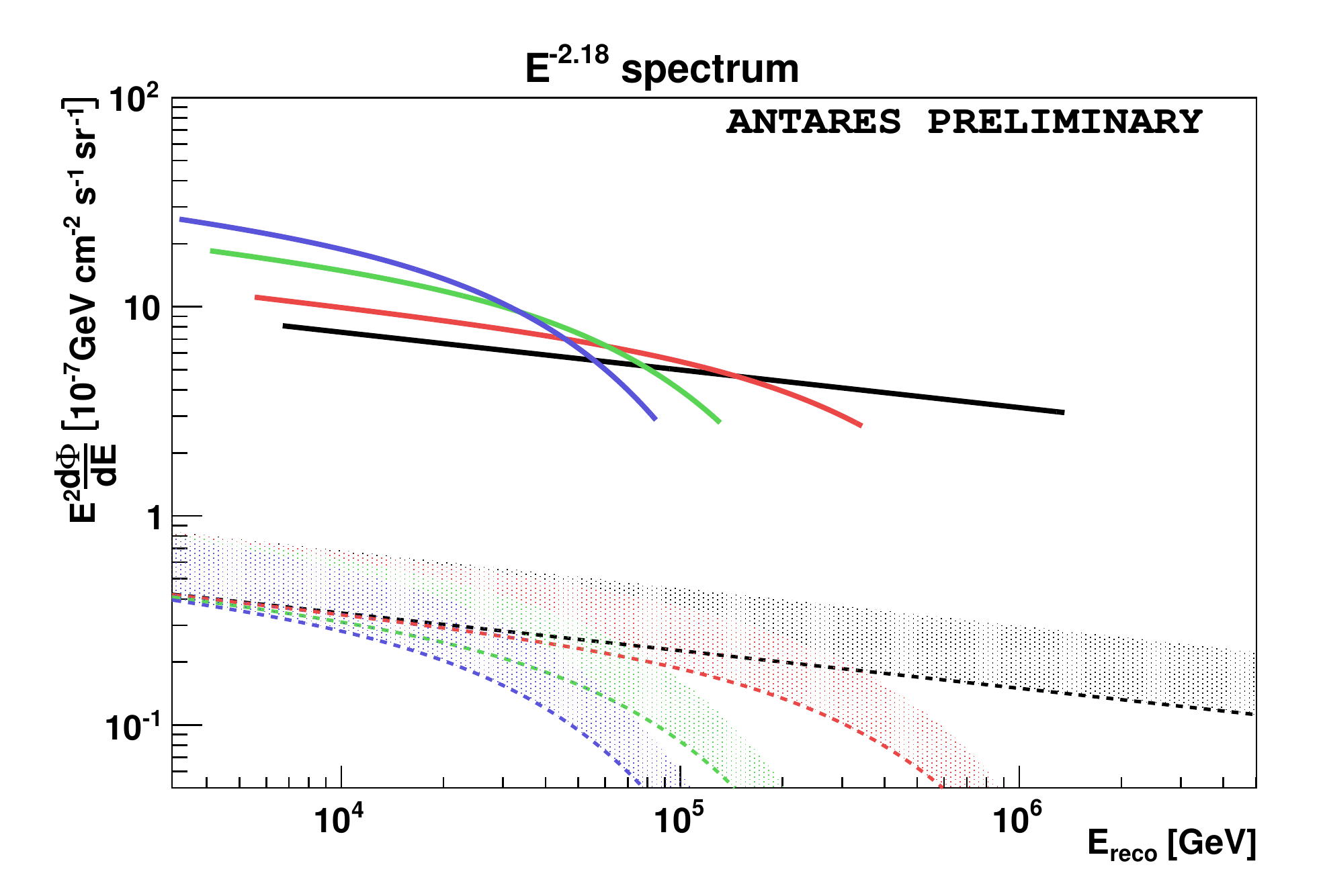}
	\end{center}
	\caption{Upper limits on the neutrino flux from the Fermi Bubbles for different cutoffs (black: no cut-off, red: 500~TeV, green: 100~TeV, blue: 50~TeV) assuming a purely hadronic emission scenario. The shaded areas are the corresponding flux predictions. The limits are drawn for the energy range where 90\% of the signal is expected.}
	\label{fig:limits}	
\end{figure}
\section{Conclusions and outlook}\label{sec:Conclusions}
In six years of \antares data the number of events observed in the Fermi Bubbles region shows yet no statistically significant excess over the background expectation. In the first 806 days, 16 events were found in the on-region with respect to 33 in the three off-regions, corresponding to an excess of $1.2\sigma$. Adding the new data set of 366 days, the number of events in the signal region increases to 22, and the background to 39/3, with an excess of $\approx 1.9\sigma$.

This analysis used track-like event signatures coming from charged current muon neutrino interactions. In contrast, charged current interactions with an electron in the final state and neutral current interactions produce showers of light with a much shorter extension in the forward direction. Recently developed methods provide an angular resolution of $5^\circ$ and below~\cite{tinoShowers} with \antares. This makes extended sources like the Fermi Bubbles an ideal target for a combined analysis using track- and shower-like interaction channels. Also, in future, the KM3NeT detector as successor of \antares will improve the sensitivity to the neutrino flux from the FB by at least one order of magnitude~\cite{FBinKM3NeT}.





\setcounter{figure}{0}
\setcounter{table}{0}
\setcounter{footnote}{0}
\setcounter{section}{0}
\setcounter{equation}{0}
\newpage
\id{id_michael}
\addcontentsline{toc}{part}{\textcolor{blue}{\arabic{IdContrib} - {\sl T. Michael} : Neutrino Point Source Search including Cascade Events with the ANTARES Neutrino Telescope
}%
}
%
%
%
%

\newlength{\figsize}
\setlength{\figsize}{0.49\textwidth}
\newlength{\bigfigsize}
\setlength{\bigfigsize}{0.7\textwidth}

\title{\arabic{IdContrib} - Neutrino Point Source Search including Cascade Events with the
       ANTARES Neutrino Telescope}

\shorttitle{\arabic{IdContrib} - ANTARES Point Source Search}

\authors{Tino Michael} 
        \afiliations{Nikhef}
\email{tino.michael@nikhef.nl}

\abstract{

    ANTARES is the largest neutrino telescope in the Northern Hemisphere. It has been taking data since 2007. One of the prime objectives is the detection and identification of cosmic neutrino sources in the TeV to PeV energy regime. ANTARES has established excellent pointing resolution for muon neutrinos (0.4 deg). Recently, we achieved good pointing capabilities also for contained cascade events ($\approx 2^{\circ}$), which opens up the possibility for all-flavour neutrino point source searches. Together with its geographical location, this makes ANTARES an excellent/competitive tool to test for the presence of cosmic sources in the Southern Hemisphere, including the area around the Galactic Centre, where IceCube reports a slight excess.

In this contribution, we briefly discuss the method to measure the shower energy and direction, which yields degree-level resolutions. We also present the latest time-integrated point source search results, which incorporate cascade events alongside the muon-neutrino events, and the impact on the interpretation of the IceCube signal.}

%
%
\maketitle
\section{Introduction}

The ANTARES neutrino telescope has been operating in the
Mediterranean sea since 2007. The clarity of the sea water allows for 
an excellent timing measurement of the Cherenkov light induced by
charged particles, so that an angular resolution better than 0.4 degrees could be established for up-going muon neutrinos. 
This allows Antares to be competitive to IceCube in the search for point sources in the Southern Hemisphere despite its small size.\cite{lastPS}

Adding sensitivity to cascade events provides access to $\nu_\mathrm e$ charged current interactions (and from there, estimate $\nu_\tau\rightarrow\tau\rightarrow\mathrm e$ contributions) and all flavour neutral current interactions and therefore increases the sensitivity for cosmic neutrino sources even further.

\section{Cascade Reconstruction}

Cascade events are reconstructed using a novel algorithm, which was developed for the purpose of point source searches. The
reconstruction proceeds in two stages:\\
Assuming a spherically expanding shell of photons, the shower 
      mean position (which is close to, but not equal to the neutrino 
      interaction) and the time of occurrence are fitted using the detected photon
      arrival times. The optical background present in ANTARES is mitigated
      by the use of a robust so-called M-estimator\footnote{The M-estimator is a modified $\chi^2$-test that is less sensitive to outliers: $M_\text{est}=2\cdot\sqrt{1+\chi^2/2}-2$.}.\\
The shower direction is determined from the intensities of the
      detected Cherenkov light. While the timing structure of the light is
      spherical to a good approximation, cascade events cause most light
      to be emitted under the Cherenkov angle. The likelihood fit uses a
      tabulated probability density function (PDF) of the expected number of photons as a function of the
      emission angle, the arrival direction of the photon with respect to
      the photomultiplier tube (PMT), and the distance of the shower vertex to the PMT. PMTs 
      that count zero photons are also considered in the Poisson likelihood.
The likelihood used in the direction fit is:
\begin{align}\label{eq:fullLLHood}
    \mathcal{L} \hspace{8pt} = \hspace{8pt}&\sum_{i=1}^{N_\text{selected Hits}} \log\left\{ P_{q>0}(q_i | E_\nu, d_i, \phi_i, \alpha_i) + P_\text{bg}(q_i) \right\} \nonumber \\
                                          +&\sum_{i=1}^{N_\text{unhit PMTs}} \log\left\{ P_{q=0}(E_\nu, d_i, \phi_i) \right\} 
\end{align}
with:\\
\hspace*{12pt} $q_i$, the charge of hit $i$,\\
\hspace*{12pt} $P_{q>0}$, the probability for a hit PMT to measure its observed charge,\\
\hspace*{12pt} $P_{q=0}$, the probability for a PMT to not being hit,\\
\hspace*{12pt} $P_\text{bg}$, the probability for that hit to be caused by random background,\\ 
\hspace*{12pt} $E_\nu$, the neutrino energy,\\
\hspace*{12pt} $d_i = |\vec r_{\text{PMT},i} - \vec r_\text{shower}|$, the distance between the shower mean and PMT $i$,\\
\hspace*{12pt} $\phi_i$, the angle between $(\vec r_{\text{PMT},i} - \vec r_\text{shower})$ and the neutrino direction,\\
\hspace*{12pt} $\alpha_i$, the angle between $(\vec r_{\text{PMT},i} - \vec r_\text{shower})$ and the direction the PMT is facing,\\
\hspace*{12pt} $\vec r_\text{shower}$, the position of the shower mean.\\

The shower position can be reconstructed very reliably. Figure~(\ref{fig:PosPerform}) shows the longitudinal (left) and perpendicular (right) offset of the position fit with respect to the Monte Carlo neutrino axis. For electromagnetic showers (red data points), the reconstructed position along the shower axis corresponds to the mean of the shower's light emission spectrum (purple line in the figure). Hadronic showers (blue data points) have a different emission profile and are usually reconstructed a bit further along the shower axis. The feature in the em-shower channel just below $E_\nu = 10^7$ GeV is due to the Glashow-Resonance. Here, an anti electron neutrino interacts with an electron from the ambient water and produces a $W^-$ Boson. If this $W^-$ decays hadronically, it produces a hadronic shower that carries the whole energy of the original neutrino (in contrast to neutral current interactions where the hadronic shower only takes a fraction of the neutrino energy). The observed longitudinal offset, therefore, corresponds to a high energetic hadronic shower and is expected to lie further away than the ones for pure em-showers.
The median perpendicular distance to the neutrino axis is as low as 1m in either case over a wide energy range.

\begin{figure}
\centerline{\includegraphics[width=0.5\textwidth]{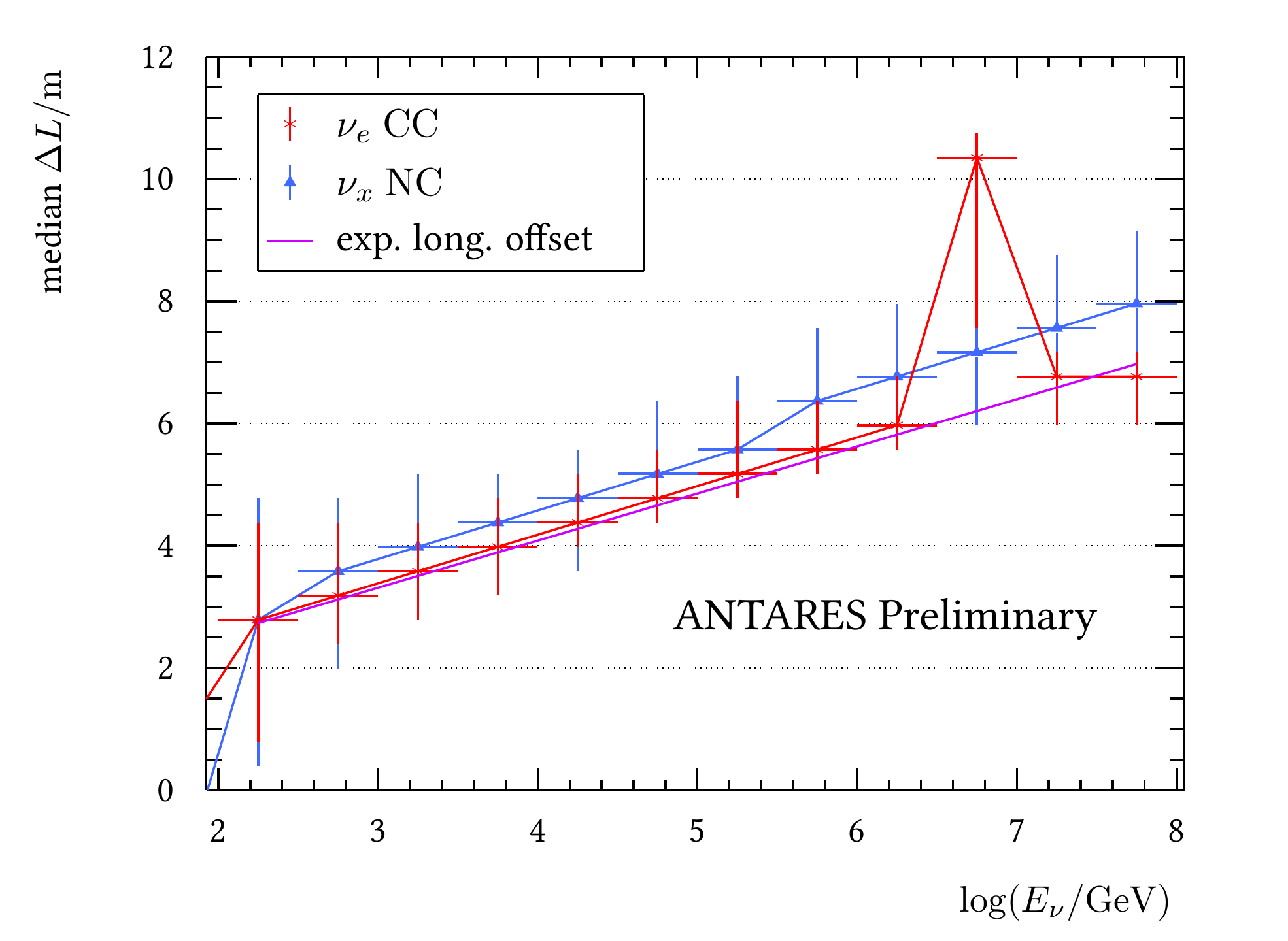}\includegraphics[width=0.5\textwidth]{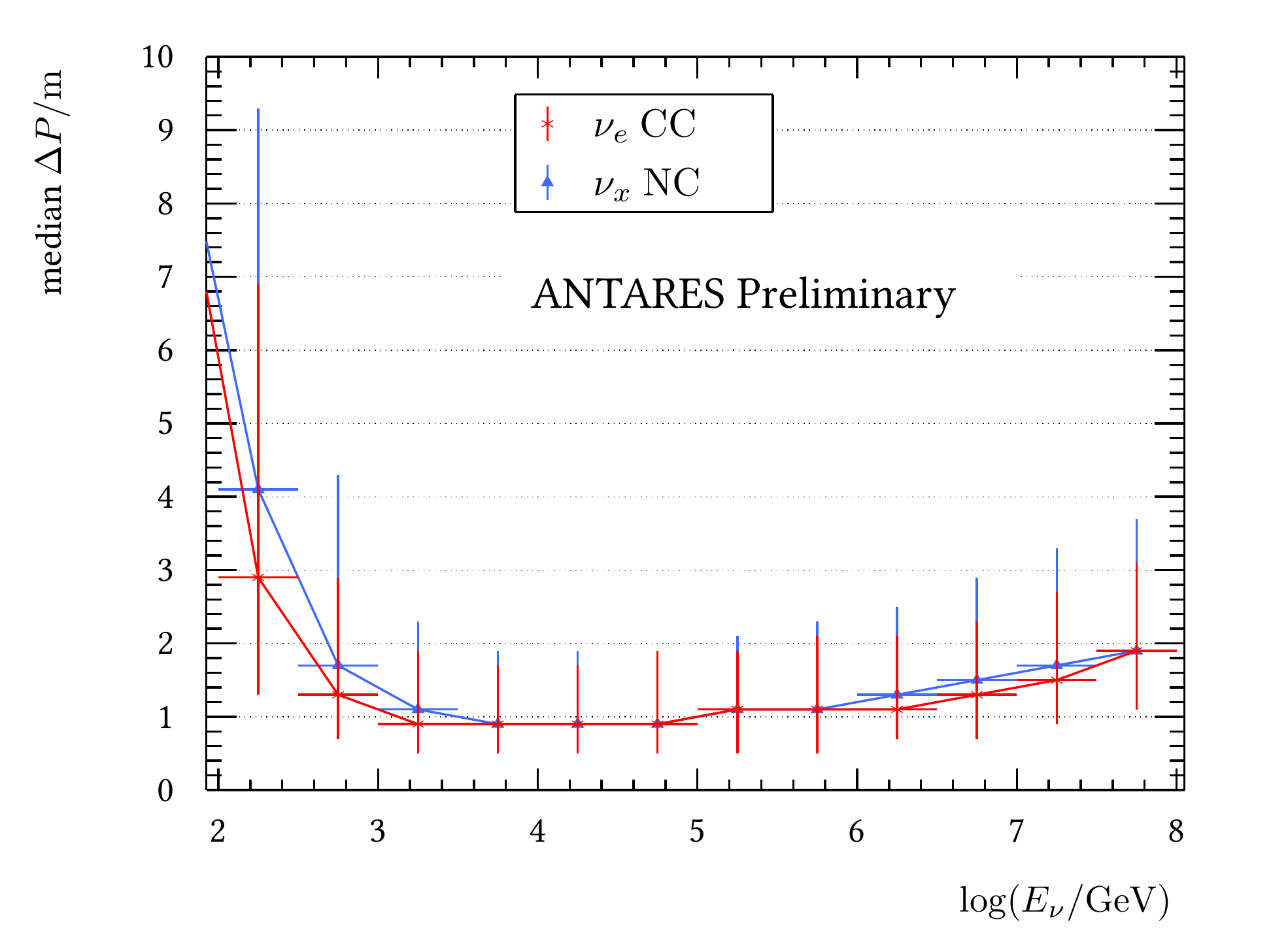}}
   \caption{\footnotesize Performance of the shower position reconstruction, red for electromagnetic showers, blue for hadronic showers, both after containment and error estimator cut (see section~\ref{sec:selection}), the purple line is the mean of the light emission spectrum for em-showers -- \textbf{Left:} The distance between the position of the neutrino interaction vertex and the reconstructed shower position along the neutrino axis. \textbf{Right:} The distance of the reconstructed shower position perpendicular to the neutrino axis.}
    \label{fig:PosPerform}
\end{figure}
\begin{figure}
\centerline{\includegraphics[width=0.5\textwidth]{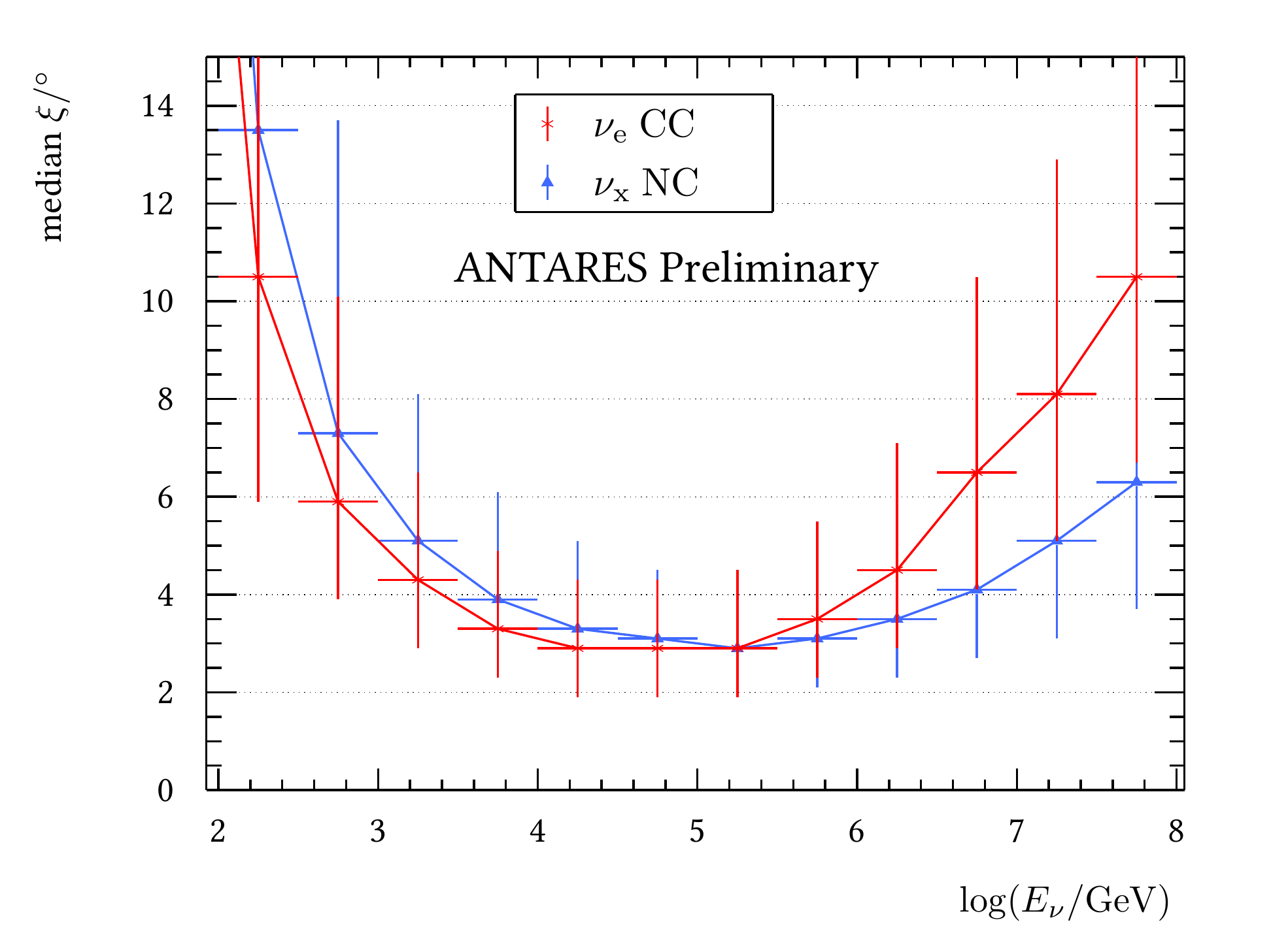}\includegraphics[width=0.5\textwidth]{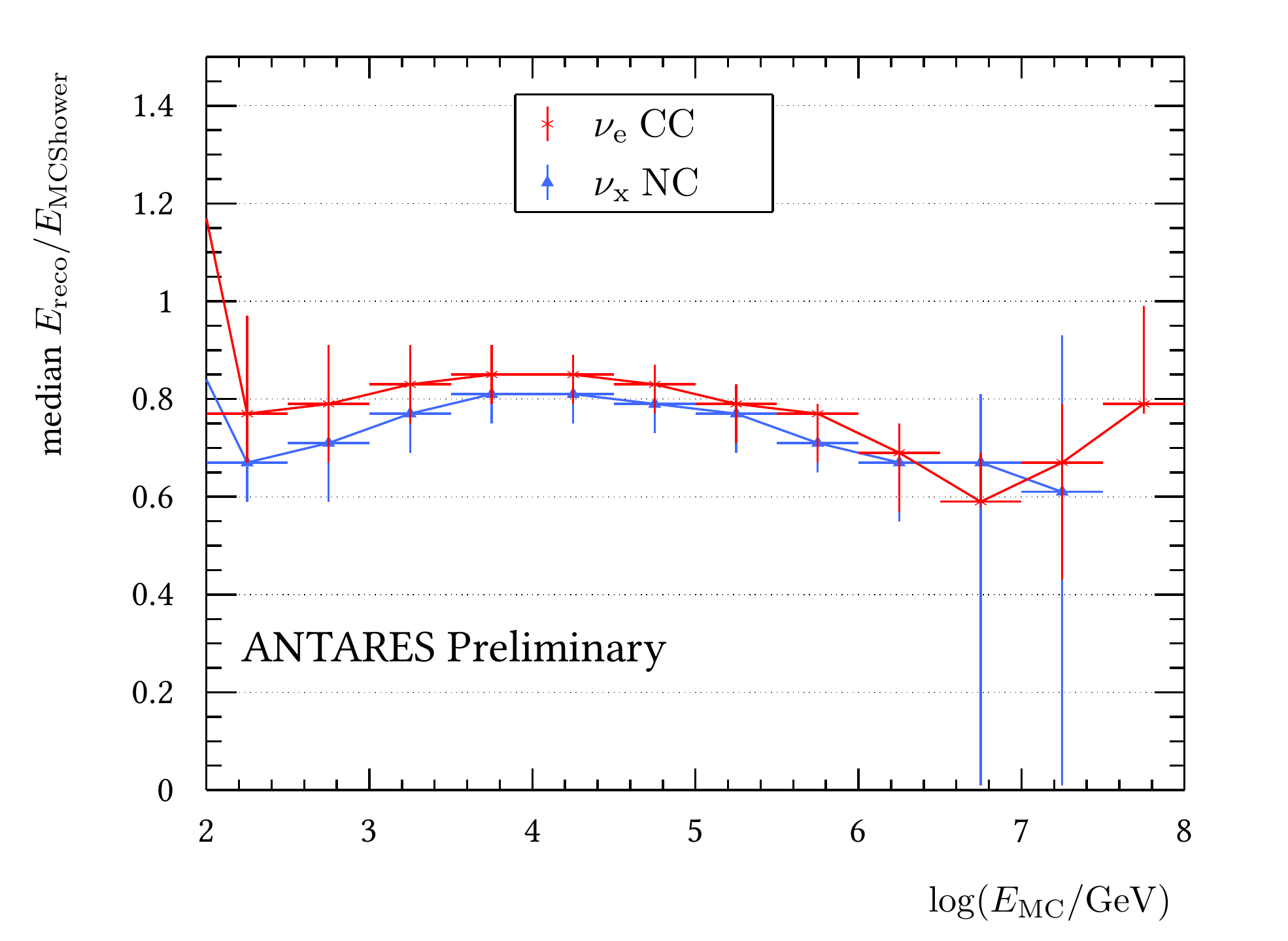}}
  \caption{\footnotesize Performance of the shower energy-direction reconstruction, red for electromagnetic showers, blue for hadronic showers, both after containment and error estimator cut (see section~\ref{sec:selection}) -- \textbf{Left:} The angle between the directions of the reconstructed shower and the Monte Carlo neutrino. \textbf{Right:} The ratio between the reconstructed energy and the Monte Carlo shower energy.}
    \label{fig:EnDirPerform}
\end{figure}


The angular resolution of the shower reconstruction is highly energy dependent. For energies $10^4 \lesssim E / GeV \lesssim 10^6$ it reaches median resolutions as low as $3^{\circ}$ with a $1\sigma$ lower spread below $2^{\circ}$. Below this energy range, not enough light is produced to illuminate sufficient PMTs for a proper reconstruction and above, most of the PMTs are saturated and the limited size of the ANTARES detector prevents us from accessing higher energies with proper resolutions. While not presently used in the point source search, it is worth mentioning that the statistical energy resolution of about $5\,\%$ has been achieved. A systematic underestimation of about $20\,\%$ can be observed over the whole energy range which is easily corrected post-reconstruction.
See figure~(\ref{fig:EnDirPerform}) for the performance of the direction (left) and energy (right) reconstruction.

 The angular resolution of the cascade reconstruction can also
 be measured in data using a sample of atmospheric muons which 
 also have a reconstructed cascade. If the reconstructed
 cascade corresponds to a true EM-shower which originates from
 the stochastic muon energy loss, the shower will have the same
 direction as the muon to a good approximation. As the muon is 
 accurately reconstructed by the track fit, a sample of EM 
 cascades of known direction can be isolated. Figure (\ref{fig:res_in_data})
 shows the result for a loose selection. A clear population of
 well reconstructed showers is visible; with a resolution of 
 two to three degrees (maximum of the distribution). This peak is well modelled
 in simulations of atmospheric muons\cite{MuPara}, which implies
 the Monte Carlo can be reliably used to determine the resolution
 for cascades of cosmic origin.

\begin{figure}
    \begin{minipage}{1.1\figsize}
        \begin{overpic}[width=1.1\figsize]
            {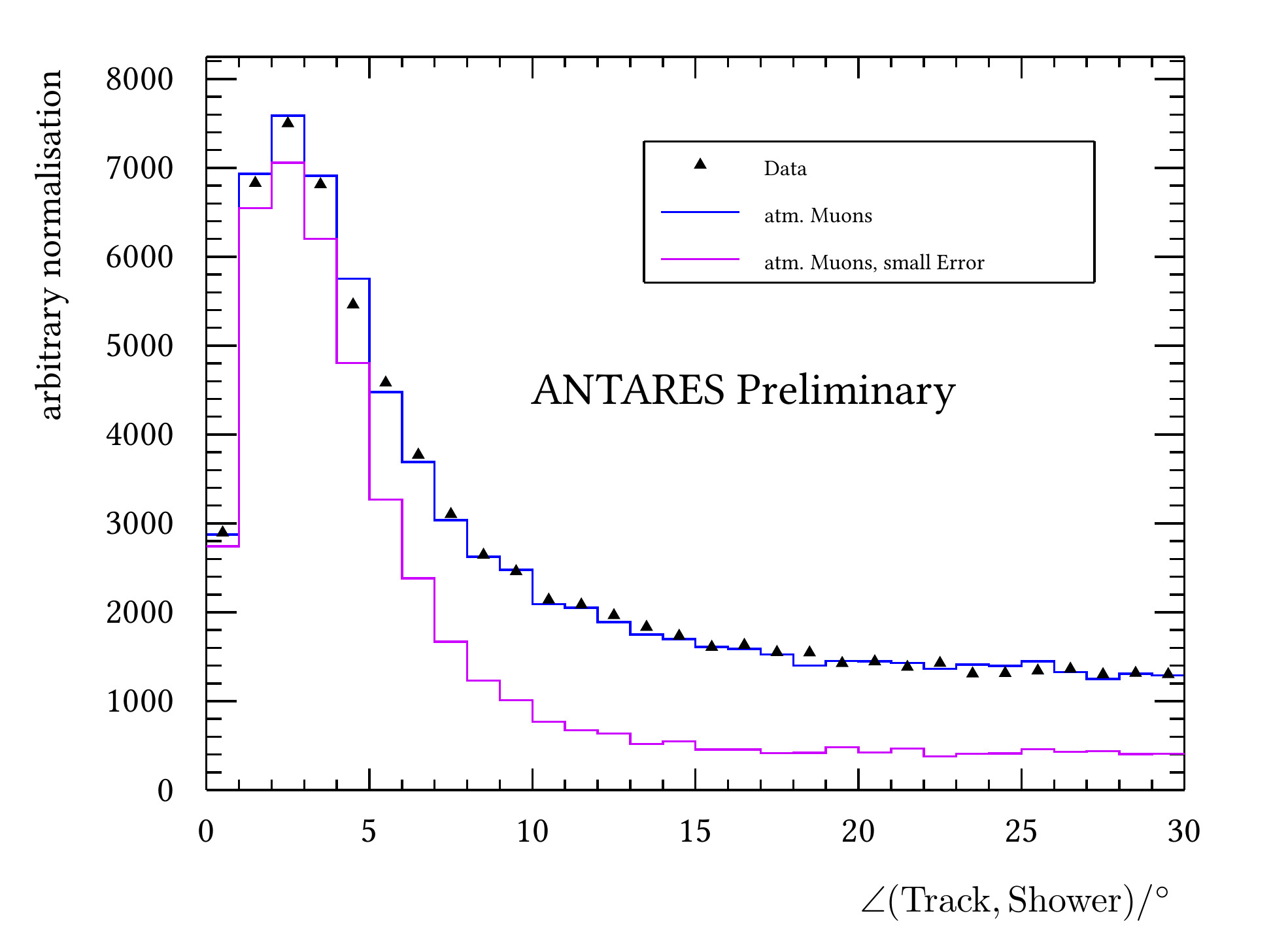}
        \end{overpic}
    \end{minipage}
    \begin{minipage}{.29\textwidth}
        \caption{\footnotesize The angular separation of the reconstructed directions using track and shower hypotheses applied to the same atmospheric muon events -- black: data, blue: atmospheric muons, violet: muons with a track reconstructed better than $5^{\circ}$.}
        \label{fig:res_in_data}
    \end{minipage}
\end{figure}

 The median energy-integrated angular resolution for an $E^{-2}$ charged-current $\nu_e$ signal 
 for the selection adopted in the analysis is $3$ degrees.

 A very similar shower reconstruction algorithm is being used for KM3NeT, which also achieves angular resolutions of $\mathcal O(1^{\circ})$.

\subsection{Selection and data sample}
\label{sec:selection}
 The selection of tracks (i.e. $\nu_\mu$ candidates) is identical to \cite{lastPS}. It requires tracks to be up-going ($\cos(\vartheta) > -0.1$), with a small estimated angular error ($\beta < 1^{\circ}$) and with a minimum reconstruction quality parameter ($\Lambda > -5.2$).

 Cascade candidates are selected using a set of criteria aimed at rejecting background from atmospheric muons, which are misreconstructed as up-going cascades -- too many to describe them all in detail here. The selection requires:\\

 {\centering
 \begin{tabular}{l}
    \textbullet\ the event not to be selected by the track channel,\\
    \textbullet\ reconstructed as up-going ($\cos(\vartheta) > -0.1$)\\
    \textbullet\ the shower position to be close to the detector ($\varrho < 300m, |z| < 250m$),\\
    \textbullet\ a maximal angular error estimate ($< 10^{\circ}$),\\
    \textbullet\ passing a combined cut on the GridFit Ratio\cite{EVisser} and number of selected hits,\\
    \textbullet\ passing a muon/em-shower likelihood discrimination specifically developed for this analysis,\\
    \textbullet\ a sufficiently low ratio between ``early'' and ``on-time'' charge\\
\end{tabular}\\}
\ \\
The used data period from 2007 to 2013 with a life time of 1622 days contains 6261 muon track candidates, $10\,\%$ of which is estimated to be atmospheric muons. A total of 156 cascade events are selected; this sample is estimated to consist to $90\,\%$ of atmospheric neutrinos, while the rest are atmospheric muons.

For an $E^{-2}$ signal flux with 1:1:1 flavour composition, the selected cascade events are expected to increase in signal event rate by 30\%.

\section{Search method}
The signature of a point source is a cluster of events.
The distribution of the angle of deviation between
the reconstructed signal event and the location of the 
source is described by the point spread function ${\cal F}(\gamma)$, 
which is the probability density of reconstructing 
an event at an angular distance $\gamma$ from the true source. 
In order to distinguish this signature from random 
clusters of background events, we use a likelihood
ratio. 
It is convenient to express the intensity of the
source in terms of the mean number of detected events
that the source produces: $\mu_{\rm sig}$. The likelihood
of the data is given by: 
\begin{equation}
    \log {\cal L}_{\rm s+b} = \sum_i \log [ \mu_{\rm sig} 
    \times {\cal F}(\gamma_i)\times \mathcal N_\mathrm{sig}(N^\mathrm{Hits}_\mathrm{i}) + {\cal B}(\delta_{\rm i})\times {\cal N}_\mathrm{backg}(N^\mathrm{Hits}_\mathrm{i}) ] - \mu_{\rm tot}, 
    \label{eq:lik1}
\end{equation} 
 where $\gamma_i$ is the angle between the reconstructed
 direction and the assumed source coordinates. $\cal N$ is the distribution for the number of selected hits for the signal / background case.
 ${\cal B}_i$ is the rate of background events at the coordinates of event $i$. For simplicity,
 we consider the background rate to be a function of
 declination. 
 The term $\mu_{\rm tot}$ represents the total number of expected events.
 The sum in the likelihood takes muon track as well as shower events 
 into account and uses the proper ingredients for $\mathcal F_i$ and $\mathcal B_i$.
 Since events that are very far away from the source position 
 yield a constant contribution and will not influence 
 maximum likelihood estimates, the sum can be restricted to a reasonably small 
 cluster of events around the hypothesized source position.
 An analogous argument allows replacing $\mu_{\rm tot}$ with  $\mu_{\rm sig}$
 in equation~(\ref{eq:lik1}).

 The first step to compute the likelihood ratio is to fit 
 the three free parameters ($\mu_{\rm sig}, \delta_s,\alpha_s$)
 in the signal hypothesis 
 to the cluster.
 In case of a fixed-point search, the coordinates are fixed
 and the fit has only $\mu_{\rm sig}$ as a free parameter.
 A selection of IceCube muon candidates has been adopted as point source candidates\cite{ICTracks}. Since those events have angular error estimators between one and two degrees, we do not treat them like the usual point source candidates with a fixed position but also fit the direction within a cone of $2^{\circ}$.

 Finally, to distinguish signal-like clusters from clusters
 produced by background, we compute the likelihood ratio $Q$:
\begin{equation}
Q = \log {\cal L}_{\rm s+b}^{\rm max} -  \log {\cal L}_{\rm b},
\end{equation}
 where the first term is the likelihood evaluated for the
 best-fit parameters and the second term is equation~(\ref{eq:lik1})
 evaluated for $\mu_{\rm sig}=0$. As we will use $Q$ to differentiate
 between signal and background, it is also called the \emph{test statistic}.

\section{Sensitivity and discovery potential}
The detector sensitivity and discovery potential can be determined with pseudo experiments. For this, various numbers of signal events are injected at a fixed position (distributed according to the point spread function ${\cal F}$) on top of a background as found in real data.
\subsection{Full sky search}
Figure~(\ref{fig:full_fitdir}) shows the fitted right ascension for various numbers of injected signal at a fixed position in the sky ($\alpha = 100^{\circ}, \delta = -70^{\circ}$).
Figure~(\ref{fig:full_disc}) shows the flux needed in a full sky search to have a $5\sigma$ discovery in $50\,\%$ of hypothetical, equivalent experiments.

\subsection{Candidate List Search}
Figure~(\ref{fig:fit_nsig}) shows the number of signal event found by the likelihood fit for different numbers of injected signal events. The fit tends to slightly overestimate the amount of injected signal by about half an event.
Figure~(\ref{fig:fix_disc}) shows the flux needed in a candidate list search to have a $5\sigma$ discovery in $50\,\%$ of hypothetical, equivalent experiments.
Figure~(\ref{fig:fix_sens}) shows the flux that can be excluded with a confidence level of $90\,\%$ in case no signal events could be found. The expected sensitivity for the fixed point search is $10^{-8}\cdot E^{2}$GeV/cm$^2$/s for declinations below $-40^{\circ}$.

\section{Results}
No discovery can be claimed -- neither in the full sky nor in the candidate list search. The following subsections show the significances for the different search methods -- figure~(\ref{fig:mostsigclus}) shows their respective most significant clusters.

\subsection{Full Sky Search}
The most significant cluster in the full sky search is very close to the one in the previous track-only analysis. It is located at $\alpha=-48.3^{\circ}, \delta=-64.6^{\circ}$ (old track-only analysis: $\alpha=-46^{\circ}, \delta=-65^{\circ}$).
    Within $3^{\circ}$ 16 tracks were found and 1 shower within $10^{\circ}$. The fitted number of signal events is $N_\mathrm{Sig} = 5.5 + 0.8$ (Tracks + Showers). The measured p-value is  18.5\,\% which corresponds to a significance of $1.33\sigma$. In the previous analysis, this cluster had a significance of $2.17\sigma$ which suggests that adding the shower channel exposes this cluster as a mere over-fluctuation in the track channel.

\newpage
\begin{figure}[h]
\centering
    \begin{subfigure}{.4\textwidth}
        \resizebox{\textwidth}{!}{
            \includegraphics[width=\textwidth]{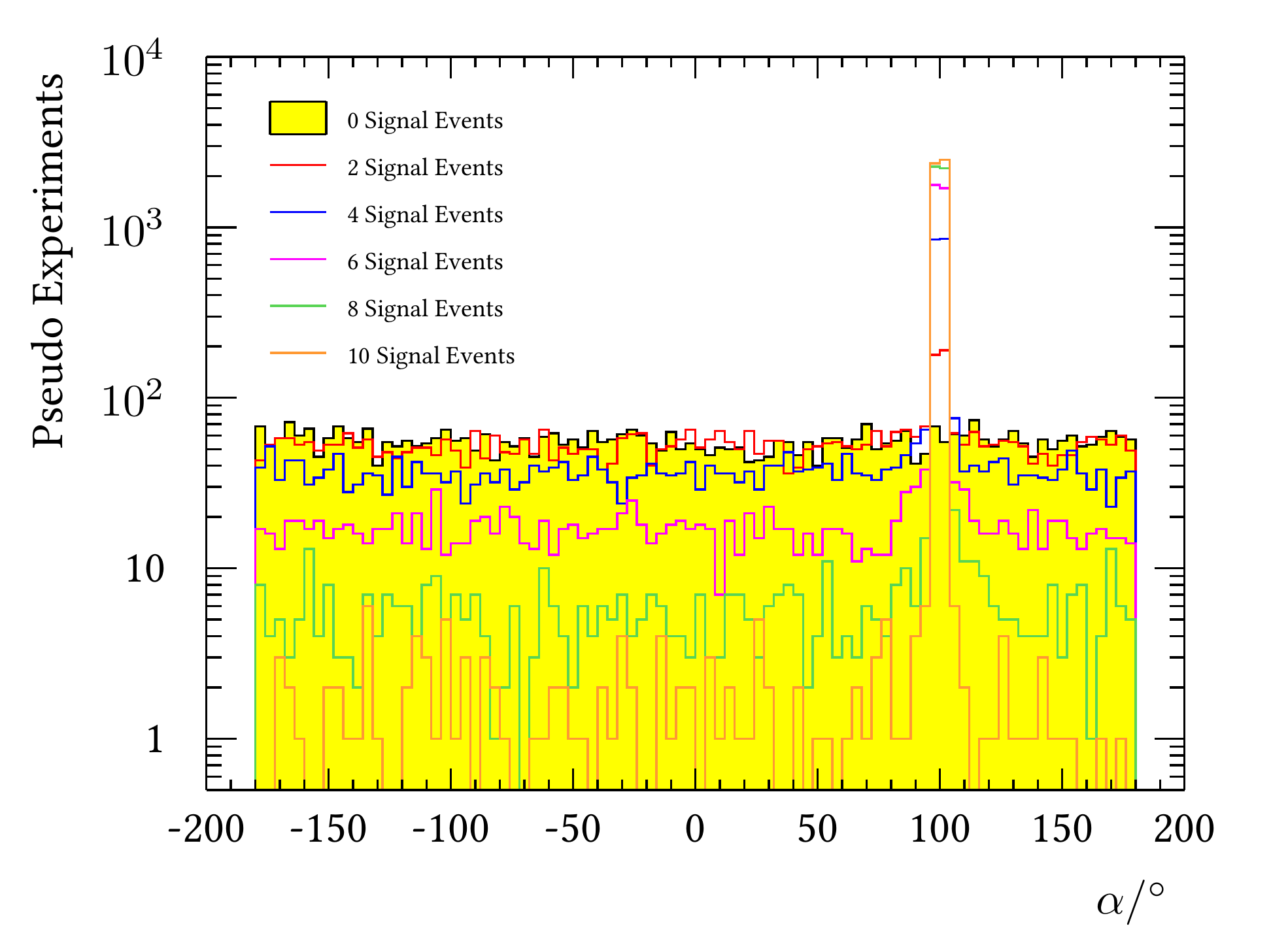}
        }
        \put(-110,110){\footnotesize ANTARES preliminary}
        \caption{\footnotesize fitted right ascension for pseudo experiments}
        \label{fig:full_fitdir}
    \end{subfigure}
    \begin{subfigure}{.4\textwidth}
        \resizebox{\textwidth}{!}{
            \includegraphics[width=\textwidth]{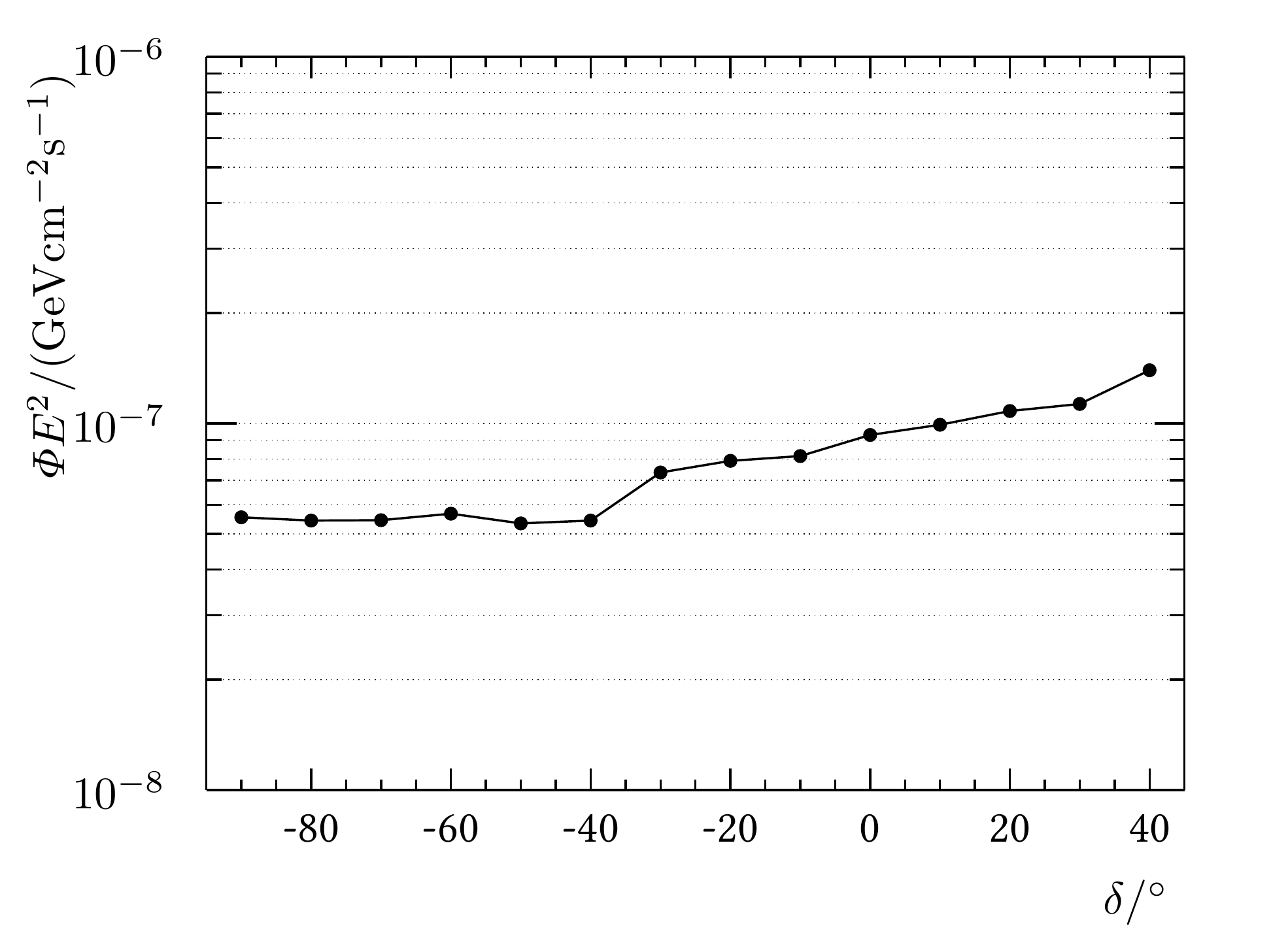}
        }
        \put(-150,43){\footnotesize ANTARES preliminary}
        \caption{\footnotesize discovery flux for full sky search}
        \label{fig:full_disc}
    \end{subfigure}
    \caption{\footnotesize (a) Fitted right ascension for pseudo experiments for various numbers of signal events injected at $\delta=-70^{\circ}$ and $\alpha=100^{\circ}$.
        (b) The flux necessary for a $50\,\%$ probability for a $5\sigma$ discovery in a full sky search.}
\end{figure}
\begin{figure}[h]
    \begin{minipage}{.4\textwidth}
        \centering
        \resizebox{\textwidth}{!}{
           \includegraphics[width=\textwidth]{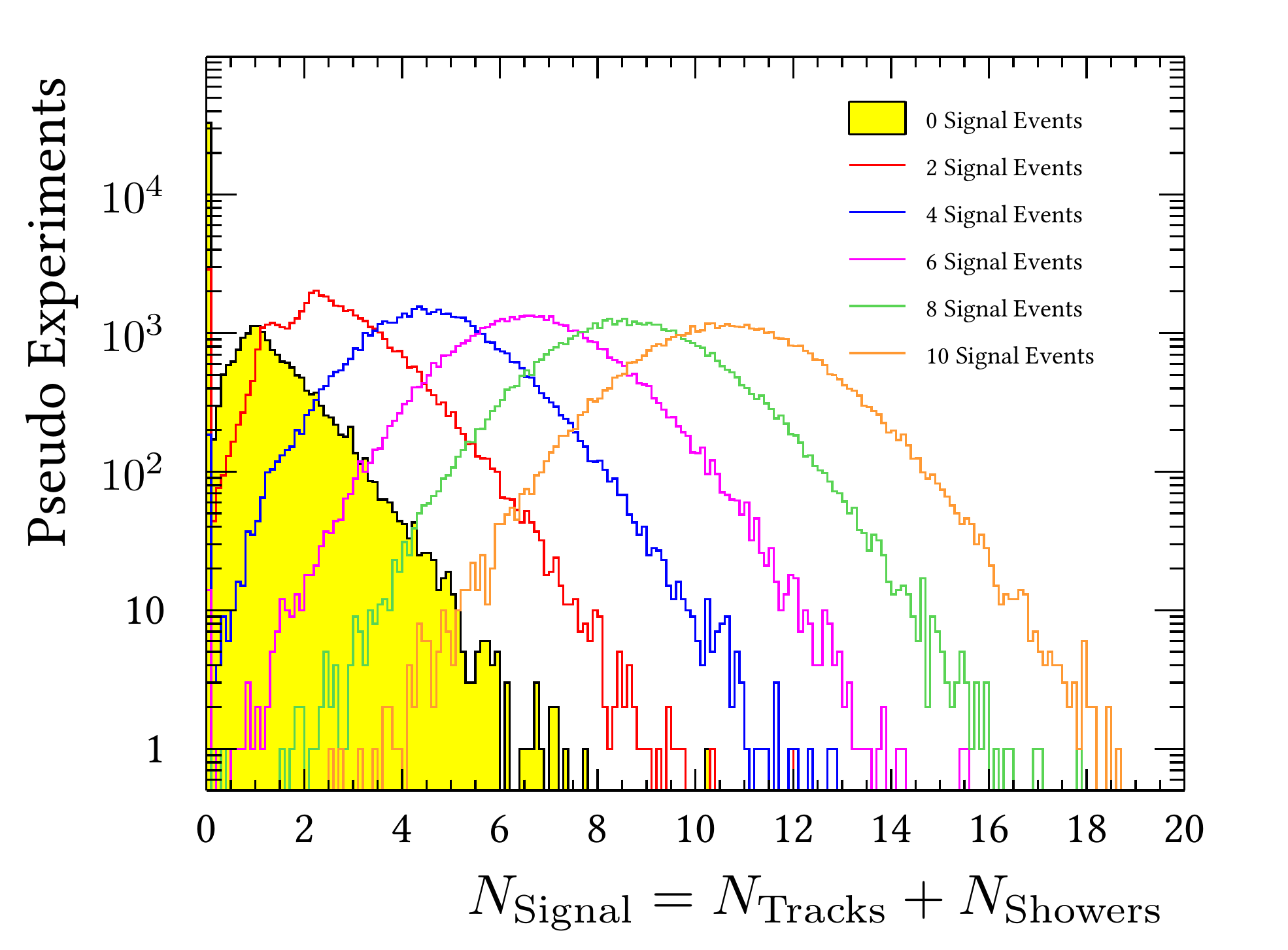}
        }
        \put(-150,110){\footnotesize ANTARES preliminary}
        \subcaption{\footnotesize number of fitted signal events}
        \label{fig:fit_nsig}
    \end{minipage}
    \begin{minipage}{.4\textwidth}
        \centering
        \resizebox{\textwidth}{!}{
            \includegraphics[width=\textwidth]{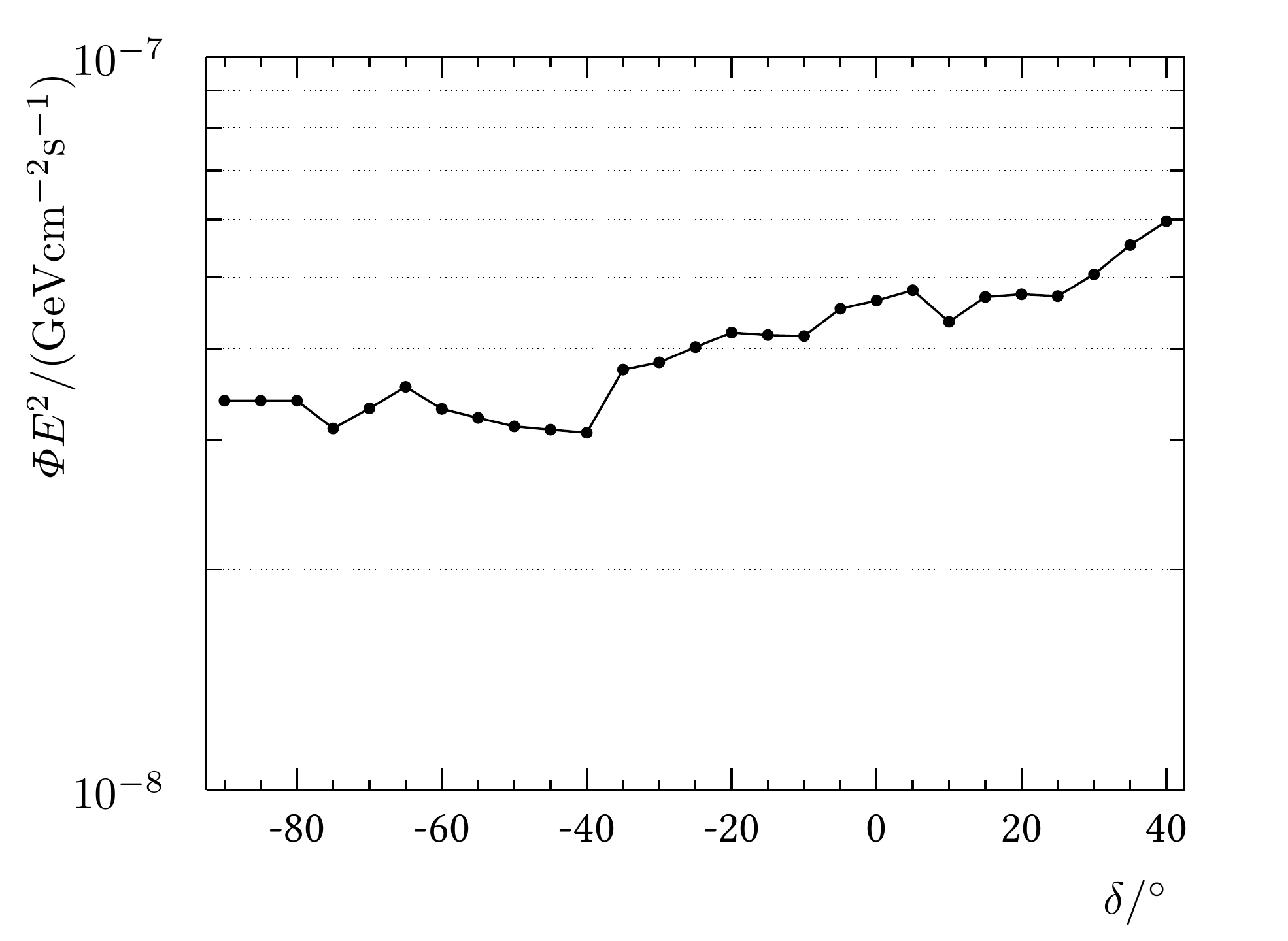}
        }
        \put(-150,43){\footnotesize ANTARES preliminary}
        \subcaption{\footnotesize discovery flux for fixed search}
        \label{fig:fix_disc}
    \end{minipage}
    \begin{minipage}{.5\textwidth}    
        \centering
        \resizebox{\textwidth}{!}{
            \includegraphics[width=\textwidth]{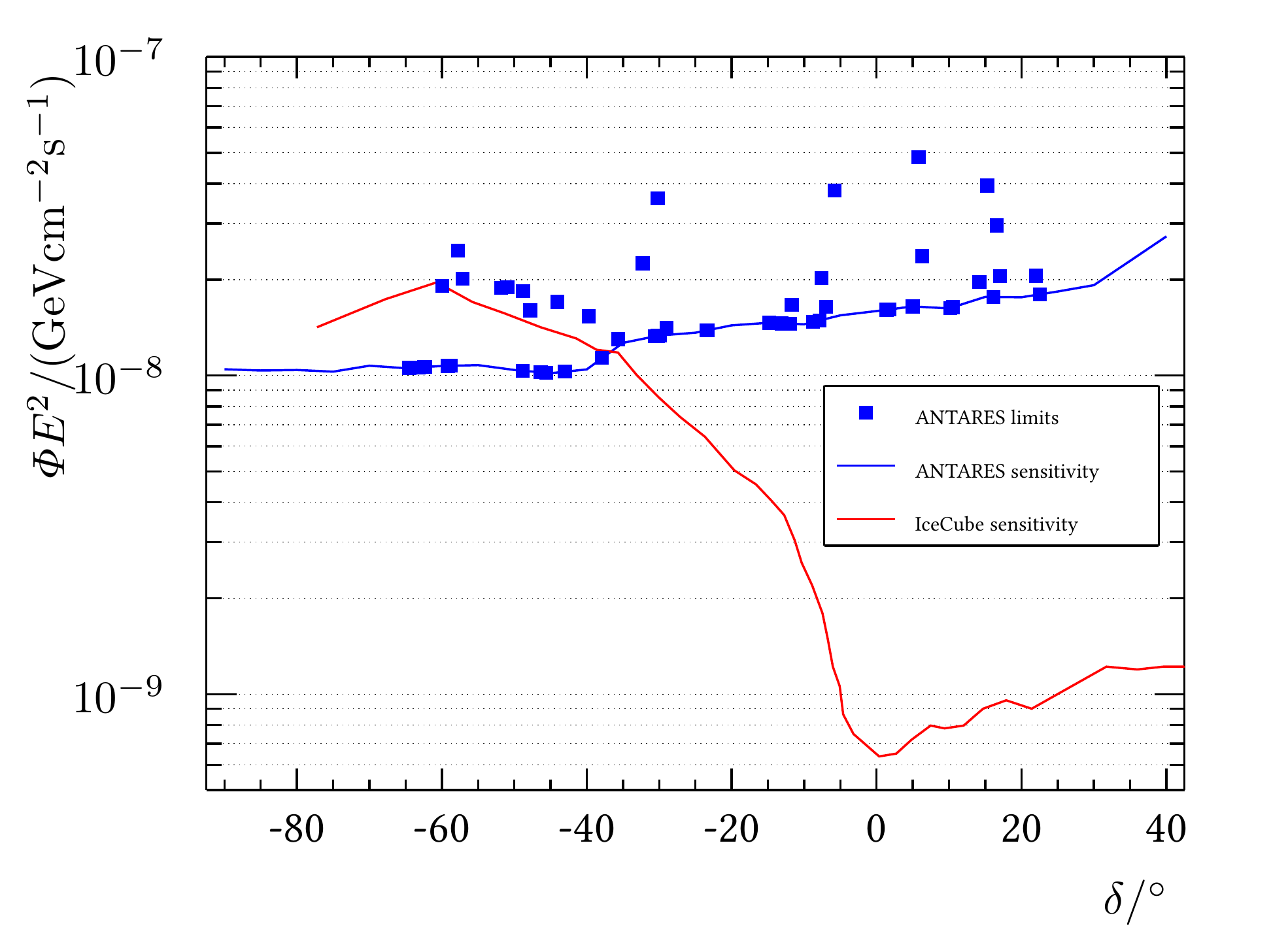}
        }
        \put(-180,43){\footnotesize ANTARES preliminary}
        \subcaption{\footnotesize sensitivity for the fixed search}
        \label{fig:fix_sens}
    \end{minipage}
    \hfill
    \begin{minipage}{.4\textwidth}    
        \caption{\footnotesize (a) The number of fitted events (Tracks + Showers) for different numbers of injected signal at $\delta=-70^{\circ}$ and $\alpha=100^{\circ}$ -- (b) The flux needed to claim a $5\sigma$ discovery in $50\,\%$ of the cases -- (c) The sensitivity for the fixed point search: blue for ANTARES, red for IceCube as comparison.}
    \end{minipage}
    \hfill {}
\end{figure}
                
                \newpage
\subsection{Candidate List Search}
The cluster with the highest significance in the candidate list search is HESSJ0632+057 ($\alpha_\mathrm{s}=98.24^{\circ},\delta_\mathrm{s}=5.81^{\circ}$) -- the same source as in the last analysis using only tracks.
With 36 tracks and 0 showers within $10^{\circ}$ around the source, the fit found $N_\mathrm{Sig} = 1.2 + 0.2$ (Tracks + Showers) signal events, corresponding to a significance of $0.75\sigma$.

\subsection{IceCube Candidate Search}
The IceCube muon track candidate with the highest significance is the event with the IceCube ID 28 ($\alpha_\mathrm{IC}=164.8, \delta_\mathrm{IC}=-71.5, \beta_\mathrm{IC}=1.3$\footnote{$\beta\rightarrow$ angular error estimate}).
7 tracks have been found within $3^{\circ}$ and 0 showers within $10^{\circ}$. The fitted signal is $N_\mathrm{Sig} = 0.005 + 0.001$ with a significance of $0\sigma$.
    
\begin{figure}[h]
    \begin{minipage}{.5\textwidth}
        \centering
        \includegraphics[width=\textwidth]{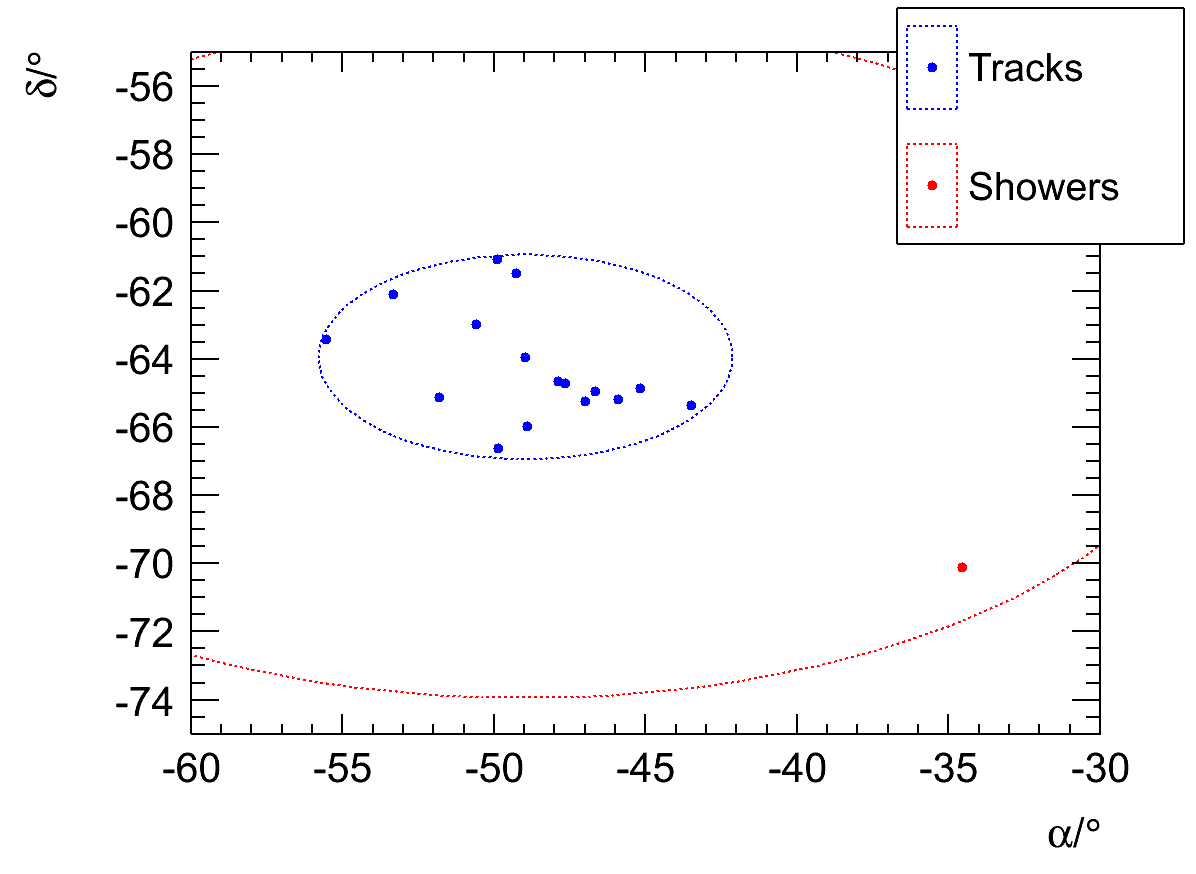}
        \put(-170,130){ANTARES preliminary}
        \subcaption{\footnotesize full sky search}
    \end{minipage}
    \begin{minipage}{.5\textwidth}
        \centering
        \includegraphics[width=\textwidth]{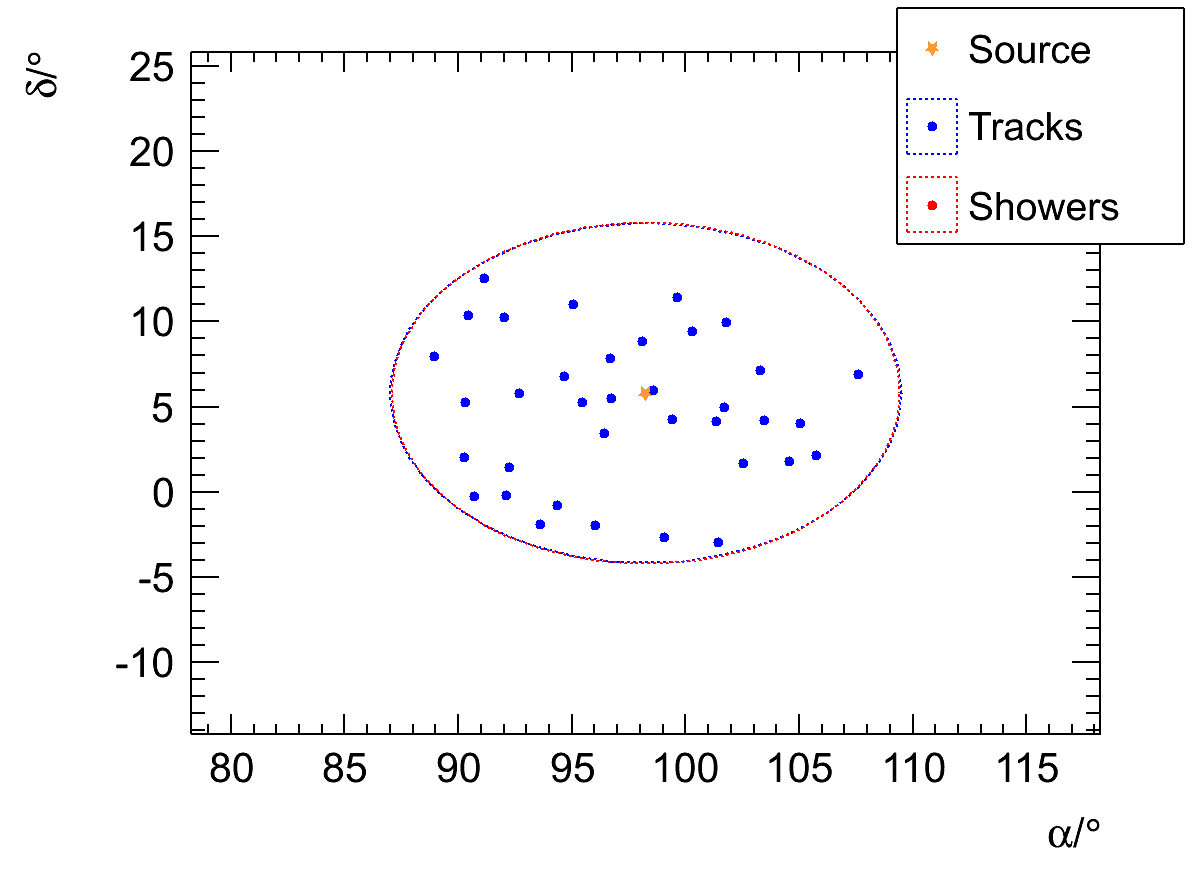}
        \put(-170,130){ANTARES preliminary}
        \subcaption{\footnotesize candidate list search}
    \end{minipage}
    \begin{minipage}{.5\textwidth}
        \centering
        \includegraphics[width=\textwidth]{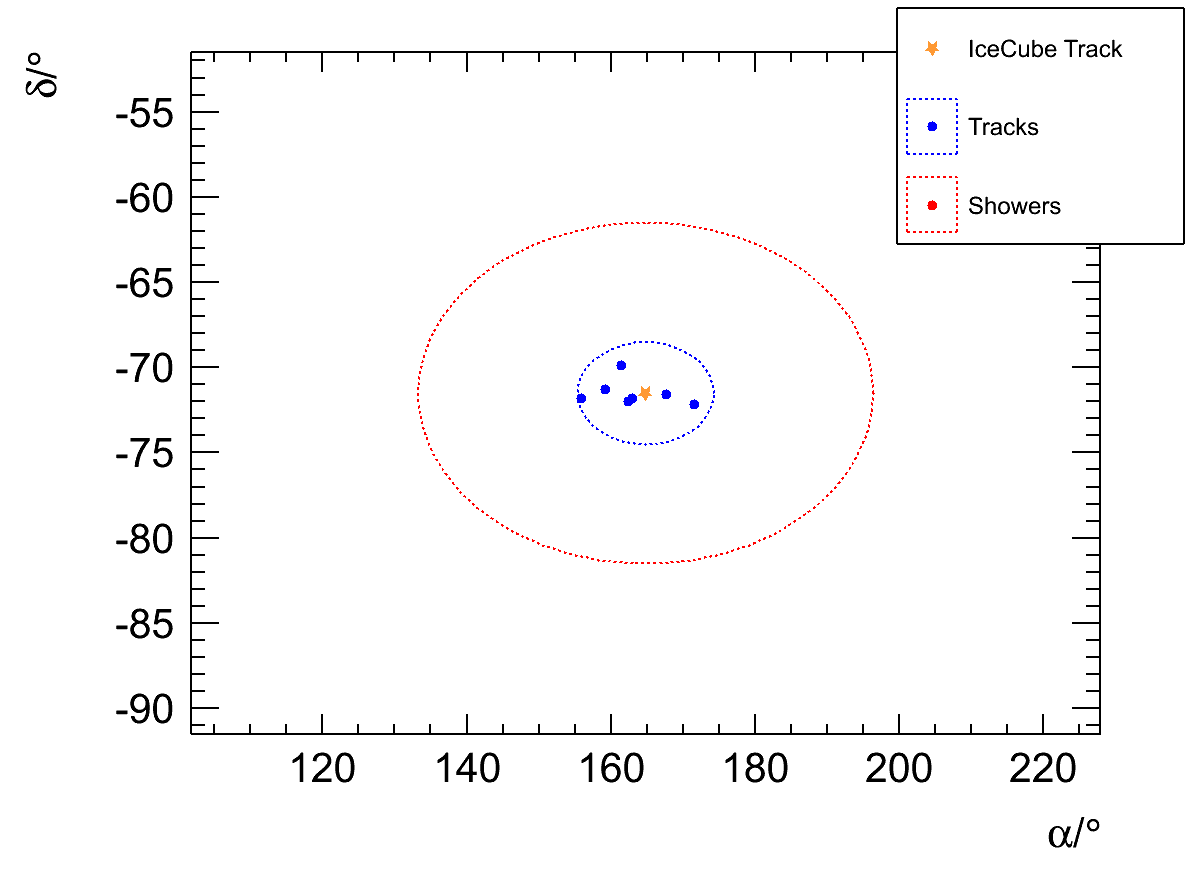}
        \put(-170,130){ANTARES preliminary}
        \subcaption{\footnotesize IceCube candidate search}
    \end{minipage}
    \hfill
    \begin{minipage}{.4\textwidth}
        \caption{\footnotesize Most significant clusters for different search methods: (a) full sky search (b) candidate list search (c) IceCube candidate search. The dashed lines mark the range in which events are considered for the current cluster.}
        \label{fig:mostsigclus}
    \end{minipage}
    \hfill {}
\end{figure}


\setcounter{figure}{0}
\setcounter{table}{0}
\setcounter{footnote}{0}
\setcounter{section}{0}
\setcounter{equation}{0}

\newpage
\id{id_cperrina}
\addcontentsline{toc}{part}{\textcolor{blue}{\arabic{IdContrib} - {\sl C. Perrina} : Search for point-like neutrino sources above the horizon with the ANTARES Neutrino Telescope}%
}



\title{\arabic{IdContrib} - Search for point-like neutrino sources above the horizon with the ANTARES Neutrino Telescope}

\shorttitle{\arabic{IdContrib} - Search for point-like neutrino sources above the horizon with ANTARES}

\authors{Chiara Perrina}
   \afiliations{"La Sapienza" University of Roma and INFN, Italy}
\email{chiara.perrina@roma1.infn.it}


\abstract{Installed in the Mediterranean Sea, at a depth of $\sim 2.5$ km, ANTARES is the largest undersea neutrino telescope currently operating. The search for point-like sources with neutrino telescopes is normally limited to a fraction of the sky, due to the selection of events where the direction of the neutrino candidate has been reconstructed as coming from below the horizon, usually referred to as ``up-going'' events, in order to significantly reduce the atmospheric muons background. Here we demonstrate that the background can be effectively suppressed through an energy and direction dependent event selection so that a part of the region above the horizon can be included in the search. The strategy for the study of a ``down-going'' neutrino flux is described and the ANTARES sensitivity for two candidate sources is presented.}


\maketitle

\section{Introduction}
\label{section:Introduction}
ANTARES, placed on the bottom of the Mediterranean Sea, $\thicksim 40$ km south-east from the coast of Toulon (France), is the first undersea neutrino telescope and the only one currently operating. Its main purpose is the search for neutrino fluxes from astrophysical objects. Its observation is based on the detection of the Cherenkov radiation induced by the passage in water of superluminal charged particles produced by the interaction of cosmic neutrinos near the detector by means of 885 photomultiplier tubes. For detailed information about the detector, refer to \cite{Collaboration:2011nsa}. 
 
The search for a point-like source of cosmic neutrinos consists in the search for a directional clustering of events. A source can be identified as a significant excess of muon tracks from a given location compared to the surrounding region dominated by the isotropic background of atmospheric neutrinos and muons. The sensitivity depends on the suppression of the background to a level at which event accumulations for expected source fluxes are visible over the statistical background fluctuations. With an assumption on the spectral shape of a given source it is possible to use the estimated energy of events as a parameter to separate signal from background, since the signal spectrum is expected to be harder than the atmospheric background one. 
In this contribution the analysis of down-going events, \textit{i.e.} events coming from above the ANTARES horizon is presented. A big challenge in this analysis is offered by atmospheric muons which can penetrate through several kilometres of water to the detector, providing the major component of the background. To retain sensitivity to a neutrino signal flux, it is thus necessary to boost the rejection power. This can be achieved by using a good energy estimator (see Sec. \ref{sec:Event selection and simulation}) and a good signal/background separation technique (see Sec. \ref{subsec:Event selection}). The search for neutrino candidates in the resulting final sample will be based on spatial information in order to derive significance for event clusters, as discussed in Sec. \ref{sec:Search method}. A candidate-list search, looking for events in the direction of two candidate sources which are known gamma-ray emitters and potential sites for hadronic acceleration, has been performed. The sensitivity of the detector to a neutrino flux $\propto E_{\nu}^{-2}$ coming from the sources has been computed (see Sec. \ref{sec:Search method}). 

\section{Data and simulation}
\label{sec:Event selection and simulation}

The analysis presented here has been developed using the data collected by ANTARES between June 2009 and June 2011. This measurement period corresponds to a total live-time of 366.6 days. Triggered events are reconstructed using the time and position information of the hits by means of a maximum likelihood (ML) method. The algorithm consists of a multi-step procedure to fit the direction of the reconstructed muon by maximizing the ML-parameter, $\Lambda$, which describes the quality of the track reconstruction (\cite{AdrianMartinez:2012rp}). Neutrinos and atmospheric muons are simulated with the GENHEN and MUPAGE (\cite{Carminati:2008qb, Bazzotti:2010zz}) packages, respectively. Furthermore, the propagation of the muon tracks is simulated with the KM3 package (\cite{Becherini:2006bf}). Two candidate sources have been considered (see Table \ref{tab:sources}). A neutrino flux coming from their directions with an $\propto$ $E_{\nu}^{-2}$ (\cite{Vissani:2011ea}) has been simulated.

\begin{table}[h]
\centering
\hspace*{-1cm}
\begin{tabular}{l l l l l}
\hline
\textbf{Object} & \textbf{$b$ (deg)} & \textbf{$l$ (deg)} & \textbf{$\delta$ (deg)} & \textbf{$\alpha$ (deg)} \\
\hline

Tycho              &1.45  &120.11     &64.18          &6.36                 \\
CTA 1              &10.40 &119.60     &72.98           &1.61                 \\

\hline
\end{tabular}
\hspace*{-1cm}
\label{tab:sources}
\caption{Candidate source list. From the second to the fifth column the galactic latitude ($b$), the galactic longitude ($l$), the declination ($\delta$) and the right ascension ($\alpha$) in decimal degrees.}
\end{table}

The distribution of data and Monte Carlo signal (generated from CTA 1) and background events for the $\Lambda$ parameter can be seen in Figure \ref{fig:l_all}.

\begin{figure}[!h]
\begin{center}
\includegraphics[width=0.5\columnwidth]{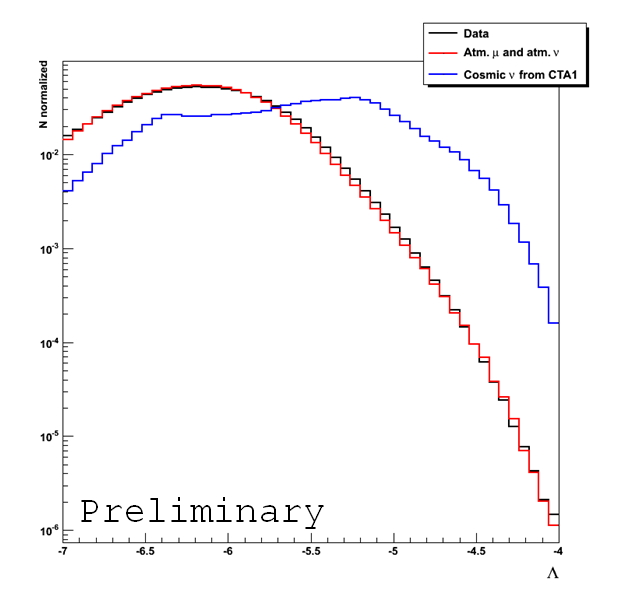}
\caption{Data and Monte Carlo events distribution for the track reconstruction quality parameter, $\Lambda$. Only down-going tracks have been considered. The  simulation of atmospheric neutrinos uses the Bartol flux. Larger values of the $\Lambda$ parameter indicate a better track reconstruction.}
\label{fig:l_all}
\end{center}
\end{figure}

The directional reconstruction resolution can also be characterized in terms of the width of the two-dimensional distribution of the angular deviation of reconstructed track directions from the true track direction. This so-called ``point-spread function'', expressed in spherical detector coordinates (Zenith and Azimuth) such that all bins span equal solid angles, is shown in Figure \ref{fig:psf}.

\begin{figure}[!h]
\begin{center}
\includegraphics[width=0.5\columnwidth]{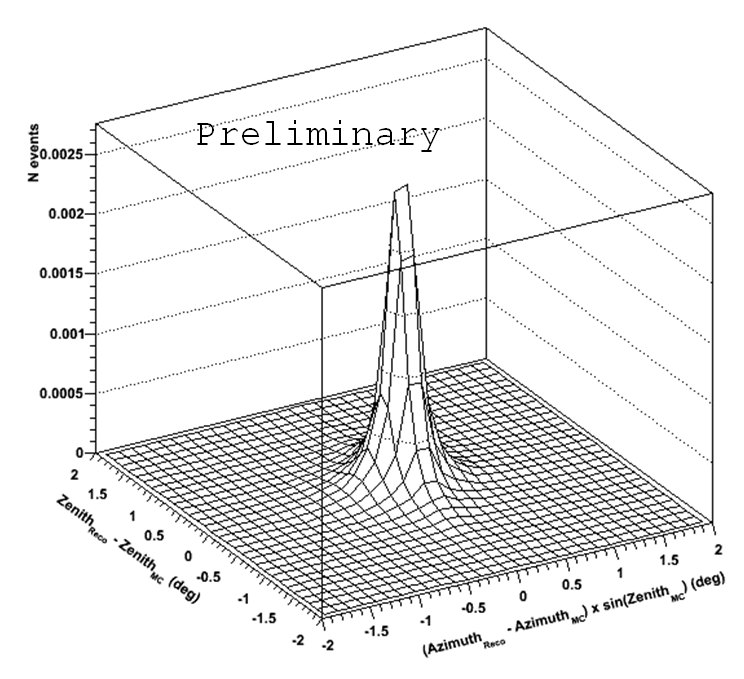}
\caption{Point-spread function in detector coordinates. The full Monte Carlo signal event sample of neutrino-induced muons from CTA 1 was used after applying the $\Lambda > -6.0$ cut.}
\label{fig:psf}
\end{center}
\end{figure}

In Figure \ref{fig:una} the neutrino energy estimator for Monte Carlo signal events as a function of the true neutrino energy is shown.

\begin{figure}[!h]
\begin{center}
\includegraphics[width=0.5\columnwidth]{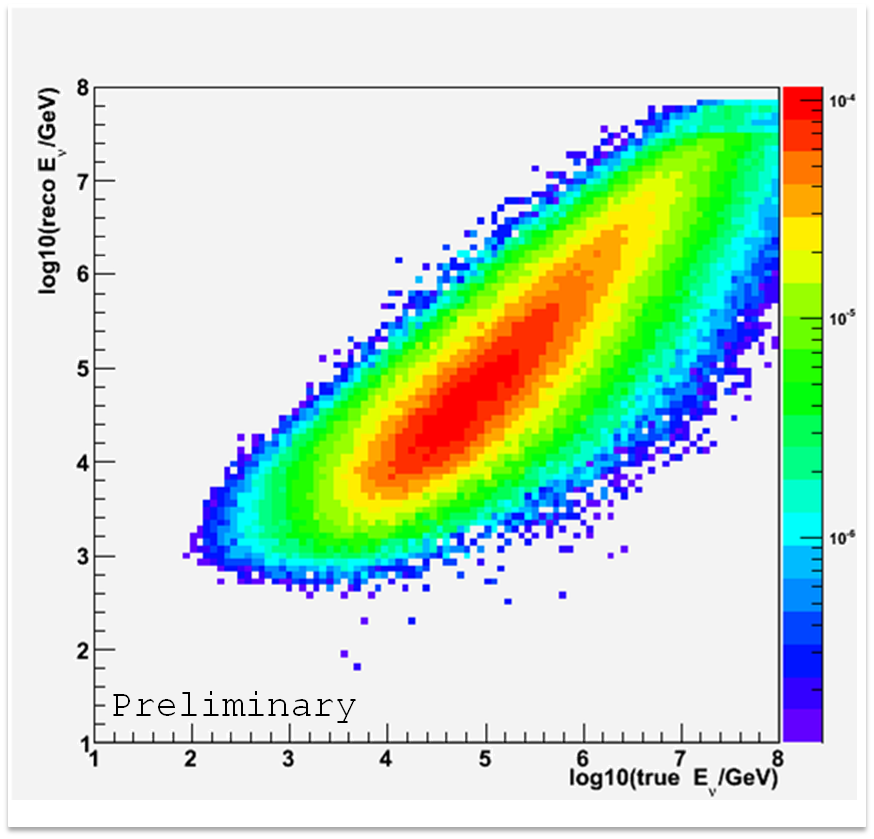}
\caption{Monte Carlo-generated signal events distribution for the neutrino energy estimator as a function of the true neutrino energy.}
\label{fig:una}
\end{center}
\end{figure}

\subsection{Event selection}
\label{subsec:Event selection}

In order to achieve the goal of this analysis a good rejection of the background is fundamental. For this purpose a multivariate analysis based on the BDT (Boosted Decision Tree) technique has been implemented. The variables used for the BDT training are: the $\Lambda$ parameter, the zenith angle and the reconstructed energy (reco $E_{\nu}$) of an event.

Fig. \ref{fig:variables} shows the distribution of the down-going Monte Carlo-generated signal events and the atmospheric background for the three variables which have been used for the BDT training. 

\begin{figure}[h]
\centering\includegraphics[width=0.7\linewidth]{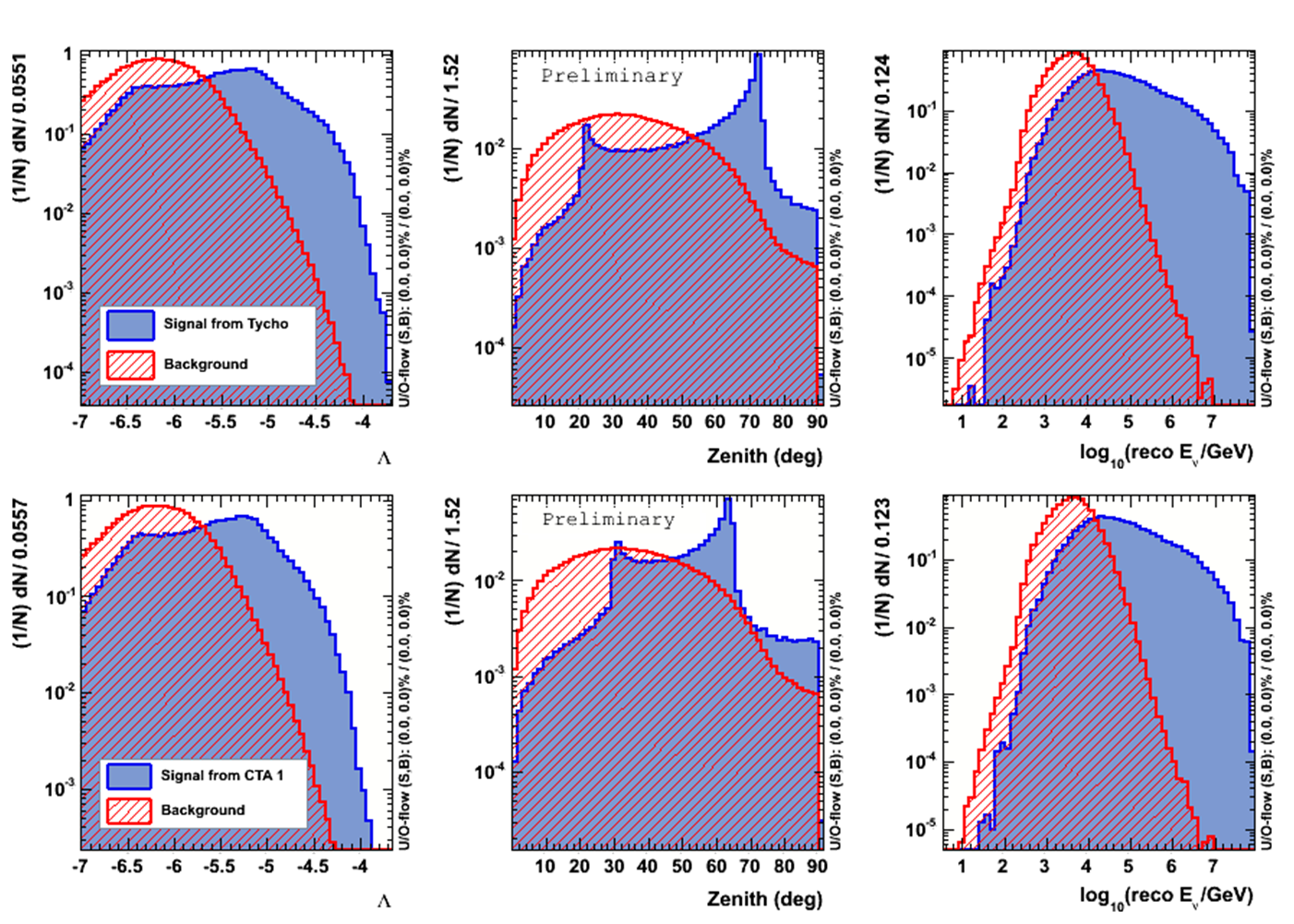}
\caption{Distribution of the down-going Monte Carlo-generated signal events (events generated from Tycho (top) and CTA 1 (bottom)) and background atmospheric events for the variables used in the BDT training: the ML-parameter ($\Lambda$), the zenith angle and the reconstructed neutrino energy (reco $E_{\nu}$.)}
\label{fig:variables}
\end{figure}

\section{Search method}
\label{sec:Search method}

A binned point source search has been performed. It consists in the search for a spatial cluster of events from a given point of the sky by counting the events occurred in small solid angles around that given point. Feldman and Cousins have proposed a method to quantify the ``sensitivity'' of an experiment independently of experimental data by calculating the average upper limit, $\bar{\mu}$, that would be obtained in absence of a signal (\cite{Feldman:1997qc}). It is calculated from the mean number of expected background events, $<n_{b}>$, by averaging over all limits obtained from all possible experimental outcomes. The average upper limit is the maximum number of events that can be excluded at a given confidence level (CL). That is, the experiment can be expected to constrain any hypothetical signal that predicts at least $<n_{s}>=\bar{\mu}$ signal events.
From the $90\%$ CL average upper limit we define the ``Model Rejection Factor'' (MRF) for an arbitrary source flux $\Phi_{\textnormal{test}}$ predicting $<n_{s}>$ signal events, as the ratio of the average upper limit to the expected signal (\cite{Hill:2002nv}). The average flux limit $\bar{\Phi}_{\nu}^{90CL}$ is found by scaling the normalization of the flux model $\Phi_{\textnormal{test}}$ such that the number of expected events equals the average upper limit:

\[
\bar{\Phi}_{\nu}^{90CL} = \Phi_{\textnormal{test}} \times (\frac{\bar{\mu}_{90}(<n_{b}>)}{<n_{s}>}) \equiv \Phi_{\textnormal{test}} \times \textnormal{MRF}.
\]
In correspondence with the minimum value of MRF, we have the best sensitivity:
\[
\Phi_{\nu}^{90CL} = \Phi_{\textnormal{test}} \times \textnormal{MRF}_{min} = \textnormal{MRF}_{min} \times 10^{-8} E_{\nu}^{-2} \textnormal{ GeV} \textnormal{ cm}^{-2} \textnormal{ s}^{-1}.
\]

Solid cones of different amplitude around the positions of the two sources have been considered. The number of signal Monte Carlo-generated events ($<n_{s}>$) and the number of background events estimated from the data ($<n_{b}>$) occurred inside each cone have been evaluated and the MRF computed. In this analysis the right ascension of the data is kept blind so that the selection procedure is as unbiased as possible.

\section{Conclusions}
\label{Conclusions}
The sensitivity of the ANTARES detector for a ``down-going'' neutrino flux coming from two candidate sources (Tycho and CTA 1) has been computed. Figure \ref{fig:sensitivity} shows the sensitivity for the two point-sources with an $E_{\nu}^{-2}$ spectrum as a function of the declination.

\begin{figure}[htbp]
\centering\includegraphics[width=0.6\linewidth]{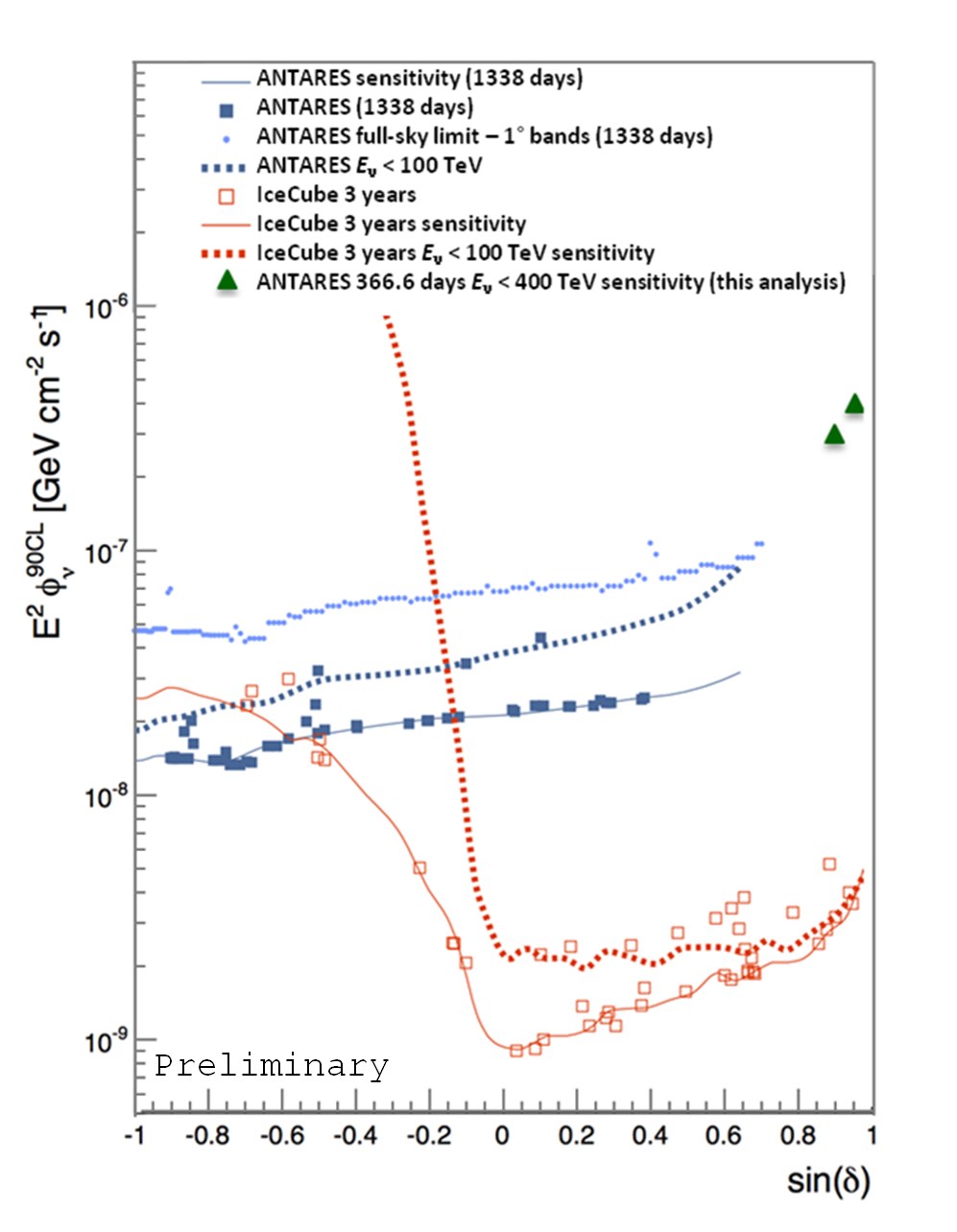}
\caption{Sensitivity for a point-sources with an $E_{\nu}^{-2}$ spectrum as a function of the declination, in green the results for the analysis of down-going events. In blue the $90 \%$ C.L. flux upper limits and sensitivities for six years of ANTARES data (\cite{Adrian-Martinez:2014wzf}). In red the IceCube results shown for comparison (\cite{Aartsen:2013uuv}).}
\label{fig:sensitivity}
\end{figure}


\setcounter{figure}{0}
\setcounter{table}{0}
\setcounter{footnote}{0}
\setcounter{section}{0}
\setcounter{equation}{0}

\newpage
\id{id_sanguineti2}
\addcontentsline{toc}{part}{\textcolor{blue}{\arabic{IdContrib} - {\sl M. Sanguineti} : Moon shadow observation with the ANTARES neutrino telescope
}%
}
\title{\arabic{IdContrib} - Moon shadow observation with the ANTARES neutrino telescope}

\shorttitle{\arabic{IdContrib} - Moon shadow observation with the ANTARES neutrino telescope }

\authors{\color{color01}Matteo Sanguineti}
\afiliations{Universit\`a\ degli\ Studi\ di\ Genova,\ INFN\ Genova}
\email{matteo.sanguineti@ge.infn.it}


\abstract{The ANTARES detector is the largest neutrino telescope currently in operation in the North Hemisphere.
One of the main goals of the ANTARES telescope is the search for point-like neutrino sources. For this reason both the pointing accuracy and the angular resolution of the detector are important and a reliable way to evaluate these performances is needed.
One possibility to measure the angular resolution and the pointing accuracy is to analyse the shadow of the Moon, i.e. the deficit in the atmospheric muon flux in the direction of the Moon induced by absorption of cosmic rays.
Analysing the data taken between 2007 and 2012, the Moon shadow is detected with about $3\sigma$ significance in the ANTARES data.
The first measurement of the ANTARES angular resolution and absolute pointing for atmospheric muons using a celestial calibration source is obtained. The presented results confirm the good pointing performance of the detector as well as the predicted angular resolution.}

%
%
\maketitle
\section{Introduction}
The neutrinos are a unique probe for the investigation of the Universe, they are chargeless, weakly interacting particles that can cross dense matter or radiation fields without being absorbed for cosmological distance.
The neutrino detection can provide more information on the nature of far Universe and the interior of the astrophysical sources, their observation can be also combined with multi-wavelength light and charged cosmic measures.

The ANTARES neutrino telescope \cite{Gen} is the largest neutrino telescope currently in operation in the North hemisphere. It is designed for the detection of high energy cosmic neutrinos and in particular the identification of point-like sources, like starburst galaxies, GRBs, Supernova remnants and AGNs.The pointing accuracy and the angular resolution of the detector are really important for the detection of point-like sources and a proper way to evaluate these performances is needed.
Several experiments, like CYGNUS \cite{CYG}, TIBET \cite{TIB}, CASA \cite{CAS}, MACRO \cite{MAC}, SOUDAN \cite{SOU} , ARGO \cite{ARG} and IceCube \cite{ICE}, used the so-called Moon shadow effect to test the pointing performance of the detector.

The Moon absorbs part of the cosmic rays, so a deficit in the event density of the atmospheric muon flux corresponding to the direction of the Moon disk is expected.
In this work we exploit this technique to measure the ANTARES angular resolution for atmospheric down-going muons and the detector absolute pointing capability.

\section{Monte Carlo simulations}

The simulation of the atmospheric muon events was performed with the MUPAGE code \cite{Mupage}, where the geo-magnetic deflection is not taken into account in the simulation code.
In order to take in account this effect a study of the deflection effect has been previously conducted by the collaboration using Corsika code \cite{corsika}. 
The correction of the muons trajectory is negligible at detector level because only low energy muons that are absorbed before reaching the detector are strongly deflected \cite{Carla}, so the geo-magnetic effect can be neglected in this analysis. 

Muon bundles were generated on the surface of a cylinder-shaped volume of water, called the {\it can},  containing the detector. It is the volume sensitive to the light and it is 200 m  larger than the instrumented volume.
The generation of Cherenkov light emitted by the muon tracks is simulated.
The simulation includes also optical background caused by bioluminescence and radioactive isotopes present in sea water.
The detector response is then simulated \cite{Ju}, the charge of the analogue pulse being evaluated according to the number of photons arriving on each PMT and the charge of consecutive pulses being integrated in a time window of 25 ns. The hit time is defined as the arrival time of the first photon. 
Finally the standard ANTARES reconstruction algorithm uses the hits detected by the PMT to reconstruct the direction of atmospheric muon tracks. The algorithm is a robust track fitting procedure based on a maximisation likelihood method.

Two different Monte Carlo simulation sets were performed: one considering the shadowing effect of the Moon and the other without this effect. The shadowing effect is simulated rejecting the muons generated within the Moon disk ($R_{Moon}=0.259^\circ$). 
The live time of each simulation is the 2080 days period considered in this data analysis (years 2007-2012).
The experimental conditions of each data run (PMT status, detector configuration, actual environmental conditions, optical background) are simulated like in the official ANTARES run-by-run simulation \cite{MC} .
The systematic uncertainties of the primary muon flux and of the detector lead to a discrepancy around 6\% between Monte Carlo simulation and data, this behaviour was already shown in other ANTARES analysis \cite{flux}.
The Monte Carlo simulations were therefore renormalized in order to reproduce the muon data rate in the region were the shadowing effect is expected to be negligible. 

The optimization of the selection criteria used in this data analysis will be described in the next section.

\section{Detection of the Moon shadow}
\label{sec:hypot}
In order to measure the deficit of muons in the direction of  the Moon, the region of the sky  around the Moon centre is divided in concentric rings with increasing radius. 
We define the  event density of each ring as the number of events detected in that sector over the surface of the ring. 
The ring size is $0.2^\circ$, so an appropriate investigation of the Moon shadow with sufficient statistics in each annular ring can be performed.
Obviously event tracks detected when the Moon is above the Horizon and reconstructed as down-going are selected.

A test statistic function $t$ is defined as:
\begin{equation}
t=\sum_{rings} \frac{(n_{m}-n_{exp,\color{bianco}{o}\color{color01}{NO}\color{bianco}{o}\color{color01}{Moon}})^2}{n_{exp,\color{bianco}{o}\color{color01}{NO}\color{bianco}{o}\color{color01}{Moon}}}-\frac{(n_{m}-n_{exp,\color{bianco}{o}\color{color01}{Moon}})^2}{n_{exp,\color{bianco}{o}\color{color01}{Moon}}} ,
\label{eq:tst}
\end{equation}
where the sum is over all the rings around the Moon centre; $n_m$ is 
the number of events detected in a ring, $n_{exp,Moon}$  
is the expected number of events in ``Moon shadow'' hypothesis and $n_{exp, NO Moon}$ is the expected number of events in ``no Moon shadow'' hypothesis.
A million of toy experiments were generated to derive the test statistic distribution in the two different hypotheses ("Moon shadow" or "No Moon shadow").

The significance of the Moon shadow deficit was estimated optimising the event selection using  the statistical tools previously described.
In this analysis quality cuts on the log-likelihood per degree of freedom  $\Lambda<\Lambda_{\hbox{cut}}$ was applied.  The maximisation of the significance is found for $\Lambda_{\hbox{cut}}= -5.9 $  as shown in Fig.~\ref{fig:lambdabeta}. 

\begin{figure}[htbp]
\begin{center}
\includegraphics[width =9cm]{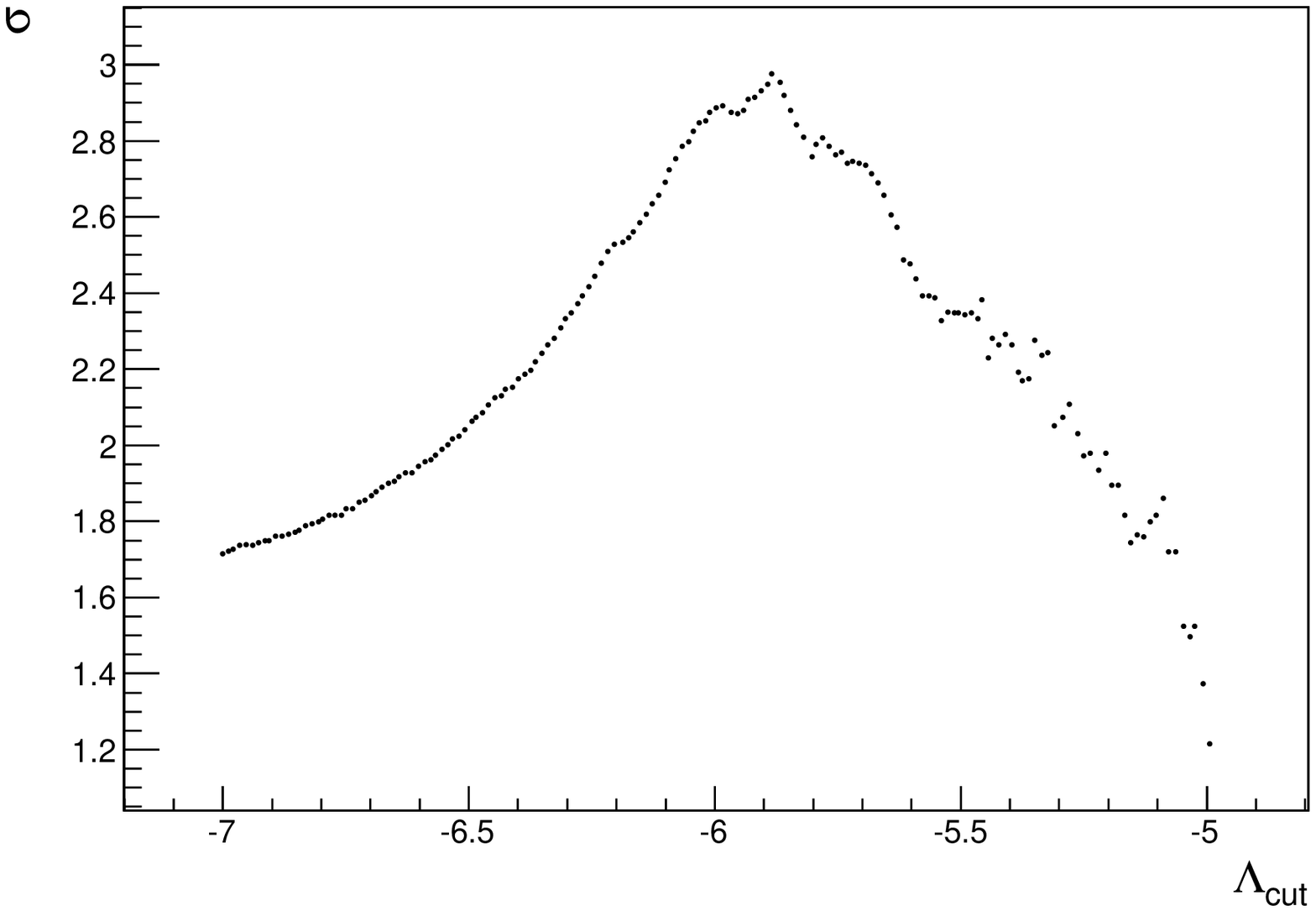}
\caption{Expected significance (expressed as number of $\sigma$) as a function of $\Lambda_{\hbox{cut}}$.} 
\label{fig:lambdabeta}
\end{center}
\end{figure}

The corresponding test function distributions are plotted in  Fig.~\ref{fig:distr}.
The shaded area gives the fraction of the toy experiments where the Moon shadow hypothesis will be correctly identified as evidence of the shadowing effect; this fraction is fixed to 50\%. The value of $t=6.15$ corresponding to this fraction of the ``Moon shadow'' toy experiments is the decision boundary of the test statistic.
The orange area corresponds to the fraction of ``No Moon shadow'' toy experiments that will be wrongly identified as evidence of shadowing effect.
In other words, this area quantifies the minimum significance of the Moon shadow discovery for experiments with  $t>6.15$. The minimum significance is here $2.9 \sigma$.

\begin{figure}[htbp]
\begin{center}
\includegraphics[width =9cm]{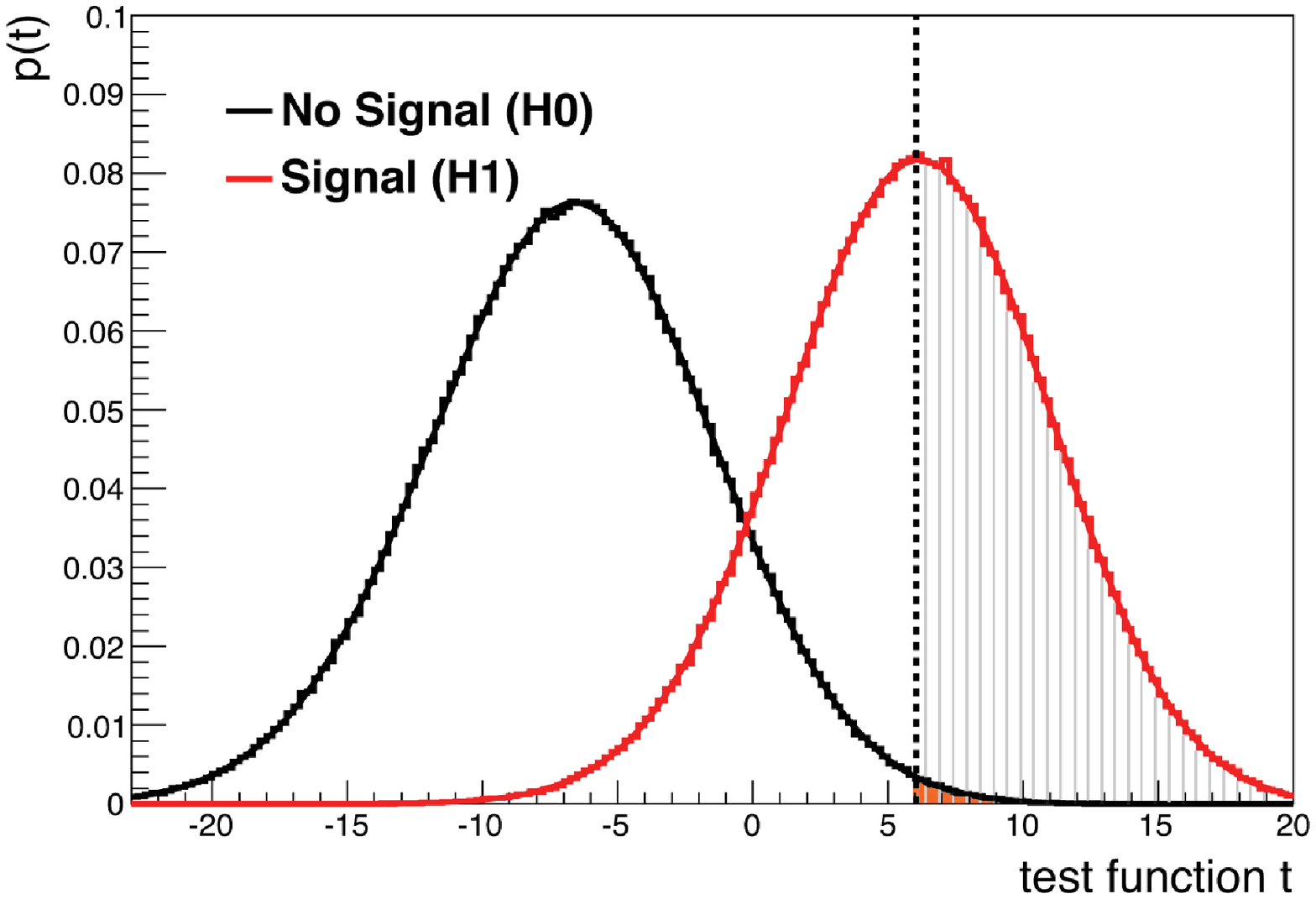}
\caption{The test function {\itshape t} distribution for ``Moon shadow'' hypothesis 
(red curve) and ``no Moon shadow'' hypothesis (black curve). The shaded area  is the fraction of the toy experiments where the Moon shadow hypothesis will be correctly identified as evidence of the shadowing effect. The orange area quantifies the minimum significance (here $2.9 \sigma$) to observe the Moon shadow.}
\label{fig:distr}
\end{center}
\end{figure}

The same quality cut $\Lambda_{\hbox{cut}}= -5.9 $  was applied to the data set.
The value of test statistic function defined in Eq. \ref{eq:tst} was then computed for data resulting in $t=7.12$.  
The ``No Moon shadow''  hypothesis can be therefore rejected with a significance of $3.1\sigma$. 

\section{Angular resolution and absolute pointing}

The angular resolution of a neutrino telescope is usually estimated through the Monte Carlo simulations, because there is not an immediate way to estimate this parameter with data.
The Moon shadow study represents an unique way to estimate the pointing performance of the detector.
The plot of event density for selected muons as a function of the angular distance from the Moon centre is shown in  Fig.~\ref{fig:datagaussfit}  

\begin{figure}[htbp]
\begin{center}
\includegraphics[width=9cm]{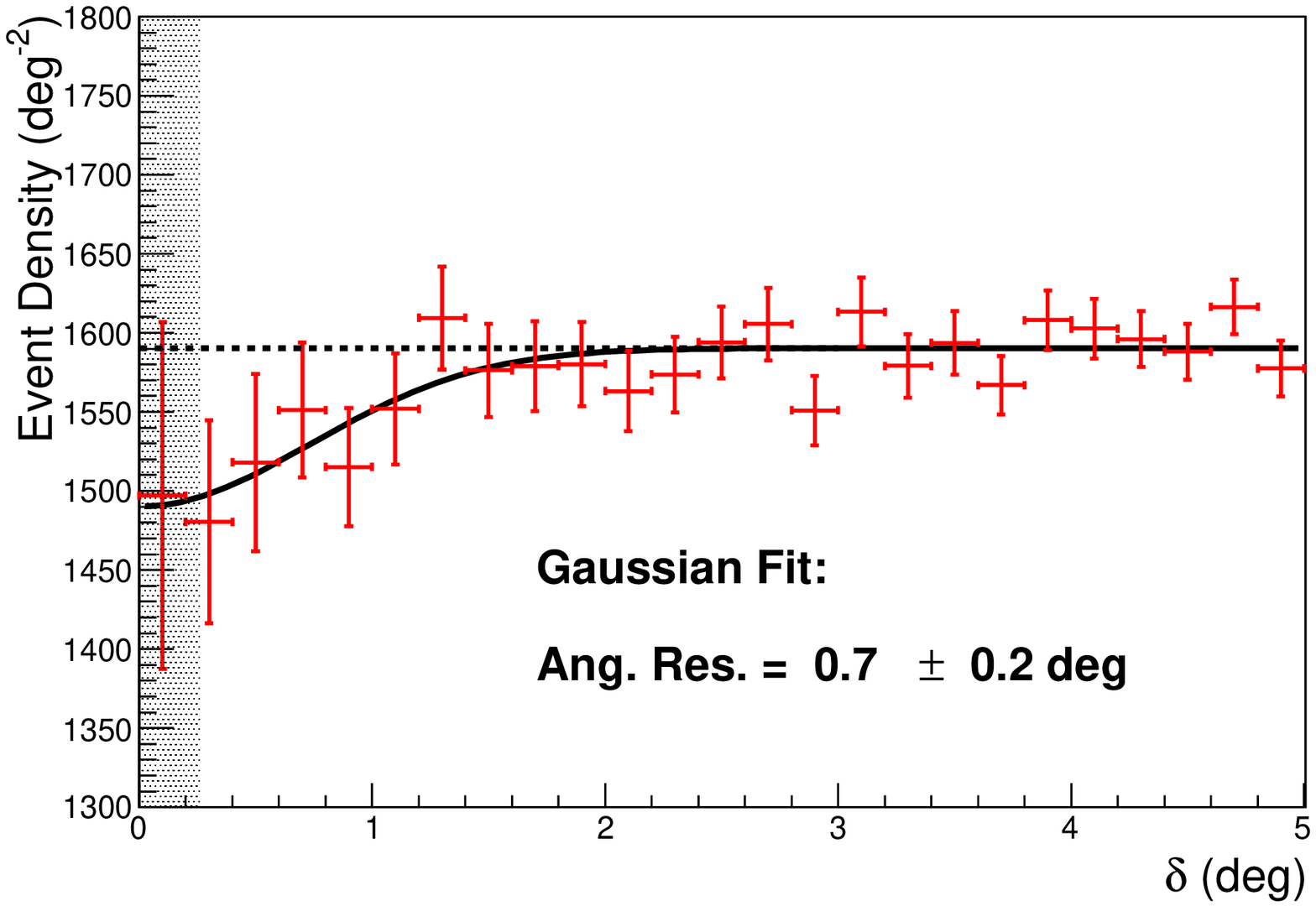}
\caption{Event density of muons after selection cut versus the angular distance from the Moon centre. The shadow is fitted assuming a Gaussian shape for the detector point spread function.  The resulting angular resolution is $\varsigma=0.7^\circ\pm0.2^\circ$ for atmospheric muons. The shaded area represents the Moon radius ($R_{Moon}=0.259^\circ$).}
\label{fig:datagaussfit}
\end{center}
\end{figure}
It is possible to evaluate the detector angular resolution fitting the event density with the formula:

\begin{equation}
\frac{\hbox{d}n}{\hbox{d}\delta^2}=k\left(1- \frac{R^2_{Moon}}{2\varsigma^2} e ^{\displaystyle -\frac{\delta^2}{2\varsigma^2}}\right),
\end{equation}
where $R_{Moon}=0.259^\circ$ is the Moon radius and $\delta$ is the angular distance from the Moon centre. The fit free parameters $k$ and $\varsigma$  are respectively the off-source density level and the detector angular resolution. 
We have assumed a Gaussian shape for the detector point spread function \cite{Formula}.
From the fit we can estimate the  angular resolution: $\varsigma=0.7^\circ\pm0.2^\circ$.

Finally the ANTARES absolute pointing performance was evaluated. It is possible that if the detector orientation is affected by a systematic error, the Moon shadow will appear shifted respect to the expected position.
In order to investigate this possibility, the concentric rings around the Moon centre are shifted (see Section \ref{sec:hypot}). In this way the detector will be "pointed'' in a wrong direction were we expect a fainter shadowing effect.

It is expected that the significance would be around $3\sigma$ for small shifts ($\le 0.1^\circ$), then it would decreases significantly while increasing the shift as we expected.  The study is ongoing, but relevant systematic errors are not expected in the absolute pointing of the ANTARES detector.

\section{Conclusions}

The Moon shadow in the atmospheric muon flux has been observed with the ANTARES neutrino telescope. 
The optimization of event selection has been performed with a dedicated Monte Carlo simulation and an opportune test statistic function has been defined to evaluate the deficit significance.
The 2007-2012 data sample has been then analysed showing a $3.1\sigma$ evidence of the effect. 
The Moon shadow profile has been fitted assuming a Gaussian shape for the detector point spread function, in this way we derived the angular resolution for the atmospheric muon flux: $0.7^\circ\pm0.2^\circ$. 

The results reported in this work are the first Monte Carlo independent measure of the angular resolution and the first study of the pointing systematics of the ANTARES detector exploiting a celestial calibration source.


\setcounter{figure}{0}
\setcounter{table}{0}
\setcounter{footnote}{0}
\setcounter{section}{0}
\setcounter{equation}{0}

\newpage
\id{id_schnabel}
\addcontentsline{toc}{part}{\textcolor{blue}{\arabic{IdContrib} - {\sl J. Schnabel} : Search for a diffuse cosmic neutrino flux with ANTARES using track and cascade events
}%
}

\title{\arabic{IdContrib} - Search for a diffuse cosmic neutrino flux with ANTARES using track and cascade events}

\shorttitle{\arabic{IdContrib} - Search for a cosmic neutrino flux with ANTARES}

\authors{Jutta Schnabel$^a$, Steffen Hallmann$^b$ (speaker)}
   \afiliations{Erlangen Centre for Astroparticle Physics, Erwin-Rommel Str. 1, 91052 Erlangen, Germany
   }
\email{$^a$jutta.schnabel@fau.de, $^b$steffen.hallmann@fau.de}
%

 
\abstract{The ANTARES neutrino telescope has since its final deployment in 2008 contributed to the searches for high-energy neutrino sources. In this work, prior ANTARES searches for the diffuse events from track-like charged-current muon neutrinos as well as cascade-like interaction from all neutrino flavours are integrated into a new comprehensive all-flavour search. The method employs a multivariate analysis approach on six years of ANTARES data optimizing for the discovery of a cosmic neutrino flux as observed by the IceCube experiment. This analysis reaches at its first stage a sensitivity of $\Phi_{IC2.5} E^{2.5} = 5.4 \times 10^{-6} GeV^{1.5}\, cm^{-2}\, sr^{-1}\, s^{-1}$ and observes a slight excess of events over the background estimation.}

%
%
\maketitle
\section{Introduction}
The search for neutrinos of cosmic origin has evolved greatly in the last few years. As decay products of, among others, $\pi$ and $K$ mesons, neutrino production is expected to occur in astrophysical sources through interaction of hadrons. At cosmic acceleration sites, the interaction of protons accelerated through shock acceleration are expected to lead to a cosmic neutrino flux that follows the distribution of the cosmic ray spectrum \cite{WB}. As hadrons from cosmic rays also lead to air showers in Earth's atmosphere, this cosmic neutrino component needs to be distinguished from an atmospheric background of neutrinos from both conventional atmospheric neutrinos \cite{AtmoFlux} and especially high-energy neutrinos emitted from prompt decays of hadrons containing charm quarks in the atmosphere \cite{Enberg}.\\

ANTARES has already set a limit on this diffuse flux of cosmic neutrinos from charged-current interactions of $\nu_{\mu}$ \cite{Antares11}. An excess of diffuse cosmic neutrinos was recently measured for all neutrino flavours by the IceCube experiment \cite{IC15}. After several more years of data taking and further development of reconstruction techniques for events from all neutrino flavours, the ANTARES sensitivity towards the cosmic neutrino flux has increased significantly, although the approach to the measurement of the cosmic neutrino flux must differ to that of IceCube due to the different technical conditions. In this work the first combined search for neutrinos of all flavours is presented by applying a new methodology which focuses on multivariate techniques in order to incorporate the different event topologies.

\section{Neutrino measurement with ANTARES}
At the ANTARES \cite{AntPaper} site at about 2.5 km below sea level off the French Mediterranean coast, the measurement of neutrinos is challenged by two main factors. On the one hand, the 12 detection lines are not only subject to the sea current and varying environmental conditions, but also detect photons from ambient light emitters like $^{40}K$ decays and, to a larger extent, bioluminescent sea life. To handle this, effective event selection and triggering schemes are in place, of which only the more stringent ones are used in this analysis to ensure a low influence of sea conditions on the event selection. On the other hand, muons produced in atmospheric air showers penetrate the overburden of water such that at the detector level they outnumber neutrino-induced events by about $1:10^6$. 

\subsection{Event Simulation and Data Selection}
Due to the varying environmental conditions, event simulation in ANTARES \cite{mupage}, \cite{simulation} is done on a run-by-run basis, accounting for changing bioluminescence rates within run periods of a few hours. Due to the complex environmental conditions, the agreement between data and simulation naturally varies, which is accounted for in the analysis procedure by restricting the optimization on simulation to runs which show a good agreement between data and simulation for all relevant parameter distributions. Consequently, an effective livetime of 913 days is selected from the data taking period between 2007 and 2013. 
A large amount of the remaining data in the same period has also good quality, but lacks an according run by run simulation. The total amount of available data is 1700 days including the previously described selection, for which the analysis was optimized. The remaining part will be included in a consecutive step which is still in progress at the time of this presentation.

\subsection{Event identification and reconstruction}
Event topologies seen in the ANTARES detector are divided into charged-current $\nu_{\mu}$ interactions which mainly produce Cherenkov emission along the extensive muon track, and cascades of short-lived secondary particles producing photon emission at the interaction point of $\nu_{e}$ and neutral-current $\nu_{\mu}$. For these track-like and cascade-like events specialized event reconstruction methods have been developed, including likelihood-based directional reconstruction from photon hit patterns and various track energy estimators. Although no special reconstruction of $\nu_{\tau}$ events was used in this work, their topology varies between cascade-like events for neutral current interactions and short track-like events for charged-current interactions producing a quickly decaying $\tau$ lepton resulting in a track-like $\mu$ or cascade, making it possible to reconstruct $\nu_{\tau}$ events with existing track and cascade reconstruction techniques.\\
In order to incorporate all event signatures in a search for a diffuse cosmic neutrino flux, multivariate techniques \cite{TMVA} were employed to identify the relevant features from both track-like and cascade-like events. As the search for cosmic neutrino events in ANTARES can roughly be divided into firstly distinguishing the atmospheric muon events from neutrino-induced events and secondly extracting the cosmic signal from the atmospheric neutrino background, two multivariate tools were used to fulfil these tasks.

\begin{figure}
 \centering
  \begin{subfigure}[b]{0.53\textwidth}
   \includegraphics[width=\textwidth]{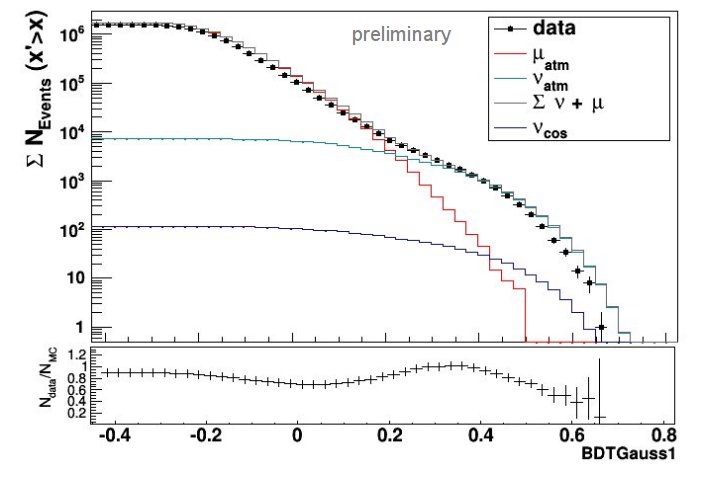}
   \caption{BDT for $\mu_{atm}$ suppression}
   \label{fig:dmc_tmva1}
  \end{subfigure}
  \begin{subfigure}[b]{0.53\textwidth}
   \includegraphics[width=\textwidth]{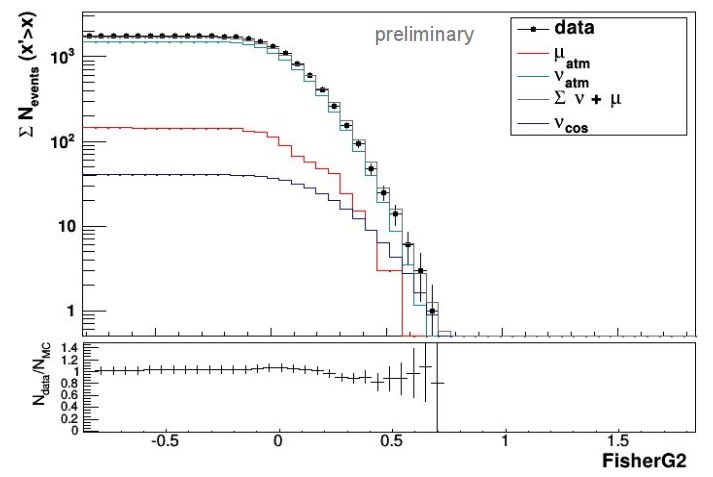}
   \caption{Fisher discriminant for $\nu_{cos}$ identification}
   \label{fig:dmc_tmva2}
  \end{subfigure}
 \caption{Agreement between data and simulation for 913 days for a) the BDT method for atmospheric muon suppression and b) the Fisher discriminant for cosmic neutrino identification after a cut on $BDT>0.345$}
 \label{Fig:DMC}
\end{figure}

\subsection{Atmospheric muon suppression}
The distinction between atmospheric muons entering the detector from above and high-energy neutrino-induced events coming from all directions can most effectively be accomplished by a combination of event angular estimates and the quality of cascade and track reconstruction methods with energy-related variables.\\
In order to find the most effective parameter combination for this task, candidate parameters and multivariate methods were tested in an optimization process employing the signal-background separation $S=(\mu(x_{sig})-\mu(x_{bkg}))/(RMS(x_{sig})-RMS(x_{bkg}))$ as optimization parameter, with $\mu$ denoting the mean and $RMS$ the root mean square of the parameter $x$ in signal and background events. Of several multivariate methods, Boosted Decision Trees (BDT) \cite{TMVA} ranked among the best performing. Following a parameter scanning procedure, nine parameters were selected as input parameters. These included two track zenith angle estimates, one track and one cascade reconstruction quality parameter, a track energy estimate and the number of photon hits measured in all PMTs in a cascade, one atmospheric muon suppression parameter and two geometrical parameters describing the extension of the event within the detector and the time residual distribution of the photons. The behaviour of the resulting BDT can be seen in Figure \ref{fig:dmc_tmva1}, where the excess of atmospheric neutrinos, weighted according to the Honda \cite{AtmoFlux} atmospheric neutrino flux model over the background of atmospheric muons can be seen at high BDT values.

\begin{figure}
 \centering
  \begin{subfigure}[b]{0.49\textwidth}
   \includegraphics[width=\textwidth]{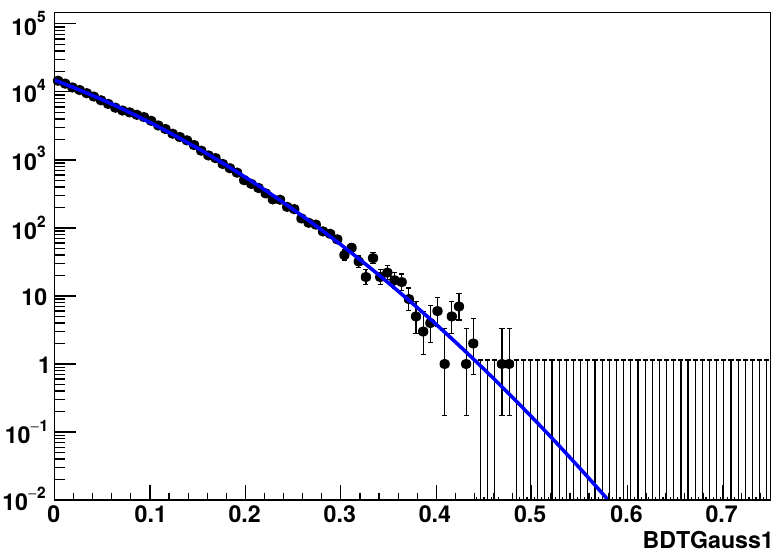}
   \caption{Muon fit for BDT}
   \label{fig:muon_tmva1}
  \end{subfigure}
  \begin{subfigure}[b]{0.49\textwidth}
   \includegraphics[width=\textwidth]{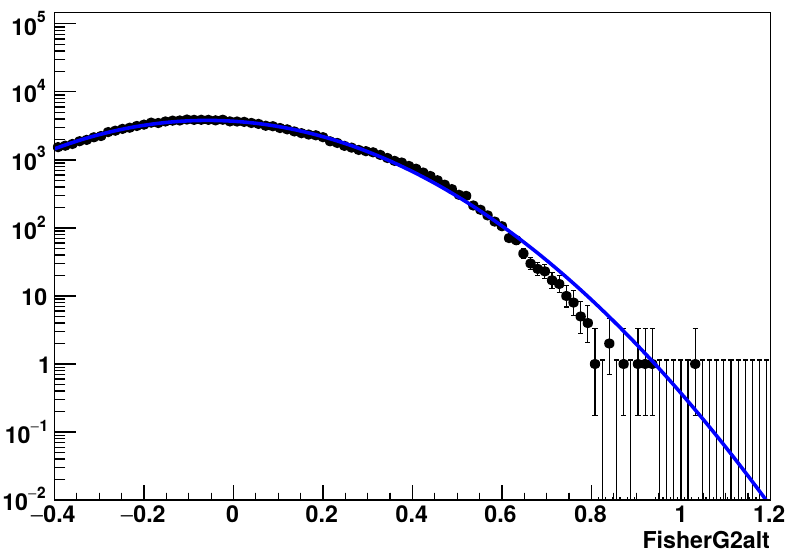}
   \caption{Muon fit for Fisher discriminant}
   \label{fig:muon_tmva2}
  \end{subfigure}
 \caption{Projections of the atmospheric muon distribution and the according extrapolation for the BDT and Fisher discriminant.}
 \label{Fig:Muons}
\end{figure}

\subsection{Cosmic neutrino identification}
The distinction between atmospheric and cosmic neutrino events is to the largest extent achieved through determining the neutrino energy, as the cosmic neutrino flux is expected to follow a harder spectrum than the background of atmospheric events. The additional energy deposited in the detector from neutrino interactions is seen as additional light yield originating from photons from either the Cherenkov emission from secondary particles at the interaction vertex or as radiation from energy loss processes along the muon track. Therefore, the number of photons, measured as charge collected on the photomultipliers, gives the simplest representation of the energy information. \\
As various sophisticated energy estimators were developed within ANTARES for the different event types, another multivariate technique was employed to arrive at a common estimate for the signal-likeness of any neutrino event. Here, the signal efficiency $\epsilon$ at very small background was employed as optimization parameter in the search for best parameters and multivariate methods, as the task of signal extraction demands a high purity of the final event sample. The following testing showed simple linear estimators to perform well for this task, leading to the use of a Fisher discriminant \cite{TMVA}, which combined three different energy estimates for tracks and cascades, three photon counts from different event-type specific photon hit selections, as well as a cascade zenith angle estimate, one track and one cascade reconstruction quality parameter and the number of storeys used for the cascade reconstruction, which adds geometrical information to the estimator. The behaviour of the Fisher discriminant can be seen in Figure \ref{fig:dmc_tmva2}, using a prior cut on the BDT parameter to reduce the contamination of the event sample by atmospheric muons to $\approx 10 \%$.

\section{Analysis procedure}
Having obtained tools for the suppression of both the atmospheric muon and atmospheric neutrino background, the analysis procedure can be reduced to a simple search for the optimal combination of parameter cuts on these two multivariate parameters. As the sensitivity of ANTARES is, by extrapolation from previous results, expected to come close to the flux of cosmic neutrinos observed by IceCube, the selection of the optimal cuts should both fulfil the requirements of a model discovery and a model rejection technique \cite{MRF}.

\begin{figure}
 \centering
 \hspace{-2cm}
  \begin{subfigure}[b]{0.53\textwidth}
   \includegraphics[width=\textwidth]{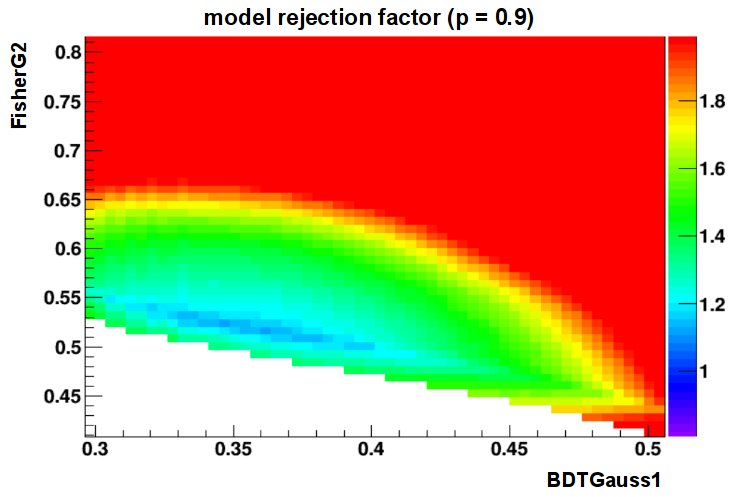}
   \label{fig:MRF}
  \end{subfigure}
  \begin{subfigure}[b]{0.53\textwidth}
   \includegraphics[width=\textwidth]{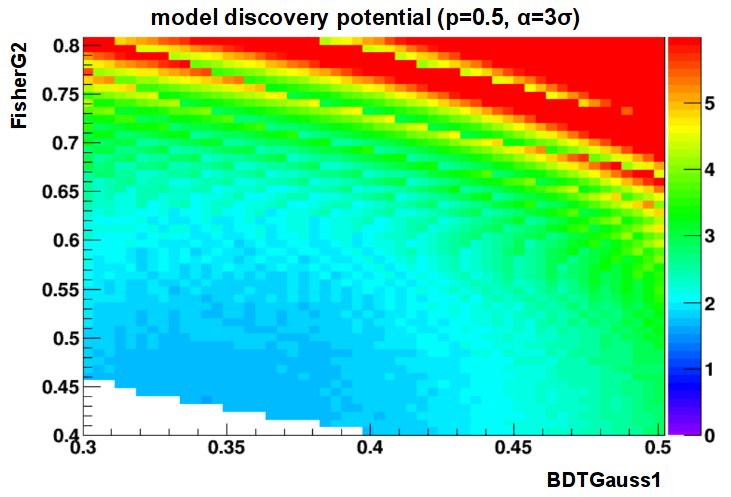}
   \label{fig:MDP}
  \end{subfigure}
   \hspace{-2cm}
 \caption{Model rejection factor (left) and discovery potential (right) (for $3\sigma$ at $50\%$) for various event cut configurations, employing both TMVA methods with a Gaussian preprocessing \cite{TMVA}. The compromise discussed below was set for the best MRF at $BDT>0.345$ and $Fisher>0.52$}
 \label{Fig:Cutoptimization}
\end{figure}

\subsection{Signal optimization}
In order to perform the signal optimization as accurately as possible, a fit on the distribution of the atmospheric muon component was introduced as well as a prompt atmospheric neutrino flux following \cite{Enberg}. The fit is necessary due to the limited statistics of the simulated atmospheric muon sample, which only accounts for 1/3 of the total data taking time. Here, a two-dimensional Gaussian function was fit to the atmospheric muon distribution for both multivariate parameters (blue lines in Figure \ref{Fig:Muons}), introducing the uncertainty of the fit parameter propagated to the muon number as error on the estimated atmospheric muon number. The contribution of $\nu_{\tau}$ events was estimated from a small simulation and not included in the optimization procedure. The procedure was therefore performed for $\nu_{e}$ and $\nu_{\mu}$ events from a cosmic signal according to the IceCube measurement \cite{IC15}, assuming $\Phi_{IC2.5} E^{2.5} = 4.1 \times 10^{-6} GeV\, cm^{-2}\, sr^{-1}\, s^{-1}$,  with atmospheric neutrinos simulated using the conventional flux from \cite{AtmoFlux} and including the extrapolated muon number. Intending to ultimately use this analysis on the full data sample of 1700 days, event numbers were scaled to this livetime for the event selection optimization. As can be seen in Figure \ref{Fig:Cutoptimization}, a model rejection optimization then leads to an optimal result that still exhibits a good model discovery potential, as both minimal regions overlap.

\subsection{Error estimates}

In order to account for simulation uncertainties in the standard ANTARES simulation, the uncertainty of water propagation properties, i.e. the water absorption and scattering length, was estimated on a simulation including 12 days data taking and the difference in event numbers after final cuts from variation of these properties by 10$\%$ is taken into account.\\
Also, a small simulation of $\nu_{\tau}$ events equivalent to 12 days was produced to estimate the behaviour of these events in the analysis. As the event topology does to a large extent agree with that of cascade events, the $\nu_{\tau}$ contribution could also be extrapolated from cascade simulations as done in \cite{Folger}. As both methods have limited accuracy, the $\nu_{\tau}$ simulation was used to estimate this contribution, while the difference between event numbers from both methods was introduced as error. The errors drawn from these estimates are, together with the final event numbers, shown in Table \ref{Tab:Event_numbers}.

\begin{table}
\centering
\begin{tabular}{|l|r|r||l|r|r|}
\hline
Signal & $N_{events}$ & error & Background & $N_{events}$ & error \\  \hline
$N^{2.5} \nu_{\mu, CC}$ & 1.4 & 0.37$^b$+0.57$^c$ & $N_{Honda} \nu_{\mu, CC}$ & 5.3 & 2.29$^b$\\ 
$N^{2.5} \nu_{\mu, NC}, \nu_{e}$ & 2.6 & 0.63$^b$+0.06$^c$ & $N_{Honda} \nu_{\mu, NC}, \nu_{e}$ & 2.4 & 0.9$^b$+0.1$^c$\\ 
$N^{2.5} \nu_{\tau}$ & 0.9 & 0.53$^d$ & $N_{Enberg} \nu_{\mu, CC}$ & 0.2 & 0.08$^b$+0.45$^c$\\ 
$N^{2.0} \nu_{\mu, CC}$  & 1.8 &  & $N_{Enberg} \nu_{\mu, NC}, \nu_{e}$ & 0.6 & 0.14$^b$+0.01$^c$\\
$N^{2.0} \nu_{\mu, NC}, \nu_{e}$  & 2.2 & & $N_{Enberg} \nu_{\tau}$ & 0.01 & 0.0\\ 
$N^{2.0} \nu_{\tau}$   & 0.6  & &  $N \mu_{atm}$ & 1.0 & 0.15$^a$\\ \hline
$\Sigma$   & $5.0^{(2.5)}$/$4.5^{(2.0)}$ & $\pm$ 1.1 & $\Sigma$   & 9.5 & $\pm$ 2.5\\ \hline
\end{tabular}
\caption{Signal and background expectation including error estimates for 913 days of ANTARES lifetime. As cosmic flux, the IceCube measurement \cite{IC15} is used assuming either a spectral index $\lambda = 2.5$ or $\lambda = 2.0$. Error estimates are drawn from a) error on muon fit parameters, b) water absorption length uncertainty, c) water scattering length uncertainty, d) difference between $\tau$ estimate and toy simulation.}
\label{Tab:Event_numbers}
\end{table}

\section{Results}
\begin{figure}
 \centering
   \includegraphics[width=0.7\textwidth]{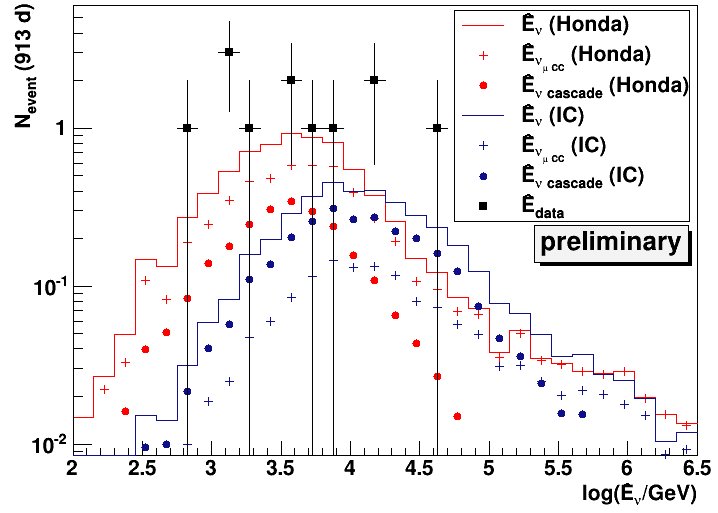}
 \caption{Energy distribution for the events found in 913 days, giving the reconstructed vertex energy $\hat{E}$ by a cascade reconstruction \cite{Folger} for data and simulated $\nu_{e}$ and $\nu_{\mu}$ contributions}
 \label{Fig:Energy of events}
\end{figure}
The search for a diffuse flux of cosmic neutrinos in ANTARES leads to the expected event numbers given in Table \ref{Tab:Event_numbers} over the background of conventional \cite{AtmoFlux} and prompt \cite{Enberg} neutrino flux. Assuming a spectral index of $-2.5$ and the cosmic neutrino flux per flavour as $\Phi_{IC2.5} E^{2.5} = 4.1 \times 10^{-6} \mathrm{GeV^{1.5}}\, \mathrm{cm}^{-2}\, \mathrm{sr}^{-1}\, \mathrm{s}^{-1}$, a sensitivity of $\Phi_{90\% IC2.5} = 1.33\, \Phi_{IC2.5}$ between 6.8 TeV and 1.1 PeV is reached for 913 days. Accordingly, a harder spectrum of $\Phi_{IC2.0} E^{2.0} = 1.1 \times 10^{-8} \mathrm{GeV}\, \mathrm{cm}^{-2}\, \mathrm{sr}^{-1}\, \mathrm{s}^{-1}$ following the spectral index of \cite{WB} and the magnitude of \cite{IC15} reaches a sensitivity per flavour of $\Phi_{90\% IC2.0} = 1.6 \times 10^{-8} \mathrm{GeV}\, \mathrm{cm}^{-2}\, \mathrm{sr}^{-1}\, \mathrm{s}^{-1}$, valid within 18 TeV to 7.5 PeV.\\
In 913 days of ANTARES data, 12 events were found, which is a slight excess over the background expectation of 9.5 events. The events studied in this analysis generally exhibit similar event topology which allows each to be reconstructed as both track and cascade events. As the events are found to be either interacting close to the detector or inside the instrumented volume, the number of photons measured by the detector is generally large. However, the various energy reconstruction methods vary in the interpretation of the neutrino energy depending on their event signature assumption, as e.g. track energy reconstructions generally interpret the energy deposition as one of several catastrophic energy losses and therefore assign a higher primary neutrino energy. In Figure \ref{Fig:Energy of events}, the cascade vertex energy is shown for the final events. Including error estimates according to \cite{pole}, upper limits on the respective fluxes can be set as $\Phi_{90\% u.l. IC2.5} = 2.4\, \Phi_{IC2.5}$ and $\Phi_{90\% u.l. IC2.0} = 2.6\, \Phi_{IC2.0}$. These results are compared to previous analyses and the flux measured by IceCube \cite{IC15} in Figure \ref{Fig:fluxlimits}.\\
This first analysis step shows the capability of ANTARES to combine the former separate searches for a diffuse cosmic neutrino flux through multivariate methods into an effective analysis of all neutrino event types. As the analysis presented here only incorporates a little more than half of the data taken by the ANTARES experiment until end 2013, a full analysis can be expected to reach a sensitivity similar to the flux measured by IceCube.

\begin{figure}
 \centering
   \includegraphics[width=0.78\textwidth]{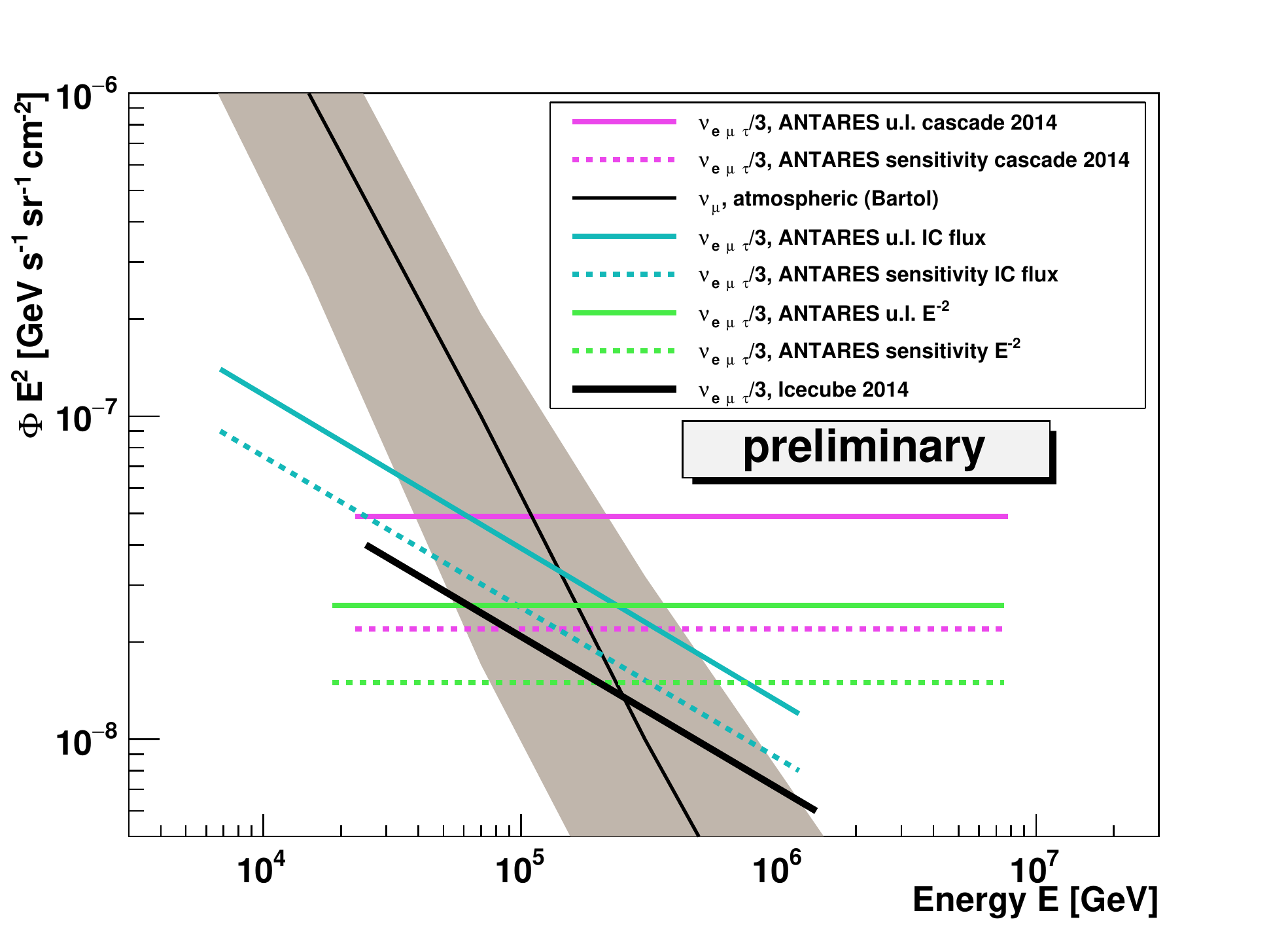} 
 \caption{Flux limits for the last track and cascade analyses in ANTARES, the Icecube result and limits and sensitivity of this most recent analysis}
 \label{Fig:fluxlimits}
\end{figure}


\setcounter{figure}{0}
\setcounter{table}{0}
\setcounter{footnote}{0}
\setcounter{section}{0}
\setcounter{equation}{0}

%

%
%

%

%
%

%
\newpage
\addcontentsline{toc}{part}{{\sc Multi-Messenger searches}%
\vspace{-0.5cm}
}
\id{id_ddornic}
\addcontentsline{toc}{part}{\textcolor{blue}{\arabic{IdContrib} - {\sl D. Dornic} : Time-dependent search of neutrino emission from X-ray binaries with the ANTARES telescopes}%
}

\title{\arabic{IdContrib} - Time-dependent search of neutrino emission from X-ray binaries with the ANTARES telescopes}

\shorttitle{\arabic{IdContrib} - X-ray binaries with the ANTARES}

\authors{Damien Dornic$^a$, A. S\'{a}nchez-Losa$^b$}
        \afiliations{$^a$Aix Marseille Universit\'e, CNRS/IN2P3, CPPM UMR 7346, 13288, Marseille, France\\
        $^b$IFIC - Instituto de F\'{i}sica Corpuscular, Edificios Investigaci\`{o}n de Paterna, CSIC - Universitat de Valencia, Apdo. de Correos 22085, 46071 Valencia, Spain}
\email{$^a$dornic@cppp.in2p3.fr, $^b$agustin.sanchez@ific.uv.es}
%


\abstract{ANTARES is currently the largest neutrino telescope operating in the Northern Hemisphere,
aiming at the detection of high-energy neutrinos from astrophysical sources. By design, 
neutrino telescopes constantly monitor at least one complete hemisphere of the sky
and are thus well set to detect neutrinos produced in transient astrophysical sources. The
flux of high-energy neutrinos from transient sources is expected to be lower than the one expected from steady
sources, but the background originating from interactions of charged cosmic rays in the Earth's atmosphere can be drastically reduced by
requiring a directional and temporal coincidence of the astrophysical
phenomenon detected by a satellite. The time-dependent point-source search has been applied to a list 
of 34 x-ray binary systems while observed in high flaring activities in the 2008-2012 satellite data, 
RXTE/ASM, MAXI and Swift/BAT. The results of this search are presented together with the comparison 
between the neutrino flux upper-limits with the measured gamma-ray spectral energy distribution and 
the prediction from astrophysical models.}

%
%
\maketitle

\section{Introduction}
Neutrinos are unique messengers for studying the high-energy Universe as they are neutral, 
stable, interact weakly, and travel directly from their sources without absorption or deflection. 
Therefore, the reconstruction of the arrival directions of cosmic neutrinos would allow both the 
sources of the cosmic rays - supernova remnant shocks, active galactic nuclei jets, x-ray binary jets,  
gamma-ray bursts, etc.~\cite{bib:reviewNeutrinosources} - and the relevant acceleration mechanisms 
acting within them to be identified. 

X-ray binaries (XRB) are a class of binary stars that are luminous in x-rays. The x-rays are produced by matter falling from the donor (usually a relatively normal star), to 
the accretor, which is compact: a white dwarf, neutron star (NS), or black hole (BH). These systems 
are usually classified as low-mass x-ray binary (LMXB) and high-mass x-ray binary (HMXB), depending on the mass of the donor. In very few cases, the 
presence of relativistic jets has been confirmed by radio measurements. The jet signature may be present in all the XRB sources. However, the composition of the jets is still unknown.
Their spectral energy distribution can be described by two components: a low-energy one from radio to X-rays and a high-energy one from X-rays to very 
high-energy gamma rays. The non-thermal emission is probably dominated by leptonic processes of accelerated eletrons but a hadronic component could also be present.
In hadronic models, associated with the very high-energy gamma rays from $\pi^{0}$ decays, the decay of the charged pions gives rise to a correlated neutrino emission.
Up to now, in only three cases, a hadronic component has been identified by spectroscopy (detection of iron or nikel lines)~\cite{bib:hadronic_origin1,bib:hadronic_origin2}. Several 
authors have estimated the flux of high-energy neutrino coming from XRB, resulting in very different shapes and normalisations~\cite{bib:hadronic_model1,bib:hadronic_model2,bib:hadronic_model3}. To cover the majority of the range allowed by the models accessible to the ANTARES sensitivity , three 
neutrino-energy spectra are tested in this analysis: $E^{-2}$, $E^{-2}\exp(-E/100~\rm{TeV})$ and $E^{-2}\exp(-E/10~\rm{TeV})$, where $E$ is the neutrino 
energy. 

In the ANTARES telescope~\cite{bib:Antares}, events are primarily detected underwater by observing the Cherenkov light induced 
by relativistic muons in the darkness of the deep sea. Owing to their low interaction probablility, only neutrinos have the ability to cross the Earth. 
Therefore, an upgoing muon is an unambiguous signature of a neutrino interaction close to the detector. To distinguish astrophysical neutrino events from background events (muons and 
neutrinos) generated in the atmosphere, energy and direction reconstructions have been used in several searches~\cite{bib:PointSource}~\cite{bib:Diffuse}. To 
improve the signal-to-noise discrimination, the arrival time information can be used, significantly reducing the effective background~\cite{bib:MDP}. 

In this paper, the results of a time-dependent search for cosmic neutrino sources using the ANTARES data taken from 2008 to 2012 is presented. 
This extends a previous ANTARES analysis~\cite{bib:AntaresMicroQ} where only five sources and the first three years were considered. The analysis 
is applied to a list of promising x-ray binaries candidates detected by various satellites such as Swift, RXTE, MAXI and Fermi. Section 
2 and 3 present the algorithms to identify the outburst periods and the statistical method adopted for this analysis, respectively. Section 4 summarised the results of this search. 
Conclusions are drawn in Section 5.

\section{Selection of outburst periods}
The time-dependent analysis described in the following section is applied to a list of x-ray binaries 
exhibiting outburst periods in their light-curves. The light curves are obtained mainly using the Swift/BAT telescope\footnotemark[1]. 
These data are complemented by the data from others instruments: RXTE/ASM\footnotemark[2], MAXI\footnotemark[3] and Fermi/GBM\footnotemark[4]. A maximum likelihood block 
(MLB) algorithm~\cite{bib:scargle} is used to remove noise from the light curve by iterating over 
the data points and selecting periods during which data are consistent with a constant flux within statistical errors. This algorithm 
is applied independently to all the light curves from all the satellites. Depending on the time period and the avaliability of the 
different instruments, the outbursts are more defined in one sample compare to the others. As the energy range and the sensitivity of these 
telescopes are different, it is not easy to merge the flares of each sources. The value of the steady state (i.e. baseline, BL) and its 
fluctuation ($\sigma_{BL}$) are determined with a Gaussian fit of the lower part of the distribution of the flux data points. The baseline 
is removed from the light curve and the amplitude is converted to a relative amplitude using the sigma of the baseline fluctuations. Finally, 
the relative light curves from different instruments are merged. 

The flaring periods are defined in three main steps. Firstly, seeds are identified by searching for points with an amplitude, or blocks 
with a fluence above $BL~+~8\sigma_{BL}$. Then, each period is extended forward and backward up to an emission compatible with $BL~+~1\sigma_{BL}$. 
An additional delay of 0.5 days is added before and after the flare in order to take into account that the precise time of the flare is not known 
(one-day binned light curve). Finally, spurious flares are discarded if they are not visible by at least one other intrument.  The final list includes 
34 x-ray binaries: 1 HMXB (BH), 12 HMXB (NS), 8 HMXB (BH candidate), 10 LMXB (NS), 3 XRB (BH candidate). The main characteristics of these XRB are reported in 
Table~\ref{table:Sources}. 

\footnotetext[1]{$http://swift.gsfc.nasa.gov/results/transients$}
\footnotetext[2]{$http://xte.mit.edi/ASM_lc.html$}
\footnotetext[3]{$http://maxi.riken.jp$}
\footnotetext[4]{$http://heasarc.gsfc.nasa.gov/W3Browse/fermi/fermigdays.html$}

\begin{table}[ht!]
\caption{List of 34 X-ray binaries with significant flares selected for this analysis.}	       
\label{table:Sources}	 
\centering
\begin{tabular}{|c ||c |c |c|}
\hline Name  & Class & {RA [$^\circ$]} & {Dec [$^\circ$]}  \\
\hline
\hline{Cyg X-1}  	& HMXB (BH) &  230.170 & -57.167   \\
\hline{1A0535p262}  	& HMXB (NS) &  84.727  & 26.316   \\ 
\hline{1A1118-61}  	& HMXB (NS) &  170.238 & -61.917   \\
\hline{Ginga 1843p00}  	& HMXB (NS) &  281.404 & 0.863   \\
\hline{GS 0834-430}  	& HMXB (NS) &  128.979 & -43.185   \\
\hline{GX 304-1}  	& HMXB (NS) &  195.321 & -61.602   \\
\hline{H 1417-624}  	& HMXB (NS) &  215.300 & -62.70   \\
\hline{MXB 0656-072}  	& HMXB (NS) &  104.572 & -7.210  \\
\hline{XTE J1946p274}  	& HMXB (NS) &  296.414 & 27.365  \\ 
\hline{Cyg X-3}  	& HMXB (NS) &  308.107 & 40.958  \\
\hline{GX 1p4}  	& HMXB (NS) &  263.009 & -24.746   \\
\hline{MAXI J1409-619}  & HMXB (NS) &  212.011 & -61.984   \\
\hline{GRO J1008-57}  	& HMXB (NS) &  152.433 & -58.295  \\
\hline{GX 339-4}  	& LMXB (BHC) & 255.7   & -48.8 \\
\hline{4U 1630-472}  	& LMXB (BHC) & 248.504 & -47.393  \\
\hline{IGR J17091-3624}  & LMXB (BHC) & 257.282 & -36.407  \\
\hline{IGR J17464-3213}  & LMXB (BHC) & 266.565 & -32.234  \\
\hline{MAXI J1659-152}  & LMXB (BHC) & 254.757 & -15.258   \\
\hline{SWIFT J1910.2-0546} & LMXB (BHC)& 287.595 & -5.799  \\
\hline{XTE J1752-223}  	& LMXB (BHC) &  268.063 & -22.342  \\
\hline{SWIFT J1539.2-6227} & LMXB (BHC) & 234.800 & -62.467   \\
\hline{4U 1954p31}  	& LMXB (NS) & 298.926 & 32.097   \\
\hline{Aql X-1}  	& LMXB (NS) &  287.817 & 0.585   \\
\hline{Cir X-1}  	& LMXB (NS) &  230.170 & -57.167   \\
\hline{EXO 1745-248}  	& LMXB (NS) &   267.022 & -24.780  \\
\hline{H 1608-522}  	& LMXB (NS) &  243.179  & -52.423  \\
\hline{SAX J1808.4-3658} & LMXB (NS) &   272.115 & -36.977   \\
\hline{XTE J1810-189}  	& LMXB (NS) &   272.586 & -19.070  \\
\hline{4U 1636-536}  	& LMXB (NS) &   250.231 & -53.751  \\
\hline{4U 1705-440}  	& LMXB (NS) &  257.225 & -44.102  \\
\hline{IGR J17473-2721}  & LMXB (NS) &   266.825 & -27.344   \\
\hline{MAXI J1836-194}  & XRB (BHC) &   278.931 & -19.320  \\ 
\hline{XTE J1652-453}  	& XRB (BHC) &   253.085 & -45.344   \\
\hline{SWIFT J1842.5-1124} & XRB (BHC) &  280.573 & -11.418   \\
\hline
\end{tabular}
\end{table}

\section{Time-dependent analysis}

The ANTARES data collected between 2008 and 2012, corresponding to 1044 days of livetime, are analysed to search for
neutrino events around the selected sources, in coincidence with the time periods defined in the previous section. The 
statistical method adopted to infer the presence of a signal on top of the atmospheric neutrino background, or alternatively 
set upper limits on the neutrino flux is an unbinned method based on a likelihood ratio test statistic. The likelihood, 
$\mathcal{L}$, is defined as:
\begin{equation}
\ln \mathcal{L} = \left(\sum_{i=1}^{N} \ln[\mathcal{N}_{\rm S}\mathcal{S}_{i}+\mathcal{N}_{\rm B}\mathcal{B}_{i}]\right)-[\mathcal{N}_{\rm S}+\mathcal{N}_{\rm B}]
\label{eq:EQ_likelihood2}
\end{equation}
\noindent where $\mathcal{S}_{i}$ and $\mathcal{B}_{i}$ are the probabilities for signal and background for an event i, 
respectively, $\mathcal{N}_{\rm S}$ (unknown) and $\mathcal{N}_{\rm B}$ (known) are the number of expected signal and 
background event in the data sample. To discriminate the signal-like events from the background ones, these probabilities 
are described by the product of three components related to the direction, energy, and timing of each event. For an event 
\textit{i}, the signal probability is:
\begin{equation}
\mathcal{S}_{i} = \mathcal{S}^{\rm space}(\Psi_{i}(\alpha_{s},\delta_{s}))\cdot \mathcal{S}^{\rm energy}(dE/dX_{i})\cdot \mathcal{S}^{\rm time}(t_{i}+lag)
\label{eq:EQ_likelihood3}
\end{equation}
\noindent where $\mathcal{S}^{\rm space}$ is a parameterisation of the point spread function, i.e., 
$\mathcal{S}^{\rm space}(\Psi_{i}(\alpha_{s},\delta_{s}))$ the probability to reconstruct an 
event \textit{i} at an angular distance $\Psi_{i}$ from the true source location ($\alpha_{s}$,$\delta_{s}$). 
The energy PDF $\mathcal{S}^{\rm energy}$ is parametrised with the normalised distribution of the muon energy 
estimator, dE/dX, of an event according to the studied energy spectrum. The shape of the time PDF, 
$\mathcal{S}^{\rm time}$, for the signal event is extracted directly from the gamma-ray light curve assuming 
the proportionality between the gamma-ray and the neutrino fluxes. A possible lag of up to $\pm$5 days has been 
introduced in the likelihood to allow for small lags in the proportionality. This corresponds to a possible shift 
of the entire time PDF. The lag parameter is fitted in the likelihood maximisation together with the number of fitted 
signal events in the data. The background probability for an event \textit{i} is:
\begin{equation}
\mathcal{B}_{i} = \mathcal{B}^{\rm space}(\delta_{i})\cdot \mathcal{B}^{\rm energy}(dE/dX_{i})\cdot \mathcal{B}^{\rm time}(t_{i})
\label{eq:EQ_likelihood4}
\end{equation}
where the directional PDF $\mathcal{B}^{\rm{space}}$, the energy PDF $\mathcal{B}^{\rm{energy}}$ and the time PDF 
$\mathcal{B}^{\rm{time}}$ for the background are derived from data using, respectively, the observed declination 
distribution of selected events in the sample, the measured distribution of the energy estimator, and the observed 
time distribution of all the reconstructed muons. 

The goal of the unbinned search is to determine, in a given direction in the sky and at a given
time, the relative contribution of each component, and to calculate the probability to have a signal
above a given background model. This is done via the test statistic, $\lambda$, defined as the ratio of the 
probability for the hypothesis of background and signal ($H_{\rm sig+bkg}$) over the probability of only 
background ($H_{\rm bkg}$):
\begin{equation}
\lambda=\sum_{i=1}^{N} \ln\frac{\mathcal{P}(x_{i}|H_{\rm{sig+bkg}}(\mathcal{N}_{\rm S}))}{\mathcal{P}(x_{i}|H_{\rm{bkg}})} 
\label{eq:TS}
\end{equation}
where $N$ is the total number of events in the considered data sample and $x_i$ are the observed event properties ($\delta_i$, $RA_i$, $dE/dX_i$ and $t_i$). 
The evaluation of the test statistic is performed by generating pseudo-experiments simulating background and 
signal in a 30$^{\circ}$ cone around the considered source according to the background-only and background plus 
signal hypotheses. The performance of the time-dependent analysis is computed with a toy experiment with a source 
assuming a square-shaped flare with a varying width assuming a flat background period. For time ranges characteristic of 
flaring activity, the time-dependent search presented here improves the discovery potential by on-average a factor 2-3 
with respect to a standard time-integrated point-source search~\cite{bib:PointSource} under the assumption that the neutrino 
emission is correlated with the gamma-ray flaring activity. 

\section{Results}

The results of the search is summarised in Table~\ref{table:results}. Only two sources, GX1+4 and IGRJ17091-3624, have a 
pre-trial p-value lower than 10\%. The lowest p-value, 4.1\%, is obtained for GX1+4 where one (three) event is coincident 
in a cone of 1(3) degres with large outbursts detected by Fermi/LAT. Figure~\ref{fig:GX14results} shows the light curve 
of GX 1+4 with the time of the neutrino events, the estimated energy distribution, and the angular distribution of the events 
around the position of this source. The post-trial probability, computed by taking into account the 34 searches, is 72$\%$, 
and is thus compatible with background fluctuations.

\begin{table}[ht!]
\caption{Results of the search for neutrinos in coincidence with XRB outbursts. The total duration of all
identified flares $\Delta t$, the optimised $\Lambda_{opt}$ cuts, the number of required events for a $3\sigma$ 
discovery ($N_{3\sigma}$) pre-trial, the number of fitted signal events by the likelihood ($N_{fit}$), the fitted time 
lag (Lag) and the corresponding pre-trial (post-trial) probability are given together with the energy spectra.}	       
\label{table:results}	 
\centering
\begin{tabular}{|c|c|c|c|c|c|c|c|c|}
\hline{Source}              &  	$\Delta t$	 &   $\Lambda_{opt}$ 	&    $N_{3\sigma}$ & $N_{fit}$ & Lag & P-value & Post-trial & Spectrum \\	      
\hline
\hline{GX1+4}               &  660 d  		 &    -5.2          	&    2.45   		&    0.69 & -5 d &  0.041 & 0.72 &      cutoff 100TeV              	\\
\hline{IGRJ17091-3624}         &  62 d 		 &    -5.4          	&    1.75   		&    0.31 & +4 d &  0.065 & 0.94 &      cutoff 10TeV             	\\
\hline
\end{tabular}
\end{table}

\begin{figure}[ht!]
\centering
\includegraphics[width=0.8\textwidth]{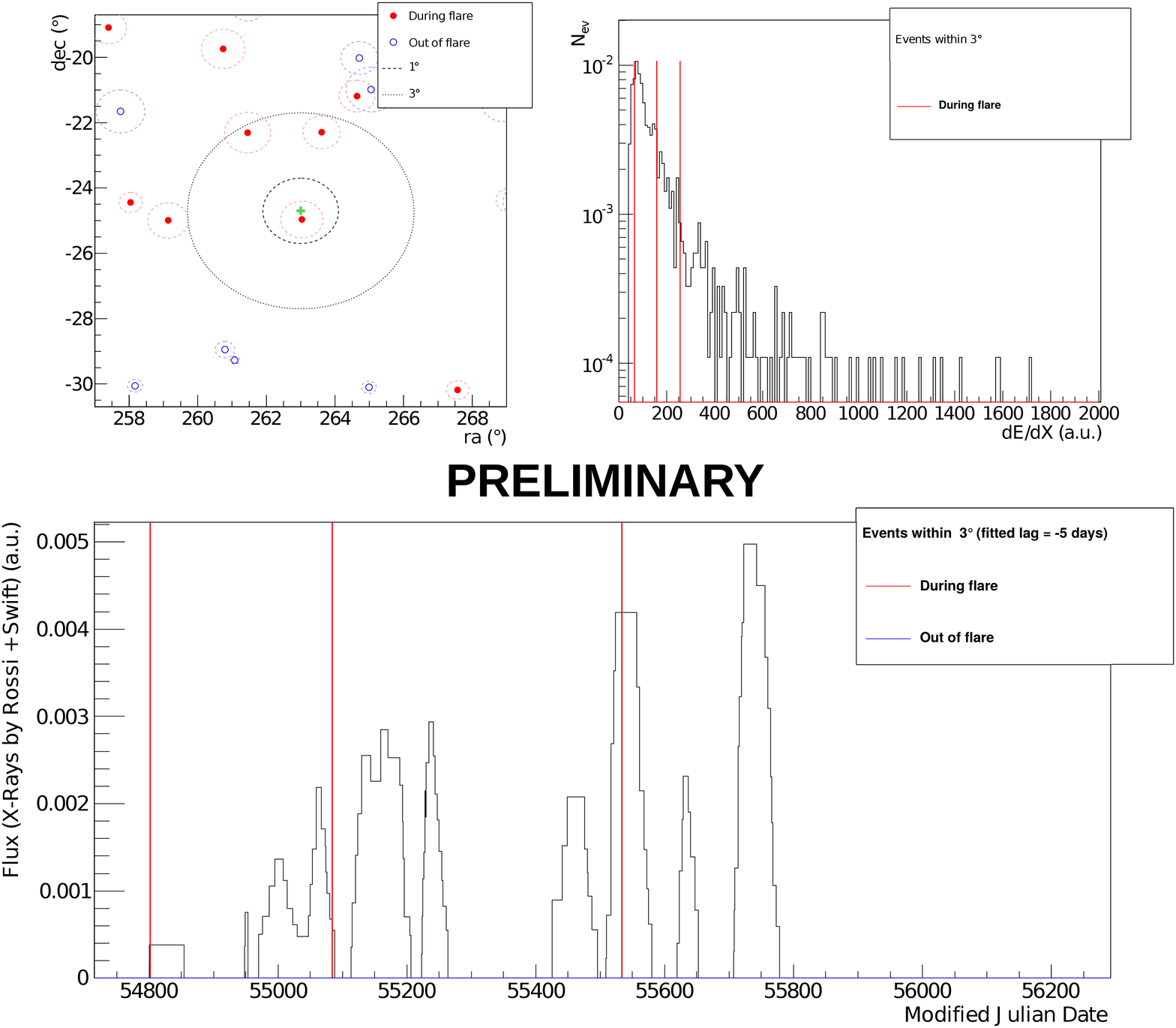}
\caption{Results for GX 1+4. (Top left) Event map around the direction of GX1+4 indicated by the green cross. The full red
(hollow blue) dots indicate the events (not) in time coincidence with the selected flares. The size of the circle around the dots is proportional 
to the estimated angular uncertainty for each event. (Top right) Distribution of the energy estimator dE/dX in a $\pm10^{\circ}$ declination band around the 
source direction. The red line displays the value of the event in coincidence with the flare in a 1$^\circ$ cone around the source direction. (Bottom) Time PDF for the signal simulation (proportional to
the x-ray light curve). The red line displays the times of the ANTARES events associated with the source during a flaring state in a 3$^\circ$ box around the source position.
}
\label{fig:GX14results}
\end{figure}

In the absence of a discovery, upper limits on the neutrino fluence, $\mathcal{F}_{\nu}$, at 90\% confidence 
level are computed using 5-95\% of the energy range and the total effective flare duration.  The limits are 
calculated according to the classical (frequentist) method for upper limits~\cite{bib:Neyman}. Figure~\ref{fig:limit} 
displays these upper limits. Systematic uncertainties of 15\% on the angular resolution and 15\% on the detector acceptance 
have beenincluded in the upper limit calculations.
 
\begin{figure}[ht!]
\centering
\includegraphics[width=0.53\textwidth]{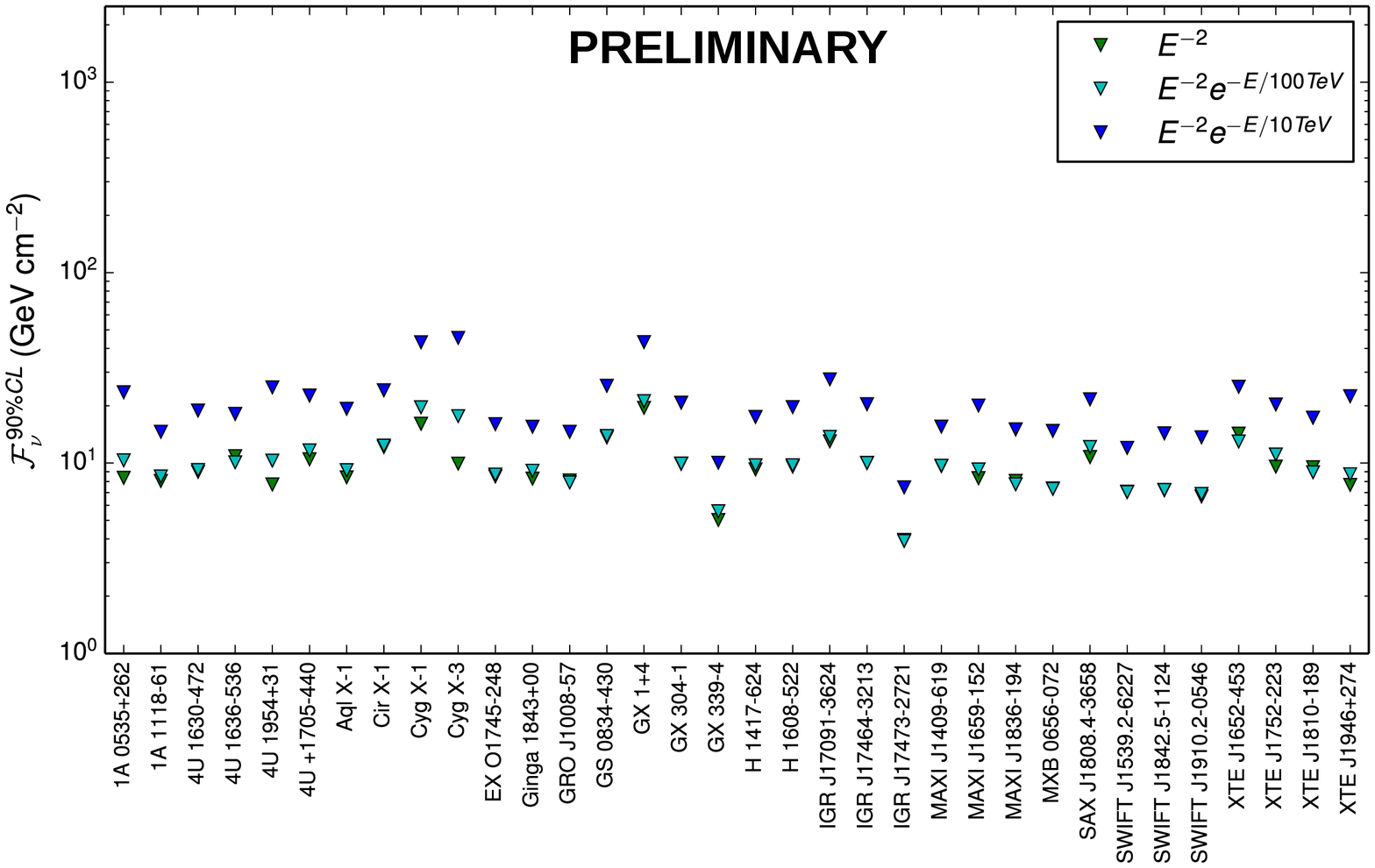}\hfill
\includegraphics[width=0.47\textwidth]{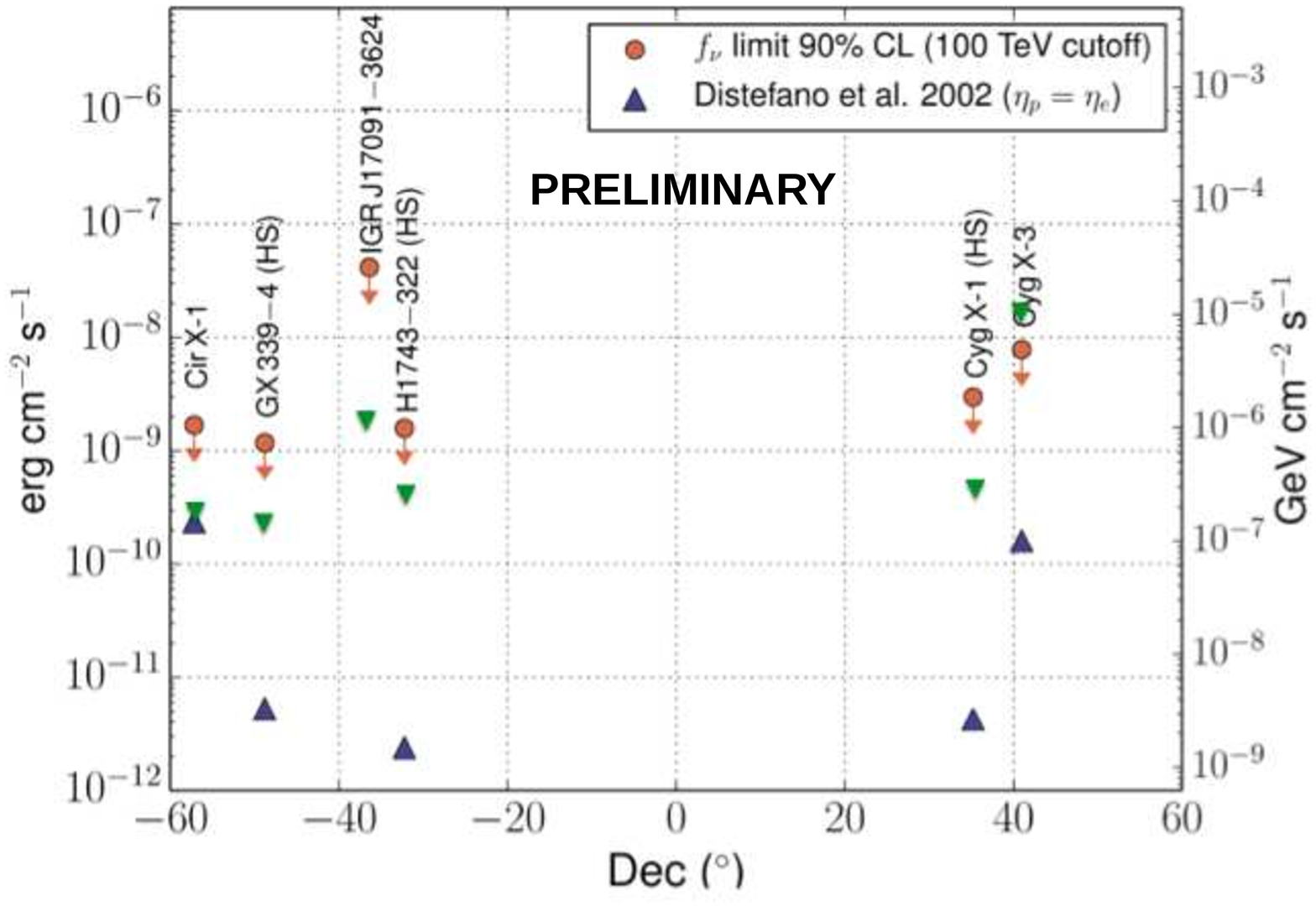}
\caption{Left: upper limits on the neutrino fluence for the 34 studied XRB in the case of  $E^{-2}$ (green triangle), $E^{-2}\exp(-E/100~\rm{TeV})$ (cyan triangle), 
$E^{-2}\exp(-E/10~\rm{TeV})$ (blue triangle) neutrino energy spectra. Right: upper limits at 90\% C.L. on the energy flux in neutrinos obtained in this analysis considering 
a flux $E^{-2}\exp(-\sqrt(E/100~\rm{TeV}))$(circles), compared with the expectations by Ref.~\cite{bib:hadronic_model1} in the case equipartition between electrons and 
protons (triangles).
}
\label{fig:limit}
\end{figure}

The neutrino flux prediction for five microquasars have been computed according to the model~\cite{bib:hadronic_model1} using the 
latest measurements of the distance and of the jet parameters of the microquasars. Figure~\ref{fig:Results_model1} (left) displays these predictions togethers with the upper limits 
computed for this analysis. For Cir X-1, the prediction is less than a factor 2 bellow the ANTARES upper limit. In Ref.~\cite{bib:hadronic_model4}, the authors have 
provided a calculation of the high-energy neutrino emission from GX339-4 in the hypothesis that the primary spectrum of the injected 
particles in the jets has spectral indexes  = -1.8; -2.0 and that the ratio between proton and electron energy is equal to 1 and 100, 
respectively (Figure~\ref{fig:Results_model1} (middle)). The model with a ratio equal to 100 is excluded by the present limit. Finally, Figure~\ref{fig:Results_model1} (right)
shows the comparison between the neutrino flux expectations from Cyg X-3 provided by \cite{bib:hadronic_model5} and \cite{bib:hadronic_model6} and the computed upper limits. The upper limit does not
allow to constrain these types of models.

\begin{figure}[ht!]
\centering
\includegraphics[width=0.5\textwidth]{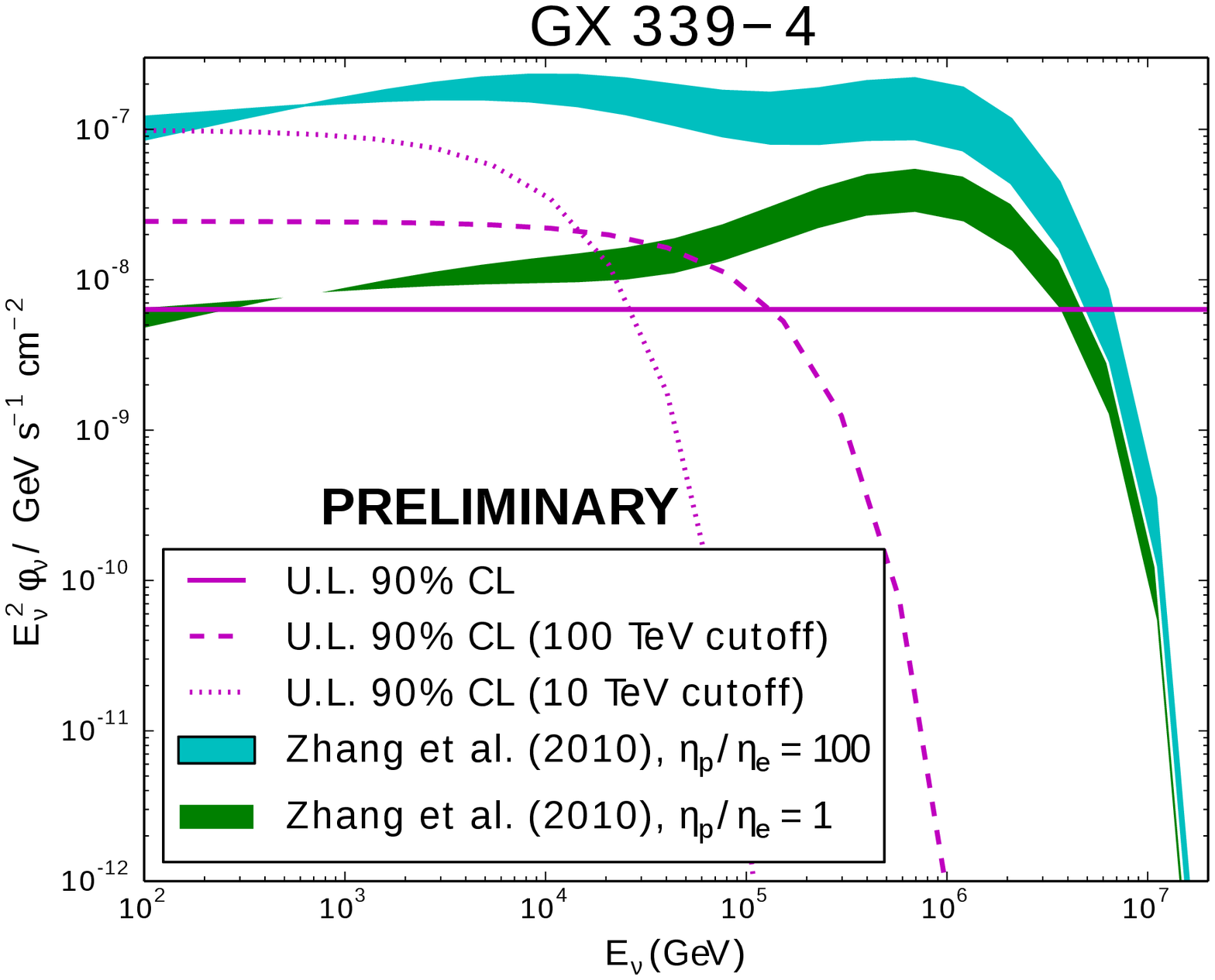}\hfill
\includegraphics[width=0.5\textwidth]{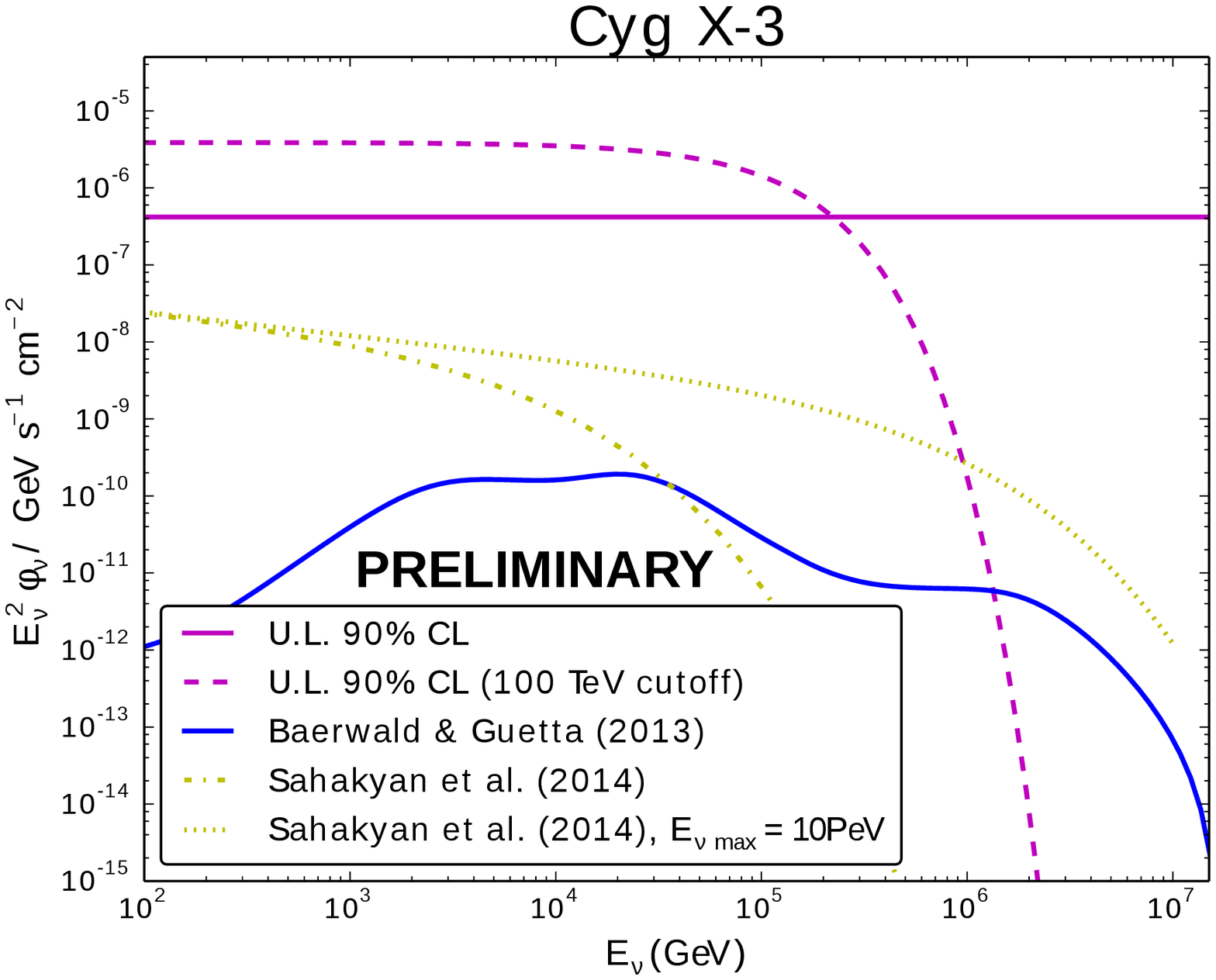}
\caption{Left: upper limits on the neutrino flux for GX 339-4 in both the 
hypotheses without and with the cutoff at 100 TeV, compared to the prediction by the authors of Ref.~\cite{bib:hadronic_model4} for a spectral index of the injected particles 
$-1.8<\alpha<-2.0$ and the ratio $n_{p}$=$n_{e}$ equal to 1 and 100, respectively. Right: Upper limits on the neutrino flux for Cyg X-3 in both the hypotheses without and with the 
cutoff at 100 TeV, compared to the predictions by Ref.\cite{bib:hadronic_model5} and \cite{bib:hadronic_model6}.}
\label{fig:Results_model1}
\end{figure}

\section{Conclusion}
This paper discusses the time-dependent search for cosmic neutrinos from x-ray binaries using the data taken with the full ANTARES 
detector between 2008 and 2012. These searches have been applied to a list of 34 XRB sources. The searches did not result in a statistically significant excess above the expected background from
atmospheric neutrino and muon events. The most significant correlation was found for the source GX 1+4 for which few neutrino events was detected in time/spatial coincidence with the 
x-ray emission. However, the post-trial probability is of 72\%, thus compatible with the background fluctuations. The comparison with predictions from several models have shown that for some sources, the upper limits are 
closed from the expectations. Therefore, with additional data from ANTARES and with the order of magnitude sensitivity improvement expected from the next generation neutrino 
telescope, KM3NeT~\footnotemark[5], the prospects for future searches for neutrino emission from x-ray binaries are very promising.

\footnotetext[5]{$http://www.km3net.org$}


\setcounter{figure}{0}
\setcounter{table}{0}
\setcounter{footnote}{0}
\setcounter{section}{0}
\setcounter{equation}{0}

\newpage
\id{id_amathieu}
\addcontentsline{toc}{part}{\textcolor{blue}{\arabic{IdContrib} - {\sl A. Mathieu} : Follow-up of high energy neutrinos detected by the ANTARES telescope}%
}

\title{\arabic{IdContrib} - Follow-up of high energy neutrinos detected by the ANTARES telescope}

\shorttitle{\arabic{IdContrib} - Follow-up of high energy neutrinos detected by the ANTARES telescope}

\authors{Aurore Mathieu, on behalf of the ANTARES, TAROT, ROTSE, MASTER and SWIFT Collaborations} 
       \afiliations{ Aix Marseille Universit\'e, CNRS/IN2P3, CPPM UMR 7346, 13288, Marseille, France}
        \email{amathieu@cppm.in2p3.fr}


\abstract{ANTARES is currently the largest neutrino telescope operating in the Northern Hemisphere, aiming at the detection of high energy neutrinos from astrophysical sources. Such observations would provide important clues about the processes at work in those sources, and possibly contribute to discover the sources of high energy cosmic rays. Transient sources such as gamma-ray bursts (GRBs) and core-collapse supernovae (CCSNe) are promising candidates, and multi-messenger programs offer a unique opportunity to detect these transient sources. In this way, a method based on optical and X-ray follow-ups of high energy neutrino alerts has been developed within the ANTARES Collaboration. This program, denoted as TAToO (Telescopes-ANTARES Target-of-Opportunity), triggers a network of robotic optical telescopes (TAROT, Zadko, MASTER) and the \textit{Swift}-XRT within a delay of few seconds after the neutrino detection. 
In this contribution, the analysis of optical and X-ray follow-up observations to search for GRBs and CCSNe is presented.
}


\maketitle
\section{Introduction}

High energy neutrinos could be produced in the interaction of charged cosmic rays with matter or radiation surrounding astrophysical sources. Even with the recent detection of extraterrestrial high energy neutrinos by the IceCube experiment \cite{IceCube_neutrinos}, no astrophysical neutrino source has yet been discovered. Such a detection would be a direct evidence of hadronic acceleration mechanisms and would therefore provide important information on the origin of very high energy cosmic rays.

High energy neutrinos are thought to be produced in several kinds of astrophysical sources, such as GRBs \cite{GRBs}, CCSNe \cite{CCSNe} or active galactic nuclei (AGN) \cite{AGN}, in which the acceleration of hadrons may occur. These sources also show a transient behavior covering a large range in the time domain, from seconds for GRBs to weeks for CCSNe or AGN. By combining the information provided by the ANTARES neutrino telescope \cite{ANTARES} with information coming from other observatories, the probability of detecting a source is enhanced since the neutrino background is significantly reduced in the time window of the transient event.

Based on this idea, a multi-wavelength follow-up program, TAToO, operates within the ANTARES Collaboration since 2009 \cite{TAToO}. It relies on optical and X-ray follow-ups of selected high energy neutrino events very shortly after their detection. This online search is mostly motivated by models of neutrinos from long duration GRBs and CCSNe. Both are thought to produce a jet, which is highly relativistic in case of long GRBs, but only mildly relativistic in case of choked jet CCSNe. Follow-up observations have the potential to reveal the electromagnetic counterpart of these transient candidate neutrino sources.

\section{Neutrino alerts}

After the selection of up-going events, which largely removes the huge background of atmospheric muons, the ANTARES neutrino sample consists mainly of atmospheric neutrinos. To select the events which might trigger an alert, a fast and robust algorithm is used to reconstruct the data \cite{BBfit}. This algorithm uses an idealized detector geometry and is independent of the dynamical positioning calibration. This reconstruction and subsequent quality selections allow the rate of events to be reduced from few Hz down to few mHz. The remaining events are then passed to a more precise reconstruction tool which allows the up-going direction of the event to be confirmed and the angular resolution to be improved.

To select neutrino candidates with an increased probability to be of cosmic origin, three online neutrino trigger criteria are currently implemented in the TAToO alert system:
\begin{itemize}
\item High energy trigger: the detection of a single high energy neutrino.
\item Directional: the detection of a single neutrino for which the direction points toward (< 0.5$^{\circ}$) a local galaxy (< 20 Mpc).
\item Doublet trigger: the detection of at least two neutrinos coming from similar directions (< 3$^{\circ}$) within a predefined time window (< 15 min).
\end{itemize}

The main performances of these three triggers are described in Table \ref{tab:perf_tatoo}. Until now, no doublet trigger has been sent to the network. 
	\begin{table*}
	\centering
	\begin{threeparttable}[b]
	\begin{scriptsize}
	\caption{Performances of the three alert criteria. The third column corresponds to the fraction of events inside a
	$2^\circ \times 2^\circ$ field of view.}           
	\label{tab:perf_tatoo}    
	\begin{tabular}{c c c c c}
	\hline\hline
	\noalign{\smallskip}
	Trigger 	 & Angular Resolution (median)   & Fraction of events in FoV     & Muon contamination 	& Mean energy\tnote{a}  	 \\  
	\noalign{\smallskip}    
	\hline
	\noalign{\smallskip}  	
	Doublet              &  $\leq~0.7^{\circ}$  &                           &    0~\%       & $\sim$ 100 GeV  \\
	single HE       	 &  0.25-0.3$^{\circ}$  &  96~\% (GRB), 68~\% (SN)  &  $<0.1~\% $   &  $\sim$ 7 TeV  \\
	single directional   &  0.3-0.4$^{\circ}$   &  90~\% (GRB), 50~\% (SN)  &  $\sim$ 2~\%  &  $\sim$ 1 TeV  \\
	\noalign{\smallskip}
	\hline
	\end{tabular}
	\begin{tablenotes}
    	\item[a]Neutrino energy weighted assuming the atmospheric muon neutrino spectrum.
    \end{tablenotes}
	\end{scriptsize}
 	\end{threeparttable}
	\end{table*}

The trigger criteria are inspired by the features expected from astrophysical sources and are tuned to comply with the alert rate to send to the telescope network. An agreement between ANTARES and the optical telescope collaborations allows a rate of $\sim$ 25 alerts per year to be sent to each optical telescope, while an agreement to send 6 alerts per year to the \textit{Swift} satellite have been accepted. Due to this reduced rate, a subset of the high energy trigger with more restrictive requirements on the neutrino energy, provides a dedicated trigger for the \textit{Swift} satellite.

The TAToO alert system is able to send alerts within few seconds ($\sim$ 3-5 s) after the neutrino detection with an angular resolution better than 0.5$^{\circ}$. 
Since 2009, around 150 and 7 neutrino alerts have successfully been sent to the optical telescope network and the \textit{Swift} satellite, respectively.

\section{The optical and X-ray follow-up system}

The network is composed of small robotic optical telescopes such as TAROT~\cite{TAROT}, Zadko~\cite{Zadko} and MASTER~\cite{MASTER}, and has been extended in June 2013 to the \textit{Swift}-XRT telescope \cite{XRT} for X-ray follow-up. TAROT is a network of two identical 0.25 m telescopes with a field of view (FoV) of 1.86$^{\circ}$ x 1.86$^{\circ}$ located in Calern (France) and La Silla (Chile). These telescopes reach a limiting magnitude of $\sim$ 18.5 mag with an exposure time of 180 s.
Zadko is a one meter telescope located at the Gingin observatory in Western Australia. It covers a FoV of about 0.15 square degrees and can reach a limiting magnitude 1.4 mag deeper compared to the TAROT telescopes with only 60 s of exposure. Recently, 5 telescopes from the MASTER network have also joined the TAToO program. These telescopes, located in Russia and in South Africa, consist of 5 pairs of tubes with a diameter of 0.40 m covering a FoV of up to 8 square degrees for each pair of telescopes. Until the end of 2014, the network also comprises the four optical telescopes ROTSE~\cite{ROTSE}, which have progressively stopped their activity. These 0.45 m telescopes had a FoV of 1.86$^{\circ}$ x 1.86$^{\circ}$ and a sensitivity of $\sim 18.5$ mag with 60 s of exposure.
The wide FoV and the fast response of these telescopes (images can be taken less than 20 s after the neutrino detection) are well suited to the search for transient sources. For each alert, the optical observation strategy is composed of an early follow-up (within 24 hours after the neutrino detection), to search for fast transient sources such as GRB afterglows, complemented by several observations during the two following months, to detect for example the rising light curves of CCSNe. For TAROT telescopes, 6 images of 180 s exposure are taken for each observation, while for ROTSE, 30 and 8 images of 60 s are taken for each observation of the early and long follow-up, respectively.

The \textit{Swift} satellite with its XRT provides a unique opportunity to observe X-ray counterparts to neutrino triggers. The detection sensitivity of the XRT is $5\times10^{-13}$ erg cm$^{-2}$ s$^{-1}$ in 1 ks, and an energy band from 0.3 to 10 keV is covered. Due to the small FoV of the XRT (radius $\sim 0.2^{\circ}$) and the typical error radius of an ANTARES alert ($\sim$ 0.3-0.4$^{\circ}$), each observation of a neutrino trigger is composed of 4 tiles of 2 ks each. This mapping covers about 72\% of the ANTARES PSF for a high energy neutrino. The observation strategy is composed of an automatic response to the neutrino trigger with observations starting as soon as possible. There is an online analysis of the data and in the case where an interesting candidate to be the counterpart is found, further observations are scheduled.

Images provided by follow-up observations must be processed. Optical images are analyzed with a dedicated pipeline based on the image subtraction method\footnote{The MASTER Collaboration analyzes its images with its own reduction pipeline.}, while X-ray data are automatically analyzed by detection algorithms at the UK Swift Science Data Centre.

\section{Results}


\subsection{Early follow-up}

    42 neutrino alerts from January 2010 to January 2015 with early optical images (< 24 h after the neutrino alert) have been analyzed. No optical counterpart associated to one of the 42 neutrinos has been found. Upper limits on the magnitude of possible transient sources which could have emitted the neutrino have thus been derived and are listed in Table \ref{tab:table_optical_alerts}. These limits correspond to the limiting magnitude of images, which is the faintest signal that can be detected. As we are looking for rapidly-fading sources, the signal is supposed to be more important in the first image of the observation, so the upper limits are the limiting magnitude of each first image computed at 5$\sigma$ and corrected for Galactic extinction~\cite{Schlegel}.

Concerning X-ray follow-up, the \textit{Swift}-XRT responded to 7 neutrino triggers between mid 2013 and the beginning of 2015. 22 X-ray sources have been found in the tiled analysis and only 2 sources were already catalogued. Although 20 new X-ray sources have been detected, none of them can be clearly associated with the neutrino trigger. Upper limits on the flux density one may expect from an X-ray counterpart have thus been derived (see Table \ref{tab:table_alerts_swift}). These limits correspond to the sensitivity reached for each 4-tile observation, which lasted from 0.8 to 1.9 ks for the 7 alerts.

\subsection{Discussion on GRB association}

GRBs are the major candidates as sources of high-energy neutrinos among the population of fast transient sources. Because in this study no optical and X-ray counterpart has been observed in coincidence with the 42 and 7 neutrino alerts respectively, the probability to reject a GRB association to each neutrino alert can be directly estimated. To do so, a comparison is done between upper limits obtained for each neutrino alert with optical and X-ray detected afterglow light curves, as shown in Fig. \ref{fig:GRB_limits}.

\begin{figure}[!ht]
	\includegraphics[width=0.45\textwidth]{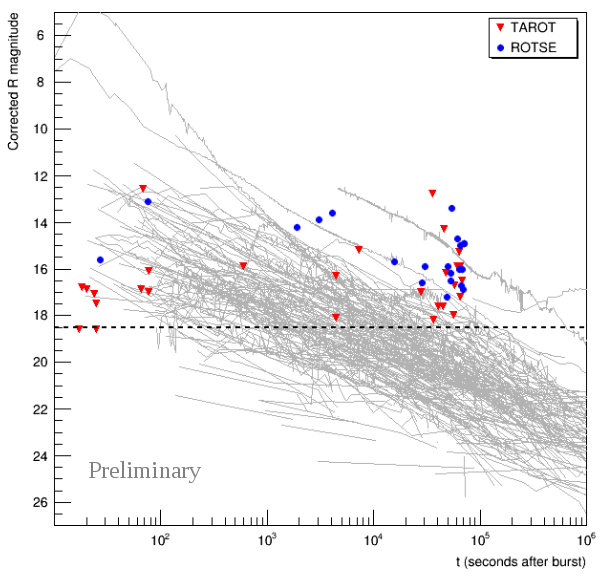} \hfill
	\includegraphics[width=0.5\textwidth]{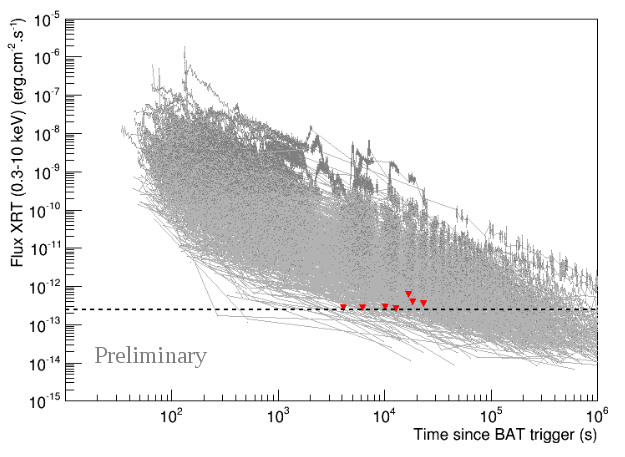}
	\caption{\textit{Left}: optical afterglow light curves observed from 1997 to 2014 by optical telescopes with upper limits on GRB afterglow magnitude for neutrino alerts followed by TAROT (red triangles) and ROTSE (blue circles). Each point represents the first image of the observation which corresponds to an exposure of 180 s for TAROT and 20 s or 60 s for ROTSE images. The horizontal dashed line corresponds to the sensitivity of these telescopes. \textit{Right}: 689 X-ray afterglow light curves detected by the \textit{Swift}-XRT from 2007 to 2015. Upper limits on GRB fluxes for 7 neutrino alerts are represented by red triangles. The horizontal dashed line corresponds to the sensitivity reached with a 2 ks exposure.}
	\label{fig:GRB_limits}
\end{figure}

The cumulative distribution functions (CDFs) of optical afterglow magnitudes and X-ray afterglow fluxes are computed at times coincident with the first optical and X-ray observation of the neutrino alerts, respectively. Figure \ref{fig:cumul_proba} shows these CDFs at typical times after the GRB in the two wavelengths.
\begin{figure}[!ht]
	\includegraphics[width=0.45\textwidth]{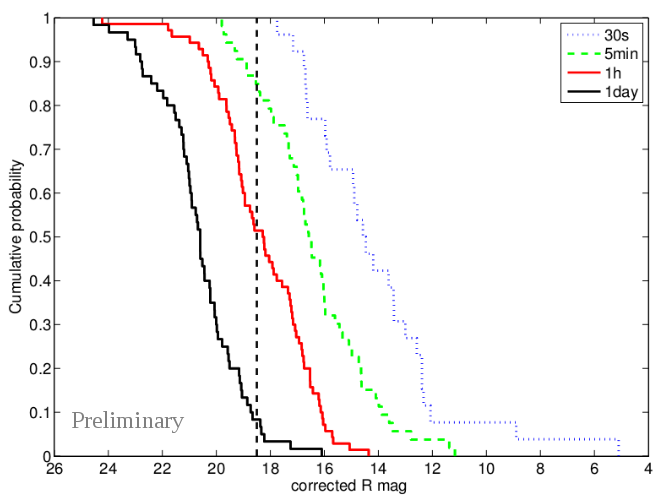}\hfill
	\includegraphics[width=0.47\textwidth]{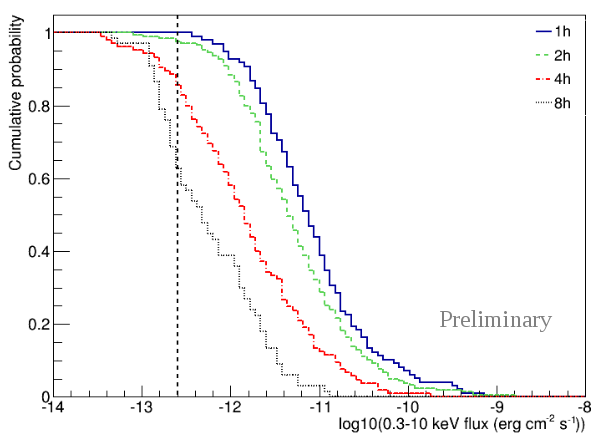}
	\caption{Cumulative distribution of optical (\textit{left}) and X-ray (\textit{right}) afterglow magnitudes for 301 and 689 detected GRBs, respectively. Each curve represents different time after burst. The vertical dashed lines represent the sensitivity of the optical telescopes and the XRT.}
	\label{fig:cumul_proba}
\end{figure}
Under the hypothesis that each detected neutrino comes from GRB, the probability to reject this hypothesis, $P_{reject}^{GRB,\nu}$, can be directly extracted from the CDFs and by considering that the GRB occurred in the field of view of the telescopes. These probabilities are listed in Table \ref{tab:table_optical_alerts} and \ref{tab:table_alerts_swift} for the optical and X-ray follow-ups. 
For most of the really early optical observations (< 1 min after the neutrino trigger), a GRB association is rejected with $\sim$ 90\% probability. With X-ray follow-up, a GRB origin of neutrino alerts can be excluded with $\sim$ 70\% probability for observations made no later than $\sim$ 2 h after the trigger.

	\begin{table*}  
		\centering
		\begin{threeparttable}[b]
		\begin{scriptsize}
		\caption{Details of the 42 neutrino alerts for which early optical images have been taken.}
		\label{tab:table_optical_alerts}  
		\begin{tabular}{cccccccc}
		\hline
		\hline
		\noalign{\smallskip}
		Alert name & Telescope & Analyzed & Exposure \tnote{a} & Delay \tnote{b} & M$_{lim}$ \tnote{c} & $A_v$ \tnote{d} & P \tnote{e} \\
		           &            &  images  & (sec)    &                                           &   (mag)   & (mag) & \\
		\noalign{\smallskip}
		\hline
		\noalign{\smallskip}
		ANT100123A & TAROT & 6  & 180 & 17h47m  & 15.3 & 0.2 & 0    \\ 
		ANT100725A & TAROT & 6  & 180 & 1m17s   & 16.1 & 0.3 &  0.50 \\ 
				   & ROTSE & 30 & 20  & 1m15s   & 13.1 & 0.3 &  0.12 \\ 
		ANT100913A & TAROT & 6  & 180 & 11h24m  & 17.6 & 0.0 &  0.06 \\ 
		ANT100922A & ROTSE & 26 & 20  & 1h08m   & 13.6 & 0.5 &  0    \\ 
		ANT110305A & ROTSE & 29 & 60  & 4h19m   & 15.7 & 0.1 &  0.06 \\ 
		ANT110409A & TAROT & 6  & 180 & 1m08s   & 12.6 & 5.6 &  0.04 \\ 
		ANT110531A & TAROT & 6  & 180 & 12h34m  & 17.6 & 0.1 &  0.06 \\ 
		ANT110923A & TAROT & 7  & 180 & 9h58m   & 12.8 & 3.9 &  0    \\ 
		ANT110925B & TAROT & 6  & 180 & 2h01m   & 15.2 & 1.8 &  0.10 \\ 
				   & ROTSE & 30 & 60  & 50m58s  & 13.9 & 1.8 &  0    \\ 
		ANT111008A & TAROT & 5  & 180 & 12h53m  & 14.3 & 2.5 &  0    \\ 
		ANT111019A & ROTSE & 8  & 60  & 18h22m  & 16.7 & 0.1 &  0.02 \\ 
		ANT111019B & ROTSE & 8  & 60  & 19h09m  & 16.9 & 0.1 &  0.02 \\ 
		ANT111101A & ROTSE & 8  & 60  & 13h33m  & 17.2 & 0.1 &  0.02 \\ 
		ANT111205A & TAROT & 6 & 180 & 10h05m  & 18.2 & 0.4 &  0.16 \\ 
		ANT111228A & TAROT & 6 & 180 & 7h44m   & 17.0 & 0.1 & 0.04 \\ 
				   & ROTSE & 8  & 60  & 7h53m   & 16.6 & 0.1 & 0.04 \\ 
		ANT120102A & TAROT & 4 & 180 & 1m17s   & 17.0 & 0.1 &  0.60 \\ 
		ANT120105A & ROTSE & 8  & 60  & 17h39m  & 16.0 & 0.4 & 0.02 \\ 
		ANT120730A & TAROT & 26 & 180 & 20s     & 16.9 & 0.4 & 0.88 \\ 
		ANT120907A & TAROT & 14 & 180 & 9m53s   & 15.9 & 0.2 & 0.31 \\ 
		ANT120907B & TAROT & 11 & 180 & 18h15m  & 17.2 & 0.2 & 0.02 \\ 
				   & ROTSE & 27 & 60  & 8h28m   & 15.9 & 0.2 & 0.02 \\ 
		ANT120923A & TAROT & 6  & 180 & 15h43m  & 18.0 & 0.1 & 0.03 \\ 
		ANT121010A & TAROT & 24 & 180 &  25s   & 18.6 & 0.0 & 0.90 \\ 
		ANT121012A & TAROT & 6 & 180 &  19h06m & 16.5 & 0.7 & 0.02 \\ 
		ANT121027A & ROTSE & 8 & 20  & 14h56m  & 13.4 & 2.6 & 0    \\ 
		ANT121206A & ROTSE & 27 & 60  & 27s     & 15.6 & 1.1 & 0.62 \\ 
		ANT130210A & ROTSE & 8 & 60  & 14h46m  & 16.5 & 0.1 & 0.02 \\ 
		ANT130724A & TAROT & 3  & 180 & 18h04m  & 15.9 & 0.1 & 0.02 \\ 
		ANT130928A & ROTSE & 8 & 60  & 13h49m  & 15.9 & 0.1 & 0.02 \\ 
		ANT131027A & ROTSE & 8  & 20  & 18h14m  & 15.0 & 0.7 & 0    \\ 
		ANT131209A & TAROT & 6  & 180 & 1h14m   & 16.3 & 0.1 & 0.14 \\ 
		ANT131221A & TAROT & 2 & 180 & 18s     & 16.8 & 0.5 & 0.83 \\ 
		ANT140123A & TAROT & 23 & 180 & 13h21m  & 16.2 & 1.3 & 0.02 \\ 
		ANT140125A & TAROT & 6  & 180 & 1h14m   & 18.1 & 0.0 & 0.43 \\ 
		ANT140203A & ROTSE & 8 & 60  & 19h43m  & 14.9 & 0.1 & 0    \\ 
		ANT140223A & TAROT & 3 & 180 & 17h08m  & 15.9 & 0.1 & 0.02 \\ 
				   & ROTSE & 3  & 60  & 31m29s  & 14.2 & 0.1 & 0.02 \\ 
		ANT140304A & TAROT & 18 & 180 & 25s     & 17.5 & 0.6 & 0.92 \\ 
		ANT140309A & TAROT & 16 & 180 & 24s     & 17.1 & 0.1 & 0.88 \\ 
		ANT140323A & ROTSE & 8 & 60  & 14h47m  & 16.2 & 0.2 & 0.02 \\ 
		ANT140408A & TAROT & 6 & 180 & 16h11m  & 16.7 & 0.1 & 0.02 \\ 
				   & ROTSE & 8  & 60  & 19h07m  & 16.0 & 0.1 & 0.02 \\ 
		ANT140505A & ROTSE & 2  & 60  & 17h11m  & 14.7 & 0.1 & 0    \\ 
		ANT140914A & TAROT & 13 & 180 &  1m05s  & 16.9 & 0.5 & 0.62 \\ 
		ANT150122A & TAROT & 8 & 180 &  17s     & 18.6 & 0.1 & 0.90 \\ 
		\noalign{\smallskip}
		\hline
		\end{tabular}
		\begin{tablenotes}
    	\item[a] Exposure of each image.
    	\item[b] Delay in hours, minutes and/or seconds between the neutrino trigger and the first image.
    	\item[c] Limiting magnitude of the first image computed at 5$\sigma$ and corrected for the galactic extinction.
    	\item[d] Galactic extinction from \cite{Schlegel}.
    	\item[e] Probability to reject an association between the neutrino trigger and a GRB.
   	 	\end{tablenotes}
		\end{scriptsize}
 		\end{threeparttable}	
	\end{table*}
	
		\begin{table*}
		\centering
		\begin{threeparttable}[b]
		\begin{scriptsize}
		\caption{Details of the 7 ANTARES triggers observed by the \textit{Swift}-XRT since 2013.}
		\label{tab:table_alerts_swift}
		\begin{tabular}{cccccccc}
		\hline
		\hline
		\noalign{\smallskip}
		Trigger name &   Error radius   &  Delay\tnote{a}  & Mean exposure  & Sensitivity & 
		 New sources (total)\tnote{b} & Counterpart & P\tnote{c} \\
		(ANTyymmddA) &  ($^\circ$)  &  (hours)  &  (ks)  &  ($\times 10^{-13}$ erg cm$^{-2}$ s$^{-1}$)  & 
		 & candidates &  \\
		\noalign{\smallskip}
		\hline
		\noalign{\smallskip}
		ANT130722A	 & 	0.4	  &  1.1  &  1.8  & 2.74  &  4 (5) &  0  &  0.71   \\ 
		ANT130915A 	 & 	0.3	  &  6.5  &  1.4  & 3.48  &  2 (2) &  0  &  0.60   \\ 
		ANT130927A   & 	0.4   &  5.1  &  1.3  & 3.84  &  0 (1) &  0  &  0.60   \\ 
		ANT140123A   & 	0.35  &  4.7  &  0.8  & 5.99  &  1 (1) &  0  &  0.55   \\ 
		ANT140311A   & 	0.35  &  2.8  &  1.7  & 2.88  &  3 (3) &  0  &  0.68   \\ 
		ANT141220A   & 	0.4	  &  3.5  &  1.9  & 2.63  &  4 (4) &  0  &  0.67   \\ 
		ANT150129A   & 	0.35  &  1.7  &  1.9  & 2.67  &  6 (6) &  0  &  0.69   \\ 
		\noalign{\smallskip}
		\hline
		\end{tabular}
		\begin{tablenotes}
    	\item[a] Delay between the neutrino trigger and the first observation by the \textit{Swift}-XRT.
    	\item[b] Number of uncatalogued sources among the total number of detected sources in each 4-tile observation.
    	\item[c] Probability to reject an association between the neutrino trigger and a GRB.
   	 	\end{tablenotes}
		\end{scriptsize}
 		\end{threeparttable}
	\end{table*}

\subsection{Long term follow-up}

71 alerts from October 2009 to January 2015 with optical follow-up observations (at least three night of observations) have been processed. No slowly varying transient optical counterpart was found in association with a neutrino trigger. This null result is consistent with the small expectation value of 0.2 accidentally
discovered SNe for 71 alerts.

From this result, constraints on the Ando \& Beacom model \cite{CCSNe} parameters will be set. In this model, the production of high energy neutrinos from mildly relativistic jets of CCSNe is proposed, depending on the jet energy $E_{jet}$, the Lorentz boost factor $\Gamma$ and the rate of CCSNe with such jets $\rho$. To test this model, a test statistic depending on an ANTARES term and an optical follow-up term will be used to check the compatibility of the measurement with the model expectations. If a set of model parameters predicts a significant larger amount of neutrinos and SN counterparts than measured in the data sample, the model can be excluded.

\section{Conclusion}

The optical and X-ray follow-ups of the ANTARES neutrino alerts have been running stably since 2010 and mid 2013, respectively. About 150 and 7 alerts have been sent to the optical telescope network and to the \textit{Swift}-XRT. The main advantage of the ANTARES program is that it is able to send alerts within few seconds after the neutrino detection with a precision better than 0.5$^{\circ}$ for high energy neutrinos.
The analysis of 42 and 7 early follow-up observations in optical and X-ray has not yet permitted to discover any transient sources associated to the neutrino events. Upper limits on the magnitude of possible transient sources have been derived. Compared to the state-of-the-art of the GRB afterglow detected light curves, the very rapid response time of the optical telescopes has allowed stringent constraints on the GRB origin of individual neutrinos to be placed. Even with the larger response time of the XRT follow-up, early observations have allowed the GRB origin for the 7 neutrino alerts to be excluded with a high probability.



\setcounter{figure}{0}
\setcounter{table}{0}
\setcounter{footnote}{0}
\setcounter{section}{0}
\setcounter{equation}{0}
\newpage
\id{id_asanchezlosa}
\addcontentsline{toc}{part}{\textcolor{blue}{\arabic{IdContrib} - {\sl A. S\'anchez-Losa} : Time-dependent search of high energy cosmic neutrinos from variable Blazars with the ANTARES telescope}%
}
%

\title{\arabic{IdContrib} - Time-dependent search of high energy cosmic neutrinos from variable Blazars with the ANTARES telescope}

\shorttitle{\arabic{IdContrib} - Time-dependent search of Blazar cosmic neutrinos with ANTARES}

\authors{Agust\'in S\'anchez-Losa$^a$, Damien Dornic$^b$}
        \afiliations{$^a$IFIC - Instituto de F\'isica Corpuscular, Edificios Investigaci\'on de Paterna, CSIC - Universitat de Val\`encia, Apdo. de Correos 22085, 46071 Valencia, Spain\\
       $^b$Aix Marseille Universit\'e, CNRS/IN2P3, CPPM UMR 7346, 13288, Marseille, France }
\email{$^a$agustin.sanchez@ific.uv.es, $^b$dornic@cppm.in2p3.fr}
%

\abstract{ANTARES, the largest neutrino telescope operating in the Northern Hemisphere, performs multiple analyses in the search for neutrino point-source candidates. In a time-dependent search, the background is drastically reduced, and the point-source sensitivity improved, by selecting a narrow time window around the assumed neutrino production period. Blazars are particularly attractive potential neutrino point sources, since they are among the most likely sources of the observed very-high-energy cosmic rays. Neutrinos and gamma rays may be produced in hadronic interactions with the surrounding medium. Moreover, blazars generally show large time variability in their light curves at different wavelengths and on various time scales. For the time-window selection, their gamma ray emission measured by the LAT instrument on-board the Fermi satellite is derived, and the resulting light curves are characterised by a time series analysis. The studied periods are determined by applying a threshold on the fluence on the light curves. In addition, the flares reported at TeV energies by the IACTs HESS, MAGIC and VERITAS have been included in a second dedicated analysis. The sensitivities reached with this method improve by a factor 2-3 with respect to a standard time-integrated point source search. The results of the two searches, using data from the years 2008 up to 2012, will be presented.}


\maketitle


\section{Introduction}

High-energy neutrino source detection would yield to identify the cosmic ray sources~\cite{bib:reviewNeutrinosources} and provide answer to the responsible acceleration mechanisms hosted on them. Active galactic nuclei (AGN) are among these candidates, although it remains unclear if their gamma ray emissions are due to leptonic~\cite{bib:AGNleptonic} or hadronic processes~\cite{bib:hadronic1,bib:hadronic2}. In the later case, the emission is attributed to $\pi^{0}$ decays; the corresponding production of charged pions implies a correlated neutrino emission.

Flat-Spectrum radio quasars (FSRQs) and BL Lacs, together classified as blazars~\cite{bib:Blazars}, are among the most likely sources of the observed very high-energy cosmic rays~\cite{bib:hadronic3,bib:hadronic4}. Multiple models predict different neutrino fluxes from AGNs with different normalisations and shapes~\cite{bib:AGNhadronic1,bib:AGNhadronic2,bib:AGNhadronic3,bib:AGNhadronic4,bib:model3,bib:model2,bib:model1}. The $E^{-2}$ spectrum is generally the most expected, yet some authors estimate more optimistic spectral indexes up to one~\cite{bib:model4,bib:model5}. Additionally, in most gamma ray sources is observed an energy cutoff. To cover the wide range of possibilities, four neutrino spectra are tested in this analysis: $E^{-1}$, $E^{-2}$, $E^{-2}\exp(-E/10~\rm{TeV})$ and $E^{-2}\exp(-E/1~\rm{TeV})$, with $E$ is the neutrino energy.

The ANTARES telescope~\cite{bib:Antares} detects events through the Cherenkov light emitted by muons in the deep sea. To distinguish astrophysical neutrino events from background ones (atmospheric muons and neutrinos), energy and direction reconstruction of events have been used in several searches~\cite{bib:PointSource,bib:diffuse,bib:AntaresICRC2013}. To improve this discrimination, the arrival time information can be used reducing significantly the effective background. Blazars present time variable emissions through different wavelengths at different time scales~\cite{bib:FermiLATAGNvariability,bib:AGNvariability1,bib:AGNvariability2}. This variability would take place also in the corresponding neutrino emission. The use of this information in the time-dependant methods improve the detection probability with respect to time-integrated approaches.

The results of a time-dependent search for cosmic neutrino sources in the sky visible to the ANTARES telescope using data taken from 2008 to 2012 are presented. This extends a previous ANTARES analysis~\cite{bib:flare} where only the last four months of 2008 were considered. The analysis is applied to a list of promising AGN candidates detected flaring by the FERMI satellite, and to a list of flares reported by TeV-range experiments (H.E.S.S., MAGIC and VERITAS).


\section{Time-dependent search method}

An unbinned likelihood-ratio maximisation method is used to perform the analysis. Data are parametrised as a two-component mixture of signal and background. The probability density function $\mathcal{P}_{i}$ and the likelihood $\mathcal{L}$ are:
\begin{equation}
\mathcal{P}_{i} = \mathcal{N}_{\rm S}\mathcal{S}_{i} + \mathcal{N}_{\rm B}\mathcal{B}_{i}
\label{eq:EQ_likelihood1}
\end{equation}
\begin{equation}
\ln \mathcal{L} = \sum_{i=1}^{N} \ln[\mathcal{N}_{\rm S}\mathcal{S}_{i}+\mathcal{N}_{\rm B}\mathcal{B}_{i}]-[\mathcal{N}_{\rm S}+\mathcal{N}_{\rm B}]
\label{eq:EQ_likelihood2}
\end{equation}
where $\mathcal{N}_{\rm S}$ and $\mathcal{N}_{\rm B}$ are the expected number of signal (unknown) and background (known) events, with probability distributions (PDFs) for an event $i$, $\mathcal{S}_{i}$ and $\mathcal{B}_{i}$, respectively. These PDFs are the product of three components that describe the event direction, energy and timing probabilities.

For an event $i$, the signal PDF is:
\begin{equation}
\mathcal{S}_{i} = \mathcal{S}_{i}^{\rm space}(\Psi_{i}(\alpha_{s},\delta_{s})) \cdot \mathcal{S}_{i}^{\rm energy}(dE/dX_{i}) \cdot \mathcal{S}_{i}^{\rm time}(t_{i}+lag)
\label{eq:EQ_likelihood3}
\end{equation}
where $\mathcal{S}_{i}^{\rm space}$ represents the point spread function, $\mathcal{S}_{i}^{\rm energy}$ is the energy PDF according to the studied energy spectrum and $\mathcal{S}_{i}^{\rm time}$ is the time PDF, extracted from the gamma ray emission of the studied source. A possible lag of up to $\pm$5 days is implemented in the likelihood to allow small lags in the gamma ray emission and the neutrino signal. This parameter is maximised in the likelihood together the number of signal events ($\mathcal{N}_{\rm S}$). For the signal simulation, the correlation between $\mathcal{S}_{i}^{\rm space}(\Psi_{i}(\alpha_{s},\delta_{s}))$ and $\mathcal{S}_{i}^{\rm energy}(dE/dX_{i})$ is taken into account.

The background PDF is:
\begin{equation}
\mathcal{B}_{i} = \mathcal{B}_{i}^{\rm space}(\delta_{i})\cdot \mathcal{B}_{i}^{\rm energy}(dE/dX_{i})\cdot \mathcal{B}_{i}^{\rm time}(t_{i})
\label{eq:EQ_likelihood2}
\end{equation}
where the directional PDF $\mathcal{B}_{i}^{space}$, the energy PDF $\mathcal{B}_{i}^{energy}$ and the time PDF $\mathcal{B}_{i}^{time}$ for the background are derived from data. 

The significance of the analysis is determined through a likelihood ratio test statistic, $\lambda$, defined as:
\begin{equation}
\lambda=\sum_{i=1}^{N} \ln\frac{\mathcal{P}(x_{i}|H_{\rm{sig+bkg}}(\mathcal{N}_{\rm S}))}{\mathcal{P}(x_{i}|H_{\rm{bkg}})} 
\label{eq:TS}
\end{equation}
where $\mathcal{N}_{\rm S}$ and $N$ are respectively the unknown number of signal events and the total number of events in the considered data sample. Its evaluation is performed through pseudo-experiment simulations.


Tests on the performance of the time-dependent analysis shows on-average a factor 2-3 of improvement with respect to the time-integrated case~\cite{bib:PointSource,bib:AntaresICRC2013}.


\section{Gamma-ray flares}

\subsection{GeV flares: Fermi LAT}

The time-dependent analysis described is applied to bright and variable Fermi blazar sources reported in the second Fermi LAT catalogue~\cite{bib:Fermicatalogue} and in the LBAS catalogue (LAT Bright AGN sample~\cite{bib:FermicatalogueAGN}). From there, are selected the sources visible by ANTARES ($\delta~<~35^\circ$) with a gamma ray flux greater than $10^{-9}~\rm{photons}\cdot \rm{cm}^{-2}\cdot \rm{s}^{-1}$ above 1~GeV, a detection significance $TS~>~25$ and a significant time variability. This list is completed up to a total of 154 sources by including sources reported as flaring in the Fermi Flare Advocates in 2011 and 2012~\cite{bib:FermiAdvocates}.

The gamma ray light curves are produced using the Fermi Public Release Pass 7 data with the source class event selection (evclass=2) and the Fermi Science Tools v9r35p1 package~\cite{bib:FermiData}, processing the photon counting data above 100~MeV, from August 2008 to December 2012, in a 2$^\circ$ cone radius around the studied source direction. Sources close to the galactic plane (galactic latitude |l|~<~10$^\circ$) or with other sources within a 2$^\circ$ cone (or 3$^\circ$ for very bright sources) are excluded due to different origin gamma ray contamination. 

A maximum likelihood block (MLB) algorithm~\cite{bib:scargle, bib:scargle1, bib:scargle2} is used to remove noise from the light curves by iterating over the data points and selecting periods during which data are consistent with a constant flux within statistical errors. The flaring periods are defined through a threshold on the fluence on these denoised light curves, based on the gamma ray emission baseline and flare significance. The final list which includes any gamma ray flaring source reduces to 41 blazars: 33 Flat Spectrum Radio Quasars, 7 BL-Lacs and 1 unknown identification.

\subsection{TeV flares: IACTs}

Imaging air Cherenkov telescopes (IACTs) such as H.E.S.S., MAGIC and VERITAS cannot monitor sources continuously. These telescopes detect photons with energies in the GeV-- TeV range that can be better correlated with high energy neutrinos. These observatories often emit alerts reporting flares to Astronomer's Telegram or directly in a dedicated paper. From them, the flaring periods are extracted, assuming a single square-shaped flare. The sources are chosen for this analysis according to the same visibility criteria as for Fermi/LAT observations, comprising 7 blazars. The same analysis as described previously is performed assuming the same four energy spectra.

~

Figure~\ref{fig:skymap} shows the position of the Blazars analysed, on top of the ANTARES visibility.

\begin{figure}[t]
\centering
\includegraphics[width=0.9\textwidth]{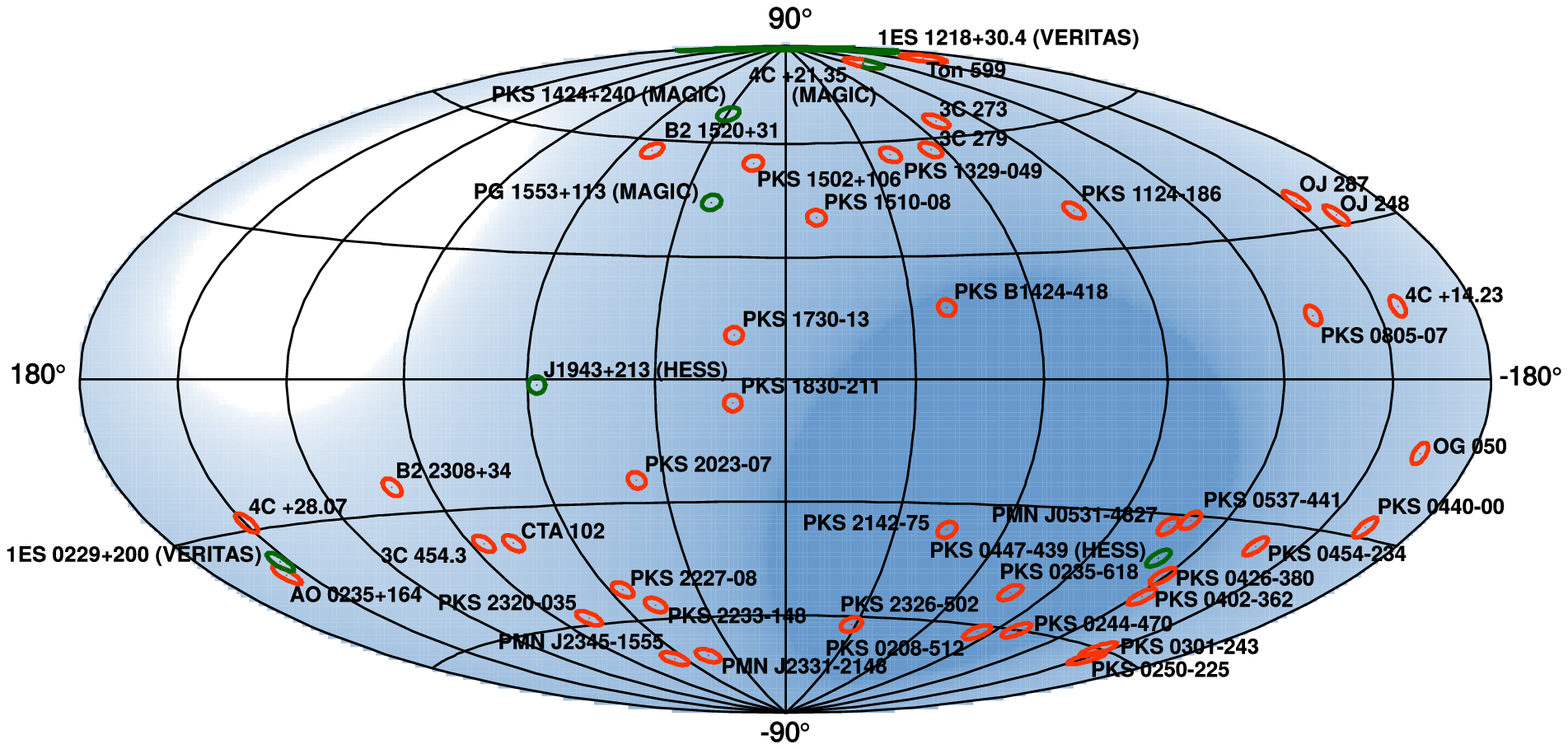}
\caption{Skymap in galactic coordinates showing the position of the 41 selected Fermi blazars (red circle) and the 7 TeV blazars (green circle) on top of the ANTARES visibility of the analyses ($\cos(\theta$)~$>$~-0.15).}
\label{fig:skymap}
\end{figure}

\begin{figure}[t]
\centering
\includegraphics[width=0.9\textwidth]{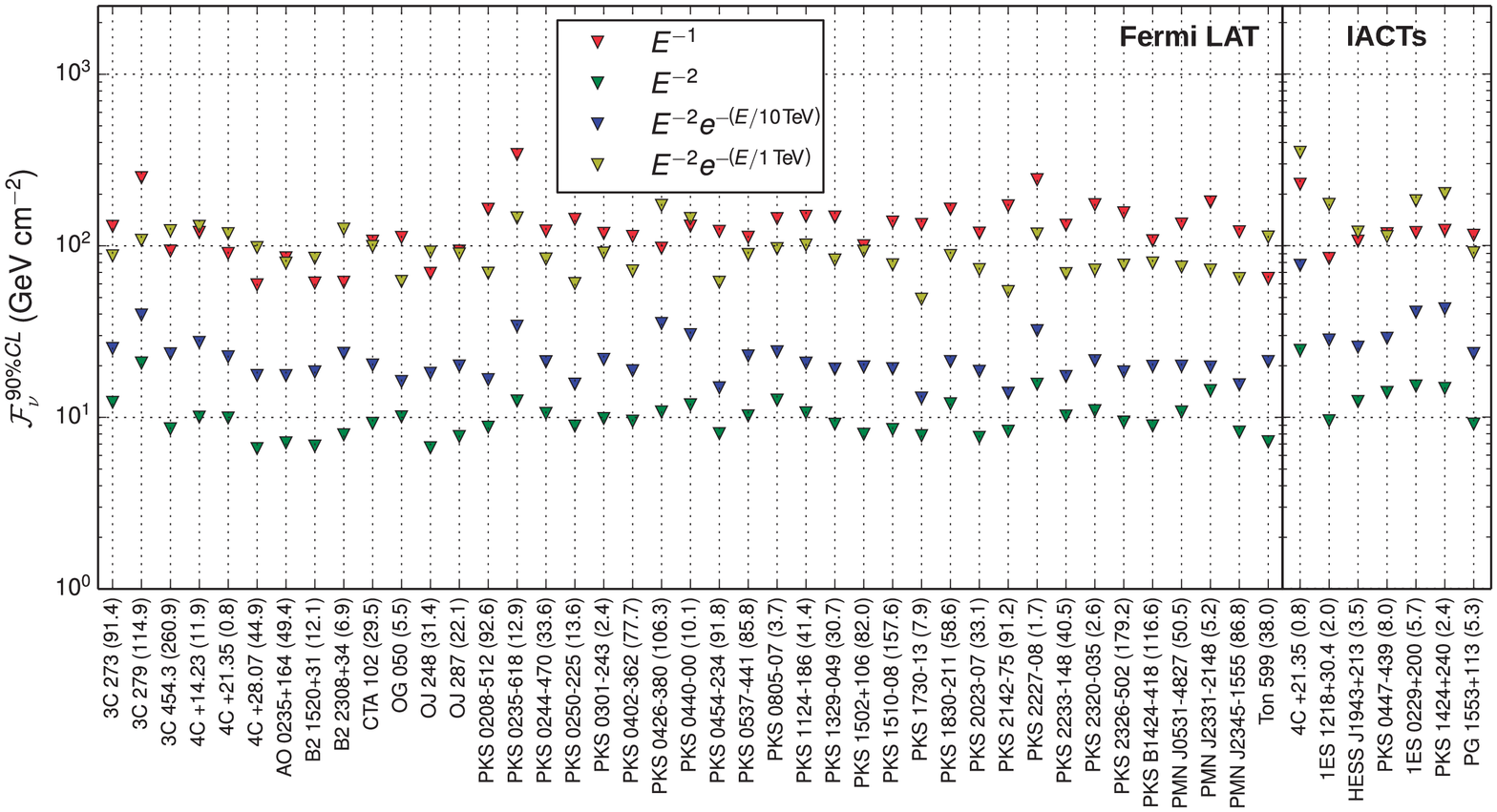}
\caption{Upper-limits on the neutrino fluence for the 41 studied Fermi blazars (left) and the 7 studied TeV blazars (right) in the case of  $E^{-2}$ (green), $E^{-2}\exp(-E/10~\rm{TeV})$ (blue), $E^{-2}\exp(-E/1~\rm{TeV})$ (yellow) and $E^{-1}$ (red) neutrino energy spectra. The number in paranthesis after the name of the source in the x-axis indicates the total flare length during the studied period.}
\label{fig:limit}
\end{figure}

\section{Results \& discussion}

Of the GeV flares, only three sources, 3C279, PKS10235-618 and PKS1124-186, have a pre-trial p-value lower than 10\%. The lowest p-value, 3.3\%, is obtained for the source 3C279 where one event is coincident with a large gamma ray flare detected by Fermi/LAT in November 2008. 
This event has already been reported in a previous analysis~\cite{bib:flare}. The post-trial probability, computed by taking into account the 41 searches, is 67$\%$, and is thus compatible with background fluctuations.

Among the seven tested flares reported by IACTs, only the blazar PKS0447-439 shows a pre-trial p-value lower than 10\% in the case of the assumed $E^{-2}\exp(-E/1~\rm{TeV})$ energy spectrum. The corresponding post trial p-value is 55\%, being also consistent with background fluctuations.

In the absence of a discovery, upper limits on the neutrino fluence $\mathcal{F}_{\nu}$ at 90\% confidence level are computed using 5-95\% of the energy range as:
\begin{equation}
  \mathcal{F_\nu} = \int_{t_{min}}^{t_{max}} \mathrm{d}t \int_{E_{min}}^{E_{max}} \mathrm{d}E \times E \frac{\mathrm{d}N}{\mathrm{d}E} = \Delta t \int_{E_{min}}^{E_{max}} \mathrm{d}E \times E \frac{\mathrm{d}N}{\mathrm{d}E}
\end{equation}

The emission duration, $\Delta t$ is computed using the effective livetime during the flare. The limits include systematic errors and are calculated according to the classical (frequentist) method for upper limits~\cite{bib:Neyman} (see Figure~\ref{fig:limit}). IceCube has performed a similar time-dependent analysis~\cite{bib:IceCubeflare} using data from 2008 to 2012 with similar results. 19 sources are in common with the Fermi-analysed sources. 
For sources in the Southern Hemisphere, the limits computed by IceCube are on the same order of magnitudes whereas the ANTARES limits are a factor 10 worse for the sources in the Northen hemisphere.

\begin{figure}[t]
\centering
\includegraphics[width=0.9\textwidth]{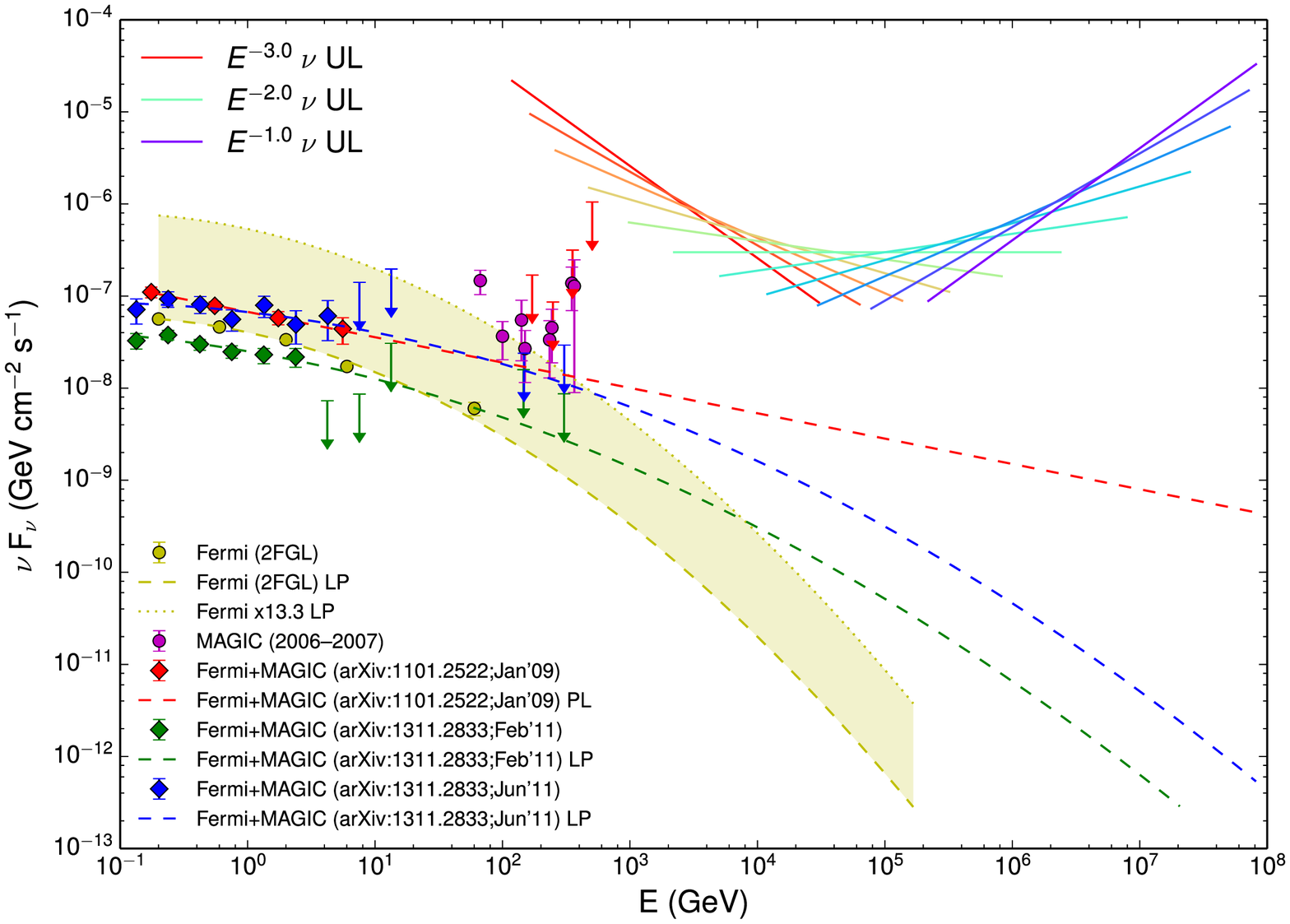}
\caption{Gamma-ray SED of 3C279 observed by Fermi/LAT at different epoches: red, green and blue dots for measurements in January 2009, January 2011 and June 2011 respectivelly~\cite{bib:3c279Magic1,bib:3c279Magic2}. The yellow dots are the average flux with 2008-2010 data (2FGL). The observed data points are corrected for absorption by the extragalactic background light assuming z = 0.536. The dashed lines represents the extrapolation fits of the Fermi data, using log-parabola (LP) and power-law (PL) functions. The shaded yellow area represents an extrapolation of the flux enhancing during the studied flares. Finally, the coloured solid lines indicates the neutrino upper-limits for different spectral indexes (from $E^{-3}$ in red to $E^{-1}$ in blue), with the ranges corresponding to the 5-95\% range of the energy sensitive to ANTARES.}
\label{fig:hybridSED_3C279}
\end{figure}

Hadronic interactions predict neutrino emission in the TeV-PeV range associated with a flux of gamma rays. The prediction that the total neutrino energy flux $F_{\nu}$ is approximately equal to the total high-energy photon flux $F_{\gamma}$ is relatively robust, at least when attributing this emission to a 100\% hadronic origin~\cite{bib:Kelner,bib:KA}. Using spectral energy distributions (SEDs) and VHE data from the literature is possible to see how gamma ray emissions compare with the neutrino upper limits for different spectral indexes, from -3 to -1. These limits are extrapolated from the $E^{-2}$ spectrum case by considering the proper change in the acceptance. In Figure~\ref{fig:hybridSED_3C279} is shown the hybrid SED for the blazar 3C279. With this simple criteria of the energy budget, the limit set by ANTARES for the blazar 3C279 is on the same order of magnitude as the gamma ray flux measured during the flares. This encourages the search for a neutrino signal during outburst periods. With more data, ANTARES will be able to significantly constrain a 100$\%$ hadronic origin of the high-energy gamma ray emission. Fermi has reported some very intense outbursts between mid 2013 and end of 2014 for 3C279~\cite{bib:3c279atel1,bib:3c279atel2}, periods not considered in this paper.

\section{Conclusions}

In this contribution are shown the results of the extended time-dependent search for cosmic neutrinos from blazars using the data taken with the full 12 line ANTARES detector between 2008 and 2012. This search is supported on the time constriction from the gamma ray variable emission of the sources as seen by Fermi satellite and IACTs. Multiple neutrino spectra has been tested and a lag between the neutrino and gamma ray emission has been considered in the analysis. The most significant correlation was found with a GeV flare of the blazar 3C279 for which one neutrino event was detected in time/spatial coincidence with the gamma ray emission. However, this event had a post-trial probability of 67$\%$ and is thus background compatible. Upper-limits were obtained on the neutrino fluence for the selected sources and compared with the gamma ray observed fluxes. These comparisons show that for the brighter blazars, the neutrino flux limits are in the same order of magnitude as the high-energy gamma ray fluxes. Given this consideration, these searches are quite promising with further years of ANTARES data and for the future KM3NeT detector. A paper with these results is in the final stages of preparation.

%



\setcounter{figure}{0}
\setcounter{table}{0}
\setcounter{footnote}{0}
\setcounter{section}{0}
\setcounter{equation}{0}
\newpage
\id{id_sanguineti1}
\addcontentsline{toc}{part}{\textcolor{blue}{\arabic{IdContrib} - {\sl M. Sanguineti} : Search for GRB neutrino emission according to the photospheric model with the ANTARES telescope
}%
}
\title{\arabic{IdContrib} - Search for GRB neutrino emission according to the photospheric model with the ANTARES telescope}

\shorttitle{\arabic{IdContrib} - ANTARES research of GRB neutrino according to the photospheric model }

\authors{\color{color01}Matteo Sanguineti}
  \afiliations{Universit\`a\ degli\ Studi\ di\ Genova,\ INFN\ Genova}
\email{matteo.sanguineti@ge.infn.it}


\abstract{The ANTARES detector is the largest neutrino telescope currently in operation in the North Hemisphere.
One of the main goals of the ANTARES detector is the search for cosmic neutrino sources including transient sources like GRBs.
In the so-called photospheric model for the emission from GRBs the interaction of the radiation field with the leptonic component of the outflow could lower the expected energy spectrum of the associated neutrino emission from GRBs.
In coincidence with a GRB alert from a satellite, ANTARES stores a window of few minutes of unfiltered data.
A dedicated directional filtering and reconstruction is applied offline to enhance the sensitivity in the lower energy range of the ANTARES detector (50 GeV - 10 TeV).
The expected improvement as derived
from Monte Carlo simulations will be presented.}

%
%
\maketitle
\section{Introduction}

Gamma ray bursts (GRB) belong to one of the most energetic phenomena of the Universe, but their origin remained a mystery for many years.
A milestone in the GRB detection was the launch of the Compton Gamma-Ray Observatory, in particular the Burst and Transient Experiment (BATSE) detected over 2700 bursts \cite{BATSE}.
BATSE showed that GRBs are distributed isotropically in the sky without any dipole or quadrupole moments, indicating an extragalactic origin later confirmed by redshift measurement \cite{Meegan}.

The fireball shock model is the best-known scenario that has been put forth to explain the gamma ray emission mechanism associated with GRBs. It predicts that different shock waves will be traveling at different relativistic speeds, and it is the interaction between these different shock fronts that cause the energetic gamma-ray emissions.

A new widely discussed scenario is the photospheric model \cite{GAO}\cite{MURASE}, which predicts a neutrino emission nearer to the central engine where the relativistic jet is still optically thick.
This model is interesting because it predicts some features of the gamma ray spectrum of GRBs that are not foreseen by the fireball scenario, like the Amati correlation \cite{AMATI}.
The photospheric model predicts a lower energetic neutrino spectrum respect to internal shock model and it has already been investigated by the IceCube collaboration \cite{ICECUBE2}.
The fireball model has already been tested by different ANTARES analyses \cite{ANT1} \cite{ANT2}. We will focus on photospheric model, so we will exploit a special data set that could offer a better sensitivity in the lower energy range. Also a low-energy optimized reconstruction algorithm and a directional filter have been used to additionally improve the sensitivity in the interesting energy range.

\section{The photospheric model}

Like in the fireball model the presence of a jet-like relativistic outflow is assumed.
The photospheric model predicts the conversion of a fraction of the bulk kinetic energy into radiation energy through a dissipation mechanism in the neighbourhood  of the photosphere.
The photosphere occurs in the acceleration phase $r<r_{sat}$ if the outflow is magnetically dominated, on the other hand in the barionic case the photosphere occurs in the coasting phase ($r>r_{sat}$), where $r_{sat}$ is the saturation radius \cite{GAO}.

In the barionic case, two different mechanisms can lead to dissipation: the dissipation of magnetohydrodynamic (MHD) turbulence or semi-relativistic shocks \cite{Thompson}  with Lorentz factor $\Gamma_r \sim 1$ as in internal shocks.
In the magnetically dominated case, before the dissipation occurs the total jet luminosity consists of a toroidal magnetic field component and a proton bulk kinetic energy component \cite{GAO}.
Calculations and simulations of such baryonic and magnetic dissipative photospheres predict a spectrum similar to the observed characteristic ``Band" spectrum \cite{Band}, parameterized  as

\begin{center}
$\frac{dN_{\gamma}}{dE}\propto(\frac{E}{E_{br}})^{x_{ph}}$
\end{center}
where a burst with z=2 redshift shows $E_{br}$ around 300 keV and 
\begin{itemize}
\item $ x_{ph}=-1$ for $E > E_{br}$
\item $ x_{ph}=-2$ for $E > E_{br}$
\end{itemize}

In the barionic photosphere scenario protons and electrons are assumed to be  accelerated through a Fermi-first order  acceleration mechanism in the surrounding  magnetic fields.
A similar process is also expected in the magnetic photosphere scenario.

Neutrinos are mainly produced though charged pion and kaon decays; these charged mesons come from p$\gamma$ and pp interactions.
For energies below 1 GeV the cross section is dominated by resonances while at higher energies multi-pion production prevails.

The pions decay is fully understood and neutrinos are produced mainly from this channel:
\begin{center}
$ \pi^{-}\rightarrow \mu^{-}+ \overline{\nu_{\mu}}\rightarrow \overline{\nu_{\mu}} + e^{-}+\overline{\nu_{e}}+\nu_{\mu} $
\end{center}
and the charge conjugate particles for the $\pi^+$.

High-energy pions lose most of their energy through synchrotron radiation. For muons with their   longer mean lifetime and smaller mass, synchrotron cooling is more severe than that of charged pions.

As previously mentioned the photospheric model can predict two different scenarios: the barionic dominated jets or magnetic fields dominated jets.
These two possibilities lead to different macroscopic acceleration rates, different proper densities in the jet rest-frame, and imply a different role for magnetic dissipation in the process of particle acceleration.

In Fig.~\ref{fig:neutrinoflux} two different estimates of the neutrino flux from a GRB according to the photospheric model are shown. 

\begin{figure}[htbp]
\begin{center}
\includegraphics[width =7.5cm, height=4.5cm]{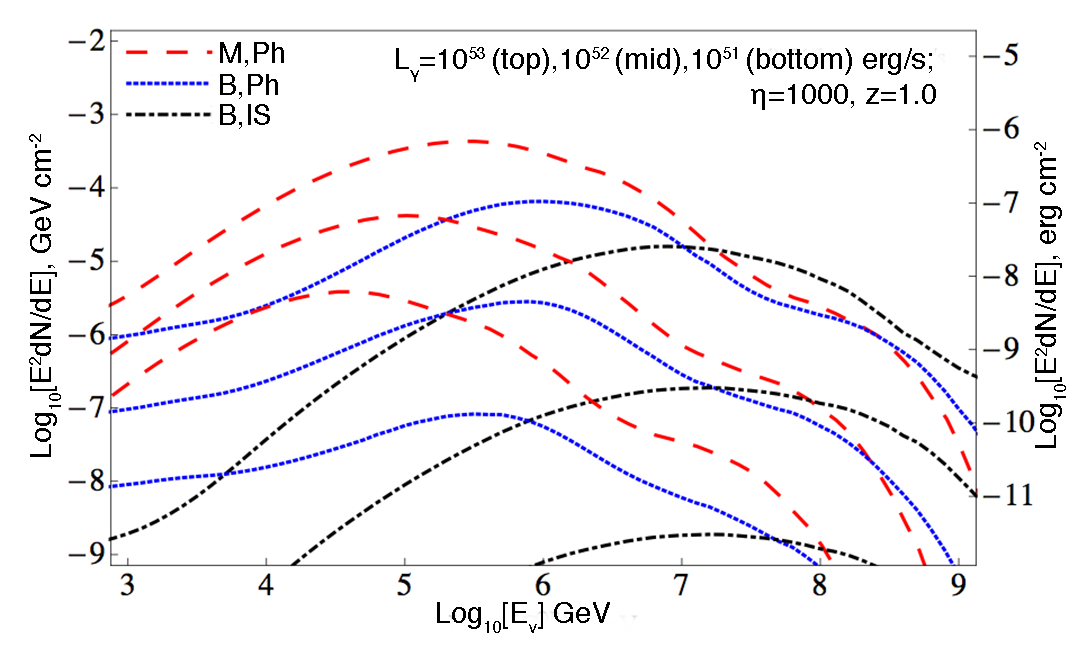}
\includegraphics[width =7.5cm, height=4.5cm]{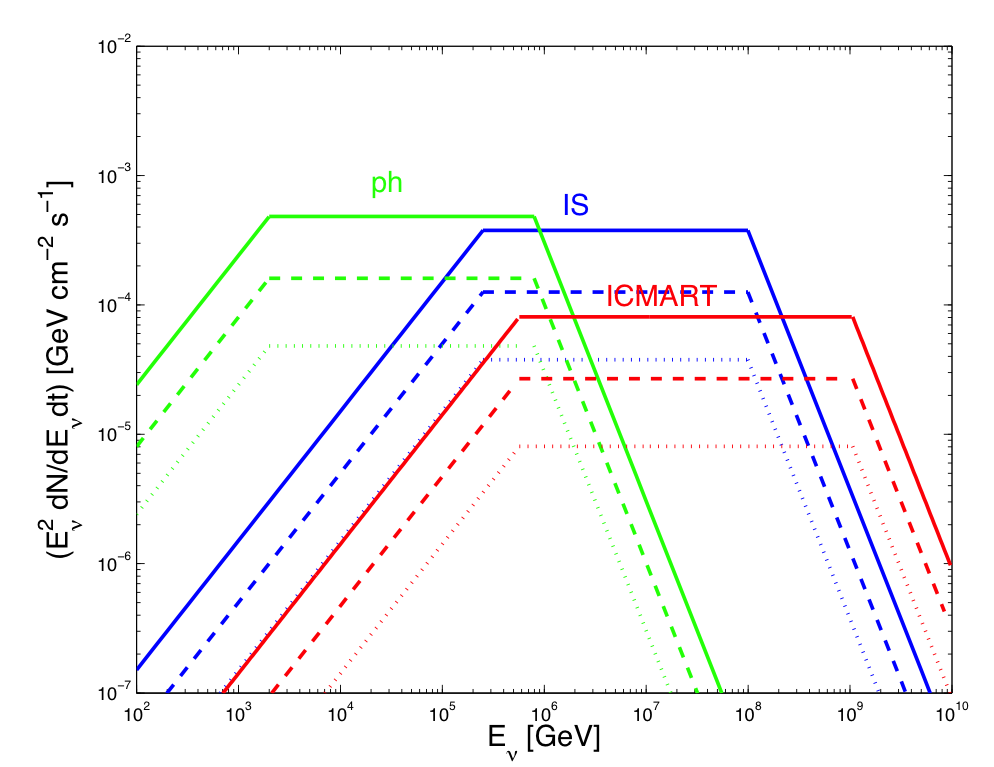}
\caption{Left: Neutrino fluence from a single GRB assuming different dissipation models. Red, dashed: magnetic photosphere; blue, dotted: baryonic photosphere; Dot-dash: baryonic internal shock \cite{GAO}. Model parameters: luminosity $L_{\gamma}$=$10^{53}$ (top curve), $10^{52}$ (mid curve), $10^{51}$ (bottom curve), redshift z=1. Right: Neutrino flux from a single GRB assuming different dissipation models: "ph" (green): dissipative photosphere model; "IS" (blue): internal shock model; "ICMART" (red): internal-collision-induced magnetic reconnection and turbulence model. Model parameters: normalized luminosity $L_{\gamma,52}$= 1, variability time scale observed in the GRB light curve $\delta$t=0.1s, redshift z=1, Lorentz factor $\gamma$ = 250 \cite{MURASE}. Three values of the ratio between photon luminosity and non-thermal proton luminosity are adopted: 0.1 (solid), 0.3 (dashed), and 1 (dotted). } 
\label{fig:neutrinoflux}
\end{center}
\end{figure}

The photospheric model predicts a neutrino flux at lower energies than expected from the internal shock model.

In the next section we will describe the tools which we will exploit to increase  the ANTARES sensitivity in the energy range between 50 GeV and 10 TeV to to address the possible neutrino flux at lower energies. 
Other GRB analyses have been performed previously by ANTARES \cite{ANT1}\cite{ANT2} and IceCube \cite{ICECUBE}\cite{ICECUBE2}. In particular the features regarding the optimization of the model discovery potential (MDP) used in this analysis have been developed and applied in \cite{ANT1} on the ANTARES data from end of 2007 to end of 2011.
We will focus first on a generic burst with a neutrino flux ($E_{\nu}^2 dN/dE_{\nu} dt$) of $5\cdot10^{-4} GeV cm^{-2} s^{-1}$ and cut-offs at $2\cdot10^{3}$ GeV and $8\cdot10^{5}$ GeV with the future goal to study two promising candidates (GRB110918A and GRB130427A).

\section{Data sample}
\label{sec:datasample}

For follow-up observations of bursts it is important to receive an alert in coincidence with a GRBs; for this purpose a global alert network has been created. The alert is distributed 
to many telescopes around the world when one of the satellites of the network detects 
a GRB. All satellites capable of GRB detection that were launched since BATSE (1991) are part of this network.
All interested telescopes can subscribe to the system to be updated promptly with the most recent GRB information.  ANTARES subscribed to the alert system, even if it is not a follow-up telescope, because we want to keep all raw data around
the alert.
In Fig.\ref{fig:mieke} the delay is shown between the detection of a GRB by the satellite and the time of the alert message distributed, in 90\% of the cases the delay is below 200 s. 
\begin{figure}[htbp]
\begin{center}
\includegraphics[width =7.5cm, height=4.5cm]{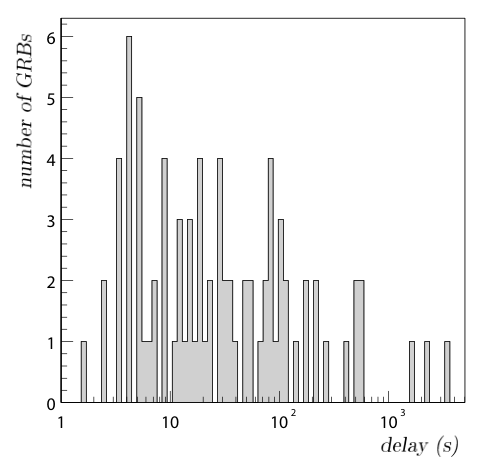}
\caption{Distribution of the delay between the detection of a GRB by the satellite and the time of the alert message is distributed.} 
\label{fig:mieke}
\end{center}
\end{figure}

The ANTARES Data Acquisition (DAQ) system is designed around the all-data-to-shore concept, which entails the transport of all photon signals recorded by the optical modules to the shore station where filtering is performed.
The filtering algorithms are operating in coincidence with a GRB alert, but in this case also raw data are saved on disks. A couple of minutes of unfiltered data (L0 data) buffered before the alert message are also available to be stored.
This configuration is maintained for a couple of minutes, in fact even long GRBs have a typical duration typical duration less than two minutes.
The filtered data are known as L1 data. A L1 hit is defined either as a local coincidence on the same storey (ANTARES consists of 295 storey with 3 PMTs each) within 20 ns, or as a single L0 hit with a large amplitude, typically 3 photoelectrons. When a muon event is triggered, all PMT pulses are recorded over 4 $\mu$s in a snapshot.
On the other hand the L0 data sample contains every signal detected above the 0.3 photoelectrons threshold for the whole alert duration (i.e. couple of minutes).

\section{Analysis principle}
\label{sec:antec}

All the filtering on the L0 data is performed offline when the data sample is analysed 
and the position of the GRB is known with the best possible accuracy.
A dedicated filtering algorithm has been developed for this data sample.
In the case of the GRB, the direction of the potential neutrino events is known, so the algorithm can look for space-time correlations with a less strict filter condition because only one direction is considered.
The direction of the muons that originate from GRB neutrinos is on average comparable 
with the direction of the burst, the angular spread depending on the neutrino 
energy.
Using this algorithm more events are expected to be detected in the interesting direction, 
which would be lost using the standard ANTARES filtering system, which is more stringent and looks for correlations in all the directions of the sky.

The hits that satisfy the filtering condition are used for the reconstruction of the track direction.
In this analysis  we will use a special reconstruction algorithm (known as GridFit \cite{GRID}) that is optimized for low energy (below  $10^3$ GeV).
As input all L0-hits (raw data) are taken and three hit selections (with different criteria) are performed. After the hits collection three reconstruction steps are implemented, the final one is based on a likelihood maximisation method. The reconstruction algorithm provides the direction of the neutrino and other useful parameters like the reconstruction quality X, which will be used later in the optimization of track selection criteria.
The parameter X is defined as $X=nhits-1.1 \cdot rLogL$ where nhits is the number of hits used in the reconstruction and rLogL is the reduced log-likelihood of the track hypothesis.

Using raw data, the special filtering algorithm, the optimized reconstruction algorithm and applying the search method developed in \cite{ANT1} we obtained a larger sensitivity for this analysis at lower energy with respect to the standard analysis.
In this analysis the number of triggered events (before quality cuts) has been doubled for energies
above $10^5$ GeV with respect to the standard ANTARES analysis. At lower energy the increase is more significant up to a factor 5 at energies around 50 GeV.

A simulation of a neutrino flux has been performed with Genhen \cite{BRUNNER} with an energy spectrum rescaled according to photospheric model predictions.
We assumed a simplified neutrino spectrum with a flux ($E_{\nu}^2 dN/dE_{\nu} dt$) of $5\cdot10^{-4} GeV cm^{-2} s^{-1}$ with cut-offs at $2\cdot10^{3}$ GeV and $8\cdot10^{5}$ GeV.

Using this simulation we derived the point spread function of the reconstructed neutrino according to the GRB photospheric model.
This function is used for building the signal probability density function (PDF), this function is called $S(\alpha)$, where $\alpha$ is the angle between the reconstructed track direction and the true MC neutrino direction in degree. The background PDF $B(\alpha)$ is assumed uniform in the search window (10 degrees).
Background events come from atmospheric neutrinos and from atmospheric muons which were misreconstructed as upwards going. In \cite{ANT1} , the sum of these two is estimated from data. The small duration of runs in our case (2mins instead of a few hours) and the higher dependence on biolumination conditions makes the direct application of this strategy impossible, in fact statistics prevents a solid direct estimate for a single 2-minute data sample.
To overcome this we have simulated a small sub-sample of raw data (a tenth of runs distributed in ANTARES life, run duration is around 2 hours) and looked for a relationship between them and the corresponding sample of official ANTARES Run by Run (RbR) Monte Carlo simulation which takes in account the experimental conditions of each data run, such as the status of each PMT,  the detector configuration, the actual environmental conditions and optical background \cite{MC}.
The rate of upgoing muons can be evaluated introducing a ratio C as follows

\begin{center}
$ C=\frac{\frac{\mu_{\uparrow}(data)}{\mu_{\downarrow}(data)}}{\frac{\mu_{\uparrow}(MC)}{\mu_{\downarrow}(MC)}}  $
\end{center}
The ratio C is constant in time and weakly depends on the detector condition or bioluminescence background, so it can be used to evaluate the number of expected upgoing muons as 
\begin{center}
$ \mu_{\uparrow}(data)=C\cdot\mu_{\downarrow}(data)\frac{\mu_{\uparrow}(MC)}{\mu_{\downarrow}(MC)} $,
\end{center}
where $\mu_{\uparrow}/_{\downarrow}(data)$ is the number of upgoing/downgoing muons in a raw data files and $\mu_{\uparrow}/_{\downarrow}(MC)$ is the number of upgoing/downgoing muons in the corresponding Run by Run Monte Carlo simulation.
The random background due to random coincidences has been also simulated, but it is irrelevant compared to the muon background. It will be neglected in our background estimation.

The muon background estimation has been verified using a reduced number of raw data files that are associated to false GRB alarm.
The estimation is compatible with the data especially for tracks of good quality. 
 We also checked the dependence of this background estimation on the zenith of the event. The ratio $raw data/RbR MC$ does not change dramatically  considering different zenith angle of our search window.
In order to be more conservative an additional safety factor 2 is added to our muon background estimation to take in account the zenith dependence of the muon background.

\section{Sensitivity study}
\label{sec:sensstud}

The sensitivity study will be performed on a generic GRB with neutrino flux ($E_{\nu}^2 dN/dE_{\nu} dt$) of $5\cdot10^{-4} GeV cm^{-2} s^{-1}$ and cut-offs at $2\cdot10^{3}$ GeV and $8\cdot10^{5}$ GeV. The GRB is assumed to be located in the part of sky where the ANTARES visibility is maximal.

According to the signal and background PDFs, as previously defined, pseudo-experiments are produced to derive the distribution of the log-likelihood  ratio Q, obtained by maximizing the log-likelihood for the signal and comparing with the background only value.
The extended maximum likelihood Q is defined as
\begin{center}
$$ Q=max_{\mu_{sig}}\sum^{n_{tot}}_{event\color{bianco}{o}\color{color01} {i}=1}log\frac{\mu_{sig}\cdot S(\alpha_i)+\mu_{bg}\cdot B(\alpha_i)}{\mu_{bg}\cdot B(\alpha_i)}-(\mu_{sig}+\mu_{bg}),$$
\end{center}

where S and B represent the signal and background PDF as previously defined, i is the index of the event with space angle $\alpha_i$ with respect to the GRBs direction, $\mu_{bg}$ is the expected number of background events and $\mu_{sig}$ is the signal contribution.
In Fig.~\ref{fig:Q_distr} the distribution of the log-likelihood ratio Q is shown.

\begin{figure}[htbp]
\begin{center}
\includegraphics[width =7.5cm]{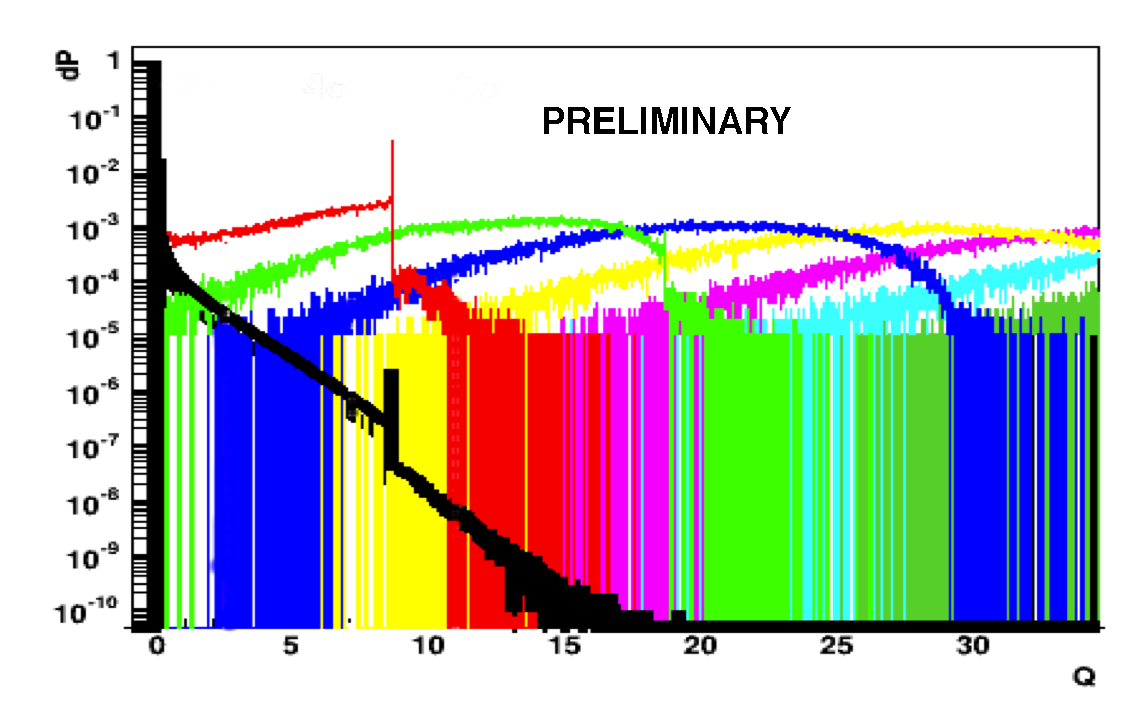}
\caption{Log likelihood ratio (Q) distributions. Background rate: 0.01. Yellow: background only. Red (green, blue, ...): background+1,2,3,... signals. } 
\label{fig:Q_distr}
\end{center}
\end{figure}

Evaluating the different curves of Fig.~\ref{fig:Q_distr}, we can compute the model discovery potential (MDP) and the expected sensitivity as a function of the expected number of events.
This strategy is repeated for several values for the cut on the quality parameter  X in order to find the selection that maximizes the model discovery potential. 

Using this optimal parameter we can derive the sensitivity as the 90\% Confidence Limit that can be put on the flux considering the median background Q value.
In the case of the spectrum considered in this analysis the sensitivity is ($E_{\nu}^2 dN/dE_{\nu} dt$)  $3\cdot10^{-1} GeV cm^{-2} s^{-1}$.

We present in Fig.\ref{fig:comp} the expected efficiency improvement factor with respect to the same analysis applied on classical filtered data.

\begin{figure}[htbp]
\begin{center}
\includegraphics[width =7.5cm]{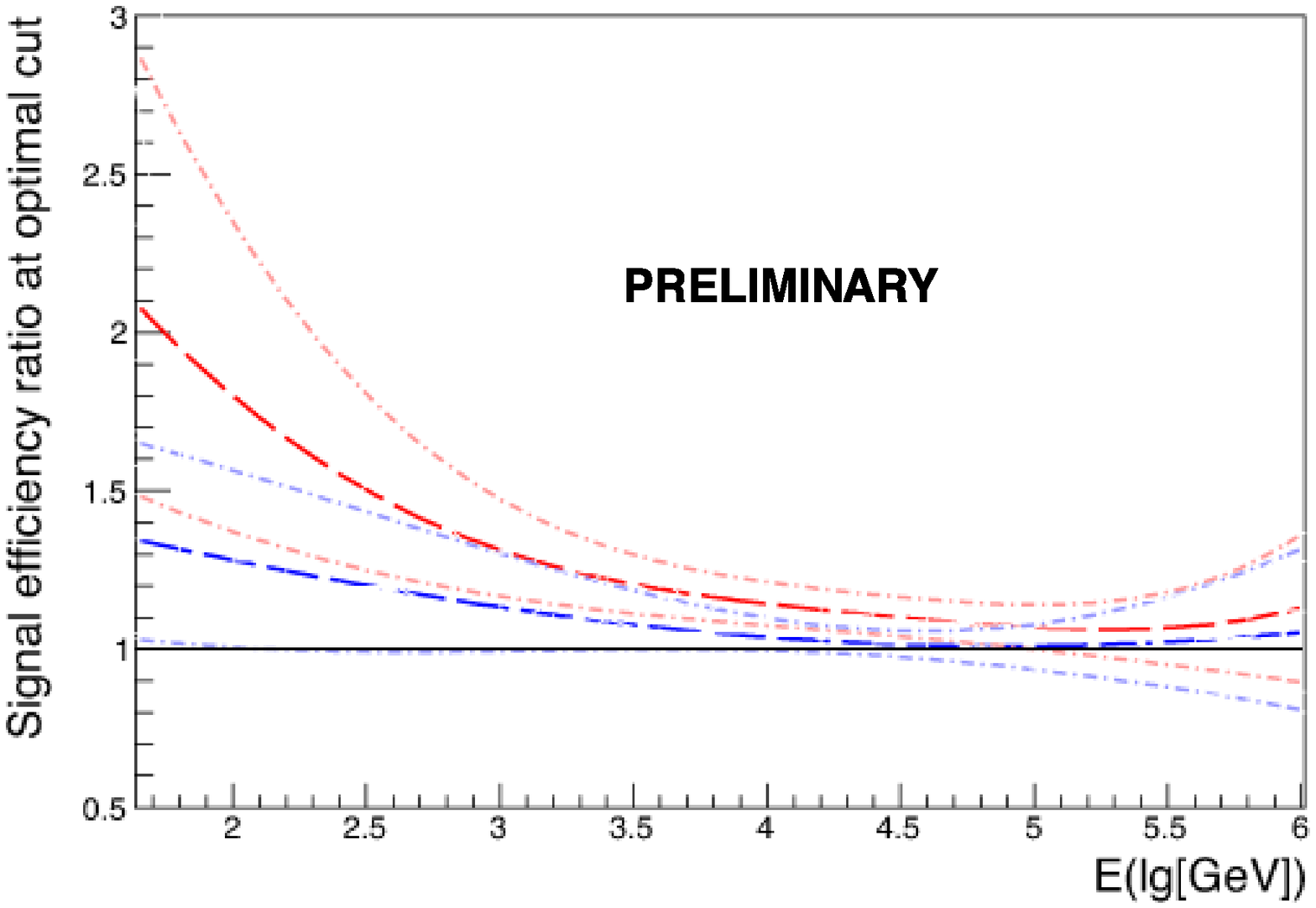}
\caption{Signal efficiency ratio at optimal cut between the proposed analysis and the standard ANTARES analysis. Red: raw data filtered with the directional trigger and reconstructed with GridFit. Blu: standard data reconstructed with GridFit. Pale red and blu: corresponding 90\% confidence interval. All results are normalized to the standard ANTARES analysis.} 
\label{fig:comp}
\end{center}
\end{figure}

As expected the proposed analysis has a better efficiency respect to the standard analysis at low energies, in particular the effectiveness is almost doubled at energies of a few hundred of GeV.
Finally the sensitivity on the expected flux from GRB 130427A according to photospheric model \cite{MURASE} has been derived applying the same quality cut on the quality parameter as in the previous case (Fig.\ref{fig:Limit}). 

\begin{figure}[htbp]
\begin{center}
\includegraphics[width =7.5cm]{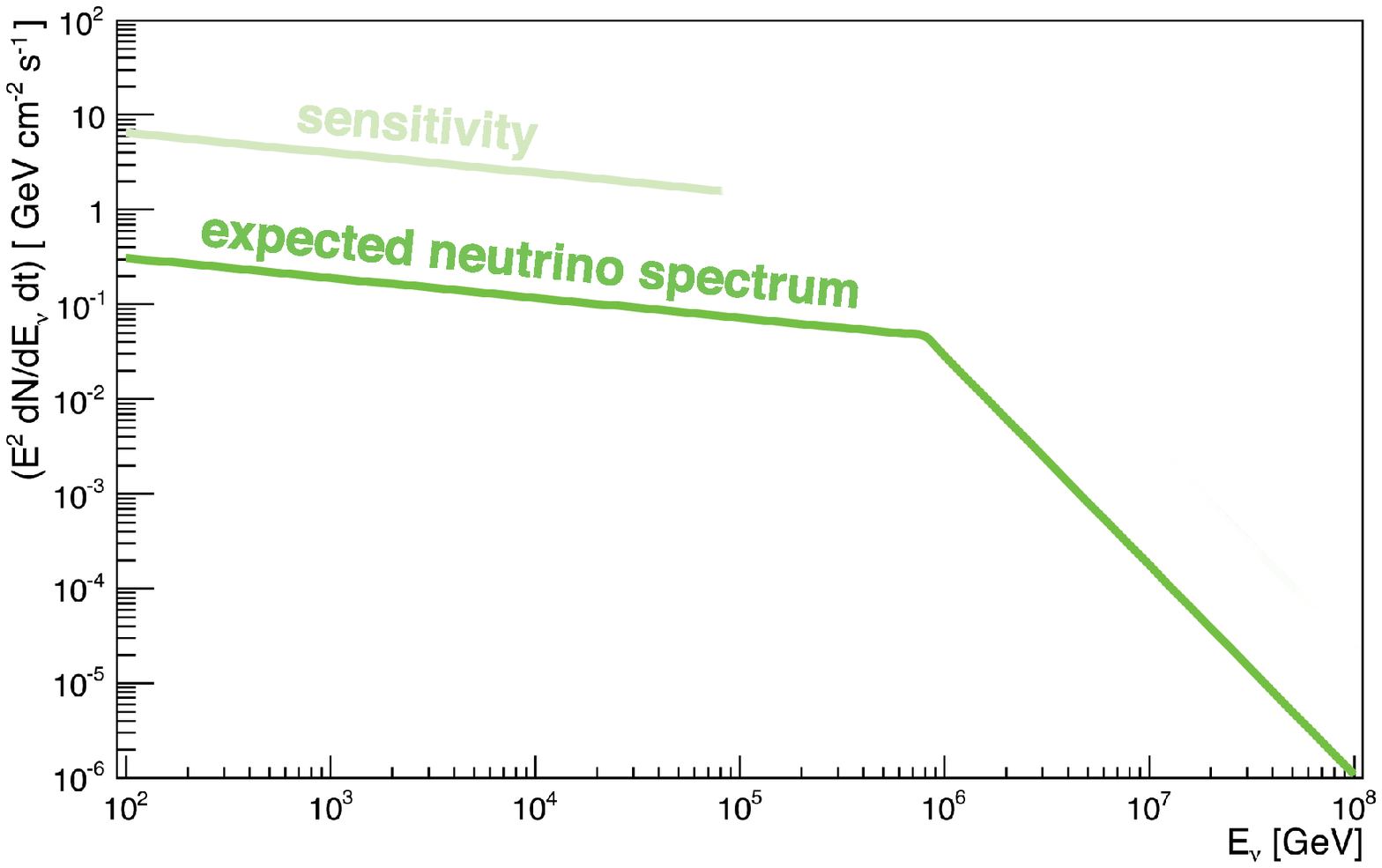}
\caption{Expected neutrino spectrum (dark green) and sensitivity  (pale green) according to photospheric model for GRB 130427A.} 
\label{fig:Limit}
\end{center}
\end{figure}

\section{Conclusions}

Adapting the strategy used to search for neutrinos from GRBs based on the widely used  internal shock model, we have studied the sensitivity of ANTARES to the concurrent GRB photospheric model. This has been done using special ANTARES data and tools enhancing the sensitivity  between 50GeV and 10TeV.
In order to enhance the sensitivity of the ANTARES in the range between 50 GeV and 10 TeV some dedicated tools have been used for this analysis: a special data sample of raw data, a directional trigger and a reconstruction algorithm optimized for this energy range.
A solid way to estimate the upgoing muon has been found and the optimization of the selection criteria has been performed and a generic sensitivity obtained.
This analysis  will be applied on the data collected in correspondence  of the two ANTARES best candidate for GRB detection of the last years (GRB110918A and GRB130427A).


\setcounter{figure}{0}
\setcounter{table}{0}
\setcounter{footnote}{0}
\setcounter{section}{0}
\setcounter{equation}{0}

\newpage
\id{id_turpin}
\addcontentsline{toc}{part}{\textcolor{blue}{\arabic{IdContrib} - {\sl D. Turpin} : Searches for neutrinos from Gamma-ray bursts with ANTARES}%
}



\title{\arabic{IdContrib} - Searches for neutrinos from Gamma-ray bursts with ANTARES}

\shorttitle{\arabic{IdContrib} - Searches for GRB-neutrinos with ANTARES}

\authors{Julia Schmid$^a$,Damien Turpin$^b$}
\afiliations{$^a$ Friedrich-Alexander-Universit\"at Erlangen-N\"urnberg, Erlangen Centre for Astroparticle Physics, Erwin-Rommel-Str. 1, 91058 Erlangen, Germany 
	\footnote{now at Laboratoire AIM, CEA-IRFU/CNRS/Universit\'e Paris Diderot, Service d'Astrophysique, CEA Saclay,
91191 Gif sur Yvette, France}\\
$^b$ Aix Marseille Universit\'e, CNRS/IN2P3, CPPM UMR 7346, 13288, Marseille, France
}
\email{$^a$julia.schmid@fau.de,$^b$damien.turpin@irap.omp.eu}

\abstract{ANTARES is the largest high-energy neutrino telescope in the Northern Hemisphere. Its main scientific purpose is the search for astrophysical muon neutrinos that are detected via their charged-current interaction in Earth and the subsequent Cherenkov emission of the secondary muon in the water of the Mediterranean Sea. Gamma-ray bursts are among the most promising candidates for the experiment as they are thought to accelerate not only electrons - leading to the observed gamma rays - but also protons, which would yield the emission of EeV neutrinos.
Compelling evidence of a high-energy cosmic neutrino signal correlated with any astrophysical source would, for the first time, prove the acceleration of hadrons beyond any doubt, a hypothesis that cannot unambiguously be put to the test by pure electromagnetic observation. However, to explain the origin of cosmic rays at ultra-high energies, it is absolutely crucial to identify those processes in the universe that are capable of accelerating baryons to such energies.
The recent searches for muon neutrinos from gamma-ray bursts using data of the ANTARES telescope will be presented, including constrains that can be put on individual model parameters and a scan for possibly time-shifted neutrino signals. 
}

%
%
\maketitle
\section{Introduction}
The detection of a high-energy neutrino signal from Gamma-ray bursts (GRB) would unambiguously probe them as powerful accelerators of hadrons. 
In the prevailing fireball model as proposed for example by M{\'e}sz{\'a}ros and Rees \cite{Meszaros93a}, the observed electromagnetic radiation is explained by synchrotron radiation and subsequent inverse Compton scattering of relativistic shock-accelerated electrons. 
Waxmann $\&$ Bahcall  \cite{Waxmann/Bahcall1997} first suggested that there could be a significant baryonic loading in GRB jets (mainly protons). If these protons are sufficiently accelerated, they interact with the ambient photon field and produce neutral and charged pions. 
Subsequent decay of the latter would yield a high-energy neutrino signal associated with the electromagnetic GRB signal. 
Neutrino astronomy can therefore provide an unique tool to probe the nature and dynamics of GRB's jets and could also serve to explain the origin of the  cosmic-ray flux at ultra-high energies. 

The underwater neutrino telescope ANTARES \cite{Ageron2011} is primarily designed to detect cosmic muon-neutrinos in the TeV-PeV range below the local horizon. 
In these proceedings, we present recent searches for muon-neutrino emission from GRBs using the ANTARES data. 





\section{The NeuCosmA model}
\label{section2}
Neutrino predictions are based on the photohadronic interactions between the accelerated protons and the ambient photon field. 
The first commonly used models of Waxmann $\&$ Bahcall \cite{Waxmann/Bahcall1997} and Guetta \cite{Guetta2001} have already been ruled out by the IceCube collaboration \cite{Abbasi2012}. 
The NeuCosmA model \cite{Hummer2010},\cite{Hummer2012} is one of the up-to-date models that takes into account the full proton-photon cross section, including $\Delta^+$ resonances, multiple pion and Kaon production which contributes to the highest energy part of the neutrino spectrum. 
The predicted neutrino spectrum depends on a set of 10 parameters describing the $\gamma$-ray prompt spectrum and the dynamics of the GRB jet: 
 $F_\nu = f(z,\alpha_\gamma,\beta_\gamma,E_p,F_\gamma,\frac{\epsilon_e}{\epsilon_B},\Gamma,f_p,t_{var},T_{90})$\footnote{The commonly-used default values for these parameters are : $z^{\rm def}=2.15$, $\alpha^{\rm def}=-1$, $\beta^{\rm def}=-2$, $E_p^{\rm def}=200\: {\rm keV}$, $F_\gamma^{\rm def}=10^{-5}\: {\rm erg.cm}^{-2}$, $\epsilon_e^{\rm def}=0.1$, $\epsilon_B^{\rm def}=0.1$, $\Gamma^{\rm def}=316$, $f_p^{\rm def}=10$, $t_{var}^{\rm def}=0.01\: {\rm s}$, $T_{90}^{\rm def}=30\:  {\rm s}$ (our choice here)} where $z$ is the cosmological redshift of the burst , $\alpha_\gamma$ and $\beta_\gamma$ are the low and high energy spectral indexes of the $\gamma$-ray spectrum, $E_p$ is the peak energy of the observed $\nu F_\nu$ $\gamma$-ray spectrum, $F_\gamma$ is the $\gamma$-ray fluence, $\epsilon_e$ and $\epsilon_B$ are the fraction of the internal jet's energy in electrons and in the magnetic field, $\Gamma$ is the bulk Lorentz factor, $f_p$ is the baryonic loading, $t_{var}$ is the minimum variability timescale of the $\gamma$-ray prompt emission and $T_{90}$ gives the burst duration.

\section{Searches for neutrinos from GRB110918A and GRB130427A}
\label{section3} 
The search methodology follows the one developed and applied in \cite{antares2013} on the ANTARES data from end of 2007 to end of 2011. It relies on the optimisation of the model discovery potential (MDP) applying per-GRB selection cuts on the track reconstruction quality parameter. In doing so, the likelihood ratio of signal (derived from Monte Carlo simulations) to background (based on data) is maximised. 
In a sample of 296 long GRBs from 2007 to 2011, no neutrino events were detected within the accumulated coincident search duration of 6.6 hours, where 0.06 neutrino events where predicted from the NeuCosmA model on a background of 0.05. 
An upper limit at 90$\%$ confidence has hence been derived \cite{antares2013}. 
The total predicted neutrino flux was mainly dominated by the very energetic and relatively close burst GRB110918A. This means that the detection of a neutrino signal from an individual GRB is very unlikely except for particular energetic bursts as GRB110918A. The very nearby burst GRB130427A was also in the ANTARES field of view and was considered as a promising candidate for a neutrino detection. Thus a specific search for a neutrino signal was performed, where the spectral and temporal properties were collected from \cite{gcnFermi}.
From the NeuCosmA model, $6.2\times 10^{-3}$ events were expected, yielding a $3\sigma$ MDP of 0.86$\%$. Equivalently for GRB110918A,  a $3\sigma$ MDP of 3.25$\%$ has been derived using NeuCosmA. No coincident neutrino signal has been observed, consequently derived upper limits after non-detection are shown in Fig. \ref{figure2}.

\section{ANTARES constraints on the physics of GRB110918A and GRB130427A}
\label{section4}
One should note that the NeuCosmA predictions were determined with standard values for the non-measured physical parameters, i.e. $\Gamma=316$, $f_p=10$ and $\frac{\epsilon_e}{\epsilon_B}=1$ and could strongly bias the final result. In order to evaluate how the unknown parameters impact the NeuCosmA expectations, the entire parameter space was scanned within expectations for long GRBs. For each parameter set, standard values were assumed for the fixed parameters and the expected numbers of neutrinos $\mu_s^i$ were derived accounting for the time-average ANTARES effective area from 2007 to 2011. The influence of each parameter on the neutrino expectations is given by the ratio between the maximum and the minimum number of predicted neutrinos: $\delta \mu_s = {\rm max}(\mu_s^i)/ {\rm min}(\mu_s^i)$. As shown in Table \ref{table1}, the bulk Lorentz factor $\Gamma$ and, to a smaller extent,  the baryonic loading $f_p$, crucially influence the neutrino predictions. For instance, a GRB with a high Lorentz factor ($\Gamma\sim900$) would exhibit $\sim 10^6$ less neutrinos than the same GRB with a low $\Gamma\sim60$ according to NeuCosmA. On the other hand, the ratio $\frac{\epsilon_e}{\epsilon_B}$ has only minor influence in the neutrino expectations. Hence it is possible to constrain regions in the $\Gamma$ and $f_p$ parameter space by excluding models that would predict a detectable signal at the 90$\%$ confidence level ($\mu_s\ge2.3$).
\begin{figure}
\centering
\includegraphics[width=0.7\textwidth]{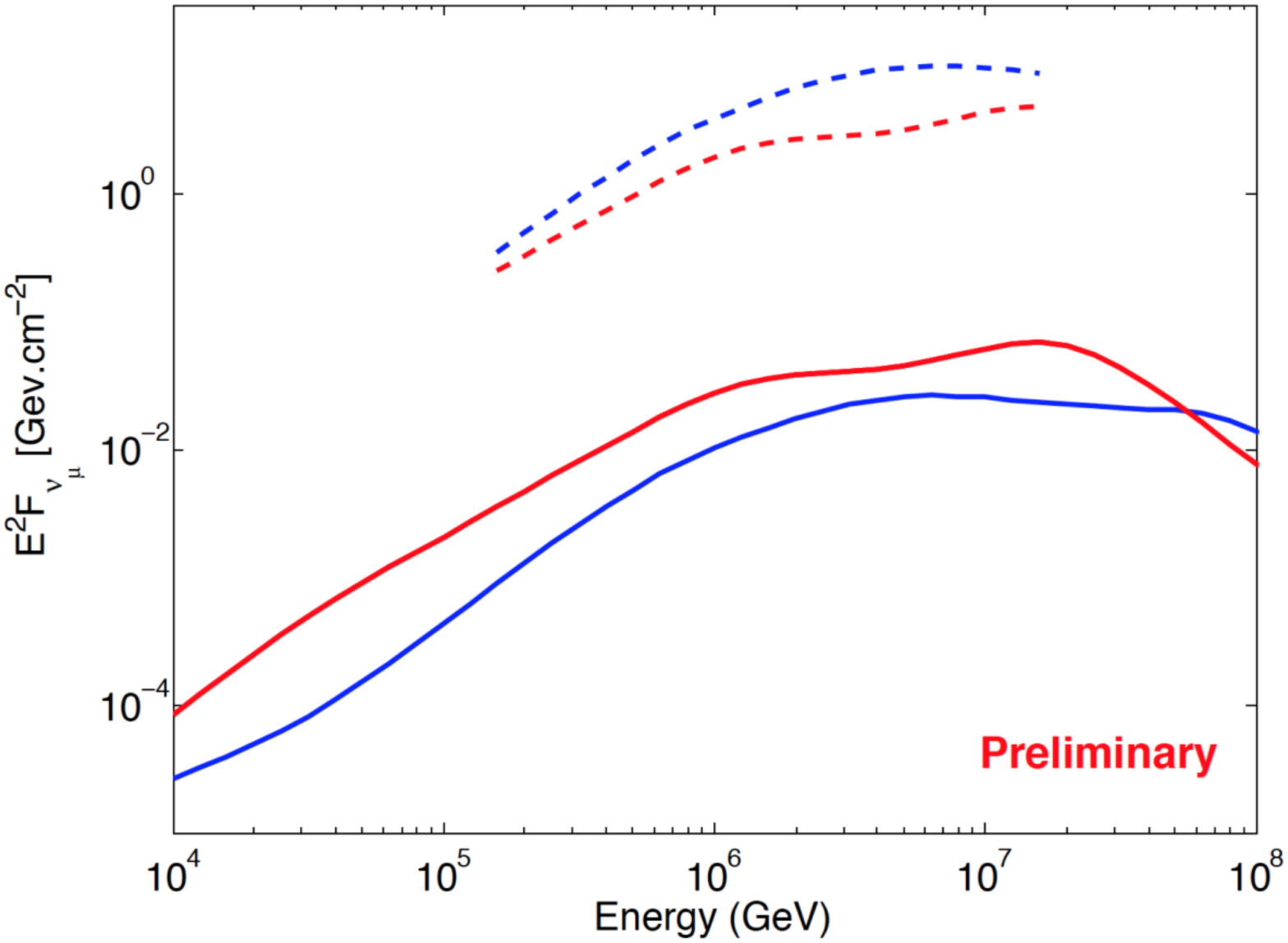}
\caption{\label{figure2}NeuCosmA spectra ($\nu_\mu+\overline{\nu}_\mu$) for GRB110918A (red) and GRB130427A (blue). The dashed lines indicate the derived limits on the coincident neutrino emission with GRB110918A (red) and GRB130427A (blue) in the energy range where 90$\%$ of the signal is expected to be detected.}
\end{figure}

\begin{table}
\begin{center}
\begin{tabular}{c|c|c|c}
\hline
\hline
{Scanned parameter}&{$\mu_s^{\rm min}$}& {$\mu_{s}^{\rm max}$}&{$\delta \mu_s$}\\
\hline
{$\Gamma$} & {$5.6\times 10^{-8}$ ($\Gamma=900$)} &{0.12 ($\Gamma=60$)} &{$2.1\times 10^6$}\\
{$f_p$} & {$4.9\times 10^{-6}$ ($f_p=0.5$)} &{$9.7\times 10^{-4}$ ($f_p=200$)} &{198.0}\\
{$\frac{\epsilon_e}{\epsilon_B}$} & {$3.3\times 10^{-5}$ ($\frac{\epsilon_e}{\epsilon_B}=0.01$)} &{$5.5\times 10^{-5}$ ($\frac{\epsilon_e}{\epsilon_B}=100$)} &{1.7}\\
\hline
\end{tabular}
\end{center}
\caption{Results of the parameter scans. GRB standard values were used for the fixed parameters, see section \ref{section2}. The minimum $\mu_s^{\rm min}$ and maximum $\mu_s^{\rm max}$ numbers of neutrinos obtained during the different scans are indicated with the associated parameter value. $\delta \mu_s$ measures the absolute variation of the expected number of neutrinos inside the parameter space of $\Gamma$, $f_p$ and $\frac{\epsilon_e}{\epsilon_B}$ according to the NeuCosmA predictions for long GRBs.}
\label{table1}
\end{table}
10000 NeuCosmA spectra were generated for each burst in order to cover the whole range of $\Gamma \in [10;900]$ and $f_p \in [0.5;200]$ ($\frac{\epsilon_e}{\epsilon_B}$ was fixed at its standard value). For each simulated spectrum, the expected number of neutrinos $\mu_s$ was calculated by taking into account the ANTARES effective area for GRB110918A and GRB130427A. The ANTARES constraints on $\Gamma$ and $f_p$ for these two burst do not strongly challenge the standard predictions of the internal shocks model as shown in Fig.\ref{figure4}.

\begin{figure}
\includegraphics[height=0.26\textheight]{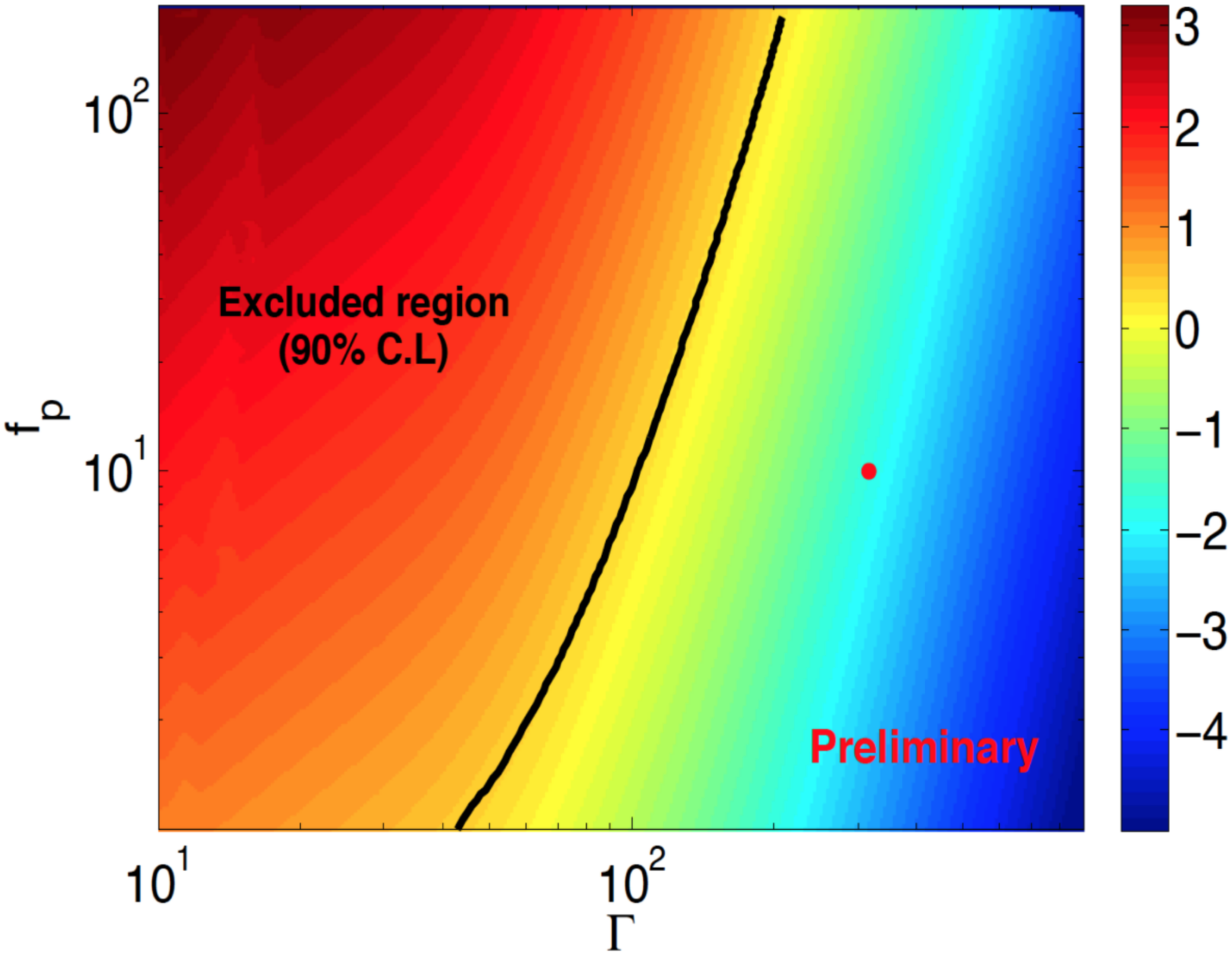}
\includegraphics[height=0.26\textheight]{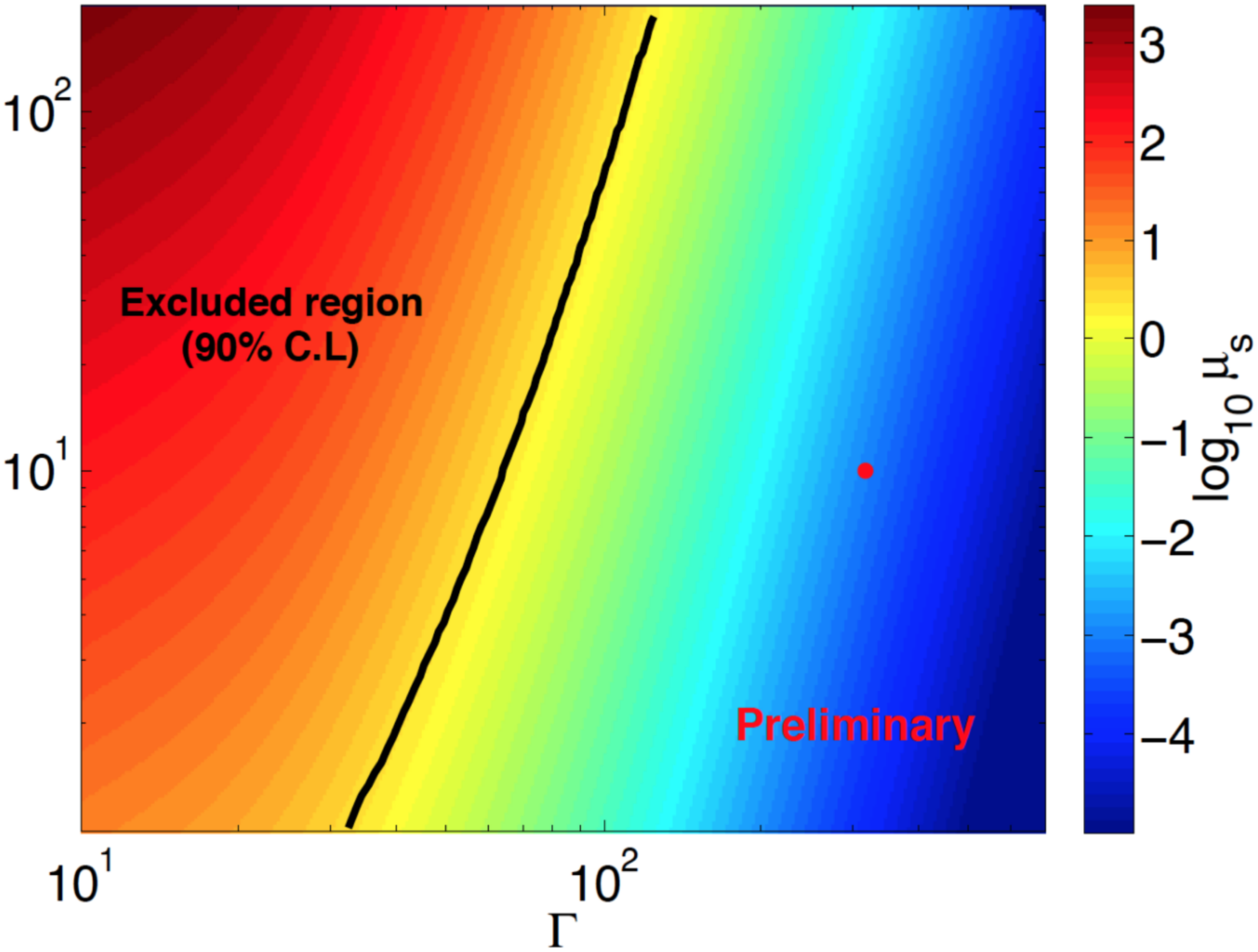}
\caption{Expected number of neutrinos (color coded) as function of $\Gamma$ and $f_p$ for GRB110918A (left-hand panel) and GRB130427A (right). The black line indicates the region excluded by ANTARES at 90$\%$ confidence level ($\mu_s\ge2.3$), the red dots shows the standard values of $\Gamma$ and $f_p$ for long GRBs. 
}
\label{figure4}
\end{figure}


\section{Search for time-shifted neutrino emission from gamma-ray bursts}
\label{section5}
Up to now no neutrino signal could be identified above the background in the data from any neutrino detector during the prompt emission phases, and the first optimistic analytical models have already been challenged by IceCube \cite{Abbasi2012}. 
Even though the search for a signal of neutrinos coincident with the emission of high-energy photons is the most generic ansatz, there are many models that predict time-shifted neutrino signals, 
such as neutrino precursors \cite{razzaque03} or afterglows \cite{waxman00}, or different Lorentz Invariance Violation (LIV) effects for photons and neutrinos on their way to Earth \cite{amelinocamelia}. Thanks to their cosmic distances and transient nature, gamma-ray bursts provide unique test environments to study and verify such effects. 
%
A novel model-independent technique was developed to distinguish a time-shifted neutrino signal from the expected background, which allows even faint signals to be detected using a large sample of GRBs. 
The search relies on stacked time profiles of neutrinos spatially coincident with GRBs in a wide time window which would enable to detect a systematic shift of neutrinos (from emission or propagation) with respect to the electromagnetic emission. 
Any neutrino emission associated with the GRBs, even if faint, would give rise to a cumulative effect in these stacked profiles, which can then be identified by its discrepancy from randomised data. 

The neutrino candidate sample for the search for neutrino point-sources \cite{ANTARES_PS} provides naturally suited data for this approach. The stringent quality cuts guarantee low muon background contamination and excellent angular resolution. This data sample consists of 5516 neutrino candidate events from March 2007 to the end of 2012. A suitable gamma-ray-burst sample was consolidated similarly to the one used in \cite{antares2013}.

Several observables sensitive to different potential origins of such a shift were considered. The simplest one is a delayed detection time $\tau = t_{\nu} - t_{\rm GRB}$ of neutrinos with respect to the detected gamma-rays. Another one, sensitive to potential shifted emission times, is corrected for the redshift $\tau_z=\frac{\tau}{1+z}$. 
Potential LIV effects are, on first order, supposed to be linearly dependent on energy\cite{amelinocamelia}, so we define the measure $\tau_{\rm LIV}=\frac{\tau}{E_{est.}\cdot D(z)}$,  with the estimated neutrino energy $E_{\rm est}$ and the luminosity distance of the GRB, $D(z)$. 
From the stacked histograms of these observables, we construct a test statistic \cite{eindhoven}:
\begin{equation}
\psi 	=  -10 \left[   \log_{10} n! + \sum_{k=1}^{m} n_k \log_{10} p_k  - \log_{10} n_k!  \right] \; ,
\end{equation}
with the total number of events $n$ being distributed in the $k \in [1 \dots m]$ bins with probability $p_k$. 

An optimal choice of the search cone size $\delta_{\rm max}$ naturally depends on the gamma-ray burst's position accuracy and the neutrino pointing uncertainty of the detector. 
We chose a per-GRB coincidence cone size by optimising the ratio of signal to square root of noise of a two-dimensional gaussian signal on flat background\cite{Alexandreas93a}:
\begin{equation}
\delta_{\rm cut} = 1.58 \cdot  \max (\sigma_{\nu}, \Delta_{\rm err, \, GRB}, \theta_{\rm lim}) \label{eq:deltacut} . 
\end{equation}
where $\sigma_{\nu}$ is the neutrino sample median resolution and $\Delta_{\rm err, \, GRB}$ the size of the GRB error box. The cone size was limited by $\theta_{\rm lim}$ such that no single GRB contributes more than an order of magnitude more background than another. 


The size of the probed time window $\tau_{\rm max}$ should be defined as the largest shift predicted by any of the models.  The largest arrival time delays between neutrinos and gamma-rays could be introduced by LIV effects. Considering  the most recent limit on the potential LIV energy scale  \cite{Vasileiou13} and the highest measured GRBs' redshift so far (z$\sim$9), we limited the maximum considered time shift to 40 days.
These choices reduced the initial sample to 563 GRBs occurring below the local ANTARES horizon (and 150 with measured redshift). 

\subsection{Sensitivity and results}
To investigate the performance of the proposed technique to identify hypothetical neutrinos from GRBs, a test signal was mimicked by associating neutrino candidates artificially with part of the GRBs at an (hypothetical) intrinsic time shift of five days.
That is, taking into account the cosmological redshift $z$, a simulated signal delayed by $t_\nu = t_{\rm GRB} + 5 {\rm d} \cdot (1+z)$. 
The sensitivity, defined as the $90\%$ confidence-level upper limit that can be placed on the number of GRBs that produced an associated neutrino signal in the ANTARES data when observing the median background, is $m(f_{{\rm all}}^{90\% {\rm CL}})=0.6\%$, as shown in Fig. \ref{figure5}. 
Considering only the sub-sample of bursts with determined redshift, the method is even sensitive to a signal in only $1.1\%$ of the bursts, which corresponds to $0.3\%$ of the entire sample.

When applying the search to actual ANTARES data, no events were found in coincidence with the GRBs where 4.4 were expected from purely randomised data (0.7 for GRBs with measured redshift) which is an under-fluctuation of 1.2\% probability (51.4\%). 
This low probability prevents us from putting a limit at  the standard $90\%$ confidence level. However, we can exclude any signal in more than $0.06\%$ of the bursts with a $99\%$ confidence.\\ 

We have performed the same search with the IceCube public IC40 point source search sample \cite{IC40PS}. It consists of 12877 neutrino collected between April 2008 and May 2009 that have been searched for associations with 40 GRBs (12 with redshift measurement). In total, 42 neutrinos (8) are found in coincidence where 35 (4) were expected from randomised data. These slight excesses with $p$-value of $13.5\%$ ($5.1\%$) are still compatible with the background expectations. 
This result derived on a complementary GRB sample as well as numerous cross-checks testing different coincidence selections confirms the fact that the observed underfluctation in the ANTARES data sample is not introduced by systematic effects of the method or the software, but is indeed inherent in the considered data sample. 
%
\begin{figure}[h!]
\centering
\includegraphics[height=0.3\textheight]{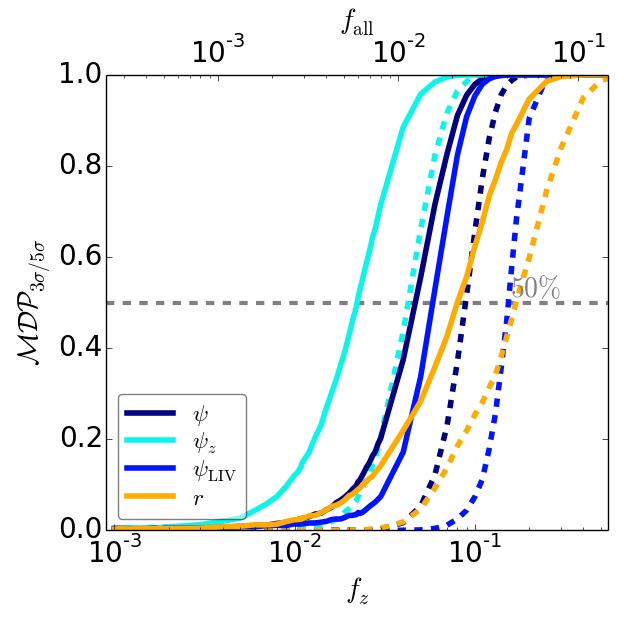}
\includegraphics[height=0.3\textheight]{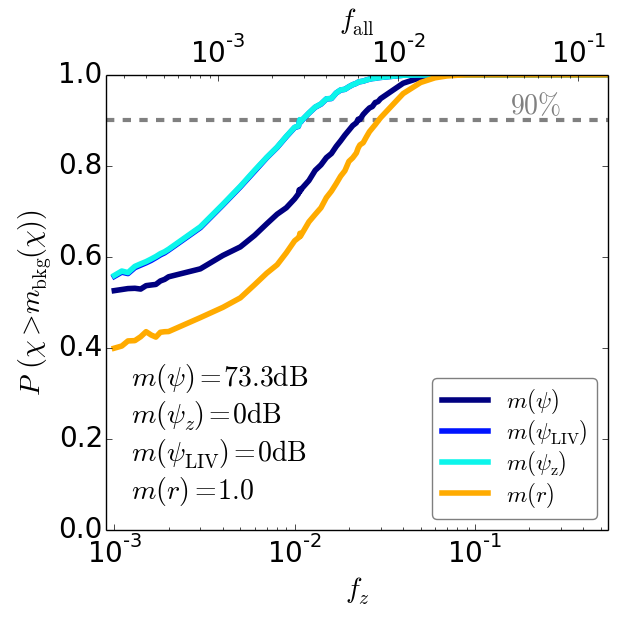}
\caption{
Efficiencies or detection probability $P$ at $3\sigma$ (solid) and $5\sigma$ (dashed lines) for the test statistics and the ratio $r$ of events before and after the GRB alert as a function of the mean fraction $f$ of GRBs with one associated signal neutrino at $t_\nu = t_{GRB} + 5 {\rm d} \cdot (1 + z)$ (left). The fraction $f_z$ denotes the fraction of GRBs with one associated signal neutrino in the ANTARES data with determined redshift $z$, whereas $f_{all}$ gives the fraction of the whole GRB sample. On the right: Probabilities $P$ to measure values of the test statistics above the median value from the background-only realizations. The sensitivity is given by the signal fraction $f$ where the curves reach 90$\%$ probability (gray dashed line). Probabilities were derived using the ANTARES data from 2007-2012. 
}
\label{figure5}
\end{figure} 
\section {Discussion and conclusion} 
Following the stacked search for neutrinos during the prompt emission of GRBs in the data of ANTARES between late 2007 and 2011, the individual analysis of the exceptionally nearby GRB130427A has also not revealed any significant signal excess. Consequently, upper limits on the neutrino fluxes have been derived. Given the ANTARES sensitivity, only the most extreme physical parameters can be excluded.
Nevertheless, the non-observation of a neutrino signal from these bursts by the ANTARES and IceCube collaborations is a strong indication that the bulk Lorentz factor of the most energetic GRBs could be very high. It has been shown that the expected NeuCosmA neutrino signal for a GRB with a high $\Gamma$ value can be up to six orders of magnitude lower than for a GRB with low $\Gamma$. Thus, the ideal GRB to produce high-energy neutrino emission would be a very luminous burst with a moderate Lorentz factor. Note, however, that these results are interpreted in the framework of internal shock models of GRBs. Other models predict high-energy neutrino emission with more or less efficiency at different places in the jet. For instance, the photospheric model \cite{Zhang/Kumar2013}, \cite{Gao2012} predicts higher neutrino flux at lower energy and will be included in future investigations. 

We also performed a model-independent search for time-shifted neutrinos with respect to the prompt emission of GRBs and could not identify any excess. 
This enables us to put a limit on the average fraction of GRBs that might produce a detectable neutrino signal in the ANTARES data, even if shifted in time, to about $1\%$. The future larger neutrino telescope KM3NeT with significantly increased sensitivity of up to a factor of $\sim50$ will be able to challenge the neutrino predictions in the framework of the GRB fireball model. In the meanwhile, the collection of more ANTARES data is still ongoing and will help to improve the neutrino flux limit, providing at the same time a continuous monitoring of GRB neutrino emission in the Southern Hemisphere. 

\section*{Acknowledgements}
We would like to thank Mauricio Bustamante for helpful discussions and making it possible to use the NeuCosmA model. J. Schmid would like to thank the Studienstiftung des Deutschen Volkes for their financial support. D. Turpin would also gratefully acknowledge financial support from the OCEVU LabEx, France.


\setcounter{figure}{0}
\setcounter{table}{0}
\setcounter{footnote}{0}
\setcounter{section}{0}
\setcounter{equation}{0}

\newpage
\id{id_vvanelewyck}
\addcontentsline{toc}{part}{\textcolor{blue}{\arabic{IdContrib} - {\sl V. Van Elewyck} : Joint search for gravitational waves and high-energy neutrinos with the ANTARES, LIGO and Virgo detectors}%
}

\def\Journal#1#2#3#4{{#1} {\bf #2}, #3 (#4)}

\def\NCA{\em Nuovo Cimento}
\def\NIM{\em Nucl. Instrum. Methods}
\def\NIMA{{\em Nucl. Instrum. Methods} A} 
\def\NPB{{\em Nucl. Phys.} B}
\def\PLB{{\em Phys. Lett.}  B}
\def\PRL{\em Phys. Rev. Lett.}
\def\PRD{{\em Phys. Rev.} D}  
\def\ZPC{{\em Z. Phys.} C}
\def\Science{\em Science}  
\def\RPP{\em Rep. Prog. Phys.}  
\def\RMP{\em Rev. Mod. Phys.}  
\def\CQG{\em Class. Quant. Grav.}
\def\JCAP{\em J. of Cosm. and Astrop. Phys. }

\title{\arabic{IdContrib} - Joint search for gravitational waves and high-energy neutrinos with the ANTARES, LIGO and Virgo detectors}

\shorttitle{\arabic{IdContrib} - ANTARES. LIGO and Virgo joint searches for GW and HEN}

\authors{V. Van Elewyck for the ANTARES Collaboration, the LIGO Scientific Collaboration and the Virgo Collaboration}
\afiliations{APC, Universit\'e Paris Diderot, CNRS/IN2P3, CEA/Irfu, Obs. de Paris, Sorbonne Paris Cit\'e, France}
\email{elewyck@apc.univ-paris7.fr}

%

\abstract{Cataclysmic cosmic events can be plausible sources of both gravitational waves (GW) and high-energy neutrinos (HEN), alternative cosmic messengers carrying information from the innermost regions of the astrophysical engines. Possible sources include long and short gamma-ray bursts (GRBs) but also low-luminosity or choked GRBs, with no or low gamma-ray emissions. 
Combining directional and timing informations on HEN events and GW bursts through GW+HEN coincidences provides a novel way of constraining the processes at play in the sources. It also enables to improve the sensitivity of both channels relying on the independence of backgrounds in each experiment. A first search was performed with concomitant data from 2007, when ANTARES was half its final size. This contribution focuses on the second, optimised search performed with data taken in 2009-2010, during the Virgo VSR2-3 and LIGO L6 science runs (with improved sensitivity) and with ANTARES in its final configuration. While the 2007 search has allowed to place the first upper limits on the density of joint GW+HEN emitters, the 2009-2010 analysis will provide a significant improvement in sensitivity.
}
%
%
%
\maketitle

\section{Introduction}
Multimessenger astronomy is at a turning point with the first cosmic High-Energy Neutrinos (HEN) detection by the IceCube experiment~\cite{IceCubeHESE} and the very probable detection of Gravitational Waves (GW)  with the advanced generation of the LIGO~\cite{Ligo} and Virgo~\cite{Virgo} detectors. In this context, a new window is about to open for the observation of the Universe with cosmic messengers conserving timing and directionality, complementary to electromagnetic observations. Both HEN and GW are expected to provide important information about the processes taking place in the core of astrophysical production sites. They could even reveal the existence of electromagnetically dark sources, that would have remained undetected so far, such as the putative "choked GRBs" which could constitute the missing link between core-collapse Supernovae and GRBs. A detailed discussion of potential GW+HEN emitters can be found in~\cite{colloquium}. 

The first concomitant data-taking phase with the whole Virgo/LIGO network, VSR1/S5, was carried out in 2007, while ANTARES~\cite{Antares} was operating in a five-line configuration. The strategy chosen for the 2007 GW+HEN joint search consisted in an event-by-event search for a GW signal correlating in space and time with a given HEN event considered as an external trigger~\cite{gwhen2007}. This approach allowed to make use of existing GW analysis pipelines developed e.g. for GRB searches. The list of 2007 HEN triggers was obtained by applying on ANTARES data a standard reconstruction algorithm (BBFit~\cite{bbfit}) and quality requirements similar to those selecting the well-reconstructed events that are used for the standalone searches for HEN point sources. The list of HEN triggers included their arrival time, direction on the sky, and an event-by-event estimation of the angular accuracy, which was used to define the angular search window for the GW search. 

This list was then processed by the X-pipeline~\cite{xpipeline}, an algorithm which performs coherent searches for unmodelled bursts of GWs on the combined data stream coming from all interferometers. The background estimation and  the optimization of the selection strategy were performed using time-shifted data from the off-source region in order to avoid contamination by a potential GW signal. Once the search parameters were tuned, the analysis was applied to the on-source dataset, consisting of data recorded within a time window of $[-500s, +500s]$ around the time of each HEN trigger. This time interval was chosen on basis of conservative estimations of the time delay between the HEN and GW signals expected for long GRBs, based on BATSE, {\it Swift} and {\it Fermi} observations~\cite{timedelay}.  No GW candidate was observed in coincidence with the selected HEN events from the 2007 data sample. A binomial test was also performed to look for an accumulation of weak GW signals, with negative results. This allowed to extract GW exclusion distances for typical source scenarios. Converting this null observation into a density of GW+HEN emitters yielded a limit ranging from $10^{-2}$ Mpc$^{-3}$\,yr$^{-1}$ for short GRB-like signals down to $10^{-3}$ Mpc$^{-3}$\,yr$^{-1}$ for long GRB-like emissions~\cite{gwhen2007}. 
\begin{figure}[ht!]
\begin{center}
\includegraphics[width=11.4cm]{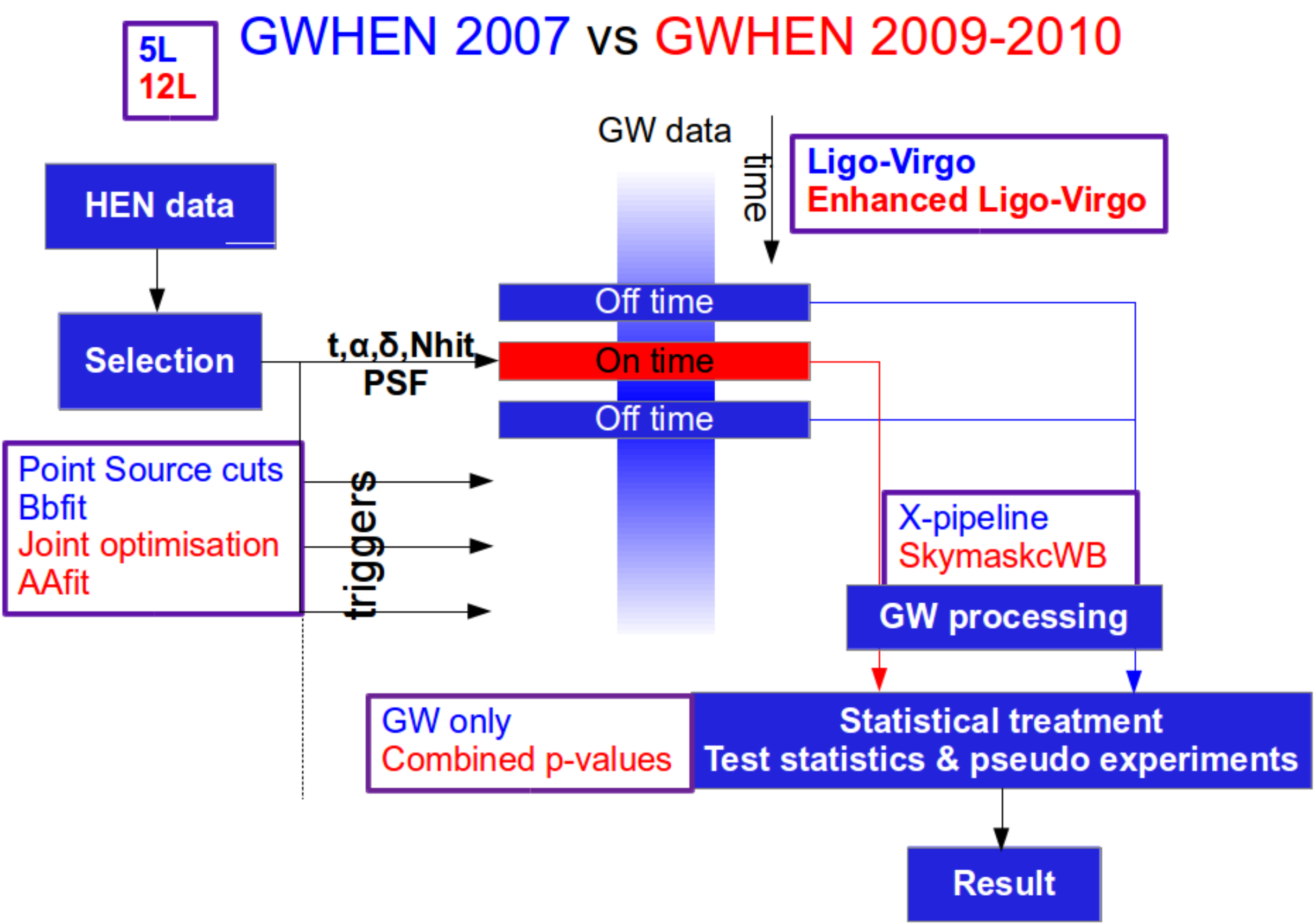}
\caption{\footnotesize Schematic flow diagram of the ANTARES/Virgo/LIGO GW+HEN analysis strategies used for the 2007 and 2009-2010 joint searches. In both cases, the neutrino candidates (with their time and directional information) act as external triggers for a given GW analysis pipeline, which searches the combined GW data flow from all active interferometers (ITFs) for a possible concomitant signal. The background
estimation and the optimization of the selection strategy are
performed using time-shifted data from the off-source region in
order to avoid contamination by a potential GW signal. Once the
search parameters are tuned, the box is opened and the analysis is
applied to the on-source dataset.}\label{fig:flow}
\end{center}
\end{figure}

This contribution focuses on the second search that is being finalized with data taken with the full ANTARES detector in 2009-2010, concomitant with the Virgo VSR2/VSR3 and LIGO S6 joint science runs, with upgraded GW detectors.  Building on the experience of the pioneering 2007 search, and following the joint 2009 IceCube-LIGO-Virgo analysis~\cite{IceCubegwhen} which introduced a more complete and symmetrical characterisation of the GW and HEN events \cite{method_Imre}, a new strategy has been adopted for the optimisation of the HEN trigger list in order to maximise the number of sources detectable by the search. A new HEN reconstruction algorithm (AAfit) has been used in order to reduce the angular error~\cite{antares_ps}. A different GW pipeline, the skymask coherent WaveBurst (s-cWB), has also been developed to allow the analysis with only 2 interferometers taking data, and the realisation of joint simulations - a necessary step to optimise the joint analysis~\cite{tez_boutayeb}. Figure~\ref{fig:flow} presents a flowchart of the analysis highlighting the main differences and  improvements between the 2007 and the 2009-2010 joint searches. 

Section~\ref{sec:det} describes the detector configuration and datasets used for the 2009-2010 joint search,  and Section~\ref{sec:opt} presents the strategy and statistical tools used for the joint optimisation procedure. Perspectives on the expected sensitivity of the search are discussed in Section~\ref{sec:exp}.

\section{Detectors and associated datasets}
\label{sec:det}
\subsection{The ANTARES neutrino telescope and associated dataset}

The ANTARES telescope~\cite{Antares} is located at a depth of 2475m in the Mediterranean Sea off the coast of Toulon, at $42^{\circ}48'~N, 6^{\circ}10'~E$. It comprises 885 optical modules consisting in 17'' glass spheres, each of them housing one 10'' photomultiplier, and installed on 12 vertical strings.

\begin{figure}[ht!]
\begin{center}
\includegraphics[width=7.5cm]{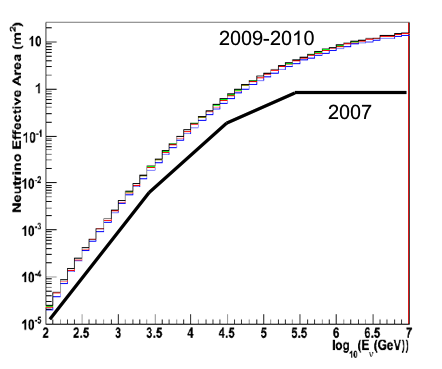}\hfill
\includegraphics[width=7.5cm]{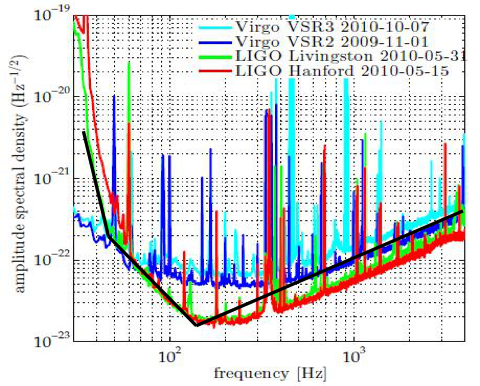}
\caption{\footnotesize \textbf{Left:} ANTARES effective area $A_{eff}$ for the two detector configurations corresponding to the datasets used in GW+HEN searches: 2007 and 2009-2010 (where the colors  correspond to different sets of quality cuts on the event reconstruction). \textbf{Right:} Detector noise spectra for LIGO and Virgo, showing typical sensitivities for the S6-VSR2/3 datasets (2009-2010). Also shown in black, to guide the eye, is the sensitivity representative of the first LIGO science run, S5 (2007).}\label{fig:aeff}
\end{center}
\end{figure}

The dataset used in this analysis covers the period from July 7th 2009 to October 20th 2010 for a total observation time of 266 days. The sample consists in events originating from muon neutrino charged-current interactions, which produce a muon that leaves a track-like signal in the detector. It is the most suited to this kind of directional searches, as the current reconstruction algorithms for this class of events achieve a sub-degree angular resolution (defined as the median angle between the neutrino and the reconstructed muon). The effective area $A_{eff}$ of the detector is plotted against energy in  Figure~\ref{fig:aeff} (left). It represents the detector response function as a function of the neutrino energy, and yields the detection rate for a given neutrino flux. The figure shows a clear increase between the 2007 datasample (5-line detector) used for the first GW+HEN search, and the 2009-2010 datasample (full, 12-line detector) used for this analysis. 

\subsection{The LIGO and Virgo gravitational wave interferometers and associated data set}
LIGO~\cite{Ligo}, with two sites in the United States, and Virgo~\cite{Virgo}, with one site in Italy, consist of perpendicular km-size Fabry-Perot cavities forming a Michelson interferometer tuned to the dark fringe. Any gravitational wave passing through the detector would induce a difference of path length in the two arms, thus changing the interference pattern.  The direction of an event is reconstructed by time-of-flight techniques which imply the use of at least two detectors.  Figure~\ref{fig:aeff} (right) shows the typical sensitivity for the LIGO and Virgo science runs taken in 2009 and 2010, compared to the one achieved in 2007. 

The GW data used in this search are the S6-VSR2/3 LIGO-Virgo data, which were collected between July 07, 2009 and October 21, 2010 by three detectors: LIGO-Livingston, LIGO-Hanford and Virgo. The concomitant data taking period between S6-VSR2/3 and ANTARES comprises all periods during which at least two out of the three interferometers were in science mode; the total duration of the joint dataset used for this analysis is $\tau\equiv$128.7 days.

\section{Joint optimisation of the common dataset}
\label{sec:opt}

\subsection{Definition of the joint figure of merit}\label{definition_of_FOM}
The approach adopted here is to optimise the HEN and GW selection cuts in order to maximize the number ${\cal N}_{\mathrm{GWHEN}}$ of detectable sources emitting both GW and HEN. A trade-off should therefore be found between two competing trends. Relaxing the cuts on the HEN sample will enhance efficiency to HEN signal, thereby increasing the number of suitable  candidates; but this will require harder cuts on the GW candidate sample in order to maintain the False Alarm Rate (FAR) below a fixed value.

Let us assume here that the sources are all identical and radiate an energy $E_{GW}$ in GW and emit a fluence $\varphi_\nu$ in HEN, and that their population is isotropic, i.e. characterised by a constant density per unit time and volume, $R$. The number of detectable sources is then given by 
\begin{equation}
{\cal N}_{\mathrm{GWHEN}}(\mathrm{cuts})= \int{dt d^3\Omega \: {\cal R}(r,t) \epsilon_\nu (\mathrm{cuts}) \epsilon_{GW} (\mathrm{cuts};E_{GW},r)}
\label{eq:N_opt1}
\end{equation}
where ${\cal R}(r,t)=R\: \mathcal{P}(N_{\nu}>0 | \frac{\varphi_\nu}{4 \pi r^2})$ is the density of detectable sources.
From Poisson statistics, we get $\mathcal{P}(N_{\nu}>0 | \frac{\varphi_\nu}{4 \pi r^2}) \propto \frac{1}{r^2}$ in the limit of small fluxes.
The optimisation is performed by varying the cut thresholds applied to the two following parameters: the quality of the muon track reconstruction $\Lambda$ for the HEN event sample,  and a proxy to the signal-to-noise ratio $\rho$ for the GW event sample, respectively. We obtain
\begin{equation}
{\cal N}_{\mathrm{GWHEN}} (\Lambda,\rho_{threshold}) \propto \int_0^\infty{4\pi r^2 dr \: \frac{1}{r^2} \epsilon_\nu (\Lambda) \epsilon_{GW} (\rho_{threshold};E_{GW},r)}
\label{eq:N_opt2}
\end{equation}
where $\epsilon_{GW}$ and $\epsilon_{HEN}$ are the respective detector efficiencies to signal. $\epsilon_{GW}$ can be reasonably well approximated by a step-like function with the edge placed at the maximum distance $D(\rho_{threshold})$  at which a GW source is detectable, defined as the GW \textit{horizon}; therefore,
\begin{equation}
{\cal N}_{\mathrm{GWHEN}} (\Lambda,\rho_{threshold}) \propto \epsilon_\nu (\Lambda) \int_0^{D(\rho_{threshold})} dr
\label{eq:N_opt3}
\end{equation}
For a GW ``standard candle'', $\rho_{threshold}$ is inversely proportional to $D(\rho_{threshold})$, leading to
\begin{equation}
{\cal N}_{\mathrm{GWHEN}} (\Lambda,\rho_{threshold}) \propto  \epsilon_\nu (\Lambda)/\rho_{threshold}
\label{FOM}
\end{equation}
The procedure then consists in tuning the HEN selection cuts in order to maximise the GWHEN figure of merit given by the ratio $\epsilon_\nu(\Lambda)/\rho_{threshold}$. 
As can be seen from Figure~\ref{fig:sky} (left), the optimal cut 
leads to 1986 neutrino candidates, each of them characterized by its arrival time, sky direction, energy and  associated error box. 
The energy estimator is the number of hits (or $n^{hit}$) used in the track fit. The error box, which depends on the track energy, is defined as the 90\% percentile of the distribution of space angles $\psi$ between the reconstructed muon and the incident neutrino direction, as estimated from Monte Carlo simulations. To each neutrino \mbox{candidate $i$} is associated a p-value $p_i^{\rm HEN}$ representing the probability that the atmospheric neutrino background would produce an event with at least the same number of hits as the considered event. Figure~\ref{fig:sky} (right) displays the skymap of the selected events, together with their error box, or angular search window (ASW90\%) used for the subsequent GW search.

\begin{figure}[h!]
\begin{center}
\includegraphics[width=6cm]{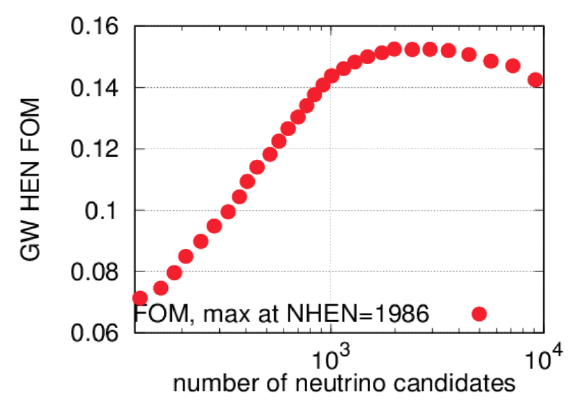}\hfill
\includegraphics[width=9cm]{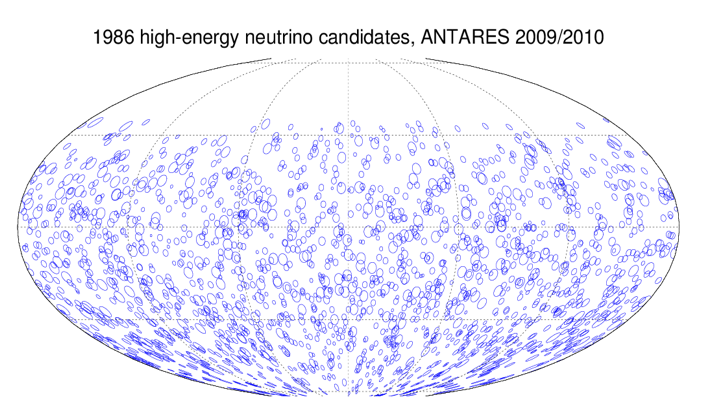}
\caption{\footnotesize \textbf{Left:} Joint GW-HEN figure of merit as a function of the number of selected HEN candidates (as determined by the value of the threshold cut on $\Lambda$). \textbf{Right:} Skymap of the 1986 selected HEN events with their associated ASW90\% angular error box.}\label{fig:sky}
\end{center}
\end{figure}

For each of the selected neutrino events, the adapted pipeline skymask coherent WaveBurst \mbox{(s-cWB)}~\cite{tez_boutayeb} performs a search for GW around the neutrino time. Among the 1986 candidate HEN, 773 are associated with 2 or more GW interferometers taking data, and are therefore usable for the purposes of the joint search. 
The whole sky is not scanned but only the region corresponding to ASW90\% centered on the reconstructed arrival direction of the neutrino $\overrightarrow{\textbf d_0 }$. For each candidate, s-cWB provides the GW skymap labeled hereafter $\mathcal{F}_i^{\rm GW}(\overrightarrow{\textbf d })$ within ASW90\%. These "sky-maps" are made of pixels of $0.4^\circ\times 0.4^\circ$, each associated with the probability that a GW is coming from it. The reconstruction pipeline also provides the value of $\rho$ for each GW candidates. This latter will correspond to a false alarm rate ${\rm  FAR_i}(\rho_{i})$ which in turn can be associated to a GW p-value indicating the probability that coherently combined background from different GW interferometers produces an event with at least this value of $\rho_i$, defined as: 
\begin{equation}
p^{\rm GW}_{i}=1-P(0| \tau_{i}\times {\rm FAR_i}(\rho_{i}))
\label{eq:pvalGW}
\end{equation}
where $\tau_i$ is the duration of the GW interferometers run in a certain configuration ({\it i.e.} combination of active detectors) during which event $i$ was recorded. The distributions are computed using $O(10^3)$ background realisations obtained with time shifts of the data stream. 
\subsection{Statistical characterisation of the joint candidates}\label{joint_ts}
The direction of the joint candidate event can be defined as the one maximizing the convolution of the GW skymaps and HEN point-spread functions (PSFs) $\mathcal{F}_i^{\rm GW}$ and $\mathcal{F}_i^{\rm HEN}$. 

The joint directional test statistic relies on the marginalized likelihood of the joint event, defined as:
\begin{equation}
\label{eq:marginalized}
\ln(\mathcal{L}_i) = \ln\left(\int{\mathcal{F}_i^{\rm GW}(\vec{x}) \times  \mathcal{F}_i^{\rm HEN}(\vec{x}) d\vec{x}}\right)
\end{equation}
and the p-value corresponding to the combined PSF-likelihood is given by: 
\begin{equation}
\label{eq:psky}
p^{sky}_i=\int_{\mathcal{L}_i}^\infty P_{bg}(\ln(\mathcal{L}))d\mathcal{L}
\end{equation}


\begin{figure}[h!]
\begin{center}
\includegraphics[width=9.5cm]{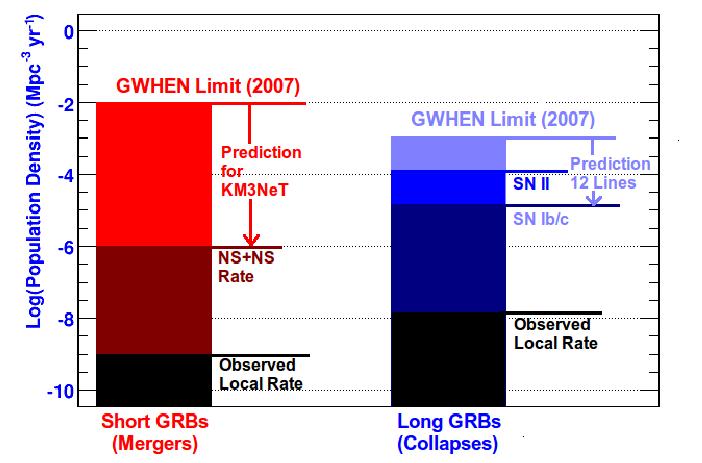}
\caption{\footnotesize GWHEN 2007 astrophysical limits as compared with local short/long GRB rates, merger rates, and SN II and SN Ib/c rates. Also shown is the expected reach of ongoing (2009-2010) and future analyses. }\label{fig:lim}
\end{center}
\end{figure}

\subsection{Final test statistic}
\label{sec:Finalteststat}
The three obtained p-values can be combined using Fisher's method~\cite{Fisher} to construct a test statistic for each event $i$:
\begin{equation}
\label{eq:combined}
X^2_i=-2\ln(p^{sky}_i\times p_i^{\rm GW} \times p_i^{\rm HEN})
\end{equation}

The final result of the search is the p-value of its most significant event $i$ defined as:

\begin{equation}
\label{eq:pgwhen}
p^{\rm GWHEN}=\int_{Max(X^2_i)}^\infty P_{bg}(max(X^2))dX^2
\end{equation}

The background probability density function $P_{bg}(max(X^2))$ is estimated by a Monte Carlo simulation of \mbox{$10^4$ pseudo-experiments} of 773 joint triggers (the remaining of the 1986 neutrinos coincident with data taking periods of GW interferometers) obtained by applying the analysis on time-shifted GW data. It will determine the significance of the loudest event once the box will be opened and the real, non-time-shifted data will be scrutinized. An accumulation of weaker signals can also be looked for, as was performed for the 2007 joint search.

\section{Perspectives and expected sensitivity}
\label{sec:exp}

The pioneering GW+HEN searches developed in~\cite{gwhen2007} and~\cite{IceCubegwhen} have opened the way towards a new multimessenger astronomy. Beyond the benefits of a potential high-confidence discovery, future analyses could be able to constrain the density of joint sources down to astrophysically meaningful levels. Figure~\ref{fig:lim} shows the upper limits on the population density of common HEN and GW emitters obtained from the ANTARES/Virgo/LIGO 2007 joint search, together with the potential reach of the ongoing (and future) searches. 

The previous discussion and the flow diagram of Figure~\ref{fig:flow} help understand the sources of the global improvement expected on these limits. The equivalent live time of the analysis is increased by 40\% with respect to the 2007 search, a gain which is also related to the possibility offered by the s-CWB pipeline to exploit data with only two interferometers active. The effective area of ANTARES has been multiplied by $\sim 3$ above 100 TeV during this data taking period. Combined with the enhanced sensitivity of the GW interferometers, and with the improvements in reconstruction and optimisation algorithms, a net gain by a factor $\sim 8$ can be expected with respect to what was achieved in the 2007 search. 

This new search should for example allow to constrain the population of events of core-collapse type at the order of $10^{-4}~\rm{Mpc^{-3} yr^{-1}}$ which is the observed rate of core-collapse supernovae. It also opens the path for the future with the advanced version of GW interferometers aLigo and aVirgo which will have ten-fold sensitivity~\cite{bibthierry:advitf} and will be operated at the same time as kilometric-scale neutrino detectors IceCube and KM3NeT~\cite{km3net}.


\setcounter{figure}{0}
\setcounter{table}{0}
\setcounter{footnote}{0}
\setcounter{section}{0}
\setcounter{equation}{0}

\newpage
\addcontentsline{toc}{part}{{\sc Exotic Physics}%
\vspace{-0.5cm}
}
\id{id_ardid}
\addcontentsline{toc}{part}{\textcolor{blue}{\arabic{IdContrib} - {\sl M. Ardid} : Constraining Secluded Dark Matter models with the ANTARES neutrino telescope}%
}

\title{\arabic{IdContrib} - Constraining Secluded Dark Matter models with the ANTARES neutrino telescope}

\shorttitle{\arabic{IdContrib} - Secluded Dark Matter models with ANTARES}

\authors{Miquel Ardid$^a$, Christoph T\"onnis$^b$ (speaker)}
  \afiliations{$^a$IGIC Universitat Polit\`ecnica de Val\`encia Paranimf 1, E-46730 Gandia, Spain\\
  $^b$ IFIC - Instituto de F\'isica Corpuscular, CSIC - Universitat de Val\`encia \\
Edificios Investigaci\'on de Paterna, Apdo. de Correos 22085, E-46071 Valencia, Spain
  }
\email{$^a$ mardid@fis.upv.es,$^b$ctoennis@ific.uv.es}


\abstract{In this work we describe the search for Secluded Dark Matter (SDM) annihilation in the Sun with ANTARES. SDM is a
special scenario where DM, which would gravitationally accumulate in astrophysical objects like the Sun, is annihilated
into a pair of non-Standard Model mediators, which subsequently decay into SM particles. It was suggested to explain
some experimental observations, such as the positron-electron ratio observed by satellite detectors. Three different
cases are studied: a) direct detection of di-muons from the mediator decay, or neutrino detection from: b) the mediator
that decays into di-muons and, in turn, into neutrinos, and c) the mediator that directly decays into neutrinos. The
ANTARES results obtained for SDM models --the first experimental limits established directly in neutrino telescopes--
are presented. \ The limits imposed to these models are much more restrictive than those derived in direct detection
searches for the case of spin-dependent interaction for a wide range of lifetimes of the meta-stable mediator.}

%
%
\maketitle

\section{Introduction}
In this paper we present the results of the analysis of ANTARES data in order to search for
signatures of Secluded Dark Matter (SDM) annihilation in the Sun. There is strong cosmological and astrophysical
evidence about the existence of Dark Matter (DM) in the Universe. There is as well a large consensus that this kind of
matter, about 83\% of the total, has the properties of being non-baryonic, non-relativistic and inert to
electromagnetic interactions, being the Weakly Interacting Massive Particles (WIMPs) hypothesis the favourite scenario
for the nature of DM. Then, DM would be embedded in the visible baryonic part of galaxies forming a halo. In the most
common scenario, WIMPs can scatter elastically with matter and become trapped in massive astrophysical objects like the
Sun. There, DM particles could self-annihilate reaching a balance between capture and annihilation rates over the age
of the Solar System. Usually, the products of DM annihilation are Standard Model (SM) particles, which interact with
the interior of the Sun and are largely absorbed. However, during this process, high-energy neutrinos may be produced,
which could scape and can be observed by neutrino detectors, such as ANTARES. In this sense, limits on WIMP DM
annihilation in the Sun have been reported already in ANTARES \cite{ardid1}, and in other neutrino telescopes: Baksan \cite{ardid2},
Super-Kamiokande \cite{ardid3} and IceCube \cite{ardid4}. Another hypothesis is based on the idea that DM will be Secluded from SM
particles, being the annihilation only possible through a metastable mediator ($\varphi $), which subsequently decays
into SM states\cite{ardid4,ardid5,ardid6,ardid7,ardid8,ardid9}. In all these models, the thermal relic WIMP DM scenario is considered as usual while there is also
the potential to explain some astrophysical observations, such as the positron-electron ratio observed by PAMELA \cite{ardid10}
or FERMI \cite{ardid11}, measured recently by AMS-II with much more accuracy \cite{ardid12}. In the Secluded Dark Matter scenario, the
presence of a mediator, as a communication way between DM and SM, can dramatically change the annihilation signature of
DM captured in the Sun. If the mediators live long enough to escape the Sun before decaying, they can produce
detectable charged-particle, $\gamma $-ray or neutrinos \cite{ardid13,ardid14} that could reach the Earth and be detected. In many of
the secluded dark matter models, $\varphi $ can decay into leptons near the Earth. Some differences appear in the
signature of leptons created by the neutrino interaction and leptons arising from $\varphi $ decays. In the latter case
as the DM mass (\~{}1 TeV) is greater than the $\varphi $ mass (\~{}1 GeV) the leptons may be boosted and parallel. If
these leptons are muons the signature in the vicinity of the detector would be two muon tracks almost parallel. Meade
et al. \cite{ardid15} discuss this possibility and calculate the expected sensitivity for the Icecube neutrino telescope to these
cases. It is worth also to mention that even in the case that the di-muon signature could be interpreted as a single
muon, the different energy deposition can help to better discriminate this case from the atmospheric neutrino signal
\cite{ardid16}. Even for short-lived mediators that decay before reaching the Earth, neutrinos from the products of mediator
decays could be detected in neutrino telescopes. Another possibility is that mediators may decay directly into
neutrinos, as discussed by \cite{ardid17}. In this case, the neutrino signal could be enhanced significantly compared to the
standard scenario even for quite short-lived mediators, since they will be able to escape the dense core of the Sun
where high energy neutrinos can interact with nuclei and be absorbed. The fact that the solar density decreases
exponentially with radius facilitates that the neutrinos injected by mediators at larger radii propagate out of the Sun
because they undergo much less absorption. 

In this work an indirect search for SDM using the 2007-2012 data recorded by the ANTARES
neutrino telescope is reported by looking at the different mediator decay products: a) direct detection of di-muons b)
neutrinos from decays of di-muons produced by mediators that decay before reaching the Earth and c) neutrinos produced
by mediators that decay directly to neutrinos and antineutrinos. The analysis procedure is basically the same as the
previous search for dark matter annihilation in the Sun \cite{ardid1}, but optimizing the search for the expected signal in the
case of SDM.

\section{The ANTARES neutrino telescope}
A description of the ANTARES neutrino telescope can be found elsewhere in these proceedings. A
more detailed description of the telescope, subsystems and methods can be found in \cite{ardid18,ardid19,ardid20,ardid21}. In this analysis, data
recorded between the 27th of January 2007 and the 31st of October 2012 are used, corresponding to a total lifetime of
1321 days, without taking into account the visibility of the Sun. During this time, the detector consisted of 5 lines
for most of 2007 and of successively 8, 9, 10 and 12 lines from 2008 to 2012. 

\section{Signal and Background estimation}
Two main sources of background are present in ANTARES: 1) Down-going atmospheric muons
resulting from the interaction of cosmic rays in the atmosphere. These background events are strongly reduced by the
deep sea location and by the reconstruction algorithms that are tuned to up-going events. Cuts on the quality of the
tracks are also applied to reject down-going muons wrongly reconstructed as up-going. 2) Atmospheric neutrinos produced
by cosmic rays. These neutrinos can traverse the Earth, so they can be detected as upgoing tracks. This is an
irreducible background. Both kinds of background have been simulated and good agreement with data has been found \cite{ardid1}.
However, the background estimation is done using scrambled data, by randomizing the time of selected events, to reduce
the effect of systematic uncertainties (efficiency of the detector, assumed atmospheric fluxes, etc.). 

Regarding \ to the signal estimation and to be able to evaluate SDM models, a new tool for
Di-Muon signal generation (DiMugen) has been developed to evaluate the sensitivity of ANTARES to the the case a) where
dimuons are detected directly \cite{ardid22}. DiMugen generates and propagates dimuons coming from decay of mediators resulting
from dark matter annihilation. For this analysis, the mediator arrives from the Sun's direction following the zenith
and azimuth information about the Sun position during the period under study. Different DM masses in the range between
30 GeV to 10 TeV have been simulated using in most cases a typical mass of 1 GeV for the mediator ${\phi}$. Once the
muons are generated in the vicinity of the detector according to these conditions, simulations of the travel and
interactions of muons are made, as well as the detection of the Cherenkov light by the optical modules. Triggering and
reconstruction algorithms are also included in the process in order to evaluate the global efficiency for the detection
of dimuons as a function of the quality parameter, Q, and the angular deviation from the Sun direction observed,
${\Psi}$.

To determine the ANTARES sensitivity for the cases where the neutrino is the final decay
product that arrives to the Earth, we have used the ANTARES effective areas for neutrinos as functions of the Q and
${\Psi}$ according to neutrino (and antineutrino) simulations. For this, it is necessary to know the energy spectra of
neutrinos arriving to the detector. In case b) the neutrino spectra have been obtained from Michel's spectra of
neutrinos and antineutrinos from muon decay and taking into account the boost. For scenario c) and assuming long
lifetime mediators with respect to the time required to go out from Sun's core, the neutrino (and antineutrino) spectra
are almost flat in the energy region under study \cite{ardid17}. For these cases, a detailed neutrino oscillation study has not
been done, but the conservative assumption that after oscillations all neutrino flavours arrive to the Earth with the
same ratio 1:1:1 has been made.

\section{Optimization of the event selection criteria}
In order to avoid any bias in the event selection, a blinding policy has been followed. The
values of the cuts have been chosen before looking at the region where the signal is expected. The best sensitivities
for di-muon flux and cross-sections are extracted with the Model Rejection Factor (MRF) method \cite{ardid23}. It consists in
finding the set of cuts which provide, in average, the best flux upper limit taking into account the existing
background and the efficiency to a possible flux signal from simulations. \ MRF is used to optimize the half-cone angle
around the sun (${\Psi}$) and the track quality cut parameters (Q) for the different cases and the different DM masses
studied. Finally, since in most of the cases the difference in flux sensitivities between different optimisations were
not large, it was decided to limit the optimisations to 4 different cuts that were representative enough of all
possible situations. There are 3 optimisations corresponding roughly to lower, intermediate and larger DM masses for
the dimuon detection case. For the neutrino detection cases, the latter one is also used for larger DM masses and
another additional optimisation is used for lower and intermediate DM masses. 

\section{Results and discussion}
After the optimisation of the flux sensitivities using the MRF with scrambled data, we have
looked at the data coming from the Sun's direction. As an example, figure \ref{fig:ardid1}-left shows the distribution of events
detected by ANTARES for Q{\textless}1.8 as a function of the angle deviation from the Sun. Good agreement between data
and the expected background obtained from scrambled data is observed. The green line shows the angle cut selected for
this analysis. Since no significant excess is observed in any of the blind cuts proposed, the 90\% CL upper limit
values in the Feldman-Cousins approach \cite{ardid24} have been extracted and used to constrain the models. The resulting flux
limits for the different cases studied are shown in figure \ref{fig:ardid1}-right.

Following the reasoning given in ref. \cite{ardid15}, the di-muon (or neutrino) flux at Earth can be
translated into DM annihilation in the Sun through the channel \ DM+DM$->{\phi}+{\phi}->$(2${\mu}$)$+$(2${\mu}$),
considering the muon decays for the detection of neutrinos. For the case in which mediators decay directly into
neutrinos, only the situation in which the mediator life is long enough has been considered, so that the absorption of
neutrinos in the Sun becomes negligible. In this case, the neutrino spectrum is harder and the signal in a neutrino
telescope is enhanced. If the lifetime of the mediator is small, the final situation would be quite similar to the
typical hard spectrum channels \cite{ardid17}. The conservative assumption that after oscillations all
neutrino flavours arrive to the Earth with the same ratio 1:1:1 has been made. Assuming 100\% branching ratios, and
taking into account the solid angle suppression and the decay probabilities, as explained in ref. \cite{ardid22}, we can start to
constrain the models by means of exclusion plots for the annihilation rates as a function of mediator lifetime and dark
matter mass. For example, figure \ref{fig:ardid3} shows the ANTARES exclusion limits for the Secluded DM scenarios studies for DM
masses of 0.5 and 5 TeV using a typical $\phi$ mass of 1 GeV. Blue lines indicate the exclusion region in the di-muon
case, either by direct detection (dot-dashed line) or through detection of neutrinos (solid line). For large decay
length L=$\gamma$c$\tau$, (L{\textgreater}1 AU), that is long mediator lifetime, the
direct detection of di-muons is more efficient than neutrino detection for small DM masses, whereas the opposite holds
for larger masses. The transition is around 0.8 TeV in DM mass. Naturally, for small L, L{\textless}{\textless}1 AU,
neutrino detection is much more efficient for all DM masses. Green lines indicate the exclusion regions of secluded DM
into neutrinos. More stringent constrains are obtained in this scenario, mainly due to the harder neutrino energy
spectrum.

\begin{figure}
\centerline{\includegraphics[width=0.5\textwidth]{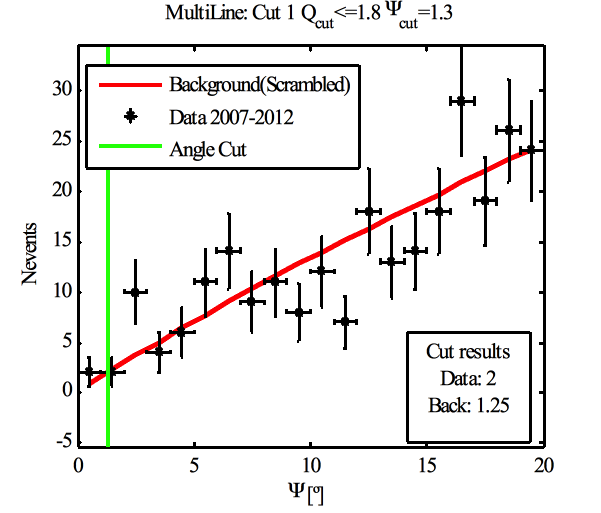}\includegraphics[width=0.5\textwidth]{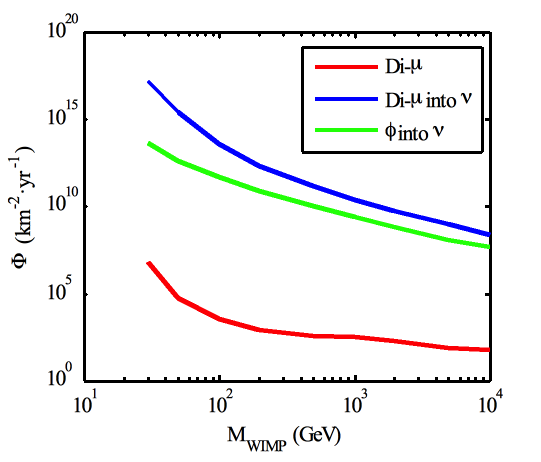}}
\caption{\label{fig:ardid1} Left: differential distribution of the angular separation of the event tracks with respect to the Sun's
direction with Q{\textless} 1.8 for data (black) and expected background (red line). Right: Limits for the flux of
di-muons and neutrinos from the SDM cases studied.}
\end{figure}


\begin{figure}
\centerline{\includegraphics[width=0.5\textwidth]{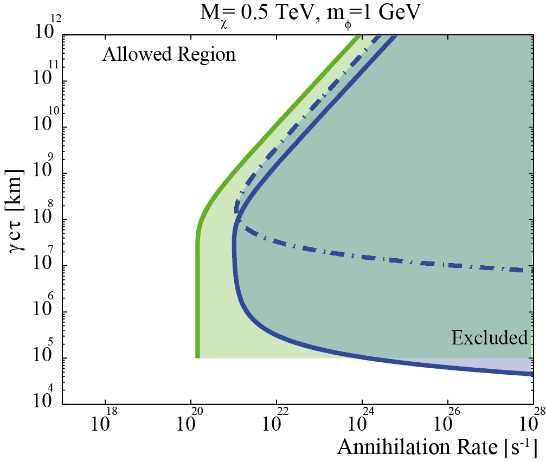}\includegraphics[width=0.5\textwidth]{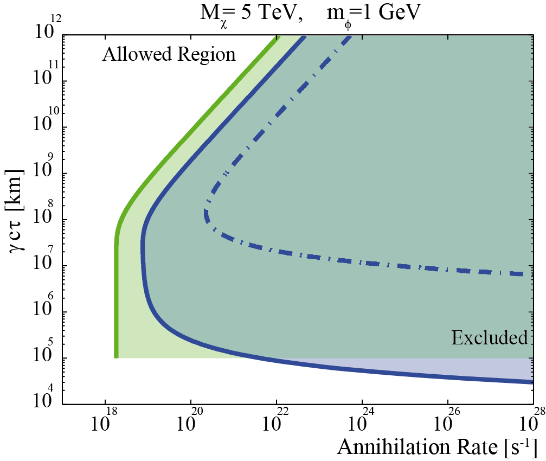}}
\caption{\label{fig:ardid2} ANTARES exclusion limits for the Secluded DM cases studied (products of DM annihilation in the Sun through
mediators decaying into: di-muons (dash-dotted blue), neutrinos from di-muons (solid blue), directly into neutrinos
(Green)) as a function of the annihilation rate ($\Gamma $) and the decay length (${\gamma}$c$\tau $) for 0.5 and 5 TeV
DM masses. The shadow regions are excluded for these models.}
\end{figure}

%

Limits on DM-nucleon interaction can also be derived for these cases. Assuming equilibrium of
the DM population in the Sun, i. e., the annihilation balances the DM, $\Gamma = C_{\textrm{CDM}}/2$, and according to \cite{ardid25} the capture is approximately:

\begin{equation}
C_{\mathit{DM}}=10^{20}s^{-1}\left(\frac{1\mathit{TeV}}{M_{\mathit{DM}}}\right)^2\frac{2.77\sigma
_{\mathit{SD}}+4270\sigma _{\mathit{SI}}}{10^{-40}\mathit{cm}^2}
\end{equation}

where, $\sigma_{\textrm{SD}}$ and
$\sigma_{\textrm{SI}}$ are the spin-dependent (SD) and
spin-independent (SI) cross-sections, respectively, and $M_{\textrm{DM}}$ is the DM mass. The limits on the SD and SI
WIMP-proton scattering cross-sections are derived for the case in which one or the other is dominant. The sensitivity
in terms of the annihilation rates depends on the lifetime of the mediator. To assess the potential to constrain these
models, lifetime values for which the sensitivities are the best possible have been assumed. For the di-muon case, the
lifetime has to be long enough to assure that the mediator reaches the vicinity of the Earth, so mediators with decay
length about Sun-Earth distance are shown. In both neutrino cases the lifetime of the mediator for best sensitivity has
to be long enough to ensure that the mediator escapes the Sun, but not too long so that it decays before reaching the
Earth. The lifetime of the mediator for the best sensitivity has been chosen, corresponding to a distance of
approximately forty times the solar radius. Figure \ref{fig:ardid3} shows the ANTARES nucleon{}-WIMP cross{}-section limits for the
SDM scenario (products of DM annihilation in the Sun through mediators decaying into: di-muons (blue) and directly into
neutrinos (green)) for the selected mediator's lifetimes. The limits are compared to those given by different
experiments of direct search for dark matter. 

\begin{figure}
\centerline{\includegraphics[width=7.853cm,height=6.585cm]{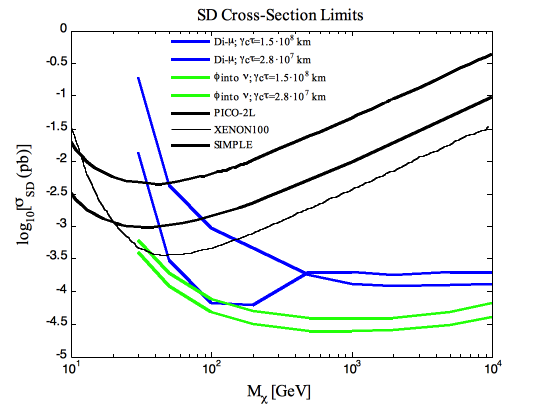}\includegraphics[width=7.105cm,height=6.586cm]{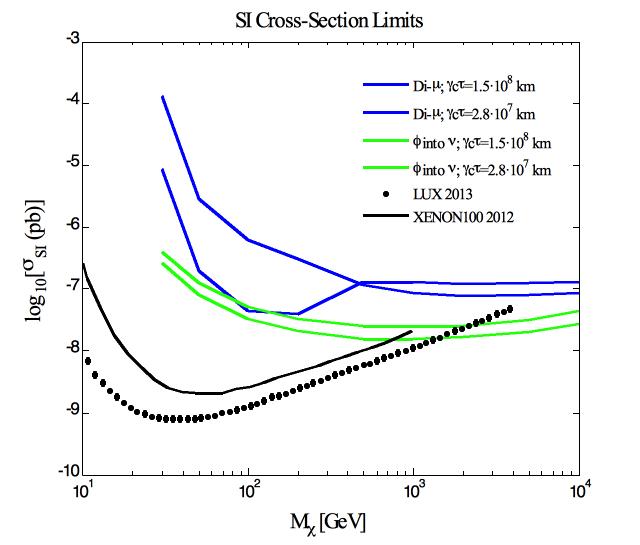}}
\caption{\label{fig:ardid3} ANTARES 90\% CL upper limits on WIMP-nucleon cross-section as a function of WIMP
mass. The left panel refers to spin-dependent and the right one to spin-independent WIMPs interactions. Two favourable
mediator lifetimes are considered. The procedure to translate from DM annihilation rate in the Sun, $\Gamma $, and
c${\tau}$ to $M_{\chi}$  and $\sigma_{\chi}$ is described in \cite{ardid15}.
Additionally, in the same panel the current bounds from SIMPLE \cite{ardid26}, PICO \cite{ardid27}, LUX \cite{ardid28} and XENON \cite{ardid29} are plotted.}
\end{figure}

%
%
%

The limits derived here are the first experimental limits on SDM models established by
neutrino telescopes. There were some previous constrains or sensitivities predicted by phenomenology physicists
\cite{ardid14,ardid15}, but naturally, the knowledge of the response of the detector in this kind of studies is quite limited, and
therefore, the results should be taken with caution. As shown in \ref{fig:ardid3}, for sufficiently long-lived, but unstable
mediators, the limits imposed to these models are much more restrictive than those derived in direct detection searches
for the case of spin-dependent interaction. In the case of spin-independent interactions, direct detection search is
more competitive for low and intermediate masses, but the SDM search becomes more competitive for larger masses
({\textgreater} 1 TeV).

Compared to other indirect detection methods, such as those using gamma-rays, the limits
derived here are in general competitive for large dark matter masses and favourable mediator lifetimes
($\gamma c{\tau} \sim 10^{11}$ m. However, the
comparison is not straightforward, since the results are usually given in terms of the
{\textless}${\sigma}$v{\textgreater} \ parameter and several astrophysical assumptions have to be made. Therefore, the
different indirect searches can be considered to provide complementary information. In that sense, this analysis
constrains in an alternative way these models that are one of the preferred solutions to explain, for example, the
energy of the positron flux measured by AMS-II \cite{ardid12}. Although one possible interpretation of this data would be the
existence of nearby pulsars, a great deal of papers study the possibility of a DM hint. In this line, the annihilation
into two mediators that results in four leptons (two di-muons, for example) is more favoured than the direct
annihilation into leptons \cite{ardid30,ardid31,ardid32}.

{\bfseries
Acknowledgements}

We acknowledge the financial support of the Spanish Ministerio de Economía y Competitividad
(MINECO) and Ministerio de Ciencia e Innovación (MICINN), Grants FPA2012-37528-C02-02, and Consolider MultiDark
CSD2009-00064, and of the Generalitat Valenciana, Grant PrometeoII/2014/079.

\setcounter{figure}{0}
\setcounter{table}{0}
\setcounter{footnote}{0}
\setcounter{section}{0}
\setcounter{equation}{0}

\newpage
\id{id_imad}
\addcontentsline{toc}{part}{\textcolor{blue}{\arabic{IdContrib} - {\sl I. El Bojaddaini} : Search for magnetic monopoles with the ANTARES neutrino telescope}%
}
%
%

\title{\arabic{IdContrib} - Search for magnetic monopoles with the ANTARES
neutrino telescope}
\shorttitle{\arabic{IdContrib} - Magnetic monopoles with ANTARES}
\authors{Imad El Bojaddaini$^a$, Gabriela Emilia Pavalas$^b$}
\afiliations{$^a$ Mohammed I University - Oujda, Morocco\\
$^b$ Institute of Space Science - Bucharest, Romania
}
\email{$^a$ elbojaddaini\_1990@hotmail.fr, $^b$gpavalas@spacescience.ro}

\maketitle

\begin{abstract}

  Magnetic monopoles are hypothetical particles predicted to be created in the early Universe in the framework of Grand Unified Theories (GUTs). The signature of the passage of relativistic magnetic monopole in a Cherenkov telescope like ANTARES (Astronomy with a Neutrinos Telescope and Abyss environmental RESearch) \cite{1} is expected to be evident and unambiguous because of the large amount of light emitted compared to that from muons.
	
A first study has been carried out in ANTARES using a limited data set of 116 days; first upper limits on the magnetic monopoles flux were established for relativistic monopoles with $\beta \ge$ 0.625. We present here an update of the analysis, using an enlarged data set (data collected from January 2008 to December 2013) and considering a wider range of values for $\beta$. No monopoles have been observed, and new sensitivity has been set, for monopoles with $\beta\ge$0.572.
\end{abstract}

\section{Introduction} \label{sec:intro}

Grand Unified Theories (GUTs) predict the creation of magnetic monopoles in the early Universe \cite{2}. Their detection in a neutrino telescope is similar to the detection of high energy muons. As for electric charges, magnetically charged particles produce Cherenkov emission when their velocity is higher than the Cherenkov threshold $\beta =1/n$, where $n$ is the phase refractive index of the medium. In this analysis, we restrict the selected sample to up-going monopoles to ensure an easy separation from atmospheric muons. However, fast monopoles can lose an energy of $10^{11}$ GeV when traversing the full diameter of the Earth, but they are expected to be accelerated in the Galactic coherent magnetic field domain to energies up to $10^{15}$ GeV. 
Thus, only monopoles in the energy range $10^{12}-10^{15}$ GeV are expected to be detectable in this analysis as up-going signals.

\section{Monte Carlo simulation and reconstruction}
Up-going magnetic monopoles have been simulated using ten equidistant ranges of velocities in the region $\beta=[0.55 ,0.995]$, and a package named $Simon$ has been used \cite{3}. It is based on Monte Carlo generators used in ANTARES to simulate neutrino interactions. This package contains two main programs, $genmon$ which is used to generate monopoles, and $geamon$ simulating the emission of light and the response of the detector. Monopoles are simulated as tracks. They are generated uniformly over the hemisphere above and below the detector. Atmospheric muons and neutrinos have been also simulated as background. The events are then reconstructed using an algorithm named $BBFit$, that is usually applied in the analysis of neutrino candidates \cite{4}. 
Indeed, the standard track reconstruction assumes that particles travel at the speed of light. In order to improve the sensitivity for magnetic monopoles traveling with lower velocities, the $BBFit$ reconstruction algorithm has been modified so as to leave the velocity $\beta$ as a free parameter to be determined by the track fit.

\section{Analysis strategy and quality cuts}

Some primary cuts are applied for the whole velocity range. The first selection cut, which is expected to remove a large part of down-going muons and neutrinos, concerns the zenith angle which must be smaller than $90\circ$ since we search for up-going monopoles. In order to further reduce the background, a second cut was applied, which consists to consider only events reconstructed on at least 2 lines of the detector $(nlines \ge 2)$. 
The other discriminative variables are based on physical properties of monopoles and the quality of reconstruction.

Two different strategies are followed in the analysis, depending on monopoles velocity. In the first range of $\beta$ the optimization is done for 6 values of $\beta$ ranging from 0.55 to 0.817. 
The discrimination of magnetic monopoles from background relies on $\beta$ reconstruction. While muons and neutrinos have approximately the speed of light, monopoles can be distinguished by their specific speed. Thus, to isolate monopoles from atmospheric muons and neutrinos a cut on the reconstructed $\beta$ will emit a large amount of direct Cherenkov light when travelling through the ANTARES detector. For $\beta$ ranging from 0.817 to 0.995 the track reconstruction algorithm is not able to discriminate the velocity and thus $\beta=1$ is assumed. The discrimination against the background relies on the number $Nhit$ of storeys used by the algorithm to reconstruct the track. Another variable named $\alpha$ containing the track fit quality parameter $t\chi^2$ and $Nhit$ is also used in this analysis.
\begin{figure}[tbh]
  \centering
  \includegraphics[width=1.0\textwidth]{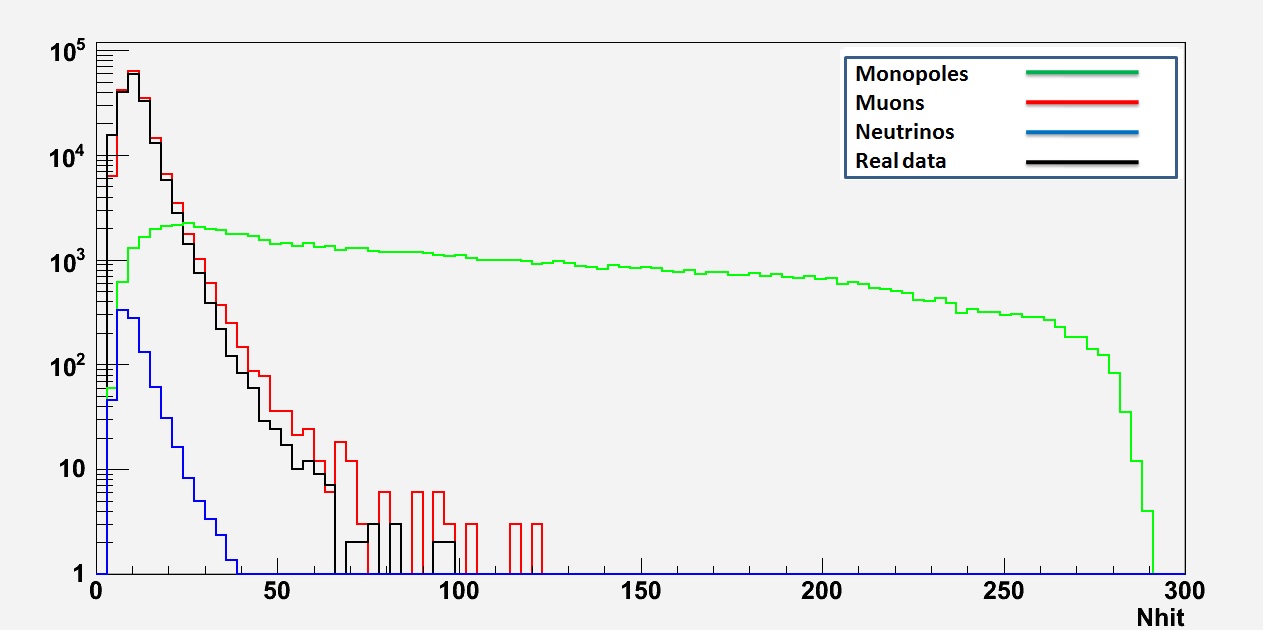}
    \caption{$Nhit$ distribution for monopoles with $\beta=0.97275$ and background, compared to the selected data set.}
    \label{fig1}
\end{figure}
\begin{figure}[h]
  \centering
  \includegraphics[width=1.0\textwidth]{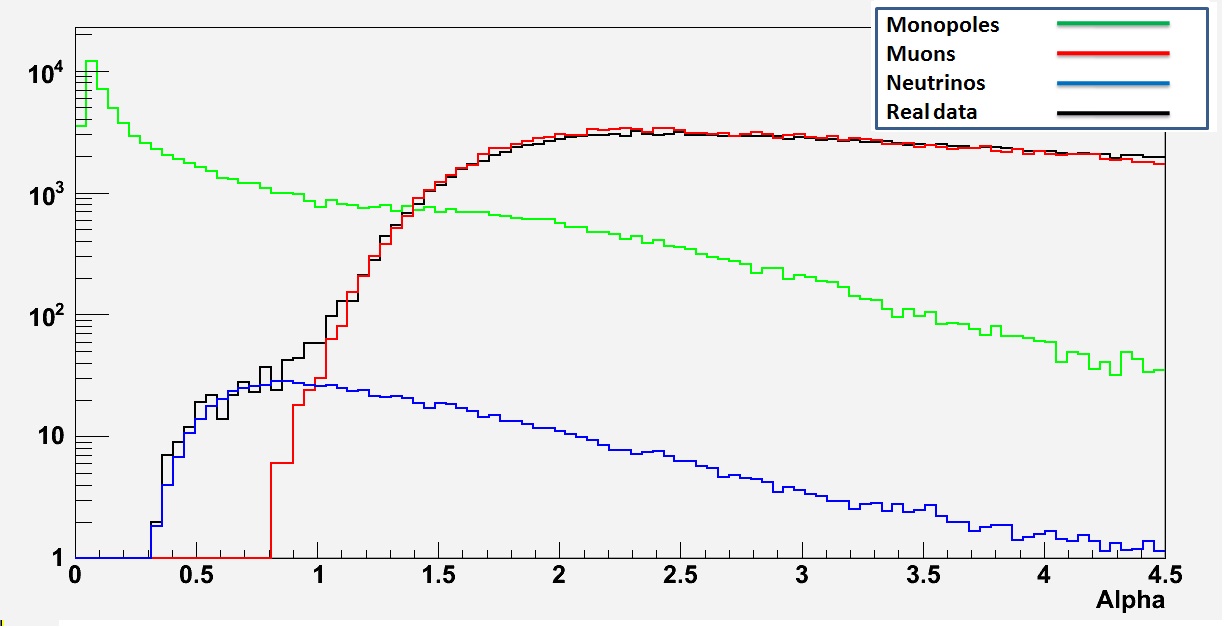}
    \caption{Alpha distribution for monopoles with $\beta=0.97275$ and background, compared to the selected data set.}
    \label{fig2}
\end{figure}

In order to avoid biases when elaborating the analysis strategy, the ANTARES Collaboration follows a $blind\ approach$: data are $blinded$ (information on the direction is $masked$) when the process of optimization of the cuts is carried out. However, in order to make comparison between real and Monte Carlo data, the collaboration allows using a sample of real data. The sample used here (Figure 1 and 2) is composed by the so-called $0\ runs$ (runs ending with a 0).

\section{ The Model Rejection Factor}

The 90\% C.L. sensitivity $S_{90\%}$ is calculated with the Feldman-Cousins formula \cite{5}, considering events which follow a Poissonian distribution:
\begin{equation}
S_{90\%}= {\overline{\mu}_{90}(n_b) \over S_{eff} \times T} \ ,
\end{equation}
where $T$ is the duration of the data taking, and where $\overline{\mu}_{90}$  and $ S_{eff} $ are defined as:
\begin{equation}
{\overline{\mu}_{90}(n_b) }= \sum^{\infty}_{n_{obs=1}} \mu_{90}(n_{obs},n_b) {n^{n_{obs}}_b\over n_{obs}! } e^{-n_b} \ ,
\end{equation}

\begin{equation}
S_{eff} ={n_{MM}\over \Phi_{MM} } ,
\end{equation}
with $ n_{MM}$ the number of monopoles remaining after cuts, and $\Phi_{MM} $ the flux of monopoles generated. 
The Model Rejection Factor consists in playing with cuts in order to get the minimum of Rejection Factor (RF) (equation 4) where the best sensitivity is obtained.
\begin{equation}
RF = {\overline{\mu}_{90}(n_b) \over n_{MM} } \ .  
\end{equation}
To optimize the 90\% C.L. sensitivity the two quantities to play with are $\alpha$ and $Nhit$. The Rejection Factor is calculated for each couple of cuts $(\alpha,Nhit)$, where $\alpha$  is varying from 0 to 4.5, and $Nhit$ varying from 0 to 300 (see figure 3).
\begin{figure}[h]
  \centering
  \includegraphics[width=1.0\textwidth]{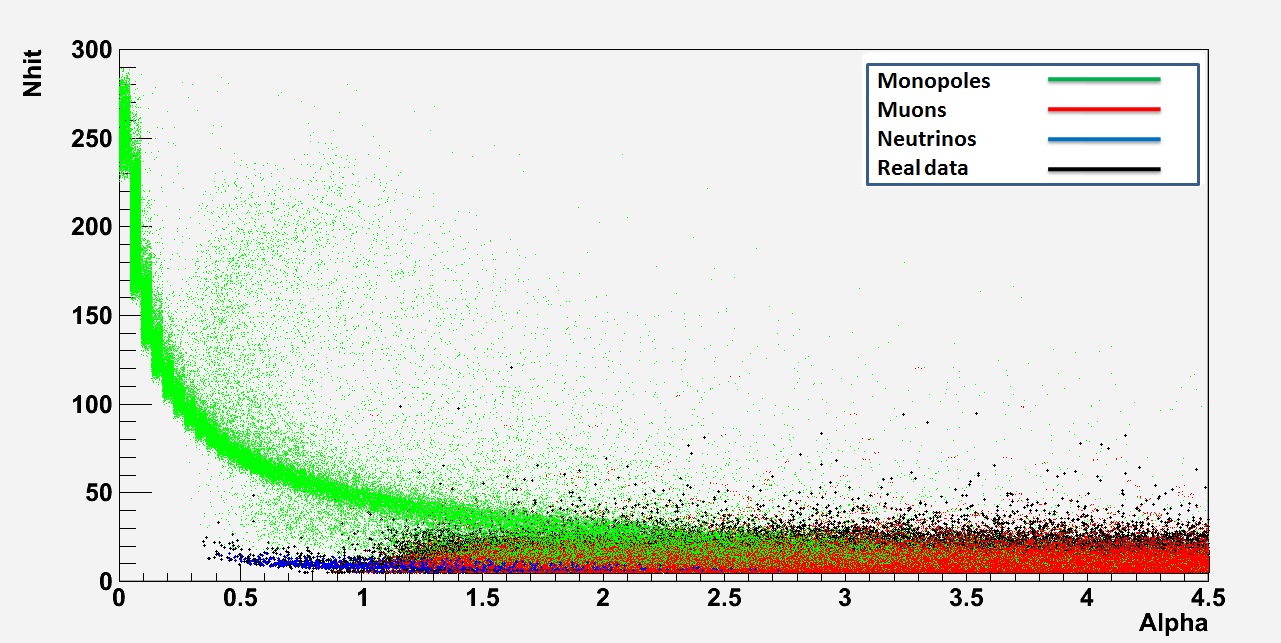}
    \caption{2D histogram representing the distribution of alpha and $Nhit$ for magnetic monopoles with a speed $\beta=0.97275$ (green points) and MC atmospheric background (red and blue points), compared to the selected data set.}
    \label{fig3}
\end{figure}
Figure 4 illustrates the variation of Rejection Factor as a function of $(\alpha,Nhit)$  cuts. In this case of $\beta=0.97275$ the minimum of RF corresponds to $5.9\times 10^{-5}$, which is then taken to calculate the sensitivity. This is done for each value of the velocity.

\begin{figure}[h]
  \centering
  \includegraphics[width=0.8\textwidth]{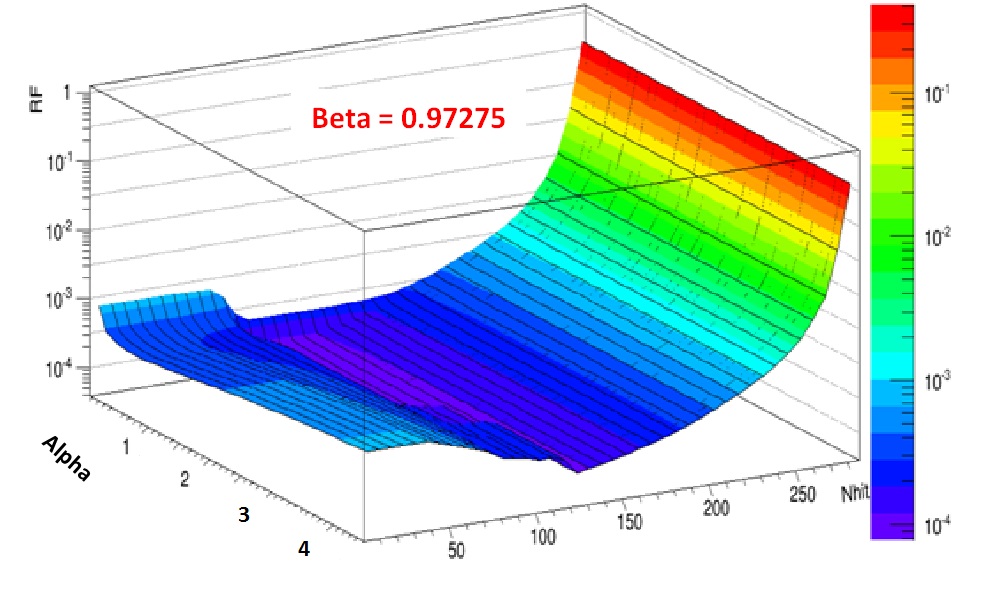}
    \caption{The rejection factor RF as a function of $\alpha$  and $Nhit$ cuts.}
    \label{fig4}
\end{figure}
\begin{figure}[h!]
  \centering
  \includegraphics[width=0.8\textwidth]{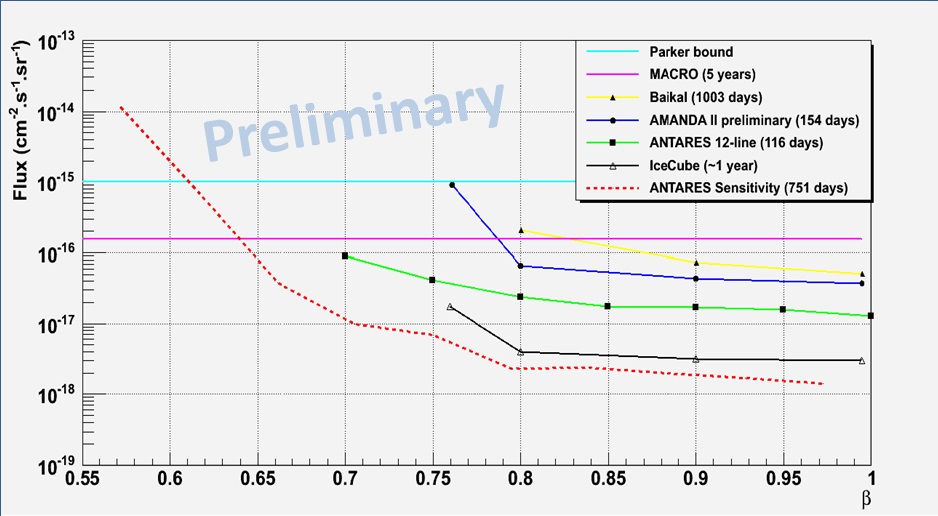}
    \caption{ANTARES sensitivity on the flux of monopoles as a function of beta, found assuming data collected during 6 years (red), compared to the upper limits on the flux obtained by other experiments.}
    \label{fig5}
\end{figure}
\section{ Sensitivity }
Figure 5 presents the ANTARES sensitivity obtained assuming data collected during 6 years when applying equation (1), compared to the upper limits on the flux found by other experiments and including the upper limit (116 days) of the previous analysis of ANTARES. As we see, despite the fluctuation of the sensitivity at lower $\beta$, it is better than all the upper limits obtained so far.

\section{Conclusion}
This paper presents a preliminary result of the analysis performed to search for up-going magnetic monopoles with velocity $\beta$  ranging from 0.55 to 0.995. The optimization of the Model Rejection Factor has led to find a new sensitivity on monopoles flux. The analysis strategy here discussed is very promising to investigate a wide range of values of $\beta$   and will be soon applied to the entire set of ANTARES data collected from January 2008 to December 2013.


\setcounter{figure}{0}
\setcounter{table}{0}
\setcounter{footnote}{0}
\setcounter{section}{0}
\setcounter{equation}{0}

\newpage
\id{id_gleixner}
\addcontentsline{toc}{part}{\textcolor{blue}{\arabic{IdContrib} - {\sl A. Gleixner} : Indirect search for dark matter towards the centre of the Earth with the ANTARES neutrino telescope}%
}

\title{\arabic{IdContrib} - Indirect search for dark matter towards the centre of the Earth
   with the ANTARES neutrino telescope}

\shorttitle{\arabic{IdContrib} - ANTARES search of DM from the Earth}

\authors{Andreas Gleixner$^a$, Christoph T\"onnis$^b$ (speaker)}
        \afiliations{$^a$Friedrich-Alexander-Universit\"at Erlangen-N\"urnberg\\
        $^b$Instituto de Fisica Corpuscular (IFIC) Valencia}
\email{$^a$Andreas.Gleixner@physik.uni-erlangen.de,$^b$ctoennis@ific.uv.es}
%


        
\abstract{The ANTARES neutrino telescope is a water Cherenkov detector and currently the largest operating neutrino telescope in the Northern Hemisphere. One of the main scientific goals of ANTARES is the indirect search for dark matter, as the Weakly Interacting Massive Particle (WIMP). WIMPs could scatter on normal matter and therefore be gravitational bound in massive astronomical objects like the Earth. Therefore an indirect search for dark matter can be performed by looking for an excess of the neutrino flux from the Earth's core. The exact spectrum of the neutrino flux from the Earth would depend on the WIMP mass, the annihilation channel, the spin independent scattering cross section and the thermally averaged annihilation cross section of the WIMPs. Such a search has been done with the data taken by ANTARES from 2007 to 2012. First limits from this search will be presented.}


\maketitle

\section{Introduction}
\label{mrf_conversion}
The hypothetical Weakly Interacting Massive Particles (WIMPs) are widely regarded as excellent dark matter (DM) candidates. WIMPs arise most prominently in supersymmetric models \cite{SDM} like the Minimal Supersymmetric Extension of the Standard Model (MSSM). In most cases the lightest supersymmetric particle (LSP) is the lightest neutralino. To ensure baryon and lepton number conservation in the MSSM it is often assumed that a multiplicative quantum number called R-parity is conserved. The LSP would then be stable, making the neutralino an excellent dark matter candidate.\\
WIMPs can be detected directly via the observation of the nuclear recoils from the scattering of WIMPs off nuclei (recent such experiments include XENON100\cite{XENON100-2}\cite{XENON100} and Lux \cite{lux_firstresults}), or indirectly via the observation of products from WIMP self-annihilations. The latter is possible for massive astrophysical objects in which WIMPs can accumulate, like the Earth \cite{SDM}\cite{icecube_theory}, the Sun \cite{antares_dmsun}\cite{icecubesun} or the Galactic Center \cite{LAT_results}. This paper deals with the indirect search for WIMPs from the center of the Earth.\\
Capturing of WIMPs in the Earth is dominated by spin-independent elastic scattering on the heavy nuclei abundant in the Earth and is kinematically suppressed if the mass of the WIMP is not close to the mass of the particle or nucleus the WIMP is scattering on. This is because the dark matter velocity dispersion is around $270\unitx{km/s}$ \cite{SDM}, but the escape velocity from Earth is only about $11.1\unitx{km/s}$ at the surface and $14.8\unitx{km/s}$ at the center. 
The WIMP annihilation rate in the Earth today can be written as \cite{SDM}:
\begin{equation}
\Gamma(t) = \frac{1}{2} C_C \tanh^2\left(\frac{t}{(C_C C_A)^{-0.5}}\right)
\label{eq:A_r}	
\end{equation}
Here $t$ is the age of the Earth, $C_A$ depends linearly on the thermally averaged annihilation cross section  times velocity $<\sigma v>$ and $C_{C}$ is the WIMP capture rate which depends linear on the  spin-independent elastic scattering cross section of the WIMP to protons $\sigma^{SI}_{p}$. The exact form of $C_A$ and $C_C$ can be found in \cite{SDM} and \cite{SRN}.\\
Assuming the annihilation cross section for dark matter in the Earth the same as during the freeze out of WIMPs, the conditions for equilibrium in the Earth ($t\geq 2 (C_C C_A)^{-1/2}$) are not generally satisfied. It would however be possible that $<\sigma v>$ becomes boosted in the case of low velocities for some reason, e.g. the Sommerfeld effect.\\
In this paper, we present limits on $\sigma^{SI}_p$, derived from data taken by ANTARES from 2007 to 2012. We consider WIMP masses between $25\unitx{GeV}$ and $1\unitx{TeV}$. The lower bound was chosen under consideration of the capability of ANTARES to reconstruct neutrinos of low energy, the upper bound was chosen roughly one order of magnitude higher than the masses of elements in the Earth. We consider WIMPs which annihilate  either into the soft  $b\overline{b}$ channel, the hard $\tau^{+}\tau^{-}$ or  $W^+W^-$ channel or the monochrome, non-SUSY $\nu_{\mu}\bar{\nu}_{\mu}$ channel. We consider both enhanced and non-enhanced scenarios for $<\sigma v>$.

\section{Simulations}
The neutrino flux from dark matter annihilation in the Earth was simulated with WimpSim \cite{WS2}\cite{WS}. It simulates WIMP pair annihilations inside the Earth without any assumptions about the dark matter model except the WIMP mass and the annihilation channel (a $100\%$ branching ratio is assumed) and the subsequent decay of the products. The resulting neutrino flux is propagated to the surface of the Earth while neutrino oscillations are taken into account in a full three flavour scenario. For an example of such fluxes, see Figure \ref{fig:WIMPSIM}.\\
\begin{figure}
\centering
\begin{minipage}{.5\textwidth}
  \centering
  \includegraphics[width=1.0\linewidth]{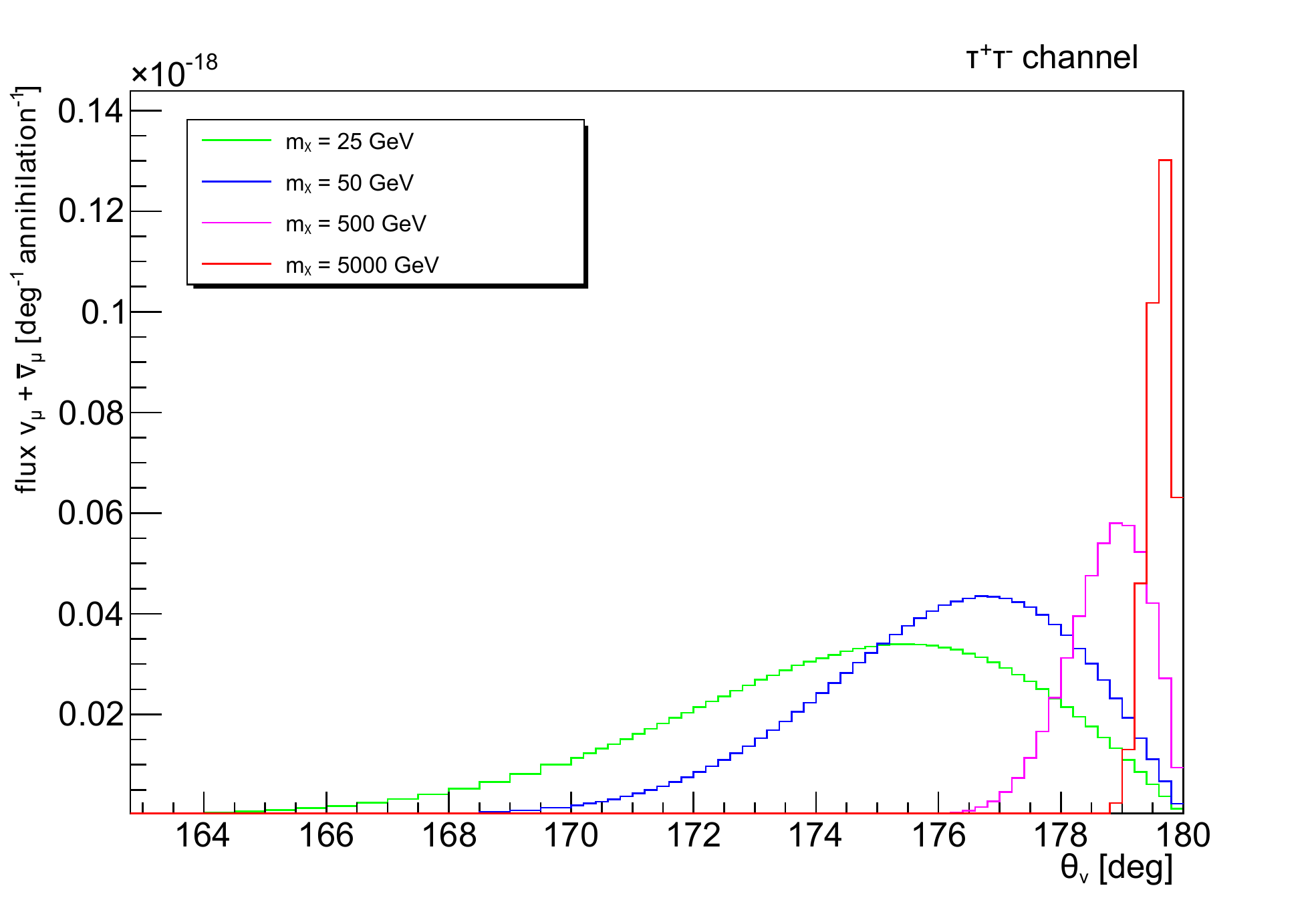}
  \label{fig:test1}
\end{minipage}%
\begin{minipage}{.5\textwidth}
  \centering
  \includegraphics[width=1.0\linewidth]{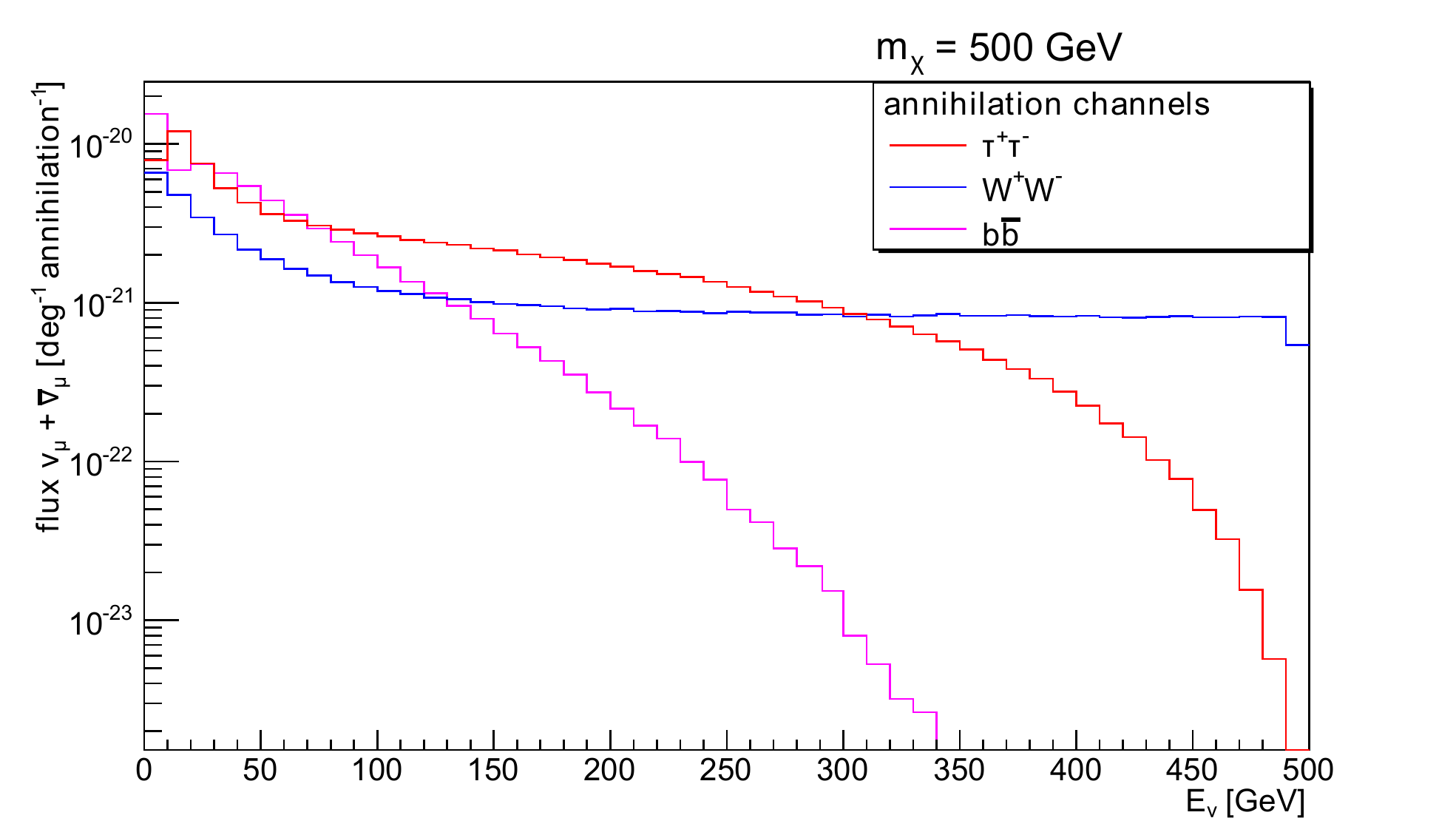}
  \label{fig:test2}
\end{minipage}
\label{fig:WIMPSIM}
\caption{Zenith and energy spectrum of the $\nu_{\mu}$ + $\overline{\nu}_{\mu}$ flux from WIMP pair annihilations for different WIMP masses (left) or annihilation channels (right) at the surface of the Earth as simulated with WimpSim.}
\end{figure}
The primary sources of background in this analysis consist of muons and muon-neutrinos, which have their origin in interactions of cosmic rays with the atmosphere of the Earth. The atmospheric muons are simulated with the MUPAGE \cite{MUPAGE} package (the parametric formulas of the fluxes of muon bundles can be found in \cite{MUPAGE-muonbundles1} and \cite{MUPAGE-muonbundles2}). For the background from atmospheric neutrinos only the contributions from charged current interactions from atmospheric muon (anti-)neutrinos contribute significant to this analysis. For the conventional neutrino flux the parametrization of \cite{bartol} is used with a prompt contribution according to \cite{enberg}.\\
The flux of particles resulting from neutrino interactions in the vicinity of the detector is simulated with the GENHEN package. For the propagation of the Cherenkov light through the sea water, both light absorption and scattering are taken into account.

\section{Event selection criteria}
\label{sec:3}
The signal neutrinos can be discriminated from the background by their zenith angle (by only selecting events which were reconstructed as up-going close to the vertical direction, i.e. with zenith angle close to $180^\circ$) and energy (by not selecting events with reconstructed energy near or higher than the WIMP mass). This analysis relies on reconstruction algorithms providing direction and energy of the neutrino candidates, and on cuts defined to select neutrinos from the direction of the Earth center, produced by WIMPs of a given mass. For the muon direction, the  BBfit \cite{BBfit}, AAfit \cite{AAfit_1} and ZAV algorithms have been used. The latter is an algorithm for verifying the reconstructed zenith angles of the former. It was designed specifically for this analysis, where all signal events reach the detector from roughly the same direction (close to the nadir). It is based on the examination of the measured light pulses and to the comparison to that expected from an up-going, vertical muon.\\
Two analysis chains were used. The first, BBchain, uses BBfit as its main method of zenith reconstruction and is more suitable for lower WIMP masses with softer annihilation channels, the second, AAchain, uses AAfit. Each analysis chain consists of several event selection criteria, derived from either BBfit, AAfit or ZAV. The selection is based on the reconstructed zenith angles of both strategies, the angular error estimate, the fit qualities, the brightness of the events in terms of its position in the detector to avoid background from edge effects. For both analysis chains, the cut parameters have been tuned individually. The event selection criteria are optimized with the approach for unbiased cut selection for optimal upper limits presented in \cite{MRF}. The WIMP annihilation rate $\Gamma(t)$ is used as scaling parameter of the source flux. The optimization is done individually for each annihilation channel and several WIMP masses in the considered mass range.\\
For higher WIMP masses harder zenith angle cuts and looser energy cuts are expected. As an example, the optimized values of the AAfit zenith angle cut $\theta_{AA,cut}$ versus WIMP mass and annihilation channels are shown in Figure \ref{fig:1}.\\
The expected background neutrino events according to simulations are shown in Figure \ref{fig:2}. Due to the limited statistics in the Monte Carlo simulations, the expected background muons events according to simulations were always 0. The structures on Figure \ref{fig:2} depend on the fact that for each mass bin, the set of cuts on the parameters defined in section \ref{sec:3} allow a different number of background events. 

\begin{figure}[t]
\begin{minipage}[t]{.45\textwidth}
\centering
\includegraphics[width=1\linewidth]{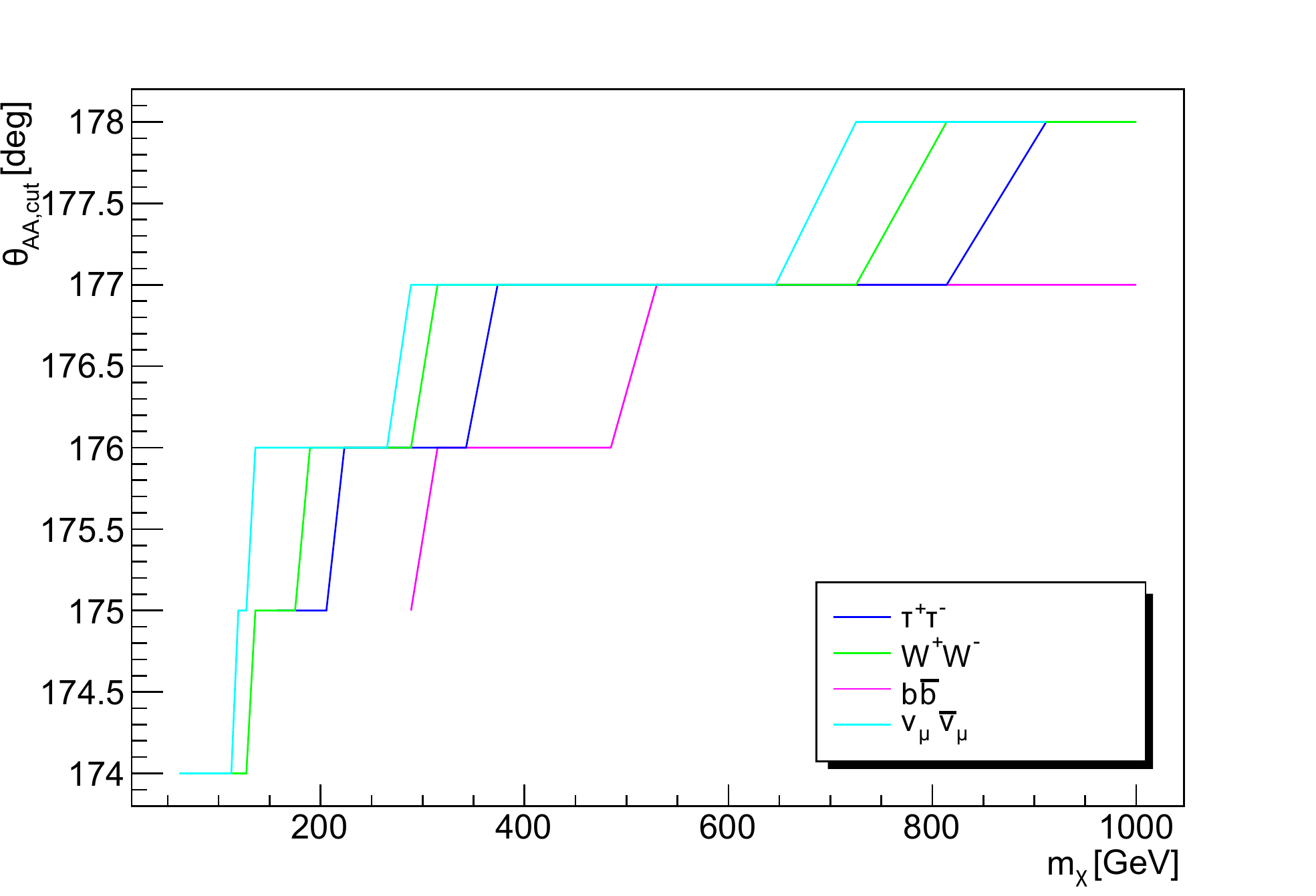}
\caption{Optimized value of $\theta_{AA, cut}$ in dependency of the WIMP mass and annihilation channel.}
\label{fig:1}
\end{minipage}%
\hfill 
\begin{minipage}[t]{.45\textwidth}
  \includegraphics[width=1\linewidth]{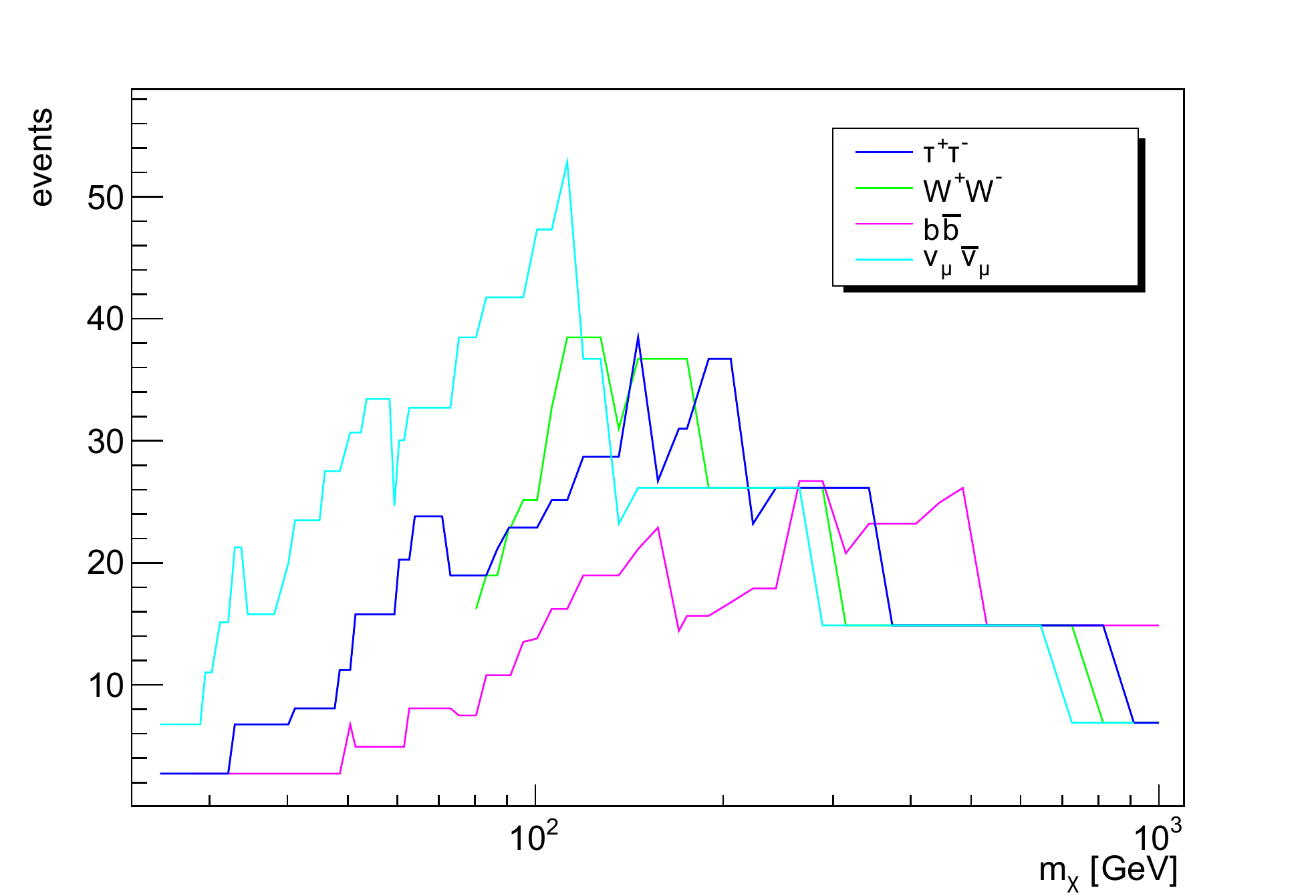}
  \caption{Number of atmospheric neutrinos expected according to simulations in dependency of the event selection criteria optimized for different WIMP masses and annihilation channels.}
  \label{fig:2}
\end{minipage}
\end{figure}

\section{Results}
After the aforementioned blinded optimization, the ANTARES data collected from 2007 to 2012 (corresponding to a livetime of 1191 days) were analysed. For each set of the cut parameters, defined in the optimization procedure for each WIMP mass interval and decay channel, the number of data events was determined. These numbers are shown in Figure \ref{fig:data}.\\
In the comparisons of the data in Figure \ref{fig:data} and the background of Figure \ref{fig:2}, no significant excess of events was observed. In particular, as shown in \cite{ant34}, the overall normalization factor for atmospheric neutrinos yielding the background given in Figure \ref{fig:2} must be increased by a factor about $25\%$. The no observation of an excess can be translated to a $90\%$ CL upper limit on the WIMP annihilation rate in the Earth (1.1). $90\%$ CL. upper limits on $\Gamma$ were calculated with the TRolke module from ROOT \cite{Trolke}, where uncertainties in the background and efficiency are considered with a fully frequentist approach \cite{Trolke} with the profile likelihood method \cite{rolke}. As a first step, a $90\%$ CL. event upper limit $\mu_{90,R}$ was calculated. Then, the limit on $\Gamma$ was then calculated as:
\begin{equation}
 \Gamma_{90}=\frac{\mu_{90,R}}{n_s}\cdot \Gamma_{0}
\end{equation}
Where $\Gamma_{0} = 1 \unitx{s^{-1}}$ and $n_s$ is the number of signal events expected for this experiment and for $\Gamma=\Gamma_{0}$.\\
It was assumed that the signal follows a Poisson distribution and that the background and efficiency can be modelled as gaussian. A systematic uncertainty of $15\%$ on the efficiency was assumed (following the studies in \cite{point_source}); a systematic uncertainty of $30\%$ was assumed for the atmospheric neutrino background (compare with \cite{point_source} and \cite{nu-uncertainty}). For the treatment of the atmospheric muon background, the most conservative approach (yielding the highest upper limit) was chosen by assuming that the atmospheric muon expectation is always 0. See Figure \ref{fig:arlimit}.\\

The limits on the annihilation rate are the main result of this analysis and the limits on $\sigma^{SI}_p$ were calculated from this result using \cite{SDM}\cite{SRN}\cite{WSC2}.
\begin{figure}
\centering
\begin{minipage}{.45\textwidth}
  \centering
  \includegraphics[width=1.0\linewidth]{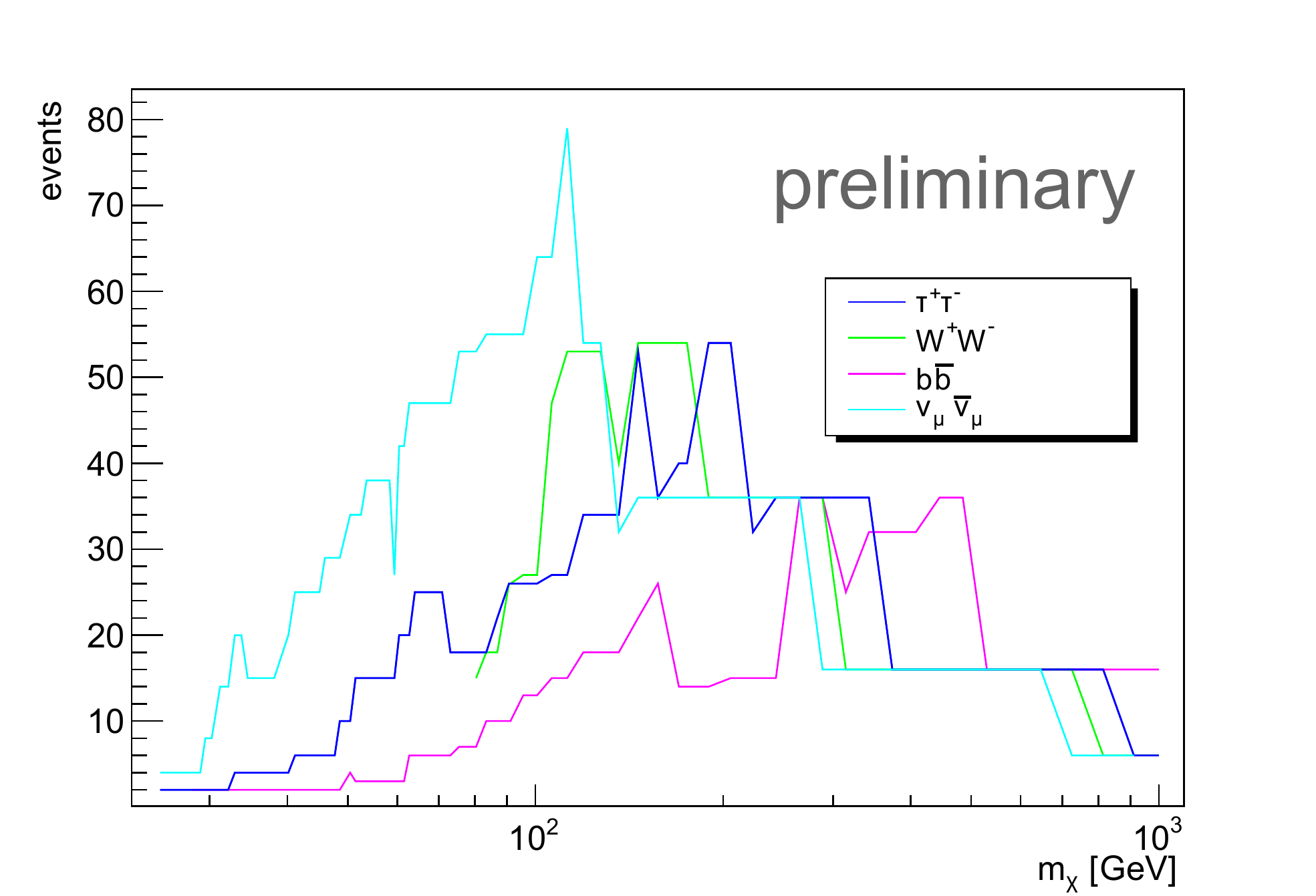}
  \caption{Number of events observed in dependency of the event selection criteria optimized for different WIMP masses and annihilation channels for ANTARES 2007 - 2011.}
  \label{fig:data}
\end{minipage}%
\hfill 
\begin{minipage}{.45\textwidth}
  \centering
  \includegraphics[width=1.0\linewidth]{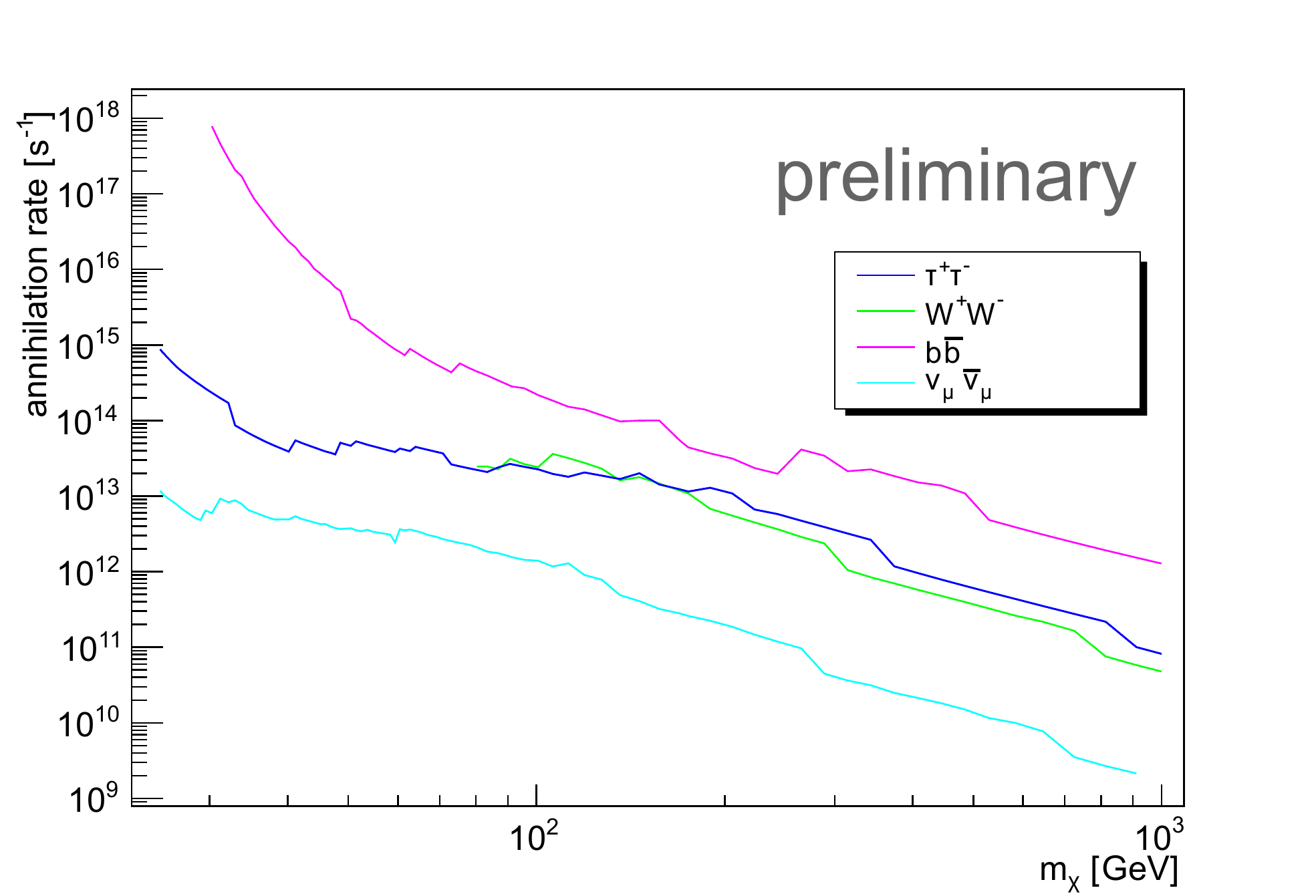}
  \caption{$90\%$ CL upper limits on $\Gamma$ as a function of the WIMP mass for WIMP pair annihilation to $100\%$ into either $\tau^{+}\tau^{-}$,  $W^+W^-$, $b\overline{b}$ or  $\nu_{\mu}\bar{\nu}_{\mu}$, for ANTARES 2007 - 2012.}
  \label{fig:arlimit}
\end{minipage}
\end{figure}
The limits on $\sigma^{SI}_p$ are shown assuming that $<\sigma v>$ for dark matter in the Earth is the same as during the freeze out ($<\sigma v> = 3 \cdot 10^{-26} cm^3 s^{-1}$) and for the annihilation channels allowed in SUSY ($\tau^+\tau^-$, $W^+W^-$ and $b\overline{b}$). The results are shown as $\sigma^{SI}_p$ versus $m_\chi$ in Figure \ref{fig:limits_s}, in comparison to the limits from other indirect and direct dark matter searches. Compared to the results from other indirect dark matter searches. This search from center of the Earth yields more stringent limits for the WIMP mass range from about 40 to 70 GeV (the mass range for which the capture rate of WIMPs would be enhanced due to the composition of the Earth). For completeness, recent limits from direct searches are shown as well. See Figure \ref{fig:limits_s}.\\
Additionally it was considered that $<\sigma v>$ of DM in the Earth is enhanced (compared to its value during the freeze out) by a boost factor. Here the $\nu_\mu \overline{\nu_\mu}$ annihilation channel is also considered. The limits are shown as $\sigma^{SI}_p$ versus the boost factor on $<\sigma v>=3 \cdot 10^{-26} cm^3 s^{-1}$ for $m_\chi=52.5$ (for which capturing of WIMPs in the Earth would be strongly enhanced due to the composition of the Earth) and $407.65$ GeV, compared to the results from Lux \cite{lux_firstresults} (which provide the most stringent limits on $\sigma^{SI}$ so far). See Figure \ref{fig:limits_s_b}.\\
\begin{figure}
\centering
  \includegraphics[width=1.0\linewidth]{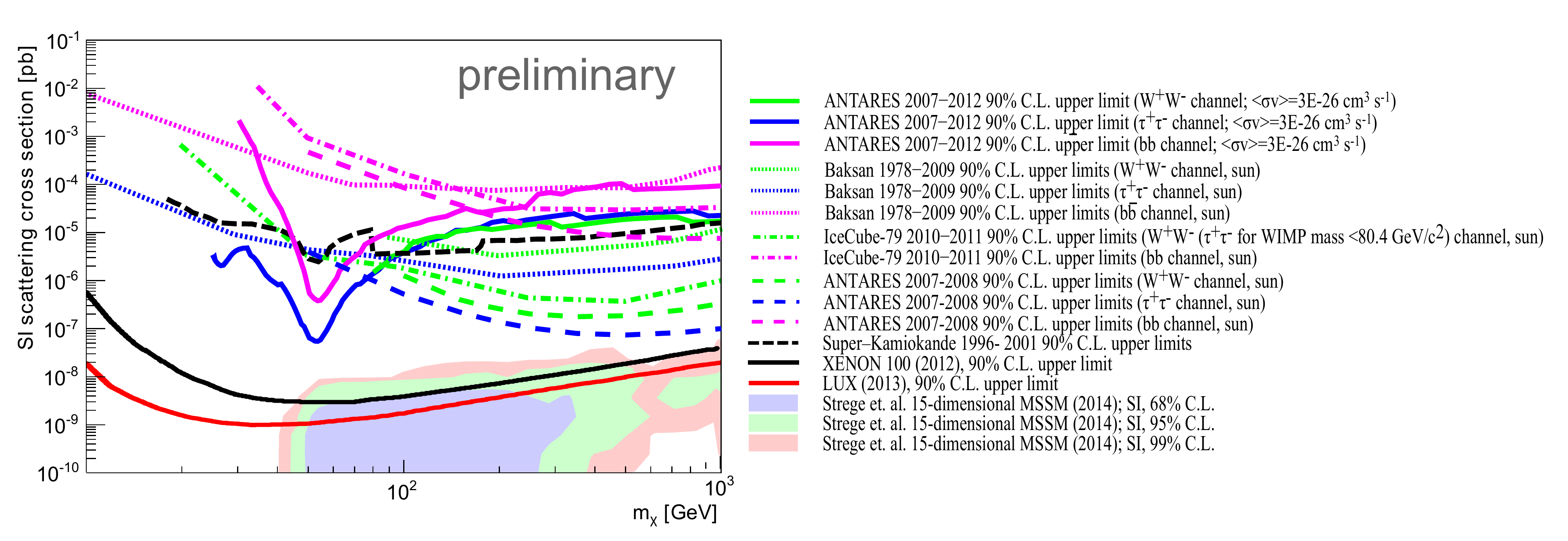}
\caption{$90\%$ CL upper limits on $\sigma^{SI}$ as a function of the WIMP mass for $<\sigma v> = 3 \cdot 10^{-26} cm^3 s^{-1}$ and WIMP pair annihilation to $100\%$ into either $\tau^{+}\tau^{-}$,  $W^+W^-$ or $b\overline{b}$, for ANTARES (Earth) 2007 - 2012, Baksan 1978 - 2009 \cite{Baksan} (from \cite{antares_dmsun}), IceCube-79 2010 - 2011 \cite{icecubesun} (from \cite{antares_dmsun}), Super-Kamiokande 1996- 2001 \cite{kamiokande}, ANTARES (Sun) 2007 - 2012 (preliminary), Xenon100 \cite{XENON100} and Lux \cite{lux_firstresults}. Also shown are the profile likelihood maps of a 15-dimensional MSSM from Strege et. al. \cite{strege}. Plot modified from \cite{plotter}.}
\label{fig:limits_s}
\end{figure}

\begin{figure}
\centering
\begin{minipage}{.5\textwidth}
  \centering
  \includegraphics[width=1.0\linewidth]{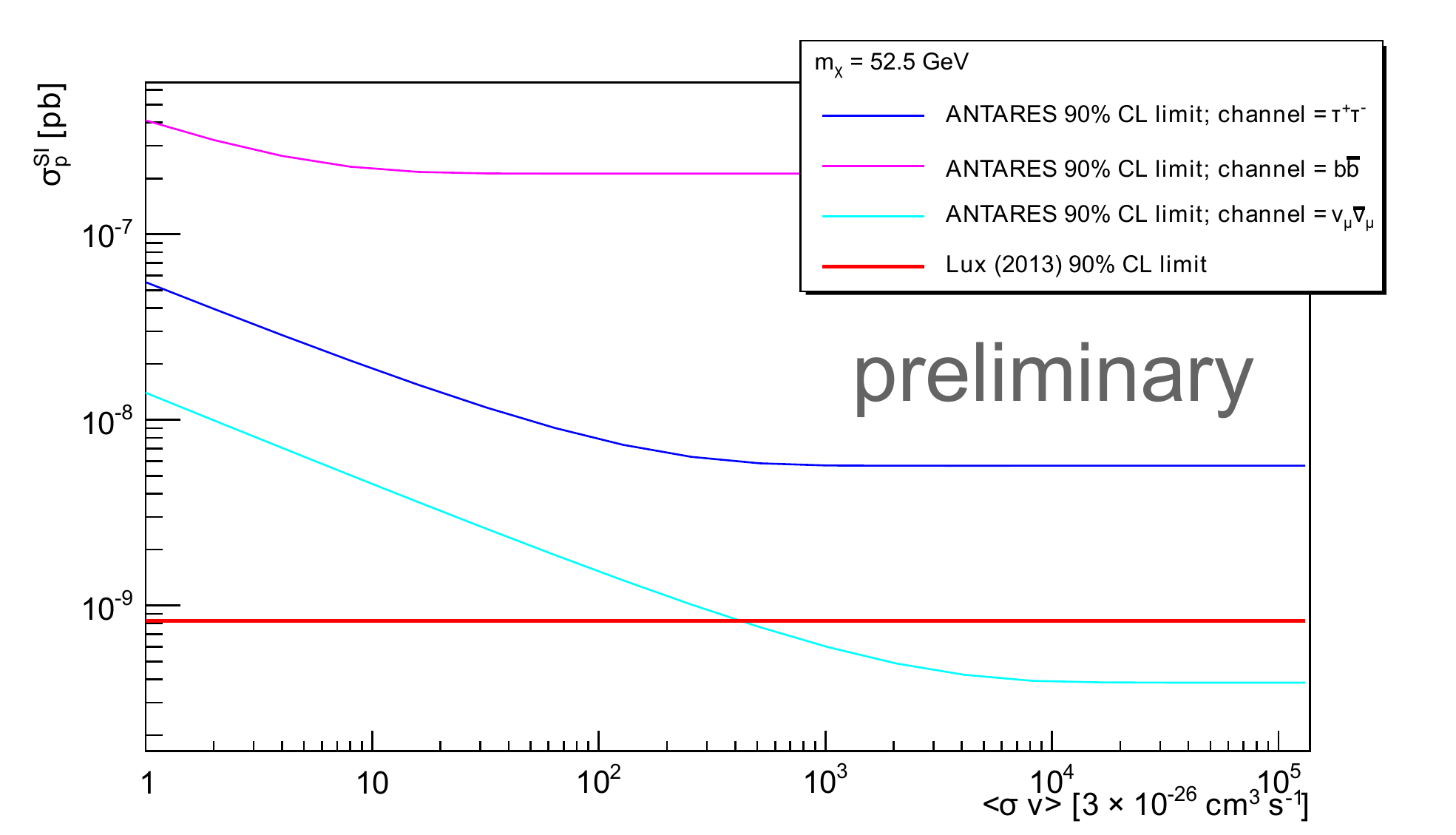}
  \label{fig:test1a}
\end{minipage}%
\begin{minipage}{.5\textwidth}
  \centering
  \includegraphics[width=1.0\linewidth]{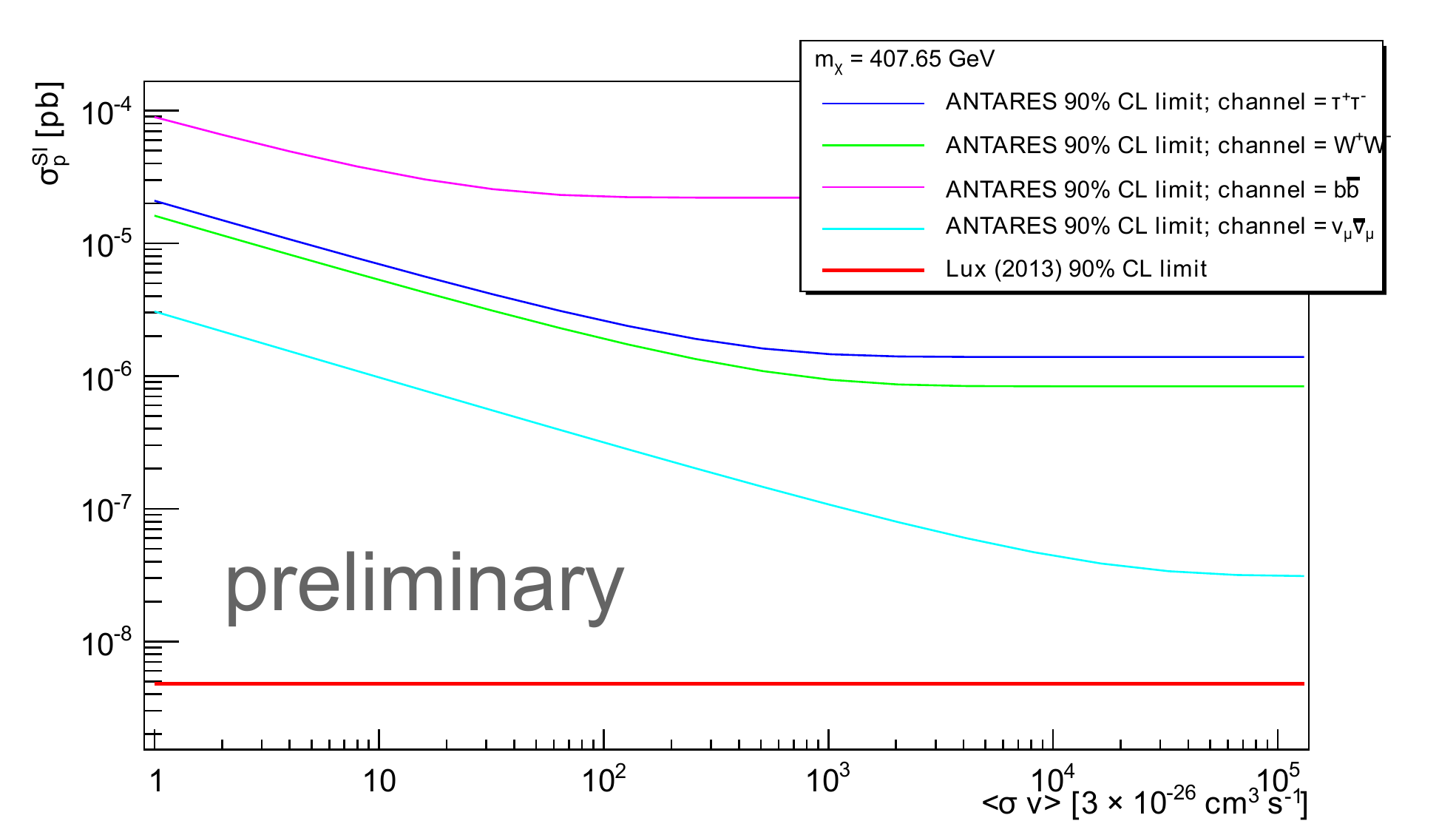}
  \label{fig:test2a}
\end{minipage}
\label{fig:limits_s_b}
\caption{$90\%$ CL upper limits on $\sigma^{SI}$ as a function of $<\sigma v>$ and WIMP pair annihilation to $100\%$ into either $\tau^{+}\tau^{-}$,  $W^+W^-$ or $b\overline{b}$, for a WIMP mass of $52.5$ GeV (left) and $407.65$ GeV (right), for ANTARES 2007 - 2011 and Lux \cite{lux_firstresults}.}
\end{figure}
Here the upper limits on $\sigma^{SI}_p$ decrease with increasing boost factor, until equilibrium would be reached. Assuming the WIMP would mainly annihilate into $\nu_\mu \bar{\nu}_\mu$ channel and $<\sigma v> \approx 1.5 10^{-23} cm^3 s^{-1}$, this search yields the so far most stringent limits on  $\sigma^{SI}_p$. It should however be noted that this scenario would not be possible if DM were mainly made up by SUSY particles or the lightest Kaluza-Klein particle.

\section{Conclusion and Outlook}
A search for dark matter from the center of the Earth has been performed with the data data collected from 2007 to 2012 by ANTARES neutrino telescope. No significant excess over the background expectation has been found. $90\%$ CL upper limits on the WIMP self annihilation rate were set as a function of the WIMP mass for WIMP pair annihilation to $100\%$ into either $\tau^{+}\tau^{-}$,  $W^+W^-$, $b\overline{b}$ or  $\nu_{\mu}\bar{\nu}_{\mu}$. These were translated to limits on the spin independent scattering cross section of WIMPs to protons. Here a scenario were the annihilation cross section for dark matter in the Earth is enhanced compared to the value during the freeze out of WIMPs was also considered. It could be demonstrated that the indirect search for dark matter towards the center of the Earth can be competitive with other types of dark matter searches, both direct and indirect. The discovery potential of such experiments strongly depends on the mass of the WIMP, its preferred annihilation channel and the thermally averaged annihilation cross section times velocity in the Earth today. A promising candidate for an improved future search is Km3Net \cite{km3net} with the ORCA extension \cite{orca}. 
\\



\setcounter{figure}{0}
\setcounter{table}{0}
\setcounter{footnote}{0}
\setcounter{section}{0}
\setcounter{equation}{0}
\newpage
\id{id_pavalas}
\addcontentsline{toc}{part}{\textcolor{blue}{\arabic{IdContrib} - {\sl G. Pavalas} : Search for nuclearites with the ANTARES neutrino telescope}%
}

\title{\arabic{IdContrib} - Search for nuclearites with the ANTARES neutrino telescope}

\shorttitle{\arabic{IdContrib} - Search for nuclearites with ANTARES }

\authors{Gabriela Emilia P\u{a}v\u{a}la\c{s}}
\afiliations{Institute of Space Science, 409 Atomistilor Street, Magurele, Ilfov, 077125, Romania}
\email{gpavalas@spacescience.ro}


\abstract{About thirty years ago, strange quark matter (SQM) was hypothesized to be the ground state of hadronic matter and was also suggested as a cold dark matter candidate. Although there is no experimental or astrophysical evidence yet for its existence, SQM may be present in the cosmic radiation as relic particles of the early Universe, or as fragments released in binary strange star collisions or supernovae.\
The ANTARES neutrino telescope is sensitive to massive and stable SQM particles, called nuclearites. Their velocity is assumed to be $\beta\sim10^{-3}$, typical of objects gravitationally trapped inside the galaxy. Nuclearites reaching the ANTARES depth would yield a large amount of light to the detector, by means of blackbody radiation emitted by the heated water molecules along their path. A dedicated analysis will be presented, as well as the ANTARES sensitivity for a flux of downgoing nuclearites, using data taken in 2009.}

%
%
\maketitle

\section{Introduction}

The nature of dark matter, representing about 85\% of the mass of the observable Universe, is an open question of today's physics. One of the hypothesized constituents of the dark matter is strange quark matter (SQM) \cite{Witten}, that may be present in cosmic radiation. Nuclearites are massive and stable lumps of SQM, composed of nearly equal numbers of up, down and strange quarks. They would interact with the ambient atoms by means of elastic or quasi-elastic collisions, displacing the atoms of matter in their path. Among relevant searches for nuclearites, the MACRO (Monopole, Astrophyics and Cosmic Ray Laboratory) experiment has set an upper limit for a downgoing nuclearite flux of $5.4\cdot 10^{-16}$ cm$^{-2}$ s$^{-1}$ sr$^{-1}$ (90\%  C.L.) \cite{Ambrosio}, for nuclearites of mass $M_{N}>10^{14}$ GeV, while the SLIM (Search for LIght Monopoles) high altitude experiment in Bolivia has set flux upper limits of $1.3\cdot 10^{-15}$ cm$^{-2}$ s$^{-1}$ sr$^{-1}$ (90\% C.L.) \cite{Cecchini}, for nuclearite masses $M_{N}>10^{10}$ GeV.

The ANTARES neutrino experiment, currently operating in the Mediterranean Sea, is also sensitive to the signal of non-relativistic nuclearites. These particles would yield a large signal inside the detector, by means of the blackbody radiation emitted along their path. Partial results on the search for nuclearites with ANTARES were reported in \cite{Gabi:2013}. 
In the following, this work describes the characteristics of nuclearites, presents the analysis performed on the selected ANTARES data and the preliminary results obtained.

\section{Nuclearites}

According to Witten \cite{Witten}, SQM lumps could be stable at zero temperature and pressure, knowing that in a three-flavor quark system, the Fermi energy and subsequently the mass of the quark bag are reduced by the third flavor added, compared to a two-flavor system (made of up and down quarks). Moreover, phenomenological models indicate that lumps of SQM are stable and metastable for a wide range of strong interaction parameters \cite{Jaffe:1984}. 
  
Nuclearites, heavy lumps of strange quark matter, would be electrically neutral; the small positive electric charge of the quark core would be neutralized by electrons, either in weak equilibrium inside the SQM, or forming an extended electronic cloud. Since direct nuclear interactions with the atoms they encounter are prevented by Coulomb repulsion, the relevant interaction mechanism of nuclearites is represented by elastic collisions \cite{Rujula}. The rate of energy loss is then given by:
  \begin{equation}
    \frac{dE}{dx} = -\sigma\rho{v^2},
    \label{en_loss}
   \end{equation}  
where $\rho$ is the density of the medium, $v $ is the nuclearite velocity and $\sigma$ its geometrical cross section:
 
   \begin{displaymath}
        \sigma = \left\{ \begin{array}{ll}
                        \pi(3M/4\pi\rho_N)^{2/3} & \mbox{for $M\geq8.4\cdot10^{14}$ GeV};\\
                       \pi\cdot10^{-16} \mbox{cm}^2 &  \mbox{for lower masses}.
                    
	             \end{array}
	             \right. 
	\label{sigma_cross}
     \end{displaymath}
with a SQM density $\rho_N=3.6\cdot10^{14}$ g cm$^{-3}$.

The propagation of nuclearites in sea water is described by the equation:
 \begin{equation}
    v(L) = v_0e^{-\frac{\sigma}{M}\int_0^L\rho dx},
    \label{velocity}
   \end{equation} 
where $\rho=1$ g cm$^{-3}$, and $v_0$ is the nuclearite speed at the Earth surface. The nuclearite collides with the atoms of water, giving them velocities of order $O (v_0)$. The temperature of the medium rises to $T\sim O (keV)$ and a hot plasma is formed that moves outwards as a shock wave. 
The luminous efficiency (defined as the fraction of dissipated energy appearing as light) was estimated, in the case of water, to be $\eta\simeq3\cdot10^{-5}$ \cite{Rujula}.

\section{Analysis}
For the present analysis, a selection of the ANTARES data from 2009 was made, considering runs that satisfy a set of quality criteria, and a well calibrated detector. The analysis complies with the {\it blinding policy} of ANTARES, that requires the optimization of the method by using simulated events and data-MC comparisons for a fraction of the selected data. For this search, a fraction equivalent to $\sim 13$ days of data acquisition was used, containing runs ending in ''0".

In what concerns the signal, nuclearite events were simulated with a dedicated Monte Carlo code, that is briefly described in the section below, with the following masses:  $10^{14}$,  $10^{15}$, $10^{16}$,  and $10^{17}$ GeV.
As for the physics background, downgoing atmospheric muons files simulated with the MUPAGE code \cite{Carminati} were used. 

Besides muons, bioluminescence is also present in the deep sea environment of the ANTARES detector. Bioluminescence background causes sporadic peaks in the singles rate of up to several MHz during periods of a few
seconds or less, that mimic at a certain extent the nuclearite signal. Short bursts can appear during data taking in relatively good conditions, on time scales of the order of a frame, i.e. $\sim 104$ ms. Programs that provide the count rate on a reduced time scale, the trigger type during particular frames and the event display are used to identify these bursts. Bioluminescence is usually localized in a region of the detector, and persists for the duration of one or more consecutive frames. 

The simulated nuclearite and atmospheric muon files were then processed with a program that uses the standard muon triggers and the charge thresholds corresponding to the considered period of operation. The optical background was added from each data run in the sample to the simulated hits from nuclearite and muon events, providing a so-called run by run simulation.

The standard muon triggers of ANTARES used in this analysis are the so-called directional trigger and cluster trigger. The directional trigger requires five local coincidences causally connected, within a time window of 2.2 $\mu$s. The cluster trigger requires two coincidences between two L1 hits\footnote{A local coincidence L1 is obtained when at least two L0 hits (hits with charge threshold > 0.3 photo-electrons) occur within 20 ns on two different PMTs in the same storey or when a large charge hit occurs. The threshold for large hits usually corresponds to 3 photo-electrons.} in adjacent or next-to-adjacent storeys. When a muon event is triggered, all PMT pulses are recorded over 4 $\mu$s in a {\it snapshot}. When two or more events have some overlapping hits, a merger of the events proceeds and a larger snapshot results.

In the following, the nuclearite simulation is briefly described, as well as the effects of the trigger processing on the simulated nuclearite events. 
Then, the reconstruction procedure and the selection conditions applied to the data and MC events are presented, followed by the preliminary results on the detector sensitivity.

\subsection{Nuclearite simulation}
Nuclearites are simulated with a Monte Carlo program, that includes the propagation of the nuclearites through the Earth's atmosphere and sea water, as well as the simulation of the expected signal at the detector level. The main assumptions of the nuclearite simulation are the following: isotropic flux above the Earth's atmosphere, galactic velocities of $\beta = 10^{-3}$ at the entrance in the atmosphere (50 km above sea level), the trajectory is a straight line, since the influence of gravity is negligible, the propagation in the atmosphere and sea water, the light yield at the detector level are calculated based on the phenomenological model proposed in \cite{Rujula}. 

Given that only nuclearites with masses larger than about $10^{22}$ GeV are able to cross the Earth, and that the nuclearite flux in cosmic rays is expected to be decreasing with increasing nuclearite mass (as for heavy nuclei), only downgoing nuclearites were considered in this analysis. 

A hemispherical volume of 548 m radius symmetrically surrounding the 12 line detector is used to generate and trace the nuclearites trajectories. The base of the hemisphere is placed on the sea bed, 100 m below the plane of the lowest ANTARES storeys, and with the pole on the ANTARES symmetry axis, 100 m above the plane of the highest storeys.
The entry point of the nuclearite trajectory is generated on the surface of the hemisphere, having the coordinates ($x_0, y_0, z_0$). The direction of the trajectory is then given by randomly generated zenith and azimuth angles. In order to simulate downgoing trajectories, a subroutine checks if the trajectory intersects the hemisphere at a point higher than the initial entry point; if true, the upper point becomes the entry point. Above the fiducial hemisphere, the path length of the nuclearite is computed by considering that the detector lies on a solid sphere with a radius equal to the Earth's radius, covered uniformly by a layer of water and by an outer layer representing the atmosphere. Both the propagation of nuclearites in the atmosphere and in the sea water are described by Equation \ref{velocity}.

In order to propagate the nuclearites through the hemisphere, the algorithm implements the energy loss mechanism presented in Section 2, and evaluates the position, the velocity $\beta$ and the number of hits on the OM in time steps of 2 ns. The procedure is repeated until the nuclearite optical energy loss (integrated over the time step) is lower than 3 eV, or the nuclearite reaches the sea floor.

Regarding the effect of the muon triggers on nuclearites, unlike the muon events that are encapsulated in one snapshot, most nuclearite events result in a series of connected snapshots of variable durations. The duration of a snapshot depends on the light yield of the particle, and on the distance to the closest optical module, ranging from muon-like snapshots ($\gtrsim 4.4 \mu$s) to large snapshots of up to few ms, produced by merging.
 
\subsection{Reconstruction and first level cuts}

The reconstruction of nuclearite trajectories uses the charge barycenter distribution as a function of time of the hits. Since the light emitted by nuclearites is isotropic, the charge barycenter gives an estimate of the position of the source at a certain moment. In case of a nuclearite passing through the detector, the displacement of the charge barycenter would indicate a downgoing track with a speed less than $10^{-3} c$. The procedure consists in retrieving the time, charge and position of OMs for the hits of each event, and distribute them in time histograms of 500 ns bins. All hits with the charge $q>0.3$ p.e. are considered in the distributions. The time histograms of the charge barycenter projected on every axis are obtained from the following ratio,  computed on each 500 ns bin: 
\begin{equation}
\frac{\sum q_i \cdot pos_i}{\sum q_i},
\end{equation}
where $pos_i=x,y,z$ is the position of the OM where the signal is detected and $i=1,2..,n$ is the number of hits in each bin.

The trajectory of the nuclearite is assumed to be linear, therefore the evolution in time of the charge barycenter distributions will be approximated by a straight line. The partial mean velocities $v_x, v_y, v_z$ at the detector level, as well as their errors, are determined from linear fits of the charge barycenter distribution on each axis as a function of time. 
Then, the total velocity and the corresponding uncertainty are obtained in a straightforward manner. The zenith angle and its uncertainty are determined as follows:
\begin{equation}
\theta=arccos(v_z/v), \\
\end{equation}
\begin{equation}
d\theta=\frac{1}{\sqrt{1-(\frac{v_z}{v})^2}}\cdot\sqrt{(\frac{dv\cdot v_z}{v^2})^2+(\frac{dv_z}{v})^2}.\\
\end{equation}

\begin{figure*}[!ht]
\centerline{\includegraphics[width=3.5in]{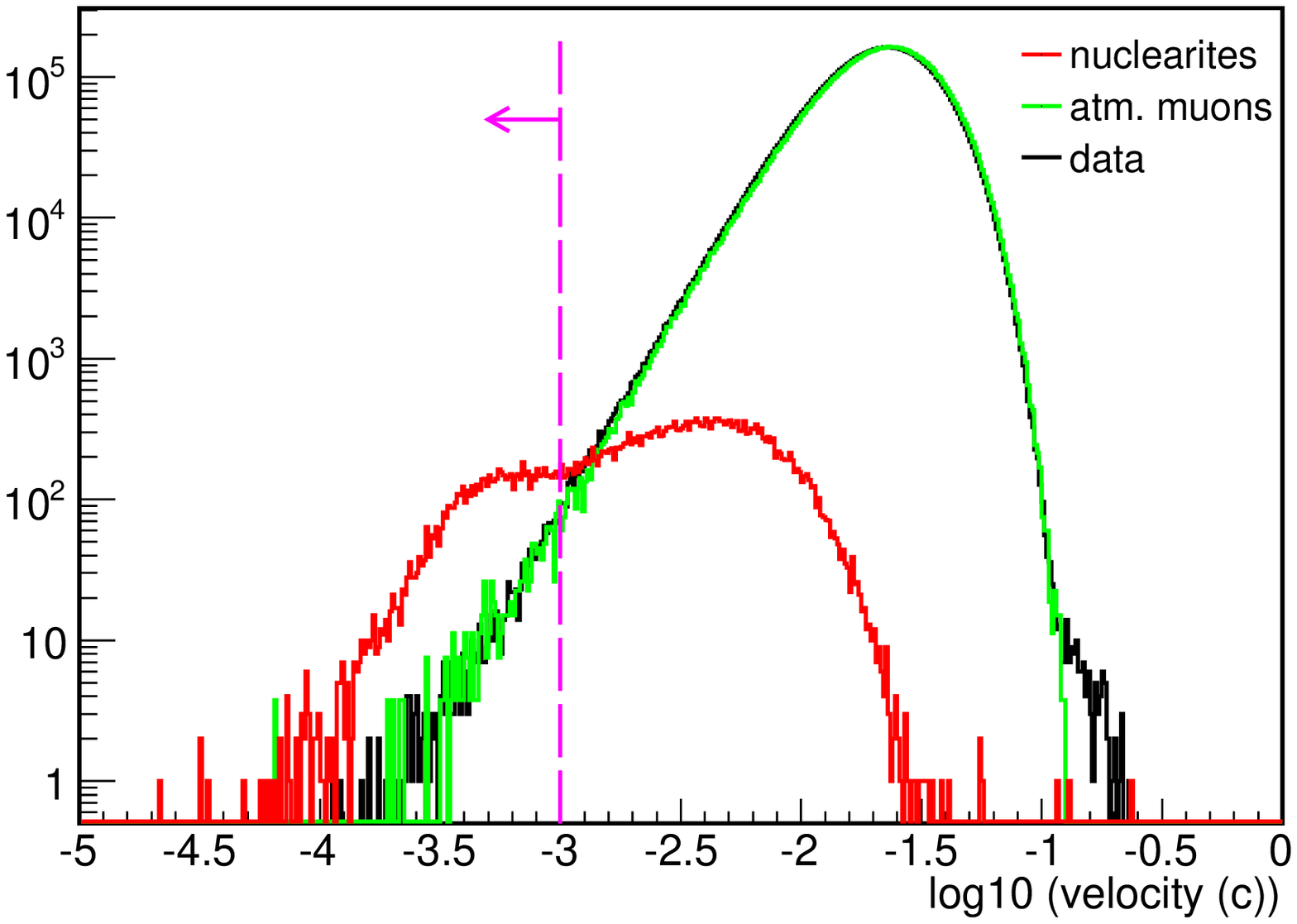}
	\hfill
         \includegraphics[width=3.5in]{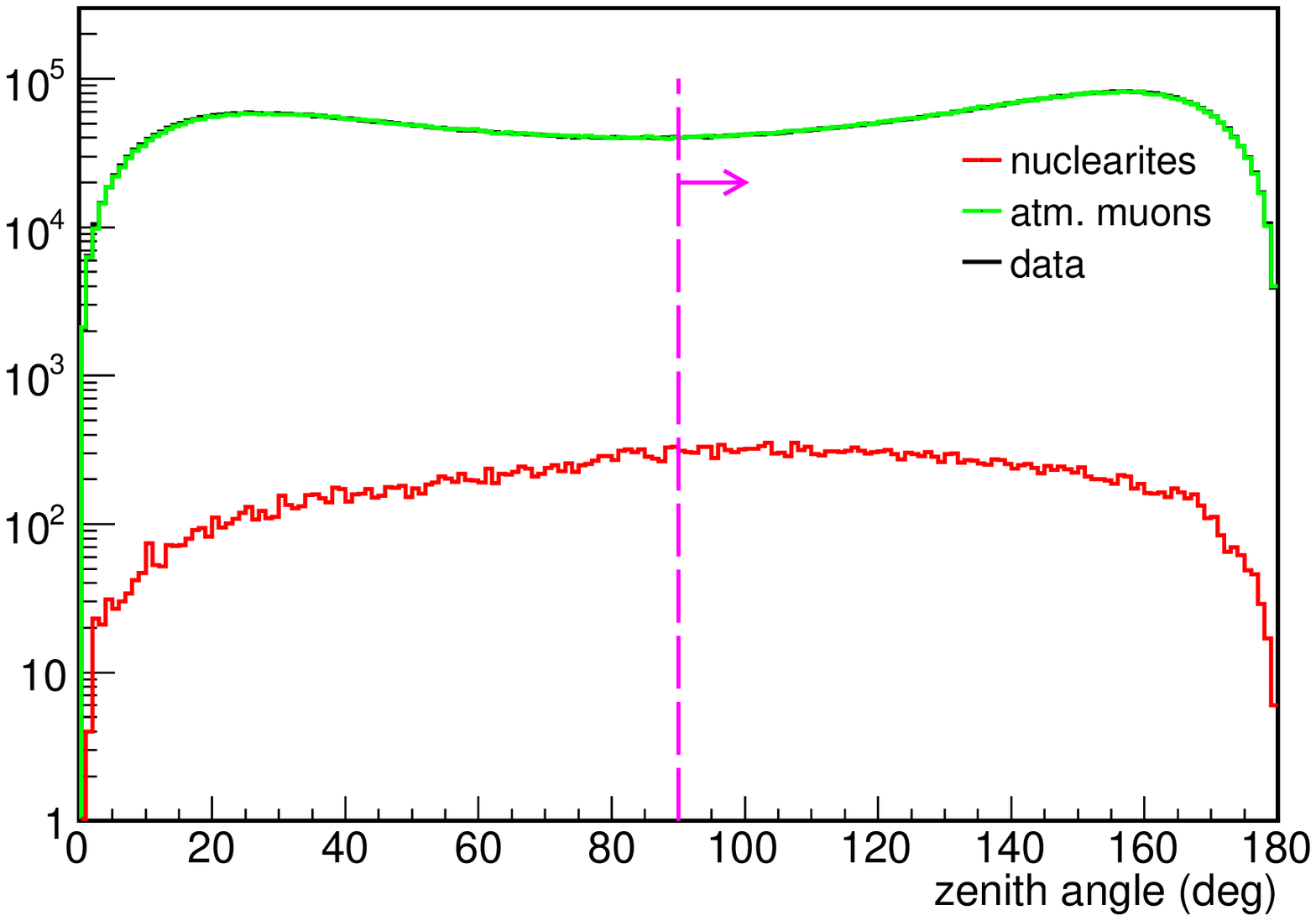}
         }
	\caption{Left: Reconstructed velocity distributions for MC nuclearite snapshots, MC muons and data sample, with first level cut corresponding to $v<10^{-3}c$. Right: Reconstructed zenith angle distributions with cut represented by the vertical line.}
   \label{fig_reco_velocity}
 \end{figure*}

The reconstruction procedure described above was applied to the selected data sample, to MC muons and nuclearite snapshots. 
The distributions of logarithmic reconstructed velocity and of zenith angle for data, MC muons and nuclearite snapshots are shown in Figure \ref{fig_reco_velocity}, with MC muon sample normalized to data.
In what concerns data-MC comparison, a reasonable agreement is observed in the velocity distribution, except for the right tail of the distribution, where an excess of events is seen in data. In the low velocity region ($v<10^{-2}c$), contribution from bioluminescence is also expected. The data-MC agreement of the zenith angle distribution is good.
First level selection conditions are then defined, requiring a reconstructed velocity $v<10^{-3}c$ and a zenith angle $\theta>90^{\circ}$, consistent with the expected characteristics of nuclearite events, i.e. non-relativistic velocities and downgoing directions.

\subsection{Second level cuts}

For the next step of the analysis, a number of discriminants was studied in order to select, from the snapshots surviving the first level cuts, the ones that might be part of a nuclearite event. These discriminants were the duration of snapshot, the number of L0 hits, and the number of L1 hits. The best discriminant for this analysis proved to be the number of L0 hits. The distributions of the logarithmic number of L0 hits for nuclearite, muon and data snapshots surviving the first level cuts  are shown in Figure \ref{double_fig_opt}, on the left-hand side, with MC muons normalized to data distribution. At this stage, muon and data distributions do not agree well, since several snapshots with large values are observed only in data. These snapshots were found to belong to several frames in two runs, 39360 and 39680, and are shown in Figure \ref{double_fig_opt} with blue line. They were investigated with tools dedicated to the bioluminescence identification, described at the begining of Section 3. The investigation indicates a bioluminescence origin of these snapshots. The number of all snapshots (including the ones triggered by one-dimensional selection algorithms) found in these frames is greater than the one usually seen in the quality runs, as shown in the right-hand side of Figure \ref{double_fig_opt}. In order to reject the suspicious snapshots, a selection cut at 200 snapshots, denoted C2a, was applied to the number of snapshots found in frames of data, as well as to the number of snapshots produced by nuclearite events. 

After the removal of the noise, the muon and data distributions are in a reasonable agreement. In order to obtain the best sensitivity for the detector, the cut on the number of L0 hits was optimized. The best sensitivity is obtained by minimizing the so-called Model Rejection Factor \cite{Hill}, MRF=$\frac{\overline{\mu}_{90\%}(n_b)}{n_s}$, where $\overline{\mu}_{90\%}$ is the "average upper limit" that would be observed by an ensemble of hypothetical experiments with no true signal and expected background $n_b$. The  $\overline{\mu}_{90\%}$ factor is taken from the Feldman-Cousins tables \cite{Feldman}. The expected background $n_b$ was determined  from the extrapolation of the L0 hits distribution of MC muons, normalized to the data distribution, while $n_s$ is given by the number of nuclearite events surviving the cuts.

\begin{figure*}[!ht]

\centerline{{\includegraphics[width=3.2in]{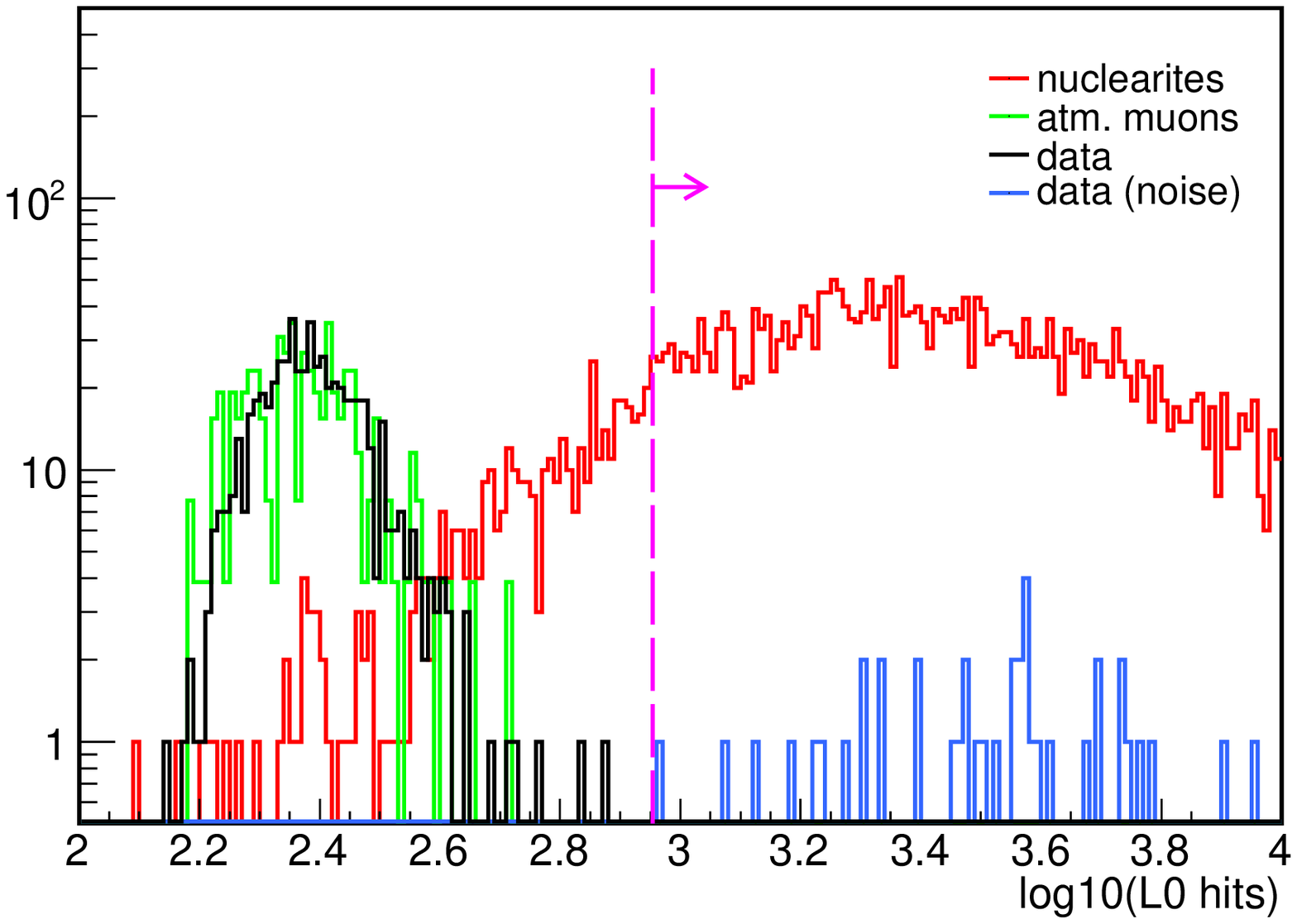}\label{L0}}
\hfill
{\includegraphics[width=3.2in]{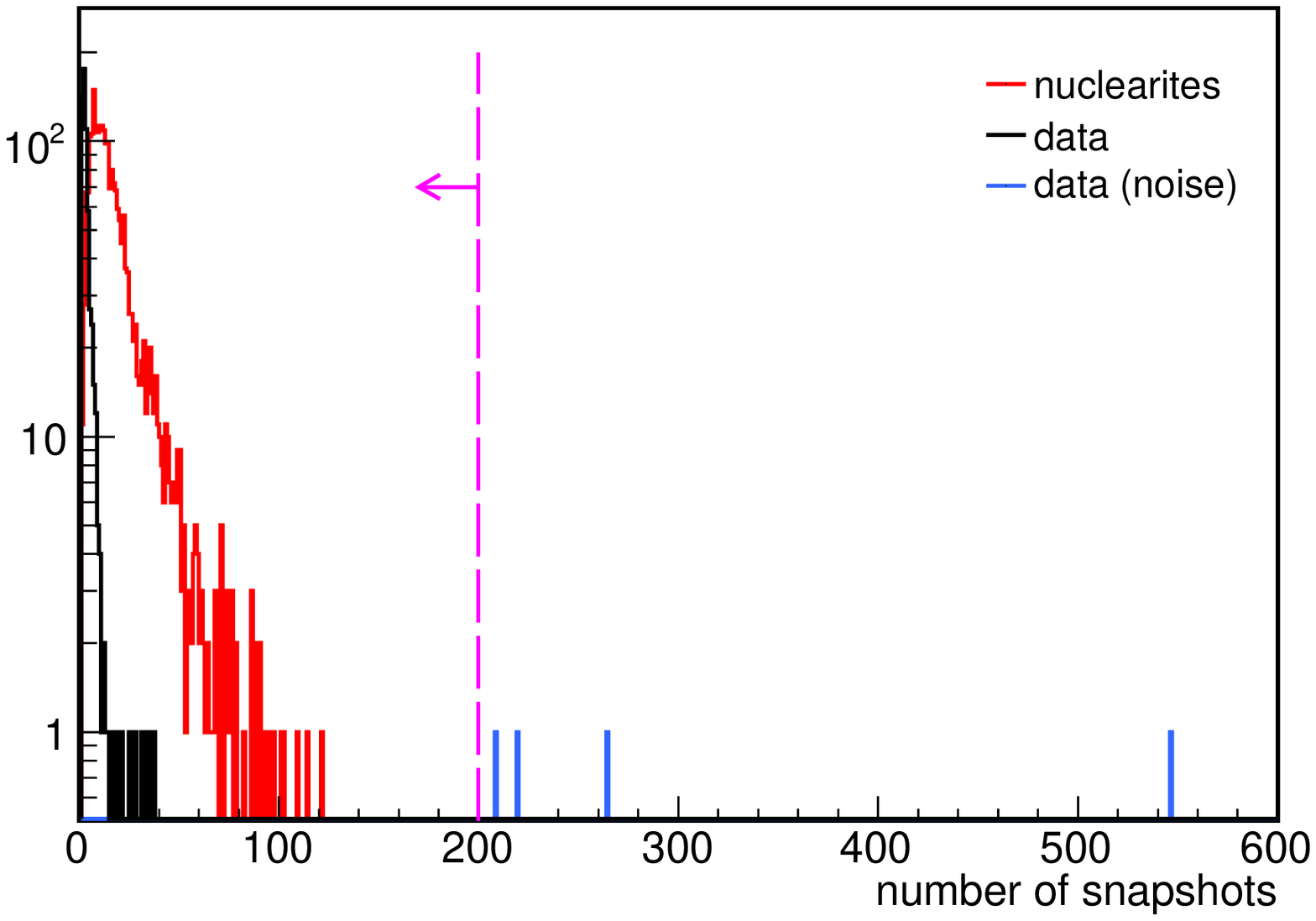} \label{sub_optim}}
}
\caption{Left: Distribution of the number of L0 hits for MC nuclearites, muons and data snapshots. Snapshots with large values in data are shown with blue line, and are rejected by the C2a cut presented in the right-hand side plot. The optimized C2b cut on the number of L0 hits is shown with a vertical dashed line. Right: Distribution of snapshots per event for nuclearites, and snapshots per frame for data; noisy frames are represented with blue line. A selection cut at 200 snapshots (C2a) rejects the bioluminescence contribution.}
\label{double_fig_opt}
\end{figure*}

The value of the cut on the number of L0 hits, denoted C2b, is chosen for the minimum MRF obtained for nuclearites. The selection condition requires that the number of L0 hits in a snapshot to be larger than 900, as shown in the left-hand side of Figure \ref{double_fig_opt}.
After applying the second level cut, no MC muon or data snapshots survived. 

As a final step in the candidate event identification, the surviving snapshots were used to look for other snaphots around them in a time interval of $\sim1$ ms, i.e. the time a particle of velocity $\beta\simeq10^{-3}$ crosses the detector. If found, the sequences of snapshots are reconstructed as events.

\section{Results}

The results of the cuts applied to the MC nuclearite and muon samples, as well as to data sample are shown in Table \ref{table_res}. 

\begin{table}[!h]
\footnotesize
\tabcolsep=0.11cm
  \centering
  \begin{tabular}{l c c c c c}
  \hline
   sample  & snapshots & after & after & after & reconstructed\\
   & & C1 cuts &  C2a cut & C2b cut & events\\
   \hline 
   nuclearites & 36403 & 5626 &  5626 & 5190 & 2254  \\
   MC muons & 2431379 & 152 & 152 &  0.0065 & 0\\
   data  &  9135988 & 628 & 587 & 0 & 0\\
   \hline
  \end{tabular}
  \caption{The number of snapshots in each sample, as well as the remaining snapshots after the first level (C1) cuts and second level (C2a and C2b) cuts were applied to the data and MC samples, are given. The last column shows the number of reconstructed events remaining in the studied samples.}
  \label{table_res}
  \end{table}
 
In order to calculate the detector sensitivity to nuclearites, the Feldman-Cousins prescription \cite{Feldman} was used,
considering events with a Poisson distribution:
\begin{equation}
\phi_{90}=\frac{\overline\mu_{90\%}}{A\times T},
\label{flux}
\end{equation}
where $A$ is the detector acceptance, and $T$ the live time. 

The effective acceptance $A$ of ANTARES to a downgoing flux of nuclearites is computed for each simulated mass as follows:
\begin{equation}
A = S\times \frac{N_{nucl}}{N_{sim}},
\label{acceptance}
\end{equation}
where $S$ is the area of the simulation hemisphere and $N_{nucl}/N_{sim}$ is the ratio of the number of nuclearite events that passed the selection cuts to the number of simulated events. 

The sensitivity expected from the analysis of $\sim 159$ days of data taken in 2009 is shown in Figure \ref{sensit}. The ANTARES preliminary sensitivity is compared with previous limits from the MACRO \cite{Ambrosio} and SLIM \cite{Cecchini} experiments and with the ANTARES upper limits obtained from the analysis of data taken in 2007 and 2008 \cite{Gabi:2013}.
\begin{figure}
\centering
\includegraphics[width=.5\textwidth]{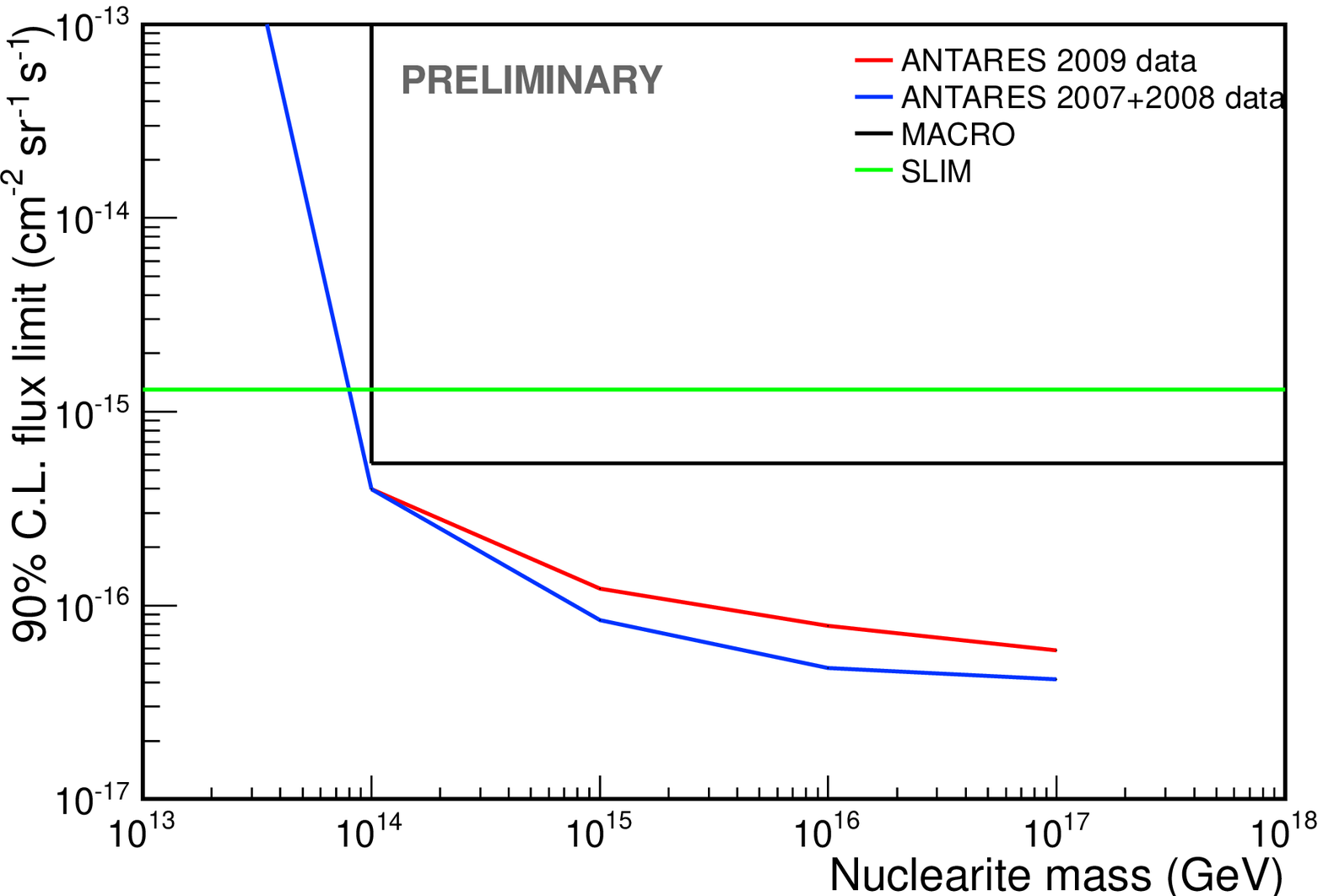}
\caption{Sensitivity of ANTARES to a flux of downgoing nuclearites, using 159 days of data taken in 2009.}
\label{sensit}
\end{figure}

\section{Conclusions}
A new analysis was developed for the search of nuclearites with the ANTARES detector, using data taken in 2009. While most of the background was removed after applying cuts on the reconstructed track parameters, hints of bursts of high bioluminescence activity were observed in the remaining data events. With these events rejected based on the noise level in  their related frames, a final optimized cut allowed to reject the background and calculate the detector sensitivity.  Preliminary results are comparable to the upper limits for a downgoing nuclearite flux, obtained in a previous analysis of the 2007 and 2008 data \cite{Gabi:2013}. Further improvement of the sensitivity and upper limits can be achieved by extending the search to the next years of ANTARES data.


\setcounter{figure}{0}
\setcounter{table}{0}
\setcounter{footnote}{0}
\setcounter{section}{0}

\newpage
\id{id_ctoennis}
\addcontentsline{toc}{part}{\textcolor{blue}{\arabic{IdContrib} - {\sl C. T\"onnis} : The indirect search for dark matter with the ANTARES neutrino telescope}%
}

 
 \title{\arabic{IdContrib} - The indirect search for dark matter with the ANTARES neutrino telescope}

\shorttitle{\arabic{IdContrib} - A summary talk on the dark matter searches with ANTARES}

\authors{Christoph T\"onnis}
 \afiliations{ IFIC - Instituto de F\'{\i}sica Corpuscular, \\
Universitat de Val\`encia--CSIC,\\ E-46100 Valencia, Spain}
\email{ctoennis@ific.uv.es}


\abstract{The indirect search for dark matter is a topic of utmost interest in neutrino telescopes. The ANTARES detector is located at the bottom of the Mediterranean Sea 40 km off the southern french coast. Results of the indirect searches for dark matter self-annihilation signals from different potential sources, including the Sun and the Galactic Center, produced with different analysis methods are presented. The specific advantages of neutrino telescopes in general and of ANTARES in particular will be explained. As an example, the indirect search for Dark Matter towards the Sun performed by neutrino telescopes currently leads to the best sensitivities and limits on the spin-dependent WIMP-nucleon cross section with respect to existing direct detection experiments.}

%
%
\maketitle

\section{Introduction}
\label{sec:1}

One of the concepts in the indirect search for dark matter is to look for annihilations of WIMPS in massive celestial objects. WIMPS can accumulate in these celestial objects due to gravitational capture or due to the formation of dark matter halos in the early universe ~\cite{indirectdm}. The annihilations of those WIMPS can produce standard model particles which can produce photons and neutrinos in secondary processes. These neutrinos and photons can then be detected in different experiments. 

In this paper, the results of the ANTARES neutrino telescope ~\cite{1ANTARES} on the searches for neutrinos from the center of the Milky Way, of the Sun and on dwarf galaxies are presented. In the following, neutrinos stands for both neutrino and antineutrinos. The search for dark matter in the Earth is presented in another contribution ~\cite{AG}. 

In the case of extended sources, as our Galaxy, galaxies and galaxy clusters as possible sources the so called J-Factor has to be calculated. The J-Factor is necessary to relate the neutrino signal flux to the thermal averaged annihilation cross section, which is a parameter that depends on the actual dark matter model employed and is customarily used to express the sensitivities and limits of the experiments, both direct and indirect, for the sake of comparison. The J-Factor is the squared dark matter density integrated along the line of sight, and can be calculated with the formula:

\begin{equation}
\rm J(\theta) =\int\limits_{0}^{l_{max}}\frac{\rho^2_{DM} \left( \sqrt{R_{SC}^{2}-2lR_{SC} \cos \left( \theta \right) + l^2} \right)}{R_{SC} \rho^2_{SC,DM}} dl
\end{equation}

\noindent $\rm R_{SC}$ is the scaling radius of the halo and $\rm \rho_{SC,DM}$ is the scaling density. The J-Factor then relates

\begin{equation}
\rm \frac{d \phi_{\nu}}{dE} = \frac{<\sigma v>}{2} J_{\Delta \Omega} \frac{R_{SC} \rho^2_{SC}}{4\pi m^2_\chi} \frac{ d N_{ \nu }}{dE}
\end{equation}

\noindent where $\rm J_{\Delta \Omega}$ is the J-Factor integrated over the observation window $\Delta \Omega$, $\rm m_\chi$ is the WIMP mass and $\rm  \frac{ d N_{ \nu }}{dE}$ is the expected signal neutrino spectrum. The dark matter halo profile $\rm \rho_{DM}$ is fitted to measurement data as for example the distribution of rotational velocities of stars in the galaxy in question. For the profile of the Dark Matter halo the NFW function is used ~\cite{clumpy}:

\begin{equation}
\rm{\rho(r) = \frac{\rho_{s}}{(r/r_{s})(1+r/r_{s})^{2}}} 
\label{nfwprofile}
\end{equation}     

\noindent with $\rm r_{s} = 21.7$ kpc. The normalisation of the profile density, $\rm \rho_{s}$, is computed by fixing the dark matter density at the Sun's position $\rm \rho(r_{Sun} = 8.5\,kpc) = 0.4\,GeV \cdot cm^{-3}$. 

In the case of the Sun a different approach has to be chosen to calculate sensitivities and limits in terms of dark matter model parameters. In this case, it is assumed that there is an equilibrium between the gravitational capture of WIMPS by their scattering with the solar plasma and the annihilation of WIMPS in the Sun. If the average number of neutrinos per WIMP annihilation is known, the total neutrino flux can be related to the total annihilation rate in the Sun, which is proportional to the capture rate. This capture rate can be expressed as ~\cite{Gould}:

\begin{equation}
\rm C_s = 3.35 \frac{1}{s} \left( \frac{\rho_{loc}}{0,3 \frac{GeV}{cm^3 }} \right) \left( \frac{270 \frac{km}{s}}{v_{rms}} \right)^3  \left( \frac{\sigma_{H,sd}+\sigma_{H,si}+0,07 \sigma_{He,si} }{10^{-6}pb} \right) \left( \frac{100GeV }{m_\chi} \right)^2
\end{equation}

\noindent The different $\rm \sigma$ are the spin-dependent and spin-independent cross section for the scattering of WIMPS with hydrogen and helium, $\rm v_{rms}$ is the root mean squared velocity of the WIMPS in the galactic halo at the Sun position and $\rm \rho_{loc}$ is the local dark matter density. 

In the analyses presented here some representative annihilation channels of the WIMPs have been chosen in order to stay model independent. For each annihilation channel a $100\%$ branching ratio for the WIMP annihilation directly into a specific pair of standard model particles is assumed. The following channels have been used:

\begin{align}
\rm WIMP + WIMP \to & b + \bar b,  W^+ + W^-, \tau^+ +\tau^-, \mu^+ + \mu^-, \nu + \bar \nu \label{bb}
\end{align}

The $\nu \bar \nu$ and $\mu^+ \mu^-$ channel have not been considered for the search for WIMP annihilations in the Sun. The $\tau^+ \tau^-$ channel is most commonly used as a benchmark for comparisons between experiments ~\cite{ANTARES_GC}. For the Sun additional effects have to be taken into account. These effects are the absorption of neutrinos and the regeneration of tau neutrinos in the solar plasma ~\cite{cirellisun,WIMPSIM}.

The ANTARES detector has obtained different limits on the flux of neutrinos from astrophysical objects. In Section 2 the result from an "unbinned" search method from the direction of the Sun is presented. 
In Section 3 a "binned" method is used for the searches for an excess of neutrinos from the direction of the Galactic center and from dwarf galaxies. In the "unbinned" method, the sensitivities and upper limits are constructed using a likelihood function. This likelihood function can be written as:

\begin{equation}
\rm log_{10}(L(n_s)) =   \sum_{i = 1} ^{N_{tot}}log_{10}\left( n_s S(\psi_i,p_i,q_i) +N_{tot}B(\psi_i,p_i,q_i) \right) - n_s -N_{tot} \label{loglik}
\end{equation}

\noindent $\rm N_{tot}$ is the total number of reconstructed events, $n_s$ is the supposed number of signal events,$\rm \psi_i$ is the angular position of the $\rm i_{th}$ event, $\rm p_i$ and $q_i$ are additional event parameters like the reconstruction quality or the estimated neutrino energy. S represents the ANTARES point spread function (PSF) for the signal and B is a function that represents the behavior of the background. 

This likelihood function is then used to analyse pseudo experiments. A pseudo experiment is a sky map filled with simulated background events, generated from a background estimate and a given number of fake signal events, using the PSF and the signal statistics. For each pseudo experiment the likelihood function is optimized with respect to $\rm n_s$. A parameter called the test statistics (TS) is then calculated as:

\begin{equation}
\rm TS =  log_{10}\left(\frac{L(n_s)}{L(0)}\right)
\end{equation}

\noindent The sensitivities in terms of detected signal events $\rm \mu_{90\%}$ are calculated from the overlap of the distribution of TS values for different numbers of inserted fake signal events. Upper limits on the number of signal events are then calculated comparing the TS value of the actual data to the TS distributions of pseudo experiments. 

The sensitivities and limits are then converted to neutrino fluxes using a quantity referred to as acceptance. The acceptance is defined as:

\begin{equation}
\rm Acc(m_{WIMP},Ch) =   \int_{E_{th}} ^{m_{WIMP}} A_{eff}(E_{\nu_\mu}) \left.\frac{dN_{\nu_\mu}}{dE_{\nu_\mu}}\right|_{Det,Ch} dE_{\nu_\mu} + \int_{E_{th}} ^{m_{WIMP}} A_{eff}(E_{\bar \nu_{ \mu}}) \left.\frac{dN_{\bar \nu_{ \mu}}}{dE_{\bar \nu_{ \mu}}}\right|_{Det,Ch} dE_{\bar \nu_{\mu}} \label{Acc1}
\end{equation}

\noindent where $\rm A_{eff}(E_{\nu_{\mu}})$ is the effective area for the muon neutrino energy $\rm E_{\nu_{\mu}}$ or muon antineutrino energy $\rm E_{\nu_{\bar \mu}}$, $\rm \left.\frac{dN_{\nu_\mu}}{dE_\nu}\right|_{Det,Ch}$ is the signal neutrino spectrum at the position of the detector for one particular annihilation channel Ch listed in equation \ref{bb}, $\rm E_{th}$ is the energy threshold of the detector and $\rm m_{WIMP}$ is the WIMP mass. The effective area, which is the size of the detector assuming a 100\% detection efficiency, is calculated from the Monte Carlo simulation. The 90\% C.L. limits and sensitivities on the fluxes are then calculated by:

\begin{equation}
\rm \bar \Phi_{\nu_\mu+\bar \nu_\mu,90\%} =  \frac{\bar \mu_{\nu_\mu+\bar \nu_\mu ,90\%}(m_{WIMP})}{Acc(m_{WIMP}) \cdot T_{live}}
\end{equation}

\noindent where $\bar \mu_{\nu_\mu + \bar \nu_\mu,90\%}$ is the 90\% C.l. sensitivity or limit and $\rm T_{live}$ is the total live time of the detector. 

\section{Indirect search for Dark Matter towards the Sun}

An indirect search for DM towards the Sun has been performed using data collected during 2007 and 2012. No excess of data has been observed in the direction of the Sun. Limits have been calculated in terms of muon  neutrino fluxes and spin-dependent WIMP-nucleon scattering cross sections, which can be seen in figure \ref{sunFLUX} and \ref{sunSDCS}, respectively. As can be seen in the flux limits the loosest cross section limits stem from the $b \bar b$ channel which is the softest of the three channels used in that analysis. The $\tau^+ \tau^-$ and the $\rm W^+ W^-$ channel lead to harder neutrino spectra and give very similar flux limits.

\begin{figure}[!h]
\begin{center}
\centering
\includegraphics[width=0.7\textwidth]{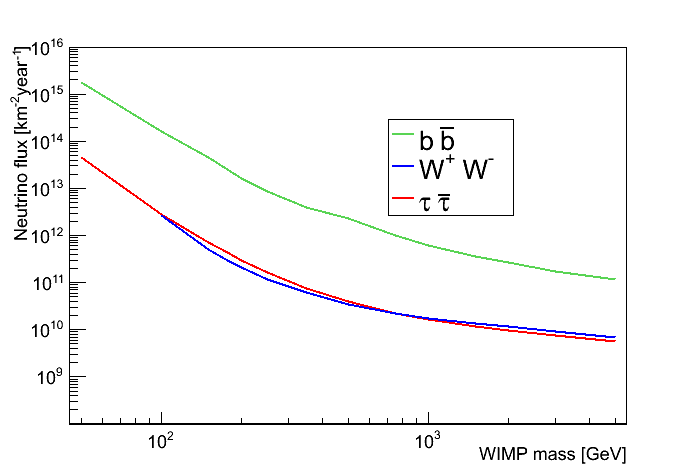}
\end{center}
\caption{$90$\% upper limits on the muon neutrino flux originating from self-annihilations of dark matter inside the Sun as a function of the WIMP mass obtained by the analysis of the data recorded by ANTARES between 2007 and 2012}
\label{sunFLUX}
\end{figure}

The spin-dependent WIMP-proton scattering cross section limits depend on the scattering of WIMPS with hydrogen in the Sun, whilst the spin-independent cross section limits depend on the scattering with helium. Since hydrogen is much more abundant in the Sun indirect searches are more sensitive to the spin-dependent scattering cross section and can surpass even direct detection experiments. Direct detection experiments, as Xenon 100 or LUX, are definitively more competitive for the spin-independent cross section. The ANTARES limits are more stringent than those of Ice Cube at higher masses (hundreds of GeV), although the instrumented volume of IceCube is significantly larger. This is due to the fact that IceCube limits are dependent on the performance of its central Deep Core ~\cite{IC_Sun}. In addition, the  angular resolution in the measurement of hundreds of GeV neutrinos is better in water than in ice.

\begin{figure}[h!]
\begin{center}
\centering
\includegraphics[width=\textwidth]{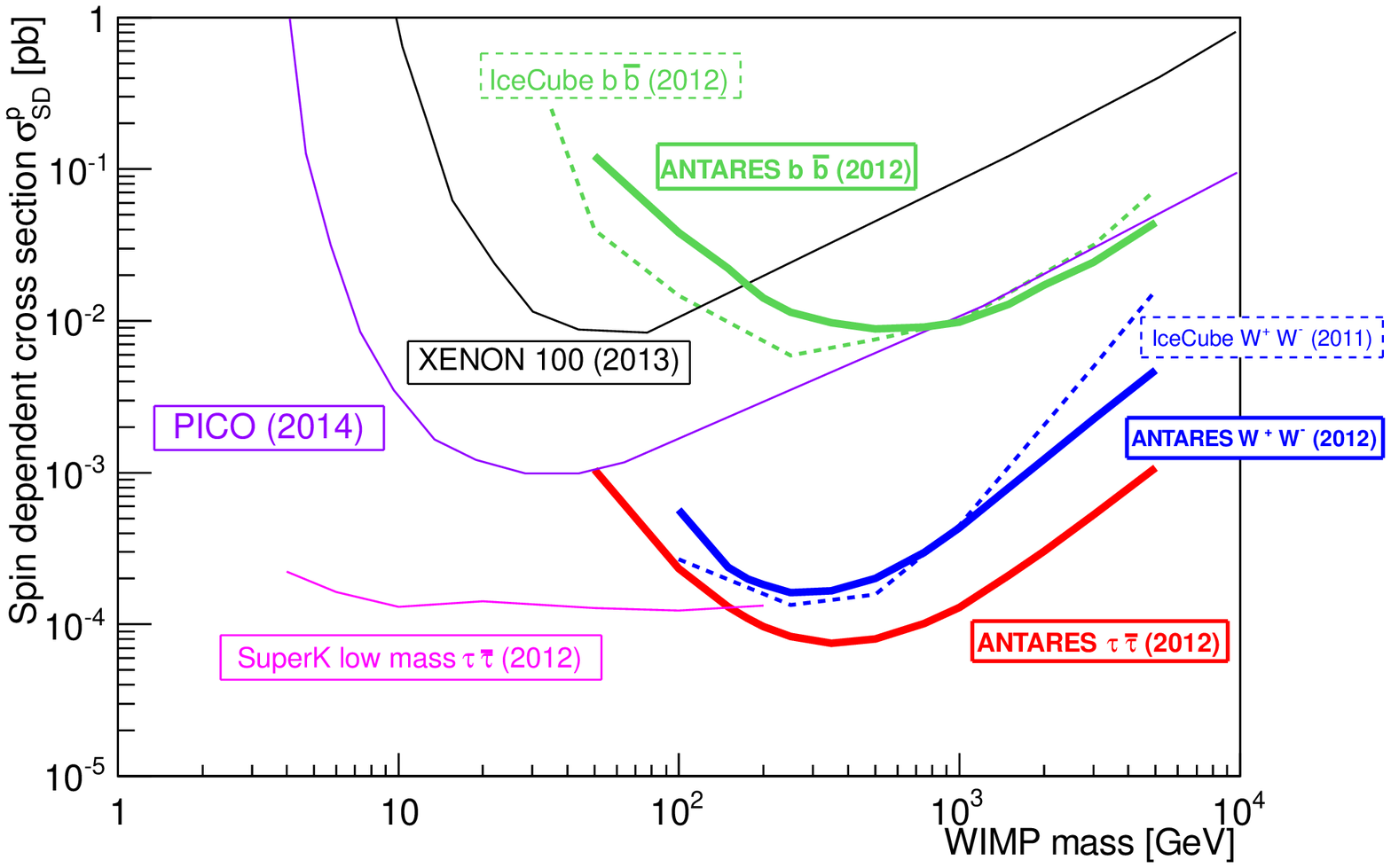}
\end{center}
\caption{$90$\% C.L. upper limits on the spin-dependent WIMP-proton scattering cross section as a function of the WIMP mass. Limits of other experiments are shown ~\cite{IC_Sun,PICO_SD,SuperK_low,XENON_SD}.}
\label{sunSDCS}
\end{figure}

\section{Indirect search for Dark Matter towards the Galactic Center and dwarf galaxies}

For the indirect search for Dark Matter towards the Galactic Center a "binned" analysis method has been used. This method calculates the amount of events within a cone around of the source and compare this to a background estimate. The sensitivities and limits are then calculated from the amount of events observed within the cone compared to those expected for the background. The size of this cone is optimized using background estimates and the sensitivities generated with it. 
No significant excess over the expected background has been found in the ANTARES data recorded between 2007 and 2012 and therefore exclusion limits have been calculated. In figure \ref{phinulimitfig} this exclusion limit in terms of neutrino signal fluxes in the direction of the GC is shown. As previously, the least stringent limit comes from the $\rm b \bar b$ channel, the $\nu_\mu \nu_{\bar \mu}$ channel lead to the most stringent limits. These limits are then converted to thermal averaged cross sections using J-Factors calculated assuming a NFW profile in equation \ref{loglik}. 

\begin{figure}[!h]
\begin{center}
\includegraphics[width=0.7\linewidth]{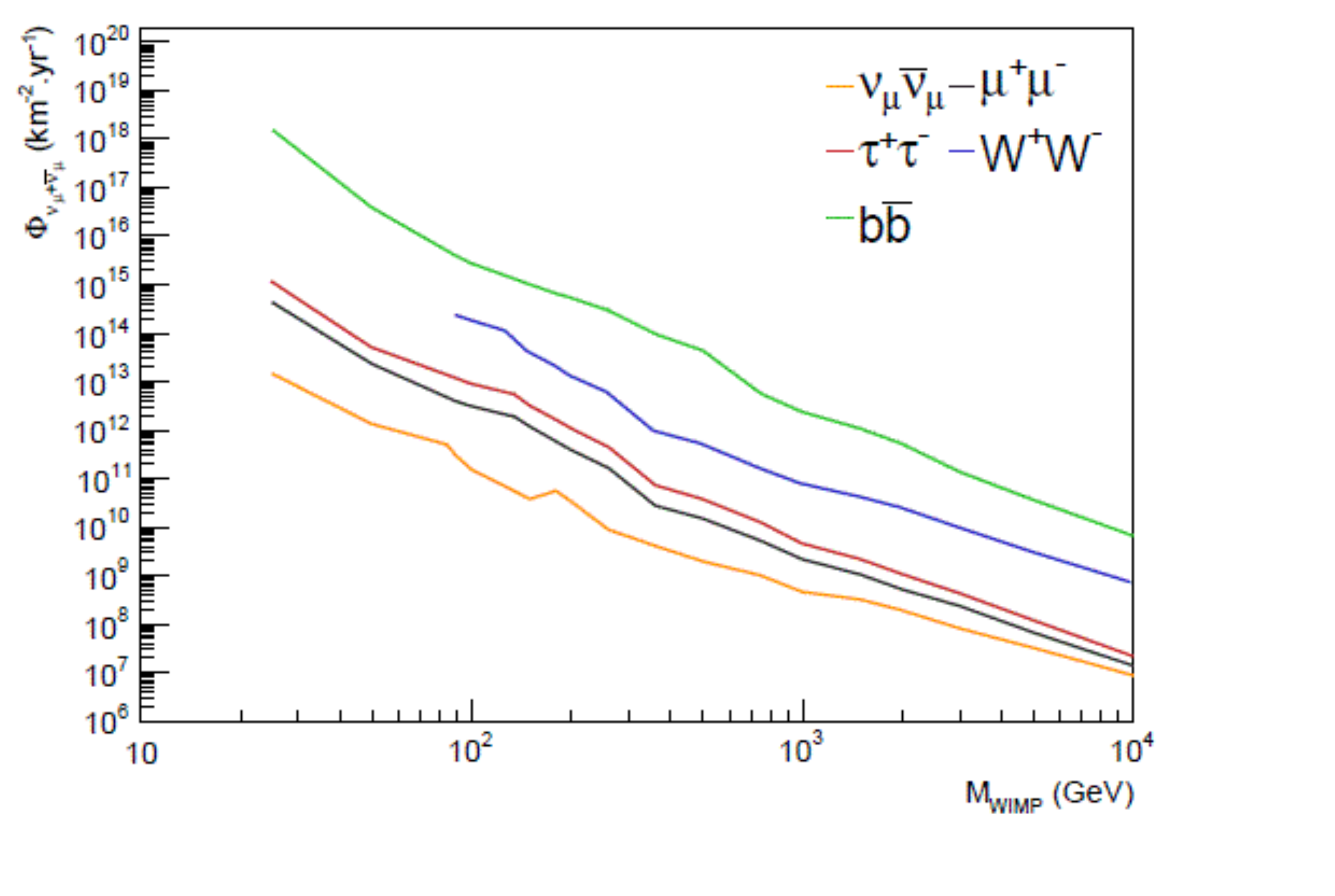}
\caption{$90$\% C.L. upper limits on the neutrino flux, $\rm \Phi_{\nu_{\mu}+\bar{\nu}_\mu}$, originating from self-annihilation of dark matter in the direction of the GC, as a function of the WIMP mass in the range $\rm 25 GeV \leq M_{WIMP} \leq 10 TeV$ for the self-annihilation channels (from top to bottom) $\rm WIMP\,WIMP \rightarrow b\bar{b} (green),W^{+}W^{-} (blue),\tau^{+}\tau^{-} (red),\mu^{+}\mu^{-} (dark grey),\nu_{\mu}\bar{\nu}_{\mu} (orange)$. }
\label{phinulimitfig}
\end{center}
\end{figure}

\begin{figure}[h!]
\begin{center}
\includegraphics[width=\linewidth]{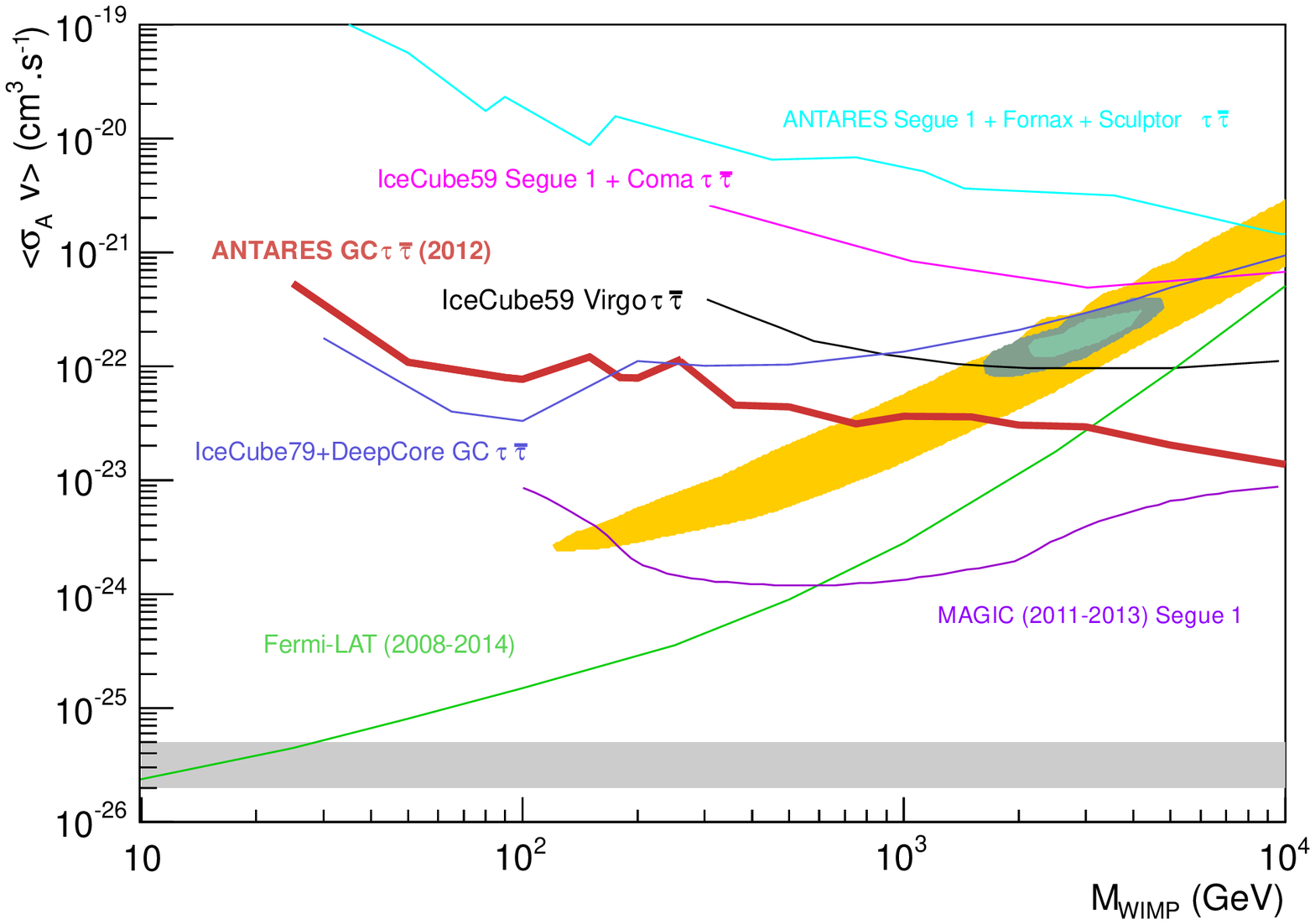}
\caption{$90$\% C.L. upper limits on the WIMP velocity averaged self-annihilation cross-section, $\rm <\sigma_{A}v>$, 
as a function of the WIMP mass in the range $\rm 10 GeV \leq M_{WIMP} \leq 10 TeV$. In this plot the IceCube limit for the Galactic Center is corrected with a constant factor for the different J-Factors used in the analysis. Limits of various experiments are shown ~\cite{IC_GC,icecube79,icecube59,fermilat,magic}.\label{bigcomparison} The allowed region of ~\cite{PAMELAint} arising from the PAMELA positron excess is also shown.}
\label{bigcomparison}
\end{center}
\end{figure}

In figure \ref{bigcomparison}, the $90\%$ C.L upper limit for the velocity averaged self-annihilation cross-section $\rm<sv>$  obtained by ANTARES is compared to that of other experiments. The $\tau^+ \tau^-$ channel has been chosen for the comparison. The original limit obtained by the IceCube experiment ~\cite{IC_GC} looking at the Galactic Center uses different halo parameters. Therefore in figure \ref{bigcomparison} a factor has been applied to the IceCube limits . This factor is the ratio of the integrated J-Factor used in the IceCube analysis to a J-Factor calculated using the halo parameters defined in section \ref{sec:1}. It is worth to notice that the limits from ANTARES reject at $90\%$ C.L. the interpretation of the PAMELA excess as a signal of leptophilic dark matter, if the constrains from HESS and Fermi-LAT ~\cite{PAMELAint} are also applied.

A similar analysis has been performed looking for a neutrino signal originating from dark matter annihilation in several dwarf spheroidal galaxies. No excess of events towards those objects has been found in the data recorded by ANTARES between 2007 and 2012. In order to derive an upper limit on the WIMP velocity averaged self-annihilation cross section, the signal of the 3 dwarf galaxies presenting the largest J-Factor and visibilities have been stacked. The resulting limit has also been included in figure \ref{bigcomparison}. 

\section{Conclusion}

As one can see the different searches for dark matter with the ANTARES neutrino telescope lead to limits, that can compete with the results of comparable experiments. Concerning especially the analysis for the Galactic Center the ANTARES limits are currently the most stringent limits from all neutrino telescopes, once the difference between the halo models used in the analyses is taken into account. Future improvements on this analysis, including the use of more advanced analysis methods, the inclusion of more recent data from ANTARES and a complementary analysis searching for neutrinos from WIMP annihilations in galaxy clusters are currently planned or in progress.


\setcounter{figure}{0}
\setcounter{table}{0}
\setcounter{footnote}{0}
\setcounter{section}{0}
\setcounter{equation}{0}

%


\end{document}